\documentclass[preprint,3p]{elsarticle}

\usepackage{amsmath,amssymb,shuffle,mathdots}
\usepackage{dsfont}
\usepackage{lineno}
\usepackage{simpler-wick}
\usepackage{enumitem}
\usepackage[hidelinks]{hyperref}
\modulolinenumbers[5]
\allowdisplaybreaks
\numberwithin{equation}{section}

\usepackage[makeroom]{cancel}

\let\Re\relax
\let\Im\relax
\DeclareMathOperator{\Re}{Re}
\DeclareMathOperator{\Im}{Im}
\DeclareMathOperator{\PT}{PT}

\def\len#1{{|#1|}}
\def\ap{{\alpha'}}
\def\be{\begin{equation}}
\def\ee{\end{equation}}
\def\ve{\varepsilon}
\def\a{{\alpha}}

\def\l{{\lambda}}

\def\b{{\beta}}

\def\g{{\gamma}}
\def\ga{{\gamma}}

\def\om{{\omega}}
\def\d{{\delta}}
\def\e{{\epsilon}}
\def\s{{\sigma}}

\def\half{{1\over 2}}
\def\p{{\partial}}

\def\t{{\theta}}

\def\bar{\overline}
\def\({\left(}
\def\){\right)}
\def\cF{ {\cal F} }

\def \bra#1{\left\langle #1\right|}
\def \ket#1{\left| #1\right\rangle}

\def \Tr{{\rm Tr}}
\def \mod{{\rm mod}}
\def \sinh{{\rm sinh}}
\def \cosh{{\rm cosh}}

\def \pbar{\bar{\partial}}

\def \be{\begin{equation}}
\def \ee{\end{equation}}

\def \la{\lambda^{\alpha}}
\def \t{\theta}

\def\psum{\mathop{\sum\nolimits'}}
\def\perm{{\rm perm}}

\newtheorem{def.}{Definition}

\newtheorem{prop.}{Proposition}
\newtheorem{lemma}{Lemma}
\newtheorem{cor}{Corollary}

\usepackage{tikz}
 \usetikzlibrary{calc} \usetikzlibrary{patterns} \usetikzlibrary{decorations.pathreplacing}
\usetikzlibrary{decorations.markings} \usetikzlibrary{decorations.pathmorphing} \usetikzlibrary{positioning}

\def\AYM{A^{\rm YM}}

\def\lie{\pounds} 
\def\lnabla{\nabla^{(L)}}
\def\eqref#1{(\ref{#1})}
\def\beq{\begin{equation}}
\def\eeq{\end{equation}}
\def\cE{{\cal E}}
\def\cF{{\cal F}}
\def\cL{{\cal L}}
\def\cW{{\cal W}}

\def\cA{{\cal A}}

\def\cK{{\cal K}}
\def\cN{{\cal N}}
\def\cR{{\cal R}}
\def\cX{{\cal X}}
\def\cZ{{\cal Z}}
\def\bA{{\mathbb A}}

\def\bW{{\mathbb W}}
\def\bF{{\mathbb F}}
\def\bL{{\mathbb L}}
\def\bH{{\mathbb H}}
\def\Wg{\bW}
\def\Fg{\bF}
\def\ce{\mathord{\hbox{$\mathfrak{e}$}}}
\def\cf{\mathord{\hbox{$\mathfrak{f}$}}}
\def\cm{\mathord{\hbox{$\mathfrak{M}$}}}
\def\cyclic#1{{\rm cyc}(#1)}
\def\frac#1#2{{#1\over #2}}

\def\Ahat{{\hat A}}
\def\What{{\hat W}}
\def\Fhat{{\hat F}}
\def\cH{{\cal H}}

\def\lnabla{\nabla^{(L)}} 

\def\Word(#1,#2,#3){W_{#1#2#3}}
\def\Wordq(#1,#2,#3,#4){W_{#1#2#3#4}}
\def\stirling{\atopwithdelims[]}

\begin{document}

\noindent{\hfill\small\texttt{UUITP-44/22}} \par\smallskip\vskip15pt

\begin{frontmatter}

\title{Tree-level amplitudes from the pure spinor superstring}

\author[1]{Carlos R. Mafra}
\address[1]{STAG Research Centre and Mathematical Sciences, University of Southampton,
Highfield, Southampton SO17 1BJ, UK}
\ead{c.r.mafra@soton.ac.uk}

\author[2]{Oliver Schlotterer}
\address[2]{Department of Physics and Astronomy, Uppsala University, 75108 Uppsala, Sweden}
\ead{oliver.schlotterer@physics.uu.se}

\begin{abstract}
We give a comprehensive review of recent developments on using the pure spinor
formalism to compute massless superstring scattering amplitudes at tree level.
The main results of the pure spinor computations are placed into the context of
related topics including the color-kinematics duality in field theory and the
mathematical structure of $\alpha'$-corrections.
\end{abstract}


\end{frontmatter}


\tableofcontents
\newpage

\section{Introduction}

Superstring theories offer ultraviolet completions of supersymmetric gauge theories
and supergravity in $D\leq 10$ spacetime dimensions (with $D=11$ and
$D=12$ in closely related M- and F-theory, respectively). Since gauge and gravity 
supermultiplets are realized through the massless vibration modes of open and closed
superstrings, respectively, their interactions are naturally unified: superstring scattering
amplitudes are computed from a topological expansion in terms of random fluctuating
surfaces dubbed {\it worldsheets} that automatically incorporate the interplay of
open and closed strings via splitting and joining. 

The worldsheet origin of string amplitudes is a rich source of structure and information. First, 
it realizes the connection between gauge theories and gravity through the Bern--Carrasco--Johansson
(BCJ) double copy in a geometrically intuitive manner. Second, the computation of string-corrections
to field-theory amplitudes from moduli spaces of punctured worldsheets reveals intriguing 
mathematical structures and cross-fertilizes with string dualities.
In order to bring these appealing implications of string amplitudes to their full fruition, it is
important to have detailed control over their explicit form and hence efficient methods to
organize their computation.

The worldsheet degrees of freedom underlying superstring theories and their amplitudes admit 
a variety of formulations. The
more recent pure spinor formalism developed by Berkovits since the year 2000 \cite{psf,Berkovits:2004px,Berkovits:2005bt} led to the first
manifestly super-Poincar\'e invariant quantization of the superstring.
The more traditional Ramond--Neveu--Schwarz (RNS) \cite{FMS, DHoker:1988pdl, DHoker:2002hof, Witten:2012bh} and Green--Schwarz (GS) \cite{GSI,GSII}
formalisms are for instance described in textbooks on string theory including
\cite{gswI, Green:1987mn, Polchinski:1998rq, Polchinski:1998rr, Zwiebach:2004tj, Becker:2006dvp, Blumenhagen:2013fgp, Kiritsis:2019npv}
and differ in the implementation of worldsheet and spacetime supersymmetry.
The equivalence
of these formalisms is widely expected based on \cite{RNSPS,Berkovits:2016xnb,Berkovits:2021xwh}
and explicitly confirmed for leading orders in string perturbation theory but in general 
a subject of ongoing research.

This is a comprehensive review of the state of the art regarding the computation of
massless superstring tree-level amplitudes in Minkowski spacetime with the 
pure spinor formalism. We will illustrate in detail how the manifest spacetime 
supersymmetry of the pure spinor
formalism simplifies computations and efficiently organizes the information on the
external gauge and gravity multiplets. The main results of this review include compact
expressions for superstring tree-level amplitudes with an arbitrary number of massless
external states revealed by a pure spinor computation in 2011 \cite{nptStringI}. These 
expressions will be shown to elegantly resonate with a web of double-copy relations between a wide
range of string- and field theories as well as number-theoretic properties of the
low-energy expansion.

\subsection{Summary of the main results}
\label{sec:1.summ}

Throughout this review, the topics are presented in an order which 
emphasizes completeness rather than brevity. As such, the topics are developed to a depth
higher than what is usually necessary for a brief application of certain parts of
the formalism.\footnote{We welcome the readers' help in spotting typos or technical mistakes.
Every correction that is firstly brought to our attention
will be rewarded with 20 Euro Cent per numbered equation,
to be paid in cash during the next in-person encounter with one of the authors.}
This is unavoidable in a comprehensive review but it can be mitigated by jumping to the 
topic of interest and choosing to pick up the minimal background as one goes along.
This section aims to give an overview of the main results in this review along with
pointers that facilitate the identification of key passages on a given topic. References to original
work can be found in the main text.

\subsubsection{Basics of the pure spinor formalism}
\label{sec:1.summ.a}

We start by summarizing the worldsheet variables in the (minimal)\footnote{See
\cite{Berkovits:2005bt} for the ``non-minimal'' pure spinor formalism with
additional worldsheet variables.} pure spinor formalism 
in ten-dimensional Minkowski spacetime, based on selected
aspects of section \ref{sec:theform}. Center stage is taken by the 
worldsheet action in (\ref{PScov})
\be
\label{summ.00}
S_{\rm PS}= \frac{1}{\pi}\int d^2z \, \Big( \frac{1}{2}\p X^m \bar{\p}X_m +p_{\a}\bar{\p}\t^{\a} -
w_\a\bar\p\l^\a \Big) 
\ee
with $\partial = \frac{\partial}{\partial z}$, $\bar \partial = \frac{\partial}{\partial \bar z}$
and $d^2z =\frac{ i}{2}dz\wedge d{\bar z}$,
and we shall now give a brief characterization of its main ingredients.
Just like the bosonic string and the RNS or GS formulation of
the superstring, the embedding coordinates $X^m$ (with vector indices 
$m,n,\ldots=0,1,\ldots,9$) enter (\ref{summ.00}) as free worldsheet bosons.
In parallel to Siegel's reformulation of the GS formalism \cite{siegel},
the matter sector of the pure spinor worldsheet action (\ref{summ.00}) also features 
a pair of anticommuting spacetime spinors $(p_\alpha,
\theta^\alpha)$ of holomorphic conformal weights $h_p=1$ and $h_\theta=0$ (with Weyl-spinor
indices $\alpha,\beta,\ldots=1,2,\ldots,16$ of $SO(1,9)$).

The main characteristic of the pure spinor formalism is the pair of commuting ghost 
variables $(w_\alpha,\lambda^\alpha)$ of holomorphic conformal weights $h_w=1$ 
and $h_\lambda =0$. They are 
spacetime spinors in contradistinction to the anticommuting scalar $(b,c)$-system 
of bosonic strings or RNS superstrings.
Cancellation of conformal anomalies and nilpotency of the BRST charge $Q_{\rm BRST}$
below requires $\l^\alpha$ to obey the pure spinor constraint
\be
\label{summ.00a}
(\lambda \gamma^m \lambda) = 0 \, , \ \ \ \ 
Q_{\rm BRST} = \oint dz \, \lambda^\alpha d_\alpha
\, , \ \ \ \ d_\alpha= p_\alpha 
- \tfrac{1}{2} \partial X^m (\gamma_m \theta)_\alpha 
- \tfrac{1}{8} (\theta \gamma^m \partial \theta) (\gamma_m \theta)_\alpha 
\ee
with $16\times 16$ Pauli matrices $\gamma^m_{\alpha \beta}=\gamma^m_{\beta \alpha}$ 
of $SO(1,9)$. Further details on the worldsheet ghosts and their contributions $N^{mn}$ 
to the Lorentz currents can be found in section~\ref{sec3.3}.

We have only displayed the left-moving spacetime spinors in (\ref{summ.00}). The
pure spinor formulation of type II superstrings involves right-moving counterparts
$(\tilde p_{\tilde \alpha},\tilde \theta^{\tilde \alpha})$ and
$(\tilde w_{\tilde \alpha},\tilde \lambda^{\tilde \alpha})$ with $\bar \partial$
in the place of $\partial$ in the action. The Weyl-spinor indices
$\tilde \alpha$ are of the same (opposite) chirality as the indices $\alpha$ of
the left-movers in (\ref{summ.00}) in case of the type IIB (type IIA) theory.
One can also construct a pure spinor version of heterotic
strings by incorporating right-moving bosons for compactified
16 extra dimensions into (\ref{summ.00}) instead of
$(\tilde p_{\tilde \alpha},\tilde \theta^{\tilde \alpha})$ and
$(\tilde w_{\tilde \alpha},\tilde \lambda^{\tilde \alpha})$, see section 
\ref{sec:7.7.4} for a brief discussion of its amplitudes.

\subsubsection{The prescription for disk amplitudes}
\label{sec:1.summ.b}

The physical spectrum of the pure spinor superstring is constructed from the
cohomology of the BRST charge in (\ref{summ.00a}). As usual in worldsheet approaches
to string theories, physical states are associated with vertex operators $V$ and $U$, 
conformal primaries of weight $h_V=0$ and $h_U=1$ in the BRST cohomology. For massless
states of the open superstring, the integrated and unintegrated representatives
of the vertex operators are
\beq
V = \lambda^\alpha A_\alpha \, , \ \ \ \ 
\int dz\, U= \int dz\, \big(\p\t^{\a}A_{\a}  + A_m \Pi^m + d_{\a}W^{\a}+ \tfrac{1}{2} N_{mn}F^{mn} \big)
\, ,
\label{summ.00b}
\eeq
see the discussion around (\ref{integrado}) and (\ref{V}).
They combine the worldsheet variables in (\ref{summ.00}), (\ref{summ.00a})
and $\Pi^m = \partial X^m + \frac{1}{2}(\theta \gamma^m \partial \theta)$ with linearized
superfields $A_\alpha,A^m,W^\alpha,F^{mn}$ of ten-dimensional super Yang--Mills (SYM) 
reviewed in section \ref{sec:10dSYM}, depending on the worldsheet variables $X^m$
and $\theta^\alpha$ but not on their derivatives.

The main subject of this review are the superstring disk amplitudes obtained
from the vertex operators in (\ref{summ.00b}) through the prescription \cite{psf}\footnote{The representation
(\ref{summ.00c}) of superstring disk amplitudes can
be derived from the gauge-fixing procedure of \cite{Hoogeveen:2007tu}.}
\be
\label{summ.00c}
{\cal A}(1,2,\ldots,n)=\!\!\!\!\! \! \! \int \limits_{-\infty<z_j <z_{j+1}<\infty}\!\!\! \! \! \! dz_2  \, dz_3 \ldots dz_{n-2} \, \langle \! \langle V_1(z_1) U_2(z_2)  U_3(z_3) {\ldots} 
 U_{n-2}(z_{n-2})  V_{n-1}(z_{n-1})  V_n(z_n)
\rangle \! \rangle \,,
\ee
see section \ref{sec:treeprescr}. The integration domain informally refers to an ordering
of the vertex-operator insertions on the disk boundary parameterized by 
$-\infty<z_1<z_2<\ldots<z_{n}<\infty$
which is associated with the Chan--Paton trace in the cyclic ordering ${\rm Tr}(t^{a_1} t^{a_2}
\ldots t^{a_n} )$. The correlators $\langle \! \langle \ldots \rangle \! \rangle$ arise from
the path integral over the worldsheet variables, and the contributions from their non-zero modes
can be evaluated from the OPEs encoded by (\ref{summ.00}). The zero modes
of the variables $\lambda^\alpha,\theta^\alpha$ with conformal weight 
$h_\theta=h_\lambda=0$ require a separate prescription 
\beq
\langle (\lambda \gamma^m \theta)  (\lambda \gamma^n \theta)
(\lambda \gamma^p \theta) (\theta \gamma_{mnp} \theta)  \rangle = 2880
\label{summ.01}
\eeq
which automatically fixes zero-mode correlators of arbitrary tensor contractions
of $\lambda^\alpha \lambda^\beta \lambda^\gamma \theta^{\delta_1} \theta^{\delta_2} 
\theta^{\delta_3} \theta^{\delta_4} \theta^{\delta_5}$ via simple group-theoretic considerations
(see also \cite{Hoogeveen:2007tu}).
Consistency conditions on (\ref{summ.00c}) and the extraction of three-point component
results for external gluons and gluinos are reviewed in detail in section~\ref{sec3.4}.
Massless $n$-point tree-level amplitudes of type II superstrings and
heterotic strings are obtained by integrating double copies of the correlator
in (\ref{summ.00c}) over the sphere, see sections \ref{sec:6.5} and \ref{sec:7.7.4}.

\subsubsection{The multiparticle formalism}
\label{sec:1.summ.c}

The driving force for the simplification of the $n$-point disk amplitude 
(\ref{summ.00c}) is the organization of the OPEs among the vertex operators
in (\ref{summ.00b}) through multiparticle superfields. The local incarnations
$A_\alpha^P,A^m_P,W^\alpha_P,F^{mn}_P$ of multiparticle superfields  
(with words or ordered sequences $P=p_1p_2\ldots$ in external-state
labels $p_i$) are obtained from recursions inspired by nested OPEs of vertex 
operators and an analysis of their equations of motion.
The construction of section \ref{sfinbcj}
in so-called BCJ gauge
leads to multiparticle superfields with generalized Jacobi identities 
under permutations of $P$, e.g.
\beq
A^m_{12} = - A_{21}^m \, , \ \ \ \ 
A^m_{123} = - A^m_{213} \, , \ \ \ \ 
A^m_{123} +A^m_{231} +A^m_{312} =0 \, .
\label{summ.02}
\eeq
More generally, the symmetries of $A^m_{1234\ldots }$ and all the other
local multiparticle superfields in BCJ gauge are those of contracted structure constants
$f^{a_1 a_2 b} f^{b a_3 c} f^{c a_4d}\ldots$ corresponding to the half-ladder
graph
\begin{figure}[h]
\begin{center}
\includegraphics[width=0.27\textwidth]{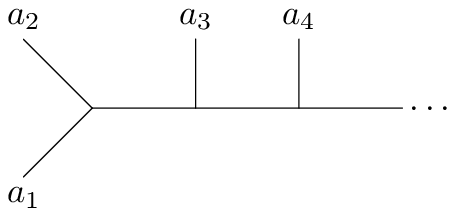}
\end{center}
\end{figure}
\vskip-15pt
\noindent Accordingly, the multiparticle
superfields inherit a diagrammatic interpretation that resonates with
the BCJ duality between color and kinematics in gauge theories 
and can be described in a combinatorial framework based on planar binary trees,
see section \ref{BGcombsec}.

By dressing the $A_\alpha^P,A^m_P,W^\alpha_P,F^{mn}_P$ with the
propagators of the associated cubic-vertex diagrams, one is led to non-local
superfields or Berends--Giele currents ${\cal A}_\alpha^P,{\cal A}^m_P,{\cal W}^\alpha_P,
{\cal F}^{mn}_P$ in {\it BCJ gauge}. This is an alternative to the construction
of Berends--Giele currents in Lorenz gauge via perturbiner methods
\cite{selivanovI,selivanovII,selivanovIII,selivanovIV}: the wave equations (\ref{nonwave})
of ten-dimensional SYM encode recursions for Lorenz-gauge currents such as
\beq
{\cal A}_\alpha^{12\ldots p} = \frac{1}{k_{12\ldots p}^2} \sum_{j=1}^{p-1} 
\big[
{\cal A}_\alpha^{j{+}1\ldots p} (k_{j{+}1\ldots p}\cdot {\cal A}_{12\ldots j})
+ {\cal A}^{m}_{j{+}1\ldots p} (\gamma_m {\cal W}_{12\ldots j})_\alpha
-(12\ldots j \leftrightarrow j{+}1\ldots p)\big]
\label{summ.03}
\eeq
with $k_{ij\ldots} = k_i+ k_j + \ldots$ and similar ones for ${\cal A}^m_P,{\cal W}^\alpha_P$ 
and ${\cal F}^{mn}_P$, see (\ref{cAalpha}). These recursions terminate with the linearized 
superfields in the vertex operators (\ref{summ.00b}) in the single-particle case, 
e.g.\ ${\cal A}^j_\alpha= A^j_\alpha$.
Berends--Giele currents in Lorenz and BCJ gauge obey the same
multiparticle equations of motion (\ref{BGEOM}) such as 
\beq
D_\alpha {\cal A}^{12\ldots p}_\beta
+ D_\beta {\cal A}^{12\ldots p}_\alpha 
= \gamma^m_{\alpha \beta} {\cal A}_m^{12\ldots p} 
+ \sum_{j=1}^{p-1} ( {\cal A}_\alpha^{12\ldots j} {\cal A}_\beta^{j+1\ldots p}
-  {\cal A}_\alpha^{j+1\ldots p} {\cal A}_\beta^{ 12\ldots j})
\label{summ.04}
\eeq
and, through their BRST-invariant combinations to be reviewed in the next sections,
both give rise to SYM and superstring tree amplitudes.

\subsubsection{SYM tree-level amplitudes}
\label{sec:1.summ.d}

Multiparticle superfields turned out to be invaluable to determine tree and
loop amplitudes in string and field theory from first principles including
locality and BRST invariance. As a simple manifestation thereof, 
color-ordered $n$-point tree amplitudes of ten-dimensional SYM 
obey the compact formula presented in (\ref{AYMgen}),
\beq
A(1,2,\ldots,n) = \sum_{j=1}^{n-2} \langle M_{12\ldots j} M_{j+1\ldots n-1} M_n  \rangle \, .
\label{summ.05}
\eeq
The $M_{12\ldots j}$ may be viewed as non-local multiparticle vertex operators
defined by the spinorial Berends--Giele currents in (\ref{summ.03}) and
whose BRST variation follows from the multiparticle equations of motion
(\ref{summ.04})
\beq
M_P= \lambda^\alpha {\cal A}_\alpha^P
\, , \ \ \ \ \ \
Q_{\rm BRST} M_{12\ldots p} = \sum_{j=1}^{p-1} M_{12\ldots j}M_{j+1\ldots p} \, .
\label{summ.06}
\eeq
Since the superfields in the zero-mode bracket of (\ref{summ.05}) are easily checked 
to be BRST invariant via (\ref{summ.06}), the component amplitudes
following from the zero-mode prescription (\ref{summ.01})
are guaranteed to be gauge invariant and supersymmetric.
An efficient Berends--Giele organization of the component 
amplitudes is described in section \ref{sec:compSYM}
which follows from (\ref{summ.05}) and a combination 
of Lorenz and Harnad--Shnider gauge for ${\cal A}_\alpha^P$.
In particular, this implies the bosonic components of (\ref{summ.05}) 
to reproduce the Berends--Giele formula \cite{BerendsME} for 
$n$-gluon tree amplitudes.

The superspace formula (\ref{summ.05}) is a convenient starting point to
prove the Kleiss--Kuijf (KK) and BCJ relations between
SYM tree amplitudes in different color orderings, see sections 
\ref{KKsec} and \ref{BCJampsec}. The KK relations
\cite{KKref} can be written as $A(P\shuffle Q,n)= 0 \ \forall \ P,Q\neq \emptyset$
with the shuffle operation defined in (\ref{shuffledef}) and follow from
the shuffle properties $M_{P\shuffle Q}= 0 \ \forall \ P,Q\neq \emptyset$ of the 
currents in (\ref{summ.06}). The BCJ relations \cite{BCJ}
take the form $A(\{P,Q\},n)=0  \ \forall \ P,Q\neq \emptyset$ with the so-called 
S-bracket $\{\cdot,\cdot\}$ defined in (\ref{smapdef}) and are derived from
multiparticle superfields in BCJ gauge using the vanishing of BRST-exact  
expressions under the zero-mode prescription (\ref{summ.01}), $\langle Q(\ldots) \rangle=0$.

\subsubsection{Superstring disk amplitudes}
\label{sec:1.summ.e}

As a key result of this review, color-ordered superstring disk amplitudes ${\cal A}(P)$
with any number of external gauge multiplets are reduced to SYM tree amplitudes 
$A(Q)$ in a BCJ basis of color orderings $Q$,
\beq
{\cal A}(1,P,n{-}1,n;\alpha') = \sum_{Q \in S_{n-3}} F_P{}^Q(\alpha') A(1,Q,n{-}1,n) \, .
\label{summ.11}
\eeq
In this simplified form of the string amplitudes, the entire $\alpha'$-dependence resides in
the disk integrals $F_P{}^Q$ (indexed by permutations $P,Q$ of legs $2,3,\ldots,n{-}2$) which are 
defined in (\ref{Fdef}) and depend on external momenta. The SYM amplitudes $A(Q)$ in turn 
carry the complete polarization dependence of (\ref{summ.11}) for any combination of external 
bosons and fermions. Remarkably, the superspace structure of string tree amplitudes
is captured by field-theory building blocks $A(Q)$ and separated from the string effects
in the scalar disk integrals $F_P{}^Q$.

As detailed in sections \ref{sec:CFTan} to \ref{theIBPsec}, the derivation
of (\ref{summ.11}) starts from the opening line (\ref{summ.00c}) and
relies on the local multiparticle superfields to perform the OPEs among the
vertex operators. After integration-by-parts reduction of the disk integrals,
the SYM amplitudes are identified through their superspace representation
(\ref{summ.05}) in BCJ gauge.

The expression (\ref{summ.11}) for $n$-point superstring disk amplitudes turns
out to line up with the Kawai--Lewellen--Tye (KLT) formula for supergravity tree
amplitudes $M_n^{\rm grav}$ once the integrals $F_P{}^Q(\alpha') $
are rewritten in a Parke--Taylor basis of $Z$-integrals defined in (\ref{Zintdef}):
\begin{align}
M_n^{\rm grav} &= -\sum_{Q,R\in S_{n-3}} \tilde A(1,Q,n,n{-}1) S(Q|R)_1 A(1,R,n{-}1,n) \label{summ.12}\\
\ \ \ \leftrightarrow \ \ \ {\cal A}(P) &= -\sum_{Q,R\in S_{n-3}} Z(P|1,Q,n,n{-}1) S(Q|R)_1 A(1,R,n{-}1,n) \, .
\notag
\end{align}
The entries of the $(n{-}3)! \times (n{-}3)!$ KLT kernel $S(Q|R)_1$ are
degree-$(n{-}3)$ polynomials in $k_i\cdot k_j$, see (\ref{kltrec}). Since the
KLT formula for $M_n^{\rm grav}$ reflects the tree-level double copy of supergravity as a square
of SYM, we interpret (\ref{summ.12}) as a field-theory double-copy construction
of the open superstring from SYM and disk integrals $Z(P|Q)$.

As detailed in section \ref{sec:6.3.3}, the disk integrals $Z(P|Q)$ at fixed
color ordering $P$ obey field-theory KK and BCJ relations
between different Parke--Taylor integrands specified by $Q$. By these
relations and their appearance in a field-theory KLT relation (\ref{summ.12}),
the $Z(P|Q)$ are proposed to furnish (single-trace) tree-level amplitudes
in a ultraviolet-completed theory of bi-colored scalars dubbed $Z$-theory.
This is furthermore supported by the emergence of doubly-partial amplitudes
of bi-adjoint scalars in the field-theory limit
\beq
\lim_{\alpha'\rightarrow 0} Z(P|Q) = m(P|Q)\, ,
\label{summ.13}
\eeq
see section \ref{subsec:biadj} and in particular (\ref{DPdefCP}) for the definition
of $m(P|Q)$.

\subsubsection{Color-kinematics duality and double copy}
\label{sec:1.summ.f}

Another main result of section \ref{sec:diskamp} is the manifestly
local $(n{-}2)!$-term representation (\ref{localFormWithZ}) of superstring disk amplitudes.
By the field-theory limit (\ref{summ.13}) of the disk integrals therein, we obtain
SYM amplitudes from a sum over permutations $Q$ of legs $2,3,\ldots,n{-}1$,
\beq
A(P) = \sum_{Q \in S_{n-2}}  m(P|1,Q,n) N_{1| Q |n}
\, .
\label{summ.14}
\eeq
The kinematic factors $N_{1|\ldots|n}$ are trilinears in local multiparticle superfields
$A_\alpha^P$ in BCJ gauge,
\beq
N_{1| P (n-1) Q |n} = (-1)^{|Q|+1} \langle V_{1P} V_{n\tilde Q} V_{n-1} \rangle
\, , \ \ \ \ \ \ 
V_P = \lambda^\alpha A_\alpha^P
\, .
\label{summ.15}
\eeq
The appearance of $N_{1|\ldots|n}$
in (\ref{summ.14}) identifies them as BCJ master numerators that manifest the 
color-kinematics duality of SYM at all multiplicities for any combination of
external bosons and fermions. More precisely, by the
discussion in section \ref{sec:6.4}, the kinematic numerators in (\ref{summ.15})
are associated with the half-ladder diagrams in
figure \ref{PSSVVV} and generate all other cubic-diagram numerators
by kinematic Jacobi identities.

In the same way as the open superstring manifests the color-kinematics duality
of $n$-point SYM tree amplitudes, section \ref{sec:6.5.4} reviews the derivation
of the gravitational double copy in its cubic-diagram formulation from closed superstrings,
\beq
M^{\rm grav}_n = \sum_{P,Q \in S_{n-2}} \tilde{N}_{1|P|n}
m(1,P,n |1,Q,n ) N_{1|Q|n}\, ,
\label{summ.16}
\eeq
which is equivalent to the KLT formula for supergravity tree amplitudes $M^{\rm grav}_n$ 
in (\ref{summ.12}). In both (\ref{summ.14}) and (\ref{summ.16}), the key to realize
the BCJ duality and double copy with manifest locality is the simplification of the correlator
in (\ref{summ.00c}) to the $(n{-}2)!$-term combination (\ref{gravsec.11}) of local multiparticle
superfields and Parke--Taylor factors.
Moreover, the construction relies
on doubly-partial amplitudes $m(P|Q)$ from the field-theory limit (\ref{summ.13})
of disk integrals and closely related sphere integrals (\ref{Jintdef}).

Similarly, we shall construct explicit BCJ numerators for the non-linear
sigma model (NLSM) of Goldstone bosons in section \ref{sec:7.7}
reflected in the amplitude representation
\beq
A_{\rm NLSM}(P) = i^{n-2} \sum_{Q \in S_{n-2}}  m(P|1,Q,n) S(Q|Q)_1
\label{summ.17}
\eeq
analogous to (\ref{summ.14}), with $S(Q|Q)_1$ the diagonal entries of
the KLT kernel in (\ref{summ.12}). Section \ref{sec:7.7.4} in turn is
dedicated to double-copy representations of Einstein--Yang--Mills
tree amplitudes similar to (\ref{summ.16}) that are derived from
the heterotic version of the pure spinor superstring.

\subsubsection{$\alpha'$-expansions of open- and closed-superstring tree amplitudes}
\label{sec:1.summ.g}

The low-energy expansion of the $n$-point disk integrals $F_P{}^Q$
and $Z(P|Q)$ in the open-string amplitudes (\ref{summ.11}) and (\ref{summ.12})
yields infinite series in dimensionless Mandelstam invariants $\alpha' k_i \cdot k_j$
with multiple zeta values (MZVs) in their coefficients,
\beq
\zeta_{n_1,n_2,\ldots,n_r} = \sum_{0<k_1<k_2<\ldots <k_r} 
k_1^{-n_1} k_2^{-n_2} \ldots k_r^{-n_r} \, , \ \ \ \ 
n_1,n_2,\ldots,n_r \in \mathbb N \, , \ \ \ \ n_r\geq 2\, .
\label{summ.21}
\eeq
After a brief review of mathematical background in section \ref{sec:7.2},
the $f$-alphabet description of (motivic) MZVs is shown to determine the entire $\alpha'$-expansion
from the coefficients of the Riemann zeta values $\zeta_w$ (i.e.\ (\ref{summ.21}) at depth $r=1$),
see (\ref{mzvsec.38}). This reflects a kind of closure of disk integrals under the motivic coaction
$\Delta$ of MZVs which has also been observed in other areas of high-energy physics and can be 
expressed in terms of another KLT formula (\ref{mzvsec.41}) for $\Delta Z(P|Q)$.

On top of these structural results, we review two recursive methods to explicitly
determine the polynomials in $k_i\cdot k_j$ within the $\alpha'$-expansion
of $n$-point disk integrals. In section \ref{sec:7.4}, matrix representations of
the Drinfeld associator relate the $(n{-}1)$-point and $n$-point versions
of the $F_P{}^Q$ basis in (\ref{summ.11}). Section \ref{sec:7.5} is dedicated
to a Berends--Giele recursion for the $Z(P|Q)$ integrals in (\ref{summ.12})
which is generated by a non-linear field equation of bi-colored scalars
in $\alpha'$-expanded form and supports the interpretation of $Z(P|Q)$ 
as $Z$-theory amplitudes.

The $\alpha'$-expansion of closed-string tree amplitudes only features the subclass
of MZVs obtained from the so-called ``single-valued'' map sv. Even though the notion
of single-valued MZVs is only well-defined in a motivic setting, we informally write
the main result of section \ref{sec:7.6.1} as
\beq
{\cal M}_n^{\rm closed}(\alpha') = - \sum_{P,Q,R \in S_{n-3}} \tilde A(1,P,n,n{-}1) S(P|Q)_1 
{\rm sv} \, F_Q{}^R(\alpha') A(1,R,n{-}1,n)\, .
\label{summ.22}
\eeq
The single-valued map acts on the MZVs order by order in $\alpha'$, for
instance ${\rm sv} \, \zeta_{2k}=0$ and ${\rm sv} \, \zeta_{2k+1}= 2 \zeta_{2k+1}$
at depth one, but leaves the external polarizations and momenta
inert. Similar to the expression (\ref{summ.11}) for open-superstring
amplitudes, the $\alpha'$-dependence of (\ref{summ.22}) is isolated in a scalar quantity ${\rm sv} \, F_Q{}^R(\alpha')$
while all the superfield-polarizations are carried by SYM amplitudes $\tilde A(P)$
and $A(R)$. With the SYM amplitudes in (\ref{summ.05}) one can access all multiplet 
components of type IIA and IIB amplitudes via (\ref{summ.22}). Moreover, the
low-energy expansion of (\ref{summ.22}) can be made fully explicit through the
single-valued map of disk integrals $F_Q{}^R$ within the reach of the
expansion methods in sections \ref{sec:7.4} and \ref{sec:7.5}.

Note that (\ref{summ.22}) only applies to closed-string amplitudes 
on the sphere. Closed-string amplitudes on the disk in turn are subleading in the string
coupling and also involve MZVs beyond the single-valued ones in the $\alpha'$-expansion.
Mixed amplitudes involving open and closed strings on the disk were studied from the
perspective of the pure spinor formalism in \cite{Alencar:2008fy, Alencar:2011tk, Bischof:2020tnf}
and can be expressed in terms of those of only open-string insertions on the disk
boundary \cite{StiebergerHQ, Stieberger:2015vya}.

\subsubsection{A web of field-theory double copies for string amplitudes}
\label{sec:1.summ.h}

There is a steadily growing web of double-copy relations among field theories of different
spins \cite{Bern:2019prr, Bern:2022wqg, Adamo:2022dcm} which can be formulated in 
terms of the KLT formula (\ref{summ.12}) with kernel $S(P|Q)_1$. In case of supergravity 
and Einstein--Yang--Mills, 
such double-copy relations can be derived from the string-theory KLT formula
reviewed in section \ref{sec:6.5.2}. It expresses closed-string 
tree-level amplitudes via bilinears in color-ordered open-string tree amplitudes 
with an $\alpha'$-dependent kernel ${\cal S}_{\alpha'}(P|Q)_1$ that depends 
trigonometrically on the external momenta.

The representations in (\ref{summ.12}) and (\ref{summ.22}) for open- and closed-string
tree-level amplitudes in turn involve the field-theory KLT kernel $S(P|Q)_1= 
\lim_{\alpha' \rightarrow 0} {\cal S}_{\alpha'}(P|Q)_1$ and are still exact
in $\alpha'$. The emergence of a field-theory double-copy in a string-theory
context can be traced back to the KLT form (\ref{KLTcorrel}) of the $n$-point 
correlation function of vertex operators in the pure spinor formalism. This correlator 
including the field-theory KLT 
kernel therein also enters the tree amplitudes of type II and heterotic superstrings
upon pairing with right movers and for instance explains the factor of $ S(P|Q)_1 $
in (\ref{summ.22}). The latter can in fact be written as
\beq
{\cal M}_n^{\rm closed} = - \sum_{P,Q \in S_{n-3}} \tilde A(1,P,n,n{-}1) S(P|Q)_1
{\rm sv} \, {\cal A}(1,Q,n{-}1,n)\, .
\label{summ.23}
\eeq
identifying type II superstrings as a field-theory double copy of SYM with the
single-valued open superstring. Field-theory KLT formulae require
BCJ relations of both double-copy constituents as a consistency condition
which is met for the ${\rm sv} \, {\cal A}(Q)$ in (\ref{summ.23})
to all orders in $\alpha'$, see the discussion in section \ref{sec:7.x.y}.
As a commonality of (\ref{summ.23}) with the KLT form (\ref{summ.12}) of
open-superstring amplitudes, SYM building blocks are
double-copied through the field-theory KLT kernel with one string-theoretic 
object -- disk integrals or single-valued open-superstring amplitudes.

Also for heterotic strings reviewed in section \ref{sec:7.7.4}, $n$-point tree amplitudes
obey a field-theory KLT formula 
\beq
{\cal M}_n^{\rm het} = - \sum_{P,Q \in S_{n-3}} \tilde A_{(DF)^2+{\rm YM}+\phi^3}(1,P,n,n{-}1) S(P|Q)_1 
{\rm sv} \, {\cal A}(1,Q,n{-}1,n)
\label{summ.24}
\eeq
with one quantum-field-theory component $ \tilde A_{(DF)^2+{\rm YM}+\phi^3}$ and 
again ${\rm sv} \, {\cal A}(Q)$ as a string-theoretic component. However, the Lagrangian
and tree amplitudes of the $(DF)^2+{\rm YM}+\phi^3$ field theory are more complicated
than those of SYM by the lack of supersymmetry and the two types of massive internal states,
see section \ref{sec:7.7.7}. On the basis of (\ref{summ.24}), the tree amplitudes
of heterotic strings with external gauge and gravity supermultiplets reveal a field-theory
double copy of $(DF)^2+{\rm YM}+\phi^3$ with single-valued open superstrings.

Together with similar field-theory KLT formulae (\ref{svhet.03}) for open- and closed-string 
amplitudes of the bosonic theories, we arrive at the web of double-copy relations 
summarized in table \ref{dcpyarray}: tree amplitudes in various perturbative string theories 
are intertwined with field-theory amplitudes and string-theoretic building blocks that share
the KK and BCJ relations of gauge theories.

\subsubsection{Example of a possible shortcut}
\label{sec:1.summ.k}

The above summary of selected main results in this review together with the pointers
to equations and sections may offer shortcuts to extract the key information on
specific topics of interest. For instance, the expression (\ref{summ.05})
for SYM tree amplitudes can already be defined through the 
Berends--Giele currents $M_P$ in Lorenz gauge. In this case, the
consistency conditions and component evaluations can already be understood
from the non-linear theory of ten-dimensional SYM using only non-local superfields, 
i.e.\ independently of string-theory methods and
the local multiparticle superfields in BCJ gauge in section \ref{sec:4.1loc}. 

However, important aspects such as the BCJ amplitude relations or the
kinematic Jacobi identities among SYM numerators require the notions of 
local multiparticle superfields and BCJ gauge as well as the associated formalism.
Therefore the theory of multiparticle superfields is given an exhaustive discussion 
in section \ref{multiSYMsec} before their applications in scattering amplitudes. Here 
and in other contexts, the reader should be aware that the years of development led to a healthy 
growth in the amount of connections between a variety of subjects which caused the review
to grow beyond the page limits envisioned in earlier stages.

\subsection{\label{sec:others} Related topics beyond the scope of this review}

There is a variety of related topics that fruitfully resonate with superstring
tree amplitudes but could not be covered in this review. As a small sample,
we shall make a few comments on the Cachazo--He--Yuan (CHY) 
formalism, string field theory, the hybrid formalism and strings
in $AdS$ spacetimes here, 
and a more detailed account on loop-level string amplitudes 
covering references up to fall 2022
can be found in section~\ref{sec:conclu}.

\subsubsection{\label{sec:others.1} The CHY formalism}

An alternative worldsheet approach to double-copy representations of field-theory amplitudes 
is offered by the CHY formalism \cite{Cachazo:2013gna, Cachazo:2013hca, DPellis}.
It may be viewed as an uplift of the Witten-RSV \cite{Witten:2003nn, Roiban:2004yf} and 
Cachazo-Skinner \cite{Cachazo:2012kg} formulae to generic spacetime dimensions
$D\neq 4$ and is underpinned by ambitwistor string theories in RNS 
\cite{Mason:2013sva, Adamo:2013tsa} and pure spinor 
\cite{Berkovits:2013xba, Adamo:2015hoa} formulations.
The reader is referred to the review \cite{Geyer:2022cey} and the white paper 
\cite{Adamo:2022dcm} for the wealth of developments in the CHY formalism
and its interplay with double copy and superstring amplitudes.

CHY formulae directly compute field-theory amplitudes from moduli-space integrals 
for punctured Riemann surfaces similar to those in superstring amplitudes. These
CHY integrals are completely localized via so-called scattering equations and, in case of Parke--Taylor
integrands seen in the main formulae for superstring tree amplitudes such as (\ref{summ.12}),
coincide with the field-theory limits of disk and sphere integrals, see for instance (\ref{summ.13}).
In fact, our main result in (\ref{KLTcorrel}) or (\ref{gravsec.11}) for the $n$-point correlation 
function of massless vertex operators in the pure spinor superstring can be readily exported
to the pure spinor version of the ambitwistor string \cite{Gomez:2013wza}.

\subsubsection{\label{sec:others.2} String field theory}

Perturbative string theories admit an alternative formulation in terms of string field theory 
where scattering amplitudes including their exact $\alpha'$-dependence are computed 
from Feynman-type rules for a string field. The wavefunction of the string field
depends on the zero and non-zero modes of the worldsheet variables and may
guide an extension of the multiparticle formalism for massless vertex operators
to the entire string spectrum. Recent lecture notes on string field theory can for
instance be found in \cite{Erler:2019loq, Erler:2019vhl, Erbin:2021smf}.

On the one hand, string field theory may face more technical complications
in a detailed evaluation of string amplitudes than the worldsheet techniques 
described in this work. On the other hand, string field theory is widely considered
more promising to describe non-perturbative features of superstring theory including 
duality symmetries or background independence. In particular, string field theory turned
out to be a successful approach to tachyon condensation 
\cite{Sen:1998sm, Sen:1999nx, Berkovits:2000hf, Schnabl:2005gv, Erler:2019fye}
or mass renormalization
\cite{Pius:2013sca, Pius:2014iaa, Sen:2016gqt} and is conjectured to 
provide an understanding of the AdS/CFT correspondence
\cite{Gopakumar:1998ki, Gaiotto:2003yb, Berkovits:2008qc, Okawa:2020llq}.

\subsubsection{\label{sec:others.3} The hybrid formalism and strings in $AdS$ spacetimes}

As an alternative to the RNS, GS and pure spinor descriptions of the superstring,
the so-called hybrid formalism admits manifestly $SO(1,3)$- or 
$SO(1,5)$-super-Poincar\'e invariant quantization. The hybrid formalism was constructed
in the 90's from a series of field redefinitions in the RNS formalism to GS-like variables 
which manifest half- or quarter-maximal spacetime supersymmetry \cite{Berkovits:1994wr, 
Berkovits:1994vy, Berkovits:1996bf, Berkovits:1999im, Berkovits:1999in}. 

Apart from manifestly supersymmetric amplitude computations in flat 
spacetime \cite{Berkovits:2001nv}, a major appeal of the hybrid formalism is its
aptitude for the description of superstrings in $AdS_3\times S^3$ 
backgrounds \cite{Berkovits:1999im} (also see \cite{Berkovits:1999zq} for $AdS_2\times S^2$).
The intricate physical-state conditions for the $AdS_3\times S^3$ superstring
are for instance discussed in \cite{Dolan:1999dc, Gaberdiel:2011vf, Gerigk:2012cq, 
Dei:2020zui, Gaberdiel:2021njm}, also see \cite{Bobkov:2002bx} for
a three-graviton amplitude.
The hybrid formulation in \cite{Berkovits:1999im} became a driving force for
recent progress on type II superstrings in $AdS_3\times S^3 \times T^4$ spacetime
with NS flux and clarified the gauge-theory dual under the AdS/CFT 
correspondence \cite{Eberhardt:2018ouy, Eberhardt:2019ywk, Dei:2020zui, Knighton:2020kuh}.

For superstrings in $AdS_5 \times S^5$ with finite radius, the RNS formulation 
faces difficulties in incorporating Ramond flux backgrounds. The pure spinor formalism
in turn preserves the full $PSU(2,2|4)$ symmetry of the coset description of
$AdS_5 \times S^5$ upon quantization \cite{Berkovits:2004xu},
though a larger amount of computations has been performed in the GS 
formalism \cite{Metsaev:1998it}. The reader is referred to the comprehensive 
review \cite{Mazzucato:2011jt} and the white paper \cite{Berkovits:2022ivl} for further 
references on both the pure-spinor and GS approach to superstrings in $AdS_5 \times S^5$;
also see the white paper \cite{Gopakumar:2022kof} for progress on relating string perturbation 
theory with conformal correlators through the AdS/CFT correspondence.

\subsection{\label{convIntrosec} Conventions and notation}

\paragraph{Ten-dimensional superspace}
The ten-dimensional superspace coordinates are denoted $\{ X^m,\t^\a\}$, where
$m=0, \ldots,9$ are the vector indices and $\a=1, \ldots,16$ denote the spinor
indices of the Lorentz group $SO(10)$. The spinor representation is based on the
$16\times 16$ Pauli matrices $\g^m_{\a\b}=\g^m_{\b\a}$ satisfying the Clifford
algebra
 $\g^{(m}_{\a\b}\g^{n)\b\g}_{\phantom\a} = 2\delta^{mn}\d_\a^\g$. In this
review, unless stated otherwise, the (anti)symmetrization of $k$ indices does not include a factor of
${1\over k!}$. For more details on gamma matrices, see \ref{sec:appA}.

\paragraph{Multiparticle index notation}
In this review we will use a notation based on words to label multiparticle states.
More precisely, let ${\mathbb N}=\{1,2,3, \ldots \}$ be the alphabet of external-particle
labels. We will consider the vector space generated by linear combinations of 
\textit{words} $P=p_1p_2 \ldots $ with \textit{letters} $p_i$ from the alphabet $\mathbb N$.
Capital letters from the Latin alphabet are used to represent words (e.g. $P=1423$) while
their composing letters are represented in lower case (e.g.\ $p=3$).
The length of a word $P=p_1p_2 \ldots p_k$ is denoted by $\len{P}=k$ and it is given by the total number of
letters contained in it. The empty word is denoted by $P=\emptyset$ and has length $\len{P}=0$.
The reversal of a word $P=p_1p_2 \ldots p_k$ is $\tilde
P=p_k \ldots p_2 p_1$. The deconcatenation of a word $P$ into two words $X$ and $Y$ 
is denoted by $\sum_{P=XY}$, and it represents all the possible ways to concatenate 
two words $X$ and $Y$ (including the empty word) such
that $XY=P$. This operation will often be used when the words are
labels of other objects (usually superfields such as $M_P$), for instance
\beq\label{decexamp}
\sum_{XY=123}M_XM_Y = M_\emptyset M_{123} + M_1 M_{23} + M_{12}M_3 + M_{123}M_\emptyset \, .
\eeq
More definitions can be found in the \ref{wordsapp}.

The multiparticle momentum $k^m_P$ for a word $P$ with letters $i$ from massless particles
$(k_i\cdot k_i)=0$ and its associated Mandelstam invariant are given by
\beq\label{mandef}
k^m_P:= k^m_{p_1}+ \cdots +k^m_{p_\len{P}}\,,\qquad s_P := \half (k_P\cdot k_P)\,.
\ee
For example $k^m_{123}:= k^m_1 + k^m_2 + k^m_3$ and $s_{123} = s_{12}+s_{13}+s_{23}$.

\section{Super Yang--Mills in ten dimensions}
\label{sec:10dSYM}

Super Yang--Mills (SYM) theory in ten dimensions is the simplest
among $D$-dimensional SYM theories; its spectrum contains just the gluon and gluino, related by sixteen supercharges \cite{SYM} that form a Majorana-Weyl spinor of $SO(1,9)$.
It is perhaps not a coincidence that it is also the theory relevant to the low-energy limit of superstring theory \cite{Green:1982sw}. Its super-Poincar\'e covariant
formulation \cite{siegelSYM,wittentwistor} is, in particular, one of the pillars supporting the
pure spinor description of massless states of the open superstring. And indeed the SYM superfields of
\cite{siegelSYM,wittentwistor}
and their multiparticle generalization \cite{EOMBBs,Gauge,genredef} reviewed in section~\ref{multiSYMsec} played an essential role in the calculation of the general $n$-point
superstring disk amplitude. It is therefore beneficial to start this review by giving a detailed account
of this beautiful field theory.

On top of the original superfields of \cite{siegelSYM,wittentwistor} we will
define additional superfields of arbitrary mass dimension and study their non-linear equations of motion. This framework simplifies the $\theta$-expansions of multiparticle superfields as
detailed in \ref{HSapp} and
the expressions of kinematic factors in higher-loop scattering amplitudes, including the
$D^6R^4$ interaction in the superstring three-loop amplitude \cite{3loop} as discussed in \cite{SYMBG}.

It is also well-known that the dimensional reduction of the simple ten-dimensional
SYM theory gives rise to various
maximally supersymmetric Yang--Mills theories in lower dimensions,
including the celebrated $\cN = 4$ theory in
$D=4$ \cite{SYM}.
Therefore a better understanding of the $D=10$ theory propagates to a variety
of applications\footnote{The dimensional reduction of the multiparticle superfields appears
to be unexplored territory so far.} to any lower dimension.

\subsection{Ten-dimensional SYM}

To describe the gluon and gluino states of ten-dimensional SYM, one introduces Lie-algebra valued superfield connections
$\Bbb A_\a = \Bbb A_\alpha(X,\t)$ and $\Bbb A_m = \Bbb A_m(X,\t)$, the
supercovariant derivatives,
\beq\label{covder}
\nabla_\a := D_\a - \Bbb A_\a\,,\qquad \nabla_m := \p_m - \Bbb A_m\,,
\ee
and imposes the constraint \cite{siegelSYM, wittentwistor}
\beq\label{SYMconst}
\big\{ \nabla_\alpha ,\nabla_\beta \big\} = \gamma^m_{\alpha \beta} \nabla_m.
\eeq
Note that $\p_m = \frac{ \partial  }{\partial X^m }$, and the superspace derivative
\beq\label{Dcov}
D_\a := {\p\over\p\t^\a} + {1\over 2}(\g^m\t)_\a\p_m
\eeq
satisfies $\{D_\a,D_\b\} = \g^m_{\a\b}\p_m$, see \ref{sec:appA.2} for
our conventions for the $16\times 16$ Pauli matrices $\g^m_{\a\b}$.

\paragraph{Non-linear equations of motion}
The constraint \eqref{SYMconst}
and the associated Bianchi identities imply the following
non-linear equations of motion \cite{siegelSYM,wittentwistor}
\begin{align}
\label{SYMeomO}
\big\{ \nabla_\a , \nabla_\b \big\} &= \g^m_{\a\b} \nabla_m\,,
& \big[ \nabla_\a , \nabla_m\big] &= - (\g_m \bW)_\a\,,\\
\big\{ \nabla_{\a} ,\bW^\beta \big\} &= {1\over 4}
(\gamma^{mn})_{\a}{}^{\beta} \bF_{mn}\,,
&\big[\nabla_{\a}, \bF^{mn} \big] &=  (\Bbb W^{[m} \gamma^{n]})_\a \,,\notag
\end{align}
where
\beq\label{WWWdef}
\Bbb F_{mn} := - \big[  \nabla_m, \nabla_n \big] \ , \qquad
\Bbb W^{\a}_m := \big[  \nabla_m, \Bbb W^\a \big]
\,,
\ee
and we recall that our conventions for (anti-)symmetrizing $k$ indices do not contain factors of
$\frac{1}{k!}$, e.g. $T^{[\mu \nu]}=T^{\mu \nu}-T^{\nu \mu }$. The superfields $\bF^{mn}$ and $\bW^\a$
are the field strengths of the gluon and gluino, respectively.

\paragraph{Gauge invariance} The equations \eqref{SYMeomO} are invariant under the infinitesimal gauge transformations of
the superfield connections under a Lie algebra-valued gauge parameter superfield $\Omega=\Omega(X,\theta)$
\begin{align}
\delta_{\Omega } \Bbb A_\a & = \big[\nabla_\a ,\Omega \big]\,,
\qquad \delta_\Omega \Bbb A_m = \big[\nabla_m ,\Omega\big] \label{NLgauge}
\end{align}
which in turn induce the gauge transformations of their field-strengths
\beq\label{gaugeF}
\delta_\Omega\Bbb W^\a =  \big[\Omega, \Bbb W^\a  \big]\, ,\qquad
\delta_{\Omega }  \Bbb F^{mn} =  \big[\Omega, \Bbb F^{mn}  \big]\, ,\qquad
\delta_{\Omega }  \Bbb W^\a_m =  \big[\Omega, \Bbb W^\a_m  \big]\,.
\eeq

\begin{lemma}
The equations \eqref{SYMeomO} imply the (massless) Dirac and Yang--Mills equations,
\beq\label{Diraceq}
\g^m_{\a\b}\big[\nabla_m , \bW^\b \big] = 0\,,\qquad{}
\big[\nabla_m, \Bbb F^{mn}\big] = \g^n_{\a\b} \big\{ \Bbb W^\a, \Bbb W^\b \big\}\,.
\eeq
\end{lemma}
\noindent{\it Proof.}
To obtain the Dirac equation, we use the constraint equation \eqref{SYMconst} to get
\begin{align}
\g^m_{\a\b}\big[\nabla_m , \bW^\b \big] &= \big[\{\nabla_\a,\nabla_\b\}, \bW^\b\big] =
- \big[\{\bW^\b,\nabla_\a\},\nabla_\b\big]
- \big[\{\nabla_\b,\bW^\b\},\nabla_\a\big] \notag \\
&=-{1\over4}(\g^{mn})_\a{}^\b\big[\bF_{mn},\nabla_\b\big]
={1\over4}(\g^{mn}\g^n \bW^m)_\a
-{1\over4}(\g^{mn}\g^m \bW^n)_\a  \notag \\
&={9\over2}\g^m_{\a\b}\big[\nabla_m,\bW^\b\big]\, ,
\end{align}
where we used $\g^{mn}\g^n = 9\g^m,\ \gamma^{mn}{}_{\b}{}^{\b}=0$ 
and \eqref{WWWdef} to arrive at the last line, implying that
$\g^m_{\a\b}\big[\nabla_m , \bW^\b \big]=0$. To obtain the Yang--Mills equation, one
evaluates the anti-commutator of the Dirac equation with $\g_n^{\a\d}\nabla_\d$ and uses
the Bianchi (or Jacobi) identity,
\begin{align}
0&=\g_n^{\a\d}\g^m_{\a\b}\big\{\nabla_\d,[\nabla_m,\bW^\b]\big\}
=
\g_n^{\a\d}\g^m_{\a\b}\big\{\bW^\b,[\nabla_\d,\nabla_m]\big\}
+ \g_n^{\a\d}\g^m_{\a\b}\big[\nabla_m,\{\bW^\b,\nabla_\d\}\big] \notag \\
&=-\g_n^{\a\d}\g^m_{\a\b}(\g_m)_{\d\s}\big\{\bW^\b,\bW^\s\big\}
+ {1\over4}\g_n^{\a\d}\g^m_{\a\b}(\g^{rs})_\d{}^\b\big[\nabla_m,\bF_{rs}\big]\notag\\
&=8\g^n_{\b\s}\big\{\bW^\b,\bW^\s\big\}-8\big[\nabla_m,\bF^{mn}\big]\, ,
\end{align}
where to arrive in the last line we used the Clifford algebra \eqref{littleclifford} and $\g^m\g_m=10$ to obtain
$-(\g^m\g^n\g_m)_{\b\s}=8\g^n_{\b\s}$ and used the trace relation \eqref{gamma_traces} to get
${1\over4}\Tr\big(\g_m\g_n\g^{rs}\big)
=4(\d^r_n\d^s_m - \d^r_m\d^s_n)$.\qed

\paragraph{Non-linear equations of motion}
The equations of motion \eqref{SYMeomO} can also be rewritten as
\begin{align}
\{\nabla_{\a},\Bbb A_{\b}\} + \{\nabla_{\b},\Bbb A_\a\}&=\gamma^{m}_{\a\b}\Bbb A_m - \{\bA_\a,\bA_\b\}\, ,
& [\nabla_{\a},\Bbb A_m]&=[\p_m,\Bbb A_\a]+(\gamma_m \Bbb W)_\a\,, \notag \\
\{\nabla_\a,\Bbb W^\b\}&={1\over4}(\g^{mn})_\a^{\phantom{\a}\beta}\Bbb F_{mn} \, ,
&[\nabla_\a, \Bbb F^{mn}]&= (\Bbb W^{[m} \g^{n]})_\a\,. \label{SYMeom}
\end{align}
After using the definitions \eqref{covder} these become
\begin{align}
\{D_{\a},\Bbb A_{\b}\} + \{D_\b,\bA_\a\} &=\gamma^{m}_{\a\b}\Bbb A_m+\{\Bbb A_\a, \Bbb A_\b\} \,,
& [D_\a,\Bbb A_m]&=[\p_m,\Bbb A_\a]+(\gamma_m \Bbb W)_\a + [\Bbb A_\a,\Bbb A_m] \,,\notag\\
\{D_\alpha,\Bbb W^\beta\}&={1\over4}(\gamma^{mn})_\a^{\phantom{\a}\b}\Bbb F_{mn}+\{\Bbb A_\a,\Bbb W^\b\}\,,
& [D_\a, \Bbb F^{mn}]&= (\Bbb W^{[m} \gamma^{n]})_\a +[\Bbb A_\a,\Bbb F^{mn}]\,,\label{solvePert}
\end{align}
which will be used later in section \ref{BGEOMsec} to obtain the Berends--Giele recursions for superfields from a perturbiner expansion. For later convenience, we use the collective notation $\Bbb K$ referring to any element of the set containing
these superfields,
\be\label{BbbKs}
\Bbb K \in \{\Bbb A_\a, \Bbb A_m, \Bbb W^\a, \Bbb F^{mn}\}\,.
\ee

\subsection{\label{linSYMsec}Linearized superfields}

The asymptotic states for external gauge multiplets in scattering amplitudes
are addressed through the {\it linearized} description of ten-dimensional SYM.
This is obtained by discarding the quadratic terms from the
equations of motion \eqref{SYMeom} and yields
\begin{align}
D_{\a} A^i_{\b} + D_{\b} A^i_{\a} & = \g^m_{\a\b} A^i_m\,, & D_\a A^i_m &= (\g_m W_i)_\a + \p_m A^i_\a\,,
\notag\\
D_\a W_i^{\b} &= {1\over 4}(\g^{mn})^{\phantom{m}\b}_\a F^i_{mn}\,, &
D_\a F^i_{mn} & = \p_{[m} (\g_{n]} W_i)_\a \,.
\label{RankOneEOM}
\end{align}
In addition, the linearized version of the gauge transformations \eqref{NLgauge} are given
by
\beq\label{linegauge}
\d_\Omega A_\a = D_\a \Omega\,,\qquad
\d_\Omega A_m = \p_m \Omega\,,
\eeq
and they will play a role in the definition of massless vertices in the pure spinor formalism in
section~\ref{sec3.4}.

In the context of scattering amplitudes, the linearized superfields are labelled by
natural numbers $i$. These numbers are the {\it single-particle labels} keeping track of
the $i^{\rm th}$ external state taking part in the scattering
process.  In addition, the linearized equations \eqref{RankOneEOM}  describe the
motion of a single SYM particle with label~$i$. More abstractly, $i$ can be thought of being a \textit{letter} from the alphabet of
natural numbers.
As we will discuss in section~\ref{multiSYMsec}, the concept of labeling superfields with a single letter $i$  has been generalized for multiparticle
states labelled by \textit{words} $P$, the {\it  multiparticle superfields}.
It will then be shown in section~\ref{SYMsec} that
SYM scattering amplitudes involving multiple particles can be compactly written
in terms of these multiparticle superfields. And in section~\ref{sec:diskamp} we will see how they
are utilized in the computation of superstring amplitudes.

\subsubsection{$\t$-expansions}
\label{thetaexpsec}

The linearized version of the gauge transformations
(\ref{NLgauge}) can be used to attain Harnad--Shnider gauge $\theta^\alpha A^i_\alpha=0$,
where the $\theta$ dependence is known in terms of fermionic power-series
expansions from \cite{harnad,tsimpis,ooguri}.
After peeling off the dependence of linearized superfields on the bosonic coordinates $X^m$
via plane waves\footnote{We absorb factors of $i$ into momentum factors
$k^m$ in order to attain plane-wave factors of $ e^{k \cdot X} $ subject to the
simple conversion $\partial_m \rightarrow k_m$ instead of the more
conventional $ e^{ik\cdot X} $ with $\partial_m \rightarrow i k_m$. 
The traditional conventions can be retrieved by replacing $k_m \rightarrow i k_m$ and $s_{ij} \rightarrow - s_{ij}$ for Mandelstam variables defined in (\ref{mandef}).} 
$e^{k_i\cdot X}$ with
on-shell momentum $k_i^2=0$, the
different orders in $\theta$ alternate between
gluino wave functions $\chi^\alpha_i$ and
gluon polarization vectors $e_i^m$, or their associated linearized field strength
\beq\label{linEXP}
f_i^{mn} = k_i^m e_i^n - k_i^n e_i^m\,.
\eeq
More precisely,
\begin{align}
A^i_{\alpha}(X,\theta)&=
 \bigg\{ {1\over 2}(\theta\gamma_m)_\alpha e^m_i
+{1\over 3}(\theta\gamma_m)_\alpha (\theta\gamma^m \chi_i)
- {1\over32}(\theta\gamma^m)^\alpha(\theta\gamma_{mnp}\theta)f_i^{np} \label{linTHEX}\\
&\quad{}
+ {1\over60}(\t\g^m)_\alpha(\t\g_{mnp}\t) k^n_i (\chi_i \g^p\t)
+ {1\over1152}(\t\g^m)_\a(\t\g_{mnp}\t)(\t\g^{p}{}_{qr}\t) k_i^n f_{i}^{  qr}  + \ldots \bigg\} \, e^{k_i\cdot X} \, ,\notag\\
A_i^m(X,\t)&=
\bigg\{e_i^m
+(\t\g^m \chi_i)
-{1\over8}(\t \g^{m}{}_{pq}\t) f_i^{pq}
+{1\over12}(\t\g^{m}{}_{np}\t) k_i^n (\chi_{i} \g^{p}\t)\notag\\
&\quad{}+{1\over 192}(\t\g^{m}_{\phantom{m}nr}\t)(\t\g^{r}_{\phantom{r}pq}\t) k_i^n f_i^{  pq}
-{1\over480} (\t\g^{m}_{\phantom{m}nr}\t)(\t\g_{\phantom{r}pq}^{r}\t)k_i^n k_i^p (\chi_{i}\g^{q}\t)  + \ldots \bigg\} \, e^{k_i\cdot X} \, , \notag\\
W_i^\a(X,\t)&=
\bigg\{\chi^\alpha_i
+{1\over 4}(\t \g_{mn})^{\alpha} f_i^{mn}
-{1\over 4}(\t \g_{mn})^{\alpha} k_i^m (\chi_i \g^{n}\t)
-{1\over48} (\t\g^{\phantom{m}q}_{m})^\alpha(\t\g_{qnp}\t) k_i^m f^{np}_i \notag\\
&\quad{}+{1\over96}  (\t\g^{\phantom{m}q}_{m})^{\alpha}(\t\g_{qnp}\t) k_i^m k_i^n (\chi_{i}\g^{p}\t)
- {1\over 1920}(\t\g^{\phantom{m}r}_{m})^{\alpha}(\t\g^{\phantom{nr}s}_{nr}\t)(\t\g_{spq}\t) k_i^m k_i^n f^{ pq}_{i} 
 + \ldots \bigg\} \, e^{k_i\cdot X} \, , \notag\\
F_i^{mn}(X,\t)&=
\bigg\{ f_i^{mn}
- k_i^{[m}(\chi_i\g^{n]}\t)+{1\over 8}(\t\g_{pq}^{\phantom{pq}[m}\t) k_i^{n]} f_i^{ pq}
 -{1\over12}(\t\g^{\phantom{pq}[m}_{pq}\t) k_i^{n]} k_i^p (\chi_{i}\g^{q}\t)\notag\\
&\quad{} -{1\over192}(\t\g^{\phantom{ps}[m}_{ps}\t) k_i^{n]} k_i^p f^{ qr}_i(\t\g^{s}_{\phantom{s}qr}\t)
+ {1\over 480}(\t\g^{[m}_{\phantom{[m}ps}\t) k_i^{n]} k_i^p k_i^q (\chi_{i}\g^{r}\t)(\t\g^{s}_{\phantom{s}qr}\t) + \ldots  \bigg\} \, e^{k_i\cdot X} \, , \notag
\end{align}
see (\ref{THEXone}) for the analogous $\theta$-expansions of the non-linear fields $\Bbb K$
in \eqref{BbbKs}. Terms in the ellipsis involve six or higher orders in $\theta$ which won't be needed for
the purpose of this review (the zero-mode prescription \eqref{summ.01}
annihilates
expressions with more than five $\theta$s) but can be obtained in closed form via expressions such as~\cite{tsimpis}
\beq
A_i^m(X,\t)=
\bigg\{ (\cosh \sqrt{{\cal O}})^m{}_q e_i^q
  + \bigg(  \frac{ \sinh \sqrt{{\cal O}} }{ \sqrt{{\cal O}} } \bigg){}^m{}_q (\theta \gamma^q \chi_i) \bigg\} \, e^{k_i\cdot X} \, ,
\eeq
where
\beq
{\cal O}^m{}_q  = \frac{1}{2} (\theta \gamma^m{}_{qn} \theta) k^n_i\, .
\eeq

\subsection{\label{sec:higherdim}Superfields of higher mass dimension}

As the loop order of SYM amplitudes increases so does the mass dimension of the associated
kinematic factors. In the pure spinor formalism the maximum mass dimension
for a four-point amplitude using only the standard SYM superfields in \eqref{BbbKs} is $k^2 F^4$ obtained from the pure
spinor superspace expression $\langle (\l\g^{mnpqr}\l)(\l\g^s W)F_{mn}F_{pq}F_{rs}\rangle$ at
genus two \cite{twoloop}.

Therefore it would be convenient to define SYM superfields of higher mass
dimension as compared to the standard ones in $\Bbb K$. The obvious candidates of
the form $\p_m\p_n \ldots \Bbb K$ are inadequate because
the ordinary derivatives $\partial_m$ do not preserve gauge covariance at a non-linear level
probed by higher-point amplitudes. So, instead, the connection $\nabla_m$ in (\ref{covder})
guides the subsequent definitions \cite{SYMBG}
\begin{align}
\bW^{m_1\ldots m_k\alpha} &:= \big[ \nabla^{m_1} , \bW^{m_2\ldots
m_k\alpha}\big]\,,\label{highmass}\\
\bF^{m_1\ldots m_k|pq} &:= \big[ \nabla^{m_1} , \bF^{m_2\ldots m_k|pq} \big]\,,\notag
\end{align}
where the vertical bar separates the antisymmetric pair of indices present in the
recursion start $\bF^{pq}$.

\subsubsection{\label{sec:hEOMs} Equations of motion at higher mass dimension}

Similarly as in the standard SYM superfields of \cite{siegelSYM,wittentwistor},
the equations of motion for the superfields of higher mass dimension \eqref{highmass}
follow from $\big[ \nabla_\alpha,\nabla_m \big] = -(\g_m \bW)_\alpha$ and
$\big[ \nabla_m,\nabla_n \big] = - \bF_{mn}$ together with
Jacobi identities among iterated brackets.
In general, one can prove by induction that
\begin{align}
\big\{ \nabla_\alpha , \Wg^{N\beta} \big\} &= \tfrac{1}{4} (\g_{pq})_\alpha{}^{\beta} \Fg^{N|pq}
-\!\!\!\! \sum_{\d(N) = R\otimes S\atop{R \neq \emptyset}}\!\!\!\!\!
\big\{ (\Wg \g)^R_\alpha , \Wg^{S \beta} \big\}\,,
\cr
\big[ \nabla_\alpha , \Fg^{N|pq} \big] &=  ( \Wg^{N[p} \g^{q]})_\alpha
- \!\!\!\sum_{\d(N) = R\otimes S\atop{R \neq \emptyset}}\!\!\! \big[ (\Wg \g)^R_\alpha , \Fg^{S|pq}\big]\,.\label{nabla47}
\end{align}
The vector indices have been gathered to a multi-index $N:= n_1n_2\ldots n_{k}$
with $(\Wg \gamma)^N:= (\Wg^{n_1\ldots n_{k-1}} \gamma^{n_{k}})$
and $\d(N)$ denotes
the deshuffle map defined in \eqref{deshuffle}.
The simplest examples of \eqref{nabla47} are given by
\begin{align}
\big\{ \nabla_\alpha ,\bW^{m\beta} \big\} &=
\tfrac{1}{4} (\g_{pq})_\alpha{}^\beta \bF^{m|pq} - \big\{ ( \bW\g^m)_\alpha, \bW^\beta\big\}\,,\\
\big[ \nabla_\alpha , \Fg^{m|pq} \big] &=( \Wg^{m[p} \g^{q]})_\alpha
- \big[ (\Wg \g^m)_\alpha, \Fg^{pq} \big]\,,\cr
\big\{ \nabla_\alpha ,\bW^{mn\beta} \big\} &=
\tfrac{1}{4} (\g_{pq})_\alpha{}^\beta \bF^{mn|pq}
- \big\{ ( \bW^m\g^n)_\alpha, \bW^\beta\big\}
- \big\{ ( \bW\g^m)_\alpha, \bW^{n\beta}\big\}
- \big\{ ( \bW\g^n)_\alpha, \bW^{m\beta}\big\}\,,\cr
\big[ \nabla_\alpha , \Fg^{mn|pq} \big] &=( \Wg^{mn[p} \g^{q]})_\alpha
- \big[ (\Wg^m \g^n)_\alpha, \Fg^{pq} \big]
- \big[ (\Wg\g^m)_\alpha, \Fg^{n|pq} \big]
- \big[ (\Wg\g^n)_\alpha, \Fg^{m|pq} \big]\,,\cr
\{\nabla_\a,\Wg^{mnp\,\b}\} &= \tfrac{1}{4} (\g_{rs})_\alpha{}^{\beta} \Fg^{mnp|rs}
-\{(\Wg^{mn}\g^p)_\a,\Wg^\b\}
-\{(\Wg^{m}\g^n)_\a,\Wg^{p\,\b}\}
-\{(\Wg^{m}\g^p)_\a,\Wg^{n\,\b}\}\cr
&\quad -\{(\Wg^{n}\g^p)_\a,\Wg^{m\,\b}\}
-\{(\Wg\g^p)_\a,\Wg^{mn\,\b}\}
-\{(\Wg\g^n)_\a,\Wg^{mp\,\b}\}
-\{(\Wg\g^m)_\a,\Wg^{np\,\b}\}\,,\notag
\end{align}
where we used $\delta(mnp)=mnp\otimes\emptyset + mn\otimes p + mp\otimes n
+ np\otimes m + p\otimes mn + n\otimes mp + m\otimes np +\emptyset\otimes mnp$.
One can also show inductively that the Dirac- and Yang--Mills equations \eqref{Diraceq}
generalize as follows at higher mass dimension:
\begin{align}
[\nabla_m, (\gamma^m \Wg^N)_\alpha] &=
\sum_{\d(N)=R\otimes S \atop{R\neq \emptyset}} \big[ \Fg^{Rm} ,
(\g_m\Wg^{S})_\a \big]\,,
\label{nabla52} \\
[\nabla_m,\Fg^{N|p m}] &= \delta_{mn}\sum_{\d(N)=R\otimes S \atop{R\neq \emptyset}}
\big[ \bF^{Rm} , \bF^{S | pn} \big]
- \sum_{\d(N) = R\otimes S} \big\{ \bW^{R\alpha} , (\g^p\bW^{S})_\a \big\}\, ,
\label{nabla53}
\end{align}
where $\Fg^{Rr}$ for non-empty $R:= Qq$ is defined as $\Fg^{Qqr}:= \Fg^{Q|qr}$.
For example,
\begin{align}\label{dirzero}
[\nabla_m, (\gamma^m \Wg^{n})_\alpha] &=
\big[ \Fg^{n r} , (\g_r\Wg)_\a \big]\,,\\
[\nabla_m, (\gamma^m \Wg^{np})_\alpha] &=
\big[ \Fg^{n|pr} , (\g_r\Wg)_\a \big]
+\big[ \Fg^{nr} , (\g_r\Wg^{p})_\a \big]
+\big[ \Fg^{pr} , (\g_r\Wg^{n})_\a \big]\,,\cr
 [\nabla_m,\Fg^{n|p m}]  &= [\Fg^{nm} , \Fg^{p}{}_m] - \{ \Wg^{n\alpha} , (\gamma^p \Wg)_\alpha \}
 - \{ \Wg^{\alpha} , (\gamma^p \Wg^n)_\alpha \}\,,\notag
\end{align}
where we used the deshuffle map $\delta(np) = np\otimes\emptyset + n\otimes p + p\otimes n + \emptyset\otimes np$.

Note that the linearized versions of higher-mass dimension superfields are simply
the outer products of derivatives
\beq
W_i^{m_1 \ldots m_k \alpha}= \partial^{m_1} \ldots \partial^{m_k} W_i^\alpha \, , \ \ \ \
F_i^{m_1 \ldots m_k| pq }= \partial^{m_1} \ldots \partial^{m_k} F_i^{pq} \, ,
\eeq
where $i$ denotes a single-particle label.
In this case, the equations of motion \eqref{nabla52} and (\ref{nabla53}) translate into
\beq
\p_m (\gamma^m W^N_i)_\a = 0 \, , \qquad  \p_m F^{N| mp}_i = 0 \, .
\eeq
In case of an empty multi-index $N\rightarrow \emptyset$, this includes the
linearized Dirac and Yang--Mills equations
$\partial_m (\gamma^m W_i)_\alpha = 0 $ and $\partial_m F^{mp}_i = 0$.

The higher-mass-dimension superfields obey further relations which
can be derived from Jacobi identities of nested (anti)commutators. For example,
\eqref{WWWdef} determines
their antisymmetrized components
\begin{align}
\Wg^{[n_1 n_2] n_3\ldots n_k \beta}_{}&= \big[  \Wg^{n_3\ldots n_k \beta} ,
\Fg^{n_1n_2} \big]\,,
 \label{nabla50} \\
\Fg^{[n_{1} n_{2}] n_{3} \ldots n_k |pq} &= \big[  \Fg^{n_3\ldots n_k |pq} , \Fg^{n_1n_2} \big]\,.
\notag
\end{align}
Similarly, more antisymmetrized indices give rise to nested commutators, for instance
\begin{align}
\label{antiWs}
\bW^{[mn]\beta} &=
[\bW^\b,\bF^{mn}]\,,\\
\bW^{[mnp]\beta} &= [\Wg^{m\b},\bF^{np}]
+ [\Wg^{n\b},\bF^{pm}]
+ [\Wg^{p\b},\bF^{mn}]\,,\notag\\
\Wg^{[mnpq]\beta} &=
[[\Wg^{\b},\Fg^{mn}],\Fg^{pq}]
-[[\Wg^{\b},\Fg^{mp}],\Fg^{nq}]
+[[\Wg^{\b},\Fg^{mq}],\Fg^{np}]\cr
&\quad +[[\Wg^{\b},\Fg^{np}],\Fg^{mq}]
-[[\Wg^{\b},\Fg^{nq}],\Fg^{mp}]
+[[\Wg^{\b},\Fg^{pq}],\Fg^{mn}]\,,\notag
\end{align}
with similar expressions at higher multiplicities.

Moreover, the definitions \eqref{highmass} via iterated commutators imply the generalized Jacobi identities of
section~\ref{genJacsec}
on the set of vector indices, of which first instances are
\begin{align}
\label{lie3}
\Fg^{[m|np]} &= 0\,, \quad \Fg^{[mn]|pq} + \Fg^{[pq]|mn}  = 0 \,.
\end{align}

\section{Pure spinor formalism and disk amplitudes}
\label{sec:theform}

The discovery of the pure spinor formalism by Berkovits in \cite{psf} led to
an efficient tool to compute superstring
scattering amplitudes in a manifestly super-Poincar\'e invariant manner. It combined numerous convenient
aspects of the Ramond--Neveu--Schwarz (RNS) \cite{FMS, DHoker:1988pdl, DHoker:2002hof, Witten:2012bh}
and Green--Schwarz (GS) formulations \cite{GSI,GSII} in a way
that allowed computations of various amplitudes previously out of reach. 

In this section we will review the basic aspects of the formalism with
a view towards the prescription to compute disk amplitudes in the superstring;
multi-loop aspects will not be covered, but a path through the recent literature can be found 
in section \ref{sec:conclu}.
The presentation will follow the ICTP lectures by Berkovits \cite{ICTP} as
well as a combination of the PhD theses of the present authors \cite{thesisCM,thesisOS}.

We will now present some of the motivations that
led to the development of the pure spinor formalism.

\subsection{Difficulties with the covariant quantization of the Green-Schwarz string}
\label{sec:gsdiff}

Type I superstrings \cite{Green:1980zg}, type II superstrings
\cite{Green:1981yb} and heterotic strings \cite{Gross:1984dd} are
supersymmetric in ten-dimensional space-time and therefore it is natural to
seek a manifestly 10d supersymmetric description of their worldsheet action.
This is traditionally achieved with the GS formalism \cite{GSI,GSII} but
unfortunately the classical action cannot be quantized while maintaining
Lorentz covariance. 

The GS action for heterotic superstrings (or a chiral half of type II superstrings) 
in conformal gauge is given by \cite{GSI}
\begin{align}
S_{\rm GS} &=
 \frac{1}{\pi}\int d^2z \left[ 
 \frac{1}{2} \Pi^m \overline{\Pi}_m+ \frac{1}{4} \Pi_m(\theta \gamma^m \bar \partial \theta)
 -  \frac{1}{4} \overline{\Pi}_m(\theta \gamma^m \partial \theta)
 \right] \label{GSact}  \\
 &=  \frac{1}{\pi}\int d^2z \left[ \frac{1}{2}\p X^m \bar{\p}X_m 
+ \frac{1}{2}  \p X_m (\theta \gamma^m \bar \partial \theta)
+ \frac{1}{8}  (\theta \gamma^m  \partial \theta) (\theta \gamma_m \bar \partial \theta)
 \right] \, ,
 \notag
\end{align}
where we employ supersymmetric momenta
\beq
\Pi^m= \partial X^m + \frac{1}{2} (\theta \gamma^m \partial \theta) \, , \ \ \ \
 \overline{\Pi}^m = \bar \partial X^m + \frac{1}{2} (\theta \gamma^m \bar \partial \theta) \, .
\label{defpibarpi}
\eeq
The dependence of $X^m,\theta$ on the worldsheet coordinates $z,\bar z$
as well as the action of the gauge sector of the heterotic string is
suppressed. Throughout this review, the integration measure is
$d^2 z = d \Re z \wedge d \Im z = \frac{i}{2} d z \wedge d \bar z$, and derivatives
are denoted by the shorthands $\partial = \partial_z$ and $\bar \partial = \partial_{\bar z}$.
Holomorphic and antiholomorphic derivatives are related to those w.r.t.\ worldsheet
coordinates $\sigma^0= \frac{1}{2}(z+\bar z)$ and $\sigma^1=\frac{1}{2}(z-\bar z)$
via $\partial_0 = \partial + \bar \partial$ and $\partial_1 = \partial - \bar \partial$.
Following the standard closed-string
conventions, we are setting $\alpha'=2$ in sections \ref{sec:theform} to \ref{SYMsec} 
(but will reinstate it in sections \ref{sec:diskamp} 
to \ref{apsec}).\footnote{The $\alpha'$-dependence of the
worldsheet CFT and the associated scattering amplitudes can be reinstated based on dimensional
analysis. For instance, demanding worldsheet actions to be dimensionless and $X^m,\sqrt{\ap}$ to
have dimensions of a length, we retrieve $S_{\rm GS} \rightarrow \frac{1}{\pi \alpha'}\int d^2z \p
X^m \bar{\p}X_m +\ldots$.}

Covariant quantization of (\ref{GSact}) is hindered by a technical challenge:
the conjugate momentum to $\t^{\a}$ 
\beq
p_{\alpha} = 2\pi \frac{ \delta S_{\rm GS} }{\delta( \partial_0 \theta^\alpha)}
= \frac{1}{2} \bigg( \Pi^m - \frac{1}{4} (\theta \gamma^m \partial_1 \theta) \bigg) (\gamma_m \theta)_\alpha
 \label{defpalpha}
\eeq
depends on $\t^{\a}$ itself, so it gives rise to the GS constraint $d_\alpha = 0$ with
\begin{equation}\label{GSconst}
d_{\alpha} = 
p_{\alpha} -  \frac{1}{2} \bigg( \Pi^m - \frac{1}{4} (\theta \gamma^m \partial_1 \theta) \bigg) (\gamma_m \theta)_\alpha \, .
\eeq
The variable $d_\a$ associated with the GS constraint satisfies the
Poisson brackets
\be
\label{ope_dd_bad}
\{d_\alpha , d_\beta \} = i \gamma^m_{\alpha \beta} \Pi_m\,.
\ee
Due to the Virasoro constraint $\Pi_m\Pi^m=0$, the relation \eqref{ope_dd_bad} mixes first- and second-class types of
constraints in a way that is difficult to disentangle covariantly \cite{Cederwall84}\footnote{Recall that first-class
(second-class)
constraints are defined by the vanishing (non-vanishing) of their
Poisson bracket \cite{diraclecture}. The constraint $\Pi^2=0$ then implies that one half of the 
Poisson brackets \eqref{ope_dd_bad} vanishes. We are grateful to 
Max Guillen for discussions on this point.}.
The standard way
to deal with this situation is to go to the light-cone gauge \cite{Green:1980zg,GSlightconeII,GSlightconeIII, Green:1981yb}, 
where the two types of constraints can be
treated separately and quantization can be achieved. However, one obviously loses
manifest Lorentz covariance in the process. These difficulties are universal to 
heterotic and type II string theories in their GS formulations.

\subsection{Siegel's reformulation of the Green--Schwarz formulation}

In 1986 Siegel \cite{siegel} proposed a new approach to
deal with the covariant quantization of the GS formalism.
His idea was to treat the conjugate momenta for
$\theta^{\alpha}$ as an independent variable, proposing the following action
for the left-moving variables
\be
\label{siegelS}
S_{\rm Siegel} = \frac{1}{\pi} \int d^2z \left[ \frac{1}{2}\p X^m \bar{\p}X_m 
    + p_{\alpha}\bar{\p} \theta^{\alpha} \right]
\ee
in which the variable $d_\alpha$
\beq
d_{\alpha} = p_{\alpha} -\frac{1}{2}\Big(\p X^m
+\frac{1}{4}(\t \g^m\p\t) \Big)(\g_m\t)_{\a}     \label{defdal}
\eeq
was assumed to be independent and not a constraint
(the difference between the expressions (\ref{GSconst}) and
(\ref{defdal}) for $d_\alpha$ is proportional to $\bar \partial \theta^\alpha$
and vanishes by the equations of motion for $p_\alpha$).
In this way, the mixing (\ref{ope_dd_bad}) of first- and second-class constraints 
of the GS formulation is not an issue in Siegel's approach.

\paragraph{Lorentz currents and energy-momentum tensor} The action \eqref{siegelS}
is easily checked to yield a Lorentz current of the spinor variables\footnote{The double-colon
notation for normal ordering of coincident operators, ${:}A(z)B(z){:}\,$, will
be left implicit in this review.}
\be
\label{sigma_siegel}
\Sigma^{mn} = -\frac{1}{2}(p\gamma^{mn}\theta)
\ee
and a holomorphic component $T:=T(z)$ of the energy-momentum tensor
\beq\label{enmom}
T = -\frac{1}{2} \p X^m \p X_m - p_{\alpha}\p \theta^{\alpha} = -\half \Pi^m\Pi_m -d_\a \p\t^\a \, .
\eeq
The supersymmetric momentum $\Pi^m = \partial X^m + \frac{1}{2}(\theta \gamma^m \partial \theta)$
is defined as in section \ref{sec:gsdiff} though its right-moving counterpart $\overline{\Pi}^m$
relevant for type II superstrings departs from (\ref{defpibarpi}) and is defined 
with separate $\theta$-variables. 
For example, under the Lorentz transformation with parameters $\ve_{mn}$,
\beq
\delta p_{\alpha}  = 
 \frac{1}{4}\ve_{mn}(\gamma^{mn})_{\a}^{\phantom{a}\beta}p_{\beta}\, , \qquad
\delta \theta^{\a}  = \frac{1}{4}\ve_{mn}(\gamma^{mn})^{\a}_{\phantom{a}\beta}\t^{\beta}\, ,
\eeq
we define the variation of \eqref{siegelS} to be
$\delta S_{\rm Siegel} = -\frac{1}{\pi} \int \frac{1}{2} \Sigma^{mn}\bar{\p}\varepsilon_{mn} $.
The calculation using Noether's method is straightforward
\begin{align}
\d S_{\rm Siegel} &= \frac{1}{\pi}\int d^2 z\,  \delta(p_{\alpha}\bar{\p}\theta^{\alpha}) =
  \frac{1}{\pi}\int d^2 z\left[
 \frac{1}{4}\ve_{mn}(\gamma^{mn})_{\a}^{\phantom{a}\beta}p_{\beta} \bar{\p}\theta^{\alpha}
+\frac{1}{4}p_{\alpha}\pbar (\ve_{mn}(\gamma^{mn}\t)^{\alpha})
   \right] \notag \\
 & = \frac{1}{\pi}\int d^2 z \left[
\frac{1}{4}\bar{\p}\ve_{mn}p_{\a}(\gamma^{mn})^{\a}_{\phantom{a}\beta}\t^{\beta}
 \right] = -\frac{1}{\pi} \int d^2 z \frac{1}{2} \Sigma^{mn}\bar{\p}\varepsilon_{mn}\, ,
\end{align}
where we used the antisymmetry $(\g^{mn})_\a{}^\b = - (\g^{mn})^\b{}_\a$, see \eqref{symgams}.

\paragraph{CFT}
The action \eqref{siegelS} defines a conformal field theory in which the holomorphic conformal 
weights of $\p X^m, p_\a$ and $\t^\a$ are $h_{\partial X} = h_p= 1$ and $h_\theta=0$, respectively.
See \cite{yellowpages} for an in-depth review of conformal field theory.
The operator product expansions (OPEs) among the variables
in $ S_{\rm Siegel} $ follow from standard path-integral methods \cite{siegel}
\begin{align}
 X^m(z,\bar{z})X^n(w,\bar{w}) &\sim - \delta^{mn}\ln|z-w|^2\, ,
& p_{\a}(z)\t^{\b}(w) &\sim \frac{\d^{\b}_{\a}}{z-w}\,, \label{pteta}\\
 d_{\a}(z)d_{\b}(w) &\sim - \frac{\g^m_{\a\b}\Pi_m(w)}{z-w}\,,
& d_{\a}(z)\Pi^m(w) &\sim  \frac{ \big(\g^m \p\t(w) \big)_{\a}}{z-w}\, ,\label{dd_ope}
\end{align}
where here and throughout this review, $\sim$ indicates that 
regular terms as $z \rightarrow w$ are dropped on the right-hand side.

\paragraph{Vertex operator}
Siegel also proposed a supersymmetric
integrated vertex operator for massless open-string states labeled by $i$ as follows
\be
\label{siegel_U}
U_i^{\rm Siegel} = \int dz\, \big(\p\t^{\a}A^i_{\a}(X,\t) +A^i_m(X,\t)\Pi^m + d_{\a}W_i^{\a}(X,\t) \big)\,,
\ee
where $\{A^i_\a,A_i^m,W_i^\a\}$ are the linearized
SYM superfields reviewed in section \ref{linSYMsec}.

\subsubsection{Difficulties with Siegel's approach}

There are three types of difficulties with Siegel's approach which will be
addressed by the pure spinor formalism to be introduced in section \ref{sec3.3} below.

\paragraph{Non-vanishing central charge}
According to the $bc$-system calculations \cite{FMS} with conformal weight 
$h_p=1$, each spinor component of the fermionic pair
$(p_{\a}, \t^{\a})$ in the energy-momentum tensor \eqref{enmom}
contributes $-3(2h_p -1)^2 + 1 = -2$ to the central charge for a total of $16\times( -2)= -32$
while the $ X^m$ contribute $+10$ \cite{Polchinski:1998rq}. Therefore the
central charge of the energy-momentum tensor \eqref{enmom} is $c_{X}+c_{p\theta}=10-32=-22$.
This non-vanishing result for the central charge leads to an anomaly when
quantizing the theory, raising a first major difficulty in Siegel's
approach to the GS formalism.

\paragraph{Inequivalence of massless vertex operators}

As emphasized in \cite{psf}, the vertex operator \eqref{siegel_U} cannot reproduce the same results
for amplitudes computed in the RNS formalism as it does not satisfy the same OPEs. More
explicitly, after using the $\t$-expansions \eqref{linTHEX} of the linearized SYM superfields,
the gluon vertex operator obtained from \eqref{siegel_U} is
\beq\label{int_siegel}
U_{i, \rm gluon}^{\rm Siegel} = \int dz\, \Big(e^m_i \p X_m - \frac{1}{4}(p\g^{mn}\t)f^i_{mn} + \ldots \Big)e^{k_i\cdot X}
\eeq
up to terms of order $\theta^3$ in the ellipsis.
The vertex operator for a gluon
with polarization vector  $e_i^m$ in the RNS formalism, on the other hand,
is given by (see (7.3.25) in \cite{gswI}\footnote{The relative factors
of the contributions from $ \p X_m $ and $\psi^m \psi^n$ to (\ref{int_rns}) depart
from most references for the following reason: for plane waves $e^{ik\cdot X}$, 
the RNS vertex operator at superghost picture zero is proportional to
$e^m (i \p X_m + (k\cdot \psi) \psi_m)$, and our conventions including
(\ref{int_rns}) are obtained by rescaling $k \rightarrow -ik$.})
\be
\label{int_rns}
U_{i, \rm gluon}^{\rm RNS} = \int dz\, \Big(e_i^m \p X_m
- \frac{1}{2}\psi^m \psi^n f^i_{mn}\Big)e^{k_i\cdot X}\, ,
\ee
where $\psi^m$ are the RNS worldsheet spinors of conformal weight $h_\psi=\frac{1}{2}$,
and $f^i_{mn} = k^i_m e^i_n - k^i_n e^i_m$ denotes the linearized field strength of the gluon.

Comparing \eqref{int_rns}
with \eqref{int_siegel} one notices
that the operator multiplying $\frac{1}{2}f^i_{mn}$ is the Lorentz current for the fermionic
variables in each formalism,
\beq\label{lors}
\Sigma^{mn}_{\rm RNS} = -\psi^m\psi^n,\qquad \Sigma^{mn}_{\rm Siegel} = -{1\over2}(p\g^{mn}\t)\,.
\eeq
The difficulty arises because their OPEs are different. On the one hand, in the RNS formalism we get
\be
\label{ope_rns}
\Sigma_{\rm RNS}^{mn}(z)\Sigma_{\rm RNS}^{pq}(w) \sim
\frac{\delta^{p[m}\Sigma_{\rm RNS}^{n]q}(w) - \delta^{q[m}\Sigma_{\rm RNS}^{n]p}(w)}{z-w} 
  + \frac{\delta^{m[q}\delta^{p]n}}{(z-w)^2}\, ,
\ee
where the double-pole term has coefficient $+1$ which can be identified with the level
of the Kac--Moody current algebra. On the other hand,
using the OPE \eqref{pteta} we get
\begin{align}
\Sigma_{\rm Siegel}^{mn}(z)\Sigma_{\rm Siegel}^{pq}(w)
         &\sim \frac{1}{4}\frac{ p_\alpha(w)(\gamma^{mn}\gamma^{pq} - \gamma^{pq}\gamma^{mn})^{\alpha}{}_{\beta}\theta^{\beta}(w)}{z-w}
	    +\frac{1}{4}\frac{\Tr(\gamma^{mn}\gamma^{pq})}{(z-w)^2}
\notag \\
&= \frac{\delta^{p[m}\Sigma^{n]q}(w)  - \delta^{q[m}\Sigma^{n]p}(w)}{z-w}
 + 4\frac{\delta^{m[q}\delta^{p]n}}{(z-w)^2}\, ,
 \label{ope_siegel}
\end{align}
where in the second line we used
$\gamma^{mn}\gamma^{pq} - \gamma^{pq}\gamma^{mn} = 2\delta^{np}\g^{mq} -
2\delta^{nq}\g^{mp}+ 2\delta^{mq}\g^{np} - 2\delta^{mp}\g^{nq}$ following from \eqref{smallex} and ${\Tr}(\gamma^{mn}\gamma_{pq}) =
16( \delta^m_q \delta^n_p -  \delta^m_p \delta^n_q)$ from \eqref{gamma_traces}.

The discrepancy in the coefficient of the double pole between \eqref{ope_rns} and
\eqref{ope_siegel} leads to analogous discrepancies in the computations of gluon scattering amplitudes
using the RNS vertex operators \eqref{int_rns} and those of Siegel in \eqref{int_siegel}.

\paragraph{Missing constraints}
Finally in Siegel's formulation \eqref{siegelS} one would need to include an appropriate set of first-class
constraints to reproduce the superstring spectrum:
the Virasoro constraint $T$ and the kappa symmetry generator $G$ of the GS formalism
\beq
T = -\frac{1}{2}\Pi^m\Pi_m - d_{\alpha}\p \theta^{\alpha}\,,\qquad
G^{\alpha} =\Pi^m(\gamma_m d)^{\alpha}
\label{TandG}
\eeq
in terms of the supersymmetric momentum and GS constraints
should certainly be elements of that set of constraints.
Even though there was a successful description of the superparticle using
Siegel's approach \cite{siegelsuperparticle,siegelsuperparticleII}, the whole set of constraints was never found for the
superstring case. Nevertheless,
Siegel's idea was not lost as it was used by Berkovits in his
proposal for the pure spinor formalism \cite{psf}.

\subsection{\label{sec3.3}Fundamentals of the pure spinor formalism}

We have seen above that while Siegel's approach circumvented the difficulties
associated to the GS constraint, the non-vanishing central charge 
$c_{X}+c_{p\theta}=-22$ and
the level $+4$ of the Lorentz current algebra presented serious challenges to this new
formulation. This motivated
Berkovits to modify Siegel's approach by introducing
{\it pure spinor} ghost variables contributing $+22$ to the central charge of the energy
momentum tensor and $-3$ to the double pole in the
OPE of the Lorentz currents, thereby fixing the most pressing issues
with the formulation by Siegel and leading to
Berkovits' {\it pure spinor formalism} \cite{psf}. Let us briefly review below some of the central elements
in this reformulation.

\paragraph{Lorentz currents for the ghosts}
Berkovits' idea was to modify the Lorentz currents \eqref{sigma_siegel}
by the addition of a contribution $N^{mn}$ coming from ghosts,
\beq\label{Mps}
M^{mn} = \Sigma^{mn} + N^{mn}.
\eeq
The newly defined $M^{mn}$ would satisfy the same OPE \eqref{ope_rns} 
as in the RNS formalism if the contribution to the double pole arising from the
ghosts $N^{mn}$ had a coefficient $-3$,\footnote{See \cite{Berkovits:2005hy} for a discussion of how to derive these OPEs from the decomposition of $N^{mn}$
in terms of $\lambda^\a$ and $w_\a$.} 
\begin{align}
N^{mn}(z)N^{pq}(w) &\sim \frac{\delta^{p[m}N^{n]q}(w)
- \delta^{q[m}N^{n]p}(w)}{z-w}
-3\frac{\delta^{m[q}\delta^{p]n}}{(z-w)^2}\, ,\notag\\
\Sigma^{mn}(z)N^{pq}(w) &\sim \text{regular}\,.
\label{ope_NN}
\end{align}
This would fix the issue with the Lorentz current OPE and set the level of the overall
Lorentz currents $M^{mn}$ to $4-3=1$, in lines with the level of the RNS currents
in (\ref{ope_rns}).

\paragraph{Energy-momentum tensor for the ghosts}
To fix the problem with the non-vanishing central charge of the energy-momentum
tensor in Siegel's approach, one would need these same ghosts to have a central
charge $c_\lambda=+22$.
Fortunately, the right solution to both
problems was found when a proposal for the BRST charge was put forward and the need
for \textit{pure spinors} became evident.

\paragraph{The BRST operator}
The next step in the line of reasoning which led to the pure spinor
formalism is the proposal of the BRST operator
\be
\label{Q_brst}
Q_{\rm BRST} = \oint  dz\, \lambda^{\alpha}(z)d_{\alpha}(z)\, ,
\ee
where $\l^{\a}$ are bosonic spinors and the Siegel variable $d_\a$ corresponding
to the GS constraint has been defined in (\ref{defdal}). The BRST
charge \eqref{Q_brst} must satisfy the consistency condition $Q^2_{\rm BRST}=0$,
otherwise the BRST charge itself would not be invariant under a variation
of the gauge constraint \cite{Polchinski:1998rq}. Using \eqref{Q_brst} and the 
OPE \eqref{dd_ope} we obtain
\beq
Q^2_{\rm BRST} = \frac{1}{2}\{Q_{\rm BRST}, Q_{\rm BRST}\} = 
            -\frac{1}{2}\oint dz \, (\lambda\gamma^m\lambda)\Pi_m\,.
\eeq
Therefore imposing that the BRST charge is nilpotent
\beq\label{Qsquared}
Q_{\rm BRST}^2=0
\eeq
implies that the bosonic fields $\l^{\a}$ must satisfy the \textit{pure spinor} constraints
\be
\label{espinor_puro}
\lambda\gamma^m\lambda=0\,,
\ee
which were first studied by Cartan from a geometrical perspective \cite{cartan}.

\subsubsection{$U(5)$ decompositions}

The formalism discovered by Berkovits is based on the properties of the pure spinor $\l^{\a}$, and
it is important to identify the number of degrees of freedom which survive
the constraints \eqref{espinor_puro}. Naively, one could think that those ten constraints
associated with $m=0,1,\ldots,9$ would imply a pure spinor of $SO(1,9)$ to 
have only $16-10=6$ degrees of freedom. However, this not the case; we will see below that
a pure spinor has eleven degrees of freedom.

\paragraph{$U(5)$ decomposition of pure spinors}
In order to see that a pure spinor has eleven degrees of freedom, it is
convenient to Wick rotate $SO(1,9)$ to $SO(10)$
and to break manifest $SO(10)$ symmetry to its $U(5)$ subgroup \cite{psf}.
The explicit calculations are shown in \ref{u5app}, with the result that a Weyl spinor
decomposes into irreducible $U(5)$ representations as 
\beq\label{u5lambdaA}
\l^\a \longrightarrow (\l^+,\l_{ab},\l^a)\,
\eeq
corresponding to ${\bf 16}\longrightarrow ({\bf 1},{\bf \bar{10}},{\bf 5})$ with $\l_{ba}=-\l_{ab}$. The solution to the
pure spinor constraint \eqref{espinor_puro} further implies that
\beq\label{u5lambda}
\l^a = {1\over 8\l^+}\e^{abcde}\l_{bc}\l_{de}\, , \quad a,b,c,d,e=1, \ldots,5
\eeq
for $\l^+\neq0$, where $\e^{abcde}$ is totally antisymmetric with $\e^{12345}=1$. In this
language, $\l_{ab}$ parameterize a $SO(10)/U(5)$ coset.
The pure spinor constraint therefore
only eliminates the ${\bf 5} \ni \l^a$ in favor of $\l^+ \in {\bf 1}$ and $\l_{ab} \in {\bf \bar{10}}$.
Hence, there remains $1+10=11$ degrees of freedom in a pure spinor of $SO(10)$.

Note that in absence of Wick rotation the
$\l_{ab}$ parameterize the compact space $SO(1,9)/(U(4)\times\cR^9)$ with $\cR^9$ representing
nine light-like boosts \cite{furlan,RNSPS,cherkis}.

\paragraph{$U(5)$ decomposition of the Lorenz currents}
To solve the pure spinor constraint \eqref{espinor_puro} it was convenient to break the manifest
$SO(10)$ symmetry to its subgroup $U(5)$, so a pure spinor is
written in terms of $U(5)=SU(5)\otimes U(1)$ variables. Consequently, the Lorentz currents must 
also be decomposed to their irreducible $U(5)$ representations
\beq
\label{Nu5s}
N^{mn} \longrightarrow (n, n^b_a, n_{ab}, n^{ab})\, ,
\eeq
with corresponding $U(1)$ charges $(0,0,{}+2,{}-2)$
in a manner specified in \ref{u5app}.
In the remainder of this section, these $SU(5)$ Lorentz currents will be constructed out of elementary
ghost variables to be denoted by $s(z)$, $u_{ab}(z)$ and their conjugate momenta $t(z),v^{ab}(z)$
such that the required condition \eqref{ope_NN} is met. 
To do this we will first state how the OPE \eqref{ope_NN} decomposes under $SO(10)\rightarrow
SU(5)\otimes U(1)$ given by \eqref{Nu5s}:
\begin{prop.}
\label{teo_lorentz}
The SO(10)-covariant OPE of the Lorentz currents
\begin{equation}
\label{ope}
N^{mn}(z)N^{pq}(w) \sim \frac{\d^{mp}N^{nq}(w) - \d^{np}N^{mq}(w) - \d^{mq}N^{np}(w) +
\d^{nq}N^{mp}(w)}{z-w}
- 3\frac{\big(\delta^{mq}\delta^{np} - \delta^{mp}\delta^{nq}\big)}{(z-w)^2}\, ,
\end{equation}
implies that the $SU(5)\otimes U(1)$ currents $(n,n_a^b,n_{ab},n^{ab})$ satisfy the
following OPEs:
\begin{align}
\label{ope1}
n_{ab}(z)n_{cd}(w) &\sim  {\rm regular}\,,
& n^{ab}(z)n^{cd}(w) &\sim {\rm regular}\, ,\\
n_{ab}(z)n^{cd}(w) &\sim {-\d_{[a}^c n^d_{b]}(w) + \d_{[a}^d n^c_{b]}(w)
- {2\over \sqrt{5}}\d^c_{[a}\d^d_{b]}n(w) \over z-w}
- 3{\d^c_b\d^d_a - \d^c_a\d^d_b\over (z-w)^2}\, , 
&n(z)n^{a}_{b}(w) &\sim {\rm regular}\,,\notag\\
n^{a}_{b}(z)n^c_d(w) &\sim {-\d^c_b n^a_d(w) + \d^a_d n^c_b(w)\over z-w} -
3{\d^a_d\d^c_b-{1\over5}\d^a_b\d^c_d\over(z-w)^2}\,,
& n(z)n_{ab}(w) &\sim + {2\over\sqrt{5}}{n_{ab}(w)\over z-w} \, ,\notag\\
n^{ab}(z)n^{c}_{d}(w) &\sim {-\d^a_d n^{bc}(w) + \d^b_d n^{ac}(w) - {2\over5}\d^c_d n^{ab}(w)\over z-w}\,,
& n(z)n^{ab}(w) &\sim -{2\over\sqrt{5}}{n^{ab}(w)\over z-w}\, ,\notag\\
n_{ab}(z)n^{c}_{d}(w) &\sim {-\d^c_b n_{ad}(w) + \d_a^c n_{bd}(w) +{2\over5}\d^c_dn_{ab}(w)\over z-w}\,,
&n(z)n(w) &\sim -{3\over (z-w)^2} \, .\notag
\end{align}
\end{prop.}
\noindent{\it Proof.}
See \ref{decsec} and also \cite{dissertacao,Hoogeveen:2010yfa}.\qed

\paragraph{$U(5)$ decomposition of spinors} There is one more consistency condition to be obeyed when constructing
the $U(5)$ Lorentz currents. The pure spinor
$\la$ must obviously transform as a spinor under the action of the total Lorentz current $M^{mn}$
in \eqref{Mps},
\beq
\delta \la = \frac{1}{2}\left[ \oint dz
\, \ve_{mn}M^{mn},\la \right] = \frac{1}{4}\ve_{mn}(\g^{mn}\l)^{\a}\, .
\label{dellambda}
\eeq
Since the OPE of $\la$ with the Lorentz currents $\Sigma^{mn}$ of \eqref{sigma_siegel} is regular we conclude that
the pure spinor must satisfy \eqref{lambda_N} given below. Given the
solution of the pure spinor constraint \eqref{u5lambda}
in $U(5)$ variables we need to know the group-theoretic
decomposition of how a $SO(10)$ spinor transforms in terms of its $U(5)$ representations.
\begin{prop.}
\label{teo_lambda}
The $SO(10)$-covariant transformation of a spinor
\beq
\label{lambda_N}
N^{mn}(z)\lambda^{\alpha}(w) \sim 
\frac{1}{2}\frac{(\g^{mn})^{\a}_{\phantom{m}\b}\l^{\b}(w)}{(z-w)}\, ,
\eeq
implies that the OPEs among the $SU(5)$ representations $(n,n^a_b,n_{ab},n^{ab})$ and
$(\lambda^+,\lambda_{cd},\lambda^c)$ are given by
\begin{align}
n(z)\lambda^+(w) &\sim -\frac{\sqrt{5}}{2}\frac{\lambda^+(w)}{z-w}\,,
& n(z)\lambda_{cd}(w) &\sim -\frac{1}{2\sqrt{5}}\frac{\lambda_{cd}(w)}{z-w}\, , \label{ope_3}\\
n(z)\lambda^c(w) &\sim \frac{3}{2\sqrt{5}}\frac{\lambda^c(w)}{z-w}\, ,
& n^a_b(z)\lambda^+(w) &\sim {\rm regular}\, , \cr
n^a_b(z)\lambda_{cd}(w) &\sim \frac{ \delta^a_d\lambda_{cb}(w)
                                     - \delta^a_c\lambda_{db}(w) }{(z-w)}
                                     -\frac{2}{5}\frac{\delta^a_b\lambda_{cd}(w)}{(z-w)} \,,
& n^a_b(z)\lambda^c(w) &\sim \frac{1}{5}{\d^a_b\l^c(w)\over(z-w)} -\frac{\d^c_b\l^a(w)}{(z-w)} \, ,\cr
n_{ab}(z)\lambda^+(w) &\sim \frac{\lambda_{ab}(w)}{z-w} \, ,
& n_{ab}(z)\lambda_{cd}(w) &\sim \frac{\epsilon_{abcde}\lambda^e(w)}{z-w}\, , \cr 
n_{ab}(z)\lambda^c(w) &\sim {\rm regular} \, ,
& n^{ab}(z)\lambda^+(w) &\sim {\rm regular}  \, ,\cr
n^{ab}(z)\lambda_{cd}(w) &\sim
                         -\frac{\delta^{[a}_c\delta^{b]}_d\lambda^+(w)}{z-w} \, ,
& n^{ab}(z)\lambda^c(w) &\sim -\frac{\epsilon^{abcde}\lambda_{de}(w)}{2(z-w)}\, . \notag
\end{align}
\end{prop.}
\noindent{\it Proof.} See \ref{u5app} and also \cite{dissertacao}.\qed

It turns out that all these OPEs
can be reproduced from an action involving the ghost variables $s(z)$, $u_{ab}(z)$, 
$t(z)$ and $v^{cd}(z)$ below that serve as the ingredients of the Lorentz
currents $(n,n_a^b,n_{ab},n^{ab})$ and pure spinor $(\l^+,\l_{cd},\l^c)$.
The pure spinor formalism crucially hinges on the existence of such a construction.

Before moving on, note the consistency between the OPE \eqref{lambda_N} and the simple pole of
\eqref{ope} arising from a twofold application of the spinorial transformation. That is, if
$[N^{mn},\l^\a] = {1\over2}(\g^{mn})^\a{}_\b \l^\b$ then
$[N^{pq},[N^{mn},\l^\a]]={1\over4}(\g^{mn})^\a{}_\b(\g^{pq})^\b{}_\d \l^\d$
which implies,
\begin{align}\label{consN}
[[N^{mn},N^{pq}],\l^\a] &= [N^{mn},[N^{pq},\l^\a]] - [N^{pq},[N^{mn},\l^\a]]
={1\over4}\big[(\g^{pq}\g^{mn})^\a{}_\b-(\g^{mn}\g^{pq})^\a{}_\b\big]\l^\b\\
&=
\d^{mp}[N^{nq},\l^\a]
- \d^{np}[N^{mq},\l^\a]
- \d^{mq}[N^{np},\l^\a]
+ \d^{nq}[N^{mp},\l^\a]\, ,
\notag
\end{align}
where we used the gamma-matrix identity $\g^{pq}\g^{mn}-\g^{mn}\g^{pq}=
2\d^{mp} \g^{nq}
- 2\d^{np} \g^{mq}
- 2\d^{mq} \g^{np}
+ 2\d^{nq} \g^{mp}$ which follows from
the product relation \eqref{gammaExpand}.
These OPEs play a crucial role in evaluating the CFT correlation functions
for string amplitudes and will for instance be used in the derivation of the
multiparticle vertex operators at multiplicity two in section~\ref{sec:4.1loc} \cite{5ptsimple,EOMBBs}.

\subsubsection{The pure spinor ghosts}

In this section we will display the solution to the above problems found by
Berkovits with the introduction of a specific $U(5)$ parameterization of pure spinors,
Lorentz currents and the energy-momentum tensor.

The action for the ghosts appearing in the pure spinor constraint is given by \cite{RNSPS,psf,ICTP}
\be
\label{acao_fantasma}
S_{\lambda} = \frac{1}{2\pi}\int d^2z \Big(
	{-}\p t\pbar s +\frac{1}{2}v^{ab}\bar{\p}u_{ab}
\Big)\,,\qquad a,b=1, \ldots,5 \, ,
\ee
where $t(z)$ and $v^{ab}(z)$ are the conjugate momenta for $s(z)$ and $u_{ab}(z)$.
Furthermore, $s(z)$ and $t(z)$ are chiral bosons, so one must impose their
equations of motions by hand, $\bar{\p}s=\bar{\p}t =0$.
The OPEs are given by
\begin{align}
\label{a1}
t(z)s(w) &\sim \ln{(z-w)} \, , \\
v^{ab}(z)u_{cd}(w) & \sim
\frac{\delta^{a}_c\delta^{b}_d-\d^a_d\d^b_c}{z-w}\,.\notag
\end{align}

\paragraph{Matching group theory with CFT} The fundamental result allowing the construction of the pure spinor formalism
is given by the explicit construction of
the $U(5)$ Lorentz currents $(n,n^a_b,n_{ab},n^{ab})$ and pure spinors
$(\l^+,\l_{ab},\l^a)$
in terms of the ghost variables $s(z)$, $t(z)$, $v^{ab}(z)$ and $u_{ab}(z)$ from the action \eqref{acao_fantasma}.
This has to be done in such a way as that their $U(5)$ OPEs among themselves satisfy
all the group-theoretic relations \eqref{ope1} and \eqref{ope_3}.
The solution found by Berkovits is given by\footnote{Note the sign flip of the Lorentz generators
and of $\l^a$ with respect to \cite{ICTP}. This ensures that the conventions of \ref{decsec} are respected.}
\cite{RNSPS,psf,ICTP}
\begin{align}
n &= -\frac{1}{\sqrt{5}}\Big( \frac{1}{4}u_{ab}v^{ab}+\frac{5}{2}\p t
- \frac{5}{2}\p s \Big) \, ,
& \! \! \! \!  \! \! \! \!  \! \! \! \! \l^+ &= e^s \, , \label{n_lorentz} \\
n^a_b & = -u_{bc}v^{ac} + \frac{1}{5}\delta^a_b u_{cd}v^{cd} \, ,
& \! \! \! \!  \! \! \! \!  \! \! \! \!  \l_{ab} &= u_{ab} \, ,\cr
n^{ab} & = -e^s v^{ab} \, ,
& \! \! \! \!  \! \! \! \!  \! \! \! \!  \l^a &= {1\over 8}e^{-s}\e^{abcde}u_{bc}u_{de} \, , \cr
n_{ab} & = -e^{-s}\Big(
           	2\p u_{ab} - u_{ab}\p t - 2u_{ab}\p s + u_{ac}u_{bd}v^{cd}
                - \frac{1}{2}u_{ab}u_{cd}v^{cd}
	   \Big)\, . \notag 
\end{align}
The unusual normalization of $n(z)$ was chosen such that the coefficient of its double pole is
$-3$.
Straightforward but long calculations show that their OPEs among themselves reproduce all
OPEs in \eqref{ope1} and \eqref{ope_3}, provided that the ghost variables
$s(z)$, $t(z)$, $v^{ab}(z)$ and $u_{ab}(z)$ satisfy the OPEs \eqref{a1}.
For instance, two sample calculations are
\begin{align}
n(z)n^{ab}(w) &= {1\over\sqrt{5}}\Big({1\over4}u_{fg}(z)v^{fg}(z)
+{5\over2}\p t(z)
- {5\over2}\p s(z) \Big)e^{s(w)}v^{ab}(w) \\
&={1\over\sqrt{5}}{1\over4}e^{s(w)}v^{fg}(z)\wick{\c1 u_{fg}(z) \c1 v^{ab}(w)}
-{\sqrt{5}\over2}\wick{\c1\p t(z) \c1 e^{s(w)}}v^{ab}(w)\notag\\
&\sim {1\over\sqrt{5}}{1\over4}e^{s(w)}v^{fg}(z){(-\d^a_f\d^b_g + \d^a_g\d^b_f)\over z-w}
-{\sqrt{5}\over2}{1\over z-w}e^{s(w)}v^{ab}(w)\notag\\
&\sim -{2\over\sqrt{5}} \frac{ n^{ab}(w)}{z-w}\notag
\end{align}
and
\begin{align}
n^{ab}(z)\l^c(w) &= -{1\over8}e^{s(z)}\epsilon^{cdefg}\big(\wick{\c1 v^{ab}(z) \c1 u_{de}(w)} u_{fg}(w)
+ u_{de}(w)\wick{\c1 v^{ab}(z) \c1 u_{fg}(w)}\big)e^{-s(w)}\\
&\sim-{1\over8}e^{s(z)}{\big(2\epsilon^{cabfg}u_{fg}(w)+2\epsilon^{cdeab}u_{de}(w)\big)\over z-w}e^{-s(w)}\notag\\
&\sim -{1\over2}\epsilon^{abcde} \frac{ \l_{de}(w)}{z-w} \, ,\notag
\end{align}
where we used the OPEs $u_{fg}(z)v^{ab}(w)\sim {-\d^a_f\d^b_g + \d^a_g\d^b_f\over z-w}$ and $\p
t(z) e^{s(w)}\sim{1\over z-w}e^{s(w)}$ that follow from \eqref{a1} and discarded
non-singular terms coming from Taylor expansions of fields at $z$ around $w$.
The above results
reproduce two of the OPEs in \eqref{ope1} and \eqref{ope_3} that were obtained from a
group-theoretic decomposition of the parental $SO(10)$-covariant OPEs. All the other OPEs can be
verified similarly. Therefore, even though the action for the ghosts $S_\l$ is not manifestly
Lorentz covariant, all OPEs involving $N^{mn}$ and $\l^\a$ descend from manifestly
$SO(10)$-covariant expressions. So the pure spinor formalism has manifest Lorentz covariance.

\paragraph{Energy-momentum tensor} We will show that the central charge
of the energy-momentum tensor for the ghosts
\beq
\label{emghost}
T_{\lambda} = \frac{1}{2}v^{ab}\p u_{ab} +\p t \p s + \p^2 s \,,
\ee
following from the ghost action \eqref{acao_fantasma}
is $+22$. This is indeed the required value for it to annihilate the
total central charge when added to Siegel's matter variables.
The derivation of \eqref{emghost} follows from
Noether's procedure using
\beq
\delta S_{\lambda} = \frac{1}{2\pi}\int d^2z  \, \big[\bar{\p}\varepsilon T_{\lambda}(z)
                     + \p\bar{\varepsilon} \bar{T}_{\lambda}(\bar{z}) \big]\, ,
\eeq
under the conformal transformations of $(v^{ab}, u_{ab}, \p s,\bar\p t)$ whose conformal weights
are $(1,0)$, $(0,0)$, $(1,0)$ and $(0,1)$, respectively,
\begin{align}
\label{t1}
\delta v^{ab} & = \p\varepsilon  v^{ab} +\varepsilon \p v^{ab} +\bar{\varepsilon}\bar{\p}v^{ab}\, ,
& \delta u_{ab} & =  \varepsilon \p u_{ab} +\bar{\varepsilon}\bar{\p}u_{ab} \, ,\\
\delta \p s & =  \p\varepsilon \p s +\varepsilon \p^2 s 
                 +\p\bar{\varepsilon} \bar{\p} s +\bar{\p}\bar{\varepsilon}\p s\, ,
& \delta \bar{\p}t & =  \varepsilon \p\bar{\p}t + \bar{\p}\varepsilon\p t 
                      +\bar{\p}\bar{\varepsilon}\bar{\p}t
                      +\bar{\varepsilon}\bar{\p}^2t \,,\notag
\end{align}
and requiring the Lorentz currents
$(n,n^a_b,n^{ab},n_{ab})$ to be primary fields \cite{psf,ICTP} (see also \cite{dissertacao} for
the explicit calculations).

\begin{prop.}
The central charge of the energy-momentum tensor for the ghosts \eqref{emghost} is $c_\lambda = 22$.
\end{prop.}
\noindent{\it Proof.}
The central charge is determined from the fourth-order
pole in $T_{\lambda}(z)T_{\lambda}(w)\sim{(c_\lambda/2)\over (z-w)^4} + \cdots $. There are two contributions
\begin{align}
{1\over4}\wick{\c1 v^{ab}(z) \c2\p u_{ab}(z)\c2v^{cd}(w) \c1\p u_{ab}(w)}
& = \frac{1}{4}\frac{\delta^{[a}_c \delta^{b]}_d\delta^{[c}_a \delta^{d]}_b}{(z-w)^4}
 	 =  \frac{10}{(z-w)^4} \, ,\\
\wick{\c1\p t(z)\c2 \p s(z) \c2\p t(w) \c1\p s(w)} & = \frac{1}{(z-w)^4} \, ,\notag
\end{align}
whose sum implies that $c_\lambda = +22$.\qed

Therefore, as there are no poles between the ghosts and matter variables,
the total central charge of the energy-momentum tensor in
the pure spinor formalism
\be
\label{tensor_total}
T_{\rm PS} = -\frac{1}{2} \p X^m \p X_m - p_{\alpha}\p \theta^{\alpha} +
       \frac{1}{2}v^{ab}\p u_{ab} +\p t \p s + \p^2 s\, ,
\ee
vanishes; $c_X+c_{p\t}+c_\lambda=10-32+22 = 0$. Therefore there will not be a conformal anomaly in the formalism.

\subsubsection{The action of the pure spinor formalism}

\paragraph{$U(5)$-covariant action}
From the discussion above we learn that adding the pure spinor ghost action of \eqref{acao_fantasma} to the
Siegel action \eqref{siegelS} implies that the energy-momentum tensor of the
theory has vanishing central charge, as $c_X+c_{p\t}=-22$ from the matter variables
is neutralized by $c_\lambda=22$ from the ghosts. Furthermore, the
Lorentz currents of the combined actions
have the same OPE as in the RNS formalism. Berkovits then proposed that
the pure spinor formalism action for the left-moving fields is given by \cite{psf}
\be
\label{acao_nathan}
S_{\rm PS}= \frac{1}{\pi}\int d^2z \, \Big( \frac{1}{2}\p X^m \bar{\p}X_m +p_{\a}\bar{\p}\t^{\a} -
\p t\pbar s +\frac{1}{2}v^{ab}\pbar u_{ab} \Big) \, .
\ee
Spacetime supersymmetry transformations are generated by
\beq\label{susyQ}
{\cal Q}_{\a} =  \oint dz \, \Big( p_{\a} +\frac{1}{2} (\gamma^m\t)_\alpha \p X_m 
+ \frac{1}{24} (\gamma^m\t)_\alpha  (\t \gamma_m \p
\t) \Big)\, ,
\eeq
and their action on the variables in the pure spinor formalism 
with Weyl-spinor parameter $\varepsilon^\alpha$ is given by
\begin{align}
\delta X^m  &= \frac{1}{2} \(\varepsilon \gamma^m \theta\),\quad
\delta \theta^{\a}  = \varepsilon^{\a} \, ,
\label{susy1}\\
\delta p_{\beta}  &=  -\frac{1}{2} (\varepsilon \gamma^m)_{\beta}\p X_m
 + \frac{1}{8}
 (\varepsilon \gamma_m \t) (\p   \t \g^m)_{\b} \, , 
 \notag\\
\delta s &= \delta t  = \delta u_{ab}  = \delta v^{ab}  = 0\, .
\notag
\end{align}
The action \eqref{acao_nathan} is found to be supersymmetric by exploiting that
the total derivatives
\begin{align}
\partial \big[ (\varepsilon \gamma_m \t) \pbar X^m \big] - 
\pbar \big[ (\varepsilon \gamma_m \t) \partial X^m \big] &=
(\varepsilon \gamma_m \partial \t) \pbar X^m - (\varepsilon \gamma_m \pbar \t) \partial X^m
\notag \\
\partial \big[ (\varepsilon \gamma_m \theta) (\theta \gamma^m   \pbar \theta) \big] - 
\pbar \big[ (\varepsilon \gamma_m \theta) (\theta \gamma^m \partial\theta)  \big]
&= 3 (\varepsilon \gamma_m \theta)  (\partial \theta \gamma^m  \pbar \theta) \label{bdyterms}
\end{align}
from the variation of $\partial X^m \pbar X_m$ and the $\theta^2$-contribution 
to $\delta p_\beta$ integrate to zero under $d^2 z$.

\paragraph{$SO(10)$-covariant action} The action \eqref{acao_nathan} in the
pure spinor formalism can be written covariantly as
\beq\label{PScov}
S_{\rm PS}= \frac{1}{\pi}\int d^2z \, \bigg( \frac{1}{2}\p X^m \bar{\p}X_m +p_{\a}\bar{\p}\t^{\a} -
w_\a\bar\p\l^\a \bigg)\, ,
\eeq
where $w_\a$ is the conjugate momentum to the pure spinor. The dependence on $\ap$ can be reinstated from the
following length dimensions of all these variables \cite{1loopH,2loopH}\footnote{We omitted all factors of $\ap$ for brevity and maximum flexibility.
For the open- and closed-string they can be restored from the conventions
$\ap=1/2$ and $\ap=2$ respectively.}
\beq\label{dim}
[\ap] = 2\, ,\quad [X^m] = 1\, , \quad [\t^\a] = [\l^\a] = \frac{1}{2} \, , \quad
[p_\a]=[w_\a] = -\frac{1}{2}\,.
\eeq
Inspired by the approach of Siegel, this action needs to be supplemented by the definitions of
the supersymmetric momentum $\Pi^m$, the GS constraint $d_\alpha$ and the
supersymmetric derivative $D_\alpha$ which we repeat here for the reader's convenience:
\begin{align}
\Pi^m &= \p X^m + \half(\t\g^m\p\t)\,, \notag \\
d_\a &= p_\a -\frac{1}{2}\Big(\p X^m +\frac{1}{4}(\t \g^m\p\t) \Big)(\g_m\t)_{\a}\,, \label{Piagain}
\\ 
D_{\a} &= \frac{\p}{\p\t^{\a}} +\frac{1}{2}(\g^m\t)_{\a}\p_m\,.
\notag
\end{align}
In addition, the BRST charge is given by\footnote{In recent years this BRST charge has
been derived from first principles \cite{Berkovits:2014aia}. For previous attempts, see
\cite{Matone:2002ft,Aisaka:2005vn,Berkovits:2007wz,Berkovits:2011gh}.} (dropping the subscript $_{\rm BRST}$ henceforth)
\beq\label{QBRST}
Q = \oint dz \, \lambda^{\alpha}(z)d_{\alpha}(z)\, .
\eeq
The $SO(10)$-covariant versions of the energy-momentum tensor
(\ref{tensor_total}) and the fermionic Lorentz currents derived from the action \eqref{PScov}
are given by
\beq\label{TandM}
T_{\rm PS} = -\frac{1}{2}\Pi^m\Pi_m - d_\a\p \t^\a + w_\a \p\l^\a\, ,\qquad
M^{mn} = -\half(p\g^{mn}\t) + \half(w\g^{mn}\l)\, .
\eeq

\subsubsection{Operator product expansions}
\label{sec:summOPE}

We shall now summarize the $SO(10)$-covariant form of the OPEs that
govern the CFT of the pure spinor formalism. The basic
worldsheet matter variables obey\footnote{Note that we have not written a $SO(10)$ covariant OPE for $w_\a(z)\l^\b(w)$ since the
pure spinor constraint implies that these are not free fields. The way around this issue is to
decompose these fields into $U(5)$ variables and to notice that the naive OPE $w_\a(z)\l^\b(w)\sim \frac{\d^\b_\a}{z{-}w}$
receives non-covariant $U(5)$ corrections needed to make the OPE of $(\l\g^m\l)$ with $w_\a$
non-singular \cite{psf}.}
\begin{align}
 X^m(z,\bar{z})X^n(w,\bar{w}) &\sim - \delta^{mn}\ln|z-w|^2 \, ,
& d_{\a}(z)\t^{\b}(w) &\sim \frac{\d^{\b}_{\a}}{z-w} \, , \label{all_opes}\\
 d_{\a}(z)d_{\b}(w) &\sim - \frac{\g^m_{\a\b}\Pi_m(w)}{z-w} \, ,
& d_{\a}(z)\Pi^m(w) &\sim \frac{\big(\g^m \p\t(w)\big)_{\a}}{z-w} \, , \notag\cr
\Pi^m(z)\Pi^n(w) &\sim - {\delta^{mn}\over (z-w)^2}\,, &&\notag
\end{align}
the OPEs involving the Lorentz currents are
\begin{align}
\label{all_opes2}
M^{mn}(z)M^{pq}(w)
         &\sim  \frac{\delta^{p[m}M^{n]q}(w)  - \delta^{q[m}M^{n]p}(w)}{z-w}
 + \frac{\delta^{m[q}\delta^{p]n}}{(z-w)^2} \, ,\\
N^{mn}(z)N^{pq}(w)
         &\sim  \frac{\delta^{p[m} N^{n]q}(w)  - \delta^{q[m} N^{n]p}(w)}{z-w}
 -3\frac{\delta^{m[q}\delta^{p]n}}{(z-w)^2}  \, , \notag\\
N^{mn}(z)\l^\alpha(w) &\sim \frac{1}{2}\frac{(\g^{mn})^{\a}_{\phantom{m}\b}\l^{\b}(w)}{(z-w)}\, , \notag
\end{align}
and generic superfields $K(X,\t)$ that do not depend on any derivatives $\p^kX^m, \ \p^k\theta^\alpha$ with $ k\ge1$ obey
\begin{align}
 d_{\a}(z)K\big(X(w,\bar w),\t(w) \big)  &\sim  \frac{D_{\a} K\big(X(w,\bar w),\t(w) \big) }{z-w} \, ,
  \label{more_opes}\\
\Pi^m(z) K\big(X(w,\bar w),\t(w) \big)  &\sim -\frac{\p^m K\big(X(w,\bar w),\t(w) \big) }{z-w}\, .\notag
\end{align}
Using these OPEs one can check that the supersymmetry currents \eqref{susyQ} satisfy the supersymmetry algebra
\be
\{ {\cal Q}_{\a}, {\cal Q}_{\beta}\} = \gamma^m_{\a\beta}\oint \p X_m\,,
\ee
and that all of $\{\p\t^\a, \Pi^m, d_\a, N^{mn}\}$ are conformal primary fields of
weight $+1$, 
\beq
\label{weights}
T_{\rm PS}(z) \big\{ \partial \theta^\alpha, \Pi^m,d_\alpha, N^{mn} \big\}(w) \sim
\frac{ \big\{ \partial \theta^\alpha, \Pi^m,d_\alpha, N^{mn} \big\}(w)}{(z-w)^2} +
\frac{ \partial \big\{ \partial \theta^\alpha, \Pi^m,d_\alpha, N^{mn} \big\}(w)}{z-w}\, ,
\eeq
a crucial fact in the construction of the integrated massless
vertex operator below.

\subsection{\label{sec3.4}Scattering amplitudes on the disk}

We shall now review the opening line for superstring disk amplitudes
along with the dictionary between superspace expressions and component amplitudes.

\subsubsection{Massless vertex operators}

In order to compute scattering amplitudes in superstring theory using conformal-field-theory
methods, first we need to describe the vertex operators containing the information about
the string states. The integrated massless vertex operator \eqref{siegel_U} proposed by Siegel
led to the discrepancy of double-pole coefficients due to the Lorentz
currents of the fermionic variables. The integrated massless vertex operator
proposed by Berkovits adds a correction to $U_{\rm Siegel}(z)$ proportional to 
the Lorentz current $N_{mn}$ of the pure spinor ghost~\cite{psf}
\beq\label{integrado}
U(z) = \p\t^{\a}A_{\a}(X,\t) + A_m(X,\t)\Pi^m + d_{\a}W^{\a}(X,\t) 
+ \half N_{mn}F^{mn}(X,\t)\,,
\ee
where the linearized SYM superfields $A_{\a}(X,\t) ,A_m(X,\t),W^{\a}(X,\t)$
and $F^{mn}(X,\t)$ were introduced in section~\ref{linSYMsec} and the dependence
$\t^\a=\t^\a(z)$ and $X^m=X^m(z,\bar z)$ on the vertex insertion points
is left implicit. The superfields have the following length dimensions \cite{1loopH,2loopH}
\beq\label{LKs}
[A_\a] = \frac{1}{2}\, , \quad [A_m] = 0\, , \quad [W^\a] = - \frac{1}{2}\, , \quad [F_{mn}] = -1\, , \quad [V(z)] = [U(z)] = 1\, ,
\eeq
and the superfields $K(X,\t)$ are decomposed into plane waves as
\beq\label{planewaves}
K(X,\t) = K(\t)e^{k\cdot X}\, .
\eeq
Using the $\t$-expansions \eqref{linTHEX} in Harnad--Shnider gauge, the gluon vertex 
operator following
from \eqref{integrado} features the complete Lorentz current $M^{mn}(z)=\Sigma^{mn}(z)+N^{mn}(z)$
of \eqref{Mps} as the coefficient of the component field strength. In this way, the issue
with the double-pole mismatch with the RNS vertex operator is absent from \eqref{integrado}.
In addition, given that $U$ has conformal weight ${}+1$, it has to
appear in the amplitude prescription
integrated over (parts of) the worldsheet boundary, i.e.\ in the conformally invariant
combination $\int dz U(z)$.

The prescription to compute tree-level amplitudes will also require a massless vertex operator with
conformal weight zero to be used at fixed locations on the Riemann surface to remove the redundancy
of the M\"obius transformations. The proposal by Berkovits for this unintegrated vertex is
\be
\label{V}
V = \l^{\a}A_{\a}(X,\t)\,.
\ee
Furthermore, the massless vertex operators represent the physical
states of gluons and gluinos and must {\it be in the cohomology}
of the BRST operator $Q$ of (\ref{QBRST}).
\begin{def.}
A state $\Psi$ is said to be in the cohomology of the BRST operator if it is {\it BRST-closed},
$Q\Psi=0$,
and not {\it BRST-exact}, $\Psi\neq Q\Omega$ for some $\Omega$.
\end{def.}
Recall that the BRST charge
satisfies $Q^2=0$ due to the pure spinor condition \eqref{espinor_puro}
and the OPE \eqref{dd_ope}.
\begin{prop.}
The unintegrated vertex operator $V(z) = \l^\a(z) A_\a(X,\t)$ for massless particles $k^2=0$ is
BRST closed $QV=0$ when the linearized superfield $A_\a(X,\t)$ is on-shell and has zero conformal weight.
\end{prop.}
\noindent\textit{Proof.}
An on-shell linearized superfield $A_\a$
satisfies the equations of motion \eqref{RankOneEOM}. In particular $D_{(\a}A_{\b)} =
\g^m_{\a\b}A_m$, so
\beq\label{QVzero}
QV(w) = \oint dz\, \l^\a(z) d_\a(z) \l^\b(w)A_{\b}\big(X(w),\t(w)\big) = \l^\a\l^\b D_\a A_\b = \frac{1}{2} (\l\g^m\l)A_m = 0\, ,
\eeq
where we used the OPE \eqref{more_opes} and the pure spinor constraint \eqref{espinor_puro}. To
show that $V$ has conformal weight zero,
first recall that in a conformal field theory the OPE of the energy-momentum
tensor with a conformal primary $\phi_h$ of weight $h$ is given 
by \cite{yellowpages, Polchinski:1998rq}
\beq\label{Tphi}
T(z)\phi_h(w) \sim {h \phi_h(w)\over (z-w)^2} + {\p_w\phi_h(w)\over (z-w)}\,.
\eeq
Using the total energy-momentum tensor $T_{\rm PS}$ from \eqref{TandM} 
and the OPEs \eqref{more_opes} we get
\beq
T(z)V(w) \sim -\half {\p^m\p_m V\over (z-w)^2} + {(\Pi^m \p_m + \p\t^\a D_\a)V+\p\l^\a A_\a\over
(z-w)} = {\p V\over (z-w)}\,,
\eeq
where we used the massless condition and the chain rule for $\p_w$
\beq\label{chaindel}
(\Pi^m \p_m + \p\t^\a D_\a)V+\p\l^\a A_\a = \l^\a \p A_\a(X,\t) + (\p\l^\a) A_\a(X,\t) = \p
V(X(w),\t(w)) = \p V(w) 
\eeq
since
\beq\label{zderivative}
( \p\t^{\b} D_{\b}  + \Pi^m \p_m )K(X,\t) = ( \p\t^{\b}\p_{\b}  + \p X^m \p_m )K(X,\t) = \p K(X,\t)
\eeq
for an arbitrary superfield $K(X,\theta)$ that is independent on $\l^\a$ and on the worldsheet derivatives
of $X^m,\theta^\alpha$, as can easily be checked
using
the expressions for $D_\a$ and $\Pi^m$ in \eqref{Piagain}.\qed

As can be seen from \eqref{linegauge}, a BRST-exact vertex operator of the form $Q\Omega$ is
interpreted as capturing the gauge variation of the super Yang--Mills fields $Q\Omega 
= \l^\a D_\a\Omega= \l^\a \d_\Omega A_\a $. In this sense, 
viewing $V$ as a representative in the cohomology of $Q$
excludes pure-gauge superfields.

The synergy between pure spinors and the SYM equations of motion seen in \eqref{QVzero}
was already anticipated by Howe and Nilsson in \cite{howe,nilsson}, also 
see \cite{Cederwall:2013vba, Cederwall:2022fwu} for a
more recent overview articles on the importance of pure spinors
for off-shell supersymmetric actions. An early application of ten-dimensional
pure spinors to the classical superstring can be found in \cite{hughston}.

\paragraph{Relating integrated and unintegrated vertices}
In the RNS formalism, the integrated vertex operator $U_{\rm RNS}$ is related to the unintegrated
vertex operator $V_{\rm RNS} = c U_{\rm RNS}$ via $Q U_{\rm RNS} = \p V_{\rm RNS}$
\cite{BigPicture}. This
can be checked by recalling that $U_{\rm RNS} = \{\oint b, V_{\rm RNS}\}$ and $T = \{Q,b\}$, 
where $(b,c)$ is the ghost system used to fix the reparametrization invariance of the worldsheet.
The proof then follows from the Jacobi identity
\be
\label{qu}
QU_{\rm RNS} = [Q,\{\oint b, V_{\rm RNS}\}] = - [ V_{\rm RNS}, \{Q, \oint b\}] - [\oint b, \{V_{\rm RNS},Q\}]
= \p V_{\rm RNS}
\ee
because the cohomology condition requires $\{V_{\rm RNS}, Q\}=0$ and the conformal
weight $h{=}0$ of $V_{\rm RNS}$ implies that $[\oint T, V_{\rm RNS}] = \p V_{\rm RNS}$ by
\eqref{Tphi}.

While the pure spinor formalism does not feature any direct analogue of the $(b,c)$ system\footnote{See
\cite{Berkovits:2005bt, Oda:2007ak, Jusinskas:2013rga,b2chandia} for a composite $b$ ghost in the
non-minimal pure spinor formalism (see also \cite{Jusinskas:2013sha}).}
-- that is why the forms of the unintegrated \eqref{V} and integrated \eqref{integrado} vertex
operators are very different --
the vertex operators $V,U$ still
satisfy the relation \eqref{qu} of their analogues in the RNS formalism (see also \cite{Chandia:2021coc}):
\begin{prop.}
The massless integrated and unintegrated vertex operators \eqref{integrado} and \eqref{V} are
related by
\beq\label{qudelv}
QU=\p V\, .
\eeq
\end{prop.}
\noindent{\it Proof.}
Using the OPEs
\eqref{all_opes} and \eqref{more_opes} as well as the equations of
motion for the linearized SYM superfields \eqref{RankOneEOM} we get
\begin{align}
Q(\p\t^{\a} A_{\a}) &= (\p \l^{\a})A_{\a} - \p\t^{\a}\l^{\b}D_{\b}A_{\a} \, ,\\
Q(\Pi^m A_m) &= (\l\g^m\p \t) A_m + \Pi^m\l^{\a}(D_{\a} A_m) \, ,\cr
Q(d_{\a}W^{\a}) &= - (\l\g^m W)\Pi_m - d_{\b}\l^{\a}D_{\a}W^{\b} \, ,\cr
Q\Big(\frac{1}{2}N_{mn}F^{mn} \Big) &= \frac{1}{4}(\g_{mn}\l)^{\a}d_{\a}F^{mn}
+ \frac{1}{2} N_{mn}\l^{\a}D_{\a}F^{mn}\,.
\notag
\end{align}
Summing them up yields
\begin{align}
QU &= (\p\l^{\a})A_{\a} - \p\t^{\b}\l^{\a}(D_{\a}A_{\b} - \g^m_{\a\b}A_m)
+ \l^{\a} \Pi^m(D_{\a}A_m - (\g_m W)_{\a})\cr
&\quad{}-\l^{\a}d_{\b}\Big(D_{\a}W^{\b}+\frac{1}{4}(\g^{mn})_{\a}^{\phantom{m}\b}F_{mn} \Big)
+N_{mn}(\l\g^n \p^m W)\cr
&= (\p\l^{\a})A_{\a} + \l^{\a}\p\t^{\b}D_{\b}A_{\a} + \l^{\a}\Pi^m\p_m A_{\a} \, ,
\label{qu3}
\end{align}
where $N_{mn}(\l\g^n \p^m W)$  vanishes due to a combination of the pure spinor condition
$(\l\g^n)_{\a}(\l\g_n)_{\b} = 0$ proven in \eqref{lagalaga}
and the linearized equation of motion $\g^m_{\a\b}\p_mW^{\b} = 0$,
\beq
N_{mn}(\l\g^n \p^m W) = \frac{1}{2}(w\g_m\g_n\l)(\l\g^n \p^m W) - (w\l)(\l\g^m\p_m W) = 0\, .
\eeq
Therefore, using \eqref{chaindel} the BRST variation $QU$ in
\eqref{qu3} becomes
\beq
QU =  (\p\l^{\a})A_{\a} + \l^{\a}( \p\t^{\b}D_{\b} A_{\a} + \Pi^m \p_m A_{\a})
= (\p\l^{\a})A_{\a} + \l^{\a} \p A_{\a} = \p (\l A) = \p V\, ,
\eeq
as we wanted to show. \qed

\begin{cor}
The integrated vertex operator $\int dz U(z)$ is
BRST invariant up to surface terms.
\end{cor}
As we will see below, surface terms
do not contribute to open- or closed-string amplitudes by the so-called
canceled-propagator argument. The cancellation of surface terms is also
used to demonstrate linearized gauge invariance of the massless vertex
operators: under the linearized variations $\delta_\Omega A_\a = D_\a \Omega$ and
$\delta_\Omega A_m = \partial_m \Omega$ of \eqref{linegauge} with some gauge-scalar superfield
$\Omega$, the variation $\delta_\Omega V = \l^\a D_\a \Omega = Q\Omega$
vanishes in the cohomology, whereas
\beq
\delta U = \partial \t^\a D_\a \Omega + \Pi^m \partial_m \Omega = \partial \Omega
\eeq
reduces to vanishing surface terms after using the chain rule (\ref{zderivative}).

Superspace vertex operators for massive open-string states $\varphi$ can be constructed
by following the same principle: identifying a BRST-closed unintegrated vertex operator $V_\varphi$
of conformal weight zero and then engineering its integrated counterpart $U_\varphi$ of weight one such
that $QU_\varphi= \partial V_\varphi$. By the conformal weight $h(e^{k \cdot X}) =-N$ at the
$N^{\rm th}$ mass level, the combinations of $ \Pi^m,d_\alpha, \ldots$
accompanying the plane waves accumulate more and more conformal weight and
Lorentz indices at growing $N$. That the pure spinor cohomology contains all massive states of the
superstring was shown in \cite{GScohomologyI,GScohomologyII} (see also \cite{Aisaka:2008vw}).
The vertex operators at the first mass level are known in superspace
from \cite{Berkovits:2002qx, Chakrabarti:2018mqd, Chakrabarti:2018bah},
and it is an open problem to pinpoint their explicit form at higher levels.

\subsubsection{Scattering-amplitude prescription at genus zero}
\label{sec:treeprescr}

The prescription to compute $n$-point tree amplitudes of
open-superstring states 
is given by the following correlation function of
vertex operators on a disk worldsheet \cite{psf}
\be
\label{treepresc}
{\cal A}(P)=\!\!\!\! \int \limits_{D(P)}\!\!\! dz_2  \, dz_3 \ldots dz_{n-2} \, \langle \! \langle V_1(z_1) U_2(z_2)  U_3(z_3) {\ldots} 
 U_{n-2}(z_{n-2})  V_{n-1}(z_{n-1})  V_n(z_n)
\rangle \! \rangle \,,
\ee
where $\langle \! \langle \ldots \rangle \! \rangle$ refers to the path integral over the
variables in the pure spinor action (\ref{PScov}) and M\"obius invariance of the 
correlator was used to fix the insertion points of the three unintegrated 
vertex operators. We adopt the particularly convenient choices to
parameterize the disk boundary by the compactified real line $z_j \in \mathbb R$
and to place the unintegrated vertex operators at
\beq\label{moebiusfix}
(z_1,z_{n-1},z_n) \rightarrow (0,1,\infty)\, .
\eeq
For a general $n$-point disk ordering characterized by a cyclic permutation 
$P := p_1 p_2 \ldots p_n$, the formal definition of the \textit{integration domain $P$} 
in (\ref{treepresc}) reads
\beq\label{domain}
D(P) := \{ (z_1,z_2,\ldots,z_n) \in \Bbb R^n\ | \ -\infty < z_{p_1} < z_{p_2} < \ldots < z_{p_n} < \infty \} \, .
\ee
For example, with the three fixed positions as in \eqref{moebiusfix}, a domain
specified by the canonical ordering $P=123 \ldots n$ amounts to the disk ordering 
$0<z_2<z_3< \cdots < z_{n-2} < 1$ of the integrated vertex operators at the boundary 
of the disk such that the cyclic ordering $1,2,\ldots,n$ in $P$ is preserved.

The prescription (\ref{treepresc}) is tailored to color-ordered amplitudes ${\cal A}(P)$
and can be color-dressed by weighting each disk ordering $D(P)$ with a trace of
Chan--Paton factors $t^{a_i}$ in the same cyclic ordering,
\begin{align}
{\cal M}(1,2,\ldots,n) = \sum_{\rho \in S_{n-1}} {\cal A}\big(1,\rho(2,3,\ldots,n)\big)
{\Tr}(t^{a_1} t^{a_{\rho(2)}} t^{a_{\rho(3)}} \ldots t^{a_{\rho(n)}} ) \, ,
\label{colordr}
\end{align}
where $\rho$ is in the set of permutations $S_{n-1}$ of the $n{-}1$ legs $2,3,\ldots,n$.
Cyclicity of the trace propagates to the color-ordered amplitudes,
${\cal A}(1,2,\ldots,n)= {\cal A}(2,3,\ldots,n,1)$, and the prescription (\ref{treepresc})
furthermore implies reflection properties ${\cal A}(1,2,\ldots,n)= (-1)^n{\cal A}(n,\ldots,2,1)$.

As will be reviewed from several perspectives, the $\frac{1}{2}(n{-}1)!$ cyclically
and reflection inequivalent permutations of ${\cal A}(1,2,\ldots,n)$ are not linearly 
independent. First, the monodromy relations \cite{BjerrumBohrRD, StiebergerHQ}
in section \ref{sec:6.5.1} only leave an $(n{-}3)!$-dimensional basis of disk orderings.
Second, these relations among the disk amplitudes can be refined according to the
multiple zeta values in the $\alpha'$-expansions, see section \ref{sec:7.3.4}: 
Parts of the string corrections obey field-theory relations of SYM tree amplitudes
\cite{Broedel:2012rc, Stieberger:2014hba} and others obey KK-like symmetries \cite{cdescent} related to
permutations in the (inverse) Solomon descent algebra
\cite{solomon1976mackey,garsia1989decomposition,schockerdescent,thibon2016lie,Reutenauer}.

\paragraph{CFT calculation and zero modes} In order to evaluate the tree-level correlation function $\langle \! \langle \ldots
\rangle \! \rangle$ in \eqref{treepresc}, one first
integrates out the non-zero modes using the OPEs \eqref{all_opes} to \eqref{more_opes} to
obtain its
dependence on the positions $z_i$ carried by the conformal-weight-one variables
$[\p\t^\a(z), \Pi^m(z), d_\a(z), N^{mn}(z)]$. As explained in \cite{Polchinski:1998rq}, this
unambiguously determines the correlator
as a function of the positions $z_i$ on a genus-zero surface.
After using the OPEs in this way, the correlation function \eqref{treepresc} will still contain the
zero modes of $\l^\a$ and $\t^\a$, as they are variables of conformal weight zero with a single
zero mode at genus zero \cite{DHoker:1988pdl}. These zero-mode correlators are denoted by
$\langle \ldots \rangle$ (as opposed to the above double brackets $\langle \! \langle \ldots
\rangle \! \rangle$ including the non-zero modes), and
one needs an ad-hoc rule to integrate them. Using the shorthand
\beq\label{la3th5}
(\l^3\t^5):= (\l\g^m\t)(\l\g^n\t)(\l\g^p\t)(\t\g_{mnp}\t)\,,
\eeq
the only non-vanishing contributions in ten-dimensional\footnote{For the dimensional reduction of the condition
\eqref{tlct} to $D=4$, see \cite{thales4d}.} zero-mode correlators is
proportional to \eqref{la3th5} \cite{psf}
\begin{equation}\label{tlct}
\langle (\l^3\t^5)\rangle = 2880\,,
\ee
where the normalization $2880$ was chosen in \cite{brenno} in order to match the
RNS tree-level amplitude conventions. As we will see in \eqref{la3th5Lie}, the scalar
\eqref{la3th5} is unique since the tensor product of three pure spinors and five
$\t$s only features a single scalar representation of $SO(10)$.
\begin{prop.} The
combination $(\l^3\t^5)$ is in the cohomology of the pure spinor
BRST operator.
\end{prop.}
\noindent\textit{Proof.}
It is BRST closed
\begin{align}\label{Qla3th5}
Q(\l\g^m\t)(\l\g^n\t)(\l\g^p\t)(\t\g_{mnp}\t) &= 3(\l\g^m \l)(\l\g^n\t)(\l\g^p\t)(\t\g_{mnp}\t)\\
&\quad{}- 2(\l\g^m\t)(\l\g^n\t)(\l\g^p\t)(\l\g_{mnp}\t) = 0\,.\notag
\end{align}
The first term vanishes by
the pure spinor constraint \eqref{espinor_puro},
while the vanishing of the second term $(\lambda \gamma^m)_{\alpha} 
(\lambda \gamma^n)_{\beta}  (\lambda \gamma^p)_{\gamma} (\l\g_{mnp}\t) =0$
can be seen by decomposing $\g^{mnp}=\g^m\g^{np}-\d^{mn}\g^p + \d^{mp}\g^n$
and using \eqref{lagalaga}, $(\l\g^m)_\a(\l\g_m)_\b =0$.
Moreover, (\ref{la3th5}) is not BRST exact,
\beq\label{la3th5not}
(\l\g^m\t)(\l\g^n\t)(\l\g^p\t)(\t\g_{mnp}\t) \neq Q\Omega(\l,\t)\,,
\eeq
because there is no scalar built from two $\l$s and six $\t$s. If there was
a $\Omega(\l,\t)$ such that $Q\Omega=(\l^3\t^5)$, it would necessarily
be a scalar function containing two $\l$s and six $\t$s since
$Q\t^\a=\l^\a$ and the $\partial_m$-admixture of $\lambda^\alpha D_{\alpha}$
can be dropped for functions of only $\l$ and $\t$. The
$SO(10)$ representation of two pure spinors $\l^{\a}$ is characterized by Dynkin labels
$(00002)$ while six antisymmetric thetas are represented
by $(01020)\oplus (20100)$, see \ref{sec:appDynkin}. However, their product \cite{Lie}
\beq\label{la2th6}
(00002)\otimes \Big( (01020) \oplus (20100)\Big) =
(00011)\oplus (00022) \oplus 2 (00120) \oplus \cdots
\eeq
has no scalar representation $(00000)$. This shows that the putative BRST ancestor $\Omega(\l,\t)$
in (\ref{la3th5not}) cannot be constructed, finishing the proof. \qed

In the formulation of the prescription \eqref{treepresc}, we have chosen legs $1,n{-}1$ 
and $n$ to be represented by an unintegrated vertex operator $V_j$ at fixed locations 
$z_1,z_{n-1},z_n$. It remains to verify that any other choice of three legs to appear in
the unintegrated picture leads to the same result for each color-ordered amplitude.
\begin{prop.} 
The disk amplitude prescription \eqref{treepresc} does not depend on which triplet 
$\{i,j,k\}$ of the external legs enters via unintegrated vertex operators $V_i(z_i)
V_j(z_j) V_k(z_k)$ at fixed punctures $z_i, z_j, z_k$ on the compactified real line.
\end{prop.}
\noindent\textit{Proof.} Following the strategy of \cite{brenno}, it is sufficient to show that
the representation $V_i, \int d z_{i+1} \,U_{i+1}$ of neighboring states $i$ and $i{+}1$ 
can always be swapped to $\int dz_i \, U_i,V_{i+1}$, i.e.
\begin{align}
&\left \langle \! \! \left\langle V_{1}(0)   \int_0^1 d z_2 \, U_2(z_2) \prod_{j=3}^{n-2}  \int^{1}_{z_{j-1}} d z_{j} \ U_{j}(z_{j})   V_{n-1}(1)  V_n(\infty)  \right \rangle \! \!  \right \rangle \notag \\
& \ \ \ \ \ \ = \left \langle \! \! \left \langle \, \int^0_{-\infty} d y \, U_{1}(y)   V_2(0) \prod_{j=3}^{n-2}   \int^{1}_{z_{j-1}} d z_{j} \, U_{j}(z_{j})   V_{n-1}(1)  V_n(\infty)   \right \rangle\!  \! \right\rangle\, .
\label{10,x1}
\end{align}
Since $Q U_j(w) = \oint  d z \, \lambda^\alpha d_\alpha(z) U_j(w) = \partial V_j(w)$, 
we can rewrite the left-hand side via
\begin{align}
V_{1}(0)    V_n(\infty) = \int^0_{-\infty} d y \, \partial V_{1}(y)   V_n(\infty)   =  \int^0_{-\infty} d y \, Q  \bigl( U_{1}(y) \bigr)     V_n(\infty) \, . \label{another10,x1}
\end{align}
In the first step, we have discarded $V_1(\infty) V_n(\infty)$ by the so-called ``canceled-propagator
argument'' \cite{Polchinski:1998rq} which states that terms with colliding vertex operators 
$V_i(z) V_j(z)$ or $V_i(z) U_j(z)$ identically vanish. As a next step, we deform the integration 
contour of the BRST current $\lambda^\alpha d_\alpha$ such that it encircles all the vertex 
operators apart from $U_1$:
\begin{align}
&\left  \langle \!\!  \left\langle  V_{1}(0)   \int_0  ^1 d z_2 \, U_2(z_2)  \prod_{j=3}^{n} \int^1 _{z_{j-1}} d z_j \, U_j(z_j)  V_{n-1}(1)   V_n(\infty)  \right  \rangle \!\!  \right \rangle \notag \\
& \ \ \ \ \ \ \ \ = - \, \left  \langle \! \!\left \langle  \int^0 _{-\infty} d y \, U_{1}(y)   \int_0  ^1 d z_2 \,  Q  \biggl[ U_2(z_2)  \prod_{j=3}^{n} \int^1 _{z_{j-1}} d z_j \, U_j(z_j) V_{n-1}(1)   V_n(\infty)  \biggr]  \right \rangle \!  \!  \right\rangle \notag \\
& \ \ \ \ \ \ \ \ = - \, \left \langle \! \!  \left\langle  \int^0 _{-\infty} d y \, U_{1}(y)   \int_0  ^1 d z_2 \, \partial V_2(z_2)  \prod_{j=3}^{n}  \int^1 _{z_{j-1}} d z_j \, U_j(z_j)  V_{n-1}(1)   V_n(\infty)  \right \rangle \! \!  \right\rangle \notag \\
& \ \ \ \ \ \ \ \ = + \, \left \langle \!  \!  \left\langle  \int^0 _{-\infty} d y \, U_{1}(y)   V_2(0)  \prod_{j=3}^{n}  \int^1 _{z_{j-1}} d z_j \, U_j(z_j)  V_{n-1}(1)   V_n(\infty)   \right \rangle \! \!  \right \rangle \label{10,x2}
\end{align}
In passing to the third line, terms where $Q$ acts on the $U_j$ vertices with $3 \leq j \leq n{-}2$
were discarded due to the canceled-propagator argument: it forces both boundary terms of
the $\int^1_{z_{j-1}} d z_j \, \partial V_j(z_j)$ integrals to vanish,
\beq
\ldots \, U_{j-1}(z_{j-1}) \, \bigl( \, V_j(1) \ - \ V_j(z_{j-1}) \, \bigr) \, \ldots \, V_{n-1}(1) \, \ldots = 0 \ .
\label{10,x3}
\eeq
On the other hand, the integral over $Q U_2 = \partial V_2$ contributes non-trivially 
to the last line of (\ref{10,x2}): while the upper integration limit $z_2 = 1$ cancels due 
to $V_2(1) \ldots V_{n-1}(1) = 0$, the lower one $z_2=0$ generically does not 
coincide with the position $y$ of $U_1$, i.e.\ $U_1(y) V_2(0) \neq 0$.
 \qed

As we have seen in \eqref{susy1}, the worldsheet action in the pure spinor formalism is 
spacetime supersymmetric. This
means that the OPEs among its worldsheet fields have the appropriate transformations under 
the generators ${\cal Q}_\a$ in (\ref{susyQ}) and will not violate supersymmetry. However,
one still needs to show that the zero-mode integration rule \eqref{tlct} for
the disk-amplitude prescription \eqref{treepresc} preserves the supersymmetric nature of the
formalism.
\begin{prop.}
The disk-amplitude prescription \eqref{treepresc} is supersymmetric \cite{psf}.
\end{prop.}
\noindent\textit{Proof.}
We will show that the supersymmetry variation of the amplitude under $\d \t^\a = \epsilon^\a$ 
vanishes, i.e.\ that $\d {\cal A}(1, \ldots,n) =0$. Note that the only possibility
of getting a non-vanishing result after the supersymmetry transformation
$\d \t^\a = \e^\a$ is if the correlator in the amplitude \eqref{treepresc} contains
a term of the form
\be
\label{notinv}
{\cal A}(P) =\!\!\! \int \limits_{D(P)}\!\!\! dz_2 \cdots   dz_{n-2}\, \langle  
 (\l\g^m \t)(\l\g^n \t)(\l\g^p \t)(\t\g_{mnp}\t)
(\t^\a \Phi_\a(z,e,\chi,k) + {\ldots} )
\rangle 
\ee
for some $\Phi_\a(z,e,\chi,k)$ depending on the worldsheet positions $z_i$ of all
open-string vertex operators as well as polarizations
$e^m$, $\chi^\a$ and momenta $k^m$. The zero-mode integration \eqref{tlct} would then 
imply the supersymmetry variation to be
$\d {\cal A} = \int dz_2 \dotsi \int dz_{n-2} \e^\a \Phi_\a(z,e,\chi,k)$.
To see why this variation must be zero, note
that the result of the OPE calculation in the amplitude prescription \eqref{treepresc}
leads to an amplitude that can be written as
$\int dz_2 \dotsi \int dz_{n-2} \langle \l^\a\l^\b\l^\g f_{\a\b\g}(\t,e,\chi,k) \rangle$ for some function $f$
depending on the zero modes of $\t$ and on the momenta and polarizations of the open-string states. However, the amplitude
must be BRST invariant, so its correlator must be such that
\be
\label{closed}
\int \limits_{D(P)} dz_2 \cdots   dz_{n-2}\,  \l^\a\l^\b\l^\g \l^\d D_\d f_{\a\b\g}(\t,e,\chi,k) = 0\,.
\ee
Using the function $f$ following from \eqref{notinv} and plugging it into \eqref{closed} we conclude
\beq
 \int \limits_{D(P)} dz_2 \cdots   dz_{n-2}\,
 \l^\a\l^\b\l^\g \l^\d \Phi_\d(z,e,\chi,k) = 0\, .
\eeq
This vanishing is only possible if $\Phi_\d$ is a total worldsheet derivative, $\Phi_\d =  \p(
\ldots)$, implying
that the supersymmetry variation of the amplitude
vanishes after integration, $\d {\cal A} = 0$. \qed

\subsubsection{The field-theory limit} 
Disk amplitudes of massless open-superstring states reduce to
$n$-point tree-level amplitudes among the supermultiplet of ten-dimensional
SYM \cite{SYM} when the dimensionless combinations $\ap k_i \cdot k_j$ are taken
to be small \cite{scherkFT,neveuscherkFT,gsoSYMft1,gsoSYMft2}. We will refer to this low-energy regime as the field-theory limit
and informally write $\alpha' \rightarrow 0$. Since the scattering energies 
are small in comparison to the inverse string-length scale, this limit can 
also be thought of as shrinking the string to a point particle.

Throughout this review, SYM tree-level amplitudes in the field-theory limit
will be denoted by $A(1, \ldots,n)$ when they contain all states in the supermultiplet, i.e.
\beq\label{aptozeroFT}
A(1,2, \ldots,n) = \lim_{\ap\to0} {\cal A}(1,2, \ldots,n)\,.
\eeq
As will be illustrated in section \ref{illustratesec} below, the superstring-amplitude 
prescription \eqref{treepresc} yields formal sums of component amplitudes with external
bosons and fermions since the pure spinor formalism is manifestly supersymmetric.
When restricted to bosonic external states, the field-theory
tree-level amplitudes will be denoted by $\AYM(1,2, \ldots,n)$. 
The construction of SYM tree-level
amplitudes $A(1,2, \ldots,n)$ using pure spinor cohomology methods will be described in section
\ref{PSSYMsec}, and the alternative derivation from the
$\ap\to0$ limit of the superstring amplitude will be reviewed in
section~\ref{FTdisksec} (see also \eqref{stringFTij}).

\subsubsection{Pure spinor superspace}

Superfield expressions containing the zero modes of three pure spinors define
{\it pure spinor superspace} \cite{psf,Berkovits:2006ik}
\begin{equation}\label{pssexpr}
\l^\a\l^\b \l^\g f_{\a\b\g}(\t,e,\chi,k)\,,
\ee
where $f_{\a\b\g}(\t,e,\chi,k)$ represents a function containing zero modes of $\t^\a$ as well as
gluon and gluino polarizations and momenta. It is easy to see that such expressions necessarily
arise from the amplitude prescription \eqref{treepresc} after integrating the non-zero modes via
OPEs as outlined above. For example, the massless three-point disk amplitude $\cA(1,2,3) = \langle V_1V_2V_3\rangle $ leads to the pure spinor superspace expression $f_{\a\b\g}(\t,e,\chi,k) =
A^1_\a(\t) A^2_\b(\t) A^3_\g(\t)$, see \eqref{linTHEX} for the $\t$-expansion of the
SYM superfields $A^i_\a(\t)$.

As seen above, the final step in the computation of string disk amplitudes boils down to integrating
out the zero modes of three pure spinors and five $\t$s using the prescription \eqref{tlct}. These zero-mode
integrations result in the component expansion of the amplitude under consideration written as
a scalar function of polarizations and momenta. Let us write the most general form of a
pure spinor superspace expression containing five $\t$s as
\be
\label{comp}
\l^{\a}\l^{\b}\l^{\g}\t^{\d_1}\t^{\d_2}\t^{\d_3}
\t^{\d_4}\t^{\d_5}f_{\a\b\g | \d_1\d_2\d_3\d_4\d_5}(e,\chi,k) \, ,
\ee
where $(e,\chi,k)$ indicates a dependence on gluon and gluino polarizations as well as their momenta.
We need to extract the Lorentz contractions of polarizations and momenta from pure 
spinor superspace
expressions like \eqref{comp}. This can be done on the basis of the group-theory
statement that there is only one scalar built from three pure spinors $\l^{\a}$ and
five unconstrained $\t$s.
\begin{lemma}
\label{la3th5lemma}
There is only one scalar representation in the decomposition of three pure spinors and five
unconstrained fermionic Weyl spinors of SO(10).
\end{lemma}
\noindent{\it Proof.}
This follows from the tensor product
of $SO(10)$ representations $(00003)$ corresponding to three pure spinors $\l^{\a}\l^{\b}\l^{\g}$
and $(00030)\oplus (11010)$ corresponding to $\t^{\d_1}\t^{\d_2}\t^{\d_3}
\t^{\d_4}\t^{\d_5}$ \cite{Lie}
\beq\label{la3th5Lie}
(00003)\otimes \big( (00030)\oplus (11010)\big) =
1\times (00000) \oplus 2\times(00011) \oplus \cdots\, ,
\eeq
where the scalar $(00000)$ occurs with multiplicity one. \qed

The above Lemma means that any expression containing three $\l$s
and five $\t$s can be reduced to its scalar component $(\l^3\t^5)$ with
proportionality constants given entirely in terms of Kronecker deltas, gamma matrices
and Levi--Civita $\e_{10}$ tensors. This will be exploited in \ref{PSSapp} to build
up a catalog of various pure spinor correlators.

For example, suppose we have the pure spinor superspace expression
$\langle (\l \g^m\t) (\l \g^n\t)(\l \g^p \t)(\t\gamma_{abc}\t)\rangle$ with free vector
indices $m,n,p$ and $a,b,c$. In order to use the rule \eqref{tlct} one needs to 
extract its scalar component $(\l^3\t^5)$. Because we know from the Lemma above that
there is only one scalar representation in this product, this
is easily done using symmetry arguments alone. The result is
\beq\label{simpleextlct}
\langle(\l\g^{m} \t) (\l\g^{n} \t)(\l\g^{p} \t)(\t\g_{abc}\t)\rangle = 24 \d^{mnp}_{abc}\,,
\eeq
where $\d^{mnp}_{abc}$ is the generalized Kronecker delta \eqref{genKronecker}.
To see this, observe that the right-hand side is the unique term that is antisymmetric
in both $[mnp]$ and $[abc]$, as required by the symmetries of the left-hand side. The
proportionality constant can be fixed by contracting the vectorial indices on both sides 
with $\d_m^a\d_n^b\d_p^c$:
On the left-hand side we get $\langle (\l^3\t^5)\rangle$, while the right-hand side
reduces to $24\times 120=2880$ (using $\d^{mnp}_{mnp}={10\choose 3}=120$, see \eqref{gKcont}).
Therefore we recover the normalization \eqref{tlct}, and the Lemma guarantees that this is the
correct tensor.

For another example, consider
$\langle (\l \g^m\t) (\l \g^n\t)(\l \g^p \t) (\chi \gamma_n \theta) (\psi \gamma_p \theta)
\rangle $, for two arbitrary Weyl spinors $\chi$ and $\psi$. Based on the Fierz identity
\eqref{Fierzlambdatheta},
$\t^\a\t^\b = {1\over 96}\g^{\a\b}_{rst}(\t\g^{rst}\t)$, we obtain
\begin{align}
\langle (\l \g^m\t) (\l \g^n\t)(\l \g^p \t) (\t \gamma_n \chi) (\t \gamma_p \psi)\rangle&=
-{1\over 96}(\chi\g_n\g^{rst}\g_p\psi)\langle(\l \g^m\t) (\l \g^n\t)(\l \g^p
\t)(\t\g_{rst}\t)\rangle\notag\\
&= -{1\over4}(\chi\g_n\g^{mnp}\g_p\psi)=18(\chi\g^m\psi)\,,\label{fermex}
\end{align}
where we used \eqref{simpleextlct} and $\g_n\g^{mnp}\g_p=-72\g^m$. For an alternative derivation,
see \cite{brenno}. And for a more in-depth excursion on the evaluation of pure spinor superspace 
zero-mode
correlators, see \ref{PSSapp}.

\subsubsection{Component expansion from pure spinor superspace}
\label{illustratesec}

As an illustration of the above steps, the supersymmetric
three-point tree amplitude following from \eqref{treepresc} is given by
\begin{align}\label{three}
{\cal A}(1,2,3) &=\langle (\l A_1)(\l A_2)(\l A_3)\rangle \\
&= {\cal A}(1_b,2_b,3_b)+{\cal A}(1_b,2_f,3_f)+{\cal A}(1_f,2_b,3_f)+{\cal A}(1_f,2_f,3_b)\,,\notag
\end{align}
where the subscripts $b$ or $f$ refer to the bosonic or fermionic component polarizations,
corresponding to the gluon or gluino at the massless level of the open superstring. 
Note that the tree-level prescription \eqref{treepresc} for less than four massless
external states does not involve any conformal fields of weight $h=1$. Hence, the
massless three-point amplitude does not receive any contributions from OPEs 
and is entirely determined by the zero modes of $\l^\a$, $\t^\a$
and their correlator \eqref{tlct}.
Component amplitudes with an odd number of fermions (say ${\cal A}(1_b,2_b,3_f)$ 
or ${\cal A}(1_f,2_f,3_f)$) are absent from (\ref{three}) due to the zero-mode prescription
\eqref{tlct}.\footnote{Contributions to $\langle (\l A_1)(\l A_2)(\l A_3)\rangle $ from an odd number 
of fermions reside at even orders in $\theta$ which are annihilated by the prescription \eqref{tlct}.}

\paragraph{Three-gluon amplitude}
Evaluating the explicit component expansion for the three-gluon amplitude is a matter of
plugging in the $\theta$-expansions \eqref{linTHEX} in Harnad--Shnider gauge and selecting the 
components with five $\t$s which contain the gluon fields. Doing this for the spinorial 
superpotential $A_\a$, the only terms in the bosonic $\t$-expansion that can contribute are
\beq
A_{\a}(X,\t) \rightarrow \Big\{ {1\over 2}e_m(\g^m\t)_\a -{1\over 32}f_{mn}(\g_p\t)_\a (\t\g^{mnp}\t) \Big\} e^{k \cdot X}\, .
\label{t5from3v}
\ee
It is easy to see that the term containing a gluon with $\t^5$ in $A_\a(X,\t)$ does not contribute
because it leads to superspace expressions of the form $\l^3\t^{p\ge 7}$
once the terms from
the other two vertex operators are taken into account, and this is
annihilated by the pure spinor bracket rule \eqref{tlct}. There are three ways to saturate $\t^5$
with bosonic contributions (\ref{t5from3v}) from each vertex, namely
$(\t^1,\t^1,\t^3)$, $(\t^1,\t^3,\t^1)$ and $(\t^3,\t^1,\t^1)$. This
results in
the three-gluon component amplitude
\begin{equation}\label{threegluon}
{\cal A}(1_b,2_b,3_b) =
- {1\over 128}e^m_1 f_2^{pq}e^n_3
\langle (\l \g^m\t) (\l \g^r\t)(\l \g^n \t)(\t\gamma_{pqr}\t)\rangle +{\rm cyc}(1,2,3)\,,
\ee
where the Koba--Nielsen factor from the plane waves $e^{k_i\cdot X}$ evaluates to a 
constant (chosen as $1$ together with an implicit momentum-conserving delta 
function $\delta^{10}(k_1{+}k_2{+}k_3)$)
due to the on-shell condition $k_i^2=0$ of massless external states.
As discussed above, symmetry arguments and the
normalization condition \eqref{tlct} fix all pure spinor correlators
and we find (using (\ref{simpleextlct}) and $\d^r_r = 10$)
\beq
\langle (\l \g^m\t) (\l \g^r\t)(\l \g^n \t)(\t\gamma_{pqr}\t)\rangle
= 24 \d^{mrn}_{pqr} = -64\d^{mn}_{pq}\,.
\label{simpl3t5}
\eeq
Hence, the three-gluon amplitude \eqref{threegluon} is given by
\begin{align}\label{tgfim}
{\cal A}(1_b,2_b,3_b)&= \half e^m_1 f_2^{mn}e_3^n + {\rm cyc}(1,2,3) \\
&= (e_1 \cdot k_2)(e_2 \cdot e_3) + {\rm cyc}(1,2,3) \,, \notag
\end{align}
where we have applied transversality $e_i \cdot k_i=0$ and momentum
conservation $k_1{+}k_2{+}k_3=0$ in passing to the second line.

\paragraph{One-gluon and two-gluino amplitude} There are three possible distributions of
$\t$ variables governing the contribution from the external states as
one gluon and two gluinos: $(1_f,2_f,3_b)$,
$(1_f,2_b,3_f)$ and $(1_b,2_f,3_f)$. For instance,
to obtain the amplitude with one gluon and two gluinos distributed as $(1_b,2_f,3_f)$,
we use the fermionic component expansion $\l^\a A^i_\a(\t)  \rightarrow  -\frac{1}{3}
(\l \gamma_m \t) (\theta \gamma^m \chi_i)e^{k_i \cdot X}$ in \eqref{linTHEX} for the external
states $i=2,3$ and $\l^\a A^1_\a(\t) \rightarrow {1\over 2}e^1_m(\l\g^m\t) e^{k_1 \cdot X}$ for
the external state $i=1$
to get
\begin{align}\label{1b2f3f}
\cA(1_b,2_f,3_f) &= {1\over18}e^m_1
\langle(\l\g_m\t)(\l\g_n\t)(\l\g_p\t)(\t\g^n\chi_2)(\t\g^p\chi_3)\rangle = e^m_1(\chi_2\g_m\chi_3) \,,
\end{align}
using (\ref{fermex}) in the last step.

\paragraph{Supersymmetric three-point amplitude}
Assembling all the different external-state contributions
to the three-point amplitude \eqref{three} yields
\beq\label{threefin}
\cA(1,2,3) =
\half e^m_1 f_2^{mn}e_3^n + e^m_1(\chi_2\g_m\chi_3) + {\rm cyc}(1,2,3)\,.
\eeq
Given the systematic nature of the above procedure,
an implementation using {\tt FORM} \cite{FORM} has been written which performs
these expansions automatically \cite{PSS} (see also \cite{Sun:2016zfa}).

In contrast to the three-gluon amplitude of the open bosonic string, the three-point
superstring amplitude
(\ref{threefin}) is independent on $\alpha'$ and therefore coincides with the SYM amplitude,
\beq\label{threefinSYM}
A(1,2,3) =\cA(1,2,3) =
\half e^m_1 f_2^{mn}e_3^n + e^m_1(\chi_2\g_m\chi_3) + {\rm cyc}(1,2,3)\,.
\eeq
For both color-ordered and color-dressed amplitudes of ten-dimensional SYM, 
\beq
A(1,2,\ldots,n) = \lim_{\alpha' \rightarrow 0} {\cal A}(1,2,\ldots,n) \, , \ \ \ \ \ \
M(1,2,\ldots,n) = \lim_{\alpha' \rightarrow 0} {\cal M}(1,2,\ldots,n) 
\label{calvsncal}
\eeq
we will use non-calligraphic letters to distinguish them from the analogous
superstring quantities, see (\ref{colordr}) for the color-dressed open-string amplitude.

\subsubsection{Preview of higher-point SYM amplitudes}

In the same way as the single term $V_1 V_2 V_3$ in the superspace expression for
${\cal A}(1,2,3)$ produces the six terms in (\ref{threefinSYM}) upon component expansion,
higher-point string and SYM amplitudes take a particularly compact form in pure spinor superspace.
For instance, we will see later that the six-point SYM tree-level amplitude
can be written in pure spinor superspace as
\beq\label{6ptsample}
A(1,2, \ldots,6) =
{1\over3}{\langle V_{12} V_{34} V_{56} \rangle \over  s_{12} s_{34} s_{56}}
+ \half \Big\langle \Big({V_{123}\over s_{12}s_{123}} + {V_{321}\over s_{23} s_{123}}\Big)
\Big({V_{45} V_6\over s_{45}} + {V_4 V_{56}\over s_{56}}\Big) \Big\rangle
+ \cyclic{1,2, \ldots,6}\,,
\eeq
in terms of multiparticle vertex operators $V_P$ subject to beautiful
combinatorial properties that will be introduced below. The expression \eqref{6ptsample} can
be checked to be gauge invariant and supersymmetric in a couple of lines with pen
and paper. Moreover, it evades BRST-exactness for purely kinematic reasons
and lines up with a recursive structure of $n$-point SYM tree amplitudes in pure spinor superspace.
Here and below, our normalization conventions for Mandelstam invariants for massless particles are
\beq
s_{12} = k_1 \cdot k_2 = \frac{1}{2}(k_1{+}k_2)^2 \, , \ \ \ \ \ \ s_{12\ldots p} = \sum_{1\leq i<j}^p s_{ij} = 
\frac{1}{2}(k_1{+}k_2{+}\ldots{+}k_p)^2\, .
\label{defmands}
\eeq
Already for the purely bosonic terms, the component expansion of \eqref{6ptsample}
in terms of single-particle polarizations and momenta produces more than 6700 terms.
Still, we will see later that the complete component expansion of \eqref{6ptsample} can 
be arranged in the compact form
\begin{align}\label{6ptcompsample}
&A(1,2, \ldots,6) =
{1\over2}\ce^m_{12}\cf^{mn}_{34}\ce^n_{56}
+ {1\over 4}\big[
\ce^m_{123}\cf^{mn}_{45}\ce^n_{6}
 + \ce^m_{45}\cf^{mn}_{6}\ce^n_{123}
 + \ce^m_{6}\cf^{mn}_{123}\ce^n_{45}
 + (4\leftrightarrow 6) \big]
\\
&\  +(\cX_{12}\g_m \cX_{34})\ce^m_{56}
+ {1\over 2} \big[ (\cX_{123}\g_m \cX_{45})\ce^m_{6}
+ (\cX_{45}\g_m \cX_{6})\ce^m_{123}
+ (\cX_{6}\g_m \cX_{123})\ce^m_{45}
+ (4\leftrightarrow 6) \big]
+ \cyclic{1,2, \ldots ,6}\notag
\end{align}
in terms of recursively defined multiparticle Berends--Giele polarizations $\ce^m_P$,
$\cX^\a_P$ of \eqref{reczero} and field strengths $\cf^{mn}_P$ of \eqref{recthree} instead of single-particle
polarizations and momenta.
Similar objects also drive compact representations of supersymmetric loop amplitudes in string
and field-theory, and they are excellently suited for numerical computations \cite{gielenumeric}.

The ten-dimensional polarization vectors in the bosonic components of expressions in pure 
spinor superspace can be straightforwardly dimensionally reduced. In this way, one obtains scalar and gluon 
amplitudes in maximally supersymmetric SYM in lower dimensions, say ${\cal N}=4$ in four dimensions. 
Upon insertion of spinor-helicity expressions, (\ref{6ptsample}) and (\ref{6ptcompsample}) then
reproduce all the six-point MHV and NMHV components of ${\cal N}=4$ SYM at tree level. Hence, pure spinor
superspace elegantly unifies all the MHV, NMHV, N$^k$MHV components upon reduction to four 
dimensions and captures all the different functional forms of color-ordered amplitudes with
particles of alike helicities in neighboring or non-neighboring legs. The number of terms in the 
pure spinor superspace representations of $n$-point SYM
amplitude grows moderately with $n$ thanks to the multiparticle formalism to be introduced below.

\section{Multiparticle SYM in ten dimensions}
\label{multiSYMsec}

OPEs among massless vertex operators of the pure spinor superstring feature rich patterns 
which led to a systematic definition of \textit{multiparticle} superfields of ten-dimensional
SYM in \cite{EOMBBs,Gauge,genredef}, in both local and non-local forms.
These multiparticle superfields encompass arbitrary numbers of single-particle gluon and gluino
states and can be constructed  independently of their OPE origins using field-theory methods,
in particular Berends--Giele recursion relations \cite{BerendsME} and perturbiner methods 
\cite{selivanovI,selivanovII,selivanovIII,selivanovIV}. The compatibility of OPE methods
and field-theory methods follows from the fact that SYM amplitudes are recovered
from the $\alpha' \rightarrow 0$ limit of open-superstring amplitudes.\footnote{The detailed
matching of multiparticle superfields constructed from OPEs and field-theory methods,
up to non-linear gauge transformations, relies on the propagator structure arising 
from the $\alpha' \rightarrow0$ limit of
the disk integrals (\ref{Zintdef}), see section \ref{subsec:biadj}.}

Over time, the definition of multiparticle superfields led to an elegant symbiosis of an ever-increasing number of related topics:
their local version is at the heart of the local BCJ-satisfying numerators, and their non-local version
is used to relate the BCJ properties of the amplitudes as originating from standard finite gauge
transformations. In addition, multiparticle superfields appear in connection with planar binary
trees leading to a combinatorial underpinning of the KLT map \cite{PScomb,flas} as well as the closely 
related $S$ bracket \cite{EOMBBs} and contact-term map \cite{genredef}. Ultimately, the use 
of multiparticle superfields simplifies the construction of expressions for scattering 
amplitudes of both SYM field theory and superstrings.

\subsection{Local superfields}
\label{sec:4.1loc}

The definition of local multiparticle superfields is inspired by OPE calculations of massless
vertex operators (\ref{integrado}) and (\ref{V}) in the pure spinor formalism. These multiparticle 
superfields generalize in a natural way the single-particle description of ten-dimensional SYM theory 
reviewed in section \ref{linSYMsec}. For each of the standard four types of superfields 
$A^i_\a(X,\t)$, $A^i_m(X,\t)$, $W_i^\a(X,\t)$ and $F^i_{mn}(X,\t)$, the single-particle label $i$ 
is generalized to labels for multiple particles, characterized either by words $P$ such as $P=1234$ 
or by nested Lie brackets as $[[[1,2],[3,4]],5]$. As such, it
will be convenient to refer to their multiparticle counterparts collectively in a set $K_P$
\beq\label{KPdef}
K_P \in \{A_\a^P(X,\t),\; A^m_P(X,\t),\; W^\a_P(X,\t),\; F^{mn}_P(X,\t)\}\,,
\ee
with obvious extension for $K_{[P,Q]}$ where $P$ and $Q$ can themselves be nested brackets.

Calculations of superstring disk correlators revealed that there is a rich set of properties obeyed by
the multiparticle superfields, reflected by the symmetry properties of their multiparticle labels.
The symmetries in turn are attained by
various gauge transformations of the individual single-particle
superfields and give rise to different definitions of multiparticle superfields, all related by
gauge transformations. Of special importance is the gauge transformation leading to Jacobi identities
within the nested brackets characterizing the multiparticle state. We will see in section \ref{sec:6.4} 
that this gauge leads to the color-kinematics duality of Bern, Carrasco and
Johansson \cite{BCJ}. At the superfield
level, this translates to non-linear gauge transformations which act on multiparticle superfields
defined recursively in the so-called Lorenz gauge.

The construction of the two-particle superfields is inspired by string-theory methods in the following way.
The insertion of a gauge-multiplet state on the boundary of an open-string worldsheet
is described by the pure spinor integrated vertex operator \eqref{integrado},
\beq
\label{Uvertex}
U_i := \p\t^\a A^i_\a + \Pi^m A^i_m + d_\a W_i^\a + \tfrac{1}{2} N^{mn}F^i_{mn}\, .
\eeq
In the computation of disk amplitudes with the prescription \eqref{treepresc},
the worldsheet fields of conformal weight one $[\p\t^\a,\Pi^m,d_\a, N^{mn}]$
contracting linearized superfields $K_i$ with particle label $i$ approach other 
linearized vertex operators describing other particle labels. This is captured by OPE singularities 
\eqref{all_opes} to \eqref{more_opes} and lead to composite superfields at their residues, dubbed
multiparticle superfields.

The first example of a multiparticle superfield appears in \cite{pureids} as the OPE residue of two 
massless vertex operators, an integrated $U_2$ describing SYM states with particle
label $2$ and an unintegrated $V_1$ with particle label $1$
\beq\label{UVs}
V_1(z_1)U_2(z_2) \sim z_{21}^{-k_1\cdot k_2} {L_{21}(z_1)\over z_{21}} \, , \ \ \ \ \ \ z_{ij} := z_i{-}z_j \,.
\eeq
In order to attain open-string conventions, we dropped the $\bar z_{ij}$ dependence of the
{\it Koba--Nielsen factor},
\beq
\langle \! \langle 
\prod_{j=1}^n e^{k_j\cdot X(z_j)}
\rangle \! \rangle = 
\prod_{1\leq i<j}^n | z_{ij} |^{-2k_i \cdot k_j}\,,
\label{introduceKN}
\eeq
i.e.\ extracted the disk correlator from the truncation
$|z_{ij}|^{-2k_i\cdot k_i} = z_{ji}^{-k_j\cdot k_i}  \bar  z_{ji}^{-k_j\cdot k_i}
\rightarrow z_{ji}^{-k_j\cdot k_i}$ of the $n$-point correlation function 
(\ref{introduceKN}) on the sphere. The Koba--Nielsen factor is most
conveniently computed via path-integral methods as in section 6.2.2 of \cite{Tong:2009np}, 
consistent with the OPE of plane waves
$e^{k_1\cdot X(z_1)} e^{k_2\cdot X(z_2)} \sim |z_{21}|^{-2k_1\cdot k_2} e^{k_{12}\cdot X(z_2)}$
and the equivalent $\Pi^m(z_1) e^{k_2\cdot X(z_2)} \sim  k_2^m z_{21}^{-1} e^{k_2\cdot X(z_2)}$ of (\ref{more_opes}).

The superfield structure of the OPE is captured by 
\beq\label{L21}
L_{21} = - A_1^m (\l\g_m W_2) - V_1(k_1\cdot A_2) + Q (A_1 W_2)
\eeq
which has a simple BRST variation 
\beq\label{QL21}
QL_{21} = ( k_{1}\cdot k_{2}) V_1V_2\,,
\eeq
where $Q=\l^\a D_\a$ denotes the action of the BRST operator \eqref{QBRST} of the pure spinor formalism on
superfields independent of $\p^{k\ge1}\t$. The BRST-exact term $Q (A_1 W_2)$ in (\ref{L21})
does not contribute to the variation (\ref{QL21}) and will be removed in passing from
$L_{21}$ to the two-particle superfield $V_{12}$ below. In the context of
$(n\geq 5)$-point disk amplitudes, replacing $L_{21} \rightarrow V_{12}$ in the
contribution from the $V_1(z_1)U_2(z_2)$ anticipates the cancellation of
terms $\langle -(A_1 W_2) Q( U_3\ldots ) \rangle$ under integration by parts in $z_j$
which are studied in detail at $n=5$ \cite{5ptsimple} and $n=6$ \cite{6ptTree} and
conjectural at higher $n\geq 7$, cf.\ section \ref{sec:CFTan}.

Proceeding recursively and defining higher-rank superfield building blocks \cite{towardsFT}
\beq\label{L2131}
L_{2131\ldots(p-1)1}(z_1) U_p(z_p) \sim z_{p1}^{-(k_{1}{+}k_2{+}\ldots k_{p-1})\cdot k_p}
{L_{2131\ldots(p-1)1p1}(z_1) \over z_{p1} }
\eeq
yields BRST transformations such as
\begin{align}
Q L_{2131} &= (k_{12}\cdot k_3) L_{21}V_3 + (k_1\cdot k_2)( L_{31}V_2 + V_1 L_{32}  )\,,\notag \\
Q L_{213141} &= (k_{123}\cdot k_4) L_{2131} V_4 + (k_{12}\cdot k_3)( L_{21} L_{43} + L_{2141} V_3)
 \label{lcl} \\
 &\quad +(k_1\cdot k_2)( L_{3141} V_2 + L_{31} L_{42} + L_{41} L_{32} + V_1 L_{3242})\notag 
\end{align}
with a suggestive recursive structure. The collection of $L_{2131\ldots p1}$
is said to be {\it BRST covariant} since their $Q$-variation is expressible in terms 
of products of lower-rank
building blocks (with $V_j$ as their rank-one version) along with factors of $k_i\cdot k_j$.
Here and below, we are using the notation
\beq
k_{12\ldots p} = k_1{+}k_2{+}\ldots{+}k_p
\label{defmultk}
\eeq
for multiparticle momenta.
However, a major shortcoming of the OPE residues $L_{2131\ldots p1}$ defined above
is their lack of symmetry under
exchange of labels $1,2,3,\ldots,p$. Luckily, the obstructions to having symmetry properties
conspire to BRST-exact terms and can be removed by redefinitions that do not affect
the desired amplitudes \cite{nptFT,nptStringI,EOMBBs}.
As a simple example of this phenomenon, the symmetric part of the rank-two OPE is BRST exact
\beq\label{treesix}
L_{21}  + L_{12} =  Q\Big\{  (A_1 W_2) + (A_2 W_1) -(A_1 \cdot A_2) \Big\} \, .
\eeq
The spinor and vector superfields $A_\alpha$ and $A^m$ of $D = 10$ SYM can be distinguished by identifying the superfields that they contract -- above these are $W^\alpha$ or $A_m$ (and we 
only use the $\cdot$ for vector-index contractions, i.e.\ not for spinor indices). 
Using the BRST transformation properties of
$L_{2131 \ldots}$, these BRST-exact admixtures have been identified in \cite{nptStringI} up to rank 
five, and their removal leads to a redefinition of the OPE residues that satisfy generalized 
Jacobi identities (see section
\ref{genJacsec} for their definition). The outcome of this removal procedure is an improved family 
of multiparticle superfields $V_{123\ldots p}$ dubbed BRST building blocks.

This approach was streamlined and further developed in \cite{EOMBBs} where all 
multiplicity-two superfields $K_{12}$ were extracted from the OPE between two integrated 
vertices as the coefficients of the conformal fields in the OPE, following earlier calculations 
in \cite{5ptsimple}
\begin{align}
U_1(z_1) U_2(z_2)&\sim z_{12}^{-k_1 \cdot k_2-1} \Big( \partial \theta^{\alpha} \big[ (k_1 \cdot A_2) A^1_{\alpha}
-(k_2 \cdot A_1) A^2_{\alpha} + D_{\alpha} A_{\beta}^2 W_1^{\beta} - D_{\alpha} A_{\beta}^1 W_2^{\beta} \big] \notag\\
&\quad\quad+ \Pi^m \big[ (k_1 \cdot A_2) A_m^1 - (k_2 \cdot A_1) A_m^2 + k_m^2 (A_2 W_1) -  k_m^1 (A_1 W_2) - (W_1 \gamma_m W_2)\big]\notag\\
&\quad\quad+ d_{\alpha} \big[ (k_1 \cdot A_2) W_1^{\alpha} - (k_2 \cdot A_1) W_2^{\alpha}
+ \tfrac{1}{ 4} (\gamma^{mn} W_1)^{\alpha} F_{mn}^2-  \tfrac{1}{ 4} (\gamma^{mn} W_2)^{\alpha} F_{mn}^1 \big]\notag\\
&\quad\quad+ {1 \over 2} N^{mn} \big[ (k_1 \cdot A_2) F^1_{mn} - (k_2 \cdot A_1) F^2_{mn} - 2 k_{m}^{12} (W_1 \gamma_n W_2)
+ 2 F_{ma}^1 F^{2}_{na} \big] \Big)\notag\\
&\quad+ (1+k_1 \cdot k_{2}) z_{12}^{-k_1 \cdot k_2-2} \big[(A_1 W_2) + (A_2 W_1) - (A_1 \cdot A_2) \big]\, .\label{UUope}
\end{align}
The superfields multiplying $z_{12}^{-k_1 \cdot k_2-2} $ in the last line 
are located at $K_i= K_i(z_i)$, 
and the position of the $K_i$ multiplying $z_{12}^{-k_1 \cdot k_2-1} $ is immaterial
in the two-particle context of (\ref{UUope}) since
less singular terms $z_{12}^{-k_1 \cdot k_2}$ are not tracked.
Using the relation $\partial K = \partial \theta^{\alpha} D_{\alpha} K + \Pi^m k_m K$ for 
superfields $K$ independent of non-zero modes $\partial \theta^{\alpha}$ and $\l^\a$, cf.\ (\ref{zderivative}),
we can absorb the most singular piece $\sim z_{12}^{-k_1 \cdot k_2-2}$ into total $z_{1},z_2$ 
derivatives and rewrite
\begin{align}
U_1(z_1) U_2(z_2) &\rightarrow
- z_{12}^{-k_1 \cdot k_2-1} \Big( \p\t^\a A^{12}_\a + \Pi^m A^{12}_m + d_\a W^\a_{12} + \half N^{mn}F^{12}_{mn} \Big)
\label{UUUope}\\
&\quad +\partial_1 \Big( z_{12}^{-k_1 \cdot k_2-1} \big[ \tfrac{1}{ 2} (A_1 \cdot A_2) - (A_1 W_2) \big] \Big) - \partial_2 \Big( z_{12}^{-k_1 \cdot k_2-1} \big[ \tfrac{1}{ 2} (A_1 \cdot A_2) - (A_2 W_1) \big] \Big)\,. \notag
\end{align}
Straightforward calculations using the linearized SYM equations of motion \eqref{RankOneEOM} yield the
following multiplicity-two superfields,
\begin{align}
A^{12}_\a &=  \tfrac{1}{2}\bigl[ A^2_\a (k_2\cdot A_1) + A_2^m (\g_m W_1)_\a - (1\leftrightarrow 2)\bigr] \, ,\cr
A_{12}^m &=  \tfrac{1}{2}\bigl[  A_2^m(k_2\cdot A_1) + A^1_p F_2^{pm} + (W_1\g^m W_2) - (1\leftrightarrow
2)\bigr] \, ,\cr
W_{12}^\a &= \tfrac{1}{4}(\g_{mn}W_2)^\a F_1^{mn} + W_2^\a (k_2\cdot A_1) - (1\leftrightarrow 2) \, ,\label{twopart} \\
F_{12}^{mn} &= F_2^{mn}(k_2 \cdot A_1)+ \tfrac{1}{2} F_{2}^{[m}{}_p F_{1}^{n]p}
+ k_1^{[m}(W_1 \gamma^{n]} W_2) - (1\leftrightarrow 2) \,,\notag
\end{align}
where we reiterate our conventions $F_{2}^{[m}{}_p F_{1}^{n]p}=F_{2}^{m}{}_p F_{1}^{np}
-F_{2}^{n}{}_p F_{1}^{mp}$ for antisymmetrization brackets. An interesting observation is 
that the two-particle field strength $F_{12}^{mn}$ admits a more conventional form
\beq\label{F12def}
F_{12}^{mn} = k_{12}^m A_{12}^n - k_{12}^n A_{12}^m - (k_1\cdot k_2)(A_1^m A_2^n -A_1^n A_2^m)
\eeq
with a non-linear extension as compared to the linearized field-strength superfield
$F_{i}^{mn} = k_{i}^m A_{i}^n - k_{i}^n A_{i}^m$.
More importantly, the covariant nature of the BRST transformations observed in \eqref{lcl} generalizes
to the whole set of superfields in $K_{12}$. In fact,
\begin{align}
\label{twoEOM}
D_{\a} A^{12}_{\b} + D_{\b} A^{12}_{\a} &= \g^m_{\a\b}A^{12}_m + (k_1\cdot k_2)(A^1_\a A^2_\b + A^1_\b A^2_\a) \, , \\
D_\a A_{12}^m &= \g^m_{\alpha \beta} W_{12}^{\beta} \!+\! k_{12}^m A^{12}_\a \!+\! (k_1\cdot k_2)(A^1_\a A_2^m \!-\! A^2_\a A_1^m) \, ,\cr
D_\a W^\b_{12}&= \tfrac{1}{4}(\g_{mn})_\a{}^\b F_{12}^{mn} + (k_1\cdot k_2)(A^1_\a W_2^\b - A^2_\a W^\b_1)\, ,\cr
D_\a F_{12}^{mn}&= k_{12}^{[m} (\g^{n]} W_{12})_\a  + (k_1\cdot k_2)\big[A^1_\a F_2^{mn}  +  A_{1}^{[n} (\g^{m]} W_2)_\a - (1\leftrightarrow 2)\big]\, .\notag
\end{align}
This set of superspace derivatives for the multiparticle 
superfields $K_{12}$ mimic the single-particle case \eqref{RankOneEOM}. The difference involves
contact-term corrections proportional to the Mandelstam invariant $k_1\cdot k_2 = s_{12}$.
We will see below that these contact terms admit a generalization and can be compactly described
by the so-called contact-term map acting on \textit{Lie polynomials}. In this review a ``Lie
polynomial'' is understood to be any linear combination of terms that can be written in terms of nested commutators.
For a more mathematical definition, see \cite{Reutenauer}.

The definition of multiplicity-two superfields can be formalized by
\begin{align}
U_{12}(z_2) &:=  - \oint  dz_1 \, (z_{1}-z_{2})^{k_1 \cdot k_2} U_1(z_1) U_2(z_2)
\label{oint}\\
&\phantom{:}=
\p\t^\a A^{12}_\a + \Pi^m A^{12}_m + d_\a W^\a_{12} + \tfrac{1}{2} N^{mn}F^{12}_{mn}\,,\notag
\end{align}
where the contour integral extracts the singular behavior of the
approaching vertex operators as $z_1\to z_2$ and annihilates the total derivatives 
w.r.t.\ $z_{1},z_{2}$ spelled out in (\ref{UUUope}). Similar to the earlier remark below
(\ref{QL21}), the OPE of $U_1(z_1) U_2(z_2)$ in the context of $(n\geq 5)$-point string
amplitudes gives rise to additional contributions where the total derivatives act 
on $z_3,z_4,\ldots$-dependent terms. By discarding BRST-exact terms and
additional total derivatives, such contributions were found to cancel from five-point
\cite{5ptsimple} and six-point \cite{6ptTree} amplitudes, and
their conjectural cancellation at higher points is used in section \ref{sec:diskamp}.

As we shall see below, these singularities
on the worldsheet translate into propagators $k_{12}^{-2} = 2 s_{12}^{-1}$ of the 
gauge-theory amplitude after performing the field-theory limit.
In other words, OPEs in string theory govern the pole structure of tree-level subdiagrams in SYM
field theory obtained from the point-particle limit. 

In addition to the 
multiplicity-two integrated vertex $U_{12}$, 
we define the multiplicity-two version of the unintegrated vertex as
\beq\label{V12}
 V_{12} = \l^\a A^{12}_\a \, .
\eeq
The two-particle equations of motion (\ref{twoEOM}) imply that
the single-particle relations $QV_1 = 0$ and $QU_1 = \p V_1$ generalize 
as follows at multiplicity two \cite{EOMBBs}:
\begin{align}\label{QUdelV}
QV_{12} &= (k_1\cdot k_2) V_1 V_2\\
QU_{12} &= \p V_{12} + (k_1\cdot k_2)(V_1 U_2 - V_2 U_1) \, .\notag
\end{align}
Note that the total derivatives in the last line of (\ref{UUUope}) are in one-to-one correspondence to
the BRST-exact difference
\beq
V_{12} = L_{21} + Q \Big\{
\tfrac{1}{2} (A_1\cdot A_2) - (A_1 W_2)
\Big\}\, .
\label{VvsL}
\eeq
The higher-multiplicity extensions of $V_{12}$ and $U_{12}$ to be constructed below also enjoy 
covariant BRST transformations among multiparticle vertex operators such as
\begin{align}\label{QUdelVpreview}
QV_{123} &= (k_1\cdot k_2)\big[V_1 V_{23} + V_{13}V_2\big]
+ (k_{12}\cdot k_3) V_{12} V_3 \, , \\
QU_{123} &= \p V_{123} + (k_1\cdot k_2)(V_1 U_{23} - V_{23} U_1 + V_{13} U_2 +  V_2 U_{13})
+ (k_{12}\cdot k_3)( V_{12} U_3 - V_3 U_{12}) \, , \notag
\end{align}
whose systematics are accurately described by the contact-term map in the subsequent section.

\subsubsection{\label{contactsec}The contact-term map}

We will see in the discussion below that many formulae simplify if we have a general
formula to associate contact terms $\sum (k_{R}\cdot k_S)( \ldots)$ with general nested brackets
of the form $[P,Q]$. The algorithm to do this is called the {\it contact-term map} and it
was defined for the first time in \cite{genredef} and further analyzed in \cite{flas}. This map
encodes in a systematic manner many properties that were implicitly used and assumed in several
papers. Among its many useful properties,
we will see that the contact-term map $C$ gives rise to the various contact terms
in the local equations of motion of multiparticle superfields. In addition, its combinatorial
properties will allow
us to prove that the later equations of motions of non-local superfields exhibit a ``deconcatenation
property'' in their non-linear terms, based on fine-tuned cancellations of the contact terms and
associated kinematic poles.

The contact-term map acting on a letter $i$ and on Lie monomials $[P,Q]$
is defined by the following recursion \cite{genredef,flas}
\begin{align}
C(i) &:= 0  \, ,\label{contactdef}\\
C([P,Q]) &:= [C(P), Q] + [P,C(Q)]
+ (k_P\cdot k_Q)\big(P\otimes Q - Q\otimes P\big)\, ,\notag
\end{align}
where the Lie bracket in the space ${\cal L}$ of all Lie
polynomials is extended canonically to $\cL\otimes \cL$
as
\begin{align}
[A\otimes B,Q] &:= [A,Q]\otimes B + A\otimes [B,Q] \, ,\label{otdef}\\
[P, A\otimes B] &:= [P,A]\otimes B + A\otimes [P,B]\,,\notag
\end{align}
and we have $k_P= k_{1}{+}k_2{+}\cdots{+}k_p$ for $P=12\ldots p$ according to
the definition (\ref{defmultk}) for multiparticle momenta.
To illustrate the definition, some examples can be worked out to give
\begin{align}
C([1,2]) &= (k_1\cdot k_2) ( 1\otimes 2 - 2 \otimes 1) \, ,
\label{Cexamples} \\
C([[1,2],3]) &= (k_1\cdot k_2)\big([1,3]\otimes 2 + 1\otimes [2,3] -
[2,3]\otimes 1 - 2\otimes [1,3]\big) \notag \\
&\quad{}+ (k_{12}\cdot k_3)\big([1,2]\otimes 3 - 3\otimes [1,2]\big) \, ,\notag\\
C([1,[2,3]]) &= (k_2\cdot k_3)\big([1,2]\otimes 3 + 2\otimes [1,3] -
[1,3]\otimes 2 - 3\otimes [1,2]\big) \notag\\
&\quad{}+ (k_{1}\cdot k_{23})\big(1\otimes [2,3] - [2,3]\otimes 1\big) \, ,\notag\\
C([[[1,2],3],4]) &=
(k_1\cdot k_2)\big([[1,3],4]\otimes 2 + [1,3]\otimes [2,4] + [1,4]\otimes [2,3] + 1\otimes [[2,3],4] \notag\\
&\qquad{}\qquad{}- [[2,3],4]\otimes 1 - [2,3]\otimes [1,4] - [2,4]\otimes [1,3] - 2\otimes [[1,3],4]\big)\notag\\
&\quad{}+(k_{12}\cdot k_3)\big([[1,2],4]\otimes 3+ [1,2]\otimes [3,4] - [3,4]\otimes [1,2] - 3\otimes [[1,2],4]\big)\notag\\
&\quad{}+(k_{123}\cdot k_4)\big([[1,2],3]\otimes 4-4\otimes [[1,2],3]\big) \, ,\cr
C([[1,2],[3,4]]) &=(k_1\cdot k_2)\big([1,[3,4]]\otimes 2 + 1\otimes [2,[3,4]] - [2,[3,4]]\otimes 1 - 2\otimes [1,[3,4]]\big)
\notag\\
&\quad{}+(k_3\cdot k_4)\big( [[1,2],3]\otimes 4 + 3\otimes [[1,2],4] - [[1,2],4] \otimes 3 - 4\otimes [[1,2],3]\big) \notag\\
&\quad{}+(k_{12}\cdot k_{34})\big( [1,2]\otimes [3,4] - [3,4]\otimes [1,2]\big) \, ,\notag
\end{align}
where the expression for $C([[1,2],3])$ for instance encodes the Mandelstam invariants
in $Q U_{123}$ previewed in (\ref{QUdelVpreview}).
By definition, the contact-term map produces the antisymmetrized combinations
$P\otimes Q - Q\otimes P$ of Lie monomials.
Therefore it is convenient to consider the image of the contact-term map
as being in the wedge product of Lie polynomials
\beq\label{PwedgeQ}
P\wedge Q := P\otimes Q - Q\otimes P\,,
\eeq
which implies that \eqref{otdef} becomes $[A\wedge B,C] = [A,C]\wedge B +
A\wedge[B,C]$.
Using this notation streamlines the output of the contact-term map, for example
\beq
C([1,[2,3]]) = (k_2\cdot k_3)\big([1,2]\wedge 3 + 2\wedge [1,3]\big)
+ (k_{1}\cdot k_{23})\big(1\wedge [2,3]\big)\,.
\eeq

\paragraph{Contact-term map and BRST charge}
A definition, implicit in \cite{genredef,flas}, extends the action of the contact-term
map to $\cL\wedge\cL$ as
\beq\label{extC}
C(P\wedge Q) = C(P)\wedge Q - P\wedge C(Q)\,.
\eeq
From this definition it follows that
\begin{lemma}
The contact-term map is nilpotent,
\beq\label{Csquared}
C^2 = 0\,.
\eeq
\end{lemma}
\noindent{\it Proof.} See \ref{Cnilap}.

The condition \eqref{Csquared} turns out to be an important consistency check,
as the contact-term map will be related to the pure spinor BRST charge in the
discussions below,
\beq\label{CQcorresp}
C \leftrightarrow Q_{\rm BRST}\,.
\eeq
In addition, when acting on a left-to-right Dynkin bracket $\ell(P)=[[ \ldots[[p_1,p_2],p_3],
\ldots],p_n]$ defined in \eqref{elldef}, it gives
rise to the deshuffle sums, as proven by induction \cite{genredef}
\beq\label{deshufflesum}
C(\ell(P))=
\!\!\!\!\sum_{XjY=P\atop \delta (Y)=R\otimes S}\!\!\!\!(k_X\cdot k_j)\big[\ell(XR)\otimes
\ell(jS)-\ell(jR)\otimes\ell(XS)\big] =
\!\!\!\!\sum_{XjY=P\atop \delta (Y)=R\otimes S}\!\!\!\!(k_X\cdot k_j)\ell(XR)\wedge
\ell(jS)
\eeq
where the deshuffle map $\d(Y)$ is defined in \eqref{deshuffle} and the effect of the swap
$X\leftrightarrow j$ is to replace $\otimes \to \wedge$ since the deshuffle map $\d(Y)=R\otimes S$ is
symmetric sum over $R$ and $S$.  For example,
\beq\label{C123}
C(\ell(123))  =  (k_1\cdot k_2)\big(1\wedge \ell(23) + \ell(13)\wedge 2\big)
+ (k_{12}\cdot k_3)\ell(12)\wedge 3\,.
\eeq
As we will see later, this is the same structure of the BRST variation of $V_{123}$ seen in
(\ref{exampOne}) and
will play an important role in motivating the correspondence \eqref{flcorresp} below.

The synergy between the contact-term map and multiparticle
superfields will become more natural once we
define how the Lie polynomials $P$ and $Q$ in the image of the
contact-term map become labels of
generic superfields\footnote{This
notation deviates from the one used in \cite{genredef}.} $K$ and $T$
\beq\label{replace}
(K\otimes T)_{P\otimes Q} := K_PT_Q\,,\qquad
(K\wedge T)_{P\wedge Q} := K_P T_Q\,,
\eeq
and extended by linearity.
For example,
\beq\label{exNota}
(A^m\otimes V)_{[[1,2],3]\otimes [4,[5,6]]} = A^m_{[[1,2],3]} V_{[4,[5,6]]}\,,\qquad
(V\wedge V)_{s_{12}1\wedge 2} = s_{12}V_1V_2\,.
\eeq

\subsubsection{Multiparticle superfield in the Lorenz gauge}

The generalization of the single-particle linearized superfields of
\eqref{RankOneEOM} to an arbitrary number of labels naturally leads to
a Lie-polynomial structure for the multiparticle labels. For a simplified
definition sufficient for this review, $P$ is a Lie polynomial if it is a linear
combination of words generated by nested Lie brackets acting on
non-commutative letters representing the particle labels. For example, $P=[[1,2],3]=
123-213-312+321$ is a Lie polynomial.

Initially defined by consistency of the resulting equations of motion in \cite{EOMBBs},
the following recursive definition of multiparticle superfields was identified
in \cite{Gauge} to correspond to a multiparticle version of superfields in
the Lorenz gauge:
\begin{align}
\hat{A}_\alpha^{[P,Q]}&=
\half\big[\hat A_\a^Q(k_Q\cdot \hat A_P) + \hat A_Q^m(\gamma_m \hat
W_P)_\a-(P\leftrightarrow Q)\big] \, ,
\label{Lorenzdef} \\
\hat{A}^m_{[P,Q]}&=\half\big[\hat A_Q^m(k_Q\cdot \hat A_P) + 
\hat A^P_n \hat F^{nm}_Q +
(\hat W_P\gamma^m \hat W_Q)-(P\leftrightarrow Q)\big] \, ,\notag\\
\hat{W}^\alpha_{[P,Q]}&=
{1\over4}\hat F_P^{rs}(\gamma_{rs}\hat W_Q)^\a
+\half(k_Q\cdot \hat A_P)\hat W^\alpha_Q
+ \half \What^{m\a}_Q \hat A_m^P
-(P\leftrightarrow Q) \, ,\notag\\
\hat F_{[P,Q]}^{mn} &=
 {1\over 2} \big[ \Fhat^{mn}_Q (k_Q \cdot \Ahat_P)
+ \Fhat_{Q}^{p|mn} \Ahat_p^P + \Fhat_Q^{[m}{}_r \Fhat_P^{n]r}
- 2 \g^{[m}_{\a\b}  \What_P^{n]\a} \What_Q^\b
- (P \leftrightarrow Q) \big]\,,\notag
\end{align}
where the momentum indexed by a Lie polynomial is understood to be stripped of brackets, for example
$k^m_{[1,[2,3]]} = k^m_{123}$.
The hat in $\hat K_{[P,Q]}$ above distinguishes this definition in the Lorenz gauge from
other definitions of multiparticle superfields in other gauges, as we will see shortly.
In order to complete (\ref{Lorenzdef}) to a recursion, we define the multiparticle instances
of the higher-mass-dimension superfields in section \ref{sec:higherdim}
\begin{align}
\hat{W}^{m\alpha}_{[P,Q]}&= k^m_{PQ}\What^\a_{[P,Q]} -
(\Ahat^m\otimes\What^\a)_{C([P,Q])}\,,\label{Wmal}\\
\Fhat^{m|pq}_{[P,Q]}&=k^m_{PQ}\Fhat^{pq}_{[P,Q]} -
(\Ahat^m\otimes \Fhat^{pq})_{C([P,Q])}\,,\notag
\end{align}
where the contact-term map $C$ is defined in \eqref{contactdef}
and we are using the notation
\eqref{replace}, for instance
\begin{align}
\hat F^{m|pq}_{[1,2]}  &= k_{12}^m \hat F^{pq}_{[1,2]}
- (k_1\cdot k_2) (\hat A_1^m  \hat F^{pq}_2 - \hat A_2^m \hat F^{pq}_1) \, , \notag \\
\hat F^{m|pq}_{[1,[2,3]]} &= k_{123}^m \hat F^{pq}_{[1,[2,3]]}
- (k_2\cdot k_3) \big(
\hat A_{[1,2]}^m \hat F_3^{pq} + \hat A_{2}^m \hat F_{[1,3]}^{pq}
- \hat A_{[1,3]}^m \hat F_2^{pq} - \hat A_{3}^m \hat F_{[1,2]}^{pq}\big)\\
&\quad -(k_{1}\cdot k_{23})\big(
\hat A_1^m \hat F_{[2,3]}^{pq}
-\hat A_{[2,3]}^m \hat F_{1}^{pq}\big)\,.\notag
\end{align}
Like in the multiplicity-two case \eqref{F12def}, the multiparticle field strength can be rewritten
in a more conventional form as
\beq\label{fsformFPQ}
\Fhat^{mn}_{[P,Q]} = k^m_{PQ}\Ahat^n_{[P,Q]}
- k^m_{PQ}\Ahat^m_{[P,Q]}
- (\Ahat^m\otimes\Ahat^n)_{C([P,Q])}\,.
\ee
The recursions \eqref{Lorenzdef} terminate with the single-particle superfields $\hat K_i = K_i$, and the
resulting two-particle superfields in Lorenz gauge turn out to match the expressions (\ref{twopart}) obtained
from OPEs, i.e.\ $\hat K_{[i,j]}=K_{ij}$.

It is important to emphasize that the above recursions apply to arbitrary
bracketing structures encompassed by $P$ and $Q$. For example,
\begin{align}
\Ahat^m_{[1,2]} &=
\half\Big[\hat A_{2}^m(k_{2}\cdot \hat A_{1}) +
\hat A^{1}_n \hat F^{nm}_{2}
+(\hat W_{1}\gamma^m \hat W_{2})
 -(1\leftrightarrow 2)\Big]\, , \label{exbra}\\
\Ahat^m_{[[1,2],3]} &=
\half\Big[\hat A_{3}^m(k_{3}\cdot \hat A_{[1,2]}) +
\hat A^{[1,2]}_n \hat F^{nm}_{3}
+(\hat W_{[1,2]}\gamma^m \hat W_{3})
 -([1,2]\leftrightarrow 3)\Big] \, ,\notag\\
\Ahat^m_{[[1,2],[[3,4],5] ]} &=
\half\Big[\hat A^m_{ [[3,4],5]}(k_{345}\cdot \hat A_{[1,2]}) +
\hat A^{[1,2]}_n \hat F^{nm}_{[[3,4],5]}
+(\hat W_{[1,2]}\gamma^m \hat W_{[[3,4],5]})
 -([1,2]\leftrightarrow [[3,4],5])\Big]\,.\notag
\end{align}
In addition, from the contact-term terms in $C([[1,2],[3,4]])$ as in \eqref{Cexamples}, namely
\begin{align}
\label{C1234}
C([[1,2],[3,4]]) &=(k_1\cdot k_2)\big([1,[3,4]]\otimes 2 + 1\otimes [2,[3,4]] - [2,[3,4]]\otimes 1 - 2\otimes
[1,[3,4]]\big)\notag\\
&\quad{}+(k_3\cdot k_4)\big( [[1,2],3]\otimes 4 + 3\otimes [[1,2],4] - [[1,2],4] \otimes 3 - 4\otimes [[1,2],3]\big) \notag\\
&\quad{}+(k_{12}\cdot k_{34})\big( [1,2]\otimes [3,4] - [3,4]\otimes [1,2]\big)\, ,
\end{align}
the field-strength \eqref{fsformFPQ} for $P=[1,2]$ and $Q=[3,4]$ becomes
\begin{align}
\Fhat_{[[1,2],[3,4]]}^{mn} &=
k^m_{1234}\Ahat^n_{[[1,2],[3,4]]}
- k^n_{1234}\Ahat^m_{[[1,2],[3,4]]}\label{Ftwotwo}\\
&\quad{}-(k_1\cdot k_2)\big(
\Ahat_{[1,[3,4]]}^m \Ahat^n_2 +
\Ahat^m_1\Ahat^n_{[2,[3,4]]}
- (1\leftrightarrow2)\big) \notag\\
&\quad{}-(k_3\cdot k_4)\big(
\Ahat^m_{[[1,2],3]}\Ahat^n_4
+ \Ahat^m_3\Ahat_{[[1,2],4]}
- (3\leftrightarrow4)\big) \notag\\
&\quad{}-(k_{12}\cdot k_{34})\big( \Ahat^m_{[1,2]}\Ahat^n_{[3,4]} -
\Ahat^m_{[3,4]}\Ahat^n_{[1,2]}\big)\,.
\end{align}
Identifying the pair of words $P$ and $Q$ for the superfields on the right-hand side
of the above examples leads to further applications of the recursions in \eqref{Lorenzdef} until
eventually all superfields are of single-particle nature. Naturally, the number of terms in the 
multiparticle superfields
increases very rapidly when expanded down to single-particle superfields. Fortunately, there is rarely
the need for doing so as even component expansions using the top cohomology 
factor (\ref{tlct}) of pure spinor superspace
can be performed efficiently at a multiparticle level, see~\ref{HSapp}.

\paragraph{OPE channels and Catalan numbers}

In presence of more than two vertex operators, different orders of performing the OPEs lead
to different multiparticle superfields. One can intuitively understand the different bracketings
of the definition (\ref{Lorenzdef}) of multiparticle superfields in the Lorenz gauge
and the associated vertex operators
\beq
\hat V_{P} := \l^\a \hat A^{P}_\a \, , \ \ \ \ \ \
\hat U_{P} :=   
\p\t^\a \hat A^{P}_\a + \Pi^m \hat A^{P}_m + d_\a\hat W^\a_{P} + \tfrac{1}{2} N^{mn}\hat F^{P}_{mn}\,,
\label{hatUVs}
\eeq
in this way: three vertex
operators $U_1(z_1)U_2(z_2)U_3(z_3)$ admit two\footnote{Of course, in
a correlation function, the ordering of operators is arbitrary and
will lead to index permutations of the multiparticle vertices. But in the end,
the \textit{BCJ-gauge} counterpart $U_{[2,[1,3]]}$ of $\hat U_{[2,[1,3]]}$ will still be
expressible in terms of $U_{[1,[2,3]]},U_{[[1,2],3]}$, see section~\ref{genJacsec}.
}
possible ways of performing two OPEs
in sequence while preserving the order of $z_i$ on the disk:
$z_2\to z_1$ and $z_3\to z_1$ or $z_3\to z_2$ and $z_2 \to z_1$.
These two possibilities lead to two
possible bracketings for the resulting multiparticle vertex at position $z_1$
\beq\label{twobrac}
\hat U_{[[1,2],3]}(z_1)\,,\qquad \hat U_{[1,[2,3]]}(z_1)\,.
\eeq
In general, for a string of $p$ vertex operators there will be $C_{p-1}$ bracketings, where $C_{p-1}$ is the
$(p{-}1)$-th Catalan number\footnote{Catalan numbers are given by $C_n = {1\over n+1}{2n\choose n}$, and
the simplest examples are $C_1=1$, $C_2=2$, $C_3=5$ and $C_4=14$.}.
For example, the $C_3=5$ bracketings corresponding to the different OPE
orderings among neighboring vertices in the correlation 
$U_1(z_1)U_2(z_2)U_3(z_3)U_4(z_4)$ give
rise to the following vertex operators at $z_1$:
\beq\label{diffbracks}
\hat U_{[[[1,2],3],4]}(z_1)\,,\quad
\hat U_{[[1,[2,3]],4]}(z_1)\,,\quad
\hat U_{[[1,2],[3,4]]}(z_1)\,,\quad
\hat U_{[1,[[2,3],4]]}(z_1)\,,\quad
\hat U_{[1,[2,[3,4]]]}(z_1)\,.
\eeq
For example, the first vertex in \eqref{diffbracks} corresponds to performing the OPEs 
as $z_2\to z_1$ first, then $z_3\to z_1$ and $z_4\to z_1$.

\subsubsection{Equations of motion of local multiparticle superfields in Lorenz gauge}

The equations of motion satisfied by the local multiparticle superfields given
in the recursive definition \eqref{Lorenzdef}
can be written in a similar fashion as their single-particle counterparts of \eqref{RankOneEOM}. To see this
we define an analogue of $\nabla_\a := D_\a - \Bbb A_\a$ at the level of local multiparticle superfields as
\beq\label{localnabladef}
\lnabla_\a \hat K_{[P,Q]} := D_\a \hat K_{[P,Q]} - (\Ahat_\a\otimes \hat K)_{C([P,Q])}
\eeq
in terms of the contact-term map \eqref{contactdef} and the notation \eqref{replace}.

With this definition, the equations of motion for the Lorenz-gauge superfields $\hat K_{[P,Q]}$
can be written as
\begin{align}
\label{localEOM}
\nabla_{(\a}^{(L)}\Ahat_{\b)}^{[P,Q]} &= \g^m_{\a\b} \Ahat_m^{[P,Q]}\, ,
& \lnabla_\a \hat A^m_{[P,Q]} &=  k^m_{PQ}  \Ahat_{\a}^{[P,Q]} + (\g^m  \What_{[P,Q]})_\a\, ,\\
\lnabla_\a \What^\b_{[P,Q]} &=  {1\over 4}(\g^{mn})_\a{}^\b \Fhat^{[P,Q]}_{mn} \, ,
& \lnabla_\a \Fhat^{mn}_{[P,Q]} &= \big(\What_{[P,Q]}^{[m}\g^{n]}\big)_\a \, ,\notag
\end{align}
which assume exactly the same form as their non-linear counterparts \eqref{SYMeom}.
After expanding out the derivatives $ \lnabla_\a$ in (\ref{localnabladef}), the 
local equations of motion for the Lorenz-gauge superfields \eqref{localEOM} are
given by
\begin{align}
\label{localEOMLorenz}
D_{(\a} \hat A_{\b)}^{[P,Q]} &= \g^m_{\a\b}\hat A_m^{[P,Q]} + (\hat A_\a\otimes \hat A_\b)_{C([P,Q])}\, ,\\
D_\a \hat A_m^{[P,Q]} &= (\g_m  \What^{[P,Q]})_\a + k_m^{PQ}  \Ahat_{\a}^{[P,Q]}
+ (\hat A_\a\otimes \hat A^m)_{C([P,Q])} \, ,\notag\\
D_\a \hat W^\b_{[P,Q]} &= {1\over 4}(\g^{mn})_\a{}^\b \Fhat^{[P,Q]}_{mn}
+ (\hat A_\a\otimes \hat W^\b)_{C([P,Q])}\, ,\notag\\
D_\a \hat F^{mn}_{[P,Q]} &= \big(\What_{[P,Q]}^{[m}\g^{n]}\big)_\a
+ (\hat A_\a\otimes \hat F^{mn})_{C([P,Q])} \, .\notag
\end{align}
In the simplest case $P=1$ and $Q=2$, the contact-term map produces a factor of $k_1\cdot k_2$
and we recover the two-particle equations of motion (\ref{twoEOM}) upon noting that $\hat K_{[i,j]}=K_{ij}$.
To illustrate the above equations, consider the equation of motion $D_\a \Ahat_m^{[P,Q]}$ for
$[P,Q]=[[1,2],3]$. Using the contact-term map $C([[1,2],3])$ from \eqref{Cexamples} leads to
\begin{align}\label{EOMex}
D_\a\Ahat^m_{[[1,2],3]} &=(\g^m  \What_{[[1,2],3]})_\a + k^m_{123}  \Ahat_{\a}^{[[1,2],3]}\\
&\quad + (k_1\cdot k_2)\big(\Ahat_\a^{[1,3]}\Ahat^m_2 + \Ahat_\a^1\Ahat^m_{[2,3]}
- \Ahat_\a^{[2,3]}\Ahat^m_1 - \Ahat_\a^2\Ahat^m_{[1,3]}\big) \notag\\
&\quad + (k_{12}\cdot k_3)\big(\Ahat_\a^{[1,2]}\Ahat^m_3 - \Ahat_\a^3\Ahat^m_{[1,2]}\big)\,,\notag
\end{align}
thus recovering equation (3.20) from \cite{EOMBBs}.

\subsubsection{Local multiparticle superfields in the BCJ gauge}

The explicit calculations of string disk amplitudes at multiplicities five
\cite{5ptsimple,towardsFT} and six \cite{6ptTree} revealed a truly fascinating
pattern arising from a conjunction of factors: first, the double poles in the
OPEs among massless vertex operators can be integrated by parts within the
full string integrand containing the Koba--Nielsen factor $\sim \prod_{1\leq
i<j}^n |z_{ij}|^{-k_i\cdot k_j}$, see (\ref{UUUope}) for a two-particle
example. This amounts to redistributing the superfields in the double-pole
terms among various single-pole terms in the OPEs of the vertex operators. Second, the
superfields in the numerators of the double poles have the precise form that,
once redistributed to single-pole terms via integration by parts, lead to
effective single-pole numerators that satisfy improved symmetry properties
within their multiparticle labels -- so-called generalized Jacobi identities
\cite{blessenohl}. This mechanism hinges on the fact that the BRST-exact terms
in (\ref{VvsL}) match the total derivatives in (\ref{UUUope}).

Unfortunately, doing these calculations in practice is tedious, and currently the best justification
for this mechanism
is the total-derivative distribution seen at the two-particle OPE \eqref{UUUope} and extensive explicit
cancellations in the six-point disk amplitude of \cite{6ptTree},
see section 3.2 and appendix B.3 of the reference.
Luckily after these patterns were understood in \cite{EOMBBs} these integration-by-parts steps 
could be bypassed by the recursive procedure to be described below. However, it 
remains a challenging open problem to rigorously prove the recursions
from the OPE approach.

\subsubsection{\label{genJacsec}Generalized Jacobi identities}

In section \ref{sfinbcj} below, we will introduce a gauge-transformed version
$K_P$ of the multiparticle superfields $\hat K_P$ in Lorenz gauge
defined in (\ref{Lorenzdef}). Before spelling out these redefinitions due 
to double-pole terms in OPEs,
we shall here describe the resulting symmetries of the multiparticle labels
in $K_P$. These symmetries can be summarized by
\beq\label{genjac}
K_{A\ell(B)C}+K_{B\ell(A)C} = 0 \, ,\quad A,B\neq\emptyset \, ,\quad\forall\,C\,,
\eeq
where $\ell(A)$ is the left-to-right Dynkin bracket,
\beq\label{ellmap}
\ell(123 \ldots n) := \ell(123 \ldots n{-}1)n - n\ell(123 \ldots n{-}1)\,,\quad
\ell(i):= i\,,\quad\ell(\emptyset):=0\,,
\eeq
for instance $\ell(12) = 12-21$ and $\ell(123) = 123-213-312+321$.
The symmetries \eqref{genjac} are known as the {\it generalized Jacobi identities\/} in the mathematics literature
\cite{blessenohl} (see also section 8.6.7 of \cite{Reutenauer}).

These are the same symmetries obtained by a string of standard structure constants \cite{1loopbb},
or equivalently, by a Dynkin bracket
\beq\label{coljacobi}
\ell(1234 \ldots p)\leftrightarrow K_{1234 \ldots p} \leftrightarrow\; f^{12 a_2} \,
f^{a_2 3 a_3} \, f^{a_3 4 a_4} \ldots f^{a_{p-1} p a_p} \, ,
\eeq
see figure~\ref{figHL}.
\begin{figure}[t]
\begin{center}
\includegraphics[width=0.5\textwidth]{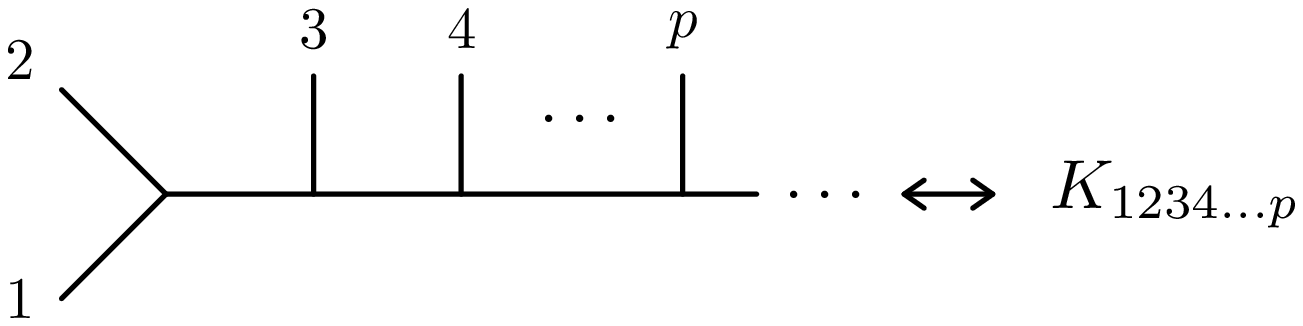}
\end{center}
\caption{The symmetry mapping between a half-ladder cubic graph and the local
SYM multiparticle superfields $K\in\{A_\a,A_m,W^\a,F^{mn}\}$.}
\label{figHL}
\end{figure}
A few examples of generalized Jacobi identities are as follows
\begin{align}
K_{12C} + K_{21C} &= 0 \, ,\quad \forall \, C \, , \notag\\
K_{123C} + K_{231C} + K_{312C} &= 0 \, ,\quad \forall \, C\, , \label{liex}\\
K_{1234C} + K_{2143C} + K_{3412C} + K_{4321C} &= 0 \, ,\quad \forall  \,C\,.\notag
\end{align}
Let $A$ be a word and $\ell(A)$
its Dynkin bracket defined in
\eqref{ellmap}.
The generalized Jacobi identities correspond to the elements in the kernel of
$\ell$. The simplest examples
\beq\label{genjacex}
\ell(12+21) = 0\,,\qquad \ell(123+231+312) = 0
\eeq
are tantamount to the antisymmetry and Jacobi identity of the Lie bracket.

Using Baker's identity $\ell(P\ell(Q)) = [\ell(P),\ell(Q)]$ \cite{Reutenauer}, it is easy to see that
$A\ell(B) + B\ell(A)$ is in the kernel of $\ell$ for any pair of words $A$ and $B$. In addition, due to
the recursive definition of $\ell$, if $\ell(P)=0$ it also follows that
$\ell(PQ)=0$ for the concatenation of $P$ with any word $Q$. Therefore the
generalized Jacobi identities can be encoded by an abstract operator $\lie_k$
\beq\label{JacobiIdentities}
\lie_k\circ K_{ABC} := K_{A\ell (B)C}+K_{B\ell (A)C}\, ,\quad\hbox{$\forall \, A,B\neq
\emptyset$ and  $\forall \, C$ such that $\len{A}+\len{B}=k\, $.}
\eeq
We emphasize the arbitrary partition of non-empty words $A$ and $B$ in the above
definition (while $C$ can be empty), leading to a non-unique operator $\lie$. For instance
\begin{align}
\lie_3\circ K_{123}&=K_{123}-K_{132}+K_{231}\,,\quad\hbox{for $A=1$, $B=23$
and $C=\emptyset$}\, , \label{nonuni}\\
\lie_3\circ K_{123}&=K_{123}+K_{312}-K_{321}\,,\quad\hbox{for $A=12$, $B=3$
and $C=\emptyset$}\,.\notag
\end{align}
However, if $\lie_2\circ K_{123}=0$ then the right-hand sides of the above
expressions agree and both are equal to $K_{123}+K_{231}+K_{312}$.

For reasons to become clear later, multiparticle superfields that satisfy $\lie_k K_P=0$ for
all $k\le \len{P}$ are said to be in the BCJ gauge.
When $K_P$ is in the BCJ gauge we use the
notation
\beq\label{BCJnotation}
K_{\ell(P)}:= K_P\,.
\eeq
For example $K_{[[1,2],3]}$ in the BCJ gauge (not to be confused with the Lorenz gauge)
is represented by $K_{123}$. In particular, with this notation we can write
Baker's identity for superfields in the BCJ gauge as
\beq\label{Baker}
K_{[P,Q]} = K_{P\ell(Q)}\,.
\eeq
For example, $K_{[12,34]} = K_{1234} - K_{1243}$. The expansion of more general bracketings
works similarly, and it amounts to rewriting an arbitrary Lie monomial $[\Gamma,\Sigma]$ 
in the Dynkin bracket basis of $\ell(1P)$, for instance:
\beq\label{hugex}
K_{[[12,34],[5,67]]} =
        K_{1234567}
       - K_{1234576}
       - K_{1234675}
       + K_{1234765}
       - K_{1243567}
       + K_{1243576}
       + K_{1243675}
       - K_{1243765}\,.
\eeq
In addition, if $K_P$ with $P=AiB$ and $i$ a single letter satisfies generalized
Jacobi identities, then it follows from \eqref{genjac} that
\beq\label{LieBasis}
K_{AiB} = - K_{i\ell(A)B}\,,\quad A\neq\emptyset\,,\,\forall  \, B\,,
\eeq
as $\ell(i)=i$ for a letter $i$. This relation
implies that there is an $(p{-}1)!$ basis of permutations $K_{i_1 \ldots i_p}$
of $K_{12\ldots p}$.

\subsubsection{\label{sfinbcj} Multiparticle superfields in the BCJ gauge}

As mentioned above, explicit calculations of superstring disk amplitudes with the pure spinor
formalism in \cite{5ptsimple,towardsFT,6ptTree} led to the discovery that the superfield numerators of single-pole integrands were
being redefined by the double-pole terms under integration by parts.\footnote{The simplicity arising
from integration by parts hinges on the fact that the double-pole terms in 
superspace are proportional to factors of $(1+s_{ijk \ldots})$. Explicit computations 
up to six points show that this is the case in the 
pure spinor formalism.} The end result of these redefinitions is an improved symmetry property of the
composite superfields, which turns out to be the same as the generalized Jacobi identities of the
previous section.

These calculations and observations led
to the definition of multiparticle superfields in the BCJ gauge.
Later computations in \cite{EOMBBs, Gauge}  suggested that these multiparticle
superfields in the BCJ gauge could
be computed by an intrinsic recursive method -- culminated
in \cite{genredef} -- independent on the (fortuitous) interference between
superstring OPEs and integration by parts when multiplied by the Koba--Nielsen
factor\footnote{The recursive construction in \cite{EOMBBs, Gauge, genredef} of
multiparticle superfields in the BCJ gauge in principle has no limitation in terms of
multiplicity. Whether the superfields in the BCJ gauge resulting from the recursive 
method agree with the result from
manipulations of superstring integrands is not yet mathematically proven
beyond six points and is taken as an
assumption in the calculations of section~\ref{sec:diskamp}.}.
The recursive method can be approached in two ways, each one convenient for
different situations. Let us now review these.

\paragraph{From a hybrid gauge to BCJ gauge}
It is convenient to encode the redefinitions needed to attain the BCJ gauge by defining
an intermediate set of multiparticle superfields in a \textit{hybrid gauge}, denoted by $\check K_{[P,Q]}$. 
The definition of multiparticle superfields in the hybrid gauge $\check K_{[P,Q]}$ assumes that all 
superfields of lower multiplicities $K_P$ and $K_Q$ have been redefined to satisfy all the generalized 
Jacobi identities, i.e.\ $\lie_k K_P = 0$ for $k\le \len{P}$ (and similarly for $Q$). We then define 
\begin{align}
\check A_\alpha^{[P,Q]}&=
\half[ A_\a^Q(k_Q\cdot  A_P) +  A_Q^m(\gamma_m  W_P)_\a-(P\leftrightarrow Q)]\, ,
\label{Hybriddef} \\
\check A^m_{[P,Q]}&=\half[ A_Q^m(k_Q\cdot  A_P) + 
 A^P_n  F^{nm}_Q +
( W_P\gamma^m  W_Q)-(P\leftrightarrow Q)] \, ,\notag\\
\check W^\alpha_{[P,Q]}&=
{1\over4} F_P^{rs}(\gamma_{rs} W_Q)^\a
+\half(k_Q\cdot  A_P) W^\alpha_Q
+ \half W^{m\a}_Q A^m_P
-(P\leftrightarrow Q) \, ,\notag\\
\check F_{[P,Q]}^{mn} &=
 {1\over 2} \big[ F^{mn}_Q (k_Q \cdot A_P)
+ F_{Q}^{r|mn} A_r^P +  F_Q^{[m}{}_{r} F_{P}^{n]r}
- 2 \g^{[m}_{\a\b} W_P^{n]\a} W_Q^\b
- (P \leftrightarrow Q) \big]\, ,\notag
\end{align}
and $\check K_i=K_i$,
where the superfields $K_P$ and $K_Q$ on the right-hand side satisfy
the generalized Jacobi identities $\lie_k K_P = 0$ for $k\le \len{P}$ and
\begin{align}
{W}^{m\alpha}_{[P,Q]}&= k^m_{PQ}W^\a_{[P,Q]} -
(A^m\otimes W^\a)_{C([P,Q])}\,,\label{WmalUnhatted}\\
F^{m|pq}_{[P,Q]}&=k^m_{PQ}F^{pq}_{[P,Q]} -
(A^m\otimes F^{pq})_{C([P,Q])}\,,\notag
\end{align}
are the local form of the multiparticle superfields of higher mass dimension defined in
\cite{Gauge} involving the contact-term map $C([P,Q])$ in \eqref{contactdef} and the notation \eqref{replace}.

In contrast to the definitions (\ref{Lorenzdef}) of Lorenz-gauge superfields $\hat K_{[P,Q]}$,
in the hybrid gauge their definitions (\ref{Hybriddef}) are {\it not} recursive:
The superfields $\check K_{[P,Q]}$ on the left-hand side of
\eqref{Hybriddef} have to be redefined $\check K_{[P,Q]}  \rightarrow
K_{[P,Q]}$ before qualifying as input on the
right-hand side in the next step of the recursion.

The hybrid gauge leads to more convenient explicit expressions to arrive at multiparticle 
superfields in the BCJ gauge. One can show that the following redefinitions
\begin{align}
K_{[P,Q]} &:= \check K_{[P,Q]} -
\!\!\!\!\sum_{P=XjY\atop \d(Y)=R\otimes S}\!\!\!\!\! (k_{X}\cdot k_j)
\bigl[ H_{[XR,Q]}\; K_{jS} - (X\leftrightarrow j)\bigr] \label{Hpdef} \\
&\quad+\!\!\!\! \sum_{Q=XjY\atop \d(Y)=R\otimes S}\!\!\!\!\! (k_{X}\cdot k_j)
\bigl[ H_{[XR,P]}\; K_{jS} - (X\leftrightarrow j)\bigr]
-
\begin{cases} D_\alpha H_{[P,Q]} &: \ K= A_\alpha \\
k_{PQ}^m H_{[P,Q]} &: \ K= A^m \\
0 &: \ K= W^\alpha
\end{cases}
\notag
\end{align}
imply that the left-hand side satisfies all generalized Jacobi identities. Note that  $\d(Y)$ in \eqref{Hpdef}
denotes the deshuffle map defined in \eqref{deshuffle} and the superfields $H$ will be defined below. 
To illustrate the above redefinitions we write down explicit examples for $A^m_{[P,Q]}$ up to multiplicity 
five (recall that $\check A^m_i:= A^m_i$ and $\check A^m_{[i,j]} := A^m_{ij}$)
\begin{align}
A^m_{[12,3]} &= \check A^m_{[12,3]}-k_{123}^mH_{[12,3]} \, ,\label{examplesH}\\
A^m_{[12,34]} &= \check A^m_{[12,34]} -k_{1234}^mH_{[12,34]}\notag\\
&\quad - (k_{1}\cdot k_{2})\Big[  H_{[1,34]}A^m_{2}-H_{[2,34]}A^m_1 \Big] \notag\\
&\quad + (k_{3}\cdot k_{4})\Big[  H_{[3,12]}A^m_{4}-H_{[4,12]}A^m_3 \Big] \, ,\notag\\
A^m_{[123,4]} &= \check A^m_{[123,4]} -k_{1234}^mH_{[123,4]}\notag\\
&\quad - (k_1\cdot k_2)\Big[ H_{[13,4]}A_2^m - H_{[23,4]}A_1^m\Big] \notag\\
&\quad -(k_{12}\cdot k_3)H_{[12,4]}A_3^m\, ,\notag\\
A^m_{[1234,5]} &= \check A^m_{[1234,5]} - k^m_{12345} H_{[1234,5]}\notag\\
&\quad - (k_1\cdot k_2) \big[
           H_{[134,5]} A^m_{2}
          + H_{[14,5]} A^m_{23}
          + H_{[13,5]} A^m_{24}
	  - (1\leftrightarrow 2)
          \big]\notag\\
&\quad- (k_{12}\cdot k_3) \big[  H_{[124,5]} A^m_{3} + H_{[12,5]} A^m_{34}  - (12\leftrightarrow 3) \big]\notag\\
&\quad- (k_{123}\cdot k_4) H_{[123,5]} A^m_{4} \, ,\notag\\
A^m_{[123,45]} &= \check A^m_{[123,45]} -k_{12345}^mH_{[123,45]}\notag\\
&\quad- (k_1\cdot k_2)\Big[ H_{[13,45]}A^m_{2} + H_{[1,45]}A^m_{23} -(1\leftrightarrow 2)\Big] \notag\\
&\quad-(k_{12}\cdot k_3)\Big[ H_{[12,45]}A^m_3 -(12\leftrightarrow 3)\Big]\notag\\
&\quad+(k_4\cdot k_5)\Big[H_{[4,123]}A^m_5-(4\leftrightarrow 5)\Big] \, .\notag
\end{align}
The explicit expressions for the new superfields $H_{[P,Q]}$
were obtained up to multiplicity five in \cite{Gauge} and for arbitrary
multiplicity in \cite{genredef}:
\beq\label{genH}
H_{[i,j]}=0\,,\qquad H_{[A,B]}=
(-1)^{\len{B}}{\len{A}\over\len{A}+\len{B}}
\sum_{XjY=\dot a\tilde B}(-1)^{|Y|}H'_{\tilde Y,j,X} - (A\leftrightarrow B)\,,
\eeq
where $\dot a$ and $\dot b$ denote the letterifications of $A$ and $B$ as defined in \eqref{letterif} and
\begin{align}
H'_{A,B,C} &:= H_{A,B,C} +\Big[\half H_{[A,B]}(k_{AB}\cdot A_C)  + {\rm
cyc}(A,B,C)\Big]\label{Hprimedef}\\
&\quad -\Big[\!\!\!\! \sum_{XjY=A\atop \delta(Y)=R\otimes S}
\!\!\!\!\!(k_{X}\cdot k_j)\big[H_{[XR,B]}H_{[jS,C]} -(X\leftrightarrow j)\big]
 + {\rm cyc}(A,B,C)\Big]\,,\notag\\
H_{A,B,C} &:= -{1\over 4}A^m_A A^n_B F^{mn}_C
+ \half (W_A\g_m W_B)A_C^m + {\rm cyc}(A,B,C)\,. \label{HABCdef}
\end{align}
It is straightforward to check that the superfields
$H_{[A,B]}$ satisfy generalized Jacobi identities within $A$ and $B$. This justifies
using the notation where nested brackets are flattened, e.g.\ $H_{[[[1,2],3],4]}=H_{[123,4]}$ 
in accordance with the notation \eqref{BCJnotation}. The combination of (\ref{genH}), 
(\ref{Hprimedef}) and (\ref{HABCdef}) reduces all the redefinitions (\ref{Hpdef}) 
from hybrid gauge to BCJ gauge to products of building blocks $H_{A,B,C}$ 
and $(k_{AB}\cdot A_C)$.

The superfields $H_{[P,Q]}$ in (\ref{genH}) up to multiplicity
seven are given by
\begin{align}
H_{[12,3]}&=\frac{1}{3}\big(H'_{1,2,3}\big) \, ,\label{explicitHs}\\
H_{[123,4]}&=\frac{1}{4}\big(H'_{12,3,4}-H'_{1,2,43}\big)\, ,\notag\\
H_{[12,34]}&=\frac{1}{4}\big(2H'_{1,2,34}-2H'_{3,4,12}\big)\, ,\notag\\
H_{[1234,5]}&=\frac{1}{5}\big(H'_{123,4,5}-H'_{12,3,54}+H'_{1,2,543}\big)\, ,\notag\\
H_{[123,45]}&=\frac{1}{5}\big(2H'_{12,3,45}-2H'_{1,2,453}-3H'_{4,5,123}\big)\, ,\notag\\
H_{[12345,6]}&=\frac{1}{6}\big(H'_{1234,5,6}-H'_{123,4,65}+H'_{12,3,654}-H'_{1,2,6543}\big)\, ,\notag\\
H_{[1234,56]}&=\frac{1}{6}\big(2H'_{123,4,56}-2H'_{12,3,564}+2H'_{1,2,5643}-4H'_{5,6,1234}\big)\, ,\notag\\
H_{[123,456]}&=\frac{1}{6}\big(3H'_{12,3,456}-3H'_{1,2,4563}-3H'_{45,6,123}+3H'_{4,5,1236} \big)\, ,\notag\\
H_{[123456,7]}&=\frac{1}{7} \big( H'_{12345,6,7}-H'_{1234,5,76}+H'_{123,4,765}-H'_{12,3,7654}+H'_{1,2,76543} \big)\, ,\notag\\
H_{[12345,67]}&=\frac{1}{7}\big(2H'_{1234,5,67}-2H'_{123,4,675}+2H'_{12,3,6754}-2H'_{1,2,67543}-5H'_{6,7,12345} \big)\, ,\notag\\
H_{[1234,567]}&=\frac{1}{7}\big( 3H'_{123,4,567}-3H'_{12,3,5674}+3H'_{1,2,56743}-4H'_{56,7,1234}+4H'_{5,6,12347}\big)\,,\notag
\end{align}
and the simplest non-vanishing instances of the primed superfields in (\ref{Hprimedef}) are
\begin{align}
H'_{1,2,3}&= H_{1,2,3} \, ,
\label{simpHprime} \\
H'_{12,3,4} &= H_{12,3,4} + \frac{1}{6} \big[ H_{1,2,3} (k_{123}\cdot A_4) - (3\leftrightarrow 4) \big]\, . \notag
\end{align}
For example, the explicit expressions for the first two superfields above are given by
\begin{align}
H_{[12,3]} &= -{1\over 12}A^m_1 A^n_2 F^{mn}_3
+ \frac{1}{6} (W_1\g_m W_2)A_3^m + {\rm cyc}(1,2,3)\, , \label{hthree}\\
H_{[123,4]} &= \frac{1}{4} (H_{12,3,4} + H_{34,1,2}) + \frac{1}{24} \big[ H_{1,2,3}(k_{123}{\cdot} A_4)
- H_{1,2,4}(k_{124}{\cdot }A_3) + H_{3,4,1}(k_{134}{\cdot} A_2) - H_{3,4,2}(k_{234}{\cdot} A_1)
\big]\, . \notag
\end{align}
It is interesting to observe that the expressions for the superfields $H_{[A,B]}$ that lead to the
BCJ gauge are not unique. In fact, simpler explicit expressions can be derived using the
Bern-Kosower formalism \cite{cristhiam}.

To complement the definition \eqref{Hpdef}, the field strength in the BCJ gauge is defined
using the contact-term map \eqref{contactdef}
\beq\label{BCJgaugeFdef}
F_{[P,Q]}^{mn}=k_{PQ}^mA_{[P,Q]}^n-k_{PQ}^nA_{[P,Q]}^m
-(A^m\otimes A^n)_{C([P,Q])}\,,
\eeq
see the definition \eqref{replace}.
Concretely, the above superfields can be explicitly checked to satisfy the generalized Jacobi identities. 
For example, using the notation \eqref{BCJnotation} as $A^m_{[1234,5]} = A^m_{12345}$, 
one can see that
\beq\label{bcjex1}
A^m_{12345} + A^m_{21435} + A^m_{34125} + A^m_{43215} = 0\, ,
\eeq
corresponding to the third identity in \eqref{liex} with $C=5$.
In addition, long calculations demonstrate that
\beq\label{ted}
A^m_{12345} - A^m_{12354} + A^m_{45123} - A^m_{45213} - A^m_{45312} + A^m_{45321} = 0\, ,
\eeq
in agreement with the expansion \eqref{Baker} applied to $A^m_{[123,45]} +
A^m_{[45,123]}=0$.

As an alternative to the method above to obtain multiparticle superfields in the BCJ gauge, one
can choose to go directly from the Lorenz gauge to the BCJ gauge. The process is more
or less the same, but the explicit formulae make it more evident that the whole process
corresponds to a finite gauge transformation of the corresponding perturbiner expansion
of Berends--Giele currents to be reviewed shortly. The discussion of these redefinitions
is left for the \ref{LorBCJapp}.

\subsubsection{Equations of motion of multiparticle superfields in the BCJ gauge}

Written in terms of the BRST charge $Q=\l^\a D_\a$, the equations of motion for the multiparticle
superfields in the BCJ gauge become ($k_\emptyset:=0$) \cite{EOMBBs}
\begin{align}
QV_{P} &= \!\!\sum_{P=XjY\atop \d(Y)=R\otimes S}\!\!(k_{X} \cdot k_j)\,V_{XR}V_{jS}\,, \label{GeneralQ}\\
QA^m_{P} &=  (\l\ga^m  W_{P}) + k^m_{P}  V_{P}\,
+\!\!\sum_{P=XjY\atop \d(Y)=R\otimes S}\!\!(k_{X} \cdot k_j)\big[V_{XR}A^m_{jS} - V_{jR}A^m_{XS}\big]\,,\notag\\
QW^\b_{P} &=  {1\over 4}(\l\ga^{mn})^\b  F^{P}_{mn}\, +\!\!
\sum_{P=XjY\atop \d(Y)=R\otimes S}\!\!(k_{X}\cdot k_j)
\big[V_{XR}W^\b_{jS} - V_{jR}W^\b_{XS}\big]\,,\notag\\
QF^{mn}_{P} &=   (\l\ga^{[n}  W^{m]}_{P})\,
+\!\! \sum_{P=XjY\atop \d(Y)=R\otimes S}\!\!(k_{X} \cdot k_j)\big[
V_{XR}F^{mn}_{jS}  - V_{jR}F^{mn}_{XS} \big]\,,\notag
\end{align}
where $V_P = \l^\a A_\a^P$ is the multiparticle unintegrated vertex operator and
the last line involves the superfield $W^{m \alpha}_P$ of higher mass dimension
defined in (\ref{WmalUnhatted}). In addition,
the sum over $P=XjY$ assembles the $|P|{-}1$ deconcatenations of the word $P$ into a 
word $X$, a single letter $j$, and a word $Y$. Moreover, $\d(Y)=R\otimes S$ denotes the
deshuffle \eqref{deshuffle} of the word $Y$ into the words $R$ and $S$.
A few examples help to illustrate the above formulae,
\begin{align}
QV_1 &= 0\,, \label{exampOne}\\
QV_{12} &= (k_1\cdot k_2)V_1V_2 \, ,\notag\\
QV_{123} & = (k_1\cdot k_2)\big[V_1 V_{23} + V_{13}V_2\big]
+ (k_{12}\cdot k_3) V_{12} V_3\,, \notag\\
Q V_{1234} &=(k_1\cdot k_2)\bigl[V_1V_{234}
+ V_{13} V_{24} + V_{14} V_{23} +  V_{134} V_2 \bigr] \notag\\
&\quad{} + (k_{12}\cdot k_3)\bigl[V_{12} V_{34}
+  V_{124} V_3\bigr]
+ (k_{123}\cdot k_4) V_{123} V_4\,,\notag\\
Q V_{12345} &=(k_1\cdot k_2)\bigl[
  V_{1} V_{2345}
+ V_{13} V_{245}
+ V_{134} V_{25}
+ V_{1345} V_{2} \notag\\
&\qquad{}+ V_{135} V_{24}
+ V_{14} V_{235}
+ V_{145} V_{23}
+ V_{15} V_{234}
\bigr]\notag\\
&\quad{} + (k_{12}\cdot k_3)\bigl[
V_{12} V_{345}
+  V_{124} V_{35}
+  V_{1245} V_{3}
+  V_{125} V_{34}
\bigr]\notag\\
&\quad{} + (k_{123}\cdot k_4)\bigl[
V_{123} V_{45}
+ V_{1235} V_{4}\bigr]\notag\\
&\quad{} + (k_{1234}\cdot k_5) V_{1234} V_{5}\,.\notag
\end{align}
Note that the instances at rank $\leq 4$ can be formally obtained
from the BRST variations (\ref{QL21}) and (\ref{lcl}) of the OPE residues upon promoting
$L_{2131\ldots p1} \rightarrow V_{123\ldots p}$. In other words, the chain of redefinitions in
(\ref{VvsL}) and generalizations to higher rank preserve the form of the covariant BRST algebra.

Recalling the notation \eqref{BCJnotation} $K_{\ell(P)}:=K_P$ for superfields in the BCJ gauge,
the BRST variations in \eqref{GeneralQ} for the left-to-right nested commutator $\ell(P)$
can be obtained as the special case $[R,S]:=\ell(P)$ of the BRST variations for general commutators
\begin{align}
\label{localEOMGeneral}
QV_{[R,S]} &= \half (V\otimes V)_{C([R,S])} \, ,\\
QA^m_{[R,S]} &= (\l\g^m  W_{[R,S]}) + k^m_{RS}  V_{[R,S]}
+ (V\otimes A^m)_{C([R,S])}\, ,\notag\\
Q W^\b_{[R,S]} &= {1\over 4}(\l\g_{mn})^\b F_{[R,S]}^{mn}
+ (V\otimes W^\b)_{C([R,S])}\, ,\notag\\
Q F^{mn}_{[R,S]} &= \big(\l W_{[R,S]}^{[m}\g^{n]}\big)
+ (V\otimes F^{mn})_{C([R,S])} \, ,\notag
\end{align}
where we are employing the notation \eqref{replace} for the contact-term map. 
The fact that
\beq\label{Qdynkin}
(V\otimes K)_{C(P)} = \sum_{P=XjY\atop \d(Y)=R\otimes S}\!\!(k_{X} \cdot k_j)\big[V_{XR}K_{jS} -
(X\leftrightarrow j)\big]
\eeq
for superfields in the BCJ gauge was proven in Lemma~1 of \cite{genredef}\footnote{The replacement $K_{\ell(P)}\to K_P$ was
left implicit in the proof of \cite{genredef}.}.

The multiparticle versions of the on-shell constraints 
$k_1^m (\gamma_m W_1)_\alpha= 0= k^1_m F^{mn}_1$
take a form similar to (\ref{GeneralQ}),
\begin{align}
k_P^m (\gamma_m W_P)_\alpha &= 
\sum_{P=XjY\atop \d(Y)=R\otimes S}\!\!(k_{X} \cdot k_j)\big[
A^m_{XR}(\gamma_m W_{jS})_\alpha 
 - A^m_{jR}(\gamma_m W_{XS})_\alpha   \big]\, ,
\label{diracloc} \\
k^P_m F^{mn}_P &=
\sum_{P=XjY\atop \d(Y)=R\otimes S}\!\!(k_{X} \cdot k_j)\big[
\gamma^n_{\alpha \beta} W^\alpha_{XR} W^\beta_{jS}
 + A_m^{XR} F_{jS}^{mn}
- (X \leftrightarrow j) \big] \, , \notag
\end{align}
for instance
\begin{align}
k_{12}^m (\gamma_m W_{12})_\alpha &= (k_1\cdot k_2) \big[ A_1^m (\gamma_m W_2)_\alpha -  A_2^m (\gamma_m W_1)_\alpha \big] \, , \label{diraclocex} \\
k_m^{12} F_{12}^{mn} &= (k_1 \cdot k_2) \big[
2 (W_1 \gamma^n W_2) + A_m^1 F^{mn}_2 -  A_m^2 F^{mn}_1 \big]\, . \notag
\end{align}
Using the multiparticle SYM superfields in the BCJ gauge,
it is natural to define the multiparticle massless vertices
in the pure spinor formalism as
\beq\label{Vhatgen}
V_{[P,Q]} = \l^\a A^{[P,Q]}_\a\,, \quad
U_{[P,Q]} = \p\t^\a A^{[P,Q]}_\a + \Pi^m A^{[P,Q]}_m + d_\a W^\a_{[P,Q]}
+ \tfrac{1}{2} N^{mn}F^{[P,Q]}_{mn}\,,
\eeq
where $[P,Q]$ denotes an arbitrary Lie monomial, e.g.\ $[[1,[2,3]],[4,5]]$. The
multiparticle equations of motion \eqref{localEOMGeneral} and the non-linear Dirac
equation (\ref{diracloc}) imply the relations
\beq
Q U_{[P,Q]} = \partial V_{[P,Q]} +  (V\otimes U)_{C([P,Q])}
\eeq
between the two types of vertex operators, in lines with $Q U_1 = \partial V_1$ and 
the rank-two example (\ref{QUdelV}).

\subsection{Non-local superfields and Berends--Giele currents}
\label{sec:42NL}

In addition to the local multiparticle superfields reviewed above, the pure spinor formalism naturally
leads to another class of multiparticle superfields containing kinematic poles in generalized Mandelstam
invariants \cite{nptFT,EOMBBs}. These non-local SYM superfields are denoted collectively by $\cK_P$ and were
dubbed \textit{Berends--Giele currents}\footnote{The story is longer than this since in \cite{nptFT} the relation with the
standard Berends--Giele current $J^m_P$ was observed from a structural similarity between the appearance of the superfield
Berends--Giele current $M_P$ in the the pure spinor cohomology formula for SYM
tree-level amplitudes and in the role played by $J^m_P$ in the standard Berends--Giele setup. It was 
much later in \cite{BGBCJ,Gauge} that a rigorous relation between the superfield version of the 
Berends--Giele currents $\cK_P$ and the standard bosonic $J^m_P$ was
established via $\cA^m(\t)$. So in fact, the non-local superfields $\cK_P$ generalize the Berends--Giele 
currents in a supersymmetric manner and may also be named Berends--Giele \textit{supercurrents}.} \cite{nptFT} due to
their relation with the standard gluonic currents $J^m_P$ defined by Berends and Giele in the 80s \cite{BerendsME}. Specifically,
they share the same shuffle symmetries, and the $\t=0$ term in the Berends--Giele superfield current $\cA^m_P(\t)$ in a suitable gauge is equal to $J^m_P$, see section \ref{SYMsec} for a review.

Each local superfield representative in $K_P \in \{ A^{P}_\alpha, A^m_{P} , W_{P} ^\alpha, F_{P} ^{mn} \}$
admits a corresponding Berends--Giele current
with multiparticle label $P=12\ldots p$ denoted by calligraphic letters
\beq\label{calK}
{\cal K}_P \!\in\! \{  \cA^{P}_\alpha, \cA^m_{P} , \cW_{P}^\alpha, \cF_{P}^{mn} \}\, ,
\eeq
starting with ${\cal K}_{1} :=  K_{1}$ and
\begin{align}
\label{BGexpl}
{\cal K}_{12} &= \frac{K_{12}}{s_{12}} \, , \\
\quad {\cal K}_{123} &= \frac{K_{123}}{s_{12}s_{123}} +\frac{K_{321}}{s_{23}s_{123}} \, , \notag\\
\cK_{1234} &= {1 \over s_{1234}} \; \bigg(\, {K_{1234}\over s_{12}s_{123} }  +  {K_{3214}\over s_{23}s_{123} }
+ {K_{3421} \over s_{34}s_{234} } + {K_{3241} \over s_{23}s_{234} }  + {K_{[12,34]} \over
s_{12}s_{34} } \,\bigg) \, ,\notag
\end{align}
with generalized Mandelstam invariants defined in \eqref{mandef}, $s_{12\ldots p} = \tfrac{1}{2} k_{12\ldots p}^2$.
In contrast to the bosonic Berends--Giele currents in \cite{BerendsME},
the supercurrents ${\cal K}_P$ also contain fermionic degrees of freedom as required by
supersymmetry,
and their construction does not
include any quartic vertices. Note that for historical reasons the Berends--Giele currents associated 
to the local multiparticle unintegrated vertex $V_P=\l^\a A^P_\a$ is denoted by $M_P$ rather than ${\cal V}_P$,
\beq\label{historic}
M_P = \l^\a \cA^P_\a\,.
\eeq
More explicitly, $M_1 = V_1$ and
\begin{align}
\label{BGexplM}
M_{12} &= \frac{V_{12}}{s_{12}}\, ,\\
\quad M_{123} &= \frac{V_{123}}{s_{12}s_{123}} +\frac{V_{321}}{s_{23}s_{123}}\, , \notag\\
M_{1234} &= {1 \over s_{1234}} \; \bigg(\, {V_{1234}\over s_{12}s_{123} }  +  {V_{3214}\over s_{23}s_{123} }
+ {V_{3421} \over s_{34}s_{234} } + {V_{3241} \over s_{23}s_{234} }  + {V_{[12,34]} \over
s_{12}s_{34} } \,\bigg)\, .\notag
\end{align}
In the early stages of unraveling the cohomology properties of multiparticle superfields with the
pure spinor formalism, the Berends--Giele currents ${\cal K}_P$ were defined case by case to encompass 
all tree subdiagrams compatible with the ordering of the external legs in $P$ in such a way as to 
transform BRST covariantly \cite{nptFT,nptStringI}. For instance,
from the equations of motion \eqref{exampOne} we get
\begin{align}
Q  M_{1} & = 0\, , \label{treethirteen}\\
Q  M_{12} & = M_1  M_2\, , \notag\\
Q  M_{123} & = M_{12}  M_3 + M_1  M_{23}\, , \notag\\
Q  M_{1234} & = M_{123}  M_4 + M_{12}  M_{34}  +  M_1  M_{234} \,.\notag
\end{align}
In contrast to $Q V_{123\ldots p}$ as given by \eqref{exampOne}, there are no explicit Mandelstam variables in \eqref{treethirteen}
as the propagators $s^{-1}_{i\ldots j}$ in \eqref{BGexpl} absorb the appearance of explicit momenta in the contact terms of the equations of motion
of the local superfields. A rigorous proof of this
statement from a combinatorial perspective can be found in \eqref{Cbdeconc}.
The generalization of
\eqref{treethirteen} to higher rank is given by \cite{nptFT}
\beq\label{treefourteen}
Q \, M_{P} = \sum_{XY=P} M_{X}  M_{Y}
\eeq
and it was proven in \cite{Gauge}.
The sum involves all the $|P|{-}1$ deconcatenations $XY{=}P$ of $P$ into
non-empty\footnote{Defining $M_\emptyset :=0$, the restriction
to non-empty words may be lifted and the general definition \eqref{decexamp}
may be applied.} words $X,Y$, e.g.\ $X=12\ldots j$ and $Y=j{+}1\ldots p$ with
$j=1,2,\ldots ,p{-}1$ in case of $P=12\ldots p$. These deconcatenations
will be later on associated with partitions of the $p$ on-shell legs on two
Berends--Giele currents while preserving the color ordering. For a
combinatorial proof of \eqref{treefourteen} from the perspective of BRST
variations of the composing local numerators $V_Q$, see \eqref{deconcproof}.

It is useful to define a BRST-exact superfield $E_P$ as
\beq\label{Edef}
E_P = \sum_{XY=P} M_{X}  M_{Y}
\eeq
which will be used in the pure spinor cohomology formula for SYM tree-level amplitudes in
section~\ref{PSSYMsec}. One can show that
$E_P$ is conditionally BRST exact
\beq\label{QEexact}
E_P= QM_P\qquad\hbox{if $s_P\neq0$} \, ,
\eeq
provided that $s_P\neq0$ is true as $M_P$ contains the propagator $1/s_P$. We
will see later that this condition of the momentum phase space is of crucial
importance.

The connection between Berends--Giele currents subject to the deconcatenation 
equation \eqref{treefourteen} and the cubic tree subdiagrams compatible with a color ordered
amplitude is supported by the following consistency check: the
number of terms or kinematic pole channels in $M_{12{\ldots}p}$ is the Catalan 
number $C_{p{-}1}$ (see \eqref{poles} below for a proof)
which counts the number of cubic diagrams in a color-ordered $(p{+}1)$-point tree amplitude.
As the relation with the Catalan number suggests, the definition of Berends--Giele currents admits a beautiful
combinatorial interpretation in terms of planar binary trees and is connected with the
KLT matrix
in many surprising ways \cite{Zfunctions,PScomb,flas}. We will return to this point later in
section~\ref{BGcombsec}.

\subsubsection{Non-linear wave equations and Berends--Giele currents}

In \cite{Gauge} the definition of Berends--Giele currents was shown to arise
from solutions of the non-linear wave equations of ten-dimensional SYM theory
in the {\it Lorenz gauge}
\beq\label{Lorenz}
[\p_m,\bA^m]=0\, .
\eeq
To show this
one first needs to derive the non-linear wave equations obeyed by the superfields $\Bbb K 
\in \{ \Bbb A_\alpha, \Bbb A^m , \Bbb W^\alpha,  \Bbb F^{mn} \}$.
This can be done starting from
\beq\label{dalemb}
\Box \Bbb K = [\p^m,[\p_m,\Bbb K]]
\eeq
and using the Jacobi identity together with $\p^m = \nabla^m + \Bbb A^m$. That is,
\begin{align}
\Box \Bbb K &= [\nabla^m + \Bbb A^m , [\partial_m,\Bbb K ]] \label{boxone} \\
&= [ [ \nabla^m , \partial_m ] , \Bbb K] + [ \Bbb A^m , [\partial_m,\Bbb K ]]
+ [ \Bbb A^m , [\nabla_m,\Bbb K ]] + [ \nabla^m , [\nabla_m,\Bbb K ]] \,.\notag
\end{align}
The first term in the second line vanishes in Lorentz gauge \eqref{Lorenz} as
$[  \partial_m , \nabla^m ] =-[  \partial_m , \Bbb A^m ] $. For the simpler
choices of superfields $\Bbb K \rightarrow \{\nabla_\alpha , \nabla_m\}$,
the last term of \eqref{boxone} can be converted to quadratic expressions in the non-linear
fields using the Dirac  and super Yang--Mills equations (\ref{Diraceq}).
In the case of $\Bbb K \rightarrow \{\Bbb W^\alpha,\Bbb F^{mn}\}$,
the analogous conversion necessitates the equations of motion\footnote{The first line of
(\ref{cors}) can be derived by applying the Clifford algebra
$ [ \nabla_m, \Bbb W^{m\a}]= \frac{1}{2} [\nabla_m,[\nabla_n, (\g^m \g^n \Bbb W)^\a
+ (\g^n \g^m \Bbb W)^\a]]$ followed by the Dirac equation, Jacobi
relations and the definition $\Bbb F_{mn} = -[\nabla_m,\nabla_n]$.
The second line of (\ref{cors}) in turn follows from
$[\nabla_m , \Bbb F^{m|pq}]=[\nabla_m,[\nabla^{[p} , [\nabla^{q]} ,\nabla^m]]]$ combined
with the super Yang--Mills equation and additional Jacobi relations.},
\begin{align}
\label{cors}
\big[ \nabla_m, \Bbb W^{m\alpha}\big]&=  {1\over 2} \big[\Bbb F_{mn} , (\g^{mn} \Bbb W)^\alpha\big]\, , \\
\big[\nabla_p , \Bbb F^{p|mn}\big] &= 2 \big[\Bbb F^{mp} , \Bbb F_p{}^n\big]
+ 2 \big\{(\Bbb W^{[m} \g^{n]})_\a , \Bbb W^{\a} \big\}
\notag
\end{align}
of the higher-dimension superfields $\bW^{m\a}:=[\nabla^m,\bW^\a]$ and
$\Bbb F^{p|mn}:= [\nabla^p,\Bbb F^{mn}]$ from \eqref{highmass}.
Upon inserting (\ref{Diraceq}) and \eqref{cors} into \eqref{boxone}, one gets \cite{Gauge}:
\begin{align}
\Box \Bbb A_{\a} &=
\big[ \Bbb A_m ,[\p^m ,\Bbb A_\a]\big] + \big[ (\gamma^m \Bbb W)_\a, \Bbb A_m \big] \, ,\label{nonwave}\\
\Box \Bbb A^m &=
 \big[ \Bbb A_p , [\p^p, \Bbb A^m]  \big]  + \big[ \Bbb F^{mp}, \Bbb A_p \big]
 +\gamma^m_{\a \beta} \{ \Bbb W^\a, \Bbb W^\beta \} \, ,\notag\\
\Box \Bbb W^{\a} &=\big[ \Bbb A_m, [ \p^m, \Bbb W^\a ]  \big] + \big[ \Bbb A^m, \Bbb W^\a_m  \big]
+ {1\over 2} \big[\Bbb F_{mn}, (\gamma^{mn} \Bbb W)^\a \big] \, ,\notag\\
\Box \Bbb F^{mn} &= [\Bbb A_p, [\p^p ,\Bbb F^{mn}]] + [\Bbb A_p,  \Bbb F^{p|mn}]
+ 2 [\Bbb F^{mp}, \Bbb F_p{}^n ] + 2 \{ (\Bbb W^{[m} \g^{n]})_\a \,, \Bbb W^\a \}\,,\notag
\end{align}
with the convention $A^{[m}B^{n]}= A^m B^n - A^n B^m$.
We will see below that these equations
are the precursors of supersymmetric Berends--Giele recursion relations. In
particular, the bosonic restriction of the equation for $\Box\Bbb A^m$ will give rise
to a derivation of the standard Berends--Giele currents of \cite{BerendsME}.

\subsubsection{\label{perturbinersec}Perturbiner solution}

To solve the wave equations \eqref{nonwave}, it is convenient to use the perturbiner method of
Rosly and Selivanov \cite{selivanovI,selivanovII,selivanovIII,selivanovIV} by expanding the superfields
$\Bbb K\in\{\Bbb A_\alpha,\Bbb A^m, \Bbb W^\alpha, \Bbb F^{mn} \}$
as a series with respect to the generators $t^{i_j}$ of a Lie algebra, summed
over all possible non-empty words $P=p_1p_2\ldots p_{|P|}$ \cite{SYMBG}
(note $t^P:= t^{p_1}t^{p_2}\cdots t^{p_{\len{P}}}$)
\begin{align}\label{series}
\Bbb K  &:=
\sum_P \cK_P t^P e^{k_P\cdot X}  = \sum_{i_1} \cK_{i_1}t^{i_1} e^{k_{i_1}\cdot X} + \sum_{i_1,i_2} \cK_{i_1i_2}t^{i_1}t^{i_2} e^{k_{i_1 i_2}\cdot X} + \cdots \\ 
&\phantom{:}= \sum_{p=1}^{\infty} \sum_{i_1,i_2,\dots,i_p} {1\over p}\ \cK_{i_1 i_2 \dots i_p}
[t^{i_1},[t^{i_2},\dots,[t^{i_{p-2}} ,[t^{i_{p-1}},t^{i_p}] ]\dots ] ] e^{k_{i_1 i_2\ldots i_p}\cdot X}  \,,\notag
\end{align}
with coefficients given by $\cK_P$, which will be identified with the Berends--Giele currents
shortly. The second line follows from the shuffle symmetry \eqref{alternal} obeyed by the Berends--Giele currents
and guarantees that
$\Bbb K$ is Lie-algebra valued, see \cite{Ree} for a proof. Note that we are implicitly
considering the generators $t^i$ to be formally nilpotent\footnote{In the original perturbiner
discussion of \cite{selivanovI}, repeated indices are avoided by adjoining nilpotent symbols
${\cal E}^i$ to each $t^i$ in the expansion (\ref{series}).}
$t^i \ldots t^i = 0$ in order to avoid
repetition of indices like in $\cK_{112}$ or $\cK_{121}$.

In order to derive recursion relations for the expansion coefficients 
$\cK_P\in\{\cA^P_\alpha,\cA^m_P,\cW^\alpha_P, \cF^{mn}_P \}$, we
insert the series \eqref{series} into \eqref{nonwave} and use the
action of Box operator $\Box e^{k_P\cdot X} = 2 s_P e^{k_P\cdot X}$
on the plane-wave factors of the superfields. By isolating the coefficient
of $t^P$ in the wave equations, one readily finds that 
\beq\label{BGdef}
\cK_P = {1\over s_{P}}\sum_{XY=P}\cK_{[X,Y]}\,,
\eeq
where the contribution from each deconcatenation of $P$ into non-empty $X,Y$ is
a non-local version of (\ref{Lorenzdef})
\begin{align}
\cA^{[P,Q]}_\a &=  \half\bigl[  \cA^{Q}_\a (k_{Q}\cdot  \cA_P)
+  \cA_{Q}^m (\g_m \cW_P)_\a - (P\leftrightarrow Q)\bigr]\,,  \label{cAalpha}\\
\cA_{[P,Q]}^m &=  \half\bigl[ \cA_{Q}^m (k_Q\cdot\cA_{P}) + \cA^{P}_n\cF_Q^{nm}
+ (\cW_{P}\g^m \cW_Q) - (P \leftrightarrow Q)\bigr]\,,\notag\\
\cW_{[P,Q]}^\a &=  {1\over 4} \cF_{P}^{rs} (\g_{rs} \cW_Q)^\a  
+ \half \cW_{Q}^\a (k_Q\cdot \cA_P) + \half \cW_Q^{m\a} \cA^m_P
- (P \leftrightarrow Q) \bigr]\,,\notag\\
\cF^{mn}_{[P,Q]} &=  {1\over 2} \big[ \cF^{mn}_Q (k_Q \cdot {\cal A}_P)
+ \cF_{Q}^{p|mn} \cA_p^P +  \cF_Q^{[m}{}_r \cF_{P }^{n]r}
- 2 \g^{[m}_{\a\b} \cW_P^{n]\a} {\cal W}_Q^\beta
- (P \leftrightarrow Q) \big]\,.\notag
\end{align}
The definition $\Bbb F^{mn} = - [\nabla^m , \nabla^n]$ and those of higher-mass-dimension superfields 
lead to the following Berends--Giele currents
\begin{align}
\cF_P^{mn} &= k_P^m  {\cal A}^{n}_P -  k_P^n  {\cal A}^{m}_P 
- \sum_{XY=P} \!\! \bigl(  {\cal A}_{X}^{m} {\cal A}^{n}_Y -{\cal A}_{Y}^{m} {\cal A}^{n}_X\bigr)\, ,
\notag \\
\cW_P^{m \alpha} &=  k_P^{m} {\cal W}_{P}^{\alpha}+ \sum_{XY=P}\!\! \bigl(  {\cal W}_{X}^{\alpha} {\cal A}^{m}_Y
-{\cal W}_{Y}^{\alpha} {\cal A}^{m}_X\bigr) \, , \label{cWm} \\
\cF_P^{m |pq} &=  k_P^{m} {\cal F}_{P}^{pq}+ \sum_{XY=P}\!\! \bigl(  {\cal F}_{X}^{pq} {\cal A}^{m}_Y
-{\cal F}_{Y}^{pq} {\cal A}^{m}_X\bigr)\,,\notag
\end{align}
and the above recursion terminates with the single-particle superfields ${\cal K}_i = K_i \in 
 \{ A^{i}_\alpha, A^m_{i} , W_{i} ^\alpha, F_{i} ^{mn} \}$.
By comparing the expressions in \eqref{BGexpl} with the
first few explicit expansions from \eqref{cAalpha}
it is possible to recognize these expansions
as the Berends--Giele currents obtained previously using BRST cohomology arguments.

\subsubsection{\label{BGEOMsec}Equations of motion of Berends--Giele currents}

By inserting the perturbiner expansions \eqref{series} of the SYM superfields in $\Bbb K$
into their non-linear equations of motion \eqref{solvePert},
one immediately obtains the equations of motion
of the Berends--Giele currents in the form
\begin{align}
D_{\alpha} \cA_{\b}^P + D_{\b} \cA_{\a}^P &=
\g^m_{\alpha \b} \cA_m^P
+ \! \! \sum_{XY=P} \! \! \bigl( \cA_\alpha^X \cA_{\b}^Y
-\cA_\alpha^Y \cA_{\b}^X\bigr)\, , \label{BGEOM}\\
D_\alpha \cA_m^P &= k^P_m \cA_{\alpha}^P + (\g_m \cW_P)_\alpha + \!\! \!
\sum_{XY=P} \! \! \bigl( \cA_\alpha^X \cA_{m}^Y -\cA_\alpha^Y \cA_{m}^X\bigr)\, ,
\notag\\
D_{\alpha} \cW^\b_P &= {1\over 4} (\g^{mn})_{\alpha}{}^{\b} {\cal F}^P_{mn}
+ \! \! \sum_{XY=P}  \! \!\bigl( \cA_\alpha^X \cW_Y^{\b} -\cA_\alpha^Y \cW_X^\b\bigr)\, ,
\notag\\
D_{\alpha} {\cal F}^{mn}_P  &= (\cW^{[m}_P \g^{n]} )_\alpha
+\!\! \sum_{XY=P} \!\!\bigl( \cA_\a^X {\cal F}_Y^{mn} -\cA_\alpha^Y {\cal F}_X^{mn}\bigr)\notag
\end{align}
by comparing coefficients of the products of gauge generators $t^P$ on both sides.
Apart from the deconcatenation sum $\sum_{XY=P}$, these
equations of motion have the same form as the linearized ones \eqref{RankOneEOM}.
For example the two- and three-particle equations of motion
of $\cA_\a^{12}$ and $\cA_\a^{123}$ read
\begin{align}
D_{\a} {\cal A}_{\b}^{12} + D_{\b} {\cal A}_{\a}^{12} &= \g^m_{\alpha \b} {\cal A}_m^{12}
+ {\cal A}_\alpha^1 {\cal A}_{\b}^{2} -{\cal A}_\alpha^{2} {\cal A}_{\b}^1\, ,
\label{simpler}
\\
D_{\a} {\cal A}_{\b}^{123} + D_{\b} {\cal A}_{\a}^{123} &=
\g^m_{\alpha \b} {\cal A}_m^{123}
+ {\cal A}_\alpha^1 {\cal A}_{\b}^{23}
+ {\cal A}_\alpha^{12} {\cal A}_{\b}^3
-{\cal A}_\alpha^{23} {\cal A}_{\b}^1
-{\cal A}_\alpha^3 {\cal A}_{\b}^{12}\,.
\notag
\end{align}
These equations lead to a simple proof of the deconcatenation property \eqref{treefourteen}, 
based on the action of the pure spinor BRST charge on superfields via $Q=\l^\a D_\a$. 
Therefore, multiplying the first equation of \eqref{BGEOM} by $\l^\a\l^\b$ and
using the pure spinor constraint $\l^\a\l^\b \g^m_{\a\b}=0$ together with
anti-commutativity of the superfields, one recovers the variation (\ref{treefourteen}). 

As we will review later,
these simple equations of motion for $\cK_P$ play a key role in various proofs of
BRST invariance  of scattering  amplitudes
in string  and field theory, see \cite{towardsFT, nptFT, nptStringI}  for examples  at
tree level and \cite{1loopbb, GreenBZA,  towardsOne, towardsTwo, oneloopIII, DHoker:2020prr} at loop level. The need for superfields that represent multi-particle contact vertices on a skeleton graph 
was also observed in the
worldline version of the pure spinor formalism \cite{BjornssonWU, BjornssonWM}.

In addition, the Lorenz gauge as well as the Dirac and super Yang--Mills equations \cite{SYMBG}
\beq
\label{extra}
\big[\p_m,\bA^m\big] = 0\,,\quad \big[ \nabla_m , (\g^m \bW)_\alpha \big] = 0\,,\quad
\big[ \nabla_m, \bF^{mn} \big]=  \g^n_{\alpha \b} \big\{ \bW^\alpha, \bW^\b\big\}
\eeq
imply, after using $\nabla_m = \p_m - \bA_m$, that
the Berends--Giele currents satisfy
\begin{align}
k^P_m \cA_P^m &= 0 \, , \label{tmpt}\\
k_m^P (\g^m \cW_P)_\alpha & =\!\!\sum_{XY=P} \! \! \big[ \cA_m^X (\g^m \cW_Y)_\a
- \cA_m^Y (\g^m \cW_X)_\a\big]\, , \label{tmpr}\\
k_m^P \cF^{mn}_P &= \!\!\sum_{XY=P} \!\!\bigl[  2(\cW_X \g^n \cW_Y)
+ \cA_m^X \cF^{mn}_Y- \cA_m^Y \cF^{mn}_X\bigr]\,.\label{tmps}
\end{align}
While (\ref{tmpr}) and (\ref{tmps}) have local counterparts (\ref{diracloc}) in BCJ gauge,
the local multiparticle-superfields $A_P^m$ subject to generalized Jacobi identities
depart from Lorenz gauge and generically obey $k^P_mA_P^m\neq  0$.

\subsubsection{Symmetry properties of Berends--Giele currents}

The symmetry properties of the ${\cal K}_P$ can be inferred from their cubic-graph expansion and
can be summarized in terms of the shuffle product $\shuffle$
\beq
\label{alternal}
{\cal K}_{A\shuffle B} = 0\, ,\quad \forall \, A,B\neq\emptyset\,,
\eeq
see \eqref{shuffles} below for the proof.
For example,
\begin{align}
0&= {\cal K}_{1 \shuffle 2} = \cK_{12} + \cK_{21}\,,\label{cKshex}\\
0&={\cal K}_{1 \shuffle 23} =
 \cK_{123} + \cK_{213} + \cK_{231}\,,\notag\\
0&={\cal K}_{12 \shuffle 3} -{\cal K}_{1 \shuffle 32}  =
\cK_{123}  -  \cK_{321} \,.\notag
\end{align}
The shuffle symmetry \eqref{alternal}
was proved for the gluonic currents $J^m_P$ of Berends and Giele in \cite{BerendsZN}, while
a proof of $\cK_{A\shuffle B} = 0$ for their supersymmetric counterparts 
$\cK\in\{\cA_\a,\cA^m,\cW^\a,\cF^{mn}\}$
can be found in the appendix of \cite{Gauge}. 
Since the $\t$-independent component of the Berends--Giele
current of the vector connection reduces to the gluonic Berends--Giele current
after setting the fermionic polarizations to zero,
$\cA^m_P(\t=0)|_{\chi_j=0}=J^m_P$, the
supersymmetric proof of \cite{Gauge} yields an alternative proof of the shuffle symmetry of
the standard Berends--Giele current $J^m_P$.

The shuffle symmetry \eqref{alternal} implies that the Berends--Giele currents admit 
a $(p{-}1)!$-element basis of permutations ${\cal K}_{i_1\ldots i_p}$ of ${\cal K}_{12\ldots p}$
which can be taken as
${\cal K}_{1\sigma(23\ldots p)}$ with $\sigma \in S_{p-1}$ via Schocker's identity \cite{schocker}
\beq
{\cal K}_{B1A} = (-1)^{|B|} {\cal K}_{1(A\shuffle \tilde B)} \, ,
\label{KlKu}
\eeq
where $\tilde B$ denotes the word reversal of $B$, see section \ref{convIntrosec}.
In particular, for $A=\emptyset$ we get the alternating parity under reversal of $P$,
\beq\label{revM}
{\cal K}_P = (-1)^{\len{P}+1}{\cal K}_{\tilde P}\,,
\eeq
for example, $\cK_{12}=-\cK_{21}$ and $\cK_{123}=\cK_{321}$,
as can be seen from \eqref{cKshex}.

\subsubsection{Berends--Giele currents and finite gauge transformations}

It was shown in \cite{Gauge,genredef} that in terms of the perturbiner series of Berends--Giele currents $\Bbb K$,
the local superfield redefinitions reviewed in the previous section correspond to a finite gauge transformation
of the superfields $\Bbb K$ satisfying the non-linear field equations \eqref{SYMeom}. To see this, one can explicitly
check that the poles in the definition of the Berends--Giele current cancel the contact terms present in the
local redefinitions from Lorenz to BCJ gauge (this will be proven in \eqref{Cbdeconc}).
More explicitly, we first define a perturbiner series of the redefining
superfields as (recall e.g. $t^{123}:= t^1t^2t^3$ etc)
\beq\label{Hseries}
\Bbb H := \sum_P \cH_P t^P e^{k_P\cdot X} \,,
\eeq
as well as Lorenz $\Bbb K^{\rm L}$ and BCJ $\Bbb K^{\rm BCJ}$ perturbiner series in which the
numerators are composed of local superfields in the Lorenz or BCJ gauge, respectively. For example,
\beq\label{BGthree}
\cK^{\rm BCJ}_{123} =
{K_{[12,3]}\over s_{12}s_{123}}
+ {K_{[1,23]}\over s_{23}s_{123}},\qquad
\cK^{\rm L}_{123} =
{\hat K_{[[1,2],3]}\over s_{12}s_{123}}
+ {\hat K_{[1,[2,3]]}\over s_{23}s_{123}} \,,
\eeq
with $\hat K_{[P,Q]} =-\hat K_{[Q,P]}$ from \eqref{Lorenzdef}. The local
redefinitions from Lorenz to BCJ gauge of the vector superpotential
\beq\label{redef3}
A^{m}_{[12,3]} = \hat A^{m}_{[[1,2],3]} - k^m_{123}\hat H_{[12,3]}\,,\qquad
A^{m}_{[1,23]} = \hat A^{m}_{[1,[2,3]]} - k^m_{123}\hat H_{[1,23]}
\eeq
(with $\hat H_{[12,3]}$ defined by \eqref{HhatDef})
imply that their perturbiner series are related by
\beq\label{BCJLorentzThree}
\cA^{m,\rm BCJ}_{123} =\cA^{m,\rm L}_{123} - k^m_{123}\cH_{123}\, , \qquad
\cH_{123} =
{\hat H_{[12,3]}\over s_{12}s_{123}}
+ {\hat H_{[1,23]}\over s_{23}s_{123}}\,,
\eeq
corresponding to the gauge transformation
\beq\label{toBCJ}
\Bbb A_m^{{\rm BCJ}} =\Bbb A_m^{{\rm L}} - [\partial_m,\Bbb H]
+ [\Bbb A_m^{{\rm L}}, \Bbb H] + \cdots \ .
\eeq
The ellipsis indicates additional terms of a finite gauge transformation (see below) that do not 
contribute to (\ref{BCJLorentzThree}) since $\cH_{1}=\cH_{12}=0$ at multiplicities one and two 
vanish identically. In fact, the calculations of \cite{genredef} using superfields up to multiplicity nine
revealed that the relation between the Lorenz  and BCJ gauges is given by a finite
gauge transformation 
\beq\label{finitegau}
\Bbb A_m^{\rm BCJ} = U\Bbb A_m^{\rm L} U^{-1} + \p_mU U^{-1}\,,\quad U =
\exp(-\Bbb H)
\eeq
whose expansion yields the omitted terms in \eqref{toBCJ} as an infinite series
\beq\label{finGauge}
\Bbb A_m^{{\rm BCJ}} =
\Bbb A_m^{{\rm L}}
+ [\Bbb H,\p_m] - [\Bbb H, \Bbb A_m^{{\rm L}}]
- \half [\Bbb H, [\Bbb H,\p_m]] + \half [\Bbb H,[\Bbb H, \Bbb A^{\rm L}_m]]
+ {1\over 3!} [\Bbb H,[\Bbb H, [\Bbb H,\p_m]]]
+ \cdots \, .
\eeq
Following \cite{SchubertFinite}, one can obtain the series \eqref{finGauge} iteratively. To see
this, define \cite{ElliotThesis}
\beq\label{Lpert}
\bL_j(\bA_m)= \bA_m - {1\over j}[\p_m,\bH] - {1\over j}[\bH,\bL_{j+1}(\bA_m)]
\eeq
and evaluate
\beq\label{LAm}
\bA_m^{\rm BCJ} = \bL_1(\bA^{\rm L}_m)\,.
\eeq
The fact that it is the gauge transformation \eqref{finGauge} that relates the superfields
$\bA_m^{\rm BCJ}$ and $\bA_m^{\rm L}$ justifies the terminology  of their corresponding local superfields
as being in the Lorenz ($\hat K_{[P,Q]}$) or BCJ gauge ($K_{[P,Q]}$).

\subsubsection{The multiparticle Berends--Giele polarizations}

In section~\ref{thetaexpsec} we have seen that the linearized superfields admit a
$\t$-expansion where each component depends on 
single-particle polarizations $e^m_i, \chi^\a_i$ and
field-strengths, $f^{mn}_i$. 
In principle, the recursive construction of multiparticle Berends--Giele currents
at the superspace level also determines the coefficients in their $\t$-expansion
in terms of single-particle polarizations. However, the tensor structure of the 
$\t$-expansion (\ref{linTHEX}) in the single-particle case is not preserved under 
the Lorenz-gauge recursion \eqref{BGdef} and (\ref{cAalpha}): generic orders in the
$\t$-expansion of multiparticle ${\cal K}_{P}$ in Lorenz gauge will receive multiple
contributions from different partitions of the $\theta$s over the lower-multiplicity 
superfields in (\ref{cAalpha}).

A notable exception arises at the zeroth order in $\theta$, where the
Lorenz-gauge recursions in superspace have an immediate echo 
at the level of components: the multiparticle polarizations
$\ce^m_P, {\cal X}^\a_P, \cf_{P}^{mn}$ defined by
setting $\theta=0$ in
\begin{equation}\label{zerocomp}
\ce^{m}_P := \cA^m_P(0) \ , \ \ \ \ \ \ {\cal X}_P^\a := \cW^\a_P(0) \ , \ \ \ \ \ \
\cf^{mn}_P := \cF_P^{mn}(0) \, ,
\ee
obey the following recursions as a consequence of \eqref{BGdef} and (\ref{cAalpha})
(with $\ce^m_i := e^m_i$ and ${\cal X}^\a_i:= \chi^\a_i$ for single-particle labels),
\begin{equation}\label{reczero}
\ce^m_P = {1 \over s_P} \sum_{XY=P} \ce^{m}_{[X,Y]} \ ,
\quad {\cal X}^\a_P = {1 \over s_P} \sum_{XY=P} {\cal X}^{\a}_{[X,Y]} \ ,
\ee
where 
\begin{align}
\ce^{m}_{[X,Y]}  &:=  {1 \over 2 }  \bigl[ \ce_{Y}^m (k_Y\cdot  \ce_{X})
+ \ce^{X}_n  \cf_Y^{nm}
+ ( {\cal X}_{X}\g^m {\cal X}_Y)
- (X \leftrightarrow Y)\bigr] \, ,\label{recone}\\
{\cal X}^{\a}_{[X,Y]} &:= 
{1 \over 2 } ( k^p_{X} + k^p_{Y} ) \gamma_p^{\a \b}
\big[ \ce_X^m ( \g_m {\cal X}_Y)_\b - \ce_Y^m (\g_m {\cal X}_X)_\b\big]  \ .\label{rectwo}
\end{align}
Moreover, the non-linear component field-strength is given by
\be\label{recthree}
\cf^{mn}_P := k_P^m \ce_P^n - k_P^n \ce_P^m
- \sum_{XY=P}\!\!\big( \ce_X^m \ce_Y^n - \ce_X^n \ce_Y^m \big)\, .
\ee
Note that the transversality $(k_i \cdot e_i) = 0$ of the gluon and the Dirac equation
$k^i_m (\g^m \chi_i)_\a=0$ of the gluino propagate as follows to the multiparticle level,
\begin{equation}\label{eoms}
(k_P \cdot \ce_P) = 0 \ , \qquad k^P_m (\g^m {\cal X}_P)_\a
=
\sum_{XY=P}\big[ \ce_X^m (\g_m \cX_Y)_\a -  \ce_Y^m (\g_m \cX_X)_\a \big]\, ,
\ee
where transversality of $\ce_P^m$ is a peculiarity of the Lorenz gauge \eqref{Lorenz} chosen in
the derivation of the corresponding superspace Berends--Giele current
$\cA^m_P(\t)$.

In Lorenz gauge, the above $\ce^m_P, {\cal X}^\a_P, \cf_{P}^{mn}$ are insufficient to
describe higher orders $\sim \theta^m$ of multiparticle ${\cal K}_P$ with $1\leq m\leq 5$ which
complicates the component expansions via (\ref{tlct}). However, one can streamline
these $\theta$-expansions by means of non-linear gauge transformation (\ref{NLgauge})
with a perturbiner expansion of both the superfields $\mathbb K$ and the gauge parameter
$\Omega$. As detailed in \ref{HSapp}, the non-linear version $\theta^\alpha
 \Bbb A_\a^{\rm HS}=0$ of Harnad--Shnider gauge reorganizes the $\theta$-expansion
 of the ${\cal K}_P$ to simple combinations of the $\ce^m_P, {\cal X}^\a_P, \cf_{P}^{mn}$.
In particular, the orders $\theta^{\leq 3}$ of ${\cal A}_\alpha^P$ relevant for $n$-point
tree-level amplitudes take the same form as in the single-particle $\theta$-expansion (\ref{linTHEX}),
see (\ref{thetaEXP}) below, which dramatically simplifies the component expansions
in section \ref{sec:compSYM}.
 
One can similarly arrange the $\theta$-expansions of local multiparticle superfields in Lorenz
or BCJ gauge such that the components relevant to tree-level amplitudes
are built from three types of multiparticle polarizations. In case of BCJ gauge,
the construction of the superfields $A^m_P,W^\alpha_P,F^{mn}_P$ in section \ref{sfinbcj}
determines the local multiparticle polarizations
\beq
e^{m}_P := A^m_P(0) \ , \ \ \ \ \ \ \chi_P^\a := W^\alpha_P(0) \ , \ \ \ \ \ \
f^{mn}_P := F_P^{mn}(0)
\label{zerolocal}
\eeq
via evaluation at $\theta =0$, for instance
\begin{align}
e_{12}^m &= e_2^m (e_1\cdot k_2) - e_1^m (e_2\cdot k_1) 
+ \frac{1}{2}(k_1^m {-}k_2^m)(e_1\cdot e_2) + (\chi_1 \gamma^m \chi_2) \, , \notag \\
\chi_{12}^\alpha &= \frac{1}{2} k_{12}^p \gamma_p^{\alpha \beta} 
\big[ e_1^m (\gamma_m \chi_2)_\alpha  - e_2^m (\gamma_m \chi_1)_\alpha  \big] \, , \label{explloc}
\\
f_{12}^{mn} &= k_{12}^m e_{12}^n - k_{12}^n e_{12}^m - (k_1\cdot k_2)(e_1^m e_2^n - e_1^n e_2^m)
\, . \notag
\end{align}
The local multiparticle polarizations (\ref{zerolocal}) obey generalized Jacobi identities
in $P$ by construction and compactly encode the components of the local BCJ
numerators to be reviewed in section \ref{sec:6.4.3}. Note that transversality
of the multiparticle polarizations $e_P^m$ at $|P|\geq 3$ is violated in BCJ gauge, e.g.
\beq
k_{123}^m e_{123}^m = s_{123} \bigg( \frac{1}{6} e_1^m e_2^n f_3^{mn} - \frac{1}{3} (\chi_1 \gamma_m \chi_2) e^m_3  + {\rm cyc}(1,2,3) \bigg) \, .
\label{noLorenz}
\eeq
Further details on local multiparticle polarizations can be found in section 4.3 of \cite{BGBCJ}.

\subsection{Combinatorial framework of Berends--Giele currents}
\label{BGcombsec}

The definition of Berends--Giele currents encompassing all the Catalan
number of poles in a color-ordered tree-level amplitude suggests a combinatorial
interpretation in terms of planar binary trees. We will see that
this point of view provides a rich mathematical framework to prove
many assertions related to Berends--Giele currents and associated topics \cite{flas}.

\subsubsection{Planar binary trees}

In the appendix of \cite{EOMBBs} a construction of Berends--Giele currents exploited the
fact that nested Lie brackets can be interpreted as planar binary trees and vice versa
\cite{garsia}. A {\it planar binary tree} is a tree embedded in a plane in which each vertex has
three edges: one {\it root} and two (left and right) {\it daughters}. An edge is called a {\it leaf} if
it has an end point. In the context of tree-level amplitudes a
planar binary tree is also called a {\it cubic graph} and we
map each
planar binary tree to a product of inverse Mandelstam invariants (the Feynman propagators) and
nested Lie brackets. In
addition, each leaf is indexed by a particle label and planarity implies that the labels are in a
fixed ordering. For example
the two planar binary trees with three leaves labelled $1,2,3$ are mapped to
\begin{center}
\includegraphics[width=0.4\textwidth]{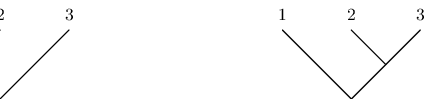}
\end{center}
\noindent
It turns out that the sum over all possible bracketings, or cubic graphs, in a color-ordered tree-level amplitude
can be generated from the Lie-polynomial valued recursion proposed in \cite{PScomb} (inspired by \cite{garsia})
\beq\label{bMap}
 b(P) := {1\over s_P}\sum_{XY=P}[b(X),b(Y)]
 \, ,\qquad b(i) := i
 \, ,\qquad b(\emptyset) := 0
 \,.
\eeq
This recursion constructs combinations $b(P)$ of non-commutative words
with inverses of Mandelstam invariants $s_P$ in (\ref{defmands}) as their coefficients,
i.e.\ the right-hand side of $b(i) = i$ is not understood as $i\in \mathbb N$, 
but as a letter in a non-commutative word.
From well-known combinatorial results, the number of terms in the
recursion above is given by the Catalan numbers
$1,2,5,14, \ldots$\footnote{This can for instance be seen from the
recursion $C_{p-1}=\sum_{x+y=p-2}C_x C_y$ for the number
of terms in $b(12\ldots p)$ with $C_0=C_1=1$ and $p\geq 3$. As detailed in
the discussion around (\ref{poles}) below, this coincides with the recursive
definition of the Catalan numbers.} and one gets, for example, the following Lie polynomials
\begin{align}
b(1) &= 1,\quad b(12) = {[1,2]\over s_{12}},
\quad b(123) = {[[1,2],3]\over s_{12}s_{123}}  + {[1,[2,3]]\over
s_{23}s_{123}} \, ,
\label{bexamp} \\
b(1234) &= {[ [ [ 1 , 2 ] , 3 ] , 4 ] \over s_{12} s_{123} s_{1234}}
+  {[ [ 1 , [ 2 , 3 ] ] , 4 ] \over s_{123} s_{1234} s_{23}}
+  {[ [ 1 , 2 ] , [ 3 , 4 ] ] \over s_{12} s_{1234} s_{34}}
+  {[ 1 , [ [ 2 , 3 ] , 4 ] ] \over s_{1234} s_{23} s_{234}}
+  {[ 1 , [ 2 , [ 3 , 4 ] ] ] \over s_{1234} s_{234} s_{34}}\,. \notag
\end{align}
The nested commutators in the numerators can be expanded in terms of formal
words in letters $12\ldots$, and the diagrammatic representation of $b(1234)$ 
can be found in figure~\ref{figBGfour}.

\begin{figure}[t]
\begin{center}
\includegraphics[width=0.9\textwidth]{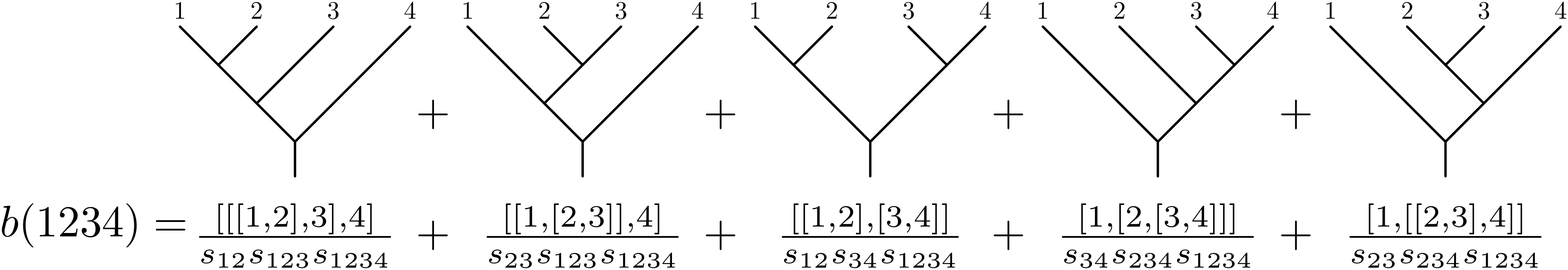}
\end{center}
\caption{The planar binary trees generated by the recursion of $b(1234)$ from \eqref{bMap}.}
\label{figBGfour}
\end{figure}

\begin{lemma} The $b$ map \eqref{bMap} is self adjoint,
\beq\label{bself}
\langle b(P), Q\rangle = \langle P, b(Q)\rangle\,,
\eeq
where $\langle A,B\rangle=\d_{A,B}$ is the scalar product of words \eqref{AdotB}.
\end{lemma}
\noindent{\it Proof.} This is easy to see when $P=i$ is a single letter with $b(i) =i$, so we will
use induction over the length the word $\len{P}:=k$ assuming that \eqref{bself}
is true for words of length up to $k{-}1$. Then,
from the definition \eqref{bMap}, the left-hand side of \eqref{bself} becomes
\beq\label{decsay}
\langle b(P), Q\rangle = {1\over s_P}\sum_{XY=P}\langle b(X)b(Y), Q\rangle - (X\leftrightarrow Y)\, .
\eeq
Using the elementary property (see (1.5.12) in \cite{Reutenauer})
\beq\label{elemen}
\langle AB, RS\rangle = \langle A,R\rangle\langle B,S\rangle\, ,\quad \len{A}{=}\len{R}\, , \ \ \len{B}{=}\len{S}\, ,
\eeq
and noting that $\len{b(X)}{=}\len{X}$ and $\len{P}{=}\len{Q}$ we get
\begin{align}
\label{indbs}
\langle b(X)b(Y),Q\rangle &= \langle b(X), q_{1}q_2 \ldots q_{\len{X}}\rangle \langle
b(Y),q_{\len{X}+1}q_{\len{X}+2} \ldots q_{\len{Q}}\rangle \\
&= \langle X, b(q_{1}q_2 \ldots q_{\len{X}})\rangle \langle Y,b(q_{\len{X}+1}q_{\len{X}+2} \ldots
q_{\len{Q}})\rangle \notag\\
&= \langle XY, b(q_{1}q_2 \ldots q_{\len{X}})b(q_{\len{X}+1}q_{\len{X}+2} \ldots q_{\len{Q}})\rangle \, ,\notag
\end{align}
where in the second line we used the induction hypothesis since $\len{X}\le k{-}1$ as the
deconcatenation \eqref{decsay} vanishes if one of the words is empty due to the definition
$b(\emptyset):=0$. Therefore,
\beq\label{upshotself}
\sum_{XY=P}\langle b(X)b(Y), Q\rangle = \sum_{XY=P}\langle P,b(q_{1}q_2 \ldots
q_{\len{X}})b(q_{\len{X}+1}q_{\len{X}+2} \ldots q_{\len{Q}})\rangle=
\sum_{XY=Q}\langle P,b(X)b(Y)\rangle\, ,
\eeq
leading to the conclusion that $\langle b(P),Q\rangle{=}\langle P,b(Q)\rangle$, finishing the
proof. \qed

Assuming linearity $b(A+B):= b(A) + b(B)$, the expansion of $b(P)$ satisfies the shuffle symmetry
$b(A\shuffle B) = 0$ for $A,B\neq\emptyset$. We will prove this in two different ways.
\begin{prop.}
The planar binary tree expansion $b(P)$ in (\ref{bMap}) satisfies the shuffle symmetry
\beq\label{bshuffle}
b(A\shuffle B) = 0\,,\qquad \forall \ A,B\neq\emptyset\,.
\eeq
\end{prop.}
\noindent\textit{Proof 1.}
We will show this by induction on the length of the word in $b(P)$
starting from $b(1\shuffle 2) = b(12) + b(21)$, which is easy to verify.
Assume that $b(A\shuffle B) = 0$ for $\len{A}{+}\len{B}=k$, and consider $b(R\shuffle S)$ for
nonempty words such that $\len{R}{+}\len{S}=k{+}1$.
The result will follow from the word identity \eqref{shuho}, the antisymmetric
nature of the deconcatenation in the definition of the $b$ map \eqref{bMap}, and
the induction hypothesis.
That is,
\begin{align}
s_{RS}b(R\shuffle S) &= \sum_{XY=R\shuffle S}[b(X),b(Y)] \notag\\
&=[b(\emptyset),b(R\shuffle S)] + [b(R\shuffle S),b(\emptyset)]
+ [b(R),b(S)] + [b(S),b(R)]\\
&\quad + \psum_{XY=R}\psum_{ZW=S} [b(X\shuffle Y), b(Z\shuffle W)]\, , \notag
\end{align}
where we used the identity \eqref{shuho} to expand the deconcatenation sum in the second line.
The second line vanishes by the antisymmetry of the Lie bracket, while the third line vanishes
since $\len{X}{+}\len{Y}{=}\len{R} \leq k$ with nonempty $X,Y$ implies, by the induction
hypothesis, $b(X\shuffle Y){=}0$. Therefore
$b(P{\shuffle} R){=}0$ for $P,R\neq\emptyset$.\qed

\noindent{\it Proof 2.} Recall
Ree's theorem \cite{Ree} that a Lie polynomial $\Gamma$
is orthogonal to shuffles with non-empty words
(see Theorem 3.1 (iv) in \cite{Reutenauer})
\beq\label{reestheo}
\langle \Gamma, R\shuffle S\rangle = 0\, , \quad R,S\neq\emptyset \, .
\eeq
Since $b(P)$ is a Lie polynomial by the definition \eqref{bMap} and $b$ is self-adjoint by
\eqref{bself}, we have
\beq\label{borth}
0=\langle b(P), R\shuffle S\rangle = \langle P, b(R\shuffle S)\rangle\, , \quad R,S\neq\emptyset\, ,
\quad\forall P\, ,
\eeq
and the result follows.\qed

\subsubsection{Berends--Giele currents from planar binary trees}

Having the planar binary tree recursion \eqref{bMap}, one can define
Berends--Giele currents in BCJ gauge $\cK_P$ or Lorenz gauge $\hat \cK_P$ as
\beq\label{BGasbMap}
\cK_P = K_{b(P)}\, , \quad \hat\cK_P = \hat K_{b(P)}\,,
\eeq
where $K_{b(P)}$ and $\hat K_{b(P)}$ are defined by linearity. 
We are here departing
from the notation in section \ref{sec:42NL}, where Berends--Giele
current in Lorenz gauge were denoted by $\cK_P$ or $\cK_P^{\rm L}$
and those in BCJ gauge by $\cK_P^{\rm BCJ}$.
For example, with $K_{[P,Q]} =
A^m_{[P,Q]}$ we get $\cA^m_1 = A^m_1$ and
\begin{align}
\cA^m_{12} &= {A^m_{[1,2]}\over s_{12}}\,,\label{BGAms}\\
\cA^m_{123} &= {A^m_{[ [ 1 , 2 ] , 3 ]} \over s_{12} s_{123}}
+  {A^m_{[ 1 , [ 2 , 3 ] ]} \over s_{123} s_{23}}\,,\notag\\
\cA^m_{1234} &=
{A^m_{[ [ [ 1 , 2 ] , 3 ] , 4 ]} \over s_{12} s_{123} s_{1234}}
+ {A^m_{[ [ 1 , [ 2 , 3 ] ] , 4 ]} \over s_{123} s_{1234} s_{23}}
+  {A^m_{[ [ 1 , 2 ] , [ 3 , 4 ] ]} \over s_{12} s_{1234} s_{34}}
+  {A^m_{[ 1 , [ [ 2 , 3 ] , 4 ] ]} \over s_{1234} s_{23} s_{234}}
+  {A^m_{[ 1 , [ 2 , [ 3 , 4 ] ] ]} \over s_{1234} s_{234} s_{34}}\,,\notag\\
\cA^m_{12345} &=
{A^m_{[ [ [ [ 1 , 2 ] , 3 ] , 4 ] , 5 ]} \over s_{12} s_{123} s_{1234} s_{12345}}
 +  {A^m_{[ [ [ 1 , [ 2 , 3 ] ] , 4 ] , 5 ]} \over s_{123} s_{1234} s_{12345} s_{23}}
 +  {A^m_{[ [ [ 1 , 2 ] , [ 3 , 4 ] ] , 5 ]} \over s_{12} s_{1234} s_{12345} s_{34}}
 +  {A^m_{[ [ [ 1 , 2 ] , 3 ] , [ 4 , 5 ] ]} \over s_{12} s_{123} s_{12345}  s_{45}}\notag\\
&\quad +  {A^m_{[ [ 1 , [ [ 2 , 3 ] , 4 ] ] , 5 ]} \over s_{1234} s_{12345} s_{23} s_{234}}
 +  {A^m_{[ [ 1 , [ 2 , [ 3 , 4 ] ] ] , 5 ]} \over s_{1234} s_{12345} s_{234} s_{34}}
 +  {A^m_{[ [ 1 , [ 2 , 3 ] ] , [ 4 , 5 ] ]} \over s_{123} s_{12345} s_{23} s_{45}}
 +  {A^m_{[ [ 1 , 2 ] , [ [ 3 , 4 ] , 5 ] ]} \over s_{12} s_{12345} s_{34}
 s_{345}}\notag\\
&\quad +  {A^m_{[ [ 1 , 2 ] , [ 3 , [ 4 , 5 ] ] ]} \over s_{12} s_{12345} s_{345} s_{45}}
 +  {A^m_{[ 1 , [ [ [ 2 , 3 ] , 4 ] , 5 ] ]} \over s_{12345} s_{23} s_{234} s_{2345}}
 +  {A^m_{[ 1 , [ [ 2 , [ 3 , 4 ] ] , 5 ] ]} \over s_{12345} s_{234} s_{2345} s_{34}}
 +  {A^m_{[ 1 , [ [ 2 , 3 ] , [ 4 , 5 ] ] ]} \over s_{12345} s_{23} s_{2345}
 s_{45}}\notag\\
&\quad +  {A^m_{[ 1 , [ 2 , [ [ 3 , 4 ] , 5 ] ] ]} \over s_{12345} s_{2345} s_{34} s_{345}}
 +  {A^m_{[ 1 , [ 2 , [ 3 , [ 4 , 5 ] ] ] ]} \over s_{12345} s_{2345} s_{345} s_{45}}\,.\notag
\end{align}
Since these expansions will be frequently used later we also write the
expansions of the Berends--Giele currents
\beq\label{MPfrombP}
M_P:=V_{b(P)}
\eeq
using this algorithm to get \eqref{bexamp}
\begin{align}
M_1 & = V_1 \, ,\quad
M_{12} = {V_{[1,2]}\over s_{12}}\,,\quad
M_{123} = {V_{[ [ 1 , 2 ] , 3 ]} \over s_{12} s_{123}}
+  {V_{[ 1 , [ 2 , 3 ] ]} \over s_{123} s_{23}}\,,\label{BGVs}\\
M_{1234} &=
{V_{[ [ [ 1 , 2 ] , 3 ] , 4 ]} \over s_{12} s_{123} s_{1234}}
+ {V_{[ [ 1 , [ 2 , 3 ] ] , 4 ]} \over s_{123} s_{1234} s_{23}}
+  {V_{[ [ 1 , 2 ] , [ 3 , 4 ] ]} \over s_{12} s_{1234} s_{34}}
+  {V_{[ 1 , [ [ 2 , 3 ] , 4 ] ]} \over s_{1234} s_{23} s_{234}}
+  {V_{[ 1 , [ 2 , [ 3 , 4 ] ] ]} \over s_{1234} s_{234} s_{34}}\,,\notag\\
M_{12345} &=
{V_{[ [ [ [ 1 , 2 ] , 3 ] , 4 ] , 5 ]} \over s_{12} s_{123} s_{1234} s_{12345}}
 +  {V_{[ [ [ 1 , [ 2 , 3 ] ] , 4 ] , 5 ]} \over s_{123} s_{1234} s_{12345} s_{23}}
 +  {V_{[ [ [ 1 , 2 ] , [ 3 , 4 ] ] , 5 ]} \over s_{12} s_{1234} s_{12345} s_{34}}
 +  {V_{[ [ [ 1 , 2 ] , 3 ] , [ 4 , 5 ] ]} \over s_{12} s_{123} s_{12345}  s_{45}}\notag\\
&\quad +  {V_{[ [ 1 , [ [ 2 , 3 ] , 4 ] ] , 5 ]} \over s_{1234} s_{12345} s_{23} s_{234}}
 +  {V_{[ [ 1 , [ 2 , [ 3 , 4 ] ] ] , 5 ]} \over s_{1234} s_{12345} s_{234} s_{34}}
 +  {V_{[ [ 1 , [ 2 , 3 ] ] , [ 4 , 5 ] ]} \over s_{123} s_{12345} s_{23} s_{45}}
 +  {V_{[ [ 1 , 2 ] , [ [ 3 , 4 ] , 5 ] ]} \over s_{12} s_{12345} s_{34}
 s_{345}}\notag\\
&\quad +  {V_{[ [ 1 , 2 ] , [ 3 , [ 4 , 5 ] ] ]} \over s_{12} s_{12345} s_{345} s_{45}}
 +  {V_{[ 1 , [ [ [ 2 , 3 ] , 4 ] , 5 ] ]} \over s_{12345} s_{23} s_{234} s_{2345}}
 +  {V_{[ 1 , [ [ 2 , [ 3 , 4 ] ] , 5 ] ]} \over s_{12345} s_{234} s_{2345} s_{34}}
 +  {V_{[ 1 , [ [ 2 , 3 ] , [ 4 , 5 ] ] ]} \over s_{12345} s_{23} s_{2345}
 s_{45}}\notag\\
&\quad +  {V_{[ 1 , [ 2 , [ [ 3 , 4 ] , 5 ] ] ]} \over s_{12345} s_{2345} s_{34} s_{345}}
 +  {V_{[ 1 , [ 2 , [ 3 , [ 4 , 5 ] ] ] ]} \over s_{12345} s_{2345} s_{345} s_{45}}\,.\notag
\end{align}
After using Baker's identity \eqref{Baker} to expand the nested brackets in
the basis of $\ell(1Q)$
and adopting the notation \eqref{BCJnotation}, e.g.,
$V_{[[ [ 1 , 2 ] , 3 ],4 ]}
= V_{1234}$, these examples reproduce the Berends--Giele expansions of $\cK_P\rightarrow \cA_P^m, M_P$
given before in (\ref{BGexpl}) and \eqref{BGexplM}.

The proof of \eqref{bshuffle} shows that any antisymmetric deconcatenation will
satisfy the shuffle symmetry, as this is a property obeyed by the underlying words.
Hence, the Berends--Giele supercurrents, defined by their antisymmetric recursion \eqref{BGdef},
and the BRST-closed superfield $E_P$, defined by \eqref{Edef}, both satisfy the shuffle symmetry.
We therefore obtain the following corollary:
\begin{cor}\label{corMEshuffle}
The Berends--Giele supercurrents $\cK_P$ \eqref{calK} and the BRST-closed superfield $E_P$ \eqref{Edef} satisfy
\beq\label{shuffles}
\cK_{R\shuffle S} = E_{R\shuffle S} = 0\,,\quad \forall \, R,S\neq\emptyset \, .
\eeq
\end{cor}

\subsubsection{\label{Sbracketsec}The S bracket}

BCJ relations for SYM amplitudes were expressed in \cite{EOMBBs,BGBCJ} using the so-called
\textit{S map} defined in \cite{EOMBBs} by its action on Berends--Giele currents. The properties
of this map provided the motivation for a more general definition in \cite{PScomb, flas} as a bracket
$\{\cdot,\cdot\}$, dubbed the \textit{$S$ bracket},
acting on words in the dual space of Lie polynomials $\cL^*$ and producing words in the dual space
$\cL^*$, i.e. $\{ \cdot , \cdot \}: \cL^*\otimes \cL^*\to\cL^*$. For our purposes, this space is defined
by the equivalence classes of words differing by proper shuffles, i.e.,
\beq\label{dualdef}
\hbox{$A\sim B\ $ if $\ A=B + \sum R\shuffle S \ $ with $\ R,S\neq\emptyset\, $.}
\eeq
For instance $\{1,2\} \sim - \{2,1\}$ because $\{1,2\}=
s_{12}12= 
- s_{12}21 + s_{12}1\shuffle 2 \sim - \{2,1\}$. See the \ref{dualLiesec} for more
information.

There are several equivalent definitions of the $S$ bracket \cite{EOMBBs,PScomb,flas}.
A recursive definition for letters $i$, $j$ and words $A$, $B$ was given in \cite{flas} as
\begin{align}
\{iAj,B\}&=i\{Aj,B\}-j\{iA,B\} \, ,\notag\\
\{B,iAj\}&=\{B,iA\}j-\{B,Aj\}i \, ,\label{smapdef}\\
\{i,j\} &= s_{ij}\,ij\,. \notag
\end{align}
Example applications are given by
\begin{align}
\{1,2\} &=s_{12}12 \, ,  \label{curlydefex}\\
\{1,23\} &= s_{12} 123 - s_{13} 132 \, ,\cr
\{12,3\} &= s_{23} 123 - s_{13} 213 \, ,\cr
\{1,234\}&= s_{12} 1234 - s_{13} 1324 - s_{13} 1342 + s_{14} 1432 \, ,\cr
\{123,4\}&= s_{34} 1234 - s_{24} 1324 - s_{24} 3124 + s_{14} 3214 \, ,\cr
\{12,34\} &= s_{23}1234 - s_{24}1243 - s_{13} 2134 + s_{14}2143 \, .\notag
\end{align}
We note that the original definition of the $S$ bracket in \cite{EOMBBs} is given in terms of a
closed formula
\beq\label{PQgen}
\{P,Q\} = \sum_{XiY=P\atop RjS = Q} k_i\cdot k_j(X\shuffle \tilde Y)ij(\tilde R\shuffle
S)(-1)^{\len{Y}+\len{R}} \, ,
\eeq
which, in particular, yields the following for one-letter words $P\rightarrow i$
\beq\label{isingle}
\{i,Q\} = \sum_{RjS=Q} k_i\cdot k_j ij(\tilde R\shuffle S)(-1)^\len{R}\,.
\eeq
Several properties of the $S$ bracket were proven in \cite{flas}:
\begin{prop.}
The $S$ bracket satisfies:
\begin{enumerate}[label=\roman*.]
\item $\{A\shuffle B,C\}=0$ for $A,B\neq \emptyset$
\item $\{\cdot, \cdot\}$ is a Lie bracket in the space of dual Lie polynomials \eqref{dualdef}
\item The binary tree map $b$ of \eqref{bMap} acting on the $S$ bracket satisfies \cite{PScomb}
\beq\label{bigclaim}
b(\{P,Q\}) = [b(P),b(Q)]\,.
\eeq
\item $\sum_{XY=P}\{X,Y\}\sim s_P P$
\end{enumerate}
\end{prop.}
The proofs of these statements can be found in \cite{flas}, we will restrict ourselves to showcasing
some examples.
As a simple illustration of the shuffle property, we consider:
\beq
\{1\shuffle 2,3\} = \{12,3\}+\{21,3\} = (s_{23}123 - s_{13}213) + (s_{13}213 - s_{23}123) = 0\, ,
\eeq
this is consistent with the fact that the $S$ bracket operates on dual words \eqref{dualdef}, where
$A\shuffle B\sim 0$.
For the Lie-bracket property we explicitly verify the simplest cases,
antisymmetry for two letters and the Jacobi identity:
\begin{align}\label{liedex}
\{1,2\}+\{2,1\}&=s_{12}12 + s_{21}21= s_{12}(1\shuffle2) \sim 0\,,\\
\{\{1,2\},3\}+\{\{2,3\},1\}+\{\{3,1\},2\} &=
 s_{12}s_{13} (
 	  3 \shuffle12
          - 2 \shuffle13
          )
       + s_{12}s_{23} (
          23 \shuffle1
          - 3 \shuffle21
          )\notag\\
&\quad       + s_{13}s_{23} (
          2 \shuffle31
          - 32 \shuffle1
          )
	\sim 0\,.\notag
\end{align}
An example for the third property is given by
\begin{align}
\label{illust}
b(\{12,3\}) &= s_{23}b(123) - s_{13}b(213)\notag \\
&= s_{23}\bigg( {[[1,2],3]\over s_{12}s_{123}}
+ {[1,[2,3]]\over s_{23}s_{123}}\bigg)
- s_{13}\bigg(
{[[2,1],3]\over s_{12}s_{123}}
+ {[2,[1,3]]\over s_{13}s_{123}}\bigg) \notag \\
& = {[[1,2],3]\over s_{12}} = [b(12),b(3)]\, ,
\end{align}
where we used the third example from \eqref{curlydefex},
$s_{12}+s_{13}+s_{23}=s_{123}$ and the Jacobi identity. Note that there is no pole $1/s_{123}$ in
the right-hand side of \eqref{illust}.
An illustration of the fourth property is the following
\begin{align}
\sum_{XY=123}\{X,Y\} &= \{1,23\}+\{12,3\} = s_{12}123-s_{13}132+s_{23}123 - s_{13}213\\
&= (s_{12}+s_{13}+s_{23})123 - s_{13}(2\shuffle 13)\sim s_{123}123\,.\notag
\end{align}
From the property \eqref{bigclaim} it is straightforward to conclude:
\begin{cor} There is no $1/s_{PQ}$ propagator in $b(\{P,Q\})$, that is \cite{flas}
\beq\label{BCJspq}
\lim_{s_{PQ}\to0} s_{PQ}b(\{P,Q\}) = 0\,.
\eeq
\end{cor}

\paragraph{BCJ amplitude relations}
We see that the $S$ bracket in $\{P,Q\}$ cancels the overall propagator $1/s_{PQ}$ from
the linear combinations in $b(\{P,Q\})$. From what we have seen in \eqref{QEexact}, this condition
implies that the superfield $E_{\{P,Q\}}$ is a BRST
exact expression, $E_{\{P,Q\}} = QM_{\{P,Q\}}$ as the divergent propagator $1/s_{PQ}$ (in the
context of an amplitude of $\len{P}{+}\len{Q}{+}1$ massless particles) is absent from $M_{\{P,Q\}}$.
This only happens when the numerators satisfy the Jacobi identity, and this fact
plays a key role in the proof of the BCJ amplitude relations using the cohomology of pure spinor
superspace, see the discussion in section~\ref{BCJampsec}. A result we will need later is the
following:
\begin{lemma} The $S$ bracket in the special case when one of the words is a letter admits
the form
\beq\label{fundiQ}
\{i,Q\} = \sum_{RS=Q}k_i\cdot k_S RiS\,.
\eeq
\end{lemma}
\noindent\textit{Proof.} We will show that the right-hand side of \eqref{fundiQ} is shuffle 
equivalent to the
expression \eqref{isingle}. Using the shuffle equivalence proven in \eqref{dualLie} we get
\begin{align}
\sum_{RS=Q}k_i\cdot k_S RiS &\sim \sum_{RS=Q}k_i\cdot k_S i(\tilde R\shuffle S)(-1)^\len{R}\notag\\
&\sim \sum_{RjkS=Q}k_i\cdot k_{kS} i(\widetilde{Rj}\shuffle kS)(-1)^{\len{R}+1}
\label{pfund} \\
&\sim\sum_{RjkS=Q}\big[k_i\cdot k_{kS}ij(\tilde R\shuffle kS)(-1)^{\len{R}+1}
+ k_i\cdot k_{kS}ik(\widetilde{Rj}\shuffle S)(-1)^{\len{R}+1}\big]\,, \notag
\end{align}
where to arrive at the second line we relabeled the summation variables as $R\to Rj$ and $S\to kS$, and
in the third line we used $(j\tilde R)\shuffle (kS) = j(\tilde R\shuffle kS) + k(j\tilde R\shuffle
S)$ from the definition of the shuffle product \eqref{shuffledef}. Now we relabel $kS\to S$ in the
first sum in the right-hand side of \eqref{pfund} and $Rj\to R$, $k\to j$ in the second one (so that
$(-1)^{\len{R}+1}\to(-1)^\len{R}$). This implies that
\begin{align}
\sum_{RS=Q}k_i\cdot k_S RiS &\sim \sum_{RjS=Q}-k_i\cdot k_S ij(\tilde R\shuffle S)(-1)^\len{R}
+ \sum_{RjS=Q}k_i\cdot k_{jS} ij(\tilde R\shuffle S)(-1)^\len{R}\\
&\sim \sum_{RjS=Q} k_i\cdot k_j ij(\tilde R\shuffle S)(-1)^\len{R}\,,\notag
\end{align}
which is the same expression as \eqref{isingle}, finishing the proof.\qed

\subsubsection{The contact-term map and the $S$ bracket}

The $S$ bracket is intimately related to the contact-term map discussed in section
\ref{contactsec}. In fact, the recursive definition of the contact-term
map \eqref{contactdef} admits an equivalent representation in terms of the $S$ bracket. If
$\Gamma$ is a Lie monomial, then \cite{flas}
\beq\label{contactAs}
\langle P\otimes Q, C(\Gamma)\rangle = \langle\{P,Q\},\Gamma\rangle\,,
\eeq
where the scalar product of words $\langle A,B\rangle $ takes values 1 for $A=B$ 
and 0 for $A\neq B$, see \eqref{AdotB}. This means that the adjoint $C^*$ of the contact-term map is the
$S$ bracket; $C^*(A\otimes B) = \{A,B\}$. Exploiting this interpretation allows one to prove that
$C^2 = 0$ as stated earlier in \eqref{Csquared}, see \ref{Cnilap}. 

Important properties of the contact-term map relevant to the description of SYM in terms of
local and non-local multiparticle superfields were proven in \cite{genredef,flas}.
For instance, the contact-term map deconcatenates the planar binary tree map involving pole
cancellations in a highly non-trivial manner:
\begin{lemma} The contact-term map \eqref{contactAs} satisfies
\beq\label{Cbdeconc}
C\big(b(P)\big) = \sum_{XY=P} \big(b(X)\otimes b(Y) - b(Y)\otimes b(X)\big):=\sum_{XY=P}b(X)\wedge b(Y)\,,
\eeq
where $A\wedge B= A\otimes B - B\otimes A$.
\end{lemma}
\noindent\textit{Proof.} From the characterization \eqref{contactAs} as the adjoint of the S
bracket we obtain
\begin{align}
\langle R\otimes S, C(b(P))\rangle &= \langle \{R,S\}, b(P)\rangle = \langle b(\{R,S\}),P\rangle\\
&= \langle [b(R),b(S)], P\rangle\notag=\langle b(R)b(S), P\rangle - (R\leftrightarrow S)\notag\\
& = \sum_{XY=P}\langle b(R),X\rangle\langle b(S),Y\rangle - (R\leftrightarrow S)\notag\\
& = \sum_{XY=P}\langle R,b(X)\rangle\langle S,b(Y)\rangle - (X\leftrightarrow Y)\notag\\
&= \sum_{XY=P}\langle R\otimes S, \big(b(X)\otimes b(Y) - (X\leftrightarrow Y)\big)\rangle\notag\,,
\end{align}
where in the first line we used that $b$ is self adjoint \eqref{bself}, in the second we used
the property \eqref{bigclaim}, and \eqref{elemen} in the third. In the fourth line we used the
self-adjoint property of the $b$ map again and finally the definition $\langle A\otimes B,
R\otimes S\rangle = \langle A,R\rangle \langle B,S\rangle$ in the last line. Since $R,S$ are
arbitrary, the result follows.\qed

For example, the simple identity $C([1,2]) = s_{12}1\wedge 2$ leads to the
deconcatenation of $b(12)={[1,2]\over s_{12}}$ (note $b(\emptyset) := 0$)
\beq\label{twob}
C\big(b(12)\big) = b(1)\wedge b(2)
= \sum_{XY=12} b(X)\wedge b(Y)\,.
\eeq
However, it is already non-trivial to explicitly check using $C([[1,2],3])$ and $C([1,[2,3]])$ given
in \eqref{Cexamples} that $C(b(123))=b(12)\wedge b(3) + b(1)\wedge b(23)$ with $b(123)$ given in
\eqref{bexamp}.

In \cite{flas,hadleigh}, the definition
\eqref{contactAs} was used to show that the $S$ bracket is in fact a \textit{Lie cobracket} as
defined in the context of Lie coalgebras \cite{Michaelis}.

\subsubsection{The KLT map}

The KLT relation was derived in \cite{KLTpaper} as a way to express
the closed-string tree-level amplitude as a sum over products of color-ordered open-string tree
amplitudes, see section \ref{sec:6.5.2} for a brief review. 
In the field-theory limit $\alpha'\rightarrow 0$, it readily implies the same type of squaring
relations between $n$-point supergravity amplitudes $M_n^{\rm grav}$ and color-ordered
SYM amplitudes $A(\ldots)$. In a modern
language, the field-theory KLT relation can be written as
\beq\label{KLTrel}
M_n^{\rm grav} = - \sum_{P,Q} A(1,P,n,n{-}1)S(P|Q)_1 \tilde A(1,Q,n{-}1,n)\, ,
\eeq
where  the $ \tilde A(\ldots)$ feature polarizations $\tilde e_i,\tilde \chi_i$ independent 
from the $ e_i, \chi_i$ in $A(\ldots)$ and $S(P|Q)_1$ is the {\it KLT matrix} or 
the {\it momentum kernel} indexed by permutations $P,Q \in S_{n-3}$ of legs 
$2,3,\ldots,n{-}2$. In a series
of papers \cite{KLTbern,KLTmatrI,KLTmatrII,dufu} the algorithm to obtain the precise 
form of the KLT matrix was sequentially simplified to 
a recursive definition \cite{Carrasco:2016ldy}
\beq\label{kltrec}
S(Aj|BjC)_i = k_j\cdot k_{iB}\, S(A|BC)_i \, ,\qquad S(\emptyset|\emptyset)_i := 1 \, ,
\eeq
where $i$ is some fixed leg, conventionally chosen $i=1$. For example,
\begin{align}\label{S0examp}
S(2|2)_1 &= k_1\cdot k_2\, ,& S(23|23)_1 &= (k_3\cdot k_{12} )(k_1\cdot k_2)\, ,\\
S(23|32)_1 = S(32|23)_1 &= (k_1\cdot k_2)( k_1\cdot k_3)\, ,& S(32|32)_1 &= (k_2\cdot k_{13} )(k_1\cdot k_3)\,.\notag
\end{align}
In the framework of twisted deRham 
theory, the entries of the inverse KLT matrix have been interpreted as intersection 
numbers \cite{Mizera:2017cqs, Mizera:2017rqa, Mizera:2019gea}, see section \ref{sec:6.5.5}
for further details.

In recent years, the KLT matrix has been found in various relations
involving the computation of string scattering amplitudes.
These relations can often be understood from a combinatorial/free-Lie-algebra perspective,
usually intimately related to planar binary trees.
For instance, we will see that the expression relating
local multiparticle superfields $V_P$ in the BCJ gauge and
Berends--Giele currents in \eqref{VSM} descends from the free-Lie-algebra relation \eqref{ellSb} below.
In addition, the KLT matrix also plays a major role in the integration-by-parts identities
used in the derivation of a closed formula for the massless $n$-point superstring disk amplitude in
section \ref{IBPsec}. Moreover, the KLT matrix is the inverse of the Berends--Giele double currents from
which the tree-level amplitudes of the bi-adjoint scalar theory are calculated. These in turn are
related to the field-theory limit of the superstring disk integrals which, as we will see in
section \ref{subsec:biadj}, admit a combinatorial interpretation.
In summary, the KLT matrix is indeed a central player
connecting various objects participating in the calculation of string scattering amplitudes.

\paragraph{Generalized KLT matrix} In the pursuit of a combinatorial framework for understanding the standard KLT matrix and its
relations to multiparticle superfields, a {\it
generalized KLT matrix} has been proposed in \cite{PScomb} and analyzed further in
\cite{flas,hadleigh} (see also \cite{frostconf}). To see this more precisely,
one defines a map that converts every Lie bracket $[\cdot, \cdot]$
in an arbitrary Lie monomial $\Gamma$ to a $S$ bracket $\{\cdot, \cdot \}$. This conversion is
denoted by~$\{\Gamma\}$
\beq\label{Lie2Dual}
\{\Gamma\}: [\cdot,,\cdot] \to \{\cdot, \cdot\}\,,
\eeq
and acts recursively in commutator depth,
transforming a Lie polynomial to a dual Lie polynomial (see
\ref{wordsapp} for the definitions). For example,
\begin{align}\label{sbracks}
\{[1,2]\} &= \{1,2\} = s_{12}12\, ,\\
\{[[1,2],3]\} &= \{\{1,2\},3\} = s_{12}\big(s_{23}123 - s_{13}213\big)\, ,\notag\\
\{[1,[2,3]]\} &= \{1,\{2,3\}\} = s_{23}\big(s_{12}123 - s_{13}132\big)\, ,\notag\\
\{[[1,2],[3,4]]\} &= \{\{1,2\},\{3,4\}\} = s_{12}s_{34}\big(
s_{23}1234 - s_{24}1243 - s_{13} 2134 + s_{14}2143
\big)\, .\notag
\end{align}
The KLT map is defined as a map between planar binary trees $\Gamma$ and its $S$ bracket version
\beq\label{KLTmap}
S:\Gamma\to\{\Gamma\}\,.
\eeq
A graphical illustration of the KLT map is given in the figure~\ref{KLTmapfig}.
\begin{figure}[t]
\begin{center}
\includegraphics[width=0.8\textwidth]{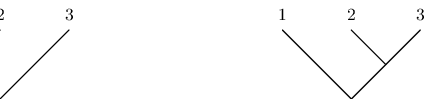}
\end{center}
\caption{Examples of the KLT map \eqref{KLTmap}, where $\{\cdot,\cdot\}$ is the $S$ bracket
\eqref{smapdef}. Each planar binary tree is mapped to the expressions given in \eqref{sbracks}.}
\label{KLTmapfig}
\end{figure}

The matrix elements of the KLT map with respect to a basis of Lie monomials $\Gamma, \Sigma$ are given
by $S(\Gamma,\Sigma) = \langle \{\Gamma \},\Sigma\rangle$ which motivate the definition of the {\it
generalized KLT matrix} for words $P$ and $Q$ \cite{PScomb}
\beq\label{genKLT}
S^\ell(P|Q) = \langle \ell\{P\},\ell(Q)\rangle\, ,
\eeq
where the dual Dynkin bracket $\ell\{P\}$ is defined as the conversion \eqref{Lie2Dual} of the
Dynkin bracket \eqref{ellmap},
\beq\label{ellS}
\ell\{P\}:=\{\ell(P)\}
\eeq
with $\ell(P)$ defined in \eqref{ellmap} and $\{\Gamma\}$ defined
in \eqref{sbracks}. Alternatively, a recursive definition is given by
\beq\label{ellsrec}
\ell\{123 \ldots n\} = \{\ell(123 \ldots n{-}1),n\}\, ,\qquad \ell\{i\} = i\,,\qquad \ell\{\emptyset\} = 0\,.
\eeq
The simplest examples of \eqref{genKLT}
include $S^\ell(12|12) = s_{12}$, and $S^\ell(12|21) = -s_{12}$, as well as
\begin{align}
S^\ell(123|123) &= s_{12}(s_{13}+s_{23})\, , & S^\ell(1234|1234) &= s_{12}(s_{13}+s_{23})(s_{14}+s_{24}+s_{34})\, ,\notag\\
S^\ell(132|123) &= s_{12}s_{13}\, , & S^\ell(1243|1234) &=s_{12}(s_{13}+s_{23})(s_{14}+s_{24})\, ,\notag\\
S^\ell(312|123) &= - s_{12}s_{13}\, , &S^\ell(3412|1234) &=-s_{12}s_{13}s_{34}\, ,\label{genkExs}\\
S^\ell(231|123) &= -s_{12}s_{23}\, , &S^\ell(3421|1234) &=s_{12}s_{23}s_{34}\,,\notag
\end{align}
and it is shown in \cite{flas} that $S^\ell(P|Q)$ is a symmetric matrix, $S^\ell(P|Q)=S^\ell(Q|P)$,
see the reference for further details. One of the main attributes of the above generalized KLT 
matrix $S^\ell(P|Q)$ is that it satisfies generalized Jacobi identities in both $P$ and $Q$,
\beq\label{genjacklt}
S^\ell(A\ell(B)|Q) + S^\ell(B\ell(A)|Q) = 0\,,\quad \forall \ Q\,.
\eeq
For example, with $A=12$ and $B=34$
\beq\label{genKLTjac}
S^\ell(1234|Q) - S^\ell(1243|Q) + S^\ell(3412|Q) - S^\ell(3421|Q) = 0\,,\quad \forall \ Q\,.
\eeq
Moreover,
the standard KLT matrix recursion \eqref{kltrec} is obtained as the special
case when the first letters in both words coincide \cite{PScomb,flas},
\beq\label{genstd}
S^\ell(iP|iQ)= S(P|Q)_i\, .
\eeq
The unrestricted nature of the first letters in the permutations of
the matrix \eqref{genKLT} is the motivation for the qualifier {\it generalized}.

An important interplay between the Lie-bracket conversion \eqref{Lie2Dual} from the space
of Lie polynomials to the space of dual Lie polynomials is summarized in:
\begin{lemma}
The planar binary tree map \eqref{bMap} satisfies
\beq\label{bdual2lie}
b(\ell\{P\}) = \ell(P)\,,
\eeq
where the dual Dynkin bracket is given by \eqref{ellsrec} and the Dynkin bracket by \eqref{ellmap}.
In fact,
\beq\label{anyGamma}
b(\{\Gamma\}) = \Gamma
\eeq
for any Lie polynomial $\Gamma$.
\end{lemma}
\noindent\textit{Proof.} Using the property \eqref{bigclaim} together
with the recursive definition
\eqref{ellsrec} yields
\begin{align}
b(\ell\{123 \ldots n\}) &= b(\{\ell\{123 \ldots n{-}1\},b(n)\}) \notag \\
&= [b(\ell\{123 \ldots n{-}1\}),n] = \ldots  
\\
&= [[ \ldots [[1,2],3], \ldots],n] = \ell(123 \ldots n)\,,\notag
\end{align}
where we used that $b(j) = j$ for a letter $j$. Since $\ell(P)$ is a basis of Lie polynomials, the
result \eqref{anyGamma} follows by a basis expansion.\qed

In effect, as $\ell\{P\} = \{\ell(P)\}$ implies $\ell\{P\}=S(\ell(P))$ in terms of the KLT map
\eqref{KLTmap}, the result \eqref{anyGamma} is the $b\circ S = Id$ part of a more general statement proven
in \cite{flas}:
\begin{prop.} The planar binary tree map $b: \cL^*\to \cL$ defined in \eqref{bMap} and the KLT map
$S: \cL\to\cL^*$ defined in \eqref{KLTmap} are inverses to each other,
\begin{align}\label{SbInv}
b\circ S: \ &\cL \to \cL\,, & S\circ b: \ &\cL^* \to \cL^*\,,\\
&\Gamma\mapsto\Gamma\, , & &P^*\mapsto P^* \, .\notag
\end{align}
\end{prop.}

\paragraph{Correspondence between multiparticle superfields and free Lie algebra}
Over the years, it became clear that relations governing multiparticle superfields
discovered in pursuit of expressions for string amplitudes had
a combinatorial flavor of the type commonly studied within the free-Lie-algebra framework.
This is particularly true in the context of the color-kinematics duality \cite{BCJ}, where the
generalized Jacobi identities played a major role in the simple form of the general massless
disk amplitude of \cite{nptStringI}.

The correspondence suggested above can be made precise with the following mapping between 
free-Lie-algebra
structures on one side and multiparticle superfields in the pure spinor formalism on the
other\footnote{The generalizations
$\ell(P)\leftrightarrow K_P$ and $b(P)\leftrightarrow \cK_P$ are immediate, where $K_P$ and $\cK_P$
are defined in \eqref{KPdef} and \eqref{calK}, respectively. We chose the
representatives $V_P$ and $M_P$ for pedagogical reason.}:
\beq\label{flcorresp}
C\leftrightarrow Q_{\rm BRST}\, ,\quad
\ell(P)\leftrightarrow V_P\, ,\quad
b(P)\leftrightarrow M_P\,,
\eeq
where the Dynkin bracket $\ell(P)$ is defined in \eqref{elldef}.
That is, the contact-term map $C$ is identified with the BRST charge $Q$ as already hinted in \eqref{CQcorresp}, the
Lie monomials encoded in the Dynkin bracket $\ell(P)$ correspond to the multiparticle 
unintegrated vertex $V_{P}$ in the BCJ gauge, and the planar binary tree
expansion $b(P)$ corresponds to the Berends--Giele current $M_P$ as in \eqref{MPfrombP}.
As an immediate consistency check, note that
both $C$ and $Q$ are nilpotent, see \eqref{Csquared} and \eqref{Qsquared}. In addition, 
the symmetries on both sides agree: generalized Jacobi identities \eqref{genjac} for both $\ell(P)$ and
$V_P$ as well as shuffle symmetries for both the planar binary tree expansion $b(P)$ in
\eqref{bshuffle} and
the Berends--Giele currents $M_P$ in \eqref{shuffles} (or more generally ${\cal K}_P$).

For a more precise relation, the identities \eqref{deshufflesum} and \eqref{GeneralQ} illustrate the correspondence
between $V_P$ and $\ell(P)$ as well as the contact-term map $C$ and the BRST charge $Q$. For
example, note the parallels of these equations at multiplicity three:
\begin{align}
C(\ell(123)) & =  (k_1\cdot k_2)\big(\ell(1)\wedge \ell(23) + \ell(13)\wedge \ell(2)\big)
+ (k_{12}\cdot k_3)\ell(12)\wedge \ell(3)\, ,\\
QV_{123} &= (k_1\cdot k_2)\big( V_1V_{23} + V_{13}V_2\big) + (k_{12}\cdot k_3)V_{12}V_3\,,\notag
\end{align}
where the fermionic property $V_PV_Q=-V_QV_P$ is mapped to the antisymmetric wedge product $\wedge$ of
\eqref{PwedgeQ}.
Moreover, using the notation defined in \eqref{replace}, the BRST variation of the unintegrated
vertex operator $V_{\Gamma}$ for an arbitrary Lie monomial $\Gamma$ can be written as
\beq\label{QVPc}
QV_{\Gamma} = (V\wedge V)_{C(\Gamma)}\,,
\eeq
which is extended to arbitrary Lie polynomials by linearity.

The precise cancellations between the contact terms in the equations of motion for local superfields and
the Mandelstam propagators featured in the definition of Berends--Giele currents constituted an
early indication of a beautiful and rigorous underlying mathematical framework. See, for example,
the discussions in
\cite{PScomb} and several proofs in \cite{flas}. For instance,  the
proof that the Berends--Giele current $M_P$ deconcatenates under the action of the BRST charge over
its local numerators $V_Q$ easily follows in a free-Lie-algebra setting,
\begin{lemma}
The Berends--Giele current $M_P$ in the BCJ gauge satisfies $QM_P = \sum_{XY=P}M_XM_Y$.
\end{lemma}
\noindent\textit{Proof.}
The Berends--Giele current $M_P$ is given by an expansion of local multiparticle vertices $V_R$
encoded in terms of planar binary trees as
$M_P = V_{b(P)}$ given in \eqref{MPfrombP}.
The deconcatenation property of $C(b(P))$ in \eqref{Cbdeconc} implies
\beq\label{deconcproof}
QM_P = QV_{b(P)} = (V\wedge V)_{C(b(P))} =\sum_{XY=P}(V\wedge V)_{b(X)\wedge b(Y)}
= \sum_{XY=P}V_{b(X)}V_{b(Y)} = \sum_{XY=P}M_XM_Y\,,
\eeq
where we used the notation \eqref{replace}.\qed

This proof sheds light on the deconcatenation property \eqref{treefourteen} of the Berends--Giele current
from a different perspective compared with the equations of motion \eqref{BGEOM} derived from the
perturbiner expansion. The result arises from the use of the equations of motion of the local
multiparticle superfields yielding contact terms that cancel the propagators present in the
planar-binary-tree expansion, demonstrating that the patterns observed in \eqref{treethirteen}
hold to all orders.
The other equations of motion for the currents in $\cK_P$ can be derived from their
local counterparts in a similar fashion \cite{genredef}.

Finally, the relation between the local multiparticle superfields $V_P$ satisfying generalized
Jacobi identities and the non-local Berends--Giele supercurrents $M_P$ in BCJ gauge
satisfying shuffle symmetries follows from the identity proven in \cite{flas}:
\begin{lemma}
The Dynkin bracket \eqref{ellmap} and the planar-binary-tree expansion \eqref{bMap} are related
by
\beq\label{ellSb}
\ell(R) = \sum_Q S^\ell(R|iQ)b(iQ)\,,
\eeq
where $S^\ell(R|iQ)$ is the generalized KLT matrix \eqref{genKLT}.
\end{lemma}
\noindent{{\it Proof.}} Consider the dual Dynkin bracket $\ell\{R\}$. Since it is a dual Lie polynomial
it can be expanded in the Lyndon basis $iQ$ of the dual Lie polynomials using the formula \eqref{dualExpansion}
\beq\label{dualexp}
\ell\{R\} = \sum_{Q}\langle\ell\{R\},\ell(iQ)\rangle iQ\,.
\eeq
Acting with the $b$ map \eqref{bMap} on both sides gives $b(\ell\{R\}) = \sum_Q\langle \ell\{R\},\ell(iQ)\rangle
b(iQ)$. The left-hand side can be rewritten using \eqref{bdual2lie}, while the scalar product on the
right-hand side is the definition of the generalized KLT matrix. Thus, $\ell(R)=\sum_Q S^\ell(R|iQ)
b(iQ)$.\qed

After setting $R=iP$ and using the definition \eqref{genstd} the generalized KLT matrix reduces to
the usual matrix $S(P|Q)_i$ in (\ref{kltrec}). Replacing $\ell(iP)\to V_{iP}$ and $b(iQ) \to M_{iQ}$ as suggested by
the correspondence \eqref{flcorresp} leads to the relation between Berends--Giele currents and
multiparticle unintegrated vertex operators:
\beq\label{VSM}
V_{iP} = \sum_Q S(P|Q)_i M_{iQ}\,.
\eeq
This identity was first explicitly mentioned in \cite{Zfunctions}, but it had already played
an implicit role in the derivation of the closed formula of the massless $n$-point
open-superstring amplitude in \cite{nptStringI}, see section \ref{IBPsec}. The inverse of
\eqref{VSM} expressing $M_{iP}$ as a linear combination of $V_{iQ}$ will be given in \eqref{VtoM}.

The relation (\ref{VSM}) can be straightforwardly adapted to reproduce the local multiparticle
superfields (see section \ref{sfinbcj}) from their respective Berends--Giele currents in BCJ gauge,
\beq\label{localfromBG}
K_{iP} = \sum_Q S(P|Q)_i {\cal K}_{iQ}\, .
\eeq
However, plugging the Berends--Giele currents in Lorenz gauge into the right-hand side of 
(\ref{VSM}) and (\ref{localfromBG}) will lead to a non-local outcome of the $S(P|Q)_i$
multiplication.\footnote{This can be anticipated from the alternative form ${\cal H}_{123}
=  \frac{ H_{1,2,3}}{3s_{123}} ( \frac{1}{s_{12}} - \frac{1}{s_{23}})$ of the 
gauge transformation (\ref{BCJLorentzThree}) due to total antisymmetry of
 $\hat H_{[12,3]}= \frac{1}{3} H_{1,2,3}$. The KLT matrix renders the 
 difference between the Lorenz- and BCJ-gauge variants of (\ref{localfromBG})
 for ${\cal A}_\alpha^{123}$ and ${\cal A}^m_{123}$ proportional to
$S(23|23)_1{\cal H}_{123}+S(23|32)_1{\cal H}_{132} = \frac{ H_{1,2,3} }{3s_{123}}(s_{13}{+}s_{23}-2 s_{12})$
and therefore non-local.}

\section{\label{SYMsec} SYM tree amplitudes from the cohomology of pure spinor superspace}

In this section we are going to review how to obtain supersymmetric expressions
for SYM tree-level amplitudes. One could in principle start by 
computing the $n$-point superstring disk amplitudes and then take its $\ap\to0$ limit
\cite{scherkFT,neveuscherkFT,gsoSYMft1,gsoSYMft2}. However, the construction in this section relies 
entirely on pure spinor cohomology considerations \cite{nptFT}, following the ideas 
of \cite{towardsFT}, and predates the calculation of the $n$-point superstring disk 
amplitude in \cite{nptStringI,nptStringII}. The alternative derivation of SYM amplitudes
from the $\ap\to0$ limit of string amplitude was done later, see section~\ref{FTdisksec}.

Inspired by progress in organizing string amplitudes, it was realized in the 1980's 
that gauge-theory amplitude calculations simplify tremendously by considering
ordered gauge invariants depending only upon kinematics --
called color-ordered or color-stripped partial amplitudes \cite{Berends:1987cv,Mangano:1987xk}.  
The full color-dressed S-matrix elements could be obtained by summing over a
product of these color-ordered amplitudes with appropriate
color-weights, either somewhat redundantly in a trace basis, or more
efficiently in the Del Duca--Dixon--Maltoni basis of \cite{KKLance}. The
advantages in considering stripped or ordered partial amplitudes are
enormous; they grow exponentially rather than factorially in local
diagram contributions. Thus here we only consider
color-ordered amplitudes.

The pure spinor cohomology formula for $n$-point SYM tree-level amplitudes turns out
to be the supersymmetrization of the standard Berends--Giele recursion relations
\cite{BerendsME}, as one might have correctly suspected from the discussion of the Berends--Giele 
currents in the previous section. Therefore we will
first review the recursive method proposed by Berends and Giele to compute Yang--Mills amplitudes.

\subsection{Berends--Giele recursion relations}

In the 80s, Berends and Giele proposed a recursive method to compute color-ordered
gluon amplitudes at tree level with the formula \cite{BerendsME}
\begin{equation}\label{BGformula}
A^{\rm YM}(1,2, \ldots,p,p{+}1) = s_{12 \ldots p}J^m_{12 \ldots p}J^m_{p+1}\,.
\ee
The \textit{Berends--Giele currents} $J^m_P$ are
defined recursively in the number of external particles
starting with the polarization vector $e^m_i$ of a single-particle gluon,
by (note $J^m_\emptyset := 0$)
\begin{align}\label{BGrecur}
J^m_i := e_i^m\,,\qquad
s_P J^m_P &:= \sum_{XY=P} [J_X, J_Y]^m + \sum_{XYZ=P}\{J_X,J_Y,J_Z\}^m\,,\\
[J_X,J_Y]^m &:=  (k_Y\cdot J_X)J_Y^m + \half k_X^m (J_X\cdot J_Y) -
(X\leftrightarrow Y)\, , \notag\\
\{J_X,J_Y,J_Z\}^m & :=  (J_X\cdot J_Z)J_Y^m - \half(J_X\cdot J_Y)J_Z^m -\half
(J_Y\cdot J_Z)J_X^m \,, \notag
\end{align}
where
the brackets $[\cdot,\cdot]^m$ and $\{\cdot, \cdot, \cdot \}^m$ are given by stripping off
one gluon field (with vector index $m$) from the cubic and
quartic vertices of the Yang--Mills Lagrangian.
The deconcatenation of the word $P$ into non-empty
words $X$ and $Y$ is denoted by $\sum_{XY=P}$,
with obvious generalization to $\sum_{XYZ=P}$, and
$P=12\ldots p$ encompasses several external particles, see section
\ref{convIntrosec} for more details on the notation.
In addition, the Mandelstam invariants $s_P$ and multiparticle momenta $k^m_P$ are defined as in \eqref{defmands} and \eqref{defmultk}.

In \cite{BerendsME} the Berends--Giele currents $J^m_P$ were shown to be conserved
\beq\label{lorGauge}
k^m_P J^m_P = 0\,,
\eeq
which can alternatively be understood as the statement that the currents are in the Lorenz gauge \cite{Gauge}.

As the simplest example of the recursion \eqref{BGrecur}, the Berends--Giele
current of multiplicity two is,
\begin{equation}\label{rankTwoJ}
s_{12} J^m_{12} = e^m_2 (e_1\cdot k_2) - e^m_1 (e_2\cdot k_1) +
\half(k_1^m-k_2^m)(e_1\cdot e_2)\,,
\ee
and leads to the well-known three-point amplitude,
\begin{equation}\label{BGthreept}
A^{\rm YM}(1,2,3) = s_{12}J^m_{12}J^m_3 = (e_1\cdot e_2)(k_1\cdot e_3)+{\rm
cyc}(123)\,,
\ee
whose manifestly cyclic form is attained after using momentum conservation $k_1{+}k_2{+}k_3=0$ and
transversality $e_j \cdot k_j=0$.
Higher-point amplitudes are generated by a straightforward application of the recursion \eqref{BGrecur}.
The multiplicity-three current
\beq\label{mul3}
s_{123}J^m_{123} = [J_{12},J_3]^m + [J_1,J_{23}]^m + \{J_1,J_2,J_3\}^m
\eeq
gives rise to the four-point amplitude $A^{\rm YM}(1,2,3,4) = s_{123}J^m_{123}e_4^m$ and so forth. These
recursion relations are a very efficient method to calculate tree amplitudes numerically, see e.g.\ \cite{BGplefka,gielenumeric}. While the Berends--Giele formula \eqref{BGformula} is not
supersymmetric -- it computes purely gluonic amplitudes -- its supersymmetrization via
uplift to pure spinor superspace will be given below.

\subsubsection{\label{KKBGsec}Kleiss--Kuijf amplitude relations}

A crucial identity satisfied by the currents was also demonstrated by Berends and Giele in \cite{BerendsZN},
the shuffle symmetries:
\begin{equation}\label{BGsym}
J^m_{A\shuffle B} = 0, \quad \forall A,B\neq\emptyset\,.
\ee
Together with the amplitude formula \eqref{BGformula}, the shuffle identity \eqref{BGsym} can be
used to demonstrate that the Yang--Mills tree amplitudes satisfy the Kleiss--Kuijf (KK) relations 
 \cite{KKref}
\beq\label{KKrelation1}
\AYM(P,1,Q,n) = (-1)^{|P|}\AYM(1,\tilde P\shuffle Q,n)\, .
\eeq
To see this one exploits the mathematical literature of free Lie algebras \cite{Ree,schocker}.
More precisely, Ree's
theorem \cite{Ree} shows that a necessary and sufficient
condition for a series of the form $\sum_{n>0}J_{i_1i_2 \ldots i_n}X^{i_1}X^{i_2} \ldots X^{i_n}$ with non-commutative
indeterminates $X^i$ to be a Lie
polynomial is the shuffle symmetry (\ref{BGsym}) of its coefficients. In the 
context of Yang--Mills tree-level amplitudes, the $X^{i_j}$ are gauge-group
generators which have to conspire to Lie polynomials and therefore contracted
structure constants $f^{abc}$ by Yang--Mills Feynman rules.
Corollary 2.4 of \cite{schocker} in turn states that 
$\sum_{n>0}J_{i_1i_2 \ldots i_n}X^{i_1}X^{i_2} \ldots X^{i_n}$ is a Lie polynomial if and only~if
\beq\label{cor2.4}
J_{PiQ} = (-1)^{|P|}J_{i\tilde P\shuffle Q}\,.
\eeq
Since both results \eqref{BGsym} and \eqref{cor2.4} are ``if and only if'' statements, they must be
equivalent (for a proof of this, see \eqref{schockerReeagain2}). As a corollary of this equivalence together with
the Berends--Giele amplitude formula \eqref{BGformula} and the shuffle relation \eqref{BGsym}, it follows that the
KK relation \eqref{KKrelation1} must be satisfied. An alternative proof of the KK relations appears in \cite{KKLance}.

We are now going to recover the Berends--Giele recursion \eqref{BGrecur} and the amplitude formula \eqref{BGformula}
from the bosonic components of
a supersymmetric formula for SYM tree amplitudes derived from pure spinor cohomology considerations.

\subsection{\label{PSSYMsec} The pure spinor superspace formula for SYM tree amplitudes}

In this subsection we will review the derivation of the recursive method in pure spinor superspace
for the computation of supersymmetric tree amplitudes of ten-dimensional SYM theory. The method first appeared
in \cite{nptFT} and it relies
on the simple cohomology properties of multiparticle superfields in pure spinor superspace
as suggested earlier in \cite{towardsFT}. The end result is a method based on the recursive
nature of the BRST variations of the supersymmetric Berends--Giele currents \eqref{treethirteen}. 
When truncated to its bosonic components, the pure spinor formula was 
later shown in \cite{BGBCJ} to reproduce the standard Berends--Giele gluonic formula of \cite{BerendsME}. 

However, there are certain beneficial novelties in the pure spinor approach worth highlighting: 
\begin{itemize}
\item the direct
derivation of multiple currents for each type of superfield 
$\Bbb K \in \{\Bbb A_\a, \Bbb A_m, \Bbb W^\a, \Bbb F^{mn}\}$ 
using their non-linear equations of motion,
\item
the usage of only cubic interactions as a result of 
identifying the natural superfields in the recursion\footnote{See \cite{duhrcubic} for a reformulation of the purely 
gluonic Berends--Giele currents using cubic vertices.} 
-- the quartic interactions appear due to the quadratic
terms in the field strength,
\item the structural relation to planar binary trees and the construction using local numerators, 
\item the derivation of the shuffle symmetry of the currents using free-Lie-algebra methods, 
\item the identification of the different gauges associated to these local numerators and the subsequent
derivation of local BCJ-satisfying numerators.
\end{itemize}

\paragraph{Reasons for the cohomology method} In intermediate states of the calculations,
the prescription to compute disk amplitudes in the pure spinor formalism yields
a superspace expression containing three pure spinors and the superfields of ten-dimensional
SYM as well as their covariant derivatives; in other words they constitute pure spinor superspace expressions.
Superstring theory dictates that the field-theory SYM tree-level amplitudes must
be recovered in the $\ap\to0$ limit of the disk amplitudes. These, in turn, are obtained
from the tree-level amplitude prescription \eqref{treepresc}, which requires multiple OPEs among
the vertex operators of schematic form $V_1 U_2 \ldots U_{n-2}V_{n-1}V_n$, see
e.g.\ section 2.3 of \cite{Berkovits:2004px}.
Given that SYM tree-level amplitudes are supersymmetric and gauge invariant, we know
from \cite{psf} that their corresponding pure spinor expressions must be in the BRST cohomology. It
is important that the amplitudes are left written in \textit{pure spinor superspace} since
integrating out the pure spinors and the fermionic theta variables via (\ref{tlct}) would lead to a multitude
of terms in polarizations and momenta \cite{tsimpis} where all the simple superspace patterns are no longer present.

Let us first review the explicit superstring calculations for low multiplicities that led to the general
method for arbitrary multiplicities, using the latest conventions for the notations.

\paragraph{Explicit results and the birth of the cohomology method} 
At low multiplicities the SYM amplitudes written in pure spinor superspace were obtained from the field-theory 
limit of the corresponding superstring disk amplitudes:
the three-point case was known from the very start \cite{psf} while
the four- and five-point amplitudes were computed in pure spinor superspace in \cite{pureids,5ptsimple}:
\begin{align}
\label{3ptSYM}
A(1,2,3) &= \langle V_1 V_2 V_3\rangle \, , \\
A(1,2,3,4) &= {1\over s_{12}}\langle V_{12}V_3V_4\rangle + {1\over s_{23}}\langle
V_{1}V_{23}V_4\rangle \, , \notag\\
A(1,2,3,4,5) &=
  {\langle V_{123}V_{4}V_{5}\rangle\over s_{12} s_{45}}
+ {\langle V_{321}V_{4}V_{5}\rangle\over s_{23} s_{45}}
+ {\langle V_{12} V_{34}V_{5}\rangle\over s_{12} s_{34}}
+ {\langle V_1 V_{234}V_{5}\rangle\over s_{23} s_{51}}
+ {\langle V_1 V_{432}V_{5}\rangle\over s_{34}s_{51}} \,.\notag
\end{align}
Furthermore, using pure spinor cohomology arguments, expressions for the six- and seven-point SYM tree amplitudes
in pure spinor superspace were proposed in \cite{towardsFT}. The six-point amplitude was later 
reproduced
from the field-theory limit of the pure spinor superstring disk amplitude in \cite{6ptTree}\footnote{The 
prior computations for five and six external bosons in the RNS formalism were
performed in \cite{Medina:2002nk} and \cite{6ptOprisa}, respectively, with considerably longer expressions
in their final results.}.
In possession of these results a general pattern was discovered
in \cite{nptFT} in terms of the Berends--Giele currents written with multiparticle
versions of the unintegrated vertices $V_P$ in the BCJ gauge\footnote{It
was later understood in \cite{Gauge} that the construction of Berends--Giele currents does not 
require any particular gauge of the associated local superfields, so the requirement of BCJ gauge 
in \cite{nptFT} was stronger than necessary.}.

The clue was to notice the composing factors in the above amplitudes satisfied a regular pattern under BRST variation
(note $s_{45}=s_{123}$ at five points)
\begin{equation}\label{patts}
QV_1 = 0\, ,\qquad{}
Q{V_{12}\over s_{12}}  = V_1 V_2\,,\qquad{}
Q\Big({V_{123}\over s_{12}s_{123}} + {V_{321}\over s_{23}s_{123}}\Big) = V_1{V_{23}\over s_{23}} + {V_{12}\over s_{12}}V_3\,.
\end{equation}
These early computations
together with six-points examples not shown led to
the pattern of the Berends--Giele currents in \eqref{treethirteen}. In addition,
the Catalan numbers govern both the number of cubic graphs in a color-ordered tree amplitude
and the
number of kinematic poles in the Berends--Giele currents $M_P$ as derived in \eqref{poles}, so 
the assumption was that tree amplitudes would be composed of $M_P$.

Given that the SYM tree amplitudes are the $\ap\to0$ field-theory limit of the superstring
\cite{scherkFT,neveuscherkFT,gsoSYMft1,gsoSYMft2},
whose correlator in the pure spinor formalism is in the cohomology of the BRST charge,
the proposal of \cite{nptFT} was based on finding
a superfield expression in the
cohomology of the BRST charge that was constructed using the Berends--Giele currents $M_P$.
Rewriting the low-multiplicity examples \eqref{3ptSYM} as
\begin{align}
\label{BG3ptSYM}
A(1,2,3) &= \langle M_1 M_2 M_3\rangle \, ,\\
A(1,2,3,4) &= \langle M_{12}M_3M_4\rangle + \langle M_{1}M_{23}M_4\rangle\, ,\notag\\
A(1,2,3,4,5) &=
  \langle M_{123}M_{4}M_{5}\rangle
+ \langle M_{12} M_{34}M_{5}\rangle
+ \langle M_1 M_{234}M_{5}\rangle\notag
\end{align}
not only simplifies their presentation but also suggests the $n$-point generalization \cite{nptFT}
\beq\label{AYMgen}
A(1,2, \ldots,n) =  \sum_{j=1}^{n-2}\langle M_{12 \ldots j}M_{j{+}1 \ldots n{-}1} \, M_n \rangle\,.
\eeq
Given that the superspace current $M_{12\ldots j}$ is associated with
the cubic tree-level subdiagrams in a color-ordered $(j{+}1)$-point
amplitude, the sum in (\ref{AYMgen}) gathers cubic $n$-point diagrams
as shown in figure \ref{fignsym}.

\begin{figure}[h]
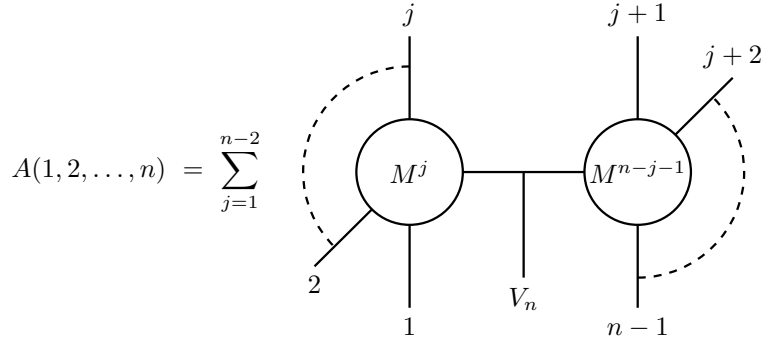

\centerline{
\tikzpicture [scale=1.0,line width=0.30mm]
\draw (0.4,0) node{$\displaystyle A(1,2,\ldots,n) \  = \ \sum_{j=1}^{n-2}$ };
\draw (4,0) -- (4,1.8) node[above]{$j$};
\draw (4,0) -- (4,-1.8) node[below]{$1$};
\draw (4,0) -- (2.75,-1.25) node[below]{$2$};
\draw[dashed] (4,1.4) arc (90:225:1.4cm);
\draw[fill=white] (4,0) circle (0.7cm);
\draw (4,0) node{$M^j$ };
\draw (4.7,0) -- (6.3,0);
\draw (5.5,0) -- (5.5,-1.4)node[below]{$V_n$};
\draw (7,0) -- (7,1.8) node[above]{$j+1$};
\draw (7,0) -- (8.25,1.25) node[above]{$j+2$};
\draw (7,0) -- (7,-1.8) node[below]{$n-1$};
\draw[dashed] (7,-1.4) arc (-90:45:1.4cm);
\draw[fill=white] (7,0) circle (0.7cm);
\draw (7,0) node{$M^{n-j-1}$ };
\endtikzpicture
}
\caption{Berends--Giele decomposition of the color ordered SYM amplitude according to (\ref{AYMgen}).}
\label{fignsym}
\end{figure}

More generally, the color-ordered field-theory tree amplitudes are given by
\beq\label{AYM}
A(P,n) = \langle E_{P} \, V_n \rangle
\eeq
in terms of the BRST-closed superfield $E_P$ in (\ref{Edef}).
This formula was later rigorously shown to match the $\ap\to0$ limit of the superstring amplitude 
in \cite{nptStringI,nptStringII} and it was also shown in \cite{BGBCJ} to reduce to the standard 
Berends--Giele gluonic formula \cite{BerendsME} reviewed in (\ref{BGformula}).
In spite of these validations, let us now give a separate proof that the expression \eqref{AYM} 
satisfies all the requirements of a color-ordered SYM tree amplitude.

\begin{prop.} In the momentum phase of $n$ massless states where $s_{12 \ldots n-1}=0$ the superfield
\beq\label{incoh}
E_{12 \ldots n-1}V_n
\eeq
is in the cohomology of the BRST charge.
\end{prop.}
\noindent\textit{Proof.} Since $QV_n=0$, to show that $E_{12 \ldots n-1}V_n$ is BRST closed
it is enough to show that $QE_{P}=0$,
\begin{align}
QE_P = \sum_{XY=P}Q(M_XM_Y) &= \sum_{XY=P}\sum_{RS=X}M_R M_S M_Y
- \sum_{XY=P}\sum_{RS=Y}M_X M_R M_S \notag \\
&= \sum_{RSY=P}\big(M_RM_SM_Y - M_RM_SM_Y\big) = 0\, ,
\end{align}
where in the last line we consolidated the sums and renamed the dummy words in the second term.

To show that \eqref{incoh} is not BRST exact
we note that the relation \eqref{Edef} depends crucially on the momentum phase space,
\begin{align}
E_{P} &= QM_{P}\,,\hbox{ if $s_P\neq0$}\label{QMP}\\
E_{P} &\neq QM_{P}\,,\hbox{ if $s_P=0$} \, .\label{notexact}
\end{align}
This is because
$M_{P} = {1\over s_{P}}\big( \ldots \big)$ contains a propagator $1/s_P$
which makes the left-hand side of \eqref{notexact} ill defined in case of $s_P=0$.
Hence, in the momentum phase space of $n$ massless particles where
$s_{12 \ldots n{-}1}=0$, the superfield
$E_{12 \ldots n-1}$ is not exact and
the expression $E_{12 \ldots n-1}V_n$ is in the cohomology of the BRST charge.\qed

\begin{prop.}
The number of kinematic poles from cubic graphs in the color-ordered $n$-point tree amplitude \eqref{AYM}
with $n\ge4$ is given by the Catalan number $C_{n-2}$.
\end{prop.}
\noindent\textit{Proof.}
The number of kinematic pole configurations in $E_{P}$ with $P$ of length $p\ge3$ and
$M_X$ of length $x\ge2$ are the Catalan
numbers $C_{p-1}$ and $C_{x-1}$, respectively\footnote{The difference in the initial lengths for $p$ and $x$
is related to the absence of the overall multiplicative pole $1/s_P$ present in $M_P$ but not in $E_P$, as
it is easy to verify.}.
To see this note that all poles on the right-hand side of
\eqref{Edef} are distinct, so it implies the recursion relation
\beq\label{poles}
C_{p-1} = \sum_{x+y=p{-}2} C_{x}C_{y}\, ,\quad C_0=C_1=1\,,\quad p\ge3\, .
\eeq
The recursion \eqref{poles} coincides with the recursive definition of the Catalan numbers
with explicit solution $C_n = {1\over n+1}{2n\choose n}$. Therefore the number of
poles in the $n$-point amplitude formula $\langle E_P V_n\rangle$ where $|P|=n{-}1$
is $C_{n-2}$.
This is the same number of cubic diagrams as
in the color ordered $n$-point SYM amplitude, see e.g.\ \cite{BCJ}. With $n=4,5,6,7$,
for instance, we get $C_{n-2}=2,5,14,42$.
And since the deconcatenation in the expression \eqref{Edef} for $E_P$ is ordered, the kinematic
poles in $E_{12 \ldots n{-}1}$ are the same as in the color-ordered $n$-point tree amplitude.\qed

\begin{prop.} The color-ordered $n$-point tree amplitude \eqref{AYM} is cyclically symmetric,
\beq\label{aymcyc}
A(1,2, \ldots,n{-}1,n) = A(2,3, \ldots,n,1)\, .
\eeq
\end{prop.}

\noindent\textit{Proof.}
By conveniently regrouping terms
of $Q M_{12\ldots j}  = \sum_{i=1}^{j-1} M_{12\ldots i} M_{i+1\ldots j}$, we can
recover the difference of cyclic images $E_{23 \ldots n}V_1$ 
and $ E_{12 \ldots n{-}1} V_n$ from
\begin{align}
Q\sum_{j=2}^{n-2} M_{12\ldots j} M_{j+1\ldots n}
&= M_1 \sum_{j=2}^{n-2} M_{2\ldots j} M_{j+1\ldots n} + \sum_{2\leq i<j}^{n-2} M_{12\ldots i}
M_{i+1\ldots j} M_{j+1\ldots n} \notag \\
&\quad -\sum_{2\leq j <k}^{n-2} M_{12\ldots j} M_{j+1\ldots k} M_{k+1\ldots n}
-  \sum_{j=2}^{n-2} M_{12\ldots j} M_{j+1\ldots n-1} M_n \label{altEoff} \\
&= M_1 \big( E_{23\ldots n} - M_{23\ldots n-1} M_n \big) - \big( E_{12\ldots n-1} -M_1 M_{23\ldots n-1}  \big) M_n
\notag \\
&=V_1E_{23 \ldots n}- E_{12 \ldots n{-}1}V_n\, .
\notag
\end{align}
The double sums in the first two lines are easily seen to cancel 
upon renaming the summation variables $i,j,k$, and the contributions
of $M_1 M_{23\ldots n-1} M_n$ in the third line drop out in the last step.
Given that all the $M_{12\ldots j} M_{j+1\ldots n}$ in (\ref{altEoff}) are
perfectly valid in the momentum phase space of $n$ massless particles 
(the highest-multiplicity currents contains non-singular $s^{-1}_{12 \ldots n{-}2}$
and  $s^{-1}_{23 \ldots n{-}1}$), we conclude that
\beq\label{banc}
\langle E_{12 \ldots n{-}1} V_n\rangle - \langle E_{23 \ldots n}V_1\rangle =
- \langle Q\big(M_{12}M_{34 \ldots n} + M_{123} M_{4\ldots n}+ \cdots + M_{12 \ldots n{-}2}M_{n{-}1n}\big)\rangle = 0 \, ,
\eeq
using the vanishing of BRST-exact expressions under the pure spinor bracket.
We have thus shown equivalence of the amplitude \eqref{aymcyc} to its cyclic image
$i \rightarrow i {-}1 \ {\rm mod} \ n$
\beq
\langle E_{12 \ldots n{-}1} V_n\rangle = \langle  E_{23 \ldots n} V_1\rangle
\eeq
which concludes the proof.\qed

For example, from the formula \eqref{AYM} and the definition \eqref{Edef},
one can also verify directly in the momentum phase space of the corresponding $n$-point amplitude
that
\begin{align}
A(1,2,3,4,5) - A(2,3,4,5,1) &= \langle M_{12}M_{34}M_5 - M_1M_2M_{345} + M_{123}M_4M_5 - M_1M_{23}M_{45}\rangle \notag\\
& = -\langle Q(M_{12}M_{345}+M_{123}M_{45})\rangle = 0\,.
\end{align}

\subsubsection{Manifesting cyclic symmetry via BRST integration by parts}

In the above discussion we have proven that the pure spinor cohomology formula \eqref{AYM} is cyclically symmetric, but
this is not manifest. We will now show how to exploit the cohomological properties of the
pure spinor formula to derive alternative expressions with manifest cyclic symmetry. In doing so,
the multiplicity of the Berends--Giele currents featured in the formulae is reduced which renders
computations including component evaluations more efficient.

In order to manifest cyclic symmetry in the pure spinor cohomology formula \eqref{AYM}, 
we exploit the decoupling of BRST-exact terms $\langle Q( \ldots)\rangle=0$
from the cohomology. Let us start with a simple example to understand the mechanism.
Consider the five point amplitude
\beq
A(1,2,3,4,5) = \langle M_1M_{234}M_5 + M_{12}M_{34}M_5 + M_{123}M_4M_5\rangle
\eeq
and note that there are BRST-exact factors of the form $M_iM_j = QM_{ij}=E_{ij}$. So it can be
rewritten as
\begin{align}\label{cycex}
A(1,2,3,4,5) & = \langle E_{51}M_{234} + M_{12}M_{34}M_5 + M_{123}E_{45}\rangle\\
&=\langle M_{51}E_{234} + M_{12}M_{34}M_5+E_{123}M_{45}\rangle\notag\\
&=\langle M_{51} (M_2M_{34}+M_{23}M_4)
+ M_{12}M_{34}M_5
+ (M_{1}M_{23}+ M_{12}M_{3})M_{45}
\rangle\notag\\
& = \langle M_{12}M_3M_{45}\rangle +{\rm cyc}(12345) \notag
\end{align}
with manifest cyclic symmetry in the last line.
Note that in the second line we integrated the BRST charge by parts;
by \eqref{QMP} this amounts to
\begin{equation}\label{cohom}
\langle E_P M_Q\rangle = \langle M_P E_Q\rangle\,,
\ee
for instance $\langle M_{123} E_{45} \rangle = \langle E_{123}M_{45}\rangle = 
\langle (M_1 M_{23}+M_{12} M_3) M_{45} \rangle$. Notice the reduction of the
highest-rank Berends--Giele currents ($M_{123}$, $M_{234}$) on the left-hand side
of (\ref{cycex}) to rank-two $M_{ij}$ on the right-hand side.

Naively, in the pure spinor cohomology formula (\ref{AYMgen}) for $n$-point SYM tree amplitudes one
needs to know all Berends--Giele currents $M_P$ with multiplicities up to $n{-}2$. For example, in the five-point
amplitude \eqref{cycex} the
first line contains Berends--Giele currents of all multiplicities up to $n{-}2=3$. However, after BRST
integration by parts, the maximum multiplicity in the last line of \eqref{cycex} is two, and its 
cyclic symmetry is manifest.

In fact, using BRST integrations by parts
it was shown in \cite{nptFT} that the highest multiplicity of currents can be
lowered to at most $\lfloor{n \over 2}\rfloor$ while at the same time
yielding superspace formulae for $n$-point
trees with manifest cyclic symmetry
\begin{align}
\label{AYMupto10}
A(1,2,\ldots,3) &= {1\over3}\langle M_1 M_2 M_3\rangle + {\rm cyc}(123)\,,\\
A(1,2, \ldots,4) &= \half \langle M_{12}E_{34}\rangle+{\rm cyc}(1234)\,,\notag\\
A(1,2, \ldots,5) &= \langle M_{12}M_3M_{45}\rangle + {\rm cyc}(12345)\,,\notag\\
A(1,2, \ldots,6)&= {1\over 2} \langle M_{123} E_{456} \rangle
+ {1\over3} \langle M_{12} M_{34} M_{56} \rangle +\cyclic{123456}\,,\notag\\
A(1,2, \ldots,7) &= \langle M_{123} M_{45} M_{67} \rangle + \langle M_{1} M_{234} M_{567} \rangle
+ {\rm cyc}(12\ldots 7)\,,\notag\\
A(1,2, \ldots,8)
&= \half \langle M_{1234} E_{5678}\rangle
+ \langle M_{123} M_{456} M_{78}\rangle
+ {\rm cyc}(12\ldots 8)\,,\notag\\
A(1,2, \ldots,9) &=
{1\over 3}\langle M_{123}M_{456}M_{789}\rangle
+ \langle M_{1234} (M_{567} M_{89} + M_{56}  M_{789} + M_{5678} M_9) \rangle
+ {\rm cyc}(12\ldots 9)\ ,\notag\\
A(1,2, \ldots,10) &= 
\half \langle M_{12345}E_{6789;10}\rangle
+ \langle M_{1234}(M_{567}M_{89;10} + M_{5678}M_{9;10})\rangle
+ {\rm cyc}(12\ldots 10)\,,\notag
\end{align}
where the fractional coefficients are introduced to avoid overcounting due to the explicit
sum over cyclic permutations. For example, after summing the cyclic permutations, the four- and
six-point instances of (\ref{AYMupto10}) can be rewritten as
\begin{align}
A(1,2, 3,4) &= \langle  M_{12} M_3 M_4 \rangle + \langle  M_{23} M_4 M_1 \rangle \, ,
\notag \\
A(1,2,3,4,5,6) &= \langle  M_{12} M_{34} M_{56} \rangle+ \langle  M_{23} M_{45} M_{61} \rangle
+  \langle  M_{123} (M_{45} M_{6}+M_{4} M_{56}) \rangle
\label{excyclic}
\\
& \ \ +  \langle  M_{234} (M_{56} M_{1}+M_{5} M_{61}) \rangle
+  \langle  M_{345} (M_{61} M_{2}+M_{6} M_{12}) \rangle\, ,
\notag
\end{align}
via BRST integration by parts and/or the fermionic nature of $M_P$,
so the factors of $\frac{1}{2}$ and $\frac{1}{3}$ in the cyclic sums in \eqref{AYMupto10} are
essential to not overcount these terms. The manifestly cyclic form of the $n$-point 
amplitudes \eqref{AYMupto10} is free of fractional coefficients when $n$ is not divisible by $2$
or $3$.\footnote{More explicitly, because of the BRST identity $\langle M_{P_1}E_{P_2}\rangle =
\langle M_{P_2}E_{P_1}\rangle$, we note that the cyclic sum of
$\langle M_{P_1}E_{P_2}\rangle$ leads to an overcounting by a factor of
two if and only if $|P_1|=|P_2|$, necessitating the symmetry factor of $\frac{1}{2}$
in the contributions $\langle M_{12\ldots k}E_{k+1\ldots 2k}\rangle$ at even
multiplicity $n=2k$.
Similarly, the cyclic sum of contributions $\langle
M_{P_1}M_{P_2}M_{P_3}\rangle$ leads to an overcounting by a factor of three if and only if
$|P_1|=|P_2|=|P_3|$ which necessitates the symmetry factors of $\frac{1}{3}$ at 
multiplicities $n \in 3\mathbb N$, e.g.\ at $n=3,6,9$ in (\ref{AYMupto10}).}

\subsubsection{Component expansion of the pure spinor SYM tree amplitude}
\label{sec:compSYM}

BRST invariance of the superfields implies gauge-invariant and
supersymmetric components, see the discussion in section \ref{sec:treeprescr}.
The gauge invariance of the SYM tree-level amplitudes allows one to choose the Harnad--Shnider
gauge at a non-linear level to perform the $\theta$-expansion of the multiparticle
supersymmetric Berends--Giele currents, see \ref{HSapp}.
After stripping off the plane-wave factor $e^{k_P \cdot X}$
as in (\ref{series}), this leads to an expansion of $\cA^P_\a$
\begin{equation}\label{thetaEXP}
{\cal A}^P_{\alpha}(\t)=  {1\over 2}(\t\g_m)_\alpha \ce^m_P
+{1\over 3}(\t\g^m)_\alpha (\t\g_m{\cal X}_P)
- {1\over 32} (\t \g^p )_\alpha (\t \g_{mnp} \t) \cf_{P}^{mn}+   \ldots 
\ee
that takes the same form as the $\theta$-expansion (\ref{linTHEX}) of the linearized
superfield $A^i_\alpha$ in the Harnad--Shnider gauge.
The Berends--Giele {\it polarization} currents $\ce^m_P,{\cal X}_P^\alpha$ and $\cf_{P}^{mn}$ 
in the component formulation of $D=10$ SYM are given by the recursions \eqref{reczero}
to (\ref{recthree}). In this way, the simple $\l^3\t^5$
correlators (\ref{simpl3t5}) and (\ref{fermex}) that govern the single-particle correlator $\langle M_1 M_2 M_3\rangle$
in the three-point amplitude \cite{psf} (see section \ref{illustratesec})
\begin{equation}\label{PSthreept}
A(1,2,3) = \langle M_1 M_2 M_3\rangle = \half \ce^m_1 \cf^{mn}_2 \ce^{n}_3 +
(\cX_1\g_m\cX_2)\ce^m_3 + \cyclic{123}
\ee
also determine the multiparticle constituents $\langle M_XM_YM_Z\rangle$ of the $n$-point amplitudes \eqref{AYM},
\begin{align}
\label{compon}
\langle M_X M_Y M_Z \rangle &= 
\half\ce^m_X\,  \cf^{mn}_Y \ce^n_Z + (\cX_X \g_m \cX_Y) \ce_Z^m + \cyclic{XYZ}
=: \cm_{X,Y,Z} \,,
\end{align}
which defines the shorthand $\cm_{X,Y,Z}$.
The component expressions for the above cohomology formulae follow easily by reducing
any $\langle M_X M_Y M_Z \rangle$ to
the combinations $\cm_{X,Y,Z}$ of $\ce^m_P$, $\cf^{mn}_P$ and $\cX^\a_P$
in \eqref{compon}. For instance, the earlier representation \eqref{AYMgen} yields components
\beq\label{cptAYMgen}
A(1,2, \ldots,n{-}1,n) =  \sum_{XY=12 \ldots n{-}1}\cm_{X,Y,n}\,,
\eeq
while the manifestly cyclic representation in \eqref{AYMupto10} gives rise to
\begin{align}
A(1,2, \ldots,4) &= {1\over 2} \cm_{12,3,4} + \cyclic{1234} \, , \label{shortMsix}\\
A(1,2, \ldots,5) &=  \cm_{12,3,45} + \cyclic{1234 5}\, ,  \cr
A(1,2,\ldots ,6) &=   {1\over3}\cm_{12,34,56}
           +{1\over 2} (\cm_{123,45,6} + \cm_{123,4,56} )    + \cyclic{12 \ldots 6}\, , \cr
A(1,2,\ldots ,7) &=  \cm_{123,45,67} + \cm_{1,234,567} + \cyclic{12\ldots 7} \, , \cr
A(1,2,\ldots ,8) &= {1\over 2} (\cm_{1234,567,8} + \cm_{1234,56,78} +\cm_{1234,5,678}  )
+ \cm_{123,456,78} +\cyclic{12\ldots 8} \, .
\notag
\end{align}
Given the recursive nature of
the definitions of the component Berends--Giele currents $\ce^m_P$, $\cf^{mn}_P$ and $\cX^\a_P$, 
the full component expansion of the above amplitudes is readily available and reproduce 
the results for SYM tree amplitudes available on the website \cite{PSSweb}. Furthermore, as
discussed in \cite{BGplefka,gielenumeric}, the Berends--Giele currents lead to fast numerical
evaluation of the amplitudes.

Note that one can obtain matrix elements of the effective operators $\ap \Bbb F^3$ and 
$\ap^2 \Bbb F^4$ of the open bosonic string from (\ref{shortMsix}) by introducing  $\alpha'$-corrections
of the gluonic components of $\cm_{X,Y,Z}$ as detailed in \cite{Garozzo:2018uzj}.

\subsubsection{Equivalence with the gluonic Berends--Giele recursion}

Using the component field-strength
\eqref{recthree}, it follows that the gluonic three-point amplitudes of
the Berends--Giele and pure spinor formulae match.

\begin{prop.}
When restricted to its gluon components the pure spinor cohomology formula
for SYM tree amplitudes \eqref{AYMgen} is equivalent to the standard Berends--Giele
formula \eqref{BGformula}.
\end{prop.}
\noindent\textit{Proof.}
The proof that the general
$n$-point amplitudes agree was written in \cite{BGBCJ}. The outline of the proof
is as follows: first one shows that
the lowest gluonic components of $\ce_P^m$ in the superfield \eqref{thetaEXP}
reproduce the standard Berends--Giele current
\begin{equation}\label{Prepcurrent}
\ce^m_P \, \big|_{\chi_j = 0} = J^m_P \ .
\ee
Then using transversality \eqref{eoms} and momentum conservation in the form of
$k^m_X+ k^m_Y + k^m_Z=0$ one rewrites \eqref{compon} as
\begin{align}
\langle M_X M_Y M_Z \rangle &= (\ce_{[X,Y]} \cdot \ce_Z)
+  \ce_X^m ({\cal X}_Y \g_m {\cal X}_Z) -  \ce_Y^m ({\cal X}_X \g_m {\cal X}_Z) \label{componZ}\\
&\quad{}+ \half\sum_{RS=Z}\big[(\ce_R\cdot \ce_X)(\ce_S\cdot \ce_Y)
- (\ce_R\cdot \ce_Y)(\ce_S\cdot \ce_X)\big]\ , \notag
\end{align}
which simplifies as follows when $Z$ is a single-particle label $p{+}1$:
\begin{equation}\label{componS}
\langle M_X M_Y M_{p+1} \rangle = (\ce_{[X,Y]} \cdot \ce_{p+1})
+  \ce_X^m ({\cal X}_Y \g_m {\cal X}_{p+1}) -  \ce_Y^m ({\cal X}_X \g_m {\cal X}_{p+1})\,.
\ee
The pure spinor superspace formula for tree-level SYM amplitudes \eqref{AYMgen} is given by
the deconcatenation sum of the correlator \eqref{componS} over $XY=12 \ldots p$ and yields
\begin{align}
A(1,2, \ldots p,p{+}1) &=
\!\!\!\! \sum_{XY=12\ldots p}\!\Bigl[ (\ce_{[X,Y]} \cdot \ce_{p+1}) +  \ce_X^m ({\cal X}_Y \g_m {\cal X}_{p+1})
-  \ce_Y^m ({\cal X}_X \g_m {\cal X}_{p+1})\Bigr]\notag\\
&= s_{12 \ldots p} (\ce_{12 \ldots p}\cdot \ce_{p+1}) + k^m_{12 \ldots p} (\cX_{12 \ldots p}\g_m \cX_{p+1})\,,\label{getAYM}
\end{align}
where in the second line the recursions \eqref{reczero} and \eqref{eoms} were used to
identify $\ce^m_{12 \ldots p}$ and $\cX^\alpha_{12 \ldots p}$. Setting the fermions to zero
and using \eqref{Prepcurrent} yields the gluonic
Berends--Giele formula \cite{BerendsME} and finishes the proof
that the pure spinor cohomology formula and the Berends--Giele formula
\eqref{BGformula} are equivalent.\qed

\paragraph{Short representations in the standard Berends--Giele formula}
Despite missing the powerful BRST cohomology manipulations,
a reduction in the multiplicities of the currents was derived in the standard gluonic Berends--Giele method in \cite{BerendsHF} to
obtain ``short'' and manifestly cyclic representations of bosonic amplitudes up to eight points.
For example, the six-point amplitude was found to be
\begin{align}
A^{\rm YM}(1,2 \ldots,6) &= \half s_{123}J^m_{123}J^m_{456} +{1\over 3}[J_{12},J_{34}]^m J^m_{56}
+\half\{J_1,J_{23},J_{4}\}^mJ^m_{56} \notag \\
&\quad+ \{J_1,J_2,J_{34}\}^m J^m_{56}
+\cyclic{123456}\,,\label{shortAsix}
\end{align}
see (\ref{BGrecur}) for the brackets $[\ldots]^m$ and $\{ \ldots \}^m$,
and similar expressions were written for the seven- and eight-point amplitudes \cite{BerendsHF}.

\subsubsection{Kleiss--Kuijf amplitude relations}
\label{KKsec}

In \cite{KKref}
the color-ordered tree amplitudes were observed to obey the KK relations
\beq\label{KKrelation}
A(P,1,Q,n) = (-1)^{|P|}A(1,\tilde P\shuffle Q,n)\,,
\eeq
which singles out legs $1$ and $n$ leading to $(n{-}2)!$ linearly independent amplitudes
(w.r.t.\ constant rather than $s_{ij}$-dependent coefficients).
As a simple example with $P=2,3$ and $Q=4$, we have
\beq
A(2,3,1,4,5) = A(1,3,2,4,5) + A(1,3,4,2,5) + A(1,4,3,2,5)\,.
\eeq
In \cite{KKref} a proof of  \eqref{KKrelation} was argued based on the shuffle symmetry
\eqref{BGsym} of the Berends--Giele currents derived
by Berends and Giele in \cite{BerendsZN}, see section~\ref{KKBGsec}. The proof that the
pure spinor cohomology formula \eqref{AYM} satisfies the KK relations is analogous, it follows
as a corollary to the equivalence in the proofs of \cite{Ree} and \cite{schocker}.
However, in this section we wish to see this equivalence more explicitly and use it
to prove the KK relations not as an indirect corollary but as a direct statement. 
The explicit equivalence between \cite{Ree} and \cite{schocker}
is given by the following lemma, first stated in \cite{BGBCJ} and
proven in \cite{ruggeroprivate} (see also equation (41) of \cite{novelli2020noncommutative}):
\begin{lemma}
Let $P,Q$ be arbitrary words and $j$ a letter, then
\beq\label{schockerRee}
\sum_{XY=P}(-1)^{|X|}\tilde X\shuffle (YjQ) = (-1)^{|P|}j(\tilde P\shuffle Q)\, ,
\eeq
where $X$ and $Y$ are allowed to be empty in the sums and $\tilde P$ denotes the reversal of $P$.
\end{lemma}
\noindent\textit{Proof.} The proof follows from an induction on the length $|P|$ of the word $P$ \cite{ruggeroprivate}.
For the base case of length zero the formula is true as it reduces to $jQ=jQ$.
Assume that the formula holds for $|P|=p$, then for $|P|=p{+}1$ set $P=Ci$ for a letter $i$ and a word $C$ of length $p$.
Using the elementary property of summations
\beq\label{elemsum}
\sum_{XY=Ci}f(X,Y) = f(Ci,\emptyset) + \sum_{XY=C}f(X,Yi)
\eeq
we get
\begin{align}
\sum_{XY=Ci}(-1)^{|X|}\tilde X\shuffle (YjQ) &= (-1)^{|Ci|}(i\tilde C)\shuffle (jQ)
+ \sum_{XY=C}(-1)^{|X|}\tilde X\shuffle (YijQ)\\
&=(-1)^{|P|}(i\tilde C)\shuffle Q' +(-1)^{|C|} i(\tilde C\shuffle Q')\notag\\
&=(-1)^{|P|}\big((i\tilde C)\shuffle (jQ) - i(\tilde C\shuffle jQ)\big)\notag\\
&
= (-1)^{|P|}j(\tilde P\shuffle Q) \, ,\notag
\end{align}
where we defined $Q'=jQ$ to use the induction hypothesis in the second line,
and we used the recursive definition of the shuffle product $(i\tilde C)\shuffle (jQ) =
i\big(\tilde C\shuffle (jQ)\big) + j\big((i\tilde C)\shuffle Q\big)$ in the fourth line. So the
induction holds true when $|P|=p{+}1$ and \eqref{schockerRee} follows.\qed

It is convenient to rewrite the identity \eqref{schockerRee} such that there are
no empty words in the sum on its left-hand side,
$\sum_{RS=P}(-1)^{|R|}\tilde R\shuffle (SjQ) + (-1)^{|P|}\tilde P\shuffle (jQ) + PjQ =
(-1)^{|P|}j(\tilde P\shuffle Q)$ where $R,S\neq\emptyset$.
That is,
\beq\label{schockerReeagain2}
PjQ
=   (-1)^{|P|}j(\tilde P\shuffle Q) 
-\sum_{RS=P}(-1)^{|R|}\tilde R\shuffle (SjQ) - (-1)^{|P|}\tilde P\shuffle (jQ)\, ,\quad
R,S\neq\emptyset\,,
\eeq
or 
\beq\label{dualLie}
PjQ=(-1)^{|P|}j(\tilde P\shuffle Q) +\hbox{shuffles}\, ,
\eeq
giving rise to an equivalence relation in
the dual of Lie polynomials \cite{flas}.

\begin{prop.}
The pure spinor cohomology formula for SYM tree amplitudes \eqref{AYM} satisfies the KK relations \eqref{KKrelation}
\beq\label{KKrelationagain}
A(P,1,Q,n) = (-1)^{|P|}A(1,\tilde P\shuffle Q,n)\,.
\eeq
\end{prop.}
\noindent\textit{Proof.} The Corollary~\ref{corMEshuffle} in \eqref{shuffles}
shows that
the superfield $E_P$ in \eqref{Edef}
also satisfies the shuffle symmetry
\beq\label{shuffleEagain}
E_{R\shuffle S} = 0\, ,\quad \forall \, R,S\neq\emptyset
\eeq
since the definition of $E_P=\sum_{XY=P}M_XM_Y$ is also over an antisymmetric deconcatenation
in view of the fermionic nature of $M_X$ and $M_Y$.
Since the words $R$ and $S$ in \eqref{shuffleEagain} must be non-empty,
the identity \eqref{schockerReeagain2} can be used to yield
\beq\label{EBG}
E_{PjQ} = (-1)^{|P|}E_{j(\tilde P\shuffle Q)}\, .
\eeq
Consequently, from the pure spinor cohomology formula \eqref{AYM} we obtain
$\langle E_{PjQ}V_n\rangle = (-1)^{|P|}\langle E_{j(\tilde P\shuffle Q)}V_n\rangle$ and
therefore the KK relation \eqref{KKrelationagain} is satisfied.\qed

\subsubsection{Bern--Carrasco--Johansson amplitude relations}
\label{BCJampsec}

The SYM amplitudes from the pure spinor cohomology formula \eqref{AYM} are almost trivially 
zero due to the fact that $E_P = QM_P$ for generic values of all the $s_{i\ldots j}$. The only 
reason why the superspace expression for
the $n$-point amplitude is not BRST exact is because $M_P$ with $P=12\ldots n{-}1$ contains a propagator
$1/s_P$ which is ill-defined in a momentum phase space of $n=|P|{+}1$ massless particles.
The BCJ amplitude relations arise when certain linear combinations of $s_{ij}E_P$
become BRST-exact expressions. Let us consider one simple example to
understand the mechanism.

\paragraph{Four point BCJ relation} Let us review the argument of \cite{BGBCJ}.
Consider the Berends--Giele current \eqref{BGVs}
\beq\label{M3bcj}
M_{123} = 
{V_{[[1,2],3]}\over s_{12}s_{123}}
+ {V_{[1,[2,3]]}\over s_{23}s_{123}}
\eeq
in the BCJ gauge, where the local superfields $V_{[P,Q]}$ satisfy the generalized Jacobi identities
\eqref{genjac}.
Now consider the linear combination $s_{23}M_{123} - s_{13}M_{213}$
\begin{equation}\label{lcM}
s_{23}M_{123} - s_{13}M_{213} = s_{23}\Big({V_{[[1,2],3]}\over s_{12}s_{123}}
+ {V_{[1,[2,3]]}\over s_{23}s_{123}}\Big)
- s_{13}\Big(
{V_{[[2,1],3]}\over s_{12}s_{123}}
+ {V_{[2,[1,3]]}\over s_{13}s_{123}}
\Big)
 = {V_{[[1,2],3]}\over s_{12}}\, ,
\eeq
where we used the generalized Jacobi identities of $V_{[P,Q]}$ and $s_{12}+s_{13}+s_{23}=s_{123}$. Note the
crucial cancellation of the overall pole in $1/s_{123}$, for which the Jacobi identity is an
essential requirement. The identity \eqref{lcM} is the Berends--Giele counterpart of the planar binary tree
relation \eqref{illust} in terms of Lie polynomials.

Therefore, while $M_{123}$ and $M_{213}$ are ill-defined objects
for a four-point amplitude where $s_{123}=0$, the linear combination $s_{23}M_{123} - s_{13}M_{213}$ 
is not! This means that it is a valid object to use as a ``BRST ancestor'' to derive $Q$-exact expressions
with vanishing components
\beq\label{Bexact}
E_{\{12,3\}}= s_{23}E_{123} - s_{13}E_{213} = Q\big(s_{23}M_{123} - s_{13}M_{213}\big) = Q \bigg( {V_{[[1,2],3]}\over s_{12}}\bigg)\, .
\eeq
Multiplying by the BRST-closed $V_4$ on the right and using the pure spinor SYM formula \eqref{AYM} we get
\beq\label{trivial}
s_{23}A(1,2,3,4) - s_{13}A(2,1,3,4) = \left \langle Q\bigg( {V_{[[1,2],3]}\over s_{12}} V_4\bigg)\right \rangle = 0\,.
\eeq
To summarize, the four-point BCJ amplitude relation \cite{BCJ}
\beq\label{4ptBCJ}
s_{23}A(1,2,3,4) - s_{13}A(2,1,3,4) = 0
\eeq
holds because the superspace expression underlying the left-hand side is BRST exact. This
is a consequence of the cancellation of the propagator $1/s_{123}$
in the linear combination \eqref{lcM} which is only true if $V_{[P,Q]}$ satisfies 
the generalized Jacobi identities. In other words, in the BCJ gauge of multiparticle 
superfields the four-point BCJ amplitude relation is obtained
due to the vanishing of BRST-exact expressions. Note, however, that the BCJ
amplitude relations are valid independently of the precise details of the
numerators by non-linear gauge invariance of the cohomology formula 
(\ref{AYMgen}) for SYM amplitudes.\footnote{The perturbiner components of the non-linear gauge variation
(\ref{NLgauge}) with Berends--Giele currents $\Omega_P$ of the gauge scalar lead to
the variation $\delta_\Omega M_P =Q \Omega_P + \sum_{P=XY}(\Omega_X M_Y - \Omega_Y M_Y)$.
The resulting non-linear gauge variation of the SYM amplitudes 
$\sum_{12\ldots n-1=XY} \langle M_X M_Y M_n \rangle$ then conspires to a BRST-exact
expression after assembling the contributions from $\delta_\Omega M_X$, $\delta_\Omega M_Y$
and $\delta_\Omega M_n$.} Hence, there
is no loss of generality in employing BCJ-gauge numerators in the
discussions above to identify BRST-exact combinations of the
superfields $E_P$ in the cohomology approach.

\paragraph{BCJ amplitude relations in general} The strategy to derive $n$-point BCJ 
amplitude relations from the pure spinor cohomology method hinges on linear 
combinations of Berends--Giele currents of
multiplicity $n{-}1$ such that the leading propagator $1/s_{12 \ldots n-1}$ is absent, 
as illustrated by the example \eqref{lcM} at $n=4$. These combinations can be
found in BCJ gauge, and their combinatorial structure is most conveniently encoded
in the $S$ bracket of section \ref{Sbracketsec} which was
used in \cite{flas} to rigorously
prove the $n$-point statements of \cite{BGBCJ}.
From the pure spinor cohomology discussion above,
the property \eqref{bigclaim} demonstrates that certain linear combinations of SYM trees vanish.
More precisely,
\begin{prop.}\textbf{(BCJ relations)}
The pure spinor cohomology formula for SYM tree amplitudes \eqref{AYM} satisfies BCJ relations
\beq\label{BCJrelations}
A(\{P,Q\},n) = 0\, ,
\eeq
for any possible distribution of the labels $\{1,2, \ldots,n{-}1\}$ between $P$ and $Q$. Moreover,
the fundamental BCJ relations in the terminology of \cite{Feng:2010my} are obtained in the 
special case when $P=1$,
$Q=23 \ldots n{-}1$ as
\beq\label{fundaBCJ}
-A(\{1,23 \ldots n{-}1\},n) = \!\!\!\! \sum_{XY=23 \ldots n{-}1}\!\!\!k_1\cdot k_X A(X,1,Y,n) = 0\,.
\eeq
\end{prop.}
\noindent\textit{Proof.} Note from the prescription \eqref{MPfrombP} and the corollary
\eqref{BCJspq} that there is no propagator $1/s_{PQ}$ in 
\beq\label{noprop}
M_{\{P,Q\}} = V_{b(\{P,Q\})}\,.
\eeq
This implies, by a similar reasoning as in \eqref{QEexact}, that
the superfield $E_{\{P,Q\}}$ for $\len{P}{+}\len{Q}=n{-}1$ is BRST exact in
the momentum phase space of $n$ massless particles
\beq\label{brstExact}
E_{\{P,Q\}} = QM_{\{P,Q\}}\,,
\eeq
even though $s_{PQ}=0$. Since $QV_n=0$, this means that the expression $E_{\{P,Q\}}V_n$ is BRST
exact; $Q(M_{\{P,Q\}}V_n)$ and therefore $\langle E_{\{P,Q\}}V_n\rangle = 0$ vanishes in the
cohomology of the pure spinor bracket. The pure spinor cohomology formula \eqref{AYM} implies that $A(\{P,Q\},n)=0$, proving
the first claim in \eqref{BCJrelations}.

To prove that the fundamental BCJ relation of \cite{Feng:2010my} is recovered as in
\eqref{fundaBCJ} we use the lemma \eqref{fundiQ}
\beq\label{SiQ}
\{i,Q\}\sim \sum_{XY=Q}k_i\cdot k_Y XiY\,.
\eeq
Therefore the BCJ relation \eqref{BCJrelations} implies
\begin{align}\label{fundaProof}
0&=-A(\{1,Q\},n)=-\sum_{XY=Q}k_1\cdot k_Y A(X,1,Y,n)\\
&= \sum_{XY=Q} k_1\cdot k_X A(X,1,Y,n) + \sum_{XY=Q} k_1\cdot k_n A(X,1,Y,n)\notag\\
&= \sum_{XY=Q} k_1\cdot k_X A(X,1,Y,n)\,,\notag
\end{align}
where we used that momentum conservation $k_Y =
- (k_X {+} k_1 {+} k_n)$ implies that $k_1\cdot k_Y = - k_1\cdot k_X - k_1\cdot k_n$ to obtain the second
line and $\sum_{XY=Q}X1Y = 1\shuffle Q$ together with $A((R\shuffle S),n)=0$ to obtain the third
line. The last equality follows from $E_{R\shuffle S}=0$ for non-empty $R,S$ as in \eqref{shuffles}.
Choosing $Q=23 \ldots n{-}1$ finishes the derivation of the fundamental BCJ relation
\eqref{fundaBCJ} and the proposition is proven.\qed

For examples of the BCJ relations (\ref{BCJrelations}) 
generated by the $S$ bracket, we consider
\begin{align}
A(\{12,3\},4) &= s_{23}A(1,2,3,4) - s_{13}A(2,1,3,4) = 0\, ,\notag\\
A(\{1,23\},4) &= s_{12}A(1,2,3,4) - s_{13}A(1,3,2,4) = 0\, ,\label{4ptBCJ2}\\
A(\{123,4\},5) &= s_{34} A(1,2,3,4,5) - s_{24} A(1,3,2,4,5) - s_{24} A(3,1,2,4,5) + s_{14}
A(3,2,1,4,5) = 0\, , \notag \\
A(\{12,34\},5)&=  s_{23}A(1,2,3,4,5) - s_{24}A(1,2,4,3,5) - s_{13} A(2,1,3,4,5) +
s_{14}A(2,1,4,3,5) = 0\, .
\notag
\end{align}
More generally, the distribution in $-A(\{1,23\ldots n{-}1\},n) = 0$ is equivalent, via
\eqref{fundaProof}, to the
fundamental BCJ relations
\beq
\label{fundBCJ}
0 = k_1\cdot k_2 A(2,1,3,\ldots,n) + k_1\cdot k_{23} A(2,3,1,4,\ldots,n)
 + \cdots + k_{1}\cdot k_{23 \ldots n{-}1}A(2,3,\ldots,n{-}1,1,n)\, ,
\eeq
whose permutations are known to leave $(n{-}3)!$ independent partial amplitudes
\cite{BCJ, BjerrumBohrRD, StiebergerHQ,Feng:2010my, Chen:2011jxa}. As will be briefly reviewed in section \ref{sec:6.4.2}, the BCJ relations
were derived in \cite{BCJ} from the color-kinematics duality, see \cite{Bern:2019prr, Bern:2022wqg} 
for reviews. The emergence of (local) BCJ-satisfying numerators from the pure spinor 
superstring will be discussed in sections~\ref{sec:6.4.3} and \ref{localMoebiussec}.

\subsection{The generating series of tree-level amplitudes}

The SYM tree-level amplitudes from the
pure spinor superspace expression \eqref{AYM} can be compactly
described by a generating function \cite{BGBCJ}.
To see this one uses the perturbiner series (\ref{series}) of the unintegrated vertex
operator expanded in terms of the Berends--Giele currents \eqref{historic}
\beq\label{sol}
\Bbb V := \lambda^\alpha \Bbb A_\alpha =  \sum_{i_1} M_{i_1} t^{i_1} e^{k_{i_1 } \cdot X} 
+ \sum_{i_1,i_2} M_{i_1 i_2} t^{i_1} t^{i_2}  e^{k_{i_1 i_2 } \cdot X} 
+\sum_{i_1,i_2,i_3} M_{i_1 i_2 i_3} t^{i_1} t^{i_2} t^{i_3} e^{k_{i_1 i_2 i_3} \cdot X} + \cdots\, .
\eeq
One can show that the generating function of color-dressed SYM amplitudes is given by
the natural generalization of the three-point amplitude $\langle V_1 V_2 V_3\rangle$ as
\beq\label{VVV}
{1\over 3}{\Tr}\langle \Bbb V \Bbb V \Bbb V \rangle =
\sum_{n=3}^{\infty} {n-2\over n}\!\!\!\! \sum_{i_1,i_2,\ldots,i_n}\!\!\!\! {\Tr}(t^{i_1} t^{i_2} \ldots t^{i_n}) 
A(i_1,i_2,\ldots,i_n)  \, .
\eeq
It is reassuring to note that the generating function of
tree-level amplitudes \eqref{VVV} reproduces the interaction term of the
ten-dimensional SYM Lagrangian of \cite{Berkovits:2001rb} evaluated on the
generating series of (non-local) Berends--Giele currents
in superspace: $\Bbb F^{mn}(X,0)$ and $\Bbb W^\a(X,0)$. To see this
note from \eqref{zerocomp} that $\ce^m_P, {\cal X}_P^\alpha$ and $\cf^{mn}_P$
are the $\theta=0$ components of the generating series
$\Bbb A^m$, $\Bbb W^\alpha$ and $\Bbb F^{mn}$.
Therefore \eqref{compon} implies that
\begin{align}
{1\over 3}{\Tr}\langle \Bbb V \Bbb V \Bbb V \rangle &=
{1\over 4}{\Tr} ( [\Bbb A_m , \Bbb A_n]\Bbb F^{mn})
+ {\Tr} (  \Bbb W \gamma^m \Bbb A_m  \Bbb W) \, \Big|_{\theta=0}  \notag\\
&=
{\Tr} \Big( {1\over 4} \Bbb F_{mn} \Bbb F^{mn}
+ (\Bbb W  \gamma^m \nabla_m \Bbb W) \Big) \, \Big|_{\theta=0}\,, \label{backtoFF}
\end{align}
where we have used the massless
Dirac equation $\nabla_m \gamma^m_{\alpha \b} \Bbb W^{\b}=0$ as
well as the field equation $\partial_m \Bbb F^{mn}=[\Bbb A_m,\Bbb F^{mn}]
+ \gamma^n_{\alpha \b} \{ \Bbb W^\alpha , \Bbb W^\b \}$ and discarded
a total derivative
to rewrite $(\partial_m \Bbb A_n) \Bbb F^{mn} =
- \Bbb A_n \big( [\Bbb A_m,\Bbb F^{mn}]
+ \gamma^n_{\alpha \b} \{ \Bbb W^\alpha , \Bbb W^\b \} \big)$.
The matching of the Lagrangian with the
resummation of all tree-level amplitudes  
is of course a strong consistency check for the manipulations
with the perturbiner series, see e.g.\ \cite{browntree}.

\section{Superstring disk amplitudes with the pure spinor formalism}
\label{sec:diskamp}

In the previous section, we have derived the elegant expression
(\ref{AYMgen}) for $n$-point SYM tree-level amplitudes in pure spinor superspace
from locality and BRST cohomology considerations. In this section, we will
review how this representation of SYM amplitudes emerges from the CFT
prescription (\ref{treepresc}) for superstring disk amplitudes through
the field-theory limit $\alpha' \rightarrow 0$. This will in fact be a corollary of the
key result of this review -- a minimal and manifestly BRST invariant form of
$n$-point open-superstring amplitudes: as will become
clear from the final and exact-in-$\alpha'$ result (\ref{nptdisk}), the entire 
polarization dependence of the string amplitude is carried by field-theory building blocks.
The CFT computation is guided by the local multiparticle superfields in section 
\ref{sec:4.1loc} and their generalized Jacobi identities. The manifestly local 
representation (\ref{stringWithTs}) of the $n$-point disk amplitude encountered in intermediate
steps will be later on identified as the origin of the color-kinematics duality, see 
section \ref{sec:6.4}.

Both the local and manifestly BRST invariant representations of the string amplitude
resonate with field-theory structures when expressed in a Parke--Taylor basis of disk
integrals with characteristic cyclic denominators $z_{i_1i_2}z_{i_2i_3}\ldots 
z_{i_{n-1}i_n} z_{i_n i_1}$. On the one hand, the full $\alpha'$-dependent open-superstring 
amplitude lines up with the field-theory KLT formula for supergravity tree amplitudes, see
(\ref{AstringP}). On the other hand, the field-theory limit $\alpha' \rightarrow0$ of
the Parke--Taylor integrals is reviewed to reproduce tree amplitudes of bi-adjoint
scalars which play a central role for the color-kinematics duality of gauge theories 
and different formulations of the gravitational double copy. In fact, as will be argued
in the present section and the following, the $\alpha'$-corrections
to Parke--Taylor integrals admit an effective-field-theory interpretation in
terms of bicolored scalars with higher-derivative interactions.

\subsection{CFT analysis}
\label{sec:CFTan}

Tree-level scattering amplitudes of open-string states are determined by iterated
integrals on the boundary of a disk worldsheet, as can be seen in the pure spinor
prescription \eqref{treepresc}.
Using the prescription to compute $n$-point disk amplitudes in the pure spinor formalism
requires the evaluation of a CFT correlator
\beq\label{nptfcttree}
\langle \! \langle V_1(z_1) \prod_{j=2}^{n-2}U_j(z_j)V_{n-1}(z_{n-1}) V_n(\infty) \rangle
\! \rangle =: \langle {\cal K}_n  \rangle
\prod_{i<j}^n |z_{ij}|^{-2\alpha' s_{ij}}\,,
\eeq
with the massless vertex operators (\ref{integrado}) and (\ref{V}).

The definition of the tree-level {\it correlators}\footnote{The context and the subscript clearly differentiates the
$n$-point correlator $\cK_n$ in \eqref{nptfcttree} and the Berends--Giele current $\cK_n$ in \eqref{calK} for the single-particle $n$ of a
generic superfield in $K_n$ of \eqref{KPdef}.} $\cK_n$ on the
right-hand side is such that it strips off
the Koba--Nielsen factor  $\prod_{i<j}^n |z_{ij}|^{-2\alpha' s_{ij}}$ from the
path integral. We have re-instated the $\alpha'$-dependence adapted to the
correlation functions on a disk worldsheet which differs from the 
Koba--Nielsen exponents on the sphere in (\ref{introduceKN}) and
section \ref{sec:6.5.2}.

\subsubsection{Double poles versus logarithmic singularities}

The computation of the correlators ${\cal K}_n$ 
from the CFT rules of the pure spinor formalism is guided
by the OPE contractions among the vertex operators in \eqref{nptfcttree}.
Since the conformal $h=1$ primaries $[\p \theta^\alpha,\Pi^m,d_\alpha , N^{mn}]$ within 
the integrated vertex (\ref{integrado}) do not have any zero modes at tree level, 
the correlator \eqref{nptfcttree} can be computed by summing all their
OPE singularities summarized in section \ref{sec:summOPE} and
placing the fields in the residues at appropriate positions, see e.g.\ section 
2.3 of \cite{Berkovits:2004px}.
As shown in \eqref{UUope}, these OPEs generically give rise to both single- and 
double-poles. However, as alluded to in \eqref{UUUope} 
and observed in explicit calculations for five \cite{5ptsimple} and six points \cite{6ptTree},
the role of the double-pole integrals is to correct the numerators of the
single-pole integrals such that any OPE residue $L_{jiki\ldots}$ as defined in \eqref{L2131} is transformed to 
the associated multiparticle vertex operator in the BCJ gauge $V_{ijk\ldots}$.
This is the consequence of a subtle interplay between
integration-by-parts identities among the disk integrals and the explicit form of 
the local superfields multiplying these integrals. In particular, the double-pole residues
feature factors of $(1+2\alpha's_{ij})$, as for instance seen in the last line of (\ref{UUope}), which
cancel the tachyon poles $(1+2\alpha's_{ij})^{-1}$ that would arise from disk 
integrals over $|z_{ij}|^{-2\alpha' s_{ij}-2}$. As a rank-three example, the
relation between $V_{ijk}$ and the object $T_{ijk}$ obtained from OPEs and
integration-by-parts corrections in \cite{6ptTree, nptStringI} 
is spelt out in \ref{Vapp}. The generating series of the Berends--Giele currents
associated with $V_{ijk}$ and $T_{ijk}$
are related by a non-linear gauge transformation that preserves BCJ gauge.


More generally, numerators $(1+2\alpha's_{ijk\ldots})$ from nested OPEs cancel tachyonic poles
of multiparticle channels in a highly nontrivial way, see for instance appendix B.3 of \cite{6ptTree}.
In this way, all the singularities of the correlator $\langle {\cal K}_n  \rangle$ in (\ref{nptfcttree})
become {\it logarithmic}: when visualizing all factors of $z_{ij}^{-1}$ by an edge between 
vertices $i$ and $j$, logarithmic singularities are characterized by obtaining a tree graph in the
frame $z_n \rightarrow \infty$. Integration by parts removes loop subgraphs associated
for instance with $z_{ij}^{-2}$
or $(z_{ij} z_{jk} z_{ki})^{-1}$, and the accompanying numerators of $(1+2\alpha's_{ij})$ or
$(1+2\alpha's_{ijk})$ due to the superfields ensure that no tachyon poles are generated in the 
coefficients of the logarithmic singularities. It follows from Aomoto's work \cite{aomoto1987gauss} that
non-logarithmic singularities can always be removed via integration by parts, but it is a
peculiarity of the superstring (as compared to bosonic or heterotic strings
\cite{Huang:2016tag, Azevedo:2018dgo}) that the coefficients of the logarithmic integrands 
become free of tachyon poles and in fact homogeneous in $\alpha'$.

\subsubsection{Lie-polynomial structure of the correlator}

After discarding total worldsheet derivatives and BRST-exact terms,
the calculation of the correlator \eqref{nptfcttree} can be summarized by an elegant pattern 
relating the symmetries of the kinematic factors and logarithmic integrands. By the superfield 
contributions from the double poles and more general non-logarithmic singularities, all
kinematic factors can be written in terms of the multiparticle vertex $V_P$ (\ref{Vhatgen})
subject to the generalized Jacobi identities (\ref{genjac}). The logarithmic singularities 
in turn are carried by the following worldsheet functions $\cZ_P$ satisfying shuffle symmetries
\beq\label{cZdef}
\cZ_{123\ldots p} := {1\over z_{12} z_{23}\ldots z_{p-1,p}}\,,\qquad
\cZ_{A\shuffle B} = 0\,,\quad \forall \ A,B\neq\emptyset \, .
\eeq
At rank $p=2,3$, shuffle symmetry is a consequence of antisymmetry 
$\cZ_{12}=z_{12}^{-1} = - z_{21}^{-1} = - \cZ_{21}$ and the partial fraction
$\cZ_{123}+\cZ_{213}+\cZ_{231} = (z_{12}z_{23})^{-1} + {\rm cyc}(1,2,3) =0$,
and a general proof can be found in Lemma 5.4 of \cite{hadleigh}.

At low multiplicities, the sum of OPEs involving one unintegrated and any number of
integrated vertices is given by (the symbol $\cong$ denotes equality up to 
total derivatives and BRST-exact terms in presence of the remaining
vertex operators $U_j(z_j)$ and $V_{n-1}(z_{n-1})V_n(\infty)$)
\begin{align}
V_1(z_1)U_2(z_2)&\cong  V_{12} {\cal Z}_{12} \, , \notag \\
V_1(z_1)U_2(z_2)U_3(z_3)&\cong   V_{123}\cZ_{123}
+  V_{132}\cZ_{132} \, , \label{experB} \\
V_1(z_1)U_2(z_2)U_3(z_3)U_4(z_4)&\cong   V_{1234}\cZ_{1234}
+ {\rm perm}(2,3,4)
\, , \notag
\end{align}
where already at rank two, we have discarded the term $ Q \big(
\tfrac{1}{2} (A_1\cdot A_2) - (A_1 W_2)\big)$ in (\ref{VvsL})
to convert the OPE residue $L_{21}$ defined by (\ref{UVs}) into the two-particle
vertex $V_{12}$. In the context of multiparticle correlators, contributions
such as $\langle (A_1 W_2) Q( U_3\ldots ) \rangle$ due to BRST integration by parts 
conspire to total derivatives which are discarded as well on the right-hand
side of the symbol $\cong$ (see the earlier comments below (\ref{QL21}) and (\ref{oint})).

Given the generalized Jacobi symmetry of $V_P$ and the shuffle symmetry of $\cZ_P$,
the sums of terms on the right-hand side of \eqref{experB} furnish Lie polynomials \cite{Ree}. 
They are in fact permutation symmetric in the labels 
of $V_1$ and all the $U_{a_i}$, say $V_{123}\cZ_{123}+  V_{132}\cZ_{132} = V_{213}\cZ_{213}
+  V_{231}\cZ_{231}$, and generalize to
\beq\label{experC}
V_1(z_1)\prod_{i=1}^{n}U_{a_i}(z_{a_i}) \cong \sum_{|A|=n}  V_{1A} {\cal Z}_{1A}\,,
\eeq
where the summation range $|A|=n$ refers to the $n!$ words $A$ formed by permutations of
$a_1a_2\ldots a_{|A|}$. The Lie-polynomial structure implies that the 
right-hand side of \eqref{experC} is permutation symmetric in $1,a_1,a_2,\ldots,a_{|A|}$
even though only the weaker symmetry in $a_1,a_2,\ldots,a_{|A|}$ is manifest.\footnote{This 
follows from the identity $\sum_{A}{1\over |A|}\cZ_{A}V_A =
\sum_{B}\cZ_{iB}V_{iB}$.}

Following this reasoning the correlator $\cK_n$ can be assembled from two factors of \eqref{experC}
corresponding to sequences of OPEs terminating on one of the unintegrated vertex
operators $V_1(0)$ or $V_{n-1}(1)$. OPE contributions involving the third unintegrated 
vertex operator $V_n(z_n)$ are suppressed in our choice of ${\rm SL}_2(\mathbb R)$
frame with $z_n\to\infty$. These selection rules for OPEs lead to $n{-}2$ deconcatenations
$AB=23\ldots n{-}2$ (including the ones with $A=\emptyset$ or $B=\emptyset$) and 
an overall permutation over $(n{-}3)!$ labels for a total of $(n{-}2)!$ terms \cite{nptStringI}:
\beq\label{experF}
{\cal K}_{n} =\!\!\!\!\!\!
\sum_{AB=23 \ldots n-2}\!\!\!\!\!\!\!\!\big(V_{1A} \cZ_{1A}\big)\big(V_{n-1,\tilde
B}\cZ_{n-1,\tilde B}\big) V_n + {\rm perm}(2,3 ,\ldots ,n{-}2)\,,
\eeq
where $\tilde B$ denotes the reversal of the word $B$.
The first few expansions of \eqref{experF} read ($\cZ_i:= 1$),
\begin{align}
{\cal K}_3 &= V_1 V_2 V_3\,, \notag \\
{\cal K}_4 &= V_{12} \cZ_{12} V_3 V_4 + V_{1} V_{32} \cZ_{32} V_4\,,\notag \\
{\cal K}_5 &=
\big(V_{123}\cZ_{123} + V_{132}\cZ_{132}\big) V_4 V_5
+ V_{1}\big(V_{423}\cZ_{423} + V_{432}\cZ_{432}\big)V_5 \label{experE} \\
&\quad{}+ \big(V_{12} \cZ_{12}\big)\big( V_{43}\cZ_{43}\big) V_5
+ \big(V_{13} \cZ_{13}\big)\big( V_{42}\cZ_{42}\big) V_5\,,\notag \\
{\cal K}_6 &=
V_{1234}\cZ_{1234} V_5 V_6+V_{123}\cZ_{123} V_{54} \cZ_{54} V_6
+V_{12}\cZ_{12} V_{543} \cZ_{543} V_6 +V_{1}  V_{5432} \cZ_{5432} V_6 + {\rm perm}(2,3,4)
\, ,\notag
\end{align}
and we reiterate that, by the Lie-polynomial structure of the correlator,
$V_{123} \cZ_{123} +  V_{132} \cZ_{132}$
is symmetric in $1,2,3$ even though only two out of $3!$ permutations are spelled out.

One can verify that \eqref{experF} can be obtained
using the following two effective rules for multiparticle OPEs
\beq\label{OPEeff}
V_A(z_a)U_B(z_b)\rightarrow {V_{[A,B]}(z_a)\over z_{ab}}\, ,\qquad
U_A(z_a)U_B(z_b)\rightarrow {U_{[A,B]}(z_a)\over z_{ab}}\, ,
\eeq
where $z_a$ and $z_b$ are the worldsheet positions corresponding to the
first letters of the words $A$ and $B$. The replacements are
again valid upon discarding BRST-exact terms and total derivatives from the 
complete correlator (\ref{nptfcttree}), see the comments below (\ref{experB}). 
The nested brackets in
$V_{[A,B]}$ or $U_{[A,B]}$ are expanded as in \eqref{Baker} by virtue 
of the generalized Jacobi identities satisfied by $V_P$ and $U_Q$.

\subsection{Local form of the disk correlator}

Using the above results the $n$-point superstring disk amplitude computed with the pure spinor
formalism becomes a sum over $(n{-}2)!$ superfield numerators along with different worldsheet 
functions (\ref{cZdef}) \cite{nptStringI},
\begin{align}
{\cal A}(\mathds{1}_n) &=
{\cal A}(1,2,\ldots,n)
\notag \\
&=(2\alpha')^{n-3} \int\limits_{D(\mathds{1}_n)} \prod_{j=2}^{n-2} dz_j \prod_{1\leq i<j}^{n-1}  |z_{ij}|^{-2\alpha's_{ij}}
\!\!\!\!\!\!\!\!\sum_{AB=23 \ldots n-2}\!\!\!\!\!\!\!\!\big\langle\big(V_{1A}
\cZ_{1A}\big)\big(V_{n-1,\tilde B}\cZ_{n-1,\tilde B}\big) V_n\big\rangle
+ {\rm perm}(23 \ldots n{-}2)\notag \\
&= (2\alpha')^{n-3}  \int\limits_{D(\mathds{1}_n)} \prod_{j=2}^{n-2} dz_j  \prod_{1\leq i<j}^{n-1}   |z_{ij}|^{-2\alpha's_{ij}}  \label{stringWithTs} \\
&\quad \times \bigg\{
\sum_{p=1}^{n-2} { \langle  V_{12 \ldots p} V_{n-1,n-2, \ldots ,p+1} V_n \rangle
\over (z_{12}  z_{23} \cdots z_{p-1,p} )( z_{n-1,n-2} \cdots z_{p+2,p+1})}  +  {\rm perm}(23 \ldots n{-}2) \bigg\}\,.\notag
\end{align}
Recall that the integration domain $D(\mathds{1}_n)=D(1,2,\ldots,n)$ defined in (\ref{domain}) 
in the present ${\rm SL}_2(\mathbb R)$-frame with $(z_1,z_{n-1},z_n)=(0,1,\infty)$
amounts to the disk ordering $0<z_2<z_3<\ldots < z_{n-2}< 1$.

In order to avoid cluttering, we adopt the notation
\beq\label{intnot}
\int d\mu^n_P := \int_{D(P)} d z_2 \, dz_3 \cdots d z_{n-2} \prod_{1\leq i<j}^{n-1} |z_{ij}|^{-2\alpha's_{ij}}\, ,
\eeq
where the superscript of the measure tracks the number $|P|{=}n$ of external labels. This shorthand
is suited for the choice of ${\rm SL}_2(\mathbb R)$-frame where the worldsheet positions $(z_1,z_{n-1},z_{n})$ 
are fixed to $(0,1,\infty)$ (or to $(1,0,\infty)$ to accommodate all the $(n{-}1)!$ cyclically inequivalent
choices of $P$), and the translation into more general ${\rm SL}_2(\mathbb R)$-frames
will be discussed in section \ref{sec:6.3.1} below. The local form of the superstring amplitude then becomes
\beq\label{localForm}
{\cal A}_n(P) = (2\alpha')^{n-3} \int d\mu^n_{P}
\!\!\!\!\! \sum_{AB=23 \ldots n-2} \!\!\!\!\!\big\langle\big(V_{1A}
\cZ_{1A}\big)\big(V_{n-1,\tilde B}\cZ_{n-1,\tilde B}\big) V_n\big\rangle
+ {\rm perm}(23 \ldots n{-}2)\,,
\eeq
where we write ${\cal A}_n(P) = {\cal A}(P)\, \big|_{|P|=n}$ whenever the 
multiplicity is not obvious from the shorthand $P$ for the color-ordering.

\subsubsection{Four-point example} 

While the three-point amplitude (\ref{three}) is completely determined
by zero modes, the simplest instance of OPE contributions occurs at four points. According to (\ref{nptfcttree}),
the four-point correlator is defined by 
\beq
\langle \! \langle V_1(z_1)  U_2(z_2)V_{3}(z_3) V_4(\infty) \rangle
\! \rangle = \langle {\cal K}_4  \rangle\,
|z_{12}|^{-2\alpha's_{12}} |z_{23}|^{-2\alpha's_{23}} 
\label{4ptex.1}
\eeq
and computed by integrating out the $h=1$ fields in $U_2(z_2)$:
\begin{align}
{\cal K}_4 &\cong {V_{[1,2]}(z_1)\over z_{12}}  V_3(z_3) V_4(\infty) + V_1(z_1) {V_{[2,3]}(z_3)\over z_{23}}   V_4(\infty) \notag \\
&\cong \frac{ V_{12} V_3 V_4 }{z_{12}} + \frac{ V_{1} V_{32} V_4 }{z_{32}} \, .
\label{4ptex.2}
\end{align}
The first line illustrates the origin of the four-point correlator from the OPE effective rules
(\ref{OPEeff}).
The second line (which is equivalent by $V_{ij} = V_{[i,j]}$) reproduces (\ref{experE}) and
results from permutations of (\ref{UVs}) and (\ref{VvsL}) while dropping BRST-exact terms and 
OPEs involving the vertex $V_4$ at infinity. The $\langle \ldots \rangle$ bracket only refers to 
the zero modes of $\lambda^\alpha,\theta^\alpha$, see (\ref{tlct}), that is why the positions of
the $V_i,V_{ij}$ are no longer displayed. The integrals in the resulting amplitude 
\begin{align}
{\cal A}(\mathds{1}_4) &=
{\cal A}(1,2,3,4) = 2\alpha' \int^1_0 dz_2 \,  \bigg( \frac{\langle V_{12} V_3 V_4 \rangle }{z_{12}} + \frac{\langle V_{1} V_{32} V_4 \rangle }{z_{32}} \bigg) |z_{12}|^{-2\alpha's_{12}} |z_{23}|^{-2\alpha's_{23}} 
 \notag \\
 &= \bigg( \frac{\langle V_{12} V_3 V_4 \rangle }{s_{12}} +  \frac{\langle V_{1} V_{23} V_4 \rangle }{s_{23}}
 \bigg) \frac{ \Gamma(1-2\alpha's_{12}) \Gamma(1-2\alpha's_{23} ) }{\Gamma(1-2\alpha's_{12}-2\alpha's_{23})}
 \label{4ptex.3} \\
 &=A(1,2,3,4) \frac{ \Gamma(1-2\alpha's_{12}) \Gamma(1-2\alpha's_{23} ) }{\Gamma(1-2\alpha's_{12}-2\alpha's_{23})}
 \notag
\end{align}
can be straightforwardly identified with the Euler beta function $\int^1_0 dx \, x^{A-1} (1-x)^{B-1} 
= \frac{\Gamma(A) \Gamma(B) }{\Gamma(A{+}B)}$ after fixing $(z_1,z_3)=(0,1)$ which
is the backbone of the famous Veneziano amplitude \cite{Veneziano:1968yb}. 
In passing to the second line, we have used the functional identity $\Gamma(A{+}1)=A\Gamma(A)$
to make the ratio $-\frac{s_{12}}{s_{23}}$ of the integrals over $z_{12}^{-1}$
and $z^{-1}_{32}$ manifest
(which is in fact a special case of the integration-by-parts identities of section \ref{IBPsec}).
As a result, the four-point SYM amplitude in \eqref{3ptSYM} has been factored out in the last
line of (\ref{4ptex.3}), and the remainder of this section is dedicated to the appearance of
SYM amplitudes from $n$-point correlators of the open superstring.
Historically, explicit four-point tree-level computations in the pure spinor formalism date
back to 2006 and 2008 \cite{tsimpis, pureids}.

\subsubsection{Five-point example} 

To illustrate the multiparticle techniques leading to the result \eqref{stringWithTs}
above, it is useful to consider the evaluation of the five-point disk correlator
via multiparticle vertex operators and the effective OPE calculations \eqref{OPEeff}.
That is, consider
\beq\label{fivecorrex}
\langle \! \langle V_1(z_1) U_2(z_2) U_3(z_3)V_{4}(z_{4}) V_5(z_5) \rangle \! \rangle = \langle {\cal K}_5  \rangle \,
|z_{12}|^{-2\alpha's_{12}} |z_{13}|^{-2\alpha's_{13}}  |z_{23}|^{-2\alpha's_{23}} |z_{24}|^{-2\alpha's_{24}}  |z_{34}|^{-2\alpha's_{34}} \, ,
\eeq
where we set $(z_1,z_4,z_5)=(0,1,\infty)$ at the end (this means that $V_5$ does not participate
in OPEs). First we eliminate $z_2$ using the OPEs of $U_2(z_2)$ to get
\beq
\cK_5 = {V_{[1,2]}(z_1)\over z_{12}}U_3(z_3)V_4(z_4)V_5(\infty)
+ V_1(z_1){U_{[3,2]}(z_3)\over z_{32}}V_4V_5(\infty)
+ V_1(z_1)U_3(z_3){V_{[4,2]}(z_4)\over z_{42}}V_5(\infty) \, ,
\label{beforeaftz3}
\eeq
followed by elimination of $z_3$ via effective OPEs (\ref{OPEeff}) of $U_3(z_3)$
\begin{align}
\cK_5 &=
{V_{[[1,2],3]}\over z_{12}z_{13}}V_4V_5
+ {V_{[1,2]}\over z_{12}}{V_{[4,3]}\over z_{43}}V_5\notag \\
&\quad
+ {V_{[1,[3,2]]}\over z_{32}z_{13}}V_4V_5
+ V_1{V_{[4,[3,2]]}\over z_{32}z_{43}}V_5 \label{aftz3} \\
&\quad
+ {V_{[1,3]}\over z_{13}}{V_{[4,2]}\over z_{42}}V_5
+ V_1{V_{[[4,2],3]}\over z_{42}z_{43}}V_5\,. \notag
\end{align}
The contributions from the first, second and third term of (\ref{beforeaftz3}) are organized
into separate lines, and we are no longer tracking the locations of the multiparticle vertex 
operators since only their zero modes remain at this point.

Using the generalized Jacobi identity \eqref{Baker} followed by the shuffle symmetry \eqref{cZdef} we get
\begin{align}
\cK_5 &=
{V_{123}\over z_{12}z_{23}}V_4V_5
+ {V_{132}\over z_{13}z_{32}}V_4V_5 
+ {V_{12}\over z_{12}}{V_{43}\over z_{43}}V_5
+ {V_{13}\over z_{13}}{V_{42}\over z_{42}}V_5
 + V_1{V_{432}\over z_{43}z_{32}}V_5
+ V_1{V_{423}\over z_{42}z_{23}}V_5\, ,
\end{align}
which reproduces \eqref{experE} and leads to
\begin{align}
\cA(\mathds{1}_5) &= (2\alpha')^2  \int^1_0  dz_3 \int^{z_3}_0 dz_2 
\, |z_{12}|^{-2\alpha's_{12}} |z_{13}|^{-2\alpha's_{13}}  |z_{23}|^{-2\alpha's_{23}} |z_{24}|^{-2\alpha's_{24}}  |z_{34}|^{-2\alpha's_{34}}  \notag\\
&\quad \times
\bigg[
{\langle V_{123}V_4 V_5\rangle\over z_{12}z_{23}}
+ {\langle V_{12}V_{43}V_5\rangle\over z_{12}z_{43}}
+ {\langle V_{1}V_{423}V_5\rangle\over z_{42}z_{23}}
 + (2\leftrightarrow3)
\bigg] \label{fivcex}
\end{align}
with $(z_1,z_4)=(0,1)$. 

In contrast to the single integral in the four-point amplitude (\ref{4ptex.3}),
the double integrals in (\ref{fivcex}) cannot be expressed in terms of Gamma
functions but instead involve a hypergeometric $_3F_2$ function \cite{Kitazawa:1987xj}
with $s_{ij}$-dependent parameters at $z=1$.
Five-point tree-level computations in the RNS formalism with external bosons include 
\cite{Medina:2002nk, Barreiro:2005hv} from the perspective of low-energy effective actions 
and \cite{Stieberger:2006bh, Stieberger:2006te, Stieberger:2007jv} in the spinor-helicity formalism 
upon dimensional reduction to $D=4$. The simplified five-point results in pure spinor
superspace \cite{5ptsimple, towardsFT} address the entire gauge multiplet and furnish
key steps towards the representation in (\ref{fivcex}).

\subsubsection{Six-point example}

The six-point instance of (\ref{stringWithTs}) is given by
\begin{align}
\cA(\mathds{1}_6) &= (2\alpha')^3  \int^1_0  dz_4 \int^{z_4}_0  dz_3 \int^{z_3}_0 dz_2 
\prod_{1\leq i<j}^5 |z_{ij}|^{- 2\alpha's_{ij}}   \label{sixcex} \\
&\quad \times
\bigg[
{\langle V_{1234}V_5 V_6\rangle\over z_{12}z_{23} z_{34}}
+ {\langle V_{123}V_{54}V_6\rangle\over z_{12} z_{23} z_{54}}
+ {\langle V_{12}V_{543}V_6\rangle\over z_{12} z_{54} z_{43}}
+ {\langle V_{1}V_{5432}V_6\rangle\over z_{54} z_{43}z_{32}}
 + {\rm perm}(2,3,4)
\bigg] \notag
\end{align}
with $(z_1,z_5)=(0,1)$ and builds upon the pure spinor computation in \cite{6ptTree}.
Earlier six-point tree-level computations in the RNS formalism have been performed
for $D$-dimensional external gluons in \cite{6ptOprisa} and in the spinor-helicity formalism
upon dimensional reduction to $D=4$ 
\cite{Stieberger:2006bh, Stieberger:2006te, Stieberger:2007jv, Stieberger:2007am}.

\subsection{Non-local form of the disk correlator}
\label{theIBPsec}

The expression \eqref{stringWithTs} for the massless $n$-point open-superstring
amplitude is characterized by its total number of $(n{-}2)!$ terms,
written in terms of local superfields $V_P$ in the BCJ gauge and $(n{-}2)!$ worldsheet integrals.
The integrands are given in terms of combinations of $\cZ_P\cZ_Q$ functions (\ref{cZdef}) 
with logarithmic singularities and with a distinctive pattern of label distributions among 
the words $P$ and $Q$. We will now see how this form can be streamlined and 
rewritten using only $(n{-}3)!$ terms.

\paragraph{Rearranging worldsheet functions}
The driving force in this rewriting is the judicious use of worldsheet integration by parts in the
presence of the Koba--Nielsen factor \cite{nptStringI}.
To do this, we will first introduce a new set of worldsheet functions $X_P$, indexed by a word $P$,
whose integration-by-parts relations involve {\it constant} rather than $s_{ij}$-dependent coefficients. 
For reasons to become clear below, it is convenient
to define $X_{iP}$ for a fixed label $i$ as
\beq\label{Xdef}
X_{iP} = \sum_Q S(P|Q)_i \cZ_{iQ}\,,
\eeq
where $S(P|Q)_i$ is the KLT matrix (\ref{kltrec}) and $\cZ_{iQ}$ is the shuffle-symmetric worldsheet
function \eqref{cZdef}. For example, $X_{12}=\frac{s_{12}}{z_{12}}$ and
\begin{align}\label{XthreeEx}
X_{123} &= S(23|23)_1\cZ_{123} + S(23|32)_1\cZ_{132} = s_{12}(s_{13}+s_{23}){1\over z_{12}z_{23}} +
s_{12}s_{13}{1\over z_{13}z_{32}}\\
&= s_{12}s_{13}\bigg({1\over z_{12}z_{23}} + {1\over z_{13}z_{32}}\bigg)  + {s_{12}s_{23}\over z_{12}z_{23}} =
{s_{12}s_{13}\over z_{12}z_{13}} + {s_{12}s_{23}\over z_{12}z_{23}}\notag\\
&= {s_{12}\over z_{12}}\bigg({s_{13}\over z_{13}} + {s_{23}\over z_{23}} \bigg)\, ,\notag
\end{align}
where we used partial fractions in the second line. In general, one can show that after using
partial-fraction identities the $X_P$ functions can be written recursively as
\beq\label{genX}
X_{Pi} = X_P (X_{p_1i}+X_{p_2i} + \cdots + X_{p_k i})\, ,\qquad X_i = 1\,,\; k=|P|{-}1\, ,
\eeq
where the base case for a letter $i$ is set to one for later convenience. Solving the recursion
leads to the simplest instances
\begin{align}
\label{Xdefex}
X_1 = 1\, ,\quad X_{12} = {s_{12}\over z_{12}}\, ,\quad
X_{123} = {s_{12}\over z_{12}}\Big({s_{13}\over z_{13}} + {s_{23}\over z_{23}} \Big)
\, ,\quad
X_{1234} = {s_{12}\over z_{12}}\Big({s_{13}\over z_{13}} + {s_{23}\over z_{23}} \Big)
\Big({s_{14}\over z_{14}} + {s_{24}\over z_{24}} + {s_{34}\over z_{34}}\Big)
\end{align}
and more generally to
\beq
X_{P} = \prod_{j=2}^{|P|} \sum_{i=1}^{j-1} \frac{ s_{p_i p_j} }{z_{p_i p_j}} \, .
\label{explicX}
\eeq
One can also describe the worldsheet functions (\ref{Xdef}) in terms of the
\textit{generalized KLT matrix} (\ref{genKLT}) \cite{PScomb}
satisfying generalized Jacobi identities in $P$ and $Q$ such that
$S^\ell(iA|iB)=S(A|B)_i$. The definition \eqref{Xdef} then generalizes to 
arbitrary words (not necessarily starting with $i$) as
\beq\label{Xdefgen}
X_{P}:= {1\over \len{P}}\sum_Q S^\ell(P|Q) \cZ_{Q}
\eeq
such that the factor ${1\over\len{P}}$ compensates the higher number of permutations
being summed over objects that satisfy shuffle symmetry and generalized Jacobi identities.

The following property of (\ref{Xdef}) has been first experimentally
observed in \cite{1loopbb} and
later proved in \cite{PScomb,flas} from the properties of $S^\ell(P|Q)$ in (\ref{Xdefgen}):
\begin{lemma}
The worldsheet functions $X_P$ satisfy the generalized Jacobi identities
\beq\label{XjacsA}
X_{P\ell(Q)} + X_{Q\ell(P)} = 0\,,
\eeq
\end{lemma}
for instance
\beq\label{XjacsB}
X_{12}=-X_{21}\,,\quad
X_{123}+X_{231}+X_{312} = 0\,,\quad
X_{1234}-X_{1243}+X_{3412}+X_{3421}=0\,  .
\eeq

\subsubsection{\label{IBPsec}Integration by parts}

As observed in \cite{nptStringI}, the chain of $\tfrac{s_{ij}}{z_{ij}}$ factors that appear
in $X_P$ is ideally suited for integration by parts (IBP) when multiplied by the Koba--Nielsen 
factor of the disk.
The key idea is to exploit the vanishing of boundary terms in the total worldsheet derivatives 
\beq\label{tot}
\int_{z_a}^{z_b} d z_k \; {\partial \over \partial z_k} { \prod_{1\leq i<j}^{n-1} |z_{ij}|^{-2\alpha' s_{ij}} \over z_{i_1 j_1} \cdots
z_{i_{n-4} j_{n-4}} } = 0 
\eeq
relevant to arbitrary orderings $D(\ldots,a,k,b,\ldots)$ of $n$-point disk amplitudes.
The absence of boundary terms follows from the contributions $|z_k{-}z_b|^{-2\alpha' s_{bk}}$ 
and $|z_k{-}z_a|^{-2\alpha's_{ak}}$ to the Koba--Nielsen factor which evidently vanish as 
$z_k \rightarrow z_b$ and $z_k \rightarrow z_a$ if $\Re(s_{bk}), \Re(s_{ak})<0$.
Analytic continuations in the $s_{ij}$ then imply the validity of (\ref{tot}) for generic complex kinematics 
which has already been used in the context of the canceled-propagator argument in (\ref{10,x3}).

Particularly simple instances of (\ref{tot}) arise if
$z_k$ does not appear in the denominator, i.e.\ if $k \notin \{ i_l,j_l \}$. In these
cases, the derivative ${\partial \over \partial z_k}$ only acts on the Koba--Nielsen factor
and the resulting IBP identity is homogeneously linear in Mandelstam invariants,
\beq\label{total}
\int_{D(P)} d z_2 \, dz_3 \cdots d z_{n-2} { \prod_{1\leq i<j}^{n-1} |z_{ij}|^{-2\alpha's_{ij}} \over z_{i_1 j_1} \cdots z_{i_{n-4} j_{n-4}} } 
\sum_{m=1 \atop{ m \neq k}}^{n-1} {s_{mk} \over z_{mk}} = 0\, ,
\eeq
with an arbitrary permutation $P=p_1 p_2\ldots p_n$ of $12 \ldots n$ characterizing the 
integration domain $D(P)$ in (\ref{domain}). The simplest examples at $n=4$ 
\begin{align}
 \int_{D(P)}  dz_2 \, |z_{12}|^{-2\alpha's_{12}}   |z_{23}|^{-2\alpha's_{23}}{ s_{12} \over z_{12} }
 &= \int_{D(P)} dz_2 \, |z_{12}|^{-2\alpha's_{12}}   |z_{23}|^{-2\alpha's_{23}}{ s_{23} \over z_{23} }
\label{4ptIex}
\end{align}
reproduces the ratio $-\frac{s_{12}}{s_{23}}$ between the four-point integrals
in the first line of (\ref{4ptex.3}) without invoking any Gamma-function identity.
At five points, IBP implies
\begin{align}
 \int_{D(P)}  dz_2 \,dz_3 \prod_{1\leq i<j}^4 |z_{ij}|^{-2\alpha's_{ij}} { s_{12} \over z_{12} }
\left( { s_{13} \over z_{13}} + { s_{23} \over z_{23}} \right)  &=
\int_{D(P)}dz_2 \,dz_3  \prod_{1\leq i<j}^4 |z_{ij}|^{-2\alpha's_{ij}} { s_{12} \over z_{12} }  { s_{34} \over z_{34}}\, .
\label{5ptIex}
\end{align}
Using the notation \eqref{intnot}, the four- and five-point instances of IBP identities 
 \eqref{total} relevant to the local correlators
(\ref{4ptex.3}) and (\ref{fivcex}) are obtained by
\begin{align}
\int d\mu^4_P \, X_{32}  &= -\int d\mu^4_P \, X_{12} \, , \notag \\
\int d\mu^5_P \, X_{12}X_{43}  &= - \int d\mu^5_P \, X_{123}\,, \label{4ptIex5ptIex} \\
\int d\mu^5_P \, X_{432}  &= \int d\mu^5_P  \, X_{123}\notag
\end{align}
and relabelings $2\leftrightarrow 3$ of the five-point cases.
At six points, the set of master IBPs for the correlator in (\ref{sixcex}) is given by
\begin{align}
\int d\mu^6_P \, X_{123}X_{54}  &= -\int d\mu^6_P \, X_{1234}\,, \notag \\
\int d\mu^6_P \, X_{12}X_{543}  &= \int d\mu^6_P \, X_{1234}\,, \label{6ptIexs} \\
\int d\mu^6_P \, X_{5432}  &= -\int d\mu^6_P \, X_{1234} \notag
\end{align}
and permutations in $2,3,4$,
while at seven points we get
\begin{align}
\int d\mu^7_P \, X_{1234}X_{65}  &= -\int d\mu^7_P \, X_{12345}\,, &\int d\mu^7_P \, X_{12}X_{6543}  &= -\int d\mu^7_P \,X_{12345}\,, \notag \\
\int d\mu^7_P \, X_{123}X_{654}  &= \int d\mu^7_P \,X_{12345}\,,  &\int d\mu^7_P \, X_{65432}  &= \int d\mu^7_P \, X_{12345}\,, \label{7ptIexs}
\end{align}
and permutations in $2,3,4,5$.
In general, these IBP identities can be written as
\beq\label{generalIBP}
\int d\mu^n_P \, X_{1A}X_{(n{-}1)\tilde B} = (-1)^\len{B}\int  d\mu^n_P \, X_{1AB}\,,
\eeq
where $\tilde B$ denotes the reversal of the word $B$, see the notation in section~\ref{convIntrosec}.
Recalling that we defined $X_i=1$, the general form \eqref{generalIBP} is
valid even when one of $A$ and $B$ is empty. Note again that the labels $i=1$ and $i=n{-}1$ are
singled out, reflecting the ${\rm SL}_2(\mathbb R)$-frame implicit in the shorthand notation
\eqref{intnot}. Still, the IBP identity (\ref{generalIBP}) holds universally for any disk ordering $P$
since any cyclically inequivalent domain $D(P)$ in (\ref{domain}) is compatible with the 
${\rm SL}_2(\mathbb R)$ frames $(z_1,z_{n-1},z_{n})=(0,1,\infty)$ or $(1,0,\infty)$ employed in (\ref{intnot}).

Moreover, the $|B|$-dependent minus signs cancel out when
the identity \eqref{generalIBP} is used together with the
reflection property (\ref{revM}) of the Berends--Giele supercurrents. This leads to
the important identity,
\beq\label{IBPwithM}
\int d\mu^n_P \, (M_{1A}X_{1A})(M_{n{-}1\tilde B}X_{(n{-}1)\tilde B}) = \int d\mu^n_P \, X_{1AB}\,M_{1A}M_{B(n{-}1)}\,,
\eeq
which will be used in the derivation of the non-local and manifestly BRST invariant form 
of the superstring $n$-point scattering amplitude on the disk in the next section.

\subsubsection{\label{sec:trading}The trading identity}

The IBP identity \eqref{IBPwithM}
will ultimately allow us to derive the non-local $(n{-}3)!$-term representation of the massless
$n$-point superstring disk amplitude in an elegant manner. Before we do this there is
one final important identity to prove, the so-called {\it trading identity} \cite{nptStringI}.
\begin{prop.}
The local superfields $V_P$ satisfying generalized Jacobi identities and the worldsheet functions
$\cZ_P$ satisfying shuffle symmetries are related by
\beq\label{tradingid}
\sum_A V_{iA}\cZ_{iA}
= \sum_A M_{iA} X_{iA}\,
\eeq
to the Berends--Giele supercurrents $M_P$ satisfying shuffle symmetries and the worldsheet
functions $X_P$ satisfying generalized Jacobi identities\footnote{As a side note,
the interplay of the generalized Jacobi identity and shuffle symmetry in
the trading identity \eqref{tradingid} gives
rise to a Lie-series interpretation of the string disk correlator. This same structural behavior
was argued to be present in the genus-one string correlator
and exploited to derive the genus-one correlators up to
seven points (with partial results at eight points) in \cite{oneloopIII}. 
In fact, similar Lie-polynomial structures are expected for correlators at all genera.}.
\end{prop.}
\noindent{\it Proof.} Starting from $V_{iA} = \sum_B S(A|B)_i M_{iB}$ (see \eqref{VSM} and \cite{Zfunctions,flas}), we have
\beq\label{tradingidproff}
\sum_A V_{iA}\cZ_{iA} = \sum_{A,B} M_{iB} S(A|B)_i \cZ_{iA}
= \sum_B M_{iB} X_{iB}\, ,
\eeq
where the second step is based on the symmetry $S(A|B)_i = S(B|A)_i$ of the KLT matrix 
and the definition of $X_{iB}$ in \eqref{Xdef}. Similarly, we could have used the relation
$\cZ_{iA} = \sum_B \langle b(iA),iB\rangle X_{iB}$,
where the binary-tree map introduced in (\ref{bMap}) inverts the 
KLT matrix in the sense of (\ref{ellSb}), to obtain the same conclusion. \qed

\subsubsection{The $n$-point disk amplitude}

The identities derived above allow us to cast the massless
$n$-point disk amplitude into a manifestly gauge invariant form
that contains
$(n{-}3)!$ terms \cite{nptStringI}.
Let us first write down explicit examples at low multiplicities
before stating the final result.

\paragraph{Four points} 
Starting from the rewriting
\beq\label{4ptdisklocal}
{\cal A}_4(P) = 2\alpha'  \int d\mu^4_P \,  \langle V_{12} \cZ_{12} V_3 V_4 + V_{1} V_{32} \cZ_{32} V_4\rangle
\eeq
of the local form (\ref{4ptex.2}) of the four-point disk correlator,
the trading identity \eqref{tradingid} regroups the Mandelstam factors to
\beq
{\cal A}_4(P) = 2\alpha'  \int d\mu^4_P \,  \langle X_{12} M_{12} M_3 M_4 + X_{32} M_{1} M_{32} M_4\rangle\, .
\eeq
Then, IBP using \eqref{4ptIex5ptIex} and the shuffle symmetry of the Berends--Giele
current $M_{32}=-M_{23}$ yield
\begin{align}
{\cal A}_4(P) &= 2\alpha'  \int d\mu^4_P \,  X_{12}\langle M_{12} M_3 M_4 + M_{1} M_{23} M_4\rangle \notag \\
&= 2\alpha'  \int d\mu^4_P \,  X_{12}\langle
E_{123}M_4\rangle \\
& = 2\alpha'  \int d\mu^4_P \,  X_{12}A(1,2,3,4)\,, \notag
\end{align}
where we identified the four-point SYM tree amplitude \eqref{AYM}. After unfolding the
notation \eqref{intnot} we get the equivalent of (\ref{4ptex.3}),
\beq\label{string4}
{\cal A}_4(P) = 2\alpha'  \int_{D(P)} dz_2 \, |z_{12}|^{-2\alpha' s_{12}}  |z_{23}|^{-2\alpha' s_{23}}  \, {s_{12}\over z_{12}} A(1,2,3,4)\,,
\eeq
for the massless four-point superstring amplitude on the disk.

\paragraph{Five points} Similarly, starting from the local form \eqref{fivcex} of the 
five-point disk amplitude, the trading identity \eqref{tradingid} together with
the IBP identities \eqref{4ptIex5ptIex} yields
\begin{align}\label{5ptst}
{\cal A}_5(P)&= ( 2\alpha' )^2\int d\mu^5_P \, \big[ 
\langle V_{123}\cZ_{123}V_4 V_5\rangle
+ \langle V_{12} \cZ_{12}V_{43}\cZ_{43} V_5\rangle
+ \langle V_{1}V_{423}\cZ_{423}V_5\rangle
 + (2\leftrightarrow3) \big]\\
&= ( 2\alpha' )^2 \int d\mu^5_P \,  \big[  \langle X_{123}M_{123}M_4 M_5 + X_{12}X_{43}M_{12}M_{43} M_5 + X_{432}M_1M_{432}M_5\rangle + (2\leftrightarrow3) \big]\cr
 &= ( 2\alpha' )^2 \int  d\mu^5_P \, \big[   X_{123}\langle M_{123}M_4 M_5 + M_{12}M_{34} M_5 + M_1M_{234}M_5\rangle + (2\leftrightarrow3) \big]\cr
&= ( 2\alpha' )^2 \int d\mu^5_P \, \big[   X_{123}\langle E_{1234}M_5\rangle + (2\leftrightarrow3) \big]\cr
&= ( 2\alpha' )^2 \int d\mu^5_P \,  \big[  X_{123}A(1,2,3,4,5) + (2\leftrightarrow3) \big] \,.\notag
\end{align}
In passing to the third line, we have used the instances
$M_{43}=-M_{34}$ and $M_{432}=M_{234}$ of the reflection identity (\ref{revM}).
We then identified the BRST-closed superfields $E_P$ using
\eqref{Edef} and the five-point SYM amplitude from
the pure spinor cohomology formula \eqref{AYM}. Finally, restoring the integrals from the
shorthand notation \eqref{intnot}, we obtain the massless superstring five-point amplitude on the disk:
\begin{align}
{\cal A}_5(P)&= ( 2\alpha' )^2 \int_{D(P)} dz_2 \, dz_3 \prod_{1\leq i<j}^4 |z_{ij}|^{-2\alpha' s_{ij}} 
\label{A5ptstring} \\
&\quad \times \bigg[
{s_{12}\over z_{12}}\bigg({s_{13}\over z_{13}}+{s_{23}\over z_{23}}\bigg)A(1,2,3,4,5)
+ {s_{13}\over z_{13}}\bigg({s_{12}\over z_{12}}+{s_{32}\over z_{32}}\bigg)A(1,3,2,4,5)\bigg]\,. \notag
\end{align}

\paragraph{Six points} The non-local form of the massless six-point disk amplitude can
be derived in similar fashion. Starting from the local form (\ref{sixcex}),
\beq\label{loc6}
{\cal A}_6(P) = (2\alpha')^3 \int d\mu^6_P \, \bigg[ \sum_{AB=234}\!\!\big\langle\big(V_{1A}
\cZ_{1A}\big)\big(V_{5\tilde B}\cZ_{5\tilde B}\big) V_6\big\rangle
+ {\rm perm}(2,3,4) \bigg]\,,
\eeq
we use the trading identity \eqref{tradingid} to obtain
\begin{align}
\cA_6(P) = (2\alpha')^3 \int d\mu^6_P \,  \Big[ \langle\big(
&M_{1234}X_{1234}M_5
+ M_{123}X_{123}M_{54}X_{54} \label{nloc6a}\\
+ &M_{12}X_{12}M_{543}X_{543}
+ M_1 M_{5432}X_{5432}
\big) M_6\rangle + \perm(2,3,4) \Big]\,. \notag
\end{align}
The IBP identities (\ref{6ptIexs}) of the worldsheet functions 
multiplied by the Berends--Giele currents yield
\begin{align}
\cA_6(P) &= (2\alpha')^3 \int  d\mu^6_P \, \big[ X_{1234}\langle
M_{1234}M_5M_5
+ M_{123}M_{45}M_6
+ M_{12}M_{345}M_6
+ M_1 M_{2345}M_6\rangle + \perm(2,3,4) \big] \notag \\
&= (2\alpha')^3 \int d\mu^6_P \, \big[ X_{1234}A(1,2,3,4,5,6) + \perm(2,3,4) \big]\,, 
\label{nloc6b}
\end{align}
where we easily recognize the expansion of $E_{12345}M_6$ from \eqref{Edef} in the first line, and consequently
of the tree-level six-point SYM amplitude \eqref{AYM} in the last line. So finally \cite{nptStringI},
\begin{align}
{\cal A}_6(P)&= (2\alpha')^3 \int_{D(P)}\!\!\! dz_2 \,dz_3\, dz_4\! \prod_{1\leq i<j}^5 \! |z_{ij}|^{-2\alpha' s_{ij}}  \label{A6ptstring} \\
& \quad \times \bigg[
{s_{12}\over z_{12}}\bigg({s_{13}\over z_{13}}+{s_{23}\over z_{23}}\bigg)
\bigg({s_{14}\over z_{14}}+{s_{24}\over z_{24}}+{s_{34}\over z_{34}}\bigg)A(1,2,3,4,5,6) +
\perm(2,3,4)\bigg]\,.
\notag
\end{align}

\paragraph{Higher points}
Since all the key formulae above generalize to any multiplicity --
the local version of the open string disk correlator
\eqref{stringWithTs}, the trading identity (\ref{tradingid}), the IBP relations \eqref{IBPwithM} 
and the pure spinor cohomology formula \eqref{AYM}
for SYM tree amplitudes -- we
propose the following generalization \cite{nptStringI}:
The massless $n$-point superstring disk amplitude is given by
\beq\label{nptdisk}
{\cal A}_n(P)= (2\alpha')^{n-3} \int_{D(P)} \prod_{j=2}^{n{-}2} dz_j \prod_{1\leq i<j}^{n-1} |z_{ij}|^{- 2\alpha' s_{ij}} 
\bigg[ \prod_{k=2}^{n{-}2}\sum_{m=1}^{k-1} {s_{mk} \over z_{mk}}\; A(1,2,\ldots,n)  + \perm(2,3,\dots,n{-}2) \bigg]\, .
\eeq
We therefore see that the multiparticle superfield techniques and several related combinatorial
identities, all inspired by the simplicity of the pure spinor formalism, lead to a striking simplification of
the $n$-point superstring disk amplitude: 
\begin{itemize}
\item All polarization dependence is carried by a linear combination of
$(n{-}3)!$ field-theory SYM amplitudes $A(1,Q,n{-}1,n)$ with $Q=q_2q_3\ldots q_{n-2}$ a permutation
of $2,3,\ldots,n{-}2$. These $(n{-}3)!$ permutations in fact form a basis under the BCJ relations
(\ref{BCJrelations}) or (\ref{fundBCJ}).
\item All the $\alpha'$-dependence of the $n$-point amplitude (\ref{nptdisk}) resides in the
disk integrals over permutations of $X_{12\ldots n-2}=\prod_{k=2}^{n{-}2}\sum_{m=1}^{k-1} 
{s_{mk} \over z_{mk}}$ multiplying the SYM amplitudes. Hence, all the string corrections
to SYM field-theory are carried by scalar, i.e.\ polarization independent, integrals.
\end{itemize}
Additional structures become visible when restricting the integration domains of
(\ref{nptdisk}) to the $(n{-}3)!$-family of $D(1,P,n{-}1,n)$ with $P=p_2p_3\ldots p_{n-2}$
a permutation of $2,3,\ldots,n{-}2$, see the definition in \eqref{domain}. 
This $(n{-}3)!$-vector of color-ordered string amplitudes
\beq\label{npttree}
{\cal A}(1,P,n{-}1,n;\alpha') = \sum_{Q \in S_{n-3}} F_P{}^{Q}(\alpha') A(1,Q,n{-}1,n)
\eeq
can then be organized
through the following square matrix of integrals
\begin{align}\label{Fdef}
F_P{}^{Q}(\alpha')&:= 
(2\ap)^{n-3} \! \! \! \! \! \! \! \! \! \!  \! \! \! \! \! \! \!
\int\limits_{0 < z_{p_2}< z_{p_3} < \ldots < z_{p_{n-2}} < 1}
\! \! \! \! \! \! \! \! \! \! \! \! \! \! \! \! \!  dz_2 \, dz_3\, \ldots \, dz_{n-2}
\prod_{1\leq i<j}^{n-1} |z_{ij}|^{-2\alpha' s_{ij}} \, { s_{1 q_2} \over z_{1 q_2}}
\left(  { s_{1 q_3} \over z_{1 q_3}} + { s_{q_2 q_3} \over z_{q_2 q_3}}\right) \\
& \ \ \ \ \ \times \left(  { s_{1 q_4} \over z_{1 q_4}} + { s_{q_2 q_4} \over z_{q_2 q_4}} + { s_{q_3 q_4} \over z_{q_3 q_4}}
\right) \ldots \left({s_{1 q_{n-2}} \over z_{1 q_{n-2}}} + { s_{q_2 q_{n-2}} \over z_{q_2 q_{n-2}}}
+\ldots + { s_{q_{n-3} q_{n-2}} \over z_{q_{n-3} q_{n-2}}}\right)\,,
\notag
\end{align}
indexed by permutations $P$ and $Q$ of the $n{-}3$ labels $23\ldots n{-}2$. On the one hand, there
is no obstruction to extending (\ref{npttree}) beyond the $(n{-}3)!$ disk orderings $D(1,P,n{-}1,n)$ -- more
general choices would simply place some of the integration variables of (\ref{Fdef}) into the
regions $(-\infty,0)$ and $(1,\infty)$. On the other hand, as will be elaborated in section \ref{sec:6.5.1},
the $(n{-}3)!$ color-ordered open-string amplitudes (\ref{npttree}) already form a basis of the complete
$(n{-}1)!$-family of ${\cal A}_n(Q)$ in the color-dressed amplitude (\ref{colordr}).

One can already anticipate from the symmetric footing of color-ordered
open-string and SYM amplitudes in (\ref{npttree}) that the field-theory limit
of $F_P{}^{Q}(\alpha')$ yields a Kronecker delta in the permutations $P,Q$,
\beq
F_P{}^{Q}(\alpha') = \delta_P^Q +{\cal O}(\ap^2)\, ,
\label{ftofFPQ}
\eeq
and we will study this relation and its $\alpha'$-corrections from several perspectives.

\subsection{The open superstring as a field-theory double copy}

We shall now relate the form of the disk integrand in the $n$-point
open-string amplitude (\ref{nptdisk}) to the structure of the KLT formula (\ref{KLTrel}) 
for gravitational tree amplitudes. In the same way as KLT formulae in field
theories are hallmarks of double copy, the form of the disk amplitude is
argued to identify
the interactions of massless open-superstring excitations as a
double copy of SYM with a theory of bicolored scalars dubbed $Z$-theory.

\subsubsection{Parke--Taylor factors and $Z$-integrals}
\label{sec:6.3.1}

As pointed out above, the calculation of the string disk amplitudes was carried out in the
${\rm SL}_2(\mathbb R)$ frames where $(z_{1}, z_{{n-1}} ,  z_{n})$ are fixed to one of
$(0,1,\infty)$ or $(1,0,\infty)$ to account for the residual M\"obius symmetry of the disk. 
In order to generalize the $n$-point formula \eqref{nptdisk} to arbitrary ${\rm SL}_2(\mathbb R)$ 
frames, we need to undo the above fixing of $z_{1}, z_{{n-1}} ,  z_{n}$. The task is to identify 
an {\it ${\rm SL}_2(\mathbb R)$-covariant uplift} of the worldsheet functions $\cZ_{1A}
\cZ_{n-1,\tilde B}$
or $X_{1Q}$ in the amplitude representations (\ref{stringWithTs}) or (\ref{nptdisk}). In other
words, it remains to reverse the ${\rm SL}_2(\mathbb R)$-fixing ($D(P)$ is defined in
\eqref{domain})
\beq
\int_{D(P)}  {dz_1 \ dz_2 \ \cdots
\ dz_n \over {\rm vol}({\rm SL}_2(\Bbb R))} 
= | z_{1,n-1} z_{1,n} z_{n-1,n} | 
\int_{D(P)} 
dz_{2} \, dz_3 \, \ldots \, dz_{n-2}\,,
\label{volsl}
\ee
and to identify a suitable function $f_{A,B}(z_1,\ldots,z_n)$ such that,
\beq
\lim_{(z_1,z_{n-1},z_n) \rightarrow (0,1, \infty)} | z_{1,n-1} z_{1,n}
z_{n-1,n} | \cdot f_{A,B}(z_1,\ldots,z_n) =  \cZ_{1A} \cZ_{n-1,B} \,
\Big|^{z_1=0}_{z_{n-1}=1}\,,
\label{Zuplift.1}
\eeq
or $z_1\leftrightarrow z_{n-1}$.
The Jacobian $| z_{1,n-1} z_{1,n} z_{n-1,n} |$ on the right-hand side of (\ref{volsl}) is part of
the prescription $1/{\rm vol}({\rm SL}_2(\Bbb R))$ that avoids an infinite overcount of 
$z_j$-configurations that are related by M\"obius transformations
\cite{gswI, Polchinski:1998rq, Blumenhagen:2013fgp}.\footnote{One could alternatively fix any other triplet of punctures $z_a,z_b,z_c$
and change the Jacobian and measure on the right-hand side of (\ref{volsl}) to $| z_{ab} z_{ac} z_{bc} |$
and $\prod_{j\neq a,b,c}^n dz_{j} $.} The desired uplift
$f_{A,B}$ is uniquely determined by (\ref{Zuplift.1}) and requiring  ${\rm SL}_2(\mathbb R)$-weight
two in each variable: in the same way as
\beq
\frac{1}{z_{i} - z_{j}} \rightarrow \frac{ (c z_i + d)(c z_j + d)}{z_{i}-z_j}  \ \
{\rm under} \ \ z_k \rightarrow \frac{az_k + b}{cz_k+d} \  \ {\rm with}\ \ \big(\begin{smallmatrix} a &b \\ c &d \end{smallmatrix} \big) \in {\rm SL}_2(\mathbb R)
\label{Zuplift.2}
\eeq
is said to have ${\rm SL}_2(\mathbb R)$-weight one in $z_i, z_j$, the uplift $f_{A,B}$ 
is required to transform as
\beq
f_{A,B}(z_1,\ldots,z_n) \rightarrow f_{A,B}(z_1,\ldots,z_n) \prod_{j=1}^n (c z_j + d)^2
\label{Zuplift.3}
\eeq
to yield well-defined integrals with the measure on the left-hand side of (\ref{volsl}).
The simplest (though not the only) quantities with ${\rm SL}_2(\mathbb R)$-weight two in 
all of $z_1,z_2,\ldots , z_n$ are the so-called {\it Parke--Taylor factors}
\beq
\PT(c_1,c_2,\ldots,c_n) = \frac{1}{z_{c_1c_2} z_{c_2 c_3}\ldots z_{c_{n-1}c_{n}} z_{c_n c_1}}\, .
\label{Zuplift.4}
\eeq
After adapting the permutation $C=c_1 c_2\ldots c_n$ to the target expression 
$\cZ_{1A}\cZ_{n-1,B}$, it is easy to check that the ${\rm SL}_2(\mathbb R)$-covariant 
solution to (\ref{Zuplift.1}) is given by
\beq
f_{A,B}(z_1,\ldots,z_n) = (-1)^{| B|-1} \PT(1,A,n,\tilde B, n{-}1)
\label{Zuplift.5}
\eeq
with $\tilde B$ the reversal of $B$. In other words, the functions $\cZ_{1A}\cZ_{n-1,B}$
in the local representation (\ref{stringWithTs}) of the $n$-point amplitude descend from
Parke--Taylor integrals
\beq
\int_{D(P)}  {dz_1 \ dz_2 \ \cdots
\ dz_n \over {\rm vol}({\rm SL}_2(\Bbb R))}  \PT(1,A,n,\tilde B, n{-}1)
=   (-1)^{| B|-1} 
\int_{D(P)}
dz_{2} \, dz_3 \, \ldots \, dz_{n-2} \, \cZ_{1A}\cZ_{n-1,B}\, ,
\label{volslPT}
\ee
and the simplest examples are given by
\beq
\PT(1,2,4,3) \rightarrow - \frac{1}{z_{12}}\ , \ \ \ \ \ \ 
\PT(1,4,2,3) \rightarrow  \frac{1}{z_{32}}
\eeq
as well as
\begin{align}
\PT(1,2,3,5,4) &\rightarrow - \frac{1}{z_{12} z_{23}} \,, 
&
\PT(1,2,5,3,4) &\rightarrow  \frac{1}{z_{12} z_{43}} \, ,
&
\PT(1,5,2,3,4) &\rightarrow  -\frac{1}{z_{43} z_{32}} \, ,
\notag \\
\PT(1,3,2,5,4) &\rightarrow - \frac{1}{z_{13} z_{32}} \,, 
&
\PT(1,3,5,2,4) &\rightarrow  \frac{1}{z_{13} z_{42}} \, ,
&
\PT(1,5,3,2,4) &\rightarrow  -\frac{1}{z_{42} z_{23}}\, .
\end{align}
Upon dressing with the Koba--Nielsen factor of ${\rm SL}_2(\mathbb R)$-weight zero in each
variable, the gauge-fixed integrals in (\ref{stringWithTs}) and \eqref{nptdisk} are
found to be expressible in terms of Parke--Taylor- or $Z$-integrals defined by
\beq\label{Zintdef}
Z(P|q_1,q_2,\ldots,q_n) :=(2 \ap)^{n-3}\!\!\!\!
\int \limits_{D(P)} {dz_1 \ dz_2 \ \cdots  \ dz_n \over {\rm vol}({\rm SL}_2(\Bbb R))}
 { \prod_{i<j}^n |z_{ij}|^{-2\ap s_{ij}}  \over z_{q_1 q_2} z_{q_2 q_3} \ldots z_{q_{n-1} q_n} z_{q_n q_1}} \,.
\ee
In this setting, $Z$-integrals are labelled by two permutations $P,Q \in S_{n}$ up to cyclic identifications
in $P$ or~$Q$: the permutation $P := p_1 p_2 \ldots p_n$ in the first entry encodes the integration 
domain $D(P)$ in (\ref{domain}), while the second permutation $Q := q_1 q_2 \ldots q_n$ refers
to a Parke--Taylor factor (\ref{Zuplift.4}) in the integrand.
In summary,
\begin{lemma}
The ${\rm SL}_2(\mathbb R)$-covariant uplift of the worldsheet integrals
appearing in the local form of the superstring amplitude \eqref{stringWithTs} is given by
\beq\label{ZZToZint}
(2\ap)^{n-3} \int d\mu_P^n \, \cZ_{1A}\cZ_{n{-}1\tilde B} = - (-1)^\len{B} Z(P|1,A,n,B,n{-}1)\,,
\eeq
where the measure $d\mu_P^n$ is defined in \eqref{intnot}.
\end{lemma}
Thus, the local form (\ref{localForm}) of the superstring
amplitude can be rewritten as
\beq\label{localFormWithZ}
{\cal A}_n(P) = -
\!\!\!\!\!\!\!\!\sum_{AB=23 \ldots n-2}\!\!\!\!\!\!\!\!\big\langle V_{1A}
V_{n-1,\tilde B} V_n\big\rangle (-1)^\len{B}Z(P|1,A,n,B,n{-}1)
+ \perm(23 \ldots n{-}2)\,,
\eeq
for instance
\begin{align}
{\cal A}_4(P) &= - \langle V_{12} V_3 V_4 \rangle Z(P|1,2,4,3) 
+  \langle V_{1} V_{32} V_4 \rangle Z(P|1,4,2,3) \, ,
\notag \\
{\cal A}_5(P) &= - \langle V_{123} V_4 V_5 \rangle Z(P|1,2,3,5,4) 
+\langle V_{12} V_{43} V_5 \rangle Z(P|1,2,5,3,4) 
- \langle V_{1} V_{432} V_5 \rangle Z(P|1,5,2,3,4) 
\label{exZloc} \\
&\quad - \langle V_{132} V_4 V_5 \rangle Z(P|1,3,2,5,4) 
+\langle V_{13} V_{42} V_5 \rangle Z(P|1,3,5,2,4) 
- \langle V_{1} V_{423} V_5 \rangle Z(P|1,5,3,2,4)\, .
\notag
\end{align}

\subsubsection{Open superstrings as a KLT formula}

The $F_P{}^{Q}$-functions (\ref{Fdef}) in the $n$-point disk amplitude (\ref{npttree}) are integrals 
of worldsheet functions $X_{1Q}$ as one can readily check from their expressions in (\ref{explicX}),
\beq
F_P{}^{Q}(\alpha')=
(2\ap)^{n-3}  
\! \! \! \int\limits_{D(1,P,n-1,n)} \! \! \! 
   dz_2 \, dz_3\, \ldots \, dz_{n-2}
\prod_{1\leq i<j}^{n-1} |z_{ij}|^{-2\alpha' s_{ij}}  X_{1Q}\, .
\eeq
In \eqref{Xdef}, the integrands $X_{1Q}$ were related to the chains $\cZ_{1R}$ 
of simple poles through the KLT kernel $S(Q|R)_1$ defined in \eqref{kltrec}. Accordingly,
one can represent the $F_P{}^{Q}(\alpha')$ via
\begin{align}
F_P{}^{Q}(\alpha') &= (2\ap)^{n-3} \sum_{R\in S_{n-3}} S(Q|R)_1 
\! \! \! \int\limits_{D(1,P,n-1,n)} \! \! \!   dz_2 \, dz_3\, \ldots \, dz_{n-2}
\prod_{1\leq i<j}^{n-1} |z_{ij}|^{-2\alpha' s_{ij}} \, \cZ_{1R}
\\
&= - (2\ap)^{n-3} \sum_{R\in S_{n-3}} S(Q|R)_1  
\! \! \! \int\limits_{D(1,P,n-1,n)} \! \! \!   \frac{ dz_1 \, dz_2\, \ldots \, dz_{n}}{{\rm vol} ( {\rm SL}_2(\mathbb R))} 
\prod_{1\leq i<j}^{n} |z_{ij}|^{-2\alpha' s_{ij}} \, \PT(1,R,n,n{-}1)\, ,
\notag
\end{align}
using (\ref{volslPT}) at $B=\emptyset$ in passing to the ${\rm SL}_2(\Bbb R)$-covariant 
last line. 
This identifies the integrals $F_P{}^{Q}$ to be linear combinations of
$Z$-integrals \eqref{Zintdef} selected by the KLT kernel \cite{Zfunctions},
\beq\label{KLTopen}
F_P{}^{Q} = - \sum_{R \in S_{n-3}} S(Q|R)_1 Z(1,P,n{-}1,n|1,R,n,n{-}1) \ .
\eeq
Hence, the $n$-point open-superstring amplitude \eqref{npttree}
takes the form of the field-theory KLT relations \cite{Zfunctions}
\beq\label{AstringP}
{\cal A}_n(P) = - \sum_{Q,R \in S_{n-3}} Z(P|1,R,n,n{-}1) S(R|Q)_1 A(1,Q,n{-}1,n)
\eeq
with one of the two SYM factors in the supergravity tree amplitude (\ref{KLTrel}) 
replaced by disk integrals $A(1,R,n,n{-}1) \rightarrow Z(P|1,R,n,n{-}1)$. This is
the case for any choice of the open-string color ordering $P$ in (\ref{AstringP}) 
which is a spectator in the sum over permutations $Q,R$ entering the KLT kernel.
We have dropped the restriction of the disk integrals (\ref{nptdisk}) to the $(n{-}3)!$-family of
domains $D(1,P,n{-}1,n)$ which was convenient to organize the $F_P{}^Q$ in (\ref{Fdef})
into a square matrix.

Given that the KLT formula is a central tree-level incarnation of 
the double-copy structure in perturbative gravity, it is tempting to interpret (\ref{AstringP}) 
as signaling the open superstring to be a double copy. Indeed, we support
the interpretation of the $Z$-integrals (\ref{Zintdef}) as amplitudes in a theory of
bi-colored scalars in sections \ref{subsec:biadj}, \ref{sec:7.7} and \ref{sec:7.5}.

We conclude by illustrating the KLT products (\ref{KLTopen}) and (\ref{AstringP})
through their four- and five-point examples:

\paragraph{Four points} Since the permutation sums over $S_{n-3}$ trivialize at $n=4$
and the KLT kernel becomes a scalar $S(2|2)_1=s_{12}$, we find the simple results
\begin{align}
F_2{}^{2} &= - s_{12} Z(1,2,3,4|1,2,4,3) = \frac{\Gamma(1-2\alpha' s_{12}) \Gamma(1-2\alpha' s_{23}) }{\Gamma(1-2\alpha' s_{12}-2\alpha' s_{23})} \, ,\notag \\
{\cal A}_4(P)&= -  Z(P|1,2,4,3) s_{12} A(1,2,3,4)\, ,
\end{align}
where we have imported the Gamma-function representation of the four-point 
disk integral from (\ref{4ptex.3}) in the first line.

\paragraph{Five points} At five points, the permutation-inequivalent entries of the symmetric
KLT kernel are $S(23|23)_1 = s_{12}(s_{13}{+}s_{23})$ and $S(23|32)_1 = s_{12}s_{13}$. The resulting
functions $F_P{}^Q$ and KLT representation of the disk amplitude are
\begin{align}
F_{23}{}^{23} &= - s_{12}(s_{13}{+}s_{23}) Z(1,2,3,4,5|1,2,3,5,4)
- s_{12}s_{13} Z(1,2,3,4,5|1,3,2,5,4) \\
F_{23}{}^{32} &= - s_{12}s_{13} Z(1,2,3,4,5|1,2,3,5,4)
-  s_{13}(s_{12}{+}s_{23})  Z(1,2,3,4,5|1,3,2,5,4) \notag 
\end{align}
as well as
\begin{align}
{\cal A}_5(P)&=  - \left( \begin{matrix}  Z(P|1,2,3,5,4) \\
Z(P|1,3,2,5,4)  \end{matrix} \right)^T
 \left( \begin{matrix}  s_{12}(s_{13}{+}s_{23}) & s_{12}s_{13} \\
 s_{12}s_{13} &s_{13}(s_{12}{+}s_{23}) \end{matrix} \right)
 \left( \begin{matrix}  A(1,2,3,4,5) \\ A(1,3,2,4,5) \end{matrix} \right)\, .
\end{align}

\subsubsection{KK and BCJ relations of $Z$-integrals}
\label{sec:6.3.3}

The KLT formula (\ref{KLTrel}) does not manifest the permutation symmetry of
the supergravity amplitude by the sum over $(n{-}3)!$ rather than $(n{-}1)!$ 
color-orderings of the two types of SYM amplitudes.\footnote{See \cite{KLTpaper} 
for alternative versions of the KLT formula with manifest permutation symmetry and
\cite{aomotoKLT} for an early discussion thereof in the mathematics literature.} One can verify
on the basis of the KK and BCJ relations (\ref{KKrelation}) and (\ref{fundBCJ}) 
of SYM amplitudes that different choices of the legs $(1,n{-}1,n) \leftrightarrow (a,b,c)$
excluded from the permutation sums yield equivalent KLT formulae.

In the context of disk amplitudes, the KLT formula in (\ref{AstringP}) ultimately
applies to the $n$-point correlator (\ref{nptfcttree}) in the amplitude prescription 
(\ref{treepresc}),
\begin{align}
\langle {\cal K}_n \rangle =   -
\frac{ dz_1 \, dz_{n-1} \, dz_n}{ {\rm vol}({\rm SL}_2(\Bbb R)) }
  \sum_{Q,R \in S_{n-3}} \PT(1,R,n,n{-}1) S(R|Q)_1 A(1,Q,n{-}1,n)
\ {\rm mod} \ \nabla_{z_k}\, ,
\label{KLTcorrel}
\end{align}
where the total Koba--Nielsen derivatives $\nabla_{z_k} f $ discarded in the IBP procedure 
are not tracked,
\beq
\nabla_{z_k} f = \partial_{z_k} f -2\alpha' f \sum_{j=1\atop {j\neq k}}^n \frac{s_{kj} }{z_{kj}}\, .
\label{defnabf}
\eeq
By the discussion in section \ref{sec:treeprescr}, the integrated correlator does
not depend on the distribution of external legs to integrated and unintegrated vertices,
so the KLT representation (\ref{KLTcorrel}) of the correlator is bound to be permutation invariant.
In the same way as the permutation invariance of supergravity amplitudes originates from 
KK and BCJ relations of its SYM constituents, the symmetry of the correlator requires
KK and BCJ relations of the Parke--Taylor factors (\ref{Zuplift.4}) modulo total 
Koba--Nielsen derivatives. The factors of
$dz_1  dz_{n-1}  dz_n$ on the right-hand
side of (\ref{KLTcorrel}) are merely a reminder of the ${\rm SL}_2(\Bbb R)$ frame
used to define ${\cal K}_n$ in (\ref{nptfcttree}) and lead to a permutation invariant
measure upon insertion into (\ref{treepresc}).

For a fixed choice of the integration domain $D(P)$, the $Z(P|Q)$ integrals \eqref{Zintdef}
associated with different permutations of $Q=q_1,q_2,\ldots,q_n$ indeed satisfy
the same relations as color-ordered SYM amplitudes. First, cyclic
symmetry and reflection parity immediately follow from the definition
(\ref{Zuplift.4}) of Parke--Taylor factors,
\begin{align}
Z(P|q_1,q_2,q_3 \ldots,q_n) &= Z(P|q_2,q_3, \ldots,q_n,q_1)\, ,  \label{cycZ} \\
Z(P|q_1,q_2, \ldots,q_n) &= (-1)^n Z(P|q_n,\ldots, q_2,q_1)\,.
\notag
\end{align}
Second, partial-fraction rearrangements of the integrand imply,
\begin{align}
Z(P|1,A,n,B) &=   (-1)^\len{B} 
Z(P|1,\tilde B \shuffle A,n)\,,\quad\forall \, A,B\, , \label{ZKK}
\end{align}
which is the direct $A(\cdot) \rightarrow Z(P|\cdot)$ analogue
of the KK relation (\ref{KKrelationagain}). Third, IBP
relations as in section \ref{IBPsec} take the form of the BCJ relations 
(\ref{fundBCJ}) with $A(\cdot) \rightarrow Z(P|\cdot)$
\begin{align}
0 &= \sum_{j=2}^{n-1} (k_{q_1}\cdot k_{q_2 q_3\ldots q_j}) Z(P| q_2,q_3,\ldots
,q_{j},q_1 ,q_{j+1}, \ldots ,q_n)\, , \label{ZBCJ}
\end{align}
see appendix B of \cite{Zfunctions} for a proof. By analogy with (\ref{BCJrelations}),
an equivalent system of integration-by-parts relations is furnished by
\beq
Z(P| \{Q,R\},n) = 0\, ,
\label{ZequivBCJ}
\eeq
see section \ref{Sbracketsec} for the $S$ bracket.
The fact that $Z(P|\cdot)$-integrals at fixed choice of $P$ obey direct analogues
(\ref{cycZ}) to (\ref{ZequivBCJ}) of field-theory amplitude relations supports the
interpretation of disk integrals as amplitudes in a scalar theory.

Note that the expressions for $n$-point disk amplitudes in four-dimensional MHV helicities
in \cite{Stieberger:2012rq} follow from the relabeling of (\ref{KLTcorrel}) involving
the Parke--Taylor basis ${\rm PT}(1,2,3,R)$ with permutations $R$ of $4,5,\ldots,n$
as well as the Parke--Taylor formula \cite{Parke:1986gb} for the dimensional reduction of the 
SYM amplitudes.

\subsubsection{Bi-adjoint scalars from the field-theory limit of $Z$-integrals}
\label{subsec:biadj}

The interpretation of $Z$-integrals (\ref{Zintdef}) in terms of scalar field theory is
further substantiated by their low-energy limits $\alpha' \rightarrow 0$,
where tree-level amplitudes of bi-adjoint scalars are recovered. More specifically, we encounter the
theory of bi-adjoint scalars $\Phi:=\Phi_{i|a}t^i\otimes \tilde t^a$ taking values
in the tensor product $U(N) \times U(\tilde N)$ of color groups with associated
structure constants $f_{ijk}$ and $\tilde f_{abc}$, i.e.\
$[t^i,t^j] = f_{ijk}t^k$ and $[\tilde t^a,\tilde t^b] = \tilde f_{abc}\tilde t^c$.
The Lagrangian defining the bi-adjoint theory features a cubic interaction
\beq\label{Lagphi}
{\cal L}_{\phi^3} = \half\p_m \Phi_{i|a}\p^m\Phi_{i|a}
+ {1\over 3!}f_{ijk}\tilde f_{abc}\Phi_{i|a}\Phi_{j|b}\Phi_{k|c} \, ,
\eeq
and by the two types of adjoint indices of the scalars, its
tree amplitudes can be expanded in terms of two species
of independent traces involving either $t^i$ or $\tilde t^a$,
\beq\label{DPdefCP}
M^{\phi^3}_n = \sum_{P,Q}\Tr(t^{1P})
\Tr(\tilde t^{1Q})m(1,P|1,Q)
\,,\qquad t^{1P}:=t^1t^{p_1}t^{p_2} \ldots t^{p_{n{-}1}}
\,,\qquad \tilde t^{1Q}:=\tilde t^1 \tilde t^{q_1}\tilde t^{q_2} \ldots \tilde  t^{q_{n{-}1}}\, .
\eeq
This doubles the color-decomposition of open-string and gauge-theory
tree amplitudes in (\ref{colordr}), and the color-independent building blocks
$m(A|B)$ are referred to as {\it doubly-partial} amplitudes \cite{DPellis}. From the
Feynman rules of the Lagrangian (\ref{Lagphi}), the doubly-partial amplitudes
solely depend on the $s_{ij\ldots k}$ via propagators of tree-level diagrams 
with cubic vertices or {\it cubic diagrams} for short.

Also the field-theory limit of disk integrals \eqref{Zintdef} yields
kinematic poles that correspond to the propagators of cubic diagrams 
(or planar binary trees) \cite{scherkFT,Frampton, nptStringII}. These poles appear only in the
planar channels of the associated planar binary trees, corresponding to groups of adjacent external
particles in the planar trees. Luckily, these adjacent poles in
the field-theory limits of the disk integrals\footnote{The factor of $(2\ap)^{n-3}$ in the
definition \eqref{Zintdef} of $Z(P|Q)$ guarantees that the leading term
in the low-energy expansion is of order $s_{ij}^{3-n}$, without any
accompanying factors of $\ap$.} \eqref{Zintdef} admit a nice combinatorial
expansion encoded in the doubly-partial amplitudes \cite{DPellis}
\beq\label{ftlim}
\lim_{\ap\to0} Z(P|Q) = m(P|Q)\,.
\ee
In other words, the bi-adjoint scalar theory \eqref{Lagphi} gives the
field-theory limit of $Z$-integrals, as expected from the early discussions
of \cite{scherkFT,Frampton}. Hence, bi-adjoint scalars furnish the low-energy
limit of $Z$-theory.

\paragraph{Berends--Giele double currents}
While the kinematic poles in $Z$-integrals have been systematically studied
at various multiplicities from several perspectives
\cite{Stieberger:2006te, nptStringII, Zfunctions, Brown:2019wna}, the straightforward
Feynman-diagram expansion and rich combinatorial structure of the doubly-partial amplitudes
make (\ref{ftlim}) a rewarding shortcut for the computation of field-theory limits.
In particular, we shall now introduce {\it Berends--Giele double currents}
that encode the planar-binary-tree expansion of $m(P|Q)$ and
offer a highly efficient approach to the kinematic poles of $Z$-integrals.

The field equation following from the Lagrangian \eqref{Lagphi} can be written as
\beq\label{EOMPhi}
\Box \Phi = [\![\Phi,\Phi]\!]\,,
\eeq
where we define $[\![\Phi,\Phi]\!]:= (\Phi_{i|a}\Phi_{j|b}-\Phi_{j|a}\Phi_{i|b})t^it^j \otimes 
\tilde t^a\tilde t^b$ for $\Phi:= \Phi_{i|a}t^i \otimes \tilde t^a$.
A solution to the non-linear field equation \eqref{EOMPhi} can be constructed
perturbatively in terms of {\it Berends--Giele double currents} $\phi_{P|Q}$ with
the ansatz \cite{FTlimit},
\beq\label{doubleLie}
\Phi(X) := \sum_{P,Q} \phi_{P|Q}\,t^P \otimes \tilde t^Q\, e^{k_P\cdot X},
\qquad t^P:= t^{p_1}t^{p_2} \ldots t^{p_{|P|}}
\eeq
which generalizes the perturbiner expansion (\ref{series}) in SYM to two
species of Lie-algebra generators. Since the ansatz \eqref{doubleLie}
contains the plane-wave factor $e^{k_P\cdot X}$ (as opposed to $e^{k_Q\cdot X}$),
the coefficients $\phi_{P|Q}$ must vanish unless $P$ is a permutation of $Q$
in order to have a well-defined multiparticle interpretation, i.e.
\beq\label{BGconstraint}
\phi_{P|Q}:= 0\, ,\qquad \hbox{if } P\setminus Q\neq\emptyset\, .
\eeq
Plugging the ansatz \eqref{doubleLie}
into the field equation \eqref{EOMPhi}
leads to the following recursion \cite{FTlimit}
\beq\label{BGphi}
\phi_{P|Q} = {1\over s_P}\sum_{XY=P}\sum_{AB=Q}\big(\phi_{X|A}\phi_{Y|B} -
(X\leftrightarrow Y)\big)\,,\qquad \phi_{i|j}=\d_{ij}\,,
\eeq
see (\ref{defmands}) for the definition of multiparticle Mandelstam invariants $s_P$.
The recursion terminates with the single-particle double current subject to the
linearized equation $\Box\phi_{i|j}e^{k_i\cdot X}=0$ such that $k_i^2=0$,
and we can pick normalization conventions where $\phi_{i|j}=\delta_{ij}$.
Given that the two entries $P,Q$ of the currents enter the recursion
(\ref{BGphi}) on equal footing, the symmetry $\phi_{i|j}=\phi_{j|i}$ propagates
to arbitrary rank,
\beq\label{symFTphi}
\phi_{P|Q}=\phi_{Q|P}\,.
\eeq
Since the summands on the right-hand side of \eqref{BGphi} are antisymmetric in both
$X ,Y$ and $A,B$, the shuffle symmetry
\beq\label{anotherBGsym}
\phi_{A\shuffle B|Q} = 0 \ \forall \ A,B\neq \emptyset \, , \ \ \ \ \ \
 \phi_{A|P\shuffle Q} = 0 \ \forall \ P,Q\neq \emptyset
\eeq
follows from the same type of combinatorial proof 
as given for the Berends--Giele currents of SYM below (\ref{bshuffle}).
In particular, Schocker's identity \cite{schocker} can be applied to both slots 
to infer 
\beq\label{schinf}
\phi_{AiB|Q} = (-1)^{|A|}\phi_{i(\tilde A\shuffle B)|Q}\, ,\qquad
\phi_{A|PiQ} = (-1)^{|P|}\phi_{A|i(\tilde P\shuffle Q)}
\eeq
from (\ref{anotherBGsym}), see (\ref{KlKu}) for the analogous identity for SYM currents.
Upon setting $B\rightarrow \emptyset$ in the first identity or $Q \rightarrow \emptyset$ in the
second, \eqref{schinf} specializes to the reflection identities 
\beq\label{BGrev}
\phi_{Ai|Q} = (-1)^{|A|} \phi_{i\tilde A|Q}\,,\qquad
\phi_{A|Pi} = (-1)^{|P|} \phi_{A|i\tilde P}\, .
\eeq
The symmetries \eqref{anotherBGsym} generalize the standard Berends--Giele
symmetry to both sets of color generators and guarantee that
the ansatz \eqref{doubleLie} is a (double) Lie series \cite{Ree}, thereby
preserving the Lie-algebra-valued nature of $\Phi(X)$
in \eqref{EOMPhi} w.r.t.\ both $t^i$ and $\tilde t^a$.

\paragraph{Examples of Berends--Giele double currents}

Based on $\phi_{i|j} = \d_{ij}$, the simplest application of the recursion \eqref{BGphi} 
leads to rank-two double currents:
\beq\label{simpleEx}
\phi_{12|12} ={1\over s_{12}}\bigl(\phi_{1|1}\phi_{2|2}
-\phi_{2|1}\phi_{1|2})={1\over s_{12}}\, ,\quad
\phi_{12|21} ={1\over s_{12}}\bigl(\phi_{1|2}\phi_{2|1} -\phi_{2|2}\phi_{1|1})=-{1\over s_{12}}\, .
\eeq
At rank three and four, it is straightforward to work out examples such as
\begin{align}
\phi_{123|123} &= {1\over s_{123}}\Bigl({1 \over s_{12}} + { 1 \over s_{23}}\Bigr)\, ,  &
\phi_{1234|1234} &={1\over s_{1234}}\Bigl(
{1\over s_{123}s_{12}}
+ {1\over s_{123}s_{23}}
+ {1\over s_{12}s_{34}}
+ {1\over s_{234}s_{23}}
+ {1\over s_{234}s_{34}}
\Bigr)
\,, \notag\\
\phi_{123|132} &= - { 1 \over s_{23} s_{123}}\, , &\phi_{1234|1243} &= -{1\over s_{1234}}\Bigl(
{1\over s_{12}s_{34}}
+ {1\over s_{234}s_{34}}
\Bigr)\, .\label{moreEx}
\end{align}

\paragraph{Berends--Giele double currents from planar binary trees}
As pointed out in \cite{PScomb}, the Berends--Giele double currents $\phi_{P|Q}$ can be obtained
from the planar-binary-tree expansions given by the $b$ map \eqref{bMap} as
\beq\label{phib}
\phi_{P|Q} = \langle b(P),Q\rangle = \langle P,b(Q)\rangle\,,
\eeq
where $\langle\cdot, \cdot\rangle$ denotes the scalar product of words \eqref{AdotB}, and
the symmetry $\phi_{P|Q}=\phi_{Q|P}$ of \eqref{symFTphi} is a consequence of the
self-adjoint property \eqref{bself} of the $b$ map. In addition,
the shuffle symmetry \eqref{anotherBGsym} follows from
the property $b(R\shuffle S)=0$ proven in \eqref{bshuffle}.

For an example application of \eqref{phib},  using the expansion \eqref{bexamp} for $b(123)$ we get
\beq\label{exbphi}
\phi_{123|132} = \langle b(123),132\rangle = {1\over s_{12}s_{123}}\langle [[1,2],3], 132\rangle
+ {1\over s_{23}s_{123}}\langle [1,[2,3]], 132\rangle = -{1\over s_{23}s_{123}}
\eeq
as $\langle [[1,2],3], 132\rangle = \langle 123 - 213 - 312 + 321,
132\rangle =0$ and $\langle [1,[2,3]],132\rangle =
\langle 123-132-231+321,132\rangle = -1$.

We will see later in \eqref{phiapdef} that there is a generalization of the relation \eqref{phib} between
planar binary trees and the Berends--Giele double current to a series expansion in $\ap$.

\paragraph{Berends--Giele formula for doubly-partial amplitudes}
Similar to the Berends--Giele formulae (\ref{BGformula}) in gauge theory, the double currents 
of bi-adjoint scalars yield their doubly-partial amplitudes via \cite{FTlimit}\footnote{The convention 
for the sign of the Mandelstam invariants here is such that
$m^{\rm here}(P,n|Q,n)= (-1)^{|P|}m^{\rm there}(P,n|Q,n)$ in
comparison with the normalization of \cite{DPellis}.}
\beq\label{BGamplitude}
m(P,n|Q,n) = \lim_{s_P\to0} s_P \phi_{P|Q}\,.
\eeq
We reiterate that $\phi_{i|j}=\delta_{ij}$, and $\phi_{P|Q}$ vanishes unless $P$ is
a permutation of $Q$ such that $s_P=s_Q$. By the cyclic symmetry of $m(R|S)$
in both words $R$ and $S$, there is no loss of generality in assuming their $n$-point 
instances to take the form $m(P,n|Q,n)$, where $|P|=|Q|=n{-}1$.\footnote{While cyclic symmetry of 
$m(R|S)$ is not manifest from the Berends--Giele formula (\ref{BGamplitude}),
it is built in from the definition (\ref{DPdefCP}) of doubly-partial amplitudes 
due to the cyclicity of the traces in $t^i$ and $\tilde t^a$.}

It is easy to see using the symmetries \eqref{anotherBGsym} obeyed by the 
double currents that the $m(P,n|Q,n) $ in (\ref{BGamplitude}) obey 
KK relations independently in both sets of color orderings.
Moreover, the Cachazo--He--Yuan (CHY) representation of doubly-partial 
amplitudes lead to BCJ relations in both entries \cite{DPellis},
\beq
m(\{ A,B\},n|Q) = m(A|\{P,Q\},n) = 0\, .
\label{bcjdpartial}
\eeq
As an earlier alternative to (\ref{BGamplitude}), doubly-partial amplitudes $m(R|S)$
can be determined from the algorithm described in \cite{DPellis} based on
drawing polygons and collecting the products of propagators associated to 
cubic graphs which are compatible with both of $R$ and $S$ as planar orderings. 
Their overall sign, however, requires keeping track of the polygon orientations.

As the main result of this section, by combining (\ref{ftlim}) with
(\ref{BGamplitude}) we get
\begin{prop.}
The field-theory limit of the $n$-point disk integrals \eqref{Zintdef} is given by
\beq\label{ftlimBG}
\lim_{\ap\to0} Z(P,n|Q,n) = \lim_{s_P\to0} s_P \phi_{P|Q} \,.
\ee
\end{prop.}
\paragraph{Examples of field-theory limits} Typical expressions for
doubly-partial amplitudes or field-theory limits of $Z$-integrals are
illustrated by the following examples at four points
\begin{align}\label{FTexs4pt}
\lim_{\ap\to0}Z(1234|1234) &= \frac{1}{s_{12}}+\frac{1}{s_{23}}\, , \ \ \ \ 
\lim_{\ap\to0}Z(1234|1243) &= -\frac{1}{s_{12}} \, , \ \ \ \ 
\lim_{\ap\to0}Z(1234|1423) &= -\frac{1}{s_{23}}\, ,
\end{align}
at five points
\begin{align}\label{FTexs5pt}
\lim_{\ap\to0}Z(12345|12345) &= \frac{1}{s_{12}s_{34}}+ \frac{1}{s_{23}s_{45}}
+ \frac{1}{s_{34} s_{51}}+ \frac{1}{s_{45}s_{12}}+ \frac{1}{s_{51}s_{23}}\, , \\
\lim_{\ap\to0}Z(12345|12354) &= - \frac{1}{s_{12} s_{45}}- \frac{1}{s_{23} s_{45}} \, , \ \ \ \ 
\lim_{\ap\to0}Z(12345|13524) = 0 \notag
\end{align}
and at six points
\begin{align}\label{FTexs6pt}
\lim_{\ap\to0}Z(123456|134256) &= - \frac{1}{s_{234} s_{34}} \bigg( \frac{1}{s_{56}}+ \frac{1}{s_{61}} \bigg)
\, .
\end{align}

\paragraph{Relation to the inverse KLT kernel} As another central result of \cite{DPellis},
doubly-partial amplitudes of bi-adjoint scalars are related to the {\it inverse} of the 
KLT matrix (\ref{kltrec}). More specifically, bases of $m(P|Q)$ under BCJ relations 
(\ref{bcjdpartial}) form invertible $(n{-}3)! \times (n{-}3)!$ matrices with entries
given by
\beq
 m^{-1}(1,R,n{-}1,n | 1,Q,n,n{-}1) = - S(R|Q)_1\, .
 \label{misinvklt.1}
\eeq
The particular choices of BCJ bases on the left-hand side are consistent with the fact that
the recursion (\ref{kltrec}) for the KLT matrix is tailored to the same BCJ bases of SYM
amplitudes in (\ref{KLTrel}). By (\ref{ftlim}), this implies that field-theory limits of
disk integrals can also be assembled from the KLT matrix as firstly pointed out in \cite{Zfunctions}.

As another consequence of (\ref{misinvklt.1}), the relation (\ref{KLTopen}) between the 
disk integrals $F_P{}^Q$ in terms of their Parke--Taylor analogues $Z$ can be 
inverted to give \cite{BGap} 
\beq
Z(1,P,n{-}1,n | Q ) = \sum_{R} m(Q | 1,R,n{-}1,n)  F_P{}^R\, ,
 \label{misinvklt.2}
\eeq
which is for instance instrumental to convert results on the $\alpha'$-expansions of both sides.
Moreover, the appearance (\ref{ftlim}) of doubly-partial amplitudes in the field-theory limit of 
$Z$-integrals can be derived from (\ref{misinvklt.2}) and the field-theory limit (\ref{ftofFPQ}) 
of $F_P{}^R$ on the right-hand side. This is not a circular conclusion since (\ref{ftofFPQ}) 
is a necessity for the consistent reduction of open-string amplitudes to those of SYM 
under (\ref{npttree}), and we will furthermore substantiate (\ref{ftofFPQ}) through the
method for its $\alpha'$-expansion in section \ref{sec:7.4}.

\paragraph{Inverse KLT matrix and Berends--Giele double currents} In terms of the Berends--Giele double
currents the statement in \eqref{misinvklt.1} translates
to the observation in \cite{PScomb} (see also \cite{FTlimit}) later proved in \cite{flas}:
\begin{lemma}
The matrix of Berends--Giele double currents $\phi_{iP|iR}$ of \eqref{phib} is the inverse to the standard KLT matrix
$S(R|Q)_i$ of \eqref{genstd}
\beq\label{phiSdelta}
\sum_R \phi_{iP|iR}
S(R|Q)_i = \d_{P,Q}\,.
\eeq
\end{lemma}
\noindent{{\it Proof.}} Taking the scalar product with $iP$ of the result $\ell(iR) = \sum_Q
S^\ell(iR|iQ)b(iQ)$ from lemma \eqref{ellSb}, we get
$\langle iP,\ell(iR)\rangle = \sum_QS^\ell(iR|iQ)\langle iP, b(iQ)\rangle$. That is, $\d_{P,R} =
\sum_Q S(R|Q)_i \phi(iQ|iP)$, where we used that $S^\ell(iR|iQ) = S(R|Q)_i$ in \eqref{genstd} and
$\langle iP, b(iQ)\rangle = \phi(iQ|iP)$ in \eqref{phib}.\qed

A positive aspect of the formula \eqref{phiSdelta} identifying the Berends--Giele double current
as the inverse of the KLT matrix is that there is no need to choose the
relative positions of $1, n{-}1,n$ like in \eqref{misinvklt.1} as no extraneous labels are present
in \eqref{phiSdelta}.
Moreover, this identity
allows us to invert the relation \eqref{VSM},
\beq\label{VtoM}
V_{iP} = \sum_Q S(P|Q)_i M_{iQ}\quad\Longrightarrow\quad
M_{iP} =\sum_Q \phi_{iP|iQ}V_{iQ}\,,
\eeq
directly without reference to extra labels.

\subsection{The field-theory limit of the superstring disk amplitudes}
\label{FTdisksec}

On the one hand, as reviewed in section \ref{PSSYMsec},
a closed formula for SYM tree-level amplitudes can be obtained using pure spinor
cohomology methods as
\beq\label{symnptAgain}
A(1,2, \ldots,n{-}1,n) = \langle E_{12 \ldots n{-}1}V_n\rangle = \sum_{XY=12 \ldots n{-}1}\langle
M_XM_Y M_n\rangle\, .
\eeq
On the other hand, we know that the SYM tree-level amplitudes are obtained as the limit
$\ap\to0$ of the superstring amplitude \eqref{localFormWithZ}. In the {\it non-local} 
KLT-representation (\ref{AstringP}) of the string amplitude, this follows from the field-theory 
limit (\ref{ftlim}) together with the inverse relation (\ref{misinvklt.1}) between the KLT matrix and 
a $(n{-}3)!^2$ basis of doubly-partial amplitudes of bi-adjoint scalars.

The goal of this section is to give an alternative proof and to recover the
cohomology formula (\ref{symnptAgain}) of SYM amplitudes from the {\it local}
representation of the string amplitude
\beq\label{localFormWithZAgain}
{\cal A}(1,2,\ldots,n) = -
\sum_{\rho \in S_{n-3}}
 \sum_{XY=\rho(23 \ldots n-2)}
\!\!\!\!\!\!\!\!\big\langle V_{1X}
V_{n-1,\tilde Y} V_n\big\rangle (-1)^\len{Y}Z(1,2, \ldots,n|1,X,n,Y,n{-}1)\,.
\eeq
\begin{prop.}
The field-theory limit of the pure spinor superstring amplitude
in its local representation
\eqref{localFormWithZAgain} yields
the SYM tree-level formula \eqref{symnptAgain}
\beq\label{stringFT}
\lim_{\ap\to0}{\cal A}(1,2,\ldots,n) = A(1,2,\ldots,n)\,.
\eeq
\end{prop.}
\noindent\textit{Proof.} The field-theory limit of the $Z$-integral in \eqref{localFormWithZAgain}
with the canonical ordering $P=12 \ldots n$ in the domain
is given by \eqref{ftlimBG},
\beq
\label{canoP}
\lim_{\ap\to0}Z(1,2, \ldots,n|Y,n{-}1,1,X,n) =
s_{12 \ldots n{-}1}\phi_{12 \ldots n{-}1|Y(n{-}1)1X}\, ,
\eeq
where we cyclically rotated $1,X,n,Y,n{-}1 \rightarrow Y,n{-}1,1,X,n$ to
attain the form of $Z(\ldots,n|\ldots,n)$ with matching end labels. 
Note that when the domain is the canonical ordering, the
deconcatenation formula \eqref{BGphi} for $\phi_{12 \ldots n{-}1|Y(n{-}1)1X}$ 
simplifies due to the constraint \eqref{BGconstraint} and we get
$s_{12 \ldots n-1}\phi_{12 \ldots n-1|Y(n-1)1X} = -
\phi_{1A|1X}\phi_{B(n-1)|Y(n-1)}$ where $AB=23 \ldots n{-}2$ with $\len{A}=\len{X}$, $\len{B}=\len{Y}$.
This means that we can write
\beq\label{spphi}
\lim_{\ap\to0}Z(1,2, \ldots,n|Y,n{-}1,1,X,n) = -
\sum_{AB=23\ldots n-2} \phi_{1A|1X}\phi_{(n-1) \tilde B|(n-1)\tilde Y}\, ,
\eeq
where we used the reflection property \eqref{BGrev} to rewrite
$\phi_{B(n-1)|Y(n-1)} = \phi_{(n-1)\tilde B|(n-1)\tilde Y}$, and only
a single term contributes to (\ref{spphi}) where $|A|=|X|$ and $|Y|=|B|$.
Therefore the limit of the string tree amplitude \eqref{localFormWithZAgain} as $\ap\to0$ becomes
\begin{align}
\lim_{\ap\to0}{\cal A}(1,2,\ldots,n) &=
\sum_{\rho \in S_{n-3}}
 \sum_{XY=\rho(23 \ldots n-2)}
\sum_{AB=23 \ldots n-2}
\big\langle \bigl(\phi_{1A|1X}V_{1X}\bigr)
\bigl(\phi_{(n-1)\tilde B|(n-1)\tilde Y}V_{n-1,\tilde Y}\bigr) V_n\big\rangle (-1)^\len{Y}
\notag \\
&=
\sum_{AB=23 \ldots n-2} \langle \bigg( \sum_C \phi_{1A|1C}V_{1C} \bigg)
(-1)^{|B|} \bigg( \sum_D \phi_{(n-1)\tilde B|(n-1)D}V_{(n-1)D} \bigg) V_n
\rangle
\label{manywor} \\
&=  \sum_{AB=23 \ldots n-2} \langle M_{1A}  (-1)^{|B|} M_{(n-1)\tilde B} V_n \rangle
=\!\!\!\!\!\sum_{XY=12 \ldots n-1}\!\!\!\!\!\langle M_X M_Y V_n\rangle \notag \\
&= \langle E_{12 \ldots n-1}V_n\rangle = A(1,2,\ldots,n)\,,\notag
\end{align}
where \eqref{BGconstraint} implies that $C$ and $D$ in the second line only need
to be summed over permutations of $A$ and $B$, respectively. We could therefore insert
\begin{align}
\label{permXY}
\sum_C \phi_{1A|1C}V_{1C} = M_{1A} \, , \ \ \ \ \ \ 
\sum_D \phi_{(n-1)\tilde B|(n-1)D}V_{(n-1)D} = M_{(n-1)\tilde B}
\, ,
\end{align}
from \eqref{VtoM}
followed by $M_{(n-1)\tilde B} = (-1)^{|B|} M_{B(n-1)}$ which finishes the proof.\qed

The above proof is valid for the canonical ordering $P=123 \ldots n$ due to
\eqref{canoP}. The generalization of the relation \eqref{stringFT} for
a general color ordering $P$ was proposed in \cite{FTlimit}, see \eqref{stringFTij}.


\section{String and field-theory amplitude relations}
\label{AmpRelsec}

In this section, we review the rich interplay of the results in the previous section
with amplitude relations in field and string theory. 
Section \ref{sec:6.4} is dedicated to the color-kinematics duality in gauge theory and
explicit realizations of kinematic Jacobi identities through the local representation
of disk amplitudes in the $\alpha' \rightarrow 0$ limit. We focus on gravitational
amplitudes in section \ref{sec:6.5}, briefly review the KLT relations between open- and 
closed-string tree amplitudes and extract the cubic-diagram organization of the gravitational 
double copy from different representations of closed-string amplitudes.
In section \ref{sec:6.5.1} we shall review the monodromy relations between 
open-superstring amplitudes with different disk orderings which furnish an 
elegant derivation of the BCJ relations among gauge-theory amplitudes.

The structure of disk amplitudes has implications
for field-theory double-copy relations beyond gauge theories and gravity. In 
section \ref{sec:7.7}, we shall discuss different
representations of Born--Infeld amplitudes and manifest the color-kinematics duality
of the non-linear sigma model of Goldstone bosons. Finally, the applications of the
disk correlator for heterotic string theories will be discussed in section \ref{sec:7.7.4},
along with the resulting amplitude relations for Einstein--Yang--Mills theory.

The discussions of this section does not rely on the detailed structure of the low-energy expansion
of string tree amplitudes. As will be detailed in section \ref{apsec}, the coefficients in the
$\alpha'$-expansion of disk and sphere integrals exhibit an elegant pattern of multiple zeta 
values (MZVs). By organizing the string-corrections to SYM and supergravity amplitude 
according to their MZV content, we will find echoes of the field-theory structures of this
section at all orders in $\alpha'$, see section \ref{sec:7.3.4}.

\subsection{Color-kinematics duality}
\label{sec:6.4}

This section is dedicated to an explicit realization of the color-kinematics duality
in SYM tree amplitudes, based on the $\alpha' \rightarrow 0$ limit of superstring disk amplitudes.
As we will see, this field-theory limit will naturally generate parameterizations of SYM
amplitudes in terms of cubic diagrams whose kinematic factors obey the same
Jacobi relations as their color factors. The {\it BCJ numerators} we will derive are
simple combinations of the local building blocks $\langle V_{1A} V_{n-1,B} V_n \rangle$
in pure spinor superspace descending from the $(n{-}2)!$-term representation (\ref{stringWithTs}) 
of disk amplitudes.

\subsubsection{Review of the color-kinematics duality}
\label{sec:6.4.1}

Our perspective on scattering amplitudes in gauge theories dramatically
changed due to the seminal conjecture of Bern, Carrasco and Johansson in 2008
that their kinematic dependence can be arranged to exhibit the same symmetries
as the color factors \cite{BCJ}. This {\it color-kinematics duality} holds for a variety of
tree-level amplitudes and loop integrands of gauge theories with different numbers 
of supersymmetries. Together with the closely related gravitational double copy
to be reviewed in section \ref{sec:6.5} below, the color-kinematics duality led to
a large web of connections between field and string theories, see
\cite{Bern:2019prr, Bern:2022wqg} for reviews and \cite{Adamo:2022dcm} for a white paper.

Already at tree level, the color-kinematics duality is obscured in a Feynman-diagrammatic 
computation of $(n\geq 5)$-point amplitudes. The string-theoretic approach reviewed in
this section led to the first explicit realizations of the color-kinematics duality in multiparticle
tree-level amplitudes in 2011 \cite{Mafra:2011kj}. In the first place, these results apply to
ten-dimensional SYM, but they straightforwardly propagate to dimensional reductions
including ${\cal N}=4$ SYM in four dimensions. In fact, the manifestations of
color-kinematics duality in this section also apply to pure Yang--Mills since
its gluon amplitudes are the same as in maximally supersymmetric gauge theories.

\paragraph{Cubic-diagram parameterization} The color-kinematics duality relies on an
elementary observation on tree amplitudes or loop integrands of pure or supersymmetric
YM theory: any dependence on the adjoint degrees of freedom (or {\it color dependence} in
short) of the external states occurs via contractions of the structure constants $f^{abc}$. In Feynman 
diagrams with exclusively cubic vertices, these contractions are straightforwardly determined by
dressing internal lines with $\delta^{ab}$ and vertices with $f^{abc}$.

While any non-abelian gauge-theory Lagrangian features a quartic interaction
$\sim {\Tr}( [\Bbb A^m , \Bbb A^n][\Bbb A_m,\Bbb A_n])$, its
color structure $f^{abe}f^{ecd}$ still resembles cubic diagrams. Each quartic
vertex bypasses one of the propagators of the cubic diagrams, but one can still enforce
a uniform number of propagators for all gauge-theory diagrams by inserting $1=\frac{p^2}{p^2}$
with suitably chosen momenta $p$ for each quartic vertex. As illustrated in figure \ref{fig:expandquartic}, this
amounts to expanding each quartic-vertex contribution in a channel $1=\frac{s_{i\ldots j} }{s_{i\ldots j}}$
that is compatible with the accompanying color
factors. 

\begin{figure}[h]
\begin{center}
\begin{tikzpicture}[scale=1.5]
\scope[xshift=-3.3cm]
\draw [line width=0.30mm]  (2,0.5) -- (1.5,1) node[above,left]{$2$};
\draw [line width=0.30mm]  (2,0.5) -- (1.5,0) node[below,left]{$1$};
\draw [line width=0.30mm]  (2,0.5) -- (2.5,1) node[above,right]{$3$};
\draw [line width=0.30mm]  (2,0.5) -- (2.5,0) node[below,right]{$4$};
\endscope
\scope[xshift=0.3cm]
\draw  (0.3,0.5) node{$\displaystyle = \ \  \ \frac{(k_1+k_2)^2}{(k_1+k_2)^2}$};
\draw [line width=0.30mm]  (2,0.5) -- (1.5,1) node[above,left]{$2$};
\draw [line width=0.30mm]  (2,0.5) -- (1.5,0) node[below,left]{$1$};
\draw [line width=0.30mm]  (2,0.5) -- (2.5,1) node[above,right]{$3$};
\draw [line width=0.30mm]  (2,0.5) -- (2.5,0) node[below,right]{$4$};
\endscope
\scope[xshift=4.1cm]
\draw  (0.3,0.5) node{$\displaystyle = \ \ \ (k_1+k_2)^2 $};
\scope[xshift=-0.1cm]
\draw [line width=0.30mm]  (2,0.5) -- (1.5,1) node[above]{$2$};
\draw [line width=0.30mm]  (2,0.5) -- (1.5,0) node[below]{$1$};
\draw [line width=0.30mm]  (2,0.5) -- (2.5,0.5) ;
\draw [line width=0.30mm]  (2.5,0.5) -- (3,1) node[above]{$3$};
\draw [line width=0.30mm]  (2.5,0.5) -- (3,0) node[below]{$4$};
\endscope
\endscope
%
\scope[xshift=0.3cm,yshift=-1.8cm]
\draw  (0.3,0.5) node{$\displaystyle = \ \  \ \frac{(k_2+k_3)^2}{(k_2+k_3)^2}$};
\draw [line width=0.30mm]  (2,0.5) -- (1.5,1) node[above,left]{$2$};
\draw [line width=0.30mm]  (2,0.5) -- (1.5,0) node[below,left]{$1$};
\draw [line width=0.30mm]  (2,0.5) -- (2.5,1) node[above,right]{$3$};
\draw [line width=0.30mm]  (2,0.5) -- (2.5,0) node[below,right]{$4$};
\endscope
\scope[xshift=4.1cm,yshift=-1.8cm]
\draw  (0.3,0.5) node{$\displaystyle = \ \ \ (k_2+k_3)^2 $};
\scope[xshift=-0.1cm]
\draw [line width=0.30mm]  (2,0.75) -- (1.5,1.25) node[left]{$2$};
\draw [line width=0.30mm]  (2,0.75) -- (2.5,1.25) node[right]{$3$};
\draw [line width=0.30mm]  (2,0.75) -- (2,0.25) ;
\draw [line width=0.30mm]  (2,0.25) -- (1.5,-0.25) node[left]{$1$};
\draw [line width=0.30mm]  (2,0.25) -- (2.5,-0.25) node[right]{$4$};
\endscope
\endscope
\end{tikzpicture}
\end{center}
 \caption{Two possibilities of expanding a quartic vertex: the first line is compatible with
 a color factor $f^{a_1 a_2 b}  f^{b a_3 a_4}$ while the second line captures the second
 term in rewriting the color factor as $f^{a_1 a_3 b}  f^{b a_2 a_4}-f^{a_2 a_3 b}  f^{b a_1 a_4}$ 
 via the Jacobi identity.}
\label{fig:expandquartic}
\end{figure}

Hence, it is always possible to parameterize gauge-theory trees and loop integrands
in terms of cubic diagrams $i$ whose propagators $D_{i}$ and
color factors $c_{i}$ can be straightforwardly read off from the cubic vertices
and internal lines. For color-dressed tree amplitudes, this parameterization reads\footnote{We 
depart from our notation $M_n$ for color-dressed SYM amplitudes to later on compare 
$M^{\rm gauge}_n$ in (\ref{revbcj.2}) with gravitational amplitudes $M^{\rm grav}_n$ and
those of bi-adjoint scalars in (\ref{DPdefCP}).}
\beq
M^{\rm gauge}_n = \sum_{i \in \Gamma_n} \frac{ c_i N_i }{D_i}\, .
\label{revbcj.2}
\eeq
The associated kinematic numerators $N_i$ encoding all the dependence on 
momenta and polarizations receive contributions from various Feynman diagrams
with different numbers of quartic vertices. The symbol $\Gamma_n$ in the summation 
range of (\ref{revbcj.2}) denotes the set of $(2n{-}5)!!$ cubic tree diagrams with $n$ 
external legs that are inequivalent under flips of cubic vertices. 

However, the Jacobi identity 
\beq
f^{abe}f^{ecd}+f^{bce}f^{ead}+f^{cae}f^{ebd}=0
\label{revbcj.1}
\eeq
introduces ambiguities in the alignment of quartic-vertex contributions with the propagator
structure of cubic diagrams. These ambiguities illustrated in figure \ref{fig:expandquartic} lead
to immense freedom in moving terms between the $N_i$ of different cubic diagrams. 
This freedom was initially referred to as
generalized gauge invariance \cite{BCJ, Bern:2010yg, loopBCJ} and later on related to 
non-abelian gauge transformations
of perturbiners \cite{Gauge}, for instance the transformation (\ref{finitegau}) mediating 
between Lorenz and BCJ gauge (see \cite{genredef} for an all-order expression).

\begin{figure}[h]
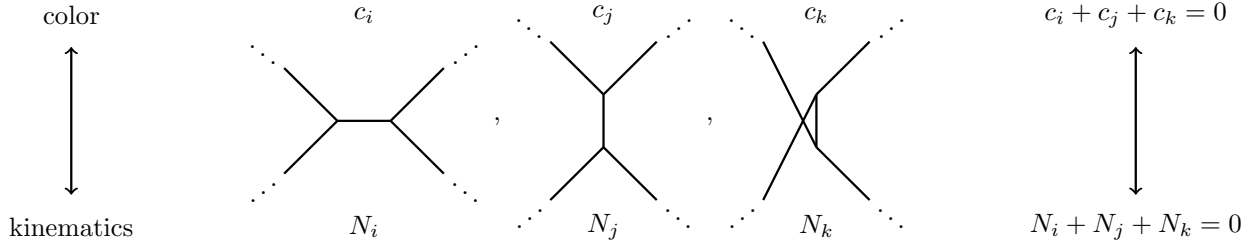

\begin{center}
 \tikzpicture [scale=1.4]
 \draw (-3,-1) node {kinematics};
 \draw (-3,1) node {color};
 \draw [line width=0.30mm,<->] (-3,0.7) -- (-3,-0.7);
 \draw (7,-1) node {$N_i + N_j + N_k= 0$};
 \draw (7,1) node {$c_i + c_j + c_k= 0$};
 \draw [line width=0.30mm,<->] (7,0.7) -- (7,-0.7);
 \scope[yshift=-0.5cm, xshift=-2.5cm]
 \draw [line width=0.30mm]  (2,0.5) -- (1.5,1) ;
 \draw (1.3,1.2) node {$\ddots$};
 \draw [line width=0.30mm]  (2,0.5) -- (1.5,0) ;
 \draw (1.3,-0.1) node {$\iddots$};
 \draw [line width=0.30mm]  (2,0.5) -- (2.5,0.5) ;
 \draw [line width=0.30mm]  (2.5,0.5) -- (3,1) ;
 \draw (3.2,1.2) node {$\iddots$};
 \draw [line width=0.30mm]  (2.5,0.5) -- (3,0) ;
 \draw (3.2,-0.1) node {$\ddots$};
 \draw (2.25,-0.5) node {$\displaystyle   N_i$}; 
 \draw (2.25,1.5) node {$\displaystyle   c_i$}; 
 \endscope
 \scope[xshift=-1.5cm]
 \draw (4.5,0) node{$,$};
 \draw [line width=0.30mm]  (5.5,-0.25) -- (5,0.75) ;
 \draw (4.8,0.95) node {$\ddots$};
 \draw [line width=0.30mm]  (5.5,0.25) -- (5,-0.75) ;
 \draw (4.8,-0.85) node {$\iddots$};
 \draw [line width=0.30mm]  (5.5,0.25) -- (5.5,-0.25) ;
 \draw [line width=0.30mm]  (5.5,0.25) -- (6,0.75) ;
 \draw (6.2,0.95) node {$\iddots$};
 \draw [line width=0.30mm]  (5.5,-0.25) -- (6,-0.75);
 \draw (6.2,-0.85) node {$\ddots$};
 \draw (5.5,-1) node {$\displaystyle  N_k$}; 
\draw (5.5,1) node {$\displaystyle  c_k$}; 
 \endscope
 \scope[xshift=-3.5cm]
 \draw (4.5,0) node{$,$};
 \draw [line width=0.30mm]  (5.5,0.25) -- (5,0.75) ;
 \draw (4.8,0.95) node {$\ddots$};
 \draw [line width=0.30mm]  (5.5,-0.25) -- (5,-0.75) ;
 \draw (4.8,-0.85) node {$\iddots$};
 \draw [line width=0.30mm]  (5.5,0.25) -- (5.5,-0.25) ;
 \draw [line width=0.30mm]  (5.5,0.25) -- (6,0.75) ;
 \draw (6.2,0.95) node {$\iddots$};
 \draw [line width=0.30mm]  (5.5,-0.25) -- (6,-0.75);
 \draw (6.2,-0.85) node {$\ddots$};
 \draw (5.5,-1) node {$\displaystyle  N_j$}; 
 \draw (5.5,1) node {$\displaystyle  c_j$}; 
 \endscope
 \endtikzpicture
\caption{Triplets of cubic graphs whose color factors $c_i$ and kinematic factors $N_i$ are
both related by a Jacobi identity if the duality between color and kinematics is manifest.
The dotted lines at the corners represent arbitrary tree-level subdiagrams and are understood
to be the same for all of the three cubic graphs.}
\label{fig:colkin}
\end{center}
\end{figure}

\paragraph{Kinematic Jacobi identities}
For all triplets of cubic diagrams $i,j,k \in \Gamma_n$ that share all propagators except for one,
see figure \ref{fig:colkin}, the Jacobi identity (\ref{revbcj.1}) implies that the associated color factors obey $c_i+c_j+c_k=0$. According to the color-kinematics duality, one can choose the numerators
$N_l$ in (\ref{revbcj.2}) such that the {\it kinematic Jacobi identity} $N_i+N_j+N_k=0$ holds
 for each such triplet $i,j,k$. Moreover, the antisymmetry $f^{abc} = f^{[abc]}$ implies that 
 color factors $c_i$ change their sign upon flipping any of the cubic vertices. Kinematic
 numerators with manifest color-kinematics duality are understood to also change
$N_i \rightarrow - N_i$ under flips of cubic vertices in diagram $i$.
 In other words,
 \begin{align}
{\rm manifest} \ \textrm{color-kinematics} \ {\rm duality}: \ \ \left\{ \begin{array}{rll}  c_i+c_j+c_k = 0
&\Longrightarrow &N_i+N_j+N_k=0 \
\forall \ i,j,k \in \Gamma_n\, ,   \\
c_i \rightarrow - c_i
&\Longrightarrow &N_i \rightarrow - N_i \
\forall \ i \in \Gamma_n \, .
\end{array} \right.
\label{revbcj.3}
\end{align}

\paragraph{Examples up to four points}
The three-point instance of the gauge-amplitude parameterization (\ref{revbcj.2})
in ten-dimensional SYM reduces to a single diagram without any propagators 
$D_i\rightarrow 1$, with color factor $c_i \rightarrow f^{123}$ and kinematic numerator
\beq
N_i \rightarrow \langle V_1 V_2 V_3 \rangle = (e_1 \cdot k_2)(e_2\cdot e_3) 
+e_1^m (\chi_2 \gamma_m \chi_3) + {\rm cyc}(1,2,3)\, .
\label{revbcj.4}
\eeq
Here and below, we use the shorthand $a_i \rightarrow i$ for the adjoint indices
of the $i^{\rm th}$ external state, e.g.\ write $f^{123}$ in the place of $f^{a_1a_2a_3}$.

The first instance of quartic-vertex contributions arises at four points.
The parameterization (\ref{revbcj.2}) comprises three diagrams
in the $s$-, $t$- and $u$-channel associated with inverse propagators
$s=s_{12}, \ t=s_{23}$ and $u=s_{13}=-s-t$,
\beq
M^{\rm gauge}_4 = \frac{ N_s c_s }{s}+  \frac{ N_t c_t }{t}+  \frac{ N_u c_u }{u} \, .
\label{revbcj.5}
\eeq
The color factors are indexed by the relevant channel, and their Jacobi identity
literally matches (\ref{revbcj.1})
\beq
\left. \begin{array}{r}
c_s = f^{12a} f^{a34} \\
c_t = f^{23a} f^{a14} \\
c_u = f^{31a} f^{a24} 
\end{array} \right\} \ \ \Longrightarrow \ \ c_s+c_t+c_u=0\, .
\label{revbcj.6}
\eeq
One admissible choice of numerators in ten-dimensional SYM reads
\begin{align}
N_s = \langle V_{12} V_3 V_4 \rangle \, , \ \ \ \ 
N_t = \langle V_{23} V_1 V_4 \rangle \, , \ \ \ \ 
N_u = \langle V_{31} V_2 V_4 \rangle\,,
\label{revbcj.7}
\end{align}
and they obey the kinematic Jacobi identity by BRST exactness of \cite{towardsFT}
\beq
N_s+N_t+N_u = \langle (V_{12}V_3+V_{23}V_1+V_{31}V_2) V_4 \rangle = - \frac{1}{s_{12}} \langle Q( V_{123} V_4) \rangle = 0
\label{revbcj.8}
\eeq
using (\ref{exampOne}) and $s_{13}+s_{23}=-s_{12}$ in the momentum phase space of
four massless particles. Still, any other choice of $\{N_s,N_t,N_u\}$ besides
(\ref{revbcj.7}) that yields the same amplitude (\ref{revbcj.5}) will obey kinematic
Jacobi identities: this can be seen by adding $0=K(\frac{s c_s}{s}+\frac{t c_t}{t}+\frac{u c_u}{u})$
to $M_4^{\rm gauge}$ with an arbitrary kinematic factor $K$ which modifies the numerators 
in (\ref{revbcj.5}) by $\delta N_s=sK,\ \delta N_t=tK$ and $\delta N_u=uK$. The modification to the triplet 
in the kinematic Jacobi identity (\ref{revbcj.8}) then vanishes by momentum conservation,
\beq 
\delta(N_s+N_t+N_u)=K(s+t+u)=0 \, .
\eeq

\paragraph{Examples at five points}

At five points, the cubic-diagram parameterization (\ref{revbcj.2}) involves $5!! = 15$ terms
\beq
M^{\rm gauge}_5 = \frac{ N_{12|3|45} c_{12|3|45} }{s_{12}s_{45}}+  
 \frac{ N_{14|3|25} c_{14|3|25} }{s_{14}s_{25}}+
  \frac{ N_{15|3|24} c_{15|3|24} }{s_{15}s_{24}}+{\rm cyc}(1,2,3,4,5) 
\label{revbcj.9}
\eeq
with color factors $c_{ab|d|gh}=f^{abi} f^{idj} f^{jgh}$ subject to Jacobi identities
$c_{ab| [d|gh]} = c_{[ab|d]|gh} = 0$. However, generic choices of kinematic 
numerators $N_{ab|d|gh}$ -- say a naive Feynman-diagram computation or
a crossing symmetric choice $N_{ab|d|gh} \rightarrow \langle V_{ab} V_d V_{gh} \rangle$ -- will 
fail to obey kinematic Jacobi identities even though they yield the correct 
color-dressed amplitude (\ref{revbcj.9}). 

Still, reparametrizations of the amplitude (\ref{revbcj.9}) will generically modify the
three-term sum of numerators that decide about kinematic Jacobi identities: adding
$0=K_{45}(\frac{s_{12} c_{12|3|45}}{s_{12} s_{45}}
+\frac{s_{23} c_{23|1|45}}{s_{23} s_{45}}
+\frac{s_{13} c_{31|2|45}}{s_{13} s_{45}})$ with some kinematic factor $K_{45}$ modifies
three of the numerator factors, $\delta N_{12|3|45} =s_{12}K_{45}$, $\delta N_{23|1|45} =s_{23}K_{45}$ 
and $\delta N_{31|2|45} =s_{13}K_{45}$, while leaving the remaining 12 inert
\cite{Bjerrum-Bohr:2010mia}. The sum of the
three numerators which vanishes
in a manifestly color-kinematics dual parameterization is modified by the above reparametrization via 
\beq 
\delta( N_{12|3|45}+N_{23|1|45}+N_{31|2|45})=(s_{12}+s_{23}+s_{13} )K_{45}=s_{45} K_{45}\neq 0\, .
\eeq
One can similarly check that, in the naive crossing-symmetric choice 
$N_{ab|d|gh} \rightarrow \langle V_{ab} V_d V_{gh} \rangle$, the 
superspace expression $(V_{12} V_3+ V_{23} V_1 +V_{31} V_2) V_{45}$
is not BRST closed. In this way, the validity of the kinematic Jacobi identity 
$N_{[12| 3] | 45} =0$ is seen to be gauge dependent. Hence, it is generically a matter of a 
suitable parameterization of the gauge-theory amplitude whether the color-kinematics 
duality is manifest or not.

\paragraph{Relation to the color decomposition}

In order to extract color-ordered gauge-theory amplitudes $A(1,2,\ldots,n)$ from
the cubic-diagram representation (\ref{revbcj.2}) of the color-dressed 
amplitude, one relies on the unique expansion of the
color factors $c_i$ in terms of traces of gauge-group generators 
${\Tr}(t^P) ={\Tr}(t^{p_1} t^{p_2} \ldots t^{p_n})$.
The above four- and five-point examples give\footnote{We are following
normalization conventions $[t^a, t^b]= f^{abc} t^c$ and ${\Tr}(t^a t^b)= \delta^{ab}$ 
which leads to ${\Tr}(t^1[t^2 , t^3]) = f^{123}$ at three points
and the coefficients $\pm 1$ in (\ref{revbcj.10}).}
\begin{align}
c_s &= {\Tr}(t^1 t^2 t^3 t^4 - t^1 t^2 t^4 t^3 - t^2 t^1 t^3 t^4 + t^2 t^1 t^4 t^3)\, ,
\label{revbcj.10} \\
c_{12|3|45} &= {\Tr}(t^1 t^2 t^3 t^4 t^5 - t^1 t^2 t^3 t^5 t^4 - t^2 t^1 t^3 t^4 t^5 + t^2 t^1 t^3 t^5 t^4) - (t^1 t^2 \leftrightarrow t^4 t^5)\, ,
\notag
\end{align}
see section 2.1 of \cite{Dixon:1996wi} for a general algorithm for arbitrary
color factors $c_i$. Hence, the color-ordered amplitudes obtained from
(\ref{revbcj.2}) are sums of diagrams
\beq
A(P)=\sum_{i \in \Gamma_n} \frac{ N_i }{D_i}  c_i  \, \big|_{ {\Tr}(t^{P}) }\, ,
\label{revbcj.11}
\eeq
where the coefficients take values in $c_i  \, \big|_{ {\Tr}(t^{P}) } \in \{0,1,-1\}$. The four- 
and five-point instances in the above notation
\begin{align}
A(1,2,3,4) &= \frac{ N_s }{s} - \frac{ N_t }{t}\, ,
\label{revbcj.12} \\
A(1,2,3,4,5) &= \frac{ N_{12|3|45} }{s_{12}s_{45} } + {\rm cyc}(1,2,3,4,5)
\notag
\end{align}
are clearly invariant under the reparametrizations $\sim K,K_{45}$
of the numerators in the color-dressed amplitude.

\paragraph{BCJ amplitude relations from the color-kinematics duality}

The BCJ relations (\ref{fundBCJ}) among color-ordered gauge-theory amplitudes were
firstly derived in \cite{BCJ} by assuming the existence of color-kinematics dual
numerators (\ref{revbcj.3}). For instance, inserting $N_u = -N_s - N_t$ into
\beq
A(1,2,3,4) = \frac{ N_s }{s} - \frac{ N_t }{t} \,, \ \ \ \ 
A(2,3,1,4) = \frac{ N_t }{t} - \frac{ N_u }{u} 
\label{revbcj.13}
\eeq
leads to the BCJ relation $A(2,3,1,4) = \frac{ s}{u} A(1,2,3,4)$. However, BCJ relations
are gauge-independent statements, i.e.\ they affect color-ordered amplitudes which
do not depend on reparametrizations of (\ref{revbcj.2}). Hence, the gauge dependent 
kinematic Jacobi relations cannot be necessary conditions for BCJ amplitude relations.
Instead, they are sufficient conditions as shown in \cite{BCJ}.

\subsubsection{DDM form of YM and bi-adjoint scalar amplitudes}
\label{sec:6.4.2}

In preparation for our proof that the $\alpha'\rightarrow 0$ limit of disk amplitudes
yields numerators $N_i$ subject to all kinematic Jacobi relations, we introduce
the so-called Del Duca--Dixon--Maltoni (DDM) representation of color-dressed gauge-theory
amplitudes. The color decomposition of $M_n^{\rm gauge}$ in terms of $(n{-}1)!$ 
color traces does not expose that the latter conspire to products of $n{-}2$ structure
constants as required by Feynman rules. Only after exhaustive use of KK relations
(\ref{KKrelation}) among color-ordered amplitudes, one can see that the color decomposition 
simplifies to contracted $f^{abc}$ in the coefficients of the KK independent
$A(1,P,n)$ with $P \in S_{n-2}$ a permutation of $2,3,\ldots,n{-}1$. This kind
of reduction by KK relations is known as the DDM form \cite{KKLance}
\beq
 \label{revbcj.14}
M_n^{\rm gauge} = \sum_{P \in S_{n-2}} c_{1|P|n} A(1,P,n)\,,\qquad
c_{1|P|n} := f^{1p_2a} f^{ap_3b}\ldots f^{yp_{n-2}z} f^{zp_{n-1}n}\,,
\eeq
where the color factor $c_{1|P|n}$ with $P=p_2p_3\ldots p_{n-1}$ corresponds
to the cubic diagram of half-ladder topology in figure~\ref{fig:masterdiag} below. In fact,
the collection of $\{ c_{1|P|n}, \ P \in S_{n-2}\}$ furnishes an $(n{-}2)!$-element basis
of all the $(2n{-}5)!!$ color factors $c_i$ under Jacobi identities.\footnote{This is a consequence of the fact that arbitrary Lie monomials 
built from non-commuting $t^{i_1},t^{i_2},\ldots, t^{i_k}$ can be expanded
in a Dynkin bracket basis of $[ [\ldots [ [ t^1,t^{\rho(2)} ],t^{\rho(3)} ], \ldots, t^{\rho(k-1)}], t^{\rho(k)} ]$, see section \ref{genJacsec}.} Accordingly the $(n{-}2)!$-family of half-ladders
in figure~\ref{fig:masterdiag} is referred to as the {\it master diagrams}.

\begin{figure}[h]
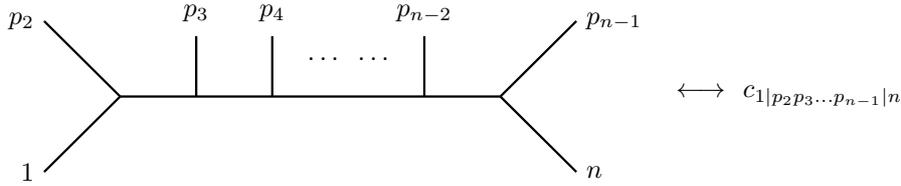

\begin{center}
 \tikzpicture[scale=1, line width=0.30mm]
 \draw(0,0) --(-1,-1) node[left]{$1$};
 \draw(0,0) --(-1,1) node[left]{$p_2$};
 \draw(0,0) -- (5,0);
 \draw(1,0) -- (1,0.8) ;
  \draw(2,0) -- (2,0.8) ;
 \draw (2.67,0.5)node{$\ldots$};
  \draw (3.33,0.5)node{$\ldots$};
  \draw(4,0) -- (4,0.8) ;
  \draw (1,1.1)node {$p_3$};
  \draw (2,1.1)node {$p_4$};
  \draw (4,1.1)node {$p_{n{-}2}$};
 \draw(5,0) --(6,-1) node[right]{$n$};
 \draw(5,0) --(6,1) node[right]{$p_{n-1}$};
  \draw(8.8,0)node{${\Large \longleftrightarrow} \ \ c_{1|p_2 p_3\ldots p_{n-1}|n}$};
 \endtikzpicture
 \end{center}
 \caption{Master diagrams whose color factors $c_{1|P|n}$ defined in
(\ref{revbcj.14}) are independent under Jacobi relations.}
\label{fig:masterdiag}
\end{figure}

In the same way as (\ref{revbcj.14}) follows from KK relations of the color-order
amplitudes in $M_n^{\rm gauge}$, one can start from the double color decomposition
(\ref{DPdefCP}) of bi-adjoint scalars and exhaustively insert KK relations of their doubly-partial
amplitudes $m(A|B)$. After expanding both entries $A$ and $B$ in $(n{-}2)!$-term KK bases,
(\ref{DPdefCP}) takes the form \cite{DPellis}
\beq
M_n^{\phi^3} = \sum_{P,Q \in S_{n-2}} c_{1|P|n} m(1,P,n | 1,Q,n) \tilde c_{1|Q|n} \label{revbcj.15}
\eeq
by analogy with (\ref{revbcj.14}), where $\tilde c_{1|Q|n}$ is the half-ladder in 
figure~\ref{fig:masterdiag} with $\tilde f^{abc}$ in the place of $f^{ijk}$.

At the same time, color-dressed $\phi^3$ amplitudes can be written as
\beq
M_n^{\phi^3} = \sum_{i \in \Gamma_n} \frac{ c_i \tilde c_i }{D_i}
\label{revbcj.16}
\eeq
as one can see from the straightforward Feynman-diagram computation with only one cubic vertex
$\sim f^{ijk}\tilde f^{abc}$ in the Lagrangian (\ref{Lagphi}). While (\ref{revbcj.16}) involves the complete $(2n{-}5)!!$-element
collection of $c_i,\tilde c_i$ with $i \in \Gamma_n$ related by Jacobi identities, the equivalent form 
(\ref{revbcj.15}) of $M_n^{\phi^3}$ only features the color factors $c_{1|P|n},\tilde c_{1|Q|n} $
of the master diagrams in figure~\ref{fig:masterdiag} under Jacobi relations. Hence, by equating
(\ref{revbcj.15}) and (\ref{revbcj.16}), the doubly-partial amplitudes $m(1,P,n | 1,Q,n)$ turn out to 
summarize the net effect of solving {\it all} Jacobi relations.

The last observation can be used to rewrite the color-dressed gauge-theory amplitude
(\ref{revbcj.2}): in a color-kinematics dual representation with $N_i$ obeying the same
Jacobi identities as $c_i$, the expansion of $M_n^{\rm gauge}$ in terms of 
color and kinematic factors of master diagrams must take the form
\beq
M_n^{\rm gauge} = \sum_{P,Q \in S_{n-2}} c_{1|P|n} m(1,P,n | 1,Q,n) N_{1|Q|n}\, . \label{revbcj.17}
\eeq
This can be understood from the fact that (\ref{revbcj.15}) follows from (\ref{revbcj.16}) solely 
by application of Jacobi identities among $c_i$, irrespective of their detailed form, and our
assumption that the $N_i$ obey the same Jacobi identities as the $\tilde c_i$.
The $(n{-}2)!$-family of $N_{1|Q|n}$ in (\ref{revbcj.17}) is again associated with the half-ladder
diagrams in figure~\ref{fig:masterdiag} and referred to as {\it master numerators}. Indeed, all the
$(2n{-}5)!!$ instances of $N_i$ in a color-kinematics dual parameterization (\ref{revbcj.2}) must be 
combinations of $N_{1|Q|n}$ with coefficients in $\{0,1,-1\}$ determined by (\ref{revbcj.17}).
In other words, any parameterization (\ref{revbcj.17}) of gauge-theory amplitudes
implies all kinematic Jacobi relations of the cubic-diagram numerators since the
same is evidently true in the $\phi^3$ case (\ref{revbcj.15}) and (\ref{revbcj.16}).

In summary, we have encountered two representations of color-dressed tree amplitudes
of gauge theories and bi-adjoint scalars: cubic-diagram expansions (\ref{revbcj.2}) 
and (\ref{revbcj.16}) related by trading kinematic numerators for another species of 
color factors $\tilde c_i \leftrightarrow N_i$.
While cubic-diagram expansions still exist if some of the kinematic Jacobi relations are
violated, the $(n{-}2)!^2$-term representations (\ref{revbcj.15}) and (\ref{revbcj.17}) are tied
to Jacobi relations reducing all of $c_i,\tilde c_i,N_i$ to an $(n{-}2)!$ basis. One can again
relate the gauge-theory amplitude (\ref{revbcj.17}) to the bi-adjoint scalar amplitude (\ref{revbcj.15})
by exchanging the kinematic master numerators with corresponding color factors,  
$N_{1|Q|n}\leftrightarrow \tilde c_{1|Q|n}$.

\subsubsection{Local BCJ numerators from disk amplitudes}
\label{sec:6.4.3}

We shall now take advantage of the representation (\ref{revbcj.17}) of color-dressed
gauge-theory amplitudes to retrieve the Jacobi relations of the kinematic 
numerators obtained from the $\alpha' \rightarrow 0$ limit of $n$-point disk amplitudes (\ref{stringWithTs}).
By matching the DDM form (\ref{revbcj.14}) of color-dressed gauge-theory amplitudes with the
expansion (\ref{revbcj.17}) in terms of master numerators, color-ordered $n$-point amplitudes
are found to take the form
\beq
A_n(P) = \sum_{Q \in S_{n-2}} m(P | 1,Q,n) N_{1|Q|n}\, .
 \label{revbcj.18}
\eeq
We also made use of the linear independence of the color factors $c_{1|P|n}$ associated
with the master diagrams in figure~\ref{fig:masterdiag} and the fact that $A(P)$ and 
$m(P|1,Q,n)$ obey the same KK relations in $P$. 

It turns out that (\ref{revbcj.18}) is precisely the form of $A(P)$ obtained in the field-theory 
limit of disk amplitudes: as detailed in section \ref{sec:6.3.1},
adapting (\ref{stringWithTs}) to a generic ${\rm SL}_2(\mathbb R)$ frame
leads to the $(n{-}2)!$ term representation (\ref{localFormWithZ}) 
of $n$-point disk amplitudes in terms
of Parke--Taylor or $Z$-integrals (\ref{Zintdef}) \cite{Zfunctions}
\beq\label{revbcj.19}
{\cal A}_n(P) =\!\!\!\!\! \sum_{XY=23 \ldots n{-}2} \!\!\!\!\!\!(-1)^{\len{Y}+1}Z(P|1,X,n,Y,n{-}1)\langle V_{1X}V_{(n-1)\tilde
Y}V_n\rangle + \perm(2,3, \ldots,n{-}2)\, ,
\eeq
for instance
\begin{align}
{\cal A}_4(P) &= - Z(P|1,2,4,3) \langle V_{12} V_{3} V_4 \rangle
+Z(P|1,4,2,3) \langle V_{1} V_{32} V_4 \rangle \, , \label{revbcj.20} \\
{\cal A}_5(P) &= - Z(P|1,2,3,5,4) \langle V_{123} V_{4} V_5 \rangle
+Z(P|1,2,5,3,4) \langle V_{12} V_{43} V_5 \rangle
-Z(P|1,5,2,3,4) \langle V_{1} V_{432} V_5 \rangle + (2\leftrightarrow 3)\, ,
\notag \\
{\cal A}_6(P) &= - Z(P|1,2,3,4,6,5) \langle V_{1234} V_{5} V_6 \rangle
+Z(P|1,2,3,6,4,5) \langle V_{123} V_{54} V_6 \rangle   \notag\\
&\quad -Z(P|1,2,6,3,4,5) \langle V_{12} V_{543} V_6 \rangle
+Z(P|1,6,2,3,4,5) \langle V_{1} V_{5432} V_6 \rangle
+ {\rm perm}(2,3,4) \, .
 \notag
\end{align}
Here and in later equations, the sum over permutations of $2,3,\ldots,n{-}2$ is
understood to not act on the labels in the integration domain $P$.

Given that the field-theory limit (\ref{ftlim}) of the $Z$-integrals yields doubly-partial
amplitudes, the SYM amplitudes resulting from (\ref{revbcj.19}) are given by
\begin{align}
A_n(P) &=
\!\!\!\!\! \sum_{XY=23 \ldots n{-}2} \!\!\!\!\!\!(-1)^{\len{Y}+1}m(P|1,X,n,Y,n{-}1)\langle V_{1X}V_{(n-1)\tilde
Y}V_n\rangle + \perm(2,3, \ldots,n{-}2)
\notag \\
&= \sum_{Q\in S_{n-2}} m(P|1,Q,n{-}1) N_{1|Q|n-1} 
\label{revbcj.21}
\end{align}
with master numerators \cite{Mafra:2011kj}
\beq
N_{1| XnY | n-1} = (-1)^{|Y|-1} \langle V_{1X}V_{(n-1)\tilde
Y}V_n\rangle \, ,
\label{revbcj.22}
\eeq
for instance
\beq
N_{1|23\ldots j n (j+1)\ldots n-2 |n-1} =  (-1)^{n-j-1} \langle V_{12\ldots j} V_{n-1,n-2\ldots j+1} V_n\rangle \, .
\label{revbcj.22a}
\eeq
The second line of (\ref{revbcj.21}) exposes the expansion of $A(P)$ in an
$(n{-}2)!$ family of $m(P|1,Q,n{-}1)$ (with $Q$ a permutation of $2,3,\ldots,n{-}2,n$)
characteristic to color-kinematic dual
representations of SYM amplitudes. Since (\ref{revbcj.21}) is related
to the color-kinematics dual form (\ref{revbcj.18}) via $n\leftrightarrow n{-}1$,
we identify the local superfields (\ref{revbcj.22}) as the BCJ master numerators
of the half-ladder diagrams in figure~\ref{fig:masterdiag} with $n{-}1$ in
the place of $n$. In fact, by the diagrammatic interpretation of $V_P$ in
figure \ref{figHL}, the right-hand side of (\ref{revbcj.22}) organizes the master diagrams 
into three subdiagrams as visualized in figure~\ref{PSSVVV}.

\begin{figure}[h]
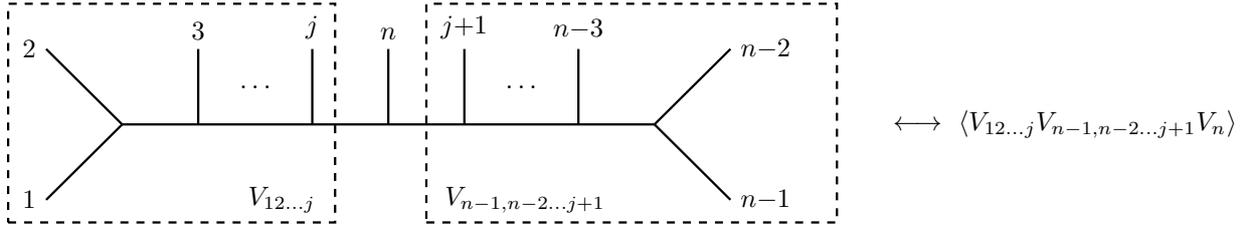

\begin{center}
\tikzpicture[line width=0.30mm]
\draw(0,0) --(-1,-1) node[left]{$1$};
\draw(0,0) --(-1,1) node[left]{$2$};
\draw(0,0) -- (7,0);
\draw(1,0) -- (1,1) node[above]{$3$};
\draw (1.75,0.5)node{$\ldots$};
\draw (2.5,0)--(2.5,1) node[above]{$j$};
\draw (3.5,0)--(3.5,1) node[above]{$n$};
\draw (4.5,0)--(4.5,1) node[above]{$j{+}1$};
\draw (5.25,0.5)node{$\ldots$};
\draw(6,0) -- (6,1) node[above]{$n{-}3$};
\draw(7,0) --(8,-1) node[right]{$n{-}1$};
\draw(7,0) --(8,1) node[right]{$n{-}2$};
\draw[dashed] (-1.5,-1.3) rectangle (2.8,1.6);
\draw[dashed] (9.4,-1.3) rectangle (4.0,1.6);
\draw(2.1,-1)node{$V_{12\ldots j}$};
\draw(5.3,-1)node{$V_{n-1,n-2\ldots j+1}$};
\draw(12.4,0)node{$\longleftrightarrow  \ \langle V_{12\ldots j} V_{n-1,n-2\ldots j+1} V_n \rangle $};
\endtikzpicture
\end{center}
\caption{The mapping between master numerators and expressions in pure spinor superspace according to (\ref{revbcj.22a}).}
\label{PSSVVV}
\end{figure}

We spell out the simplest examples at four points
\begin{align}
N_{1|24|3} = -\langle V_{12} V_3 V_4 \rangle \, , \ \ \ \  N_{1|42|3} = \langle V_{1} V_{32} V_4 \rangle
\label{revbcj.23}
\end{align}
and at five points
\begin{align}
N_{1|235|4} &= -\langle V_{123} V_4 V_5 \rangle \, , \ \ \ \  
N_{1|253|4} = \langle V_{12} V_{43} V_5 \rangle \, , \ \ \ \  
N_{1|523|4} = -\langle V_{1} V_{432} V_5 \rangle\, ,
\label{revbcj.24} \\
N_{1|325|4} &= -\langle V_{132} V_4 V_5 \rangle \, , \ \ \ \  
N_{1|352|4} = \langle V_{13} V_{42} V_5 \rangle \, , \ \ \ \  
N_{1|532|4} = -\langle V_{1} V_{423} V_5 \rangle \, . \notag
\end{align}
The superspace numerators (\ref{revbcj.22}) enter the representations (\ref{revbcj.21})
of color-ordered gauge-theory amplitudes that are hallmarks of manifest 
color-kinematics duality by the discussion of section \ref{sec:6.4.2}.
Hence, the master numerators (\ref{revbcj.22}) determine
all other cubic-diagram numerators in (\ref{revbcj.2}) by a sequence of
kinematic Jacobi identities, and the coefficients can be conveniently determined
by isolating the propagators of interest from the doubly-partial amplitudes in 
(\ref{revbcj.21}). Moreover, the master numerators are local, i.e.\ free of kinematic 
poles, by the construction of multiparticle superfields $A_\alpha^P$ in BCJ gauge that enter 
$V_P = \lambda^\alpha A_\alpha^P$, see section \ref{sec:4.1loc}. On these
grounds, the superspace expressions (\ref{revbcj.22}) are referred
to as local BCJ numerators \cite{Mafra:2011kj}.

\subsubsection{Component expansion of BCJ numerators}

In order to extract the superspace components from master numerators 
$\langle V_X V_Y V_Z \rangle$, it is convenient to combine BCJ gauge for the 
superfields with the non-linear version of Harnad--Shnider
gauge for their $\theta$-expansion, see section 4.3 of \cite{Gauge} 
and \ref{HSapp}. In this BCJ--Harnad--Shnider
gauge, the relevant orders in $\theta$ are,
\beq
V_P =
{1\over 2}(\lambda \gamma_m \theta)  e^m_P
+{1\over 3}(\lambda \gamma_m \theta) (\theta\gamma^m \chi_P)
- {1\over32}(\lambda \gamma^m \theta) (\theta\gamma_{mnq}\theta)f_P^{nq}+
{\cal O}(\theta^4)\,,\label{thetaVP.1}
\eeq
with local multiparticle polarizations $e_P^m,\chi_P^\alpha,f_P^{mn}$ 
defined by (\ref{zerolocal}) in the place of the
single-particle polarizations $e_i^m,\chi_i^\alpha,f_i^{mn}$ in (\ref{linTHEX}).
Similar to the discussion in section \ref{sec:compSYM}, this organization of the $\theta$-expansion
reduces the component expansion at all multiplicities,
\beq
\langle V_X V_Y V_Z \rangle = 
\half e^m_X\,  f^{mn}_Y e^n_Z + (\chi_X \g_m \chi_Y) e_Z^m + \cyclic{XYZ}\,,
\label{thetaVP.2}
\eeq
to the $\lambda^3 \theta^5$ correlators (\ref{fermex}) and (\ref{simpl3t5}) of
the three-point amplitude \cite{BGBCJ}.

\subsubsection{The M\"obius product}

While the SYM amplitudes $A(P,n)=\langle E_P V_n\rangle$ satisfy the BCJ amplitude relations, a
naive relabeling of $P$ does not lead to numerators that satisfy the color-kinematics duality.
As discussed in \cite{Mafra:2011kj} and reviewed above, the way string theory disk amplitudes give
rise to a local representation of numerators satisfying the color-kinematics duality is via the
field-theory limit of the pure spinor parameterization of the correlator with $(n{-}2)!$ numerators
of the form $\langle V_{1P}V_{(n{-}1)Q}V_n\rangle$ with each multiparticle vertex $V_R$ in the
BCJ gauge reviewed in section~\ref{sfinbcj}. This parameterization is generated by
\eqref{revbcj.18}, and its essential feature is the
distribution of the labels $1,n{-}1,n$ into three separate superfields $V_R$
within master numerators \eqref{revbcj.22}. This
splitting can be traced back to  the fixing of the M\"obius invariance of the disk correlator in
\eqref{treepresc}.

The field-theory limit of the disk integrals with different disk orderings, given by bi-adjoint amplitudes
\eqref{ftlimBG}, does not modify this label distribution in the numerators, while relabeling the color
ordering of $A(P,n)$ in \eqref{AYM} does. Note, however, that \eqref{AYM} and the field-theory
limit of the string disk amplitude manifestly coincide for the canonical ordering $P=12 \ldots n$
(and in fact for a $(n{-}3)!$ basis of color-orderings $P=1R(n{-}1)n$), as
demonstrated in \eqref{manywor}.
Let us illustrate the above point with an example.

The pure spinor formula \eqref{AYM} for the
ordering $P=12435$ in $A(1,2,4,3,5)=\langle E_{1243} V_5\rangle$ yields
\beq
A(1,2,4,3,5) =
 {\langle V_{124}V_3V_5\rangle\over s_{12}s_{124}}
+ {\langle V_{421}V_3V_5\rangle\over s_{24}s_{124}}
+ {\langle V_{12}V_{43}V_5\rangle\over s_{12}s_{34}}
+ {\langle V_1V_{243}V_5\rangle\over s_{24} s_{34}}
+ {\langle V_1V_{342}V_5\rangle\over s_{34}s_{234}}\,,
\eeq
while the $\ap\to0$ limit of the superstring amplitude \eqref{localFormWithZ} with the same
ordering gives, after using \eqref{ftlimBG},
\begin{align}
\lim_{\ap\to0}{\cal A}(1,2,4,3,5) &=
{\langle(V_{12}V_{43} + V_{123}V_4)V_5\rangle \over s_{12}s_{124}}
- { \langle(V_1 V_{423} + V_{13}V_{42})V_5\rangle \over s_{24}s_{124}}
+ {\langle V_{12}V_{43}V_5\rangle\over s_{34}s_{12}} \\
&\quad - { \langle V_1V_{432}V_5\rangle\over s_{34}s_{234}}
- {\langle V_1V_{423}V_5\rangle\over s_{24}s_{234}}\,.\notag
\end{align}
These two amplitudes must be equal, but the numerators differ as only the latter preserves
the superfield structure $V_{1A}V_{(n{-}1)\tilde B}V_n$ due to fixing
$(z_1,z_{n{-}1},z_n) \rightarrow (0,1,\infty)$ by M\"obius invariance.
Comparing both amplitudes we see the correspondence
\begin{align}
V_{124}V_3 &\rightarrow V_{12}V_{43} + V_{123}V_4\, ,
& V_1V_{243} &\rightarrow -V_1V_{423}\, ,\\
V_{421}V_3 &\rightarrow - V_1 V_{423} - V_{13}V_{42}\, ,
&V_1V_{342} &\rightarrow -V_1V_{432}\, .\notag
\end{align}
In \cite{FTlimit} an algorithm was
proposed that reproduces the above correspondences.
The idea is to guarantee that the two labels $i$ and $j$ (usually $i=1$ and $j=n{-}1$) always appear
in two separate vertices\footnote{The third vertex $V_n$ can
always be fixed with label $n$ by cyclic invariance.}. Therefore the algorithm redistributes the labels
$i$ and $j$ between two vertices if they originally appear inside a single vertex $V_{iAjB}$ and
does nothing otherwise. To this effect, we define the \textit{M\"obius product}
$\circ_{ij}$ as \cite{FTlimit}
\beq\label{moebius}
V_{iAjB}\circ_{ij} V_C :=\!\!\! \sum_{\d(B\dot\ell(C))=R\otimes S}\!\!\! V_{iAR}V_{jS}\, ,
\qquad V_{AiB}\circ_{ij} V_{CjD} := V_{AiB}V_{CjD}\, ,
\eeq
where $\dot\ell(C)$ denotes the letterification \eqref{letterif} of the Dynkin bracket
$\ell(C)$ of \eqref{elldef} and $\d(P)$ is the
deshuffle map \eqref{deshuffle}. Note that it is always possible to move label $i$ to the front
using \eqref{LieBasis}, $V_{PiQ} = - V_{i\ell(P)Q}$, so these two rules are sufficient. 
In summary, the mapping \eqref{moebius} ensures that the labels $i$ and $j$
are split between the two vertices $V_A$ and $V_B$ in
the product $V_A \circ_{ij}V_B$. The choice $i=1$ and $j=n{-}1$ corresponds to the
usual fixing of the M\"obius symmetry of the disk.
For example applications of \eqref{moebius} we list
\begin{align}
\label{moebiusEx}
V_{124}\circ_{14}V_3 &= V_{12} V_{43} +  V_{123} V_{4}\, , \\
V_{142}\circ_{14}V_3 &= V_{1} V_{423} +  V_{12} V_{43} + V_{123}V_4 +
V_{13}V_{42}\, ,\cr
V_{421}\circ_{14}V_3 &= -  V_{1} V_{423} -  V_{13} V_{42} \, ,\cr
V_{143}\circ_{14}V_2 &= V_1V_{432} + V_{13}V_{42}+ V_{12}V_{43} + V_{132}V_{4} \, ,\cr
V_{134}\circ_{14}V_2 &= V_{13}V_{42} + V_{132}V_{4} \, ,\cr
V_{1235}\circ_{15}V_4 &=  V_{123} V_{54} +  V_{1234} V_{5}\, ,\cr
V_{1253}\circ_{15}V_{4} &=
 V_{12} V_{534}
+  V_{123} V_{54}
+  V_{1234} V_{5}
+  V_{124} V_{53} \, ,\cr
V_{1523}\circ_{15}V_{4} &=
 V_{1} V_{5234}
+  V_{12} V_{534}
+  V_{123} V_{54}
+  V_{1234} V_{5}\cr
&\quad{} +  V_{124} V_{53}
+  V_{13} V_{524}
+  V_{134} V_{52}
+  V_{14} V_{523} \, ,\cr
V_{1235}\circ_{15}V_{46} &=  V_{123} V_{546} -  V_{123} V_{564} +  V_{12346}V_{5} -  V_{12364} V_{5} \, ,\cr
V_{152}\circ_{15} V_{34} & =
         V_{1} V_{5234}
       - V_{1} V_{5243}
       + V_{12} V_{534}
       - V_{12} V_{543}
       + V_{1234} V_{5}\cr
       &\quad{}- V_{1243} V_{5}
       + V_{134} V_{52}
       - V_{143} V_{52} \, .\notag
\end{align}
From the translation $V_A V_B\to [\ell(A),\ell(B)]$ we obtain
the free-Lie-algebra interpretation the above mapping: it is a rewriting
system of nested commutators from $[\ell(iAjB),\ell(C)]$ to a
basis\footnote{Despite appearances,
this is not a $(n-2)!$ dimensional basis of the free Lie algebra but a
$(n-1)!$ one, even after the fixing of two letters $i$ and $j$. The reason is
that the lengths of $P$ and $Q$ are not fixed. The simplest example is the
case $n=3$ where the $(n-1)!=2$ dimensional basis is $[\ell(1),\ell(23)]$ and
$[\ell(13),\ell(2)]$.}
of brackets of the form $[\ell(iP),\ell(jQ)]$. For instance, the first example
in \eqref{moebiusEx} is equivalent to
\beq\label{freeCom}
[[[1,2],4],3] = [[1,2],[4,3]] + [[[1,2],3],4]\, ,
\eeq
which can be explicitly verified by expanding the commutators. The correctness of the other examples can be checked similarly.
In \cite{FTlimit}, a similar interpretation was used to map the product
$V_{iAjB}V_CV_n$ to a multi-peripheral color factor composed from a
string of structure constants. The map \eqref{moebius}
was then shown to correspond to a closed formula to rewrite the
multi-peripheral factors in the DDM basis of Del Duca, Dixon and Maltoni
\cite{KKLance}.

\subsubsection{Local BCJ numerators from the M\"obius product}
\label{localMoebiussec}

Using the M\"obius product \eqref{moebius} it is easy to obtain local numerators for SYM tree
amplitudes satisfying the color-kinematics duality, and in fact the full tree amplitudes in BCJ form.
To this end, for an $n$-point tree amplitude, we map the planar binary trees in the expansion of
$b(P)$ in (\ref{bMap}) with $\len{P}=n{-}1$ and the root identified as the $n$-th leg to
pure spinor superspace numerators as follows \cite{FTlimit}
\beq\label{bcjmap}
[\Gamma,\Sigma] \longrightarrow \langle V_\Gamma\circ_{ij}V_\Sigma V_n\rangle
\eeq
with superfields in the BCJ gauge and
for arbitrary choices for $i,j$ (usually $i,j=1,n{-}1$).
The graphical depiction is the following:
\begin{center}
\includegraphics[width=0.17\textwidth]{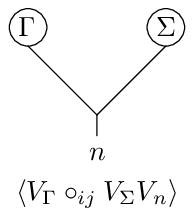}
\end{center}
where the blobs $\Gamma$ and $\Sigma$ represent arbitrary cubic trees.
For example, the expression for the amplitudes $A(1,2,3,4,5)$ and $A(1,4,3,2,5)$
are obtained from $s_{1234}b(1234)$ and $s_{1234}b(1432)$ from \eqref{bMap}
using the prescription \eqref{bcjmap} with $i,j=1,4$, see
figure~\ref{A12345fig}.
\begin{figure}[t]
\begin{center}
\includegraphics[width=\textwidth]{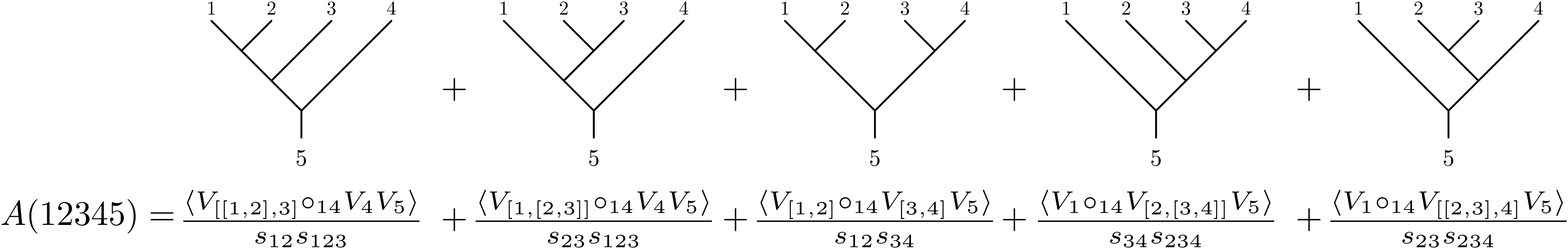}
\end{center}
\begin{center}
\includegraphics[width=\textwidth]{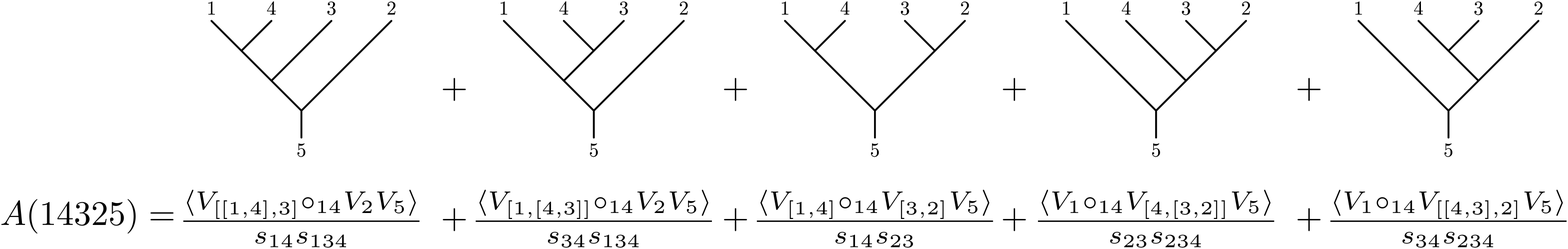}
\end{center}
\caption{The amplitudes $A(12345)$ and $A(14325)$ parameterized with BCJ numerators according to
the M\"obius map \eqref{bcjmap} with $i,j=1,4$. The expanded numerators after applying the
M\"obius product \eqref{moebius} are given in \eqref{A12345nums}. See figure \ref{figBGfour} for
the binary tree expansion of $b(1234)$.}
\label{A12345fig}
\end{figure}
More explicitly, after applying the M\"obius product to the numerators one
gets
\begin{align}
A(1,2,3,4,5) &= 
{\langle V_{123}V_4V_5\rangle\over s_{12}s_{123}}
+{\langle V_{123}V_4V_5{-}V_{132}V_4V_5\rangle\over s_{23}s_{123}}
-{\langle V_{12}V_{43}V_5\rangle\over s_{12}s_{34}}
+{\langle V_{1}V_{432}V_5\rangle\over s_{34}s_{234}}
+{\langle V_{1}V_{432}V_5{-}V_{1}V_{423}V_5\rangle\over s_{34}s_{234}}\, ,
\notag\\
A(1,4,3,2,5) &= 
       {\langle V_{1}V_{432}V_5
       + V_{12}V_{43}V_5
       + V_{13}V_{42}V_5
       + V_{132}V_{4}V_5\rangle\over s_{14}s_{134}}
       + {\langle V_{1}V_{432}V_5
       + V_{12}V_{43}V_5\rangle\over s_{34}s_{134}}\label{A12345nums}\\
&\quad + {\langle V_{1}V_{432}V_5
       - V_{1}V_{423}V_5
       - V_{123}V_{4}V_5
       + V_{132}V_{4}V_5\rangle\over s_{14}s_{23}}
+ {\langle V_{1}V_{432}V_5-V_1V_{423}V_5\rangle\over s_{23}s_{234}}
+{\langle V_{1}V_{432}V_5\rangle\over s_{34}s_{234}} \, ,\notag
\end{align}
which agree with the results of \cite{Mafra:2011kj}.
For instance, the numerator of the pole $1/(s_{34}s_{134})$ in the amplitude $A(1,4,3,2,5)$ 
is given by $V_{[1,[4,3]]}\circ_{14}V_2 V_5$, whose evaluation via \eqref{moebius} yields
\beq
\big(V_{143}\circ_{14}V_2-V_{134}\circ_{14}V_2\big)V_5 = V_1V_{432}V_5 - V_{12}V_{43}V_5\, ,
\eeq
where we used \eqref{BCJnotation} and \eqref{Baker} to rewrite $V_{[1,[4,3]]}=V_{143}-V_{134}$
followed by the examples in \eqref{moebiusEx}. Comparing with the parameterization of the
five-point numerators $n_{j=1,2,\ldots,15}$ in \cite{BCJ}
\begin{align}
A(1,2,3,4,5) &=
{n_1\over s_{12}s_{123}}
+ {n_2\over s_{23}s_{234}}
+ {n_3\over s_{34}s_{12}}
+ {n_4\over s_{123}s_{23}}
+ {n_5\over s_{234}s_{34}}\, , \\
A(1,4,3,2,5) &=
{n_6\over s_{14}s_{134}}
+ {n_5\over s_{234}s_{34}}
+ {n_7\over s_{23}s_{14}}
+ {n_8\over s_{134}s_{34}}
+ {n_2\over s_{234}s_{23}} \, ,\notag
\end{align}
it is easy to verify that the BCJ triplet
identity $n_3-n_5+n_8=0$  is satisfied:
\beq
-\langle V_{12}V_{43}V_5\rangle
-\langle V_1V_{432}V_5\rangle
+ \langle V_1 V_{432}V_5 + V_{12}V_{43}\rangle = 0\, .
\eeq
All the other BCJ numerator identities can be similarly verified.

\subsubsection{The field-theory limit of the superstring disk amplitude for arbitrary orderings}
\label{sec:otherorderings}

Finally, we define the M\"obius product of Berends--Giele currents
$M_X\circ_{ij}M_Y$ by the action on the products $V_A\circ_{ij}V_B$ arising
from the expansion \eqref{BGexplM} of $M_X$ and $M_Y$ which extends to
\beq\label{Eijdef}
E^{(ij)}_P := \sum_{XY=P}M_X\circ_{ij}M_Y\,.
\eeq
It was argued in \cite{FTlimit} that the field-theory limit of the superstring
amplitude with arbitrary color ordering can be written as
\beq\label{stringFTij}
\lim_{\ap\to0}{\cal A}(P,n) = \langle E^{(1,n{-}1)}_P V_n\rangle =:
A^{(1,n{-}1)}(P,n)\,,
\eeq
such that the right-hand side can be viewed as a closed-formula yielding field-theory amplitudes whose (local)
numerators satisfy the color-kinematics duality.

\subsubsection{Local BCJ numerators from finite gauge transformations}

In \cite{flas} a straightforward parameterization of YM tree amplitudes satisfying the
color-kinematics duality was obtained. The idea  is to  map the Lie-polynomial numerators
$\Gamma=[\Gamma_1,\Gamma_2]$ of the planar-binary-tree-expansions $b(P)$ generated by
\eqref{bMap} into kinematic numerators. This can be done using the $\t{=}0$ component of the local vector potential
$A^m_{[\Gamma_1,\Gamma_2]}$ in the BCJ gauge of \cite{Gauge,genredef} after setting the external fermions to zero. In this gauge, the vector
potential $A^m_{\Gamma}$ is associated to a cubic-tree Lie polynomial $\Gamma$ and satisfies the
same Jacobi identities of the
associated color factors but in kinematic space\footnote{More recently, mapping binary trees to
kinematic numerators was proposed in \cite{Brandhuber:2022enp,Chen:2022nei} exploiting a beautiful
connection to free Lie algebras via the
quasi-shuffle product \cite{quasi}.}. The BCJ gauge at arbitrary multiplicity
was shown in \cite{genredef} to be equivalent to a standard finite gauge transformation of the
SYM field $\bA^m$.

Starting from the binary-tree expansion $b(P)b(n)$, where $b(n){=}n$ is a single letter,
the YM tree amplitude $\AYM(P,n)$ is obtained from
the map\footnote{Similar maps were considered in \cite{genredef}.}
\beq
\label{kinalgebra}
N(\Gamma n) =  (e_\Gamma\cdot e_n)\, ,
\eeq
where $e^m_\Gamma$ is the local $\t{=}0$ component \eqref{zerolocal} of the superfield $A^m_\Gamma$ in the BCJ gauge reviewed in section~\ref{sfinbcj},
and the fermion wave functions such as the contribution $(\chi_1 \gamma^m \chi_2)$ to $e_{12}^m$ in
(\ref{explloc}) are understood to be set to zero. More precisely,
\beq
\label{symmap}
\AYM(P,n) = \lim_{s_P\to0} s_P N\big(b(P)b(n)\big)\,,
\eeq
where the expansion of the binary-tree map $b(P)$ decorates the color-kinematics dual numerators
with the cubic-diagram propagators. While the earlier expressions (\ref{thetaVP.2}) for 
the components of $n$-point BCJ numerators involve multiparticle polarizations of rank
$\leq n{-}2$, the numerators in (\ref{kinalgebra}) involve rank-$(n{-}1)$ building blocks.

For example, the four-point amplitudes in
the KK basis of color ordering following from \eqref{symmap} are given by
\begin{align}
\AYM(1234) &= \bigg({e^m_{[[1,2],3]} \over s_{12}}  +
{e^m_{[1,[2,3]]}\over s_{23}}\bigg)e^m_4\, , \label{the4ptcase}\\
\AYM(1324) &= \bigg({e^m_{[[1,3],2]} \over s_{13}}  +
{e^m_{[1,[3,2]]}\over s_{23}}\bigg)e^m_4 \, ,\notag
\end{align}
from which all BCJ kinematic numerator identities map one-to-one to the Jacobi identities of the
associated Lie polynomials. Similarly, the five-point amplitudes
\begin{align}
\label{amps5}
\AYM(12345) &=
\bigg({e^m_{[ [ [ 1 , 2 ] , 3 ] , 4 ]} \over s_{12} s_{45}}
+  {e^m_{[ 1 , [ [ 2 , 3 ] , 4 ] ]} \over s_{23} s_{51}}
+  {e^m_{[ [ 1 , 2 ] , [ 3 , 4 ] ]} \over s_{12} s_{34}}
+  {e^m_{[ [ 1 , [ 2 , 3 ] ] , 4 ]} \over s_{45} s_{23}}
+  {e^m_{[ 1 , [ 2 , [ 3 , 4 ] ] ]} \over s_{51} s_{34}}\bigg) e^m_5\,,\\
\AYM(14325) &=
\bigg({e^m_{[ [ [ 1 , 4 ] , 3 ] , 2 ]} \over s_{14} s_{25}}
+  {e^m_{[ 1 , [ [ 4 , 3 ] , 2 ] ]} \over s_{43} s_{51}}
+  {e^m_{[ [ 1 , 4 ] , [ 3 , 2 ] ]} \over s_{14} s_{32}}
+  {e^m_{[ [ 1 , [ 4 , 3 ] ] , 2 ]} \over s_{25} s_{43}}
+  {e^m_{[ 1 , [ 4 , [ 3 , 2 ] ] ]} \over s_{51} s_{32}}\bigg) e^m_5\, ,\cr
\AYM(13425) &= \bigg({e^m_{[ [ [ 1 , 3 ] , 4 ] , 2 ]} \over s_{13} s_{25}}
+  {e^m_{[ 1 , [ [ 3 , 4 ] , 2 ] ]} \over s_{43} s_{51}}
+  {e^m_{[ [ 1 , 3 ] , [ 4 , 2 ] ]} \over s_{13} s_{42}}
+  {e^m_{[ [ 1 , [ 3 , 4 ] ] , 2 ]} \over s_{25} s_{43}}
+  {e^m_{[ 1 , [ 3 , [ 4 , 2 ] ] ]} \over s_{51} s_{42}}\bigg) e^m_5\, ,\cr
\AYM(12435) &=
\bigg({e^m_{[ [ [ 1 , 2 ] , 4 ] , 3 ]} \over s_{12} s_{35}}
+  {e^m_{[ 1 , [ [ 2 , 4 ] , 3 ] ]} \over s_{24} s_{51}}
+  {e^m_{[ [ 1 , 2 ] , [ 4 , 3 ] ]} \over s_{12} s_{34}}
+  {e^m_{[ [ 1 , [ 2 , 4 ] ] , 3 ]} \over s_{35} s_{24}}
+  {e^m_{[ 1 , [ 2 , [ 4 , 3 ] ] ]} \over s_{51} s_{34}}\bigg) e^m_5\, ,\cr
\AYM(14235) &=
\bigg({e^m_{[ [ [ 1 , 4 ] , 2 ] , 3 ]} \over s_{14} s_{35}}
+  {e^m_{[ 1 , [ [ 4 , 2 ] , 3 ] ]} \over s_{24} s_{51}}
+  {e^m_{[ [ 1 , 4 ] , [ 2 , 3 ] ]} \over s_{14} s_{32}}
+  {e^m_{[ [ 1 , [ 4 , 2 ] ] , 3 ]} \over s_{35} s_{24}}
+  {e^m_{[ 1 , [ 4 , [ 2 , 3 ] ] ]} \over s_{51} s_{32}}\bigg) e^m_5\, , \cr
\AYM(13245) &=
\bigg({e^m_{[ [ [ 1 , 3 ] , 2 ] , 4 ]} \over s_{13} s_{45}}
+  {e^m_{[ 1 , [ [ 3 , 2 ] , 4 ] ]} \over s_{23} s_{51}}
+  {e^m_{[ [ 1 , 3 ] , [ 2 , 4 ] ]} \over s_{13} s_{24}}
+  {e^m_{[ [ 1 , [ 3 , 2 ] ] , 4 ]} \over s_{45} s_{23}}
+  {e^m_{[ 1 , [ 3 , [ 2 , 4 ] ] ]} \over s_{51} s_{24}}\bigg) e^m_5\notag
\end{align}
have kinematic numerators that
manifestly satisfy the color-kinematics duality. Similar expressions can be written down at
arbitrary multiplicity, and their form closely resembles the form of the amplitudes in the
Berends--Giele method, but now they arise from the planar binary tree expansion $b(P)$.

The above BCJ representations are equivalent to
\beq
\AYM(P) = \sum_{Q \in S_{n-2}} m(P|1,Q,n) (e_{1Q} \cdot e_n) \, ,
\label{equivbcj}
\eeq
where the propagators are now organized into doubly-partial amplitudes instead of $b(P)$.
This representation was studied in section 5 of \cite{Garozzo:2018uzj} and
generalized to tree-level matrix elements for the effective operators $\ap \Tr( \Bbb F^3)$ and 
$\ap^2 \Tr( \Bbb F^4)$ of the open bosonic string.

\subsubsection{An explicit solution to BCJ relations in KLT form}
\label{bcjfromklt}

The process of obtaining the field-theory limit (\ref{revbcj.21}) from the local $(n{-}2)!$-term
representation of disk amplitudes (\ref{revbcj.19}) can be repeated for the non-local form
(\ref{AstringP}) with $(n{-}3)!$ terms. From the low-energy limit \eqref{ftlim}, we arrive at an
explicit representation of the BCJ amplitude relations in terms of $(n{-}3)!$
SYM amplitudes,
\beq\label{revbcj.41}
A(P) = - \sum_{Q,R \in S_{n-3}} m(P|1,R,n,n{-}1) S(R|Q)_1 A(1,Q,n{-}1,n)\,,
\eeq
of both practical and conceptual appeal.

At the practical level, (\ref{revbcj.41}) is a closed-form solution to the entirety of
BCJ relations (\ref{BCJrelations}) or (\ref{fundBCJ}), i.e.\ for the expansion of
arbitrary color-ordered amplitudes in a prescribed $(n{-}3)!$ BCJ basis. The BCJ relations
by themselves do not offer any guidance on how to solve the huge equation
system to rewrite the $(n{-}1)!$ permutations of $A(P,n)$ in terms of the
$(n{-}3)!$ linearly independent $A(1,Q,n{-}1,n)$. Hence, it is beneficial
to have the closed formula for the expansion coefficients in (\ref{revbcj.41}),
in particular since the entries of $m(\cdot |\cdot)$ and $S(\cdot |\cdot)_1$
can be efficiently generated from the recursions (\ref{kltrec}) and (\ref{BGphi}), respectively.
For example, using \eqref{revbcj.41} to rewrite the SYM amplitude $A(24315)$ in the BCJ basis $\{A(12345), A(13245)\}$
we get
\begin{align}\label{exBCJfromString}
A(24315) &= 
- \big(m(24315|12354)S(23|23)_1
+ m(24315|13254)S(32|23)_1\big) A(12345)\\
&\quad{}- \big(m(24315|12354)S(23|32)_1
+ m(24315|13254)S(32|32)_1\big) A(13245)\notag\\
& = - {s_{12}\over s_{134}}A(12345)
- {
(s_{12}{+}s_{23})\over s_{134}}A(13245)\, ,\notag
\end{align}
where we used that $m(24315|12354) = 0$, $m(24315|13254) = 1/(s_{13}s_{134})$, as well as
\eqref{S0examp}
for the KLT matrix.

At a conceptual level, the KLT form of (\ref{revbcj.41}) leads to the
conclusion that SYM is a
double copy of bi-adjoint scalars with SYM itself. Since this statement carries over
to any other field or string theory subject to tree-level BCJ relations, bi-adjoint 
scalars can be viewed as the identity operator under taking double copies.
This can of course be anticipated from the identification (\ref{misinvklt.1}) of doubly-partial
amplitudes as the inverse KLT kernel \cite{DPellis}. The realization of SYM as a 
double copy of bi-adjoint scalars with SYM is the $\alpha' \rightarrow 0$ limit of 
the double-copy formula (\ref{AstringP}) for disk amplitudes: when interpreting
open superstrings as a double copy of $Z$-theory with SYM, bi-adjoint scalars
are recovered from the low-energy limit of the more general $Z$-theory of bi-colored
scalars, see section \ref{sec:7.5} for an $\alpha'$-expansion of their equations of motion.

We conclude by mentioning a quick consistency check of (\ref{revbcj.41}): for permutations
$P \rightarrow 1,A,n{-}1,n$ within the BCJ basis on the right-hand side, (\ref{revbcj.41})
holds trivially since $m(\cdot|\cdot)$ and $S(\cdot|\cdot)_1$ are inverse to each other by (\ref{misinvklt.1}).
For any other permutation $P$ outside the BCJ basis of $A(1,Q,n{-}1,n)$, SYM amplitudes $A(P)$
obey the same BCJ relations in $P$ as $m(P|B)$ at fixed $B$.

\subsection{String-theory KLT relations and the double-copy form of gravity numerators}
\label{sec:6.5}

This section is dedicated to gravitational amplitudes in string and field theories. We 
review the string-theory incarnation of the KLT formula, identify closed-string analogues 
of the $Z$-integrals along with their field-theory limits and deduce the local form of the
gravitational double copy with cubic-graph numerators given by perfect squares $N_i \tilde N_i$.
This is the tree-level case \cite{BCJ} of the conjecture due to Bern, Carrasco and Johansson
\cite{loopBCJ} that representations of gauge-theory amplitudes with manifest
color-kinematics duality induce explicit loop integrands in double-copy form for a variety
of gravitational theories. The BCJ double copy radically changed the computational
reach for multiloop amplitudes in supergravity and drives precision calculations
of gravitational-wave observables, 
see \cite{Bern:2019prr, Bern:2022wqg} for reviews and \cite{Adamo:2022dcm} for a white paper.

\subsubsection{String-theory KLT relations}
\label{sec:6.5.2}

The opening line for closed-string tree-level amplitudes in the pure spinor formalism is given by
\beq
{\cal M}^{\rm closed}_n = \bigg({-}\frac{ \ap}{2\pi} \bigg)^{n-3}\!\!\!\! \int \limits_{\mathbb C^{n-3} \setminus \{z_a = z_b\}} \! \! \! \! 
d^2 z_2\, d^2 z_3\,\ldots \,d^2 z_{n-2} \,
\langle \! \langle V^{\rm cl}_1(z_1) U^{\rm cl}_2(z_2) {\ldots} 
 U^{\rm cl}_{n-2}(z_{n-2})  V^{\rm cl}_{n-1}(z_{n-1})  V^{\rm cl}_n(z_n) \rangle \! \rangle\, ,
 \label{gravsec.1}
\eeq
where the integration of $z_2,z_3,\ldots,z_{n-2}$ over the Riemann sphere realizes the 
moduli-space integral over genus-zero surfaces with $n$ 
marked points in the ${\rm SL}_2(\mathbb C)$ frame with $z_1,z_{n-1},z_n$
fixed to $(0,1,\infty)$.
In comparison to the disk-amplitude prescription (\ref{treepresc}), 
the closed-string vertex operators $V^{\rm cl}_i,U^{\rm cl}_i$ are double 
copies of the open-string ones $V_i, U_i$,
\beq
V^{\rm cl}_i = | \lambda^\alpha A_{\alpha}(\theta) |^2 e^{k_i\cdot X} \, , \ \ \ \ \ \ 
U^{\rm cl}_i = \big| \p\t^{\a}A_{\a}(\t) + A_m(\t)\Pi^m + d_{\a}W^{\a}(\t) 
+ \tfrac{1}{2} N_{mn}F^{mn}(\t) \big|^2 e^{k_i\cdot X} \, ,
 \label{gravsec.2}
\eeq
where $K(\theta)$ denote the SYM superfields without their plane-wave
factor, see (\ref{planewaves}). Moreover, $| \lambda^\alpha A_{\alpha}(\theta) |^2
=  \lambda^\alpha A_{\alpha}(\theta)\tilde\lambda^{\hat \beta} \tilde A_{\hat \beta}(\theta)$
introduces right-moving counterparts $\tilde{\lambda}^{\hat \alpha},\tilde{\theta}^{\hat \beta}$
of the left-moving worldsheet variables $\lambda^\alpha ,\theta^\beta$ whose spinor indices
$\hat \alpha,\hat \beta,\ldots$ have same (opposite) chirality as $\alpha,\beta,\ldots$ in the
case of the type IIB (type IIA) theory. The $\tilde{\theta}$-expansion of
$\tilde A_{\hat \beta}(\tilde\theta)$ and all the other superfields $\tilde K(\tilde \theta)$ 
in $U^{\rm cl}_i $ again takes the form of (\ref{linTHEX}) with
independent gauge-multiplet polarizations $\tilde e_m,\tilde \chi^{\hat \alpha}$ 
in the place of $e_m, \chi^{ \alpha}$ in $A_\beta$. 

The correlator $\langle \! \langle \ldots \rangle \! \rangle$
in (\ref{gravsec.1}) is adapted to the sphere rather than the disk: apart from
the plane-wave factors $e^{k_i\cdot X}$ in (\ref{gravsec.2}),
the OPEs for the left- and right-moving parts of $V^{\rm cl}_i,U^{\rm cl}_i $ are performed
separately, and the zero-mode integral (\ref{tlct}) applies independently 
to $\lambda^{ \alpha},\theta^\beta$ and $\tilde{\lambda}^{\hat \alpha},
\tilde{\theta}^{\hat \beta}$. 
Hence, the sphere correlator in (\ref{gravsec.1}) 
factorizes into two copies of the correlators ${\cal K}_n$ on the disk defined by 
(\ref{nptfcttree}),
\beq
{\cal M}^{\rm closed}_n = \bigg({-}\frac{ \ap}{2\pi} \bigg)^{n-3}\!\!\!\!\int \limits_{\mathbb C^{n-3} \setminus \{z_a = z_b\}}
d^2 z_2 \, d^2 z_3\ldots d^2 z_{n-2}\, \langle {\cal K}_n\rangle \, \langle  \tilde{\cal K}_n\rangle
\prod_{i<j}^n |z_{ij}|^{-\alpha' s_{ij}} \, ,
 \label{gravsec.3}
\eeq
where the closed-string polarizations are obtained from the tensor products of
the superfields in $ {\cal K}_n$ and $\tilde{\cal K}_n$. The OPE singularities
in $\tilde{\cal K}_n$ are the complex conjugates $\bar z_{ij}^{-1}$ of the 
$z_{ij}^{-1}$ in $ {\cal K}_n$.

At three points, the absence of integrated punctures immediately leads to
the factorization of the string amplitudes into color-ordered open-string ones
\beq
{\cal M}^{\rm closed}_3  = {\cal A}(1,2,3) \tilde{\cal A}(1,2,3)\,,\label{gravsec.3a}
\eeq
where $\cA(1,2,3)$ is the open-string three-point amplitude \eqref{three}.
At $n\geq 4$ points, one can even decompose the closed-string Koba--Nielsen factor into products
of meromorphic and antimeromorphic functions according to
$ |z_{ij}|^{-\alpha' s_{ij}} =  (z_{ij})^{- \frac{\alpha'}{2} s_{ij}}  (\bar z_{ij})^{-\frac{\alpha' }{2}s_{ij}}$.
The integrand of (\ref{gravsec.3}) is a holomorphic square of a meromorphic but multivalued function
$\langle {\cal K}_n\rangle \prod_{i<j}^n (z_{ij})^{- \frac{\alpha'}{2} s_{ij}}$ with branch
points at the diagonals $z_a=z_b$. Hence, it requires care to extend the double-copy structure 
of (\ref{gravsec.3}) to the sphere integrals: the multivaluedness of $(z_{ij})^{- \frac{\alpha'}{2} s_{ij}}$
introduces monodromy phases $e^{\pm \frac{i \pi }{2} \alpha' s_{ij}}$ in relating different
integration contours which also take center stage in the discussion of monodromy
relations in section \ref{sec:6.5.1}.

The monodromy phases in unwinding the sphere integrals (\ref{gravsec.3})
over closed-string Koba--Nielsen factors into products of disk 
integrals (with open-string Koba--Nielsen factors at $\alpha' \rightarrow \frac{ \alpha' }{4}$)
have been firstly determined by Kawai, Lewellen and Tye (KLT) in 1986 \cite{KLTpaper}.
At four points, the phases conspire to a single trigonometric factor in
\begin{align}
{\cal M}^{\rm closed}_4 &= - \frac{ \ap}{2\pi}  \int_{\mathbb C^{n-3} \setminus \{0,1,\infty\}}
d^2 z_2 \, \langle {\cal K}_4\rangle \, \langle  \tilde{{\cal K}}_4\rangle \, 
|z_{2}|^{-\alpha' s_{12}} |1{-}z_{2}|^{-\alpha' s_{23}} \notag \\
&= -  \frac{ \ap}{2\pi} \sin \bigg( \frac{ \pi \alpha'}{2} s_{12} \bigg)
\int_0^1 dz_2 \, z_{2}^{-\frac{ \alpha'}{2} s_{12}} (1{-}z_{2})^{-\frac{ \alpha'}{2} s_{23}}  \, \langle {\cal K}_4\rangle   \label{gravsec.4} \\
&\ \ \ \ 
\times \int_0^{-\infty} d \bar z_2 \, (-\bar z_{2})^{-\frac{ \alpha'}{2} s_{12}} (1{-}\bar z_{2})^{-\frac{ \alpha'}{2} s_{23}}  \, \langle
\tilde{{\cal K}}_4\rangle \notag \\
&= - \frac{2}{\pi \alpha'} \sin \bigg( \frac{ \pi \alpha'}{2} s_{12}\bigg) {\cal A}(1,2,3,4;\tfrac{\ap}{4})
\, \tilde {\cal A}(2,1,3,4;\tfrac{\ap}{4}) \, , \notag
\end{align}
where the rescaling $\alpha' \rightarrow \frac{\alpha'}{4}$ in the open-string amplitudes on 
the right-hand side can be seen by comparison with the Koba--Nielsen exponents in $|z_{ij}|^{-2\alpha's_{ij}}$ in (\ref{Fdef}). Note that one can employ the Gamma-function 
representation (\ref{4ptex.3}) of open-string amplitudes together with $\sin(\pi x)
= \frac{ \pi x}{\Gamma(1+x) \Gamma(1-x)}$ and the field-theory KLT relations
(\ref{KLTrel}) to factor out the supergravity amplitude $M^{\rm grav}_4$
 \beq
{\cal M}^{\rm closed}_4 = M^{\rm grav}_4 \frac{ 
\Gamma(1{-}\tfrac{\ap}{2}s_{12})\Gamma(1{-}\tfrac{\ap}{2}s_{23})\Gamma(1{-}\tfrac{\ap}{2}s_{13}) }{
\Gamma(1{+}\tfrac{\ap}{2}s_{12})\Gamma(1{+}\tfrac{\ap}{2}s_{23})\Gamma(1{+}\tfrac{\ap}{2}s_{13})}\, ,
\label{4ptclosed}
\eeq
where $M^{\rm grav}_4$ is given in \eqref{KLTrel},
but this is no longer possible at five points.

The analogous trigonometric phase factors in the $n$-point KLT formula
furnish an $\alpha'$-dependent generalization of the field-theory momentum 
kernel $S(P|Q)_i$ \eqref{kltrec} involving $n{-}3$ trigonometric factors
\beq
{\cal S}_{\alpha'}(Aj|BjC)_i = \frac{2}{\pi \alpha'} \sin \bigg( \frac{ \pi \alpha'}{2}  k_j\cdot k_{iB}
\bigg)\, {\cal S}_{\alpha'}(A|BC)_i\, , \qquad {\cal S}_{\alpha'}(\emptyset|\emptyset)_1 = 1\,,
\label{gravsec.6}
\eeq
for example,
\begin{align}\label{sinKLTex}
{\cal S}_\ap(2|2)_1 &= {2\over\pi\ap}\sin\bigg({\pi\ap\over2}k_1\cdot k_2\bigg)\, ,\\
{\cal S}_\ap(23|23)_1 &= \Big({2\over\pi\ap}\Big)^2\sin\bigg({\pi\ap\over2}k_3\cdot k_{12}\bigg)\sin\bigg({\pi\ap\over2}k_1\cdot k_2\bigg)\, ,\notag\\
{\cal S}_\ap(23|32)_1 &= \Big({2\over\pi\ap}\Big)^2\sin\bigg({\pi\ap\over2}k_1\cdot k_2\bigg)
\sin\bigg({\pi\ap\over2}k_1\cdot k_3\bigg) = {\cal S}_\ap(32|23)_1 \, ,\notag\\
{\cal S}_\ap(32|32)_1 &= \Big({2\over\pi\ap}\Big)^2\sin\bigg({\pi\ap\over2}k_2\cdot k_{13}\bigg)
\sin\bigg({\pi\ap\over2}k_1\cdot k_3\bigg)\,.\notag
\end{align}
This generalizes the recursion (\ref{kltrec}) of the field-theory momentum
kernel, and the normalization factors are engineered to have
${\cal S}_{\alpha'}(P|Q)_i = S(P|Q)_i + {\cal O}(\ap^2)$.

The $n$-point KLT formula for closed-string tree amplitudes then 
takes the compact form \cite{KLTpaper, KLTmatrI, KLTmatrII}
\beq
{\cal M}^{\rm closed}_n = - \sum_{P,Q \in S_{n-3}} {\cal A}(1,P,n,n{-}1;\tfrac{\ap}{4}) 
{\cal S}_{\alpha'}(P|Q)_1  \tilde{\cal A}(1,Q,n{-}1,n;\tfrac{\ap}{4})
\label{stringkltrel} 
\eeq
and evidently reduces to the field-theory KLT relation (\ref{KLTrel})
as $\alpha' \rightarrow 0$. The KLT formula (\ref{stringkltrel})
with type I amplitudes on the right-hand side computes
tree amplitudes of the type IIB (type IIA) superstring if the chiralities of
the fermions in ${\cal A}(\ldots)$ and $\tilde{\cal A}(\ldots)$ are
the same (opposite). Similarly (\ref{stringkltrel}) relates tree amplitudes
of closed and open bosonic strings. 

As will be discussed in section \ref{sec:6.5.1},
the $(n{-}3)!$ permutations of $ {\cal A}(\ldots)$ and $ \tilde{\cal A}(\ldots)$ on the right-hand
side of (\ref{stringkltrel}) furnish bases under the monodromy relations of
color-ordered open-string amplitudes. Accordingly,
the four-point KLT relations (\ref{gravsec.4}) can be written in the alternative form
\begin{align}
{\cal M}^{\rm closed}_4 &=  
 - \frac{2}{\pi \alpha'} \sin \bigg( \frac{ \pi \alpha'}{2} s_{23}\bigg) {\cal A}(1,2,3,4;\tfrac{\ap}{4}) \,
 \tilde{\cal A}(1,3,2,4;\tfrac{\ap}{4}) \, . \label{gravsec.4a}
\end{align}
These two equivalent forms stem from different 
ways of deforming integration contours in \cite{KLTpaper}. The systematic study of the 
analogous $n$-point integration contours on the sphere led to the momentum-kernel 
formalism in \cite{KLTmatrII}.

One can also manifest the symmetry ${\cal A}  \leftrightarrow \tilde{ \cal A} $
of the KLT formula by repeated use of monodromy relations, but already the four-point
example
\begin{align}
{\cal M}^{\rm closed}_4 &=
-  \frac{2}{\pi \alpha'} \frac{ \sin \big( \frac{ \pi \alpha'}{2} s_{12}\big)
\sin \big( \frac{ \pi \alpha'}{2} s_{23}\big)}{\sin \big( \frac{ \pi \alpha'}{2} s_{13}\big)}
{\cal A}(1,2,3,4;\tfrac{\ap}{4}) \, \tilde{\cal A}(1,2,3,4;\tfrac{\ap}{4}) \label{gravsec.8}
\end{align}
shows that the locality of the KLT kernel in (\ref{gravsec.6}) is lost in this way.
This motivates the choice of asymmetric bases for ${\cal A}$ and $\tilde{ \cal A}$
in (\ref{stringkltrel}) which lead the simple and local entries 
(\ref{gravsec.6}) of the $n$-point KLT kernel.

\subsubsection{Sphere integrals and their field-theory limit}
\label{sec:6.5.3}

The derivation of the KLT formula is independent on the polarizations accompanying the
sphere integrals and the rational functions of $z_j,\bar z_j$ entering the correlators 
$ \langle {\cal K}_n\rangle , \langle  \tilde{{\cal K}}_n\rangle$ in (\ref{gravsec.3}).
Hence, one can rewrite it at the level of Parke--Taylor integrals
\beq
\label{Jintdef}
J(P|Q) := \bigg({-}\frac{ \ap}{2\pi} \bigg)^{n-3}\!\!\!\!
\int \limits_{\mathbb C^{n-3}} {d^2z_1 \ d^2z_2 \ \cdots  \ d^2z_n \over {\rm vol}({\rm SL}_2(\Bbb C))}
\, \prod_{i<j}^n |z_{ij}|^{-\ap s_{ij}}   \PT(Q) \overline{\PT(P)}
\ee
that furnish the closed-string counterparts of the $Z$-integrals (\ref{Zintdef}):
In both of $Z(P|Q)$ and $J(P|Q)$, the second entry refers to the meromorphic
Parke--Taylor factor $\PT(Q)$ in the integrand. The role of first word $P$ in turn
changes in passing from the disk to the sphere -- instead of a disk ordering
$D(P)$, it refers to an antimeromorphic (i.e.\ complex conjugate) Parke--Taylor 
factor $\overline{\PT(P)}$ in the sphere integrand of (\ref{Jintdef}) which does 
not arise in disk correlators.

The equivalent of the KLT formula (\ref{stringkltrel}) for the sphere integrals (\ref{Jintdef})
takes a universal form for any pair of Parke--Taylor factors $\PT(Q) \overline{\PT(P)}$,
\beq
J(P|Q ) = - \sum_{A,B \in S_{n-3}} Z(1,A,n,n{-}1|P ) 
{\cal S}_{\alpha'}(A|B)_1 Z(1,B,n{-}1,n|Q)
\label{gravsec.9}
\eeq
and in fact for any other pair of rational functions in $z_j , \bar z_j$ of 
the same ${\rm SL}_2(\mathbb C)$-weight. Here and below, the rescaling
$\alpha' \rightarrow \frac{ \alpha' }{4}$ within the disk integrals
$Z(1,A,n,n{-}1|P ) $, $ Z(1,B,n{-}1,n|Q)$ is implicit. This rescaling 
rule applies whenever disk integrals are imported into closed-string computations
as in (\ref{stringkltrel}) or (\ref{gravsec.9}).

The field-theory limit of (\ref{gravsec.9}) reveals another striking parallel between the $Z(P|Q) $ 
and $J(P|Q) $ integrals:
Given that the $\alpha' \rightarrow 0$ limit (\ref{ftlim}) of disk integrals introduces doubly-partial
amplitudes $m(P|Q)$ and therefore the inverse field-theory KLT matrix by (\ref{misinvklt.1}),
we conclude that \cite{Stieberger:2014hba}
\beq
\lim_{\alpha' \rightarrow 0} J(P|Q) = m(P|Q)\, ,
\label{JFTlim}
\eeq
i.e.\ the disk and sphere integrals $Z(P|Q) $ and $J(P|Q)$ have the same field-theory limit.
As will be detailed in section \ref{sec:7.6}, this parallel between $Z(P|Q) $ and $J(P|Q)$
even has an echo at all orders in their $\alpha'$-expansions.

\subsubsection{The local form of the gravitational double copy}
\label{sec:6.5.4}

The sphere integrals (\ref{Jintdef}) of Parke--Taylor type and their field-theory limit 
(\ref{JFTlim}) admit an elegant proof of the gravitational double copy at the level
of cubic tree diagrams, cf.\ section \ref{sec:6.4}. The starting point is the local
representation of the disk correlator in the form of (\ref{revbcj.19})
\begin{align}
\langle {\cal K}_n \rangle = \frac{ dz_1 \, dz_{n-1} \, dz_n}{ {\rm vol}({\rm SL}_2(\Bbb R)) }
 \sum_{P \in S_{n-2}} N_{1| P | n-1} \PT(1,P,n{-}1)\ {\rm mod} \ \nabla_{z_k}\,,
\label{gravsec.11}
\end{align}
where the superfield representation $\sim \langle V_X V_Y V_Z \rangle$
of the master numerators $N_{1| P | n-1}$ (with $P$ a permutation of $2,3,\ldots,n{-}2,n$) 
can be found in (\ref{revbcj.22}). The rescaling $\alpha' \rightarrow \frac{\alpha'}{4}$ 
in a closed-string context also applies to the expression (\ref{defnabf}) for Koba--Nielsen 
derivatives $\nabla_{z_k}$. The DDM-type formula (\ref{gravsec.11}) was already at the heart 
of deriving BCJ numerators from disk amplitudes in section \ref{sec:6.4.3}. Upon insertion
into the closed-string amplitude representation (\ref{gravsec.3}) and identifying
the $J$-integrals (\ref{Jintdef}), it leads to the $(n{-}2)!^2$-term expression
\beq
{\cal M}^{\rm closed}_n = \sum_{P,Q \in S_{n-2}} \tilde{N}_{1|P|n-1}
J(1,P,n{-}1|1,Q,n{-}1) N_{1|Q|n-1}\,,
\label{gravsec.12}
\eeq
where $P,Q$ are again permutations of $2,3,\ldots,n,n{-}2$ (our choice of ${\rm SL}_2$ frame
led to a swap $n\leftrightarrow n{-}1$ relative to the DDM-type formulae in section \ref{sec:6.4.2}).
With the field-theory limit (\ref{JFTlim}) of the sphere integrals and relabeling of
$n\leftrightarrow n{-}1$, one readily obtains gravity amplitudes in the form
\beq
M_n^{\rm grav} = \sum_{P,Q \in S_{n-2}} \tilde  N_{1|P|n} m(1,P,n | 1,Q,n) N_{1|Q|n} 
\label{gravsec.13}
\eeq
analogous to the color-dressed tree amplitudes of bi-adjoint scalars and SYM
in (\ref{revbcj.15}) and (\ref{revbcj.17}). By Jacobi identities of both color factors $c_i$
and kinematic numerators $N_i$, (\ref{revbcj.15}) and (\ref{revbcj.17}) were explained
to be equivalent to the cubic-diagram expansions (\ref{revbcj.16}) and (\ref{revbcj.2}).
Since these rewritings solely depend on the properties of the universal building block
$m(1,P,n | 1,Q,n)$, the same equivalence must hold for (\ref{gravsec.13}) and
\beq
M^{\rm grav}_n = \sum_{i \in \Gamma_n} \frac{  N_i \tilde{ N}_i }{D_i}\, ,
\label{gravsec.14}
\eeq
where both types of kinematic numerators $ N_i $ and $\tilde{ N}_i$ obey Jacobi relations.
Hence, we have derived the prescription of \cite{BCJ, loopBCJ, Bern:2010yg} 
that kinematic Jacobi relations among the numerators
are sufficient to obtain gravity amplitudes from SYM via
\beq
M^{\rm grav}_n = M^{\rm gauge}_n \, \Big|_{c_i \rightarrow \tilde{ N}_i}
= \sum_{i \in \Gamma_n} \frac{  N_i c_i }{D_i} \, \bigg|_{c_i \rightarrow \tilde{ N}_i}\, ,
\label{gravsec.15}
\eeq
i.e.\ by replacing color factors by another copy $\tilde{ N}_i$ of
kinematic numerators. In fact, the Jacobi identities of the color factors
$c_i$ imply that the cubic-diagram expansion (\ref{revbcj.2}) of gauge-theory 
amplitudes may still accommodate Jacobi-violating numerators $N_i$, see
the discussion below (\ref{revbcj.9}). Accordingly, the color-kinematics dual
representation (\ref{gravsec.14}) of gravity amplitudes is still valid if only one of
the sets of numerator $\{N_i\}$ or $\{ \tilde{ N}_i \}$ obeys Jacobi identities.

Note that (\ref{gravsec.13}) in combination with
(\ref{revbcj.18}) yields another manifestly local formulation of the
double copy
\beq
M_n^{\rm grav} = \sum_{P \in S_{n-2}} \tilde  N_{1|P|n} A(1,P,n ) \, ,
\label{gravalt.13}
\eeq
which is obtained from the DDM form (\ref{revbcj.14}) through the
same replacement $c_{1|P|n}  \rightarrow \tilde  N_{1|P|n} $
as in (\ref{gravsec.15}).

\subsubsection{Consistency check with the field-theory KLT relation}
\label{sec:6.5.9}

As exemplified in section \ref{bcjfromklt}, it is rewarding to also insert the non-local
form (\ref{KLTcorrel}) of the disk correlator into the field-theory limit of string amplitudes.
In the closed-string case, (\ref{gravsec.3}) together with a relabeling of  $n \leftrightarrow n{-}1$
in $\tilde{\cal K}_n$ leads to the amplitude representation
\beq
{\cal M}^{\rm closed}_n = \! \! \! \sum_{P,Q,A,B \in S_{n-3}} \! \! \!
 \tilde A(1,P,n,n{-}1) S(P|Q)_1 
J(1,Q,n{-}1,n|1,A,n,n{-}1) S(A|B)_1 A(1,B,n{-}1,n)\, ,
\label{gravsec.16}
\eeq
where both left- and right-moving correlators (\ref{KLTcorrel}) contribute one copy 
of the field-theory KLT kernel. In order to make contact with the
supergravity amplitude, we apply the field-theory limit (\ref{JFTlim}) of the
sphere integrals $J$ and the inverse relation (\ref{misinvklt.1}) between
$m(1,Q,n{-}1,n|1,A,n,n{-}1) $ and $S(A|B)_1$, leading to
\beq
S(P|B)_1 = - \sum_{Q,A \in S_{n-3}}
S(P|Q)_1
\lim_{\alpha' \rightarrow 0}
J(1,Q,n{-}1,n|1,A,n,n{-}1) S(A|B)_1\,.
\label{step.16}
\eeq
Upon insertion into the $\alpha' \rightarrow 0$ limit of (\ref{gravsec.16}), this
reproduces the KLT formula (\ref{KLTrel}) for gravity amplitudes.
In this way, we confirm the compatibility of the monodromy phases in
manipulating integration cycles on the sphere and disk (which among
other things led to the field-theory limit (\ref{JFTlim}) of the sphere integrals $J$) with the
expansion of disk correlators (\ref{KLTcorrel}) in a basis of Parke--Taylor factors.

\subsection{Monodromy relations}
\label{sec:6.5.1}

Color-ordered open-string amplitudes ${\cal A}(P)$ associated with different orderings
$P$ of the vertex operators
on the disk boundary obey monodromy relations \cite{BjerrumBohrRD, StiebergerHQ}. 
Similar to the KLT relations (\ref{stringkltrel}) for closed-string amplitudes, they
solely rely on analytic properties of the disk worldsheet and are therefore universal to 
the bosonic theory and type I superstrings. Monodromy relations can be equivalently formulated 
at the level of the $Z(P|Q)$-integrals (\ref{Zintdef}): while section \ref{sec:6.3.3} featured
relations between different ``integrands $Q$'' at fixed 
``integration domain $P$'', monodromy relations concern different choices
of the domain $P$ at fixed integrand $Q$. To begin with, the procedure of fixing
${\rm SL}_2(\mathbb R)$ frames in section \ref{sec:6.3.1} leads to the following
cyclicity and reflection properties,
\beq\label{cycref}
Z(p_1p_2\ldots p_n| Q) = Z(p_2 p_3\ldots p_n p_1|Q)
= (-1)^n Z(p_n \ldots p_2 p_1|Q) \, ,
\ee
yielding a naive upper bound of $\frac{1}{2}(n{-}1)!$ independent disk orderings.
However, the actual basis dimensions for color-ordered disk amplitudes identified 
by monodromy relations are considerably smaller with only 
$(n{-}3)!$ choices of $P$ at fixed $Q$ \cite{BjerrumBohrRD, StiebergerHQ}.
The proof relies on the following simple analytic property of the disk integrand and thereby
extends to integrands of suitable ${\rm SL}_2(\mathbb R)$-weight beyond Parke--Taylor factors: 
the only non-meromorphic dependence on the integration variables 
in (\ref{Zintdef}) occurs through the Koba--Nielsen factor 
$\prod^n_{1\leq i<j} |z_{ij}|^{-2\alpha' s_{ij}}$.
The latter can be related to the meromorphic but multivalued function 
$\prod^n_{1\leq i<j} (z_{ij})^{-2\alpha' s_{ij}}$ by monodromy phases $e^{\pm 2\pi i \alpha' s_{ij}}$
which differ from one ordering $P$ to another. The same type of 
monodromy phases gives rise to the trigonometric factor in the four-point
KLT relation (\ref{gravsec.4}). By applying Cauchy's theorem as 
detailed in \cite{BjerrumBohrRD, StiebergerHQ}, one obtains,
\beq\label{PBCJ}
0 = \sum_{j=1}^{n-1} \exp\! \big[2\pi i \alpha' (k_{p_1} \cdot k_{p_2 p_3 \ldots p_j}) \big]
Z(p_2p_3\ldots p_{j}p_1 p_{j+1} \ldots p_n|Q) \, ,
\ee
and the associated relation among color-ordered amplitudes of open superstrings,
\beq\label{PBCJAopen}
0 = \sum_{j=1}^{n-1} \exp\! \big[2\pi i \alpha' (k_{p_1} \cdot k_{p_2 p_3 \ldots p_j}) \big]
{\cal A}(p_2p_3\ldots p_{j}p_1 p_{j+1} \ldots p_n) \, ,
\ee
which take an identical form for open bosonic strings. Since different choices 
of branches yield identical relations with opposite phases as compared to
(\ref{PBCJ}) and (\ref{PBCJAopen}), one can take sums and differences of
both options and obtain \cite{BjerrumBohrRD, StiebergerHQ}
\begin{align}
0 &= {\cal A}(p_1  p_2p_3\ldots  p_n) 
+\sum_{j=2}^{n-1} \cos\! \big[2\pi \alpha' (k_{p_1} \cdot k_{p_2 p_3 \ldots p_j}) \big]
{\cal A}(p_2p_3\ldots p_{j}p_1 p_{j+1} \ldots p_n)  \, ,
\notag \\
0 &= \sum_{j=2}^{n-1} \sin\! \big[2 \pi \alpha' (k_{p_1} \cdot k_{p_2 p_3 \ldots p_j}) \big]
{\cal A}(p_2p_3\ldots p_{j}p_1 p_{j+1} \ldots p_n) \, . \label{sepmono}
\end{align}
For real kinematics, these are simply the real and imaginary parts of (\ref{PBCJAopen}).
At leading order in $\alpha'$, the trigonometric coefficients reduce to $\cos(\alpha' x) \rightarrow1$
and $\sin(\alpha'x) \rightarrow \alpha' x$, respectively. As a result, one obtains (special instances of) 
KK relations (\ref{KKrelation}) and the fundamental BCJ relations (\ref{fundBCJ}) as the 
low-energy limit of the first and second line of (\ref{sepmono}), respectively 
\cite{BjerrumBohrRD, StiebergerHQ}. Moreover, the fact that $\lim_{\alpha' \rightarrow 0} Z(P|Q)$ 
obeys KK and BCJ relations in $P$ at fixed $Q$ is consistent with the relations
of $m(P|Q)$ obtained in the field-theory limit.

In the canonical ordering $p_j=j$ at four points, (\ref{sepmono}) reduce to
\begin{align}
0 &= {\cal A}(1,2,3,4) + \cos(2\pi \alpha' k_1 \cdot k_2) {\cal A}(2,1,3,4) 
+ \cos(2\pi \alpha' k_1 \cdot k_{23}) {\cal A}(2,3,1,4) \, , \notag \\
0 &= \sin(2\pi \alpha' k_1 \cdot k_2) {\cal A}(2,1,3,4) 
+ \sin(2\pi \alpha' k_1 \cdot k_{23}) {\cal A}(2,3,1,4) \, .
\label{sepmono4pt}
\end{align}
The second relation together with $k_1 \cdot k_{23}=-s_{23}$ and
${\cal A}(2,3,1,4) = {\cal A}(1,3,2,4)$ establishes the equivalence between the
two forms (\ref{gravsec.4}) and (\ref{gravsec.4a}) of the four-point KLT relations
(taking the usual rescaling $\alpha' \rightarrow \frac{\alpha'}{4}$ into account).
More generally, one may view monodromy relations as a consistency condition
for permutation invariance of the $n$-point KLT formula (\ref{stringkltrel}): the 
$(n{-}3)! \times (n{-}3)!$ 
pairs of ${\cal A}(1,P,n,n{-}1) \tilde {\cal A}(1,Q,n{-}1,n)$ on the right-hand side need 
to generate bilinears ${\cal A}(X) \tilde {\cal A}(Y)$ with arbitrary orderings $X,Y$ through linear
combinations.

The monodromy relations of individual disk integrals
\begin{align}
0 &= Z(p_1  p_2p_3\ldots  p_n | Q) 
+\sum_{j=2}^{n-1} \cos\! \big[2\pi \alpha' (k_{p_1} \cdot k_{p_2 p_3 \ldots p_j}) \big]
Z(p_2p_3\ldots p_{j}p_1 p_{j+1} \ldots p_n | Q) \, ,
\notag \\
0 &= \sum_{j=2}^{n-1} \sin\! \big[2 \pi \alpha' (k_{p_1} \cdot k_{p_2 p_3 \ldots p_j}) \big]
Z(p_2p_3\ldots p_{j}p_1 p_{j+1} \ldots p_n |Q )  \label{sepmonoZ}
\end{align}
equivalent to (\ref{sepmono}) underpin our viewpoint on 
the disk integrals \eqref{Zintdef} as the doubly-partial
amplitudes of $Z$-theory. By \eqref{ZKK} and \eqref{ZBCJ}, $Z$-theory amplitudes $Z(P|Q)$ satisfy 
the color-kinematics duality in the integrand orderings $Q$ at fixed $P$ to all orders in $\alpha'$. The
converse relations \eqref{sepmonoZ} among the integration domain orderings $P$ at fixed $Q$, 
on the other hand, exhibit additional trigonometric $\alpha'$-dependence. These 
trigonometric functions imprint the monodromy properties of the disk worldsheets 
on the S-matrix of $Z$-theory upon dressing with
the relevant Chan--Paton factors $\sum_P Z(P,n|Q) {\Tr}(t^P t^n)$.

\subsubsection{The $(n{-}2)!$ form of color-dressed open-string amplitudes}
\label{sec:6.5.new}

Since the first relation of (\ref{sepmono}) deforms the KK relations of gauge-theory
amplitudes by $\cos(2\pi \alpha' k_P\cdot k_Q)$, one may wonder about the string-theory
uplift of the DDM decomposition (\ref{revbcj.14}) in gauge theory. By repeated use of 
monodromy relations, one can express the color-dressed open-superstring amplitude 
(\ref{colordr}) in terms of the $(n{-}2)!$ color orderings ${\cal A}(1,P,n)$ with $P=p_2p_3\ldots p_{n-1}$ \cite{Ma:2011um},
\beq
{\cal M}_n = \sum_{P \in S_{n-2}} \Tr(  [[ \ldots [[ t^1 , t^{p_2}]_{\alpha'}, t^{p_3}]_{\alpha'},\ldots, t^{p_{n-2}}]_{\alpha'},t^{p_{n-1}}]_{\alpha'} t^n ) {\cal A}(1,\rho(2,3,\ldots,n{-}1),n)\, .
\label{KKmono.00}
\eeq
Their coefficients generalize the
color factors in the field-theory DDM formula (\ref{revbcj.14}) 
\beq
c_{1|23\ldots n-1|n} = \Tr(  [[ \ldots [[ t^1 , t^2], t^3],\ldots, t^{n-2}],t^{n-1}] t^n )
\label{KKmono.01}
\eeq
to involve an $\alpha'$-dependent bracket instead of the conventional commutator
\beq
[t^p , t^q ]_{\alpha'} = e^{i\pi \alpha' k_p \cdot k_q} t^p t^q - e^{-i\pi \alpha' k_p \cdot k_q} t^q t^p \, .
\label{KKmono.02}
\eeq
For matrix products in the entries of $[\cdot , \cdot]_{\alpha'}$ or nested commutators, 
the exponentials are understood to involve multiparticle momenta, 
e.g.\ $[t^1 t^2 , t^3 ]_{\alpha'} = e^{i\pi \alpha' k_{12} \cdot k_3}
 t^1 t^2 t^3 - e^{-i\pi \alpha' k_{12} \cdot k_3} t^3 t^1 t^2 $. At four points, for instance
\begin{align}
{\cal M}_4 &= \Tr( [[ t^1 , t^{2}]_{\alpha'}, t^{3}]_{\alpha'} t^4 )  {\cal A}(1,2,3,4)  + (2\leftrightarrow 3)
\notag\\
&= \big[  e^{i\pi \alpha' ( k_1\cdot k_2+ k_{12}\cdot k_3)} \Tr(t^1 t^2 t^3 t^4)
- e^{i\pi \alpha' (-k_1\cdot k_2+k_{12}\cdot k_3)} \Tr(t^2 t^1 t^3 t^4) \label{KKmono.03} \\
& \ \ \ \
- e^{i\pi \alpha' (k_1\cdot k_2- k_{12}\cdot k_3)} \Tr(t^3 t^1 t^2 t^4)
+e^{i\pi \alpha' (-k_1\cdot k_2- k_{12}\cdot k_3)} \Tr(t^3 t^2 t^1 t^4) \big]
 {\cal A}(1,2,3,4) + (2\leftrightarrow 3)
 \notag \\
 &=   \Tr(t^1 t^2 t^3 t^4{+}  t^3 t^2 t^1 t^4) {\cal A}(1,2,3,4)
 - \Tr(t^2 t^1 t^3 t^4) \big[ e^{-2\pi i \alpha' s_{12}} {\cal A}(1,2,3,4) {+} e^{2\pi i \alpha' s_{13}} {\cal A}(1,3,2,4) \big] + (2\leftrightarrow 3)
 \notag
\end{align}
can be checked to reproduce the conventional form (\ref{colordr}) of the color-dressed
amplitude by means of $k_{12}\cdot k_3 = - s_{12}$ and the monodromy relation
$e^{-2\pi i \alpha' s_{12}} {\cal A}(1,2,3,4) + e^{2\pi i \alpha' s_{13}} {\cal A}(1,3,2,4)
= - {\cal A}(2,1,3,4)$. 

By isolating the coefficient of a given $\Tr(t^{p_1} t^{p_2}\ldots t^{p_n})$ on the
right-hand side of (\ref{KKmono.00}), this DDM-type decomposition of open-string
tree amplitude generates the expansion of arbitrary ${\cal A}(P)$ in a prescribed 
$(n{-}2)!$-element set of disk orderings. However, the expansion coefficients
are not unique since the ${\cal A}(1,\ldots,n)$ on the right hand side of
(\ref{KKmono.00}) are still related by monodromy relations.
As we will see in section \ref{sec:7.7}, the specialization of (\ref{KKmono.00})
to abelian Chan--Paton generators $t^{i} \rightarrow \mathds{1}$ has valuable 
applications to Born--Infeld theory and its double-copy structure.

\subsubsection{The $(n{-}3)!$ form of color-dressed open-string amplitudes}
\label{sec:6.5.5}

The next step after identifying the string-theory uplift (\ref{KKmono.00}) of 
the $(n{-}2)!$-term DDM decomposition is to reduce color-ordered string
amplitudes to an $(n{-}3)!$-element basis under monodromy relations. 
As can be anticipated from the reduction of gauge-theory amplitudes 
into a BCJ basis via (\ref{revbcj.41}), an elegant solution of the monodromy
relations is offered by the string-theory KLT kernel and its inverse.

In view of the interpretation (\ref{misinvklt.1}) of the inverse field-theory KLT
kernel as doubly-partial amplitudes of bi-adjoint scalars, the inverse of the
string-theory KLT kernel ${\cal S}_{\alpha'}$ has been firstly studied in \cite{Mizera:2016jhj}. 
Its entries w.r.t.\ the $(n{-}3)! \times (n{-}3)!$ basis of ${\cal S}_{\alpha'}$ 
in (\ref{gravsec.6}) are given by
\beq
 m_{\alpha'}^{-1}(1,R,n{-}1,n | 1,Q,n,n{-}1) = - {\cal S}_{\alpha'}(R|Q)_1\, ,
 \label{gravsec.17}
\eeq
and one can infer more general entries $m_{\alpha'}(A|B)$ by inverting
the kernel in different representations of the KLT relations (\ref{stringkltrel})
with other $(n{-}3)!$ bases ${\cal B}_1,{\cal B}_2$ of ${\cal A}(\ldots)$, $\tilde{\cal A}(\ldots)$,
\beq
{\cal M}^{\rm closed}_n = \sum_{P,Q \in {\cal B}_1,{\cal B}_2} 
{\cal A}(P) m_{\alpha'}^{-1}(P|Q) \tilde{\cal A}(Q)\, .
\label{gravsec.22}
\eeq 
At four and five points, for example, we have,
\begin{align}
m_{\alpha'}(1,2,3,4|1,2,4,3) &= -\frac{ \pi \alpha'}{2 \sin \big( \frac{ \pi \alpha'}{2} s_{12}\big)}\, ,
\label{gravsec.18}\\
m_{\alpha'}(1,2,3,4|1,2,3,4) &= \frac{\pi \alpha'}{2} \bigg\{
\cot \bigg( \frac{ \pi \alpha'}{2} s_{12}\bigg)+ \cot \bigg( \frac{ \pi \alpha'}{2} s_{23}\bigg)
\bigg\} \, , \notag\\
m_{\alpha'}(1,5,3,2,4|1,2,3,5,4) &=\Big({\pi\ap\over2}\Big)^2
{1\over \sin({\pi\ap\over2}s_{14})}\bigg\{
\cot\bigg({\pi\ap\over2}s_{23}\bigg) + \cot\bigg({\pi\ap\over2}s_{35}\bigg)\bigg\} \, ,\notag\\
m_{\alpha'}(1,5,3,2,4|1,3,2,5,4) &= -\Big({\pi\ap\over2}\Big)^2{1\over
\sin({\pi\ap\over2}s_{14})}{1\over\sin({\pi\ap\over2}s_{23})}\,,
\end{align} 
and permutations. The entries of higher-multiplicity $m_{\alpha'}$ are efficiently
generated by the {\tt Mathematica} notebook in the ancillary file of \cite{Mizera:2016jhj}.

At any multiplicity, $m_{\alpha'}(A|B)$ enjoys cyclicity and
monodromy relations of open-string amplitudes in both slots $A,B$ (while holding
the other one fixed) after performing the usual conversion $\alpha'\rightarrow 4\alpha'$
between closed- and open-string settings. This leads to the elegant formula to expand 
disk amplitudes or $Z$-integrals with an arbitrary integration cycle $P$ in a prescribed basis w.r.t.\
monodromy relations \cite{Mizera:2016jhj},
\begin{align}
{\cal A}_n(P) &= - \sum_{Q,R \in S_{n-3}} m_{4\alpha'}(P|1,R,n,n{-}1) 
{\cal S}_{4\alpha'}(R|Q)_1 {\cal A}(1,Q,n{-}1,n) \, , \label{gravsec.21}\\
Z(P| C) &= - \sum_{Q,R \in S_{n-3}} m_{4\alpha'}(P|1,R,n,n{-}1) 
{\cal S}_{4\alpha'}(R|Q)_1 Z(1,Q,n{-}1,n|C)\, ,
\notag
\end{align}
which readily follows from the logic of section \ref{bcjfromklt}. The first
line of (\ref{gravsec.21}) furnishes the string-theory uplift of the BCJ reduction
of SYM amplitudes in (\ref{revbcj.41}), and the second line is valid for
arbitrary Parke--Taylor orderings $C$ in the $Z$-integrals (see (\ref{misinvklt.2}) for the
converse formula for attaining a prescribed basis of Parke--Taylor factors at
fixed disk ordering). For example, using the expansions of $m_{\alpha'}$ 
in \eqref{gravsec.18} and the KLT matrix in \eqref{sinKLTex} we get
\begin{align}
\cA(1,5,3,2,4) &= \frac{ 
\sin (2\pi \alpha' s_{12} )\sin\big(2\pi \alpha' (s_{13}{+}s_{23}) \big)
}{  \sin(2\pi \alpha' s_{14}) \sin(2\pi \alpha' s_{23}) }
\cA(1,2,3,4,5)+\frac{ \sin(2\pi \alpha' s_{12})\sin (2\pi \alpha' s_{13}) }{
\sin(2\pi \alpha' s_{14}) \sin(2\pi \alpha' s_{23})
}
\cA(1,3,2,4,5)\notag\\
&\quad - \Big\{\cot (2\pi \alpha' s_{23} ) 
+ \cot (2\pi \alpha'  s_{35} )\Big\}
 \frac{ \sin ( 2\pi \alpha'  s_{12})\sin (2\pi \alpha' s_{13})}{ \sin(2\pi \alpha' s_{14}) }
\cA(1,2,3,4,5)\label{5ptreduce}\\
&\quad -\Big\{\cot (2\pi \alpha' s_{23} ) + \cot (2\pi \alpha' s_{35} )\Big\}
\frac{ \sin(2\pi \alpha' s_{13})\sin\big(2\pi \alpha' (s_{12}{+}s_{23})\big)
}{\sin(2\pi \alpha' s_{14})}
\cA(1,3,2,4,5) \, , \notag
\end{align}
consistent with the five-point examples in \cite{BjerrumBohrRD, StiebergerHQ}.

In fact, Mizera identified the entries of $m_{\alpha'}$ with intersection numbers
of twisted cycles \cite{Mizera:2017cqs} and thereby opened up a fascinating
connection between string perturbation theory and twisted deRham theory. In this
framework, the KLT relations are a consequence of {\it twisted period relations} 
\cite{cho1995}, and their representation (\ref{gravsec.22})
follows from elementary linear algebra in twisted homologies
and cohomologies \cite{Mizera:2017cqs}. Similarly, the field-theory KLT
relations can be understood from intersection numbers of 
twisted cocycles \cite{Mizera:2017rqa}. 

Since $m_{\alpha'}$ can be algorithmically computed
from intersection numbers, there is no circular logic in solving monodromy
relations via (\ref{gravsec.21}): the entries of $m_{\alpha'}$ beyond 
the $(n{-}3)! \times (n{-}3)!$ basis in (\ref{gravsec.17}) do not necessitate
any prior knowledge of the solutions to the monodromy relations.

\subsection{Double copies beyond gravity from string amplitudes}
\label{sec:7.7}

The color-kinematics duality and double copy apply to a much wider classes
of theories beyond gauge theories and (super-)gravity 
\cite{Bern:2019prr, Bern:2022wqg, Adamo:2022dcm}.
In this section, we will review the input of superstring tree amplitudes
on the double-copy structure of Born--Infeld theory and its supersymmetrizations.
The reasoning will be based on the KLT-form (\ref{KLTcorrel}) of the disk correlator
which implies that all tree amplitudes of Born--Infeld are double copies
involving a BCJ basis of SYM tree amplitudes. The other double-copy component
of Born--Infeld amplitudes turns out to be a non-linear sigma model (NLSM) 
of Goldstone bosons even though the latter are not part of the naive string spectrum
(but can be engineered to arise as massless excitations in the setup of \cite{Green:1995ga}).

\subsubsection{Born--Infeld and NLSM}
\label{sec:7.7.1}

The low-energy limit of abelian open-superstring tree-level interactions gives rise
to the Born--Infeld theory \cite{Metsaev:1987qp}, also see \cite{Tseytlin:1999dj} for a review
and \cite{Kallosh:1997aw, Bergshoeff:2013pia} for its supersymmetrizations to so-called 
Dirac--Born--Infeld--Volkov--Akulov theories. Tree-level amplitudes ${\cal M}_n^{\rm BI}$ 
of Born--Infeld were identified as a field-theory double copy of SYM with scalar amplitudes
in the NLSM of Goldstone bosons \cite{Cachazo:2014xea}
as can be stated through the KLT formula
\beq
{\cal M}_n^{\rm BI} = -  \sum_{Q,R \in S_{n-3}} A_{\rm NLSM}(1,R,n,n{-}1) S(R|Q)_1 A(1,Q,n{-}1,n)\, .
\label{otherDC.01}
\eeq
In contrast to the gravitational KLT formula (\ref{KLTrel}), the polarizations
of the colorless spin-one multiplets in ${\cal M}_n^{\rm BI}$ stem entirely
from those in the color-ordered SYM amplitudes $A(R)$.

The study of the NLSM \cite{Cronin:1967jq, Weinberg:1966fm, Weinberg:1968de, Brown:1967qh, Chang:1967zza} and its tree-level amplitudes \cite{Osborn:1969ku, Susskind:1970gf, Ellis:1970nt, Kampf:2013vha} has a long history. Within the modern amplitudes program, the interest
in the NLSM was fueled by the observation that its tree amplitudes obey KK and BCJ relations
\cite{Chen:2013fya} and qualify to enter field-theory double copies. Just like the
Born--Infeld amplitudes, color-ordered tree amplitudes of the NLSM vanish for odd 
multiplicity, and their simplest non-zero instances are
\begin{align}
A_{\rm NLSM}(1,2,3,4) &= s_{12}+s_{23}\, ,
\label{otherDC.02} \\
A_{\rm NLSM}(1,2,3,4,5,6) &= s_{12} - \frac{(s_{12}{+}s_{23})(s_{45}{+}s_{56}) }{2 s_{123}}+{\rm cyc}(1,2,3,4,5,6) \, .
\notag
\end{align}
In order to compute Born--Infeld amplitudes from the low-energy limit of abelian open-superstring
states, we specialize the color-dressed disk amplitude ${\cal M}_n$ in (\ref{colordr}) to $t^{i} \rightarrow \mathds{1}$
and insert the KLT formula (\ref{AstringP}) for color-ordered disk amplitudes:
\begin{align}
{\cal M}_n^{\rm BI}  &= - \lim_{\ap \rightarrow 0} \frac{1}{(2\pi \alpha')^{n-2}} {\cal M}_n \, \big|_{t^{i} \rightarrow \mathds{1}}  
= - \lim_{\ap \rightarrow 0} \frac{1}{(2\pi \alpha')^{n-2}} \sum_{Q \in S_{n-1}} {\cal A}_n(Q,n)
\label{otherDC.03}  \\
&= -  \sum_{Q,R \in S_{n-3}} \bigg( \lim_{\ap \rightarrow 0} \frac{1}{(2\pi \alpha')^{n-2}}  Z_\times(1,R,n,n{-}1) \bigg) S(R|Q)_1 A(1,Q,n{-}1,n)\, , \notag
\end{align}
where we introduce the following shorthand for symmetrized disk integrals
or abelian $Z$-theory amplitudes
\beq
Z_\times(P) = \sum_{Q \in S_{n-1}} Z(Q,n|P)\, .
\label{otherDC.04} 
\eeq
The inverse $n{-}2$ factors of $2\pi \alpha'$ in (\ref{otherDC.03}) compensate for the leading
order $\sim \alpha'^{n-2}$ in the low-energy expansion of the abelian open-string 
amplitudes ${\cal M}_n \, \big|_{t^{i} \rightarrow \mathds{1}} $ that will be exposed 
in the discussion below. Based on the reflection property
(\ref{cycref}), the symmetrization in (\ref{otherDC.04}) annihilates $Z_\times(P)$ of
odd multiplicity.

Since the disk integrals $Z_\times$ solely depend on Mandelstam invariants, 
the KLT formula (\ref{otherDC.03}) implies that Born--Infeld is a double-copy
involving SYM. For consistency with the alternative KLT formula (\ref{otherDC.01})
which identifies the NLSM as the second double-copy component \cite{Cachazo:2014xea},
the coefficients of the linearly independent $A(1,Q,n{-}1,n)$ have to agree.
Hence, the conclusion from equating (\ref{otherDC.01}) with (\ref{otherDC.03}) is
that NLSM amplitudes arise from the low-energy limit of symmetrized disk integrals
\cite{Carrasco:2016ldy},
\beq
A_{\rm NLSM}(P) = \lim_{\ap \rightarrow 0} \frac{1}{(2\pi \alpha')^{n-2}}   Z_\times(P) 
\, .
\label{otherDC.05} 
\eeq
This adds support to the interpretation of disk integrals as tree-level amplitudes in
a bi-colored scalar $Z$-theory: when abelianizing the gauge-group generators
$t^i \rightarrow \mathds{1}$
dressing the disk ordering $P$ of $Z(P|Q)$, the low-energy limit reproduces the
tree amplitudes of the NLSM, a well-known scalar field theory. The appearance
of NLSM amplitudes in the low-energy limit of abelian $Z$-theory is here deduced
from the double copy (\ref{otherDC.01}) of Born--Infeld in \cite{Cachazo:2014xea} and does not rely 
on Goldstone bosons in the superstring spectrum. On the other hand, toroidal 
compactifications of ten-dimensional superstrings along with worldsheet
boundary condensates indeed give rise to NLSM Goldstone bosons
among the massless excitations \cite{Green:1995ga}.

\subsubsection{BCJ numerators of the NLSM from disk integrals}
\label{sec:7.7.2}

The definition (\ref{otherDC.04}) of symmetrized disk integrals does not manifest
the leading term in its $\alpha'$-expansion, so it may appear surprising that the
limit (\ref{otherDC.05}) does not diverge. In order to expose the leading
order $\alpha'^{n-2}$ of the $Z_\times(Q)$, we shall employ a variant
of the DDM-type decomposition (\ref{KKmono.00}) of color-ordered open-string
amplitudes. In reading this decomposition at the level of the disk integrals and
specializing to abelian gauge-group generators, we obtain
\begin{align}
Z_\times(Q) = \sum_{P \in S_{n-2}} \Tr(  [[ \ldots [[ \mathds{1}^1 , \mathds{1}^{p_2}]_{\alpha'}, \mathds{1}^{p_3}]_{\alpha'},\ldots, \mathds{1}^{p_{n-2}}]_{\alpha'},\mathds{1}^{p_{n-1}}]_{\alpha'} \mathds{1}^n ) Z(1,P,n|Q)\, .
\label{otherDC.06} 
\end{align}
In slight abuse of notation, we have indicated through the superscripts of $\mathds{1}^j$
that these unit matrices arose from the abelianization of $t^{j}$. In this way, the information
about the momentum dependence in the $\alpha'$-deformed bracket (\ref{KKmono.02}) is preserved
and we can evaluate
\beq
\Tr(  [[ \ldots [[ \mathds{1}^1 , \mathds{1}^{p_2}]_{\alpha'}, \mathds{1}^{p_3}]_{\alpha'},\ldots, \mathds{1}^{p_{n-2}}]_{\alpha'},\mathds{1}^{p_{n-1}}]_{\alpha'} \mathds{1}^n )
= \prod_{j=2}^{n-1} ( e^{i\pi \alpha' k_{1p_2\ldots p_{j-1}} \cdot k_{p_j}} 
- e^{-i\pi \alpha' k_{1p_2\ldots p_{j-1}} \cdot k_{p_j}} ) \, .
\label{otherDC.07}
\eeq 
Upon converting the exponentials to sine functions, this implies
\begin{align}
Z_\times(Q) = (2i)^{n-2} \sum_{P \in S_{n-2}} 
Z(1,P,n|Q) \prod_{j=2}^{n-1} \sin \big( \pi \alpha'  k_{1p_2\ldots p_{j-1}} \cdot k_{p_j} \big) \, ,
\label{otherDC.08} 
\end{align}
which leads to vanishing $Z_\times(Q)$ of odd multiplicity and the following
examples at even $n$:
\begin{align}
Z_\times(q_1,q_2,q_3,q_4) &= 4 \sin^2\! \big( \pi \alpha'  k_{1} \cdot k_{2} \big) Z(1,2,3,4|q_1,q_2,q_3,q_4) + 4 \sin^2\! \big( \pi \alpha'  k_{1} \cdot k_{3} \big) Z(1,3,2,4|q_1,q_2,q_3,q_4)\, ,
\notag \\
Z_\times(q_1,q_2,\ldots,q_6) &= 16 \sum_{P \in S_4} 
\sin\! \big( \pi \alpha'  k_{1} \cdot k_{p_2} \big)
\sin\! \big( \pi \alpha'  k_{1p_2} \cdot k_{p_3} \big) \label{otherDC.ex}  \\
&\ \ \ \ \times
\sin\! \big( \pi \alpha'  k_{1p_2p_3} \cdot k_{p_4} \big) 
\sin\! \big( \pi \alpha'  k_{1p_2p_3p_4} \cdot k_{p_5} \big)
 Z(1,P,6|q_1,q_2,\ldots,q_6) \, . \notag
\end{align}
Given that the low-energy limit of $Z(1,P,n|Q)$ yields doubly-partial amplitudes
$m$ at order $\alpha'^{0}$ and each sine function introduces leading low-energy
order $\alpha'^{1}$, one can easily identify the low-energy limit of (\ref{otherDC.08}) to be
\begin{align}
Z_\times(Q) = (2i\pi \alpha')^{n-2} \bigg\{ \sum_{P \in S_{n-2}}  
m(1,P,n|Q)  \prod_{j=2}^{n-1}( k_{1p_2\ldots p_{j-1}} \cdot k_{p_j} ) + {\cal O}(\alpha') \bigg\} \, .
\label{otherDC.09} 
\end{align}
Hence, the representation (\ref{otherDC.05}) of NLSM amplitudes 
in terms of low-energy limits of symmetrized disk integrals is non-singular and 
simplifies to
\begin{align}
A_{\rm NLSM}(P) &= i^{n-2} \sum_{Q \in S_{n-2}}  m(P | 1,Q,n ) 
\prod_{j=2}^{n-1} ( k_{1q_2\ldots q_{j-1}} \cdot k_{q_j} ) \notag\\
&=  \sum_{Q \in S_{n-2}}  m(P | 1,Q,n ) N^{\rm NLSM}_{1|Q|n}\, .
\label{otherDC.10} 
\end{align}
In passing to the second line, we have manifested the formal similarity with 
the DDM form (\ref{revbcj.18}) of gauge-theory amplitudes in a BCJ form. 
Given that the coefficients in the $(n{-}2)!$-term sum (\ref{revbcj.18}) over doubly-partial amplitudes 
$m(P | 1,Q,n )$ are BCJ master numerators of SYM (cf.\ section \ref{sec:6.4.3}),
the analogous coefficients $N^{\rm NLSM}_{1|Q|n}$ in the second line of (\ref{otherDC.10})
are bound to be local BCJ numerators of the NLSM,
\begin{align}
N^{\rm NLSM}_{1|Q|n} &= i^{n-2} \prod_{j=2}^{n-1} ( k_{1q_2\ldots q_{j-1}} \cdot k_{q_j} ) 
= i^{n-2} S(Q|Q)_1\, .
 \label{otherDC.11} 
\end{align}
In the second step, we have identified the BCJ master numerators of the NLSM
as diagonal entries of the field-theory KLT matrix as initially conjectured in \cite{Carrasco:2016ldy} 
and then derived as outlined above in \cite{Carrasco:2016ygv}. A Lagrangian for the NLSM 
with manifest color-kinematics duality was presented in \cite{Cheung:2016prv} which reproduces
the numerators in (\ref{otherDC.11}) from Feynman rules. Earlier explicit BCJ numerators
for the NLSM in terms of the KLT kernel can be found in \cite{dufu}.

\subsubsection{Coupling NLSM to bi-adjoint scalars}
\label{sec:7.7.3}

The behavior of NLSM and Born--Infeld amplitudes under soft limits $k_j \rightarrow 0$
in the external momenta
singles out preferred ways of coupling Goldstone bosons to bi-adjoint scalars
and supersymmetric Born--Infeld theories to SYM \cite{Cachazo:2016njl}. These extended
theories to be referred to as NLSM$+\phi^3$ and BI$+$SYM are related by KLT formulae
\beq
A_n^{{\rm BI}+{\rm SYM}}(P) = -  \sum_{Q,R \in S_{n-3}} A_{{\rm NLSM}+\phi^3}(P|1,R,n,n{-}1) S(R|Q)_1 A(1,Q,n{-}1,n)\, ,
\label{otherDC.13}
\eeq
which can also be studied from disk amplitudes in the pure spinor formalism:

The BI$+$SYM theory of \cite{Cachazo:2016njl} results from the low-energy limit
of open-superstring amplitudes where a subset of the Chan--Paton generators 
is abelianized. This amounts to keeping some of the $t^j$ non-abelian
in (\ref{otherDC.03}) and isolating an appropriate order in $\alpha'$ as the low-energy
limit. The non-abelian $t^j$ give rise to a color-decomposition w.r.t.\ $|P| < n$
legs, and $A_n^{{\rm BI}+{\rm SYM}}(P)$ refers to the coefficient of
$\Tr(t^{p_1} t^{p_2}\ldots t^{p_{|P|}})$.

Similar to the doubly-partial amplitudes $m(A|B)$ of bi-adjoint scalars defined by
(\ref{DPdefCP}), the amplitudes $A_{{\rm NLSM}+\phi^3}(P|1,R,n,n{-}1)$ on the right-hand side
of (\ref{otherDC.13}) are the coefficients of two types of traces -- one over
generators $\tilde t^a$ shared by the bi-adjoint scalars $\Phi=\Phi_{j|a} t^j \otimes \tilde t^a$ 
and the Goldstone bosons of the NLSM as well as one over the $t^j$ exclusive 
to the $|P|$ external bi-adjoint scalars. The simplest examples that do not coincide
with pure NLSM or $\phi^3$ amplitudes occur at five points, where for instance \cite{Cachazo:2016njl}
\begin{align}
A_{{\rm NLSM}+\phi^3}(3,4,5|1,2,3,4,5) &= 1- \frac{ s_{51}{+}s_{12} }{s_{34}} - \frac{s_{12}{+}s_{23} }{s_{45}}\,, \notag \\
A_{{\rm NLSM}+\phi^3}(2,3,5|1,2,3,4,5) &= 1 - \frac{s_{45}{+}s_{51} }{s_{23}}
\, .
\label{otherDC.14}
\end{align}
Note that the coupling of Goldstone bosons to bi-adjoint scalars can be
accommodated in the NLSM Lagrangian of \cite{Cheung:2016prv} with
manifest color-kinematics duality.

In computing the SYM$+$BI amplitudes (\ref{otherDC.13}) from the low-energy
limit of the open superstring, the NLSM$+\phi^3$ amplitudes on the right-hand side 
arise via partially symmetrized
disk integrals $Z_P(Q)$ \cite{Carrasco:2016ygv}. As detailed in the reference, 
the latter can be defined by starting from Chan--Paton-dressed
$Z$-theory in the DDM-type form (\ref{KKmono.00}) 
\beq
{\cal Z}_n(Q) = \sum_{P \in S_{n-2}} \Tr(  [[ \ldots [[ t^1 , t^{p_2}]_{\alpha'}, t^{p_3}]_{\alpha'},\ldots, t^{p_{n-2}}]_{\alpha'},t^{p_{n-1}}]_{\alpha'} t^n )
Z(1,P,n|Q)
\label{otherDC.15}
\eeq
and setting a subset of the generators $t^j$ to be unit matrices. We then
obtain partially symmetrized disk integrals such as the $(|Q|=5)$-point example
\begin{align}
Z_{345}(Q) &=
{\cal Z}_5(Q)  \, \big|_{ {\rm Tr}(t^3 t^4 t^5) }
\notag \\
&= 4 \sin(\pi \alpha' k_{1}\cdot k_{2}) \sin(\pi \alpha' k_{12}\cdot k_{4}) \cos(\pi \alpha' k_{124}\cdot k_{3}) Z(1,2,4,3,5|Q) \notag \\
&\quad+4 \sin(\pi \alpha' k_{1}\cdot k_{4}) \sin(\pi \alpha' k_{14}\cdot k_{2}) \cos(\pi \alpha' k_{124}\cdot k_{3}) Z(1,4,2,3,5|Q) \notag \\
&\quad+4 \sin(\pi \alpha' k_{1}\cdot k_{4}) \sin(\pi \alpha' k_{134}\cdot k_{2}) \cos(\pi \alpha' k_{14}\cdot k_{3}) Z(1,4,3,2,5|Q) \label{otherDC.16} \\
&\quad- 4 \sin(\pi \alpha' k_{1}\cdot k_{2}) \sin(\pi \alpha' k_{12}\cdot k_{3}) \cos(\pi \alpha' k_{123}\cdot k_{4}) Z(1,2,3,4,5|Q) \notag \\
&\quad-4 \sin(\pi \alpha' k_{1}\cdot k_{3}) \sin(\pi \alpha' k_{13}\cdot k_{2}) \cos(\pi \alpha' k_{123}\cdot k_{4}) Z(1,3,2,4,5|Q) \notag \\
&\quad-4 \sin(\pi \alpha' k_{1}\cdot k_{3}) \sin(\pi \alpha' k_{134}\cdot k_{2}) \cos(\pi \alpha' k_{13}\cdot k_{4}) Z(1,3,4,2,5|Q)
\notag
\end{align}
from the color-decomposition of ${\cal Z}_n(Q)$ w.r.t.\ the non-abelian $t^j$.
In the low-energy limit where $Z(P|Q)\rightarrow m(P|Q)$ as well as
$\sin(\pi \alpha' k_{P}\cdot k_{Q})\rightarrow \pi \alpha' k_P\cdot k_Q$
and $\cos(\pi \alpha' k_{P}\cdot k_{Q})\rightarrow 1$, we recover $(2\pi \alpha')^2$
times the first line of (\ref{otherDC.14}) from (\ref{otherDC.16}). A variety
of further examples and the systematics for general numbers of abelianized
and non-abelian $t^j$ can be found in \cite{Carrasco:2016ygv}.
Among other things, the results in the reference give rise to local BCJ numerators for
the NLSM$+\phi^3$ theory from the monodromy properties of disk integrals
along the lines of section \ref{sec:7.7.2}.

By gradually converting some of the Chan--Paton dressings 
of the $Z$-amplitudes (\ref{otherDC.15}) to become abelian
$t^j \rightarrow \mathds{1}$, the low-energy limits interpolate between
pure $\phi^3$-amplitudes and pure NLSM amplitudes. In the ``semi-abelian''
case, $Z$-theory can be viewed as an ultraviolet completion of the
NLSM$+\phi^3$ theory in \cite{Cachazo:2016njl}.

\subsection{Heterotic strings and Einstein--Yang--Mills}
\label{sec:7.7.4}

The $(n{-}3)!$ form of the disk correlators in (\ref{KLTcorrel}) also has crucial implications
for massless tree amplitudes of heterotic string theories since their supersymmetric chiral
halves can be described through the left-moving modes of the pure spinor formalism.
The massless sector of the heterotic string incorporates both gauge multiplets and the
half-maximal supergravity multiplet in ten dimensions. In contrast to the type I theory, 
heterotic strings already feature gauge-gravity couplings at tree level due to worldsheets of
sphere topology. Hence, one can study Einstein--Yang--Mills amplitudes from 
the correlators of the heterotic string and the field-theory limit of the sphere integrals 
in section \ref{sec:6.5.3}.

\subsubsection{Basics of heterotic-string amplitudes}
\label{sec:7.7.5}

The opening line for tree amplitudes of the heterotic string \cite{Gross:1985rr}
\begin{align}
{\cal M}_n^{\rm het}
 = \bigg({-}\frac{ \ap}{2\pi} \bigg)^{n-3}\!\!\!\! \int \limits_{\mathbb C^{n-3} \setminus \{z_a = z_b\}} \! \! \! \! 
d^2 z_2\, d^2 z_3\,\ldots \,d^2 z_{n-2} \,
\langle \! \langle V^{\rm het}_1(z_1) U^{\rm het}_2(z_2) {\ldots} 
 U^{\rm het}_{n-2}(z_{n-2})  V^{\rm het}_{n-1}(z_{n-1})  V^{\rm het}_n(z_n) \rangle \! \rangle
 \label{hetEYM.01}
\end{align}
is almost identical to that of type II superstrings in (\ref{gravsec.1}) up to the choice
of vertex operators for the gauge and gravity multiplet: both the integrated and the 
unintegrated variant involve a chiral half from the bosonic string
\begin{align}
V^{\rm het}_i &= \lambda^{\alpha} A^i_\alpha(\theta) e^{k_i\cdot X} \bar c \cdot \left\{ \begin{array}{cl} 
\overline{ {\cal J} }^{a_i} &: \ {\rm gauge} \ {\rm multiplets} \, ,\\ 
\sqrt{\frac{ 2}{\alpha'}}\, \tilde \epsilon^m_i \bar \partial X_m&: \ {\rm gravity} \ {\rm multiplets} \, ,
\end{array} \right.
 \label{hetEYM.02}\\
U^{\rm het}_i &=\big( \p\t^{\a}A^i_{\a}(\t) + \Pi_p A_i^p(\t) + d_{\a}W_i^{\a}(\t) 
+ \tfrac{1}{2} N_{pq}F_i^{pq}(\t)  \big) e^{k_i\cdot X} \cdot \left\{ \begin{array}{cl} 
\overline{ {\cal J} }^{a_i} &: \ {\rm gauge} \ {\rm multiplets}\, , \\ 
\sqrt{\frac{ 2}{\alpha'}} \,\tilde \epsilon_i^m \bar \partial X_m&: \ {\rm gravity} \ {\rm multiplets} \, ,
\end{array} \right.
\notag
\end{align}
where $\overline{ {\cal J} }^{a_i}$ are Kac--Moody currents of antiholomorphic conformal
weight $h=1$, the $\bar c$ ghost known from bosonic strings has conformal weight $h=-1$, 
and the polarization vectors of the gravity multiplets are transversal, $\tilde \epsilon_i{\cdot} k_i=0$.
The tree-level correlators are determined by $\langle \! \langle \bar c(z_1) \bar c(z_2) 
\bar c(z_3) \rangle \! \rangle = \bar z_{12} \bar z_{13} \bar z_{23}$ and the~OPEs
\beq
\overline{ {\cal J} }^{a}(z) \overline{ {\cal J} }^{b}(w) \sim \frac{ \delta^{ab} }{(\bar z-\bar w)^2} +
\frac{ f^{abc} \overline{ {\cal J} }^{c}(w) }{\bar z-\bar w}
\label{hetEYM.03}
\eeq
as well as
\beq
\bar \partial X^m(z) e^{k \cdot X}(w) \sim -  \frac{\alpha' k^m}{2 (\bar z-\bar w)} e^{k \cdot X}(w)\, , \ \ \ \ \ \
\bar \partial X^m(z) \bar \partial X^n(w) \sim  - \frac{\alpha' \delta^{mn} }{2 (\bar z-\bar w)^2}\, .
\label{hetEYM.04}
\eeq
Similar to the organization of type II amplitudes in (\ref{gravsec.3}), 
the correlator in (\ref{hetEYM.01}) is guaranteed to comprise the
Koba--Nielsen factor on the sphere and one copy of the disk correlator
$\langle {\cal K}_n\rangle$ in (\ref{KLTcorrel}) from the supersymmetric chiral half,
\beq
{\cal M}^{\rm het}_n  = \bigg({-}\frac{ \ap}{2\pi} \bigg)^{n-3}\!\!\!\!\int \limits_{\mathbb C^{n-3} \setminus \{z_a = z_b\}}
d^2 z_2 \, d^2 z_3\ldots d^2 z_{n-2}\,  \tilde {\cal K}^{\rm bos}_n  \, \langle  {\cal K}_n\rangle
\prod_{i<j}^n |z_{ij}|^{-\alpha' s_{ij}} \, .
 \label{hetEYM.05}
\eeq
The bosonic chiral half in turn contributes the rational function 
$\tilde {\cal K}^{\rm bos}_n$ of the $\bar z_i$ that depends on the color
degrees of freedom $a_i$ of the external gauge multiplets as well as the half-polarizations
$\tilde \epsilon_i^m$ and momenta of the external gravity multiplets. The non-vanishing
three-point examples $\tilde {\cal K}^{\rm bos}_n(P|Q)$ of the bosonic correlator in (\ref{hetEYM.05})
with external gauge and gravity multiplets in the sets $P$ and $Q$ are
\begin{align}
\tilde {\cal K}^{\rm bos}_3(1,2,3|\emptyset) &= f^{a_1 a_2 a_3}  \, , \ \ \ \ \ \ \tilde {\cal K}^{\rm bos}_3(1,2|3) =  {-} \sqrt{ \frac{\ap}{2} } \delta^{a_1 a_2} (\tilde \epsilon_3\cdot k_1)
 \, , \label{hetEYM.06}  \\
\tilde {\cal K}^{\rm bos}_3(\emptyset| 1,2,3) &= \sqrt{ \frac{\ap}{2} }\,\bigg\{ \big[ (\tilde \epsilon_1\cdot \tilde \epsilon_2)  (\tilde \epsilon_3\cdot k_1) + {\rm cyc}(1,2,3) \big] - \frac{\ap}{2} (\tilde \epsilon_1\cdot k_2)
(\tilde \epsilon_2\cdot k_3)(\tilde \epsilon_3\cdot k_1) \bigg\}  \, ,
\notag
\end{align}
and the massless three-point amplitudes ${\cal M}^{\rm het}_3$ are obtained 
upon multiplication with the supersymmetric correlator $\langle  {\cal K}_3\rangle = 
\langle V_1 V_2 V_3 \rangle$, see section \ref{illustratesec} for its component expansion.
As one can see from the appearance of the color factors in these examples, the 
heterotic-string amplitudes in (\ref{hetEYM.05}) are automatically color dressed. 
In fact, the OPEs (\ref{hetEYM.03}) of the Kac--Moody currents also introduce
products of traces with a maximum of $\lfloor \frac{n}{2} \rfloor$ trace factors at $n$ points
as expected from coupling between colored gauge multiplets and uncolored 
supergravity multiplets. One can still isolate color-ordered single-trace
amplitudes by picking up the antiholomorphic Parke--Taylor factors
(\ref{Zuplift.4}) in \cite{Frenkel:2010ys}
\begin{align}
\langle \! \langle \overline{ {\cal J} }^{a_1}(z_1)   \overline{ {\cal J} }^{a_2}(z_2)
\ldots  \overline{ {\cal J} }^{a_n}(z_n)   \rangle \! \rangle \, \big|_{\Tr(t^{a_1}t^{a_2}\ldots t^{a_n})}
&= - \overline{ {\rm PT}(1,2,\ldots,n) }\, .
\label{hetEYM.07}
\end{align}
Accordingly, multi-trace amplitudes associated with $\Tr(t^{P_1})\Tr(t^{P_2})\ldots \Tr(t^{P_k})$ 
and $t^{p_1 p_2\ldots p_{|P|}} = t^{p_1} t^{p_2}\ldots t^{p_{|P|}}$
are determined by isolating the product $(-1)^k \PT(P_1) \PT(P_2)\ldots \PT(P_k)$ from
the current correlator.

\subsubsection{Heterotic double copy and Einstein--Yang--Mills}
\label{sec:7.7.6}

The expansion (\ref{KLTcorrel}) of the supersymmetric correlator $ \langle  {\cal K}_n\rangle$
in terms of SYM trees can be readily applied to the heterotic-string amplitude (\ref{hetEYM.05}):
in this way, all color-dressed tree amplitudes involving arbitrary combinations of gauge and gravity 
multiplets conspire to a field-theory double copy with one SYM constituent,
\begin{align}
{\cal M}^{\rm het}_n &= -  \sum_{Q,R \in S_{n-3}} {\cal B}(1,R,n,n{-}1) S(R|Q)_1 A(1,Q,n{-}1,n)\, ,
\label{hetEYM.08} \\
{\cal B}(1,R,n,n{-}1) &= {-} \bigg({-}\frac{ \ap}{2\pi} \bigg)^{n-3}\!\!\!\!\int \limits_{\mathbb C^{n-3} \setminus \{z_a = z_b\}}
d^2 z_2 \, d^2 z_3\ldots d^2 z_{n-2}\,  \tilde {\cal K}^{\rm bos}_n \,
{\cal Z}_{1R}\, \prod_{i<j}^n |z_{ij}|^{-\alpha' s_{ij}} \, . \notag
\end{align}
Given that the rational functions ${\cal Z}_{1R}$ on the right-hand side correspond
to ${\rm SL}_2(\mathbb R)$-fixed Parke--Taylor factors ${-}\PT(1,R,n,n{-}1)$, see (\ref{volslPT}),
one can uplift ${\cal B}(1,R,n,n{-}1)$ to a cyclic object ${\cal B}(P)$ by integrating
over $-\PT(P)$ in the place of ${\cal Z}_{1R} $.

In order to understand the significance of the $\alpha'$-dependent building 
block ${\cal B}(P)$ in the double copy (\ref{hetEYM.08}), it is instructive
to compare with the Einstein--Yang--Mills amplitudes obtained from
the low-energy limit of heterotic strings: Einstein--Yang--Mills theories
are double copies of SYM with the so-called YM$+\phi^3$ 
theory \cite{Chiodaroli:2014xia}. Similar to the NLSM$+\phi^3$ theory
in section \ref{sec:7.7.3}, YM$+\phi^3$ is characterized by a minimal
coupling of bi-adjoint scalars to pure (i.e.\ non-supersymmetric) Yang--Mills
theory such that the BCJ relations are preserved. 
More precisely, in a color-decomposition of YM$+\phi^3$ tree amplitudes w.r.t.\ the
generators $\tilde t^j$ common to the gauge bosons and the bi-adjoint scalars, 
\beq
M^{{\rm YM}+\phi^3}_n = \sum_{P \in S_{n-1}} \Tr( \tilde t^{1P})
A_{{\rm YM}+\phi^3}(1,P)\, ,
\label{hetEYM.09}
\eeq
the color-ordered amplitudes $A_{{\rm YM}+\phi^3}$ obey KK and BCJ relations.
Accordingly, they qualify to enter the following KLT formula that encodes the double
copy of Einstein--Yang--Mills:
\beq
M^{\rm EYM}_n = -  \sum_{Q,R \in S_{n-3}} A_{{\rm YM}+\phi^3}(1,R,n,n{-}1) S(R|Q)_1 A(1,Q,n{-}1,n)\,.
\label{hetEYM.10}
\eeq
Similar to the bosonic correlators $ \tilde {\cal K}^{\rm bos}_n $ and thereby
the string-theory building blocks ${\cal B}$ in (\ref{hetEYM.08}), the
$A_{{\rm YM}+\phi^3}$ still depend on the color-factors $t^j$ exclusive to
the bi-adjoint scalars because only the $\tilde t^j$ are stripped in (\ref{hetEYM.09}). 
Since the KLT formula (\ref{hetEYM.10}) is obtained from the low-energy 
limit of (\ref{hetEYM.08}) with all $\alpha'$-dependence
carried by ${\cal B}(P)$, we can identify color-ordered YM$+\phi^3$
amplitudes as its low-energy limit
\beq
{\cal B}(P) = A_{{\rm YM}+\phi^3}(P) \, \big|_{g_{\rm YM} \rightarrow \sqrt{ \frac{\ap}{2} } } \, \big(1 + {\cal O}(\alpha') \big)\, .
\label{hetEYM.11}
\eeq
As exemplified by (\ref{hetEYM.06}), the bosonic correlators $\tilde {\cal K}^{\rm bos}_n$ and
hence the heterotic-string amplitudes (\ref{hetEYM.08}) may already carry integer powers of
$\sqrt{ \tfrac{\ap}{2} }$ in their low-energy limits. These prefactors are interpreted as realizing the
gravitational coupling $\kappa$ of the heterotic string which, in the double copy
of \cite{Chiodaroli:2014xia}, translates into the
gauge coupling $g_{\rm YM}$ of the YM$+\phi^3$ theory.\footnote{In any other 
section of this review, we have stripped off the uniform
prefactors $g_{\rm YM}^{n-2}$ from $n$-point tree-level amplitudes of
SYM. In the YM$+\phi^3$ theory, on the other hand, generic tree-level amplitudes
mix different powers of $g_{\rm YM}$ according to the trace structure of the color factors: only
the gluon vertices and the minimal coupling of two scalars $\phi$ to gluons
carry powers of $g_{\rm YM}$ whereas the coefficient of the $\phi^3$ interaction is
taken to be independent on $g_{\rm YM}$.}

\subsubsection{Heterotic strings as a field-theory double copy}
\label{sec:7.7.7}

By the bosonic origin of ${\cal B}(P)$ in (\ref{hetEYM.08}), its $\alpha'$-dependence
can be further streamlined by expanding the bosonic correlator  $\tilde {\cal K}^{\rm bos}_n $
in a Parke--Taylor basis. 
Even though the computation of $\tilde {\cal K}^{\rm bos}_n$ is straightforward from the OPEs 
(\ref{hetEYM.03}) and (\ref{hetEYM.04}), the Parke--Taylor decomposition
relies on a way more intricate cascade of integrations by parts than encountered
in section \ref{theIBPsec} for supersymmetric correlators, see for instance
\cite{Huang:2016tag, Schlotterer:2016cxa, He:2018pol, He:2019drm}.
By the arguments in section 4.2 of \cite{Azevedo:2018dgo}, the coefficients
$A_{(DF)^2+{\rm YM}+\phi^3}$ in an $(n{-}3)!$-term reduction
(discarding total Koba--Nielsen derivatives 
$\nabla_{\bar z_k} f = \partial_{\bar z_k} f -\frac{ \alpha' }{2} f \sum_{j=1\atop {j\neq k}}^n \frac{s_{kj} }{\bar z_{kj}}$
on the sphere)
\beq
\tilde {\cal K}^{\rm bos}_n  = - \frac{ d\bar z_1 \, d\bar z_{n-1} \, d\bar z_n}{ {\rm vol}({\rm SL}_2(\Bbb R)) } \sum_{Q,R \in S_{n-3}} \overline{ {\rm PT}(1,R,n,n{-}1)}  S(R|Q)_1 A_{(DF)^2+{\rm YM}+\phi^3}(1,Q,n{-}1,n) \ {\rm mod} \ \nabla_{\bar z_k}
\label{hetEYM.12}
\eeq
are given by field-theory amplitudes in a massive gauge theory
dubbed $(DF)^2+{\rm YM}+\phi^3$ \cite{Johansson:2017srf}, see
the discussion below (\ref{KLTcorrel}) for the $d\bar z_i$ in the prefactor. Just like for
YM$+\phi^3$ theory, the massless states of $(DF)^2+{\rm YM}+\phi^3$ 
are bi-adjoint scalars and gauge bosons.
The massive states in the $(DF)^2+{\rm YM}+\phi^3$ theory
are tachyons $m^2 = - \frac{4}{\ap}$ as expected for the open
bosonic string, so the $A_{(DF)^2+{\rm YM}+\phi^3}$ are still rational
functions of $\alpha'$ \cite{Azevedo:2018dgo}.
Similar to (\ref{hetEYM.11}), the gauge coupling of the $(DF)^2+{\rm YM}+\phi^3$ 
theory is understood to be converted to the gravitational one in the 
double copy (\ref{hetEYM.12}), i.e.\ $g_{\rm YM} \rightarrow \kappa = \sqrt{ \frac{ \ap}{2} }$. 

The three-point amplitudes of the $(DF)^2+{\rm YM}+\phi^3$ theory
(with subscripts $\phi$ and $g$ for external scalars and gluons) reproduce
the simplest bosonic correlators in (\ref{hetEYM.06}),
\begin{align}
A_{(DF)^2+{\rm YM}+\phi^3}(1_\phi,2_\phi,3_\phi) &= f^{a_1 a_2 a_3}  \, , \ \ \ \ \ \ A_{(DF)^2+{\rm YM}+\phi^3}(1_\phi,2_\phi,3_g) =  {-} \sqrt{ \frac{\ap}{2} } \delta^{a_1 a_2} (\tilde \epsilon_3\cdot k_1)
 \, , \label{hetEYM.df3}  \\
A_{(DF)^2+{\rm YM}+\phi^3}( 1_g,2_g,3_g) &= \sqrt{ \frac{\ap}{2} }\,\bigg\{ \big[ (\tilde \epsilon_1\cdot \tilde \epsilon_2)  (\tilde \epsilon_3\cdot k_1) + {\rm cyc}(1,2,3) \big] - \frac{\ap}{2} (\tilde \epsilon_1\cdot k_2)
(\tilde \epsilon_2\cdot k_3)(\tilde \epsilon_3\cdot k_1) \bigg\}  \, ,
\notag
\end{align}
where the $\alpha'$-correction in the last line can be traced back to 
the ${\rm Tr}(F^3)$ vertex in the $(DF)^2+{\rm YM}+\phi^3$ Lagrangian \cite{Johansson:2017srf}.
External scalars $\phi$ in legs $1, 3$ and gauge multiplets $g$ in legs $2,4$ in turn give rise to
\beq
A_{(DF)^2+{\rm YM}+\phi^3}(1_\phi,2_g,3_\phi,4_g) = \delta^{a_1 a_3} \frac{\ap}{2}
\bigg\{
\frac{(\tilde \epsilon_2{\cdot} k_1)(\tilde \epsilon_4{\cdot }k_3) }{s_{12}}
+\frac{(\tilde \epsilon_2{\cdot }k_3)(\tilde \epsilon_4{\cdot} k_1) }{s_{23}}
+ (\tilde \epsilon_2{\cdot}  \tilde \epsilon_4 ) + \frac{ \ap\, \tilde f^{mn}_{2}\tilde f^{mn}_{4} }{2+ \ap s_{13}}
\bigg\} \, .
\label{hetEYM.13}
\eeq
By the double copy with SYM in (\ref{hetEYM.08}), the external states $\phi$ and $g$
in $(DF)^2+{\rm YM}+\phi^3$ amplitudes translate into gauge multiplets $g$ and gravity 
multiplets $h$ in heterotic-string amplitudes.

The decomposition (\ref{hetEYM.12})
is a useful way to disentangle the $\alpha'$-dependence of (\ref{hetEYM.08}) into the 
sphere integrals $J(P|R)$ in (\ref{Jintdef}) with all the poles of massive-state exchange
and the $A_{(DF)^2+{\rm YM}+\phi^3}$ with only massless and tachyonic poles,
\beq
{\cal B}(P)  = -  \sum_{Q,R \in S_{n-3}} J(P|1,R,n,n{-}1) S(R|Q)_1 A_{(DF)^2+{\rm YM}+\phi^3}(1,Q,n{-}1,n) \, .
\label{hetEYM.14}
\eeq
The sphere integrals $J(P|Q)$ have zeros at $s_{ij\ldots k}= - \frac{2}{\alpha'}$ that prevent the
tachyon poles of $A_{(DF)^2+{\rm YM}+\phi^3}$ from entering the heterotic-string amplitude.
At four points, this can be anticipated from the Gamma functions in the denominator 
of (\ref{4ptclosed}). By combining (\ref{hetEYM.14}) with (\ref{hetEYM.08}), 
heterotic-string amplitudes can be brought into the form
\beq
{\cal M}^{\rm het}_n = \! \! \! \! \! \! \sum_{P,Q,A,B \in S_{n-3}} \! \! \! \! \! \!
A_{(DF)^2+{\rm YM}+\phi^3}(1,P,n,n{-}1) S(P|Q)_1 
J(1,Q,n{-}1,n|1,A,n,n{-}1) S(A|B)_1 A(1,B,n{-}1,n)\, ,
\label{hetEYM.15}
\eeq
which is directly analogous to the representation (\ref{gravsec.16}) of type II amplitudes,
with $A_{(DF)^2+{\rm YM}+\phi^3}$ in the place of a second copy $\tilde A$ of SYM.
External gauge and gravity multiplets in ${\cal M}^{\rm het}_n$ are represented by
external scalars and gauge bosons in $A_{(DF)^2+{\rm YM}+\phi^3}$,
respectively, with the conversion $g_{\rm YM} \rightarrow \sqrt{\frac{\ap}{2}}$ 
of the gauge coupling as in (\ref{hetEYM.11}). 
Given that the matrix product $\sum_{Q,A \in S_{n-3}}S(P|Q)_1 
J(1,Q,n{-}1,n|1,A,n,n{-}1) S(A|B)_1$ reduces to $S(P|B)_1$ in
the field-theory limit, we arrive at the following refinement of
(\ref{hetEYM.11})
\beq
A_{(DF)^2+{\rm YM}+\phi^3}(P) = A_{{\rm YM}+\phi^3}(P) \big(1 + {\cal O}(\alpha') \big)\,.
\label{hetEYM.16}
\eeq
In the four-point example (\ref{hetEYM.13}), for instance, the last term
$ \frac{\ap}{2} \tilde f^{mn}_{2}\tilde f^{mn}_{4}$ is subleading in $\alpha'$,
and the resulting YM$+\phi^3$ amplitude entering the Einstein--Yang--Mills
double copy (\ref{hetEYM.10}) is 
\beq
A_{{\rm YM}+\phi^3}(1_\phi,2_g,3_\phi,4_g) = \delta^{a_1 a_3} g_{\rm YM}^2
\bigg\{
\frac{(\tilde \epsilon_2\cdot k_1)(\tilde \epsilon_4\cdot k_3) }{s_{12}}
+\frac{(\tilde \epsilon_2\cdot k_3)(\tilde \epsilon_4\cdot k_1) }{s_{23}}
+ (\tilde \epsilon_2\cdot  \tilde \epsilon_4 ) 
\bigg\}\, ,
\label{hetEYM.17}
\eeq
where $g_{\rm YM}^2$ translates into the prefactor $\frac{\alpha'}{2}$
of the corresponding $(DF)^2+{\rm YM}+\phi^3$ amplitude.

\subsubsection{Einstein--Yang--Mills amplitude relations from string theories}
\label{sec:7.7.8}

As a key implication of the Einstein--Yang--Mills double copy (\ref{hetEYM.10}),
any tree amplitude of Einstein--Yang--Mills (regardless on the number of external
gauge \& gravity multiplets or traces in the $t^i$) can be expanded in terms
of SYM trees. On top of the manifestly gauge invariant KLT form (\ref{hetEYM.10}), 
one can explicitly realize the double copy with manifest locality,
\beq
M^{\rm EYM}_n = \sum_{P \in S_{n-2}} \tilde  N^{{\rm YM}+\phi^3 }_{1|P|n} A(1,P,n ) \, ,
\label{hetEYM.21}
\eeq
with BCJ master numerators $\tilde  N^{{\rm YM}+\phi^3 }_{1|P|n}$ of YM$+\phi^3$. 
This DDM form of Einstein--Yang--Mills amplitudes is analogous to the 
representation (\ref{gravalt.13}) of gravitational amplitudes. Field-theoretic 
computations of the master numerators $\tilde  N^{{\rm YM}+\phi^3 }_{1|P|n} $
from gauge invariance and color-kinematics duality as well as a discussion of the
resulting Einstein--Yang--Mills amplitude relations (\ref{hetEYM.21}) can be found in
\cite{Chiodaroli:2017ngp}. We shall here review the worldsheet approach
to derive manifestly local Einstein--Yang--Mills amplitude relations that
amount to the DDM-type decomposition
\beq
\tilde {\cal K}^{\rm bos}_n =  \frac{ d\bar z_1 \, d\bar z_{n-1} \, d\bar z_n}{ {\rm vol}({\rm SL}_2(\Bbb R)) }  \sum_{P \in S_{n-2}} 
\tilde  N^{(DF)^2+{\rm YM}+\phi^3 }_{1|P|n} \overline{ \PT(1,P,n) }  \ {\rm mod} \ \nabla_{\bar z_k}
\label{hetEYM.22}
\eeq
analogous to (\ref{gravsec.11}). The YM$+\phi^3$ master numerators in (\ref{hetEYM.21})
can then be simply read off from the leading $\alpha'$-order of the $(DF)^2+{\rm YM}+\phi^3$
numerators or Parke--Taylor coefficients in (\ref{hetEYM.22}):
\begin{itemize}
\item For single-trace amplitudes ${\cal A}^{\rm het}(1,2,\ldots,n;p)$ with one external
graviton $p$ as well as $n$ external gluons and associated trace $\Tr(t^1 t^2 \ldots t^n)$,
the bosonic correlator due to (\ref{hetEYM.04}) and (\ref{hetEYM.07}) is given by
\begin{align}
\tilde {\cal K}^{\rm bos}_n &\sim \overline{ \PT(1,2,\ldots,n) }  \sum_{j=1}^n \frac{ \tilde \epsilon_p \cdot k_j }{ \overline z_{jp}} =  \overline{ \PT(1,2,\ldots,n) } \sum_{j=1}^{n-1} 
(\tilde \epsilon_p \cdot k_{12\ldots j} ) \frac{ \bar z_{j,j+1} }{ \bar z_{j,p} \bar z_{p,j+1} }\notag \\
&=  \sum_{j=1}^{n-1}  (\tilde \epsilon_p \cdot k_{12\ldots j} ) \overline{ \PT(1,2,\ldots,j,p,j{+}1,\ldots ,n{-}1,n) } \, ,
\label{hetEYM.23}
\end{align}
where we used $ \overline{ \PT(1,2,\ldots,j,p,j{+}1,\ldots ,n) }  = \frac{ \bar z_{j,j+1} }{ \bar z_{j,p} \bar z_{p,j+1} } \overline{ \PT(1,2,\ldots,j,j{+}1,\ldots,n) } $ in passing to the last line. By matching with
(\ref{hetEYM.22}), one can read off master numerators
\beq
\tilde  N^{(DF)^2+{\rm YM}+\phi^3 }_{1|23\ldots j p (j+1)\ldots (n-1) |n}
 \rightarrow  \tilde \epsilon_p \cdot k_{12\ldots j} \ \ \ \Rightarrow \ \ \ 
 \tilde  N^{{\rm YM}+\phi^3 }_{1|23\ldots j p (j+1)\ldots (n-1) |n}
 \rightarrow  \tilde \epsilon_p \cdot k_{12\ldots j}
\label{hetEYM.24}
\eeq
which result in the following amplitude relation (\ref{hetEYM.21}) (with gravity multiplet $p$ and single-trace ordering $\Tr(t^1 t^2 \ldots t^n)$)
\beq
A^{\rm EYM}(1,2,\ldots,n;p) =  \sum_{j=1}^{n-1} (\tilde \epsilon_p \cdot k_{12\ldots j} )A(1,2,\ldots,j,p,j{+}1,\ldots,n{-}1,n)\, .
\label{hetEYM.25}
\eeq
This relation has been firstly derived from disk amplitudes of type I superstrings 
with one closed-string insertion \cite{Stieberger:2016lng} and generalized to single-trace 
amplitudes with multiple external gravitons using CHY methods\footnote{The prescription 
for Einstein--Yang--Mills amplitudes in the CHY formalism has
been given in \cite{Cachazo:2014nsa, Cachazo:2014xea}.} \cite{Nandan:2016pya}
and heterotic strings \cite{Schlotterer:2016cxa}. For single-trace amplitudes with an
arbitrary number of gravity multiplets, a decomposition formula (\ref{hetEYM.21})
in terms of intersection numbers of twisted cocycles can be found in \cite{Mazloumi:2022lga}.
Note that generic $(DF)^2+{\rm YM}+\phi^3$ numerators are rational functions of $\alpha'$,
so their $\alpha'$-independent instances in (\ref{hetEYM.24}) are rather atypical.
\item For $n$-gluon double-trace amplitudes associated with $\Tr(t^1 t^P)\Tr(t^Q t^n)$ and
no external gravitons, the bosonic correlator determined by the current algebra is
\begin{align}
\tilde {\cal K}^{\rm bos}_n &\sim \overline{ \PT(1,P) } \, \overline{ \PT(Q,n) } 
=  \frac{   \alpha'   }{2+  \alpha'  s_{1P} }
\overline{ \PT(1,\{P,Q\},n) }  \ {\rm mod} \ \nabla_{\bar z_k} \label{hetEYM.26} \\
&=  {-} \frac{   \alpha'   (-1)^{|P|} }{2+  \alpha'  s_{1P} }
\sum_{i=1}^{|P|} \sum_{j=1}^{|Q|} (-1)^{i-j} s_{ij} 
\sum_{A \in p_1 p_2\ldots p_{i-1} \atop{\shuffle p_{|P|}\ldots p_{i+1}}}
\sum_{B \in q_{j+1}\ldots q_{|Q|} \atop{ \shuffle q_{j-1}\ldots q_2 q_1}}
\overline{ \PT(1,A,p_i,q_j,B,n) } \ {\rm mod} \ \nabla_{\bar z_k}
 \, ,
\notag
\end{align}
where the second line is known from \cite{Schlotterer:2016cxa}
and follows from expanding out the $S$-bracket via
(\ref{PQgen}). The master numerators (\ref{hetEYM.22})
can be read off from the Parke--Taylor coefficients, and we now have
a non-trivial $\alpha'$-dependence of the $\tilde  N^{(DF)^2+{\rm YM}+\phi^3 }_{1|P |n}$
through the geometric series
\beq
 \frac{   \alpha'   }{2+  \alpha'  s_{1P} } = \frac{\alpha'}{2} \sum_{n=0}^{\infty} \bigg( {-}\frac{\alpha'}{2} s_{1P} \bigg)^n\, .
\label{hetEYM.27}
\eeq
The master numerators of YM$+\phi^3$ are obtained from the leading terms
$ \frac{   \alpha'   }{2+  \alpha'  s_{1P} } \rightarrow \frac{\alpha'}{2}$ and
result in the following expansion of the Einstein--Yang--Mills 
double-trace amplitude $A^{\rm EYM}(1,P|Q,n)$
along with $\Tr(t^1 t^P)\Tr(t^Q t^n)$:
\beq
A^{\rm EYM}(1,P|Q,n) = \tfrac{\alpha'}{2} A(1,\{P,Q\},n)\, ,
\label{hetEYM.28}
\eeq
for instance
\begin{align}
A^{\rm EYM}(1,2|3,4) &=  \tfrac{\alpha'}{2} s_{23} A(1,2,3,4)  \, , \notag\\
A^{\rm EYM}(1,2,3|4,5) &=  \tfrac{\alpha'}{2} \big[ s_{34} A(1,2,3,4,5)  - s_{24} A(1,3,2,4,5) \big] \, ,
\label{hetEYM.29}
\\
A^{\rm EYM}(1,2,3,4|5,6) &= \tfrac{\alpha'}{2} \big[ s_{45} A(1,2,3,4,5,6)- s_{35} A(1,2,4,3,5,6) 
+(2\leftrightarrow 4) \big] \, ,
\notag \\
A^{\rm EYM}(1,2,3|4,5,6) &=  \tfrac{\alpha'}{2}
\big[ s_{34} A(1,2,3,4,5,6) - s_{24} A(1,3,2,4,5,6) - (4\leftrightarrow 5) \big] \, .
\notag
\end{align}
The prefactor is identified with the gravitational coupling in 
Einstein--Yang--Mills theory according to $\kappa^2 = \frac{\ap}{2}$
and signals two gravitational vertices as expected: each diagram contributing 
to double-trace amplitudes with external gauge multiplets involves one gravitational propagator ending on vertices with one factor of $\kappa$ each.

The double-trace relations (\ref{hetEYM.28}) have been derived and extended to one
external gravity multiplet via CHY methods \cite{Nandan:2016pya} and heterotic
strings \cite{Schlotterer:2016cxa}. Their generalizations to 
arbitrary number of traces can be found in \cite{Du:2017gnh}.
\end{itemize}
As illustrated by the tachyon pole $(2{+}\alpha' s_{1P})^{-1}$ in (\ref{hetEYM.26}), the
numerators $\tilde N^{(DF)^2+{\rm YM}+\phi^3}_{1|P|n}$ feature propagators
of the massive states in the $(DF)^2+{\rm YM}+\phi^3$ theory. Still, they are free of massless 
poles and therefore yield a local low-energy limit $\tilde N^{{\rm YM}+\phi^3}_{1|P|n}$ for 
the Einstein--Yang--Mills amplitude relations (\ref{hetEYM.21}). 

\subsubsection{Reducing heterotic-string amplitudes to the single-trace sector}
\label{sec:7.7.9}

In fact, the Einstein--Yang--Mills amplitude relations (\ref{hetEYM.21}) uplift to 
exact-in-$\alpha'$ relations for heterotic-string amplitudes in passing to the numerators
of the $ (DF)^2+{\rm YM}+\phi^3$ theory
\beq
{\cal M}^{\rm het}_n = \sum_{P \in S_{n-2}} \tilde  N^{ (DF)^2+{\rm YM}+\phi^3 }_{1|P|n} {\cal A}^{\rm het}(1,P,n ) \, ,
\label{hetEYM.31}
\eeq
where ${\cal A}^{\rm het}(Q ) $ denote the single-trace amplitudes for $|Q|$ external
gauge multiplets. This follows from (\ref{hetEYM.22}) in combination with
\begin{align}
{\cal A}^{\rm het}(1,R,n,n{-}1 ) &= {-}  \bigg({-}\frac{ \ap}{2\pi} \bigg)^{n-3}\!\!\!\!\int \limits_{\mathbb C^{n-3} \setminus \{z_a = z_b\}}
d^2 z_2 \, d^2 z_3\ldots d^2 z_{n-2}\,
{\cal Z}_{1R} \, \langle  {\cal K}_n\rangle
\prod_{i<j}^n |z_{ij}|^{-\alpha' s_{ij}}
\label{preEYM.32}
\end{align}
and its ${\rm SL}_2(\mathbb C)$ covariant uplift ${\cal Z}_{1R} \rightarrow - {\rm PT}(1,R,n,n{-}1)$ 
in (\ref{volslPT}) \cite{Stieberger:2014hba},
\begin{align}
{\cal A}^{\rm het}(P ) 
&= -  \sum_{Q,R \in S_{n-3}} J(P|1,R,n,n{-}1) S(R|Q)_1 A(1,Q,n{-}1,n)\, .
\label{hetEYM.32}
\end{align}
One may view (\ref{hetEYM.31})
as an alternative to the double copy (\ref{hetEYM.15}) with locality w.r.t.\ the massless
propagators but not w.r.t.\ the massive ones. For instance, specializing (\ref{hetEYM.31}) 
to double-trace amplitudes ${\cal A}^{\rm het}(1,P|Q,n)$ of the heterotic
string yields \cite{Schlotterer:2016cxa}
\beq
{\cal A}^{\rm het}(1,P|Q,n) =  \frac{ \alpha'}{2 + \alpha' s_{1P}}{\cal A}^{\rm het}(1,\{P,Q\},n)\, ,
\label{hetEYM.33}
\eeq
with the Einstein--Yang--Mills relation (\ref{hetEYM.28}) in its low-energy limit.
The only case where (\ref{hetEYM.31}) is free of massive propagators is
the following single-trace amplitude with one external gravity multiplet \cite{Schlotterer:2016cxa}
\beq
{\cal A}^{\rm het}(1,2,\ldots,n;p) = \sum_{j=1}^{n-1} (\tilde \epsilon_p \cdot k_{12\ldots j} )
{\cal A}^{\rm het}(1,2,\ldots,j,p,j{+}1,\ldots,n{-}1,n)\, ,
\label{hetEYM.34}
\eeq
which is the $\alpha'$-uplift of (\ref{hetEYM.25}). 

\section{\label{apsec}$\ap$-expansion of superstring tree-level amplitudes}

In the previous sections, we have reviewed the derivation and structure of
the expression (\ref{nptdisk}) for the $n$-point disk amplitude in terms of
SYM tree amplitudes. By disentangling the contributions of left- and right-moving 
worldsheet degrees of freedom, similar decompositions (\ref{gravsec.16}) 
and (\ref{hetEYM.15}) were deduced for sphere amplitudes of type II and heterotic 
strings. On the one hand, these genus-zero results only cover the leading order 
in string perturbation theory and still receive loop- and non-perturbative
corrections. On the other hand, (\ref{nptdisk}), (\ref{gravsec.16}) and (\ref{hetEYM.15}) 
are exact in $\alpha'$, i.e.\ they incorporate all orders in the low-energy expansion
at genus zero.

This section is dedicated to the $\alpha'$-expansion of $n$-point disk and sphere amplitudes, 
with a detailed review of the structure and explicit computation of the string corrections
to the field-theory limits discussed in the previous sections. These string
corrections are organized into infinite series in the dimensionless Mandelstam
invariants $\alpha' k_i {\cdot} k_j$ with rational combinations of multiple zeta 
values (MZVs) in their coefficients.
The appearance of MZVs unravels elegant mathematical properties of and striking
connections between tree-level amplitudes in different perturbative string theories.
Moreover, the interplay of MZVs with the accompanying polynomials in $\alpha' k_i {\cdot} k_j$ 
identifies several echos of field-theory structures at all orders of the low-energy expansion 
including Berends--Giele recursions, color-kinematics duality and double copy.

The study of low-energy expansions in string perturbation theory has a long history.
We focus on state-of-the-art techniques to expand the $n$-point disk integrals \eqref{Fdef}
or their $Z$-basis \eqref{Zintdef} using the Drinfeld associator \cite{drinfeld}, see section
\ref{sec:7.4}, or Berends--Giele recursions \cite{BGap}, see section \ref{sec:7.5}. Based on
these results for the disk integrals of open superstrings,
the analogous expansions of the sphere integrals in tree amplitudes of type II and heterotic
string theories will be obtained as corollaries under the so-called single-valued map,
see section \ref{sec:7.6}. At $n\leq 7$ points, a variety of
earlier calculations have been successfully carried out before the advent of
the all-multiplicity methods in sections \ref{sec:7.4} and \ref{sec:7.5}, often exploiting synergies
with hypergeometric functions \cite{Kitazawa:1987xj, Medina:2002nk, Barreiro:2005hv, 6ptOprisa, 
Stieberger:2006te, Stieberger:2007jv, Stieberger:2009rr, Schlotterer:2012ny, Zfunctions, Boels:2013jua, Puhlfurst:2015foi}. The loop-level extensions of the results in this section are
under active investigation, and a short summary of the state of the art as of fall 2022
can be found in section \ref{sec:9.2}.

Numerous developments related to the $\alpha'$-expansion of
string amplitudes have been crucially fueled by the recent number-theory
and algebraic-geometry literature. As will be detailed below, the mathematical 
references underlying the tree-level results of this section 
include \cite{Terasoma, Goncharov:2005sla, Brown:2009qja, 
Brown:2011ik, Schnetz:2013hqa, Brown:2013gia, Brown:2018omk, Brown:2019wna}. 
Parts of the results of this section can also be found in the reviews \cite{MadridOS, MadridSS}
from 2016, also see \cite{Duhr:2012fh} for a helpful introductory reference on the 
Hopf-algebra structure of genus-zero integrals in the particle-physics literature.

\subsection{Basics of $\alpha'$-expansions}
\label{sec:7.0}

This subsection aims to set the stage for the main results of this section by
reviewing four-point examples of $\alpha'$-expansions and the connection
with low-energy effective actions. 
 
\subsubsection{Four-point $\ap$-expansions}
\label{altsec:7.2.1}

Similar to the computation and simplification of the correlators, the main efforts
in determining low-energy expansions kick in at five points. The four-point
$\alpha'$-expansion in turn has been known in closed form for decades
from the simple expansion of the Gamma function
\beq
\log\Gamma(1{-}z) = \gamma z + \sum_{n=2}^\infty \frac{ z^n }{n} \zeta_n\, ,
\label{4ptmzv.0}
\eeq
where the Euler--Mascheroni constant $\gamma=\lim_{n\rightarrow \infty}( \sum_{k=1}^n \frac{1}{k}-\log(n))$ drops out from string tree-level
computations, and the Riemann zeta values are given by convergent infinite sums
\beq
\zeta_n = \sum_{k=1}^\infty k^{-n} \, , \ \ \ \ \ \ n\geq 2\, .
\label{4ptmzv.2}
\eeq
The $\alpha'$-dependence of the four-point open-superstring amplitude (\ref{4ptex.3}) 
can be written as
\begin{align}
F_2{}^2 &= \frac{ \Gamma(1{-}2\ap s_{12})\Gamma(1{-}2\ap s_{23}) }{\Gamma(1{-}2\ap s_{12}{-}2\ap s_{23})}
= \exp \bigg(
\sum_{n=2}^{\infty} \frac{ \zeta_n }{n}(2\ap)^n \big[  s_{12}^n {+}s_{23}^n - (s_{12}{+}s_{23})^n\big]
\bigg)  \notag \\
&= 1 - (2\ap)^2 \zeta_2 s_{12} s_{23} + (2\ap)^3 \zeta_3 s_{12}s_{23}s_{13} - (2\ap)^4 \zeta_4 s_{12}s_{23} 
\big(s_{12}^2 + \tfrac{1}{4} s_{12}s_{23} + s_{23}^2 \big)  \label{4ptmzv.1} \\
&\quad
- (2\ap)^5 \zeta_2 \zeta_3 s_{12}^2 s_{23}^2 s_{13} +\frac{1}{2} (2\ap)^5 \zeta_5 s_{12} s_{23} s_{13}
(s_{12}^2{+}s_{23}^2{+}s_{13}^2) + {\cal O}(\ap^6) \, ,
\notag
\end{align}
where $F_2{}^2$ is the scalar four-point instance of the $(n{-}3)! \times (n{-}3)!$
matrix $F_P{}^Q$ of $n$-point disk integrals in (\ref{Fdef}). The Gamma functions
in the numerators of $F_2{}^2$ introduce poles at $2\ap s_{12}, 2\ap s_{23}=1,2,\ldots$,
i.e.\ at center-of-mass energies $(k_i{+}k_j)^2 \in \frac{ \mathbb N }{\alpha'}$, that signal 
the exchange of massive open-string vibration modes. After $\alpha'$-expansions, say in
the exponential of (\ref{4ptmzv.1}), these poles are no longer manifest.

Also for closed superstrings, the $\alpha'$-expansion of the four-point amplitude in 
the form (\ref{4ptclosed}) can be extracted from a scalar combination of Gamma functions,
\begin{align}
 \frac{ 
\Gamma(1{-} \frac{\ap}{2} s_{12})\Gamma(1{-}\frac{\ap}{2}s_{23})\Gamma(1{-}\frac{\ap}{2}s_{13}) }{
\Gamma(1{+}\frac{\ap}{2}s_{12})\Gamma(1{+}\frac{\ap}{2}s_{23})\Gamma(1{+}\frac{\ap}{2}s_{13})}&= 
\exp \bigg( 2
\sum_{k=1}^{\infty} \frac{ \zeta_{2k+1}}{2k{+}1} \bigg( \frac{\ap}{2} \bigg)^{2k+1} \big[  s_{12}^{2k+1} +s_{23}^{2k+1}
+ s_{13}^{2k+1}\big]
\bigg) 
\label{4ptmzv.3}\\
&=
 1 +2 \bigg( \frac{\ap}{2} \bigg)^3 \zeta_3 s_{12}s_{23}s_{13}  + \bigg( \frac{\ap}{2} \bigg)^5 \zeta_5 s_{12} s_{23} s_{13}
(s_{12}^2{+}s_{23}^2{+}s_{13}^2) + {\cal O}(\ap^6)\, .
\notag
\end{align}
The coefficients in this series are still exclusively built from Riemann zeta
values (\ref{4ptmzv.2}), but there is no more reference to the even zeta
values $\zeta_{2k}$ seen in the open-string expansion (\ref{4ptmzv.1}).
While these cancellations at four points can still be understood from Gamma-function
expansions, their generalizations to $n\geq 5$ points are governed by an elaborate mathematical
structure known as the {\it single-valued map}, see sections \ref{sec:7.2.3} and \ref{sec:7.6} for details.

\subsubsection{Low-energy effective actions}
\label{sec:7.1}

One of the traditional motivations for $\alpha'$-expansions of massless string 
amplitudes is to determine the low-energy effective action of the gauge and gravity
multiplets. An expansion around $\alpha' \rightarrow 0$
amounts to integrating out the massive vibration modes which become infinitely
heavy in view of their mass-spectra $M^2 \in \mathbb N/\alpha'$ and
$M^2 \in 4 \mathbb N/\alpha'$ for open and closed strings, respectively. This can be
anticipated from the fact that the poles of the Gamma functions due to massive-state
exchange in (\ref{4ptmzv.1}) or (\ref{4ptmzv.3}) are no longer manifest in the exponentials 
encoding the respective $\alpha'$-expansions, let alone in the individual orders in $\alpha'$.

The joint effort of all massive string modes leads to effective string interactions of schematic 
form $\alpha'^{m+k-2} \Tr\{ D^{2k}\mathbb F^m\}$ and $\alpha'^{m+k-1} D^{2k} \mathbb R^m$ 
(with $m\geq 4$ powers of the non-linear
gluon field strength $\mathbb F$, Riemann curvature $\mathbb R$
and their respective gauge- and diffeomorphism-covariant derivatives $D$)
and their supersymmetrizations. Low-energy effective operators $\alpha'^{m+k-2}  
\Tr\{D^{2k}\mathbb F^m\}$ or 
$\alpha'^{m+k-1} D^{2k} \mathbb R^m$ can be extracted from
the $\alpha'$-expansion of massless string amplitudes, i.e.\ by reverse-engineering
the Feynman rules that generate a given $\alpha'$-order of the amplitude.
This approach turns out to be cumbersome in practice since\footnote{For a brief review and for the
practical struggles associated with
(i) and (ii), see \cite{Richards:2008jg}.}
\begin{itemize}
\item[(i)] field redefinitions and 
relations of the schematic form $D^2 \mathbb F \cong \mathbb F^2$ or
$D^2 \mathbb R \cong \mathbb R^2$ introduce ambiguities, 
\item[(ii)] extracting the new information on $D^{2k}\mathbb F^n$ or $D^{2k}\mathbb R^n$ interactions 
from $n$-point amplitudes necessitates the subtraction of reducible-diagram 
contributions with insertions of $D^{2k}\mathbb F^m$ or $D^{2k} \mathbb R^m$ at $m< n$,
\item[(iii)] operators with four or more field strengths $\mathbb F$ and in particular curvature tensors 
$\mathbb R$ admit a large number of Lorentz-index structures, which can be 
alleviated via manifestly supersymmetric approaches.
\end{itemize}
In the non-abelian gauge sector of the type I effective action, there are no explicit
results beyond the order of $\alpha'^4$ with all the tensor structures spelled out. 
The purely bosonic terms at leading orders are given by \cite{Tseytlin:1986ti, Gross:1986iv}
\begin{align}
S_{\rm eff}^{\rm open}= \int d^{10} x \, {\Tr}\bigg\{{-}\frac{1}{4} \mathbb F_{mn} \mathbb F^{mn}
+   \ap^2 \zeta_2 \Big[&{-} 2 \mathbb F^m{}_p  \mathbb F^p{}_n
\mathbb F^{q}{}_{m} \mathbb F^{n}{}_{q} 
 -\mathbb F^m{}_n \mathbb F^n{}_p \mathbb F^{p}{}_{q} \mathbb F^{q}{}_{m} \label{tensorseff} \\
 &+\tfrac{1}{2} \mathbb F^{mn} \mathbb F_{mn} \mathbb F^{pq} \mathbb F_{pq}
 + \tfrac{1}{4} \mathbb F^{mn}  \mathbb F^{pq} \mathbb F_{mn} \mathbb F_{pq} \Big] +{\cal O}(\ap^3) \bigg\} + {\rm fermions} \, ,
\notag
\end{align}
followed by 8-term expressions $\alpha'^3 \zeta_3 {\Tr} \{D^2 \mathbb F^4{+} \mathbb F^5\}$
and 96-term-expressions  $\alpha'^4 \zeta_4 {\Tr}\{D^4 \mathbb F^4{+}D^2 \mathbb F^5{+} \mathbb F^6\}$.
At higher orders $\alpha'^{w\geq 5}$,
one encounters multiple conjecturally $\mathbb Q$-independent 
combinations of MZVs (say $\zeta_5$ and $\zeta_2 \zeta_3$ at $\alpha'^5$,
also see table \ref{QQbases} below),
and the accompanying operators ${\rm Tr} \{ D^{2k} \mathbb F^m\}$ typically cover
powers $4\leq m \leq w{+}2$ of field strengths. A schematic overview
of effective gauge interactions up to $\alpha'^{6}$ is given in 
table \ref{eff:table} as presented in \cite{6ptOprisa}.

\begin{table}[h]
	\begin{center}
		\begin{tabular}{|c | c  | c  |c  |}\hline 
			order &MZV &eff.\ gauge interactions &eff.\ gravity interactions \\\hline \hline
			$(\alpha')^0$ &1 &$\mathbb F^2$ &$\mathbb R$ \\\hline
			$(\alpha')^1$ &$\times$ &$\cancel{\mathbb F^3}$ &$\cancel{\mathbb R^2}$ \\\hline
			$(\alpha')^2$ &$\zeta_2$ &$\mathbb F^4$ &$\cancel{\mathbb R^3}$ \\\hline
			$(\alpha')^3$ &$\zeta_3$ &$D^2\mathbb F^4+\mathbb F^5$ &$\mathbb R^4$ \\\hline
			$(\alpha')^4$ &$\zeta_4$ &$D^4 \mathbb F^4 + D^2 \mathbb F^5 +\mathbb F^6$ 
			&$\cancel{D^2\mathbb R^4}+\cancel{\mathbb R^5}$ \\\hline
			$(\alpha')^5$ &$\zeta_5$ &$D^6\mathbb F^4 + D^4\mathbb F^5+D^2 \mathbb F^6+\mathbb F^7$ 
			&$D^4 \mathbb R^4+ D^2 \mathbb R^5+\mathbb R^6$ \\
			  &$\zeta_3 \zeta_2$ &$D^6\mathbb F^4 + D^4\mathbb F^5+ D^2 \mathbb F^6+\mathbb F^7$ &$\cancel{D^4 \mathbb R^4}+ \cancel{D^2 \mathbb R^5}+\cancel{\mathbb R^6}$ \\\hline
			$(\alpha')^6$ &$\zeta_3^2$ &$D^8\mathbb F^4 +D^6\mathbb F^5 +\ldots+\mathbb F^8$ 
			&$D^6\mathbb R^4+D^4\mathbb R^5+D^2 \mathbb R^6+\mathbb R^7$ \\
			 &$\zeta_6$ &$D^8\mathbb F^4 +D^6\mathbb F^5 +\ldots+\mathbb F^8$ &$\cancel{D^6\mathbb R^4}+\cancel{D^4\mathbb R^5}+\cancel{D^2 \mathbb R^6}+\cancel{\mathbb R^7}$ 	\\ \hline
		\end{tabular}
	\end{center}
\caption{Schematic overview of gauge and gravity interactions and their MZV coefficients
in the tree-level effective action of type I and type II superstrings, respectively. Gauge-field
operators $D^{2k}\mathbb F^m$ are understood to be traced over the gauge
group, and the shown operators in both the gauge and gravity sector are 
representatives for supersymmetry multiplets of interactions. The crossed out
operators $\cancel{\mathbb F^3}$ and $\cancel{D^{2k} \mathbb R^m}$
could in principle have been expected from their mass dimensions but turn out to be
absent from superstring effective actions, either by supersymmetry
arguments or by the properties of their MZV coefficients. The column
on gauge interactions closely follows the presentation in \cite{6ptOprisa},
and the rightmost column is a subset of table I in \cite{Stieberger:2009rr}.}
\label{eff:table}
\end{table}

A state-of-the-art method to determine the tensor structures of the effective
gauge interactions in table~\ref{eff:table} can be found in \cite{Barreiro:2012aw}\footnote{The absence of tensor structures
$(e_j\cdot k_i)^n$ in $n$-point disk amplitudes, i.e.\ the appearance of
at least one factor of $(e_i\cdot e_j)$ in each summand of the gluon components,
was recognized as a valuable source of information on the open-string effective 
action \cite{Barreiro:2012aw} and properties of the amplitudes 
themselves \cite{Barreiro:2013dpa}.}, also see \cite{Kitazawa:1987xj, Barreiro:2005hv}
and \cite{Andreev:1988cb, Koerber:2002zb, 6ptOprisa,Howe:2010nu} for earlier 
results at the orders of $\alpha'^{\leq 4}$. 
 The abelian gauge sector of type I superstrings
incorporates supersymmetric Born--Infeld theory in its low-energy limit \cite{Metsaev:1987qp},
see also \cite{Tseytlin:1986ti, Tseytlin:1997csa,Cederwall:2002df}.

The closed-string counterparts of the effective interactions of gauge multiplets in
(\ref{tensorseff}) are summarized in the rightmost column of table \ref{eff:table} 
as presented in \cite{Stieberger:2009rr}. The table illustrates that gravitational
higher-derivative interactions are more sparse than gauge interactions: 
in contrast to the first higher-derivative operator $\alpha'^2{\rm Tr}\{\mathbb F^4\}$
of the open superstring, the first correction 
to the supergravity action of type II superstrings occurs at the third subleading
order in $\alpha'$, and its gravitational $\alpha'^3\mathbb R^4$ contribution was
firstly investigated in \cite{Gross:1986iv}. Moreover, the possible MZV coefficients are
constrained to be \textit{single-valued} as will be detailed in sections \ref{sec:7.2.3}
and \ref{sec:7.6}, leading to the absence of coefficients $\zeta_{2k}$ in
the rightmost column of table~\ref{eff:table} \cite{Stieberger:2009rr}.

\subsubsection{On the scope of four-point amplitudes}
\label{sec:7.0.1}

The polarization dependence of the four-point open- and closed-string
amplitudes (\ref{4ptex.3}) and (\ref{4ptclosed}) enjoys a simple tensor
structure at all orders in $\alpha'$: for the bosonic components, 
the polarization vectors in the prefactors $A(1,2,3,4)$ and $M_4^{\rm grav}$
combine to linearized field strengths contracted by the famous $t_8$-tensor
\beq
t_8(f_1,f_2,f_3,f_4) = f_1^{mn} f_2^{np} f_3^{pq} f_4^{qm} 
- \frac{1}{4} f_1^{mn}f_2^{mn} f_3^{pq} f_4^{pq} + {\rm cyc}(2,3,4)\, ,
\label{t8.01}
\eeq
namely
\begin{align}
s_{12}s_{23}A(1,2,3,4) &= -{1\over2} t_8(f_1,f_2,f_3,f_4)+ {\cal O}(\chi_j) \, ,
\label{t8.02}  \\
s_{12}s_{23}s_{13} M_4^{\rm grav} &=  -{1\over4} t_8(f_1,f_2,f_3,f_4)  
t_8(\tilde f_1,\tilde f_2, \tilde f_3, \tilde f_4)  + {\cal O}(\chi_j,\tilde \chi_j)\,,
\notag
\end{align}
where $A(1,2,3,4)$ and $M_4^{\rm grav}$  are given in \eqref{shortMsix} and \eqref{KLTrel}.
Accordingly, the closed-form expressions for the four-point $\alpha'$-expansions (\ref{4ptmzv.1})
and (\ref{4ptmzv.3}) can be used to swiftly propose operators 
$ {\Tr} \{D^{2k}\mathbb F^4\}$ or $D^{2k} \mathbb R^4$ which reproduce
the four-point amplitudes. For instance, the first $\alpha'$-correction
in the open-superstring effective action (\ref{tensorseff}) can be written as $-\alpha'^2 \zeta_2 
\Tr\big(t_8(\mathbb F, \mathbb F,\mathbb F,\mathbb F) \big)$ and readily reflects
the subleading order of
 \beq
 {\cal A}(1,2,3,4) = A(1,2,3,4) \Big( 1 - (2\ap)^2 \zeta_2 s_{12} s_{23} + {\cal O}(\ap^3) \Big) \, .
 \label{F4linearized}
 \eeq
However, the information from the four-point amplitudes does not fix any 
effective operators $ {\Tr} \{D^{2k}\mathbb F^m\}$ or $D^{2k} \mathbb R^m$ 
with $m\geq 5$ required by non-linear supersymmetry. Moreover, the Mandelstam
dependence of the four-point $\alpha'$-expansion does not fix the order of the 
non-commutative covariant derivatives $D_m$ acting on $\mathbb F^4$ or $\mathbb R^4$.
At the time of writing, it is not clear whether one can find a tensor structure analogous
to $t_8$ that governs the five-field operators $ {\Tr} \{D^{2k}\mathbb F^5\}$
and $D^{2k}\mathbb R^5$ at all orders in $\alpha'$.\footnote{See for instance
\cite{Peeters:2005tb, Liu:2022bfg} for explicit tensors $t_r$ contracting $r\geq 16$ 
indices in eight-derivative interactions related to superpartners of $\mathbb R^4$.}

For closed-string effective actions, the complexity proliferates more drastically with
the order in $\alpha'$. As an additional complication as compared to open strings, 
already the simplest $\alpha'$-correction $\alpha'^3 \zeta_3  \mathbb R^4$
to the ten-dimensional type II supergravity action goes beyond the $t_8$-tensor:
The sixteen Lorentz indices of $\prod_{j=1}^{4}  \mathbb R^{m_j n_j p_j q_j}$
are contracted with a combination of $t_8 t_8$ and two ten-dimensional
Levi--Civita tensors $\varepsilon_{10}^{abm_1n_1 m_2 n_2 m_3 n_3 m_4 n_4}
 \varepsilon_{10}^{abp_1q_1 p_2 q_2 p_3 q_3 p_4 q_4}$ in the tree-level effective action 
 of both type IIA and type IIB superstrings \cite{Grisaru:1986px,Grisaru:1986vi,Freeman:1986zh},
where the $\varepsilon_{10} \varepsilon_{10}$ terms do not contribute to four-point amplitudes.

Furthermore, the type II supergravity multiplets have a much richer structure than 
the gauge multiplet of ten-dimensional SYM. The multitude of superpartners of the 
type IIB tree-level interaction
$\alpha'^3 \zeta_3(t_8 t_8+\varepsilon_{10} \varepsilon_{10}) \mathbb R^4$ for instance
includes a sixteen-dilatino term \cite{Green:1997me}. More generally, the
operators in the type IIB effective action are organized according to their
charges w.r.t.\ the $U(1)$ R-symmetry of type IIB supergravity which is
broken by the string corrections. 

As an even more fundamental limitation of four-point open- and
closed-string amplitudes,
they fail to anticipate the effective operators whose coefficients are (conjecturally) 
indecomposable MZVs $\zeta_{n_1,\ldots, n_r}$ of depth $r \geq 2$ to be introduced
in section \ref{sec:7.2}. Their first open- and closed-string instances occur at the orders of
\cite{Stieberger:2009rr, Schlotterer:2012ny}
\begin{align}
&(\alpha')^8 \zeta_{3,5}{\rm Tr}\{ \cancel{D^{12} \mathbb F^4} + D^{10}\mathbb F^5 +\ldots+ \mathbb F^{10}\}\, ,
\\
&(\alpha')^{11} \zeta_{3,3,5}(\cancel{D^{16} \mathbb R^4} + D^{14}\mathbb R^5 +\ldots+ \mathbb R^{12})\, , \notag
\end{align}
followed by infinite families of operators $ {\Tr} \{D^{2k}\mathbb F^{\geq 5}\}$
and $D^{2k}\mathbb R^{\geq 5}$ with coefficients beyond the scope of the 
Riemann zeta values in table \ref{eff:table}. The absence of ${\Tr} \{D^{2k}\mathbb F^{4}\}$-
and $D^{2k}\mathbb R^{4}$ interactions with higher-depth MZVs as coefficients 
follows from the fact that Riemann zeta values capture all orders of the
four-point $\alpha'$-expansions (\ref{4ptmzv.1}) and (\ref{4ptmzv.3}).

We finally note that also five-point string amplitudes feature dropouts 
of MZVs starting from the $(\alpha')^{18}$-order \cite{Schlotterer:2012ny, Drummond:2013vz}.
As a result, certain combinations of MZVs firstly occur in six-point amplitudes
and therefore in effective interactions $ {\Tr} \{D^{2k}\mathbb F^{\geq 6}\}$
or $D^{2k}\mathbb R^{\geq 6}$ such as
$(\alpha')^{18} {\rm Tr}\{ \cancel{D^{32} \mathbb F^4} +\cancel{D^{30}\mathbb F^5}+ D^{28}\mathbb F^6 +\ldots+ \mathbb F^{20}\}$.

\subsubsection{Manifestly supersymmetric approaches}
\label{sec:7.0.2}

The manifestly supersymmetric form of the $n$-point disk amplitude in (\ref{nptdisk})
severely constrains the effective action of type I superstrings to all orders in $\alpha'$:
The amplitudes computed from reducible and irreducible diagrams at various orders
in $\alpha'$ must conspire to linear combinations of SYM trees. It is an open problem
to translate this property together with the all-order results on the $\alpha'$-expansion of 
disk integrals to be reviewed below into a new line of attack for the effective action.

By the success of pure spinor methods to obtain compact expressions for $n$-point 
amplitudes, one can expect that the open questions on effective actions will benefit
from superspace methods. The supersymmetrization of the $\mathbb F^4$ interaction
in (\ref{tensorseff}) has a long history \cite{Bergshoeff:1986jm, Gates:1986is, Cederwall:2001bt, Cederwall:2001td} and manifestly supersymmetric formulations of more general effective string interactions have for 
instance been discussed in \cite{Green:1998by, 
Koerber:2001uu, Collinucci:2002ac, Berkovits:2002ag, Drummond:2003ex, 
tsimpis, Howe:2010nu, Cederwall:2011vy, Wang:2015jna}.

\subsection{Multiple zeta values}
\label{sec:7.2}

The coefficients in the low-energy expansion of $n$-point string amplitudes and
the associated low-energy effective action are rational linear combinations 
of multiple zeta values (MZVs) \cite{Terasoma, Brown:2009qja}
\beq
\zeta_{n_1,n_2,\ldots,n_r} = \sum_{0<k_1<k_2<\ldots <k_r}^\infty k_1^{-n_1}
k_2^{-n_2} \ldots k_r^{-n_r} \, , \ \ \ \ \ \ n_1,n_2,\ldots,n_r \in \mathbb N \, 
, \ \ \ \ \ n_r\geq 2
\label{mzvsec.1}
\eeq
that generalize the Riemann zeta values (\ref{4ptmzv.2}) to depend on multiple integers $n_j$.
The infinite sum converges if $n_r\geq 2$, and we refer to $r$ and $n_1{+}n_2{+}\ldots{+}n_r$
as the {\it depth} and the {\it weight} of the MZV, respectively. While even zeta 
values $\zeta_{2k}$ are rational multiples of $\pi^{2k}$
(with ${\rm B}_{2k}$ denoting the Bernoulli numbers\footnote{The generating function
$\frac{ t  }{e^t{-}1} = \sum_{m=0}^{\infty} \frac{t^m}{m!} {\rm B}_{m}$ leads to even Bernoulli numbers such as ${\rm B}_{2}= \frac{1}{6}, \ {\rm B}_{4} = -\frac{1}{30}$ and ${\rm B}_{6} = \frac{1}{42}$
whereas the odd ones vanish, ${\rm B}_{2k+1}=0 \ \forall \ k \in \mathbb N$, apart from ${\rm B}_{1}= - \frac{1}{2}$.}),
\beq
\zeta_{2k} = - \frac{(2 \pi i)^{2 k} {\rm B}_{2k} }{2(2 k)!}\, ,
\label{mzvsec.2}
\eeq
the numbers $\pi,\zeta_3,\zeta_5,\zeta_7,\ldots$ are conjectured to be
algebraically independent over $\mathbb Q$.
MZVs arise from iterated integrals over
logarithmic forms $d \log(z_j{-}a_j)$ with $a_j \in \{0,1\} \ \forall \ j=1,2,\ldots,w$ and $a_1\neq 0$, 
\begin{align}
I(0;a_1a_2\ldots a_w;z) &= \int_{0<z_1<z_2<\ldots<z_w < z} \frac{ d z_1 }{z_1{-}a_1}
 \frac{ d z_2 }{z_2{-}a_2} \ldots  \frac{ d z_w }{z_w{-}a_w}  \, ,\notag \\
\zeta_{n_1,n_2,\ldots,n_r} &= (-1)^r I(0; 1\underbrace{0\ldots 0}_{n_1-1} 
1\underbrace{0\ldots 0}_{n_2-1}\ldots 1\underbrace{0\ldots 0}_{n_r-1};1)\, ,
\label{mzvsec.3}
\end{align}
i.e.\ multiple polylogarithms at unit argument. The combined set of relations following
from the iterated-integral and nested-sum
representations can be used to reduce any MZV of weight $w\leq 7$ to products of Riemann
zeta values $\zeta_n$ and leave the conjectural
bases over $\mathbb Q$ in table \ref{QQbases}. The first instances
of irreducible MZVs at depth 2 and 3 are believed to occur at weight 8 (e.g.\ $\zeta_{3,5}$)
and weight 11 (e.g.\ $\zeta_{3,3,5}$), respectively.

\begin{table}[h]
	\begin{center}
		\begin{tabular}{|c || c  | c  |c  |c  |c  |c  |c  |c  |c  |c  |c  |c  |}\hline 
			$w$ &0 &1 &2 &3 &4 &5 &6 &7 &8 &9 &10 &11 \\\hline \hline
			 &1 & &$\zeta_2$ &$\zeta_3$ &$\zeta_2^2$ &$\zeta_5$
			&$\zeta_2^3$ &$\zeta_7$ &$\zeta_2^4$ &$\zeta_9$ &$\zeta_2^5 \quad \zeta_{2} \zeta_{3,5}$ &$\zeta_{11} \quad \zeta_{3,3,5}$ \\
			&&&&&&$\zeta_2 \zeta_3$ &$\zeta_3^2$ &$\zeta_2 \zeta_5$ &$\zeta_{3,5}$
			&$\zeta_2 \zeta_7$ &$  \zeta_{3,7} \quad \zeta_2^2 \zeta_3^2$ &$\zeta_2 \zeta_9 \quad \zeta_2^2 \zeta_7$ \\
			MZV &&&&&&&&$\zeta_2^2 \zeta_3$ &$\zeta_2 \zeta_3^2$ &$\zeta_2^2 \zeta_5$ &$\zeta_{5}^2 \quad \zeta_3 \zeta_7$ &$\zeta_2^3 \zeta_5 \quad \zeta_2^4 \zeta_3$ \\
			&&&&&&&&&$\zeta_3 \zeta_5$ &$\zeta_2^3 \zeta_3$ &$\zeta_{2} \zeta_3 \zeta_5$ &$\zeta_3^2 \zeta_5 \quad  \zeta_{2} \zeta_3^3$ \\ &&&&&&&&&&$\zeta_3^3$  & &$\zeta_3 \zeta_{3,5}$
			\\ \hline
	dim$_w$	&1 &0 &1 &1 &1 &2 &2 &3 &4 &5 &7 &9	
			\\ \hline
		\end{tabular}
	\end{center}
\caption{Conjectural $\mathbb Q$-bases of MZVs at weights $w\leq 11$.}
\label{QQbases}
\end{table}

Comprehensive references on MZVs include \cite{Jianqiang, GilFresan}, 
and a datamine of $\mathbb Q$-relations with machine-readable ancillary files 
can be found in \cite{Blumlein:2009cf}. Any known relation among MZVs 
over $\mathbb Q$ preserves the weight, and the dimensions ${\rm dim}_w$
of the tentative $\mathbb Q$-bases at weight $w$ are conjectured to obey
the recursion ${\rm dim}_w={\rm dim}_{w-2}+{\rm dim}_{w-3}$
with ${\rm dim}_0=1={\rm dim}_2$ and ${\rm dim}_1=0$ \cite{Zag:rec},
see table \ref{QQbases} for possible representatives.

\subsubsection{Motivic MZVs and the f alphabet}
\label{sec:7.2.1}

The conjectural counting of $\mathbb Q$-linearly independent MZVs 
through the above recursion for ${\rm dim}_w$ can be reproduced from
a simple model, the so-called {\em f-alphabet} \cite{Brown:2011ik}: introduce
non-commutative variables $f_{3},f_5,f_7,\ldots$
for each odd integer $\geq 3$, a single commutative variable $f_2$ and assign
weight $w$ to $f_w$. It is easy to show that the number of weight-$w$ compositions
(non-commutative words in $f_{2m+1}$ along with non-negative powers of $f_2$)
is counted by ${\rm dim}_w$ in table \ref{QQbases}, e.g.\ $\{f_5 , \, f_2 f_3\}$ at weight 
five or $\{ f_2^4, \, f_2 f_3 f_3, \, f_3 f_5, \, f_5 f_3\}$ at weight eight. Note in particular
that the first instance $f_3 f_5 \neq f_5 f_3$ of non-commutativity ties in
with the first conjecturally irreducible MZV $\zeta_{3,5}$ beyond depth one.

It is tempting to map MZVs into the $f$-alphabet, i.e.\ the Hopf-algebra comodule
${\cal U} = \mathbb Q \langle f_3,f_5,\ldots \rangle \otimes_{\mathbb Q} \mathbb Q[f_2]$,
in order to manifestly mod out by their $\mathbb Q$-relations.
However, the unsettled transcendentality properties of MZVs currently 
obstruct a well-defined map to ${\cal U}$. As a workaround, one can consider
{\it motivic MZVs} $\zeta^{\mathfrak m}_{n_1,n_2,\ldots,n_r}$ instead of the 
$\zeta_{n_1,n_2,\ldots,n_r} \in \mathbb R$ in (\ref{mzvsec.1}).
By definition, motivic MZVs obey the complete set of $\mathbb Q$ relations
among $\zeta_{n_1,n_2,\ldots,n_r}$ known up to date, and their elaborate definition
in the framework of algebraic geometry can be found 
in \cite{Goncharov:2005sla, Brown:2011ik, BrownTate}.

By passing to motivic MZVs, one can set up an invertible map $\phi$ to the $f$-alphabet
starting from the normalization
\beq
\phi(\zeta^{\mathfrak m}_{2k+1}) = f_{2k+1} \, , \ \ \ \ \ \ 
\phi(\zeta^{\mathfrak m}_{2}) = f_2 \, . 
\label{mzvsec.4}
\eeq
For motivic MZVs $\zeta^{\mathfrak m}_{n_1,\ldots,n_r}$ of weight $w$ beyond 
depth one, the $\phi$-image up to adding a $\mathbb Q$-multiple of $\phi(\zeta^{\mathfrak m}_w) $ 
can be determined from the shuffle product $\shuffle$ and 
deconcatenation coaction $\Delta$ in~${\cal U}$:
\begin{align}
f_A \shuffle f_B &= \sum_{C \in A \shuffle B} f_C\, ,
\label{mzvsec.5}\\
\Delta (f_2^n f_A) &= f_2^n \sum_{A= BC} f_B \otimes f_C\, .
\notag
\end{align}
We employ the shorthand $f_A= f_{a_1} f_{a_2} \ldots f_{a_{|A|}}$ for 
$A=a_1a_2\ldots a_{|A|}$ (with $f_{\emptyset}=1$), and 
the sum over $\sum_{A= BC}$ includes the terms with $B= \emptyset$ or
$C= \emptyset$, for instance
\begin{align}
(f_3 f_5) \shuffle f_7 &= f_3 f_5 f_7 + f_3 f_7 f_5 + f_7 f_3 f_5 \, , \label{mzvex.5} \\
\Delta(f_2 f_9 f_3) &= f_2 \otimes f_9 f_3 + f_2 f_9 \otimes f_3 + f_2 f_9 f_3\otimes 1\, .
\notag
\end{align}
More specifically, the $\phi$-images of motivic MZVs are required to preserve
the product and coaction structure in the sense of
\begin{align}
\phi( \zeta^{\mathfrak m}_{n_1,\ldots,n_r} \cdot \zeta^{\mathfrak m}_{p_1,\ldots,p_s} )
&= \phi( \zeta^{\mathfrak m}_{n_1,\ldots,n_r}) \shuffle \phi( \zeta^{\mathfrak m}_{p_1,\ldots,p_s} )\, ,
\notag \\
\Delta\big(  \phi( \zeta^{\mathfrak m}_{n_1,\ldots,n_r} )  \big) &= 
\phi \big(  \Delta( \zeta^{\mathfrak m}_{n_1,\ldots,n_r} )  \big) \, .
\label{mzvsec.6}
\end{align}
The shuffle symbol in the first line is understood to act trivially on the commutative variable
$f_2$, e.g.\ $\phi(\zeta^{\mathfrak m}_{2}\zeta^{\mathfrak m}_{3}) = f_2 f_3$ and $\phi\big(\zeta^{\mathfrak m}_{2})^k \big) = f_2^k$ for any $k \in \mathbb N$.
In the second line of (\ref{mzvsec.6}), the 
coaction $\Delta( \zeta^{\mathfrak m}_{n_1,\ldots,n_r} )$ can be
obtained from \cite{Goncharov:2005sla} and leaves the freedom to shift
$\phi( \zeta^{\mathfrak m}_{n_1,\ldots,n_r} )$ at higher depth by the 
depth-one image $\phi( \zeta^{\mathfrak m}_{n_1+\ldots+n_r} )= f_{n_1+\ldots+n_r}$.\footnote{Strictly 
speaking, the second entry of the coaction $\Delta( \zeta^{\mathfrak m}_{n_1,\ldots,n_r} )$ 
involves deRham periods $\zeta^{\mathfrak dr}_{n_1,\ldots,n_r}$,
where the deRham version of $\zeta_2$ vanishes, see for instance \cite{Francislecture}.}
As the simplest examples of this procedure, the (conjecturally indecomposable) 
MZVs beyond depth one in table \ref{QQbases} are mapped to
\begin{align}
 \phi( \zeta^{\mathfrak m}_{3,5} ) &= - 5 f_3 f_5 \, , \ \ \ \ \ \ 
  \phi( \zeta^{\mathfrak m}_{3,7} )  = - 14 f_3 f_7 - 6 f_5 f_5 \, ,\notag \\
  \phi( \zeta^{\mathfrak m}_{3,3,5} )  &= - 5 f_3 f_3 f_5 - 45 f_9 f_2  - \frac{ 6}{5} f_7 f_2^2 
  + \frac{4}{7} f_5 f_2^3\, , \label{mzvsec.7}
\end{align}
where we have chosen to exclude $f_8, f_{10}$ and $f_{11}$ from 
$ \phi( \zeta^{\mathfrak m}_{3,5} ) ,\phi( \zeta^{\mathfrak m}_{3,7} ) $
and $ \phi( \zeta^{\mathfrak m}_{3,3,5} )$, respectively. The coefficients
of $f_w$ in each other $\phi( \zeta^{\mathfrak m}_{n_1,\ldots ,n_r})$ at
weight $w=8,10$ or $11$ are determined by (\ref{mzvsec.4}),
(\ref{mzvsec.6}), (\ref{mzvsec.7}) and imposing that $\phi$ preserves
the $\mathbb Q$-relations among motivic MZVs.

Examples of $\phi( \zeta^{\mathfrak m}_{n_1,\ldots,n_r} )$ at higher weight $w$
can be found in \cite{Schlotterer:2012ny}, where the MZVs beyond depth one in
the conjectural $\mathbb Q$ bases of \cite{Blumlein:2009cf} are taken to have no $f_w$
in their $\phi$-images (also see section \ref{sec:7.2.4} for comments on the conventions). 
As will be reviewed in section \ref{sec:7.3}, the $\alpha'$-expansion of the disk 
integrals $F_P{}^Q$ in (\ref{Fdef}) takes a very compact form
once the (motivic) MZVs in the coefficients are translated into the $f$-alphabet.

\subsubsection{The Drinfeld associator}
\label{sec:7.2.2}

As will be described in section \ref{sec:7.4}, the MZVs in the $\alpha'$-expansion 
of $n$-point disk integrals can be derived from the {\it Drinfeld associator}, a generating
series of MZVs. Besides the MZVs in (\ref{mzvsec.3}) obtained from convergent iterated integrals
$I(0;1\ldots 0;1)$, the Drinfeld associator also involves so-called shuffle-regularized
MZVs which descend from formally divergent integrals. For the iterated integrals
$I(0;a_1\ldots a_w;1)$ in (\ref{mzvsec.3}), we assign regularized values
\beq
I(0;0;1) = I(0;1;1) = 0
\label{mzvsec.8}
\eeq
to the divergent cases at weight 1. At higher weight, the regularized values of 
integrals $I(0;0  \ldots ;1)$ or $ I(0;\ldots  1;1)$ with endpoint divergences 
are defined by (\ref{mzvsec.8}) and by imposing them to obey the shuffle relations 
of convergent $I(0;1\ldots 0;1)$,
\beq
I(0;A;1) I(0;B;1) = \sum_{C \in A\shuffle B} I(0;C;1) \, , \ \ \ \ \ \ {\rm also} \ {\rm for} \ a_1,b_1 = 0 \ 
{\rm and} \ a_{|A|},b_{|B|} = 1\, .
\label{mzvsec.9}
\eeq
One can recursively remove leading zeros by relating $I(0; 0^k D;1)$ with $d_1=1$ to
$I(0; 0^k;1) I(0; D;1)$ minus terms with $\leq k{-}1$ leading zeros, for instance
$I(0;01;1) =I(0;0;1) I(0;1;1)- I(0;10;1)= \zeta_2$. Similarly, subtracting $I(0;  D ;1) I(0;  1^k;1)$
from $I(0;  D 1^k;1)$ with $d_{|D|}=0$ yields terms with $\leq k{-}1$ terminal ones,
see for instance \cite{Panzer:2015ida} for further details.
The resulting regularized values of $I(0;A;1)$ with $a_1=0$ and/or $a_{|A|}=1$ are
known as shuffle-regularized MZVs.

The Drinfeld associator $\Phi(e_0,e_1)$ (not to be confused with the perturbiners
for bi-adjoint scalars in section \ref{subsec:biadj}) is a generating series of
shuffle-regularized MZVs, where the coefficient of $I(0;A;1)$ is
a word $e_A= e_{a_1} e_{a_2} \ldots e_{a_w}$ of non-commutative 
variables $e_0,e_1$,
\begin{align}
\Phi(e_0,e_1) &= \sum_{A \in \{0,1\}^\times}  (-1)^{\sum_{j=1}^{|A|} a_j} I(0;A;1) e_{A}
\notag \\
&= 1 + \zeta_2 [e_0,e_1] + \zeta_3 [e_0{-}e_1,[e_0,e_1]] \label{mzvsec.10} 
\\
&\quad + \zeta_4 \Big(
[e_0,[e_0,[e_0,e_1]]] + \tfrac{1}{4} [e_1,[e_0,[e_1,e_0]]]
+[e_1,[e_1,[e_0,e_1]]] + \tfrac{5}{4} [e_0,e_1]^2
\Big) + \ldots\, ,
\notag
\end{align}
and the summation range $ \{0,1\}^\times$ denotes the set of words (of arbitrary
length $0,1,2,\ldots$) in letters $0,1$. The pairing of $I(0;A;1) e_{A}$
ensures that the length of the words in $e_0,e_1$ matches the weight of the accompanying 
shuffle-regularized MZVs, and the ellipsis in the last line of (\ref{mzvsec.10}) refers to words 
of length $\geq 5$.

In the first place, the Drinfeld associator has been introduced as the universal
monodromy of the KZ equation 
\beq
\frac{ d F(z)}{dz} = \bigg( \frac{ e_0}{z}+ \frac{e_1}{1-z} \bigg)F(z) \, ,
\label{mzvsec.11}
\eeq 
(with $e_0,e_1$ some non-commutative indeterminates)
relating its regularized boundary values $C_0,C_1$ \cite{Drinfeld:1989st, Drinfeld2}:
\beq
C_0= \lim_{z\rightarrow 0} z^{-e_0} F(z) \, , \ \ \ \ 
C_1 =  \lim_{z\rightarrow 1} (1{-}z)^{e_1} F(z)
\ \ \ \ \Rightarrow \ \ \ \ 
C_1 = \Phi(e_0,e_1)C_0 \ .
\label{drinfeld.11}
\eeq
The equivalence of this definition to the generating series (\ref{mzvsec.10}) was then shown
by Le and Murakami \cite{LeMura}. The relevance of the Drinfeld associator for open-string
amplitudes will later on be illustrated by presenting $(n{-}2)!$-component vectors $F$ subject to 
(\ref{mzvsec.11}) and related to disk integrals, with matrix representations of $e_0,e_1$ 
linear in $\alpha' s_{ij}$.

\subsubsection{Single-valued multiple zeta values}
\label{sec:7.2.3}

In comparing the $\alpha'$-expansions of the four-point disk and sphere integrals
(\ref{4ptmzv.1}) and (\ref{4ptmzv.3}), we already noted the dropout of even zeta
values from the closed-string amplitude. The $n$-point systematics of dropouts
in passing from open to closed strings is captured by the notion of {\it single-valued
MZVs} to be reviewed in this section.

The terminology is borrowed from the polylogarithms that specialize to MZVs at
unit argument, see (\ref{mzvsec.3}): while the meromorphic polylogarithms are notoriously
multivalued as the defining integration path is deformed by loops around $z=0$ or $z=1$,
one can form single-valued combinations by adjoining complex conjugates. For
instance, the real part of the multivalued $I(0;1;z)=\log(1{-}z)$ yields the single-valued
$I^{\rm sv}(0;1;z)= I(0;1;z) + \overline{I(0;1;z)} =\log |1{-}z|^2$.

At higher weight, single-valued polylogarithms $I^{\rm sv}(0;A ;z)$ can be systematically
constructed from products of $I(0;B ;z) \overline{I(0;C ;z)}$ and MZVs as detailed in \cite{svpolylog}.
The guiding principle of the reference is to preserve the holomorphic derivatives
\beq
\partial_z I(0;Ab ;z) = \frac{ I(0;A ;z) }{z{-}b}
\ \ \ \leftrightarrow \ \ \
\partial_z I^{\rm sv}(0;Ab ;z) = \frac{ I^{\rm sv}(0;A ;z) }{z{-}b}
\label{mzvsec.12}
\eeq
on the expense of more complicated expressions for the antiholomorphic
derivatives $\partial_{\bar z} I^{\rm sv}(0;Ab ;z)$. The weight-two example
$ I^{\rm sv}(0;10 ;z)=I(0;10 ;z)+I(0;0 ;z)\overline{I(0;1 ;z)} + \overline{ I(0;01 ;z)}$
illustrates that shuffle-regularized versions of polylogarithms (based on 
$I(0;0 ;z)=\log(z)$) are encountered even if the holomorphic part
(in this case $I(0;10 ;z)$) is convergent.

In the same way as meromorphic polylogarithms yield MZVs at $z=1$, see
(\ref{mzvsec.3}), we define single-valued MZVs as single-valued polylogarithms
at unit argument \cite{Schnetz:2013hqa, Brown:2013gia},
\beq
\zeta^{\rm sv}_{n_1,n_2,\ldots,n_r} = (-1)^r I^{\rm sv}(0; 1\underbrace{0\ldots 0}_{n_1-1} 
1\underbrace{0\ldots 0}_{n_2-1}\ldots 1\underbrace{0\ldots 0}_{n_r-1};1)\, .
\label{mzvsec.13}
\eeq
At depth one, this annihilates even zeta values and doubles odd ones,
\beq
\zeta^{\rm sv}_{2k} = 0\, , \ \ \ \ \ \ 
\zeta^{\rm sv}_{2k+1} = 2 \zeta_{2k+1} \, ,
\label{mzvsec.14}
\eeq
and the expressions for single-valued MZVs at higher depth are usually less straightforward, e.g.
\begin{align}
\zeta^{\rm sv}_{3,5} &= - 10 \zeta_3 \zeta_5 \, , \ \ \ \ \ \ 
\zeta^{\rm sv}_{3,7} = -28 \zeta_3 \zeta_7 - 12 \zeta_5^2\, , \label{mzvsec.15} \\
\zeta^{\rm sv}_{3,3,5} &= 2 \zeta_{3,3,5} - 5 \zeta_3^2 \zeta_5 +90 \zeta_2 \zeta_9 
+ \frac{12}{5} \zeta_2^2 \zeta_7 - \frac{ 8}{7} \zeta_2^3 \zeta_5 \, .
\notag
\end{align}
It is clear by the constituents of $I^{\rm sv}(0;A;z)$ that single-valued MZVs can
be expressed in terms of $\mathbb Q$-linear combinations of MZVs.

The above constructions are formalized through the {\it single-valued map}
that sends both meromorphic polylogarithms and arbitrary MZVs to their
single-valued versions. However, the single-valued map ${\rm sv}$ of MZVs is
only well-defined in a motivic setup, i.e.\ (\ref{mzvsec.14}) and (\ref{mzvsec.15}) 
are understood as
\beq
{\rm sv}(\zeta^{\mathfrak m}_{2k}) = 0 \, , \ \ \ \ \ \
{\rm sv}(\zeta^{\mathfrak m}_{2k+1}) = 2\zeta^{\mathfrak m}_{2k+1} \, , \ \ \ \ \ \
{\rm sv}(\zeta^{\mathfrak m}_{3,5}) = -10 \zeta^{\mathfrak m}_{3} \zeta^{\mathfrak m}_{5} \, .
 \label{mzvsec.16} 
\eeq
As a major advantage of adapting the single-valued map to motivic MZVs,
one can employ the $f$-alphabet where the single-valued
map follows a simple closed formula at arbitrary weight and depth
\cite{Brown:2013gia},
\beq
{\rm sv}(f_2^n f_{i_1} f_{i_2}\ldots f_{i_r} ) = \delta_{n,0} \sum_{j=0}^r f_{i_j} \ldots f_{i_2}  f_{i_1}
\shuffle f_{i_{j+1}}  f_{i_{j+2}} \ldots f_{i_r}  \, ,
 \label{mzvsec.17}
\eeq
where $i_1,i_2,\ldots, i_r \in 2\mathbb N{+}1$,
for instance
\begin{align}
{\rm sv}(f_{i_1}) &= 2 f_{i_1} \, , \ \ \ \ \ \ 
{\rm sv}(f_{i_1} f_{i_2}) 
= 2 f_{i_1}\shuffle f_{i_2} = 2 (f_{i_1} f_{i_2} +   f_{i_2} f_{i_1})  \, ,
 \label{mzvsec.18} \\
{\rm sv}(f_{i_1} f_{i_2} f_{i_3}) &= 2 (f_{i_1} f_{i_2} f_{i_3}+ f_{i_3}  f_{i_2} f_{i_1}
+ f_{i_2}  f_{i_1} f_{i_3}
+ f_{i_2}  f_{i_3} f_{i_1})
\,. \notag
\end{align}
In slight abuse of notation, we are employing the same notation
sv for the single-valued map of motivic MZVs and the induced
single-valued map $\phi \, {\rm sv}\, \phi^{-1}$ in the $f$ alphabet.
Since $f_{i_1} f_{i_2}$ and $f_{i_2} f_{i_1}$ (with $i_1,i_2$ odd) are indistinguishable
under the single-valued map by (\ref{mzvsec.18}), irreducible double
zetas such as $\zeta_{3,5}^{\mathfrak m} , \zeta_{3,7}^{\mathfrak m}$
in table \ref{QQbases} factorize into products of odd Riemann zeta values.
Accordingly, $\zeta^{\rm sv}_{3,3,5}$ in (\ref{mzvsec.15}) is the simplest
indecomposable single-valued MZV beyond depth one.

Note that the single-valued map preserves the product structure,
\beq
{\rm sv}( \zeta^{\mathfrak m}_{n_1,\ldots,n_r} \cdot \zeta^{\mathfrak m}_{p_1,\ldots,p_s} )
= {\rm sv}( \zeta^{\mathfrak m}_{n_1,\ldots,n_r}) \cdot {\rm sv}(\zeta^{\mathfrak m}_{p_1,\ldots,p_s} )\, ,
 \label{mzvsec.19}
 \eeq
as one can check from its $f$-alphabet representation (\ref{mzvsec.17}).
In section \ref{sec:7.6}, we will apply the single-valued map to
the $\alpha'$-expansions of disk integrals which then acts on (motivic) MZVs at
various weights.

\subsubsection{Comments on conventions}
\label{sec:7.2.4}

In comparing the material of this section with the literature on MZVs and
their $f$-alphabet description, the reader should be warned about two sources of
mismatching conventions. First, many references including 
\cite{Blumlein:2009cf, Jianqiang, GilFresan} define the
nested sum on the right-hand side of (\ref{mzvsec.1}) to be $\zeta_{n_r,\ldots,n_2,n_1}$ 
instead of $\zeta_{n_1,n_2,\ldots,n_r}$, and our ordering conventions for the arguments
of MZVs agree with those of \cite{Zag:rec, Terasoma, Brown:2009qja, BrownTate, 
Brown:2011ik, Schlotterer:2012ny, Drummond:2013vz, Schnetz:2013hqa, Brown:2013gia, 
Stieberger:2013wea, Stieberger:2014hba, MadridOS, MadridSS}. Second, our conventions 
for the motivic coaction are those of \cite{Duhr:2012fh, Drummond:2013vz, MadridOS} but differ 
from \cite{BrownTate, Brown:2011ik,	Schlotterer:2012ny, Schnetz:2013hqa, Brown:2013gia, 
Stieberger:2013wea, Stieberger:2014hba, GilFresan, MadridSS} by the swap 
$A\otimes B \leftrightarrow B \otimes A$. 

In the above references with opposite conventions for the motivic coaction,
the order of the non-commutative $f_{2k+1}$ will be reversed in comparison
to the expressions in this work. In particular, the coaction of the commutative
$f_2$ becomes $1\otimes f_2$ instead of $f_2\otimes 1$ in translating to those
references. Moreover, the single-valued map of $f_2^n f_{i_1} f_{i_2}\ldots f_{i_r}$
 becomes $ \delta_{n,0} \sum_{j=0}^r f_{i_1} f_{i_2}\ldots f_{i_j}
\shuffle f_{i_r} f_{i_{r-1}} \ldots f_{i_{j+1}}$ in the place of (\ref{mzvsec.17}) 
when changing the conventions to $A\otimes B \rightarrow B \otimes A$.

For instance, the (conjectural) $\mathbb Q$-bases of MZVs employed in the datamine
\cite{Blumlein:2009cf} are related to those of this work and \cite{Schlotterer:2012ny}
by reversing $\zeta_{n_1,n_2,\ldots,n_r} \rightarrow \zeta_{n_r,\ldots,n_2,n_1}$.
However, in order to import the $\phi$-images at weight $\leq 16$ from \cite{Schlotterer:2012ny}
into our present conventions, the order of the non-commutative letters $f_{2k+1}$ 
requires a separate reversal, i.e.\ $f_{i_1} f_{i_2} \ldots f_{i_r} \rightarrow  f_{i_r} \ldots  f_{i_2}  f_{i_1} $ 
with $i_j \in 2\mathbb N{+}1$.

\subsection{Patterns in the $\alpha'$-expansion}
\label{sec:7.3}

In this section, we review the structure of the $\alpha'$-expansion of the
$n$-point disk integrals $F_P{}^Q$ in (\ref{Fdef}) which has been firstly
described in \cite{Schlotterer:2012ny}. First of all, the MZVs contributing 
to the order of $\alpha'^w$ have total weight $w$, i.e.\ the $\alpha'$-expansion of $F_P{}^Q$
is said to enjoy uniform transcendentality, see section \ref{sec:7.4} for a proof.
 
Once the MZVs at given weight $w$ are organized in the conjectural $\mathbb Q$-bases
of table \ref{QQbases}, the coefficients of the Riemann zeta values $\zeta_w$ 
are claimed to determine all other coefficients, say those of indecomposable
higher-depth zetas $\zeta_{3,5}$ or products such as $\zeta_{a}\zeta_b $ or $\zeta_a \zeta_{b,c}$.
These intriguing patterns are checked for a variety of weights \& multiplicities and 
most conveniently described in the $f$-alphabet of section \ref{sec:7.2.1}.
They imply a remarkably simple formula for the coaction of the integrals $F_P{}^Q$ 
\cite{Drummond:2013vz} that resonates with recent studies of 
Feynman integrals \cite{Abreu:2017enx, Abreu:2017mtm, Abreu:2019wzk, Abreu:2021vhb} 
and Lauricella hypergeometric functions \cite{Brown:2019jng}.

\subsubsection{The pattern in terms of MZVs}
\label{sec:7.3.1}

For the four-point instance of the disk integrals $F_P{}^Q$ in (\ref{Fdef}),
the $\alpha'$-expansion is given in closed form by the exponential in (\ref{4ptmzv.1}).
This expression manifests that the coefficients,
\beq
M_{2k+1} \, \big|_{n=4} = (2\ap)^{2k+1} \frac{s_{12}^{2k+1}{+}s_{23}^{2k+1}{+}s_{13}^{2k+1}}{2k{+}1}
\, , \ \ \ \ \ \
P_{2k}  \, \big|_{n=4} =(2\ap)^{2k} \frac{ \zeta_{2k} (
s_{12}^{2k}{+}s_{23}^{2k}{-}s_{13}^{2k}) }{2k(\zeta_2)^k}\,,
 \label{mzvsec.31}
\eeq
of $\zeta_{2k+1}$ and $\zeta_{2k}$ determine those of products $\zeta_{a_1} \zeta_{a_2}\ldots$
by expanding the exponential. In order to generalize this observation to $n\geq 5$ points,
we shall consider the matrix-valued coefficients of Riemann zeta values
\beq
(M_{2k+1})_P{}^Q = F_P{}^Q \, \big|_{\zeta_{2k+1}} \, , \ \ \ \ \ \
(P_{2k})_P{}^Q = F_P{}^Q \, \big|_{\zeta_{2}^k}\, ,
 \label{mzvsec.32}
\eeq
where the entries of the $(n{-}3)! \times (n{-}3)!$ matrices $P_{w}$ and $M_{w}$
are homogeneous degree-$w$ polynomials in $2\alpha' s_{ij}$ with rational coefficients.
A variety of examples at $n=5,6,7$ is available for download from \cite{MZVWebsite},
where the conventions in this review are matched after
rescaling $s_{ij} \rightarrow - 2\alpha' s_{ij}$ in the dataset of the website.
At the leading orders in $\alpha'$, the expansion of $n$-point disk integrals
is found to exhibit the following multiplicity-agnostic pattern \cite{Schlotterer:2012ny}
\begin{align}
F &= \mathds{1} + \zeta_2 P_2 + \zeta_3 M_3 + \zeta_2^2 P_4 + \zeta_5 M_5  + \zeta_2 \zeta_3 P_2 M_3  \label{mzvsec.33}\\
&\quad + \zeta_2^3 P_6 + \frac{1}{2} \zeta_3^2 M_3^2 + \zeta_7 M_7 + \zeta_2 \zeta_5 P_2 M_5 
+ \zeta_2^2 \zeta_3 P_4 M_3 + {\cal O}(\ap^8)\,, \notag
\end{align}
where we have suppressed the row and column indices of the $F_P{}^Q$ in (\ref{Fdef}). 
The coefficients $P_w,M_w$ of $\zeta_w$ defined in (\ref{mzvsec.32}) turn out to determine
those of $\zeta_2 \zeta_3$ or $\zeta_3^2$ via matrix products $P_2 M_3$ or $M_3^2$.
In other words, there is only one piece of independent information $P_w$ or $M_w$ at 
each order $\alpha'^{w\leq 7}$ in (\ref{mzvsec.33}).

Starting from weights $w=8,10,11,\ldots$, the $\mathbb Q$-bases of MZVs are believed to contain
indecomposable elements of depth $\geq 2$ which can be chosen as $\zeta_{3,5}, \zeta_{3,7},
\zeta_{3,3,5},\ldots$ \cite{Blumlein:2009cf}. In the
conjectural bases of table~\ref{QQbases}, the simplest instances of depth-two 
and depth-three MZVs are accompanied by the following matrix commutators \cite{Schlotterer:2012ny}
\begin{align}
F  \, \big|_{(\ap)^8} &= \zeta_2^4 P_8 + \frac{1}{2} \zeta_2 \zeta_3^2 P_2 M_3^2 + \zeta_3 \zeta_5 M_5 M_3 + \frac{1}{5} \zeta_{3,5} [M_5,M_3] \, ,\notag \\
F  \, \big|_{(\ap)^9} &= \zeta_9 M_9 +  \zeta_2 \zeta_7 P_2 M_7 
+  \zeta_2^2 \zeta_5 P_4 M_5
+  \zeta_2^3 \zeta_3 P_6 M_3
+ \frac{1}{6} \zeta_3^3 M_3^3\, ,\notag \\
F  \, \big|_{(\ap)^{10}} &= \zeta_2^5 P_{10}
+ \zeta_2 \zeta_3 \zeta_5 P_2 M_5 M_3
+ \frac{1}{5} \zeta_{2} \zeta_{3,5} P_2 [M_5 , M_3]
+ \frac{1}{2} \zeta_2^2 \zeta_3^2 P_4 M_3^2 \notag \\
&\quad
+ \frac{1}{2} \zeta_5^2 M_5^2
+ \zeta_3 \zeta_7 M_7 M_3
+ \bigg( \frac{ 1}{14} \zeta_{3,7} + \frac{3}{14} \zeta_5^2 \bigg) [M_7, M_3]
\, , \label{mzvsec.34}\\
F  \, \big|_{(\ap)^{11}} &= \zeta_{11} M_{11} + \zeta_2 \zeta_9 P_2 M_9 + \zeta_2^2 \zeta_7 P_4 M_7 + \zeta_2^3 \zeta_5 P_6 M_5 + \zeta_2^4 \zeta_3 P_8 M_3 + \frac{1}{6} \zeta_2 \zeta_3^3 P_2 M_3^3
\notag \\
&\quad  + \frac{1}{2} \zeta_{3}^2 \zeta_5 M_5 M_3^2
+\frac{1}{5} \zeta_{3,5} \zeta_3 [M_5, M_3] M_3 + \bigg( \frac{1}{5} \zeta_{3,3,5}
+ 9 \zeta_2 \zeta_9 + \frac{6}{25} \zeta_2^2 \zeta_7 - \frac{4}{35} \zeta_2^3 \zeta_5 \bigg) [M_3, [M_5,M_3]]\, .
\notag
\end{align}
These expressions are consistent with the dropout of $\zeta_{3,5},\zeta_{3,7}$ and $\zeta_{3,3,5}$
at four points, see (\ref{4ptmzv.1}), since the $(n{-}3)! \times (n{-}3)!$ matrices $M_{2k+1}$ 
then reduce to the scalars (\ref{mzvsec.31}) with vanishing commutators. 
However, rational prefactors such as $\frac{1}{5}$ of $\zeta_{3,5} [M_5,M_3]$
or $- \frac{4}{35}$ of $\zeta_2^3 \zeta_5 [M_3, [M_5,M_3]]$ may appear surprising 
at first glance. As we will see in the next section, these rational numbers conspire to
unit coefficients once the MZVs in (\ref{mzvsec.34}) are taken to be motivic ones
and mapped into the $f$-alphabet reviewed in section \ref{sec:7.2.1}.

\subsubsection{The pattern in the $f$ alphabet}
\label{sec:7.3.2}

In preparation for a well-defined map into the $f$-alphabet, we promote the MZVs in the
$\alpha'$-expansion of the matrix $F$ to their motivic versions,
\beq
(F^{\mathfrak m})_P{}^Q = F_P{}^Q \, \big|_{ \zeta_{n_1,\ldots,n_r} \rightarrow  \zeta^{\mathfrak m}_{n_1,\ldots,n_r} }\, .
 \label{mzvsec.35}
\eeq
The image of (\ref{mzvsec.34}) in the $f$-alphabet can be assembled
from the action (\ref{mzvsec.4}), (\ref{mzvsec.6}) and (\ref{mzvsec.7}) of the $\phi$-isomorphism
on $\zeta^{\mathfrak m}_w, \zeta_{3,5}^{\mathfrak m},\zeta^{\mathfrak m}_{3,3,5}$ and products
thereof,
\begin{align}
\phi(F^{\mathfrak m})  \, \big|_{(\ap)^8} &= f_2^4 P_8 +  f_2f_3f_3 P_2 M_3^2 + f_3 f_5 M_3 M_5+f_5 f_3 M_5 M_3 \, , \notag \\
\phi(F^{\mathfrak m})  \, \big|_{(\ap)^9} &= f_9 M_9 +  f_2f_7 P_2 M_7 + f_2^2 f_5 P_4 M_5
+f_2^3 f_3 P_6 M_3+f_3 f_3 f_3  M_3^3 \, , \notag \\
\phi(F^{\mathfrak m})  \, \big|_{(\ap)^{10}} &= f_2^5 P_{10} 
+ f_2 f_3 f_5 P_2 M_3 M_5+ f_2 f_5 f_3 P_2 M_5 M_3 
+  f_2^2 f_3f_3 P_4 M_3^2  \label{mzvsec.36}  \\
&\quad + f_5 f_5 M_5^2
+f_3 f_7 M_3 M_7+f_7 f_3 M_7 M_3 \, , \notag \\
\phi(F^{\mathfrak m})  \, \big|_{(\ap)^{11}} &= f_{11} M_{11} + f_2 f_9 P_2 M_9 + f_2^2 f_7 P_4 M_7 + f_2^3 f_5 P_6 M_5 + f_2^4 f_3 P_8 M_3 + f_2 f_3 f_3 f_3 P_2 M_3^3
\notag \\
&\quad  + f_{3} f_3 f_5  M_3^2 M_5 + f_{3} f_5 f_3 M_3 M_5 M_3 + f_{5} f_3 f_3 M_5 M_3^2\, .
\notag
\end{align}
Each word in the non-commutative generators $f_{2k+1}$ is accompanied by a 
matrix product of $M_{2k+1}$ with a matching multiplication order, and powers of the commutative
$\phi$-image $f_2$ of $\zeta^{\mathfrak m}_2$ occur with left-multiplicative $f_2^kP_{2k}$.
Moreover, all the unwieldy rational prefactors of $\frac{1}{5}$ or $-\frac{4}{35}$
in (\ref{mzvsec.34}) have conspired to unit coefficients in passing to (\ref{mzvsec.36})!
By reinstating the lower-weight results (\ref{mzvsec.33}) in the $f$-alphabet\footnote{At the
orders of $\alpha'^{w\leq 7}$, the rational prefactors in (\ref{mzvsec.33})
remain unchanged in passing to $\phi(F^{\mathfrak m})$ apart from
$\phi( (\zeta^{\mathfrak m}_3)^2 )= f_3 \shuffle f_3 = 2 f_3 f_3$. Similarly, contributions
of $\frac{1}{n!} (\zeta^{\mathfrak m}_{2k+1}M_{2k+1})^n$ at higher orders that resemble 
the expansion of a matrix-valued exponential are mapped to $n$-fold concatenation 
products $f_{2k+1} f_{2k+1}\ldots f_{2k+1} M_{2k+1}^{n}$ under $\phi$.}, the 
orders of $\alpha'^{\leq 11}$
can be reconstructed from
\begin{align}
\phi(F^{\mathfrak m})  &= \big(\mathds{1} + f_2 P_2 + f_2^2 P_4 + f_2^3 P_6 + f_2^4 P_8 + f_2^5 P_{10} \big)
\notag \\
&\quad \times \big( \mathds{1} + f_3 M_3 + f_5 M_5 + f_3 f_3 M_3 M_3 + f_7 M_7 
+ f_3 f_5 M_3 M_5 + f_5 f_3 M_5 M_3  \label{mzvsec.37} \\
&\quad \quad
+ f_9 M_9 + f_3 f_3 f_3 M_3 M_3 M_3
+ f_5 f_5 M_5 M_5 + f_3 f_7 M_3 M_7 + f_7 f_3 M_7 M_3  \notag \\
&\quad \quad + f_{11} M_{11}
+ f_{3} f_3 f_5  M_3^2 M_5 + f_{3} f_5 f_3 M_3 M_5 M_3 + f_{5} f_3 f_3 M_5 M_3^2\big)
+ {\cal O}(\ap^{12}) \, .
 \notag
\end{align}
This suggests the following all-order formula for the $\alpha'$-expansion of
$n$-point disk integrals \cite{Schlotterer:2012ny},
\begin{align}
\phi(F^{\mathfrak m})  &= \bigg( \sum_{k=0}^\infty f_2^k P_{2k} \bigg)
\sum_{r=0}^{\infty} \sum_{i_1,i_2,\ldots,i_r \in 2\mathbb N+1} f_{i_1}f_{i_2}\ldots f_{i_r}
M_{i_1} M_{i_2}\ldots M_{i_r}  \notag \\
&=  \bigg( \sum_{k=0}^\infty f_2^k P_{2k} \bigg) \frac{1}{1- \sum_{n=1}^\infty f_{2n+1} M_{2n+1} }\, ,
 \label{mzvsec.38}
\end{align}
where the fraction in the last line is understood as a geometric series
$\frac{1}{1-x} = \sum_{m=0}^\infty x^m$. The only independent pieces
of information in (\ref{mzvsec.38}) are the matrices $M_{2n+1}$ and $P_{2k}$
along with $f_{2n+1}$ and $f_2^k$. The coefficients of any product $f_2^k f_{2n+1}$
or higher-depth term $f_{2n_1+1} f_{2n_2+1}\ldots$ are determined by (\ref{mzvsec.38})
in terms of matrix multiplications among the $P_{2k}$ and $M_{2n+1}$.

At multiplicities $n=5,6,7$, (\ref{mzvsec.38}) has been checked up to and
including weight $21,9,7$ \cite{Zfunctions} and is conjectural beyond this.

\subsubsection{Coaction}
\label{sec:7.3.3}

The coefficients $P_w,M_w$ of $\zeta_w$ in (\ref{mzvsec.32}) are defined w.r.t.\ a prescribed 
$\mathbb Q$-basis of MZVs at weight $w$. At weights $w\leq 7$, we have employed the
unique bases in (\ref{mzvsec.33}) that are expressible in terms of Riemann zeta values.
Starting from weight $w=8$, however, the choices of basis elements $\zeta_{3,5},\zeta_{3,7},
\zeta_{3,3,5},\ldots$ beyond depth one in table \ref{QQbases} is somewhat arbitrary and
leaves various equally natural alternatives. One could for instance
change the basis to include $\zeta_{5,3} = \zeta_3 \zeta_5 - \zeta_{3,5} - \zeta_8$
instead of $\zeta_{3,5}$ which would add commutator terms $\sim [M_3,M_5]$ to the 
coefficient $P_8$ of $\zeta_2^4 = \frac{ 175}{24} \zeta_8$ in the new basis. 
Similarly, trading $\zeta_{3,3,5}$ for a different basis elements at depth $\geq 3$ 
leads to a shift of $M_{11}$ by a rational multiple of $[M_3, [M_5,M_3]]$.

We shall illustrate the basis dependence of $M_{11}$ by rewriting the $\alpha'^{11}$-order
in (\ref{mzvsec.34}) in terms of $\zeta_{3,5,3}= - 2 \zeta_{3,3,5} + \frac{299}{2} \zeta_{11}
+ \zeta_{3} \zeta_{3,5} + \frac{8}{7} \zeta_2^3 \zeta_5 - \frac{12}{5} \zeta_2^2 \zeta_7
- 90 \zeta_2 \zeta_9$
rather than $\zeta_{3,3,5}$,
\begin{align}
F  \, \big|_{(\ap)^{11}} &= \zeta_{11} \bigg( M_{11} + \frac{299}{20} [M_3,[M_5,M_3]] \bigg)
+ \zeta_2 \zeta_9 P_2 M_9 + \zeta_2^2 \zeta_7 P_4 M_7 + \zeta_2^3 \zeta_5 P_6 M_5 + \zeta_2^4 \zeta_3 P_8 M_3 
\notag\\
&\quad  
+ \frac{1}{6} \zeta_2 \zeta_3^3 P_2 M_3^3
+ \frac{1}{2} \zeta_{3}^2 \zeta_5 M_5 M_3^2
+\frac{1}{10} \zeta_{3,5} \zeta_3 [M_5, M_3^2] 
- \frac{1}{10} \zeta_{3,5,3} [M_3, [M_5,M_3]]\, .
\label{neww11} 
\end{align}
The coefficient of $\zeta_{11}$ became $M_{11} + \frac{299}{20} [M_3,[M_5,M_3]]$ 
in the place of $M_{11}$ in (\ref{mzvsec.34}). Hence, the definition (\ref{mzvsec.32})
of $M_{2k+1}$ requires the specification of a (conjectural) $\mathbb Q$-basis of
MZVs at weight $2k{+}1$, and we will follow the choices of the 
datamine \cite{Blumlein:2009cf} as done in \cite{Schlotterer:2012ny}.
It would be interesting if alternative choices of basis MZVs at higher weight
may lead to similar shortenings as seen in the more compact form
(\ref{neww11}) of the $\alpha'^{11}$ order with $\zeta_{3,5,3}$ 
in the place of $\zeta_{3,3,5}$ in (\ref{mzvsec.34}).

These ambiguities in the definition of $P_{\geq 8}$ and $M_{\geq 11}$
are also reflected by the freedom to add $f_8 = \frac{24}{175} f_2^4$
to $\phi( \zeta^{\mathfrak m}_{3,5})$ 
in (\ref{mzvsec.7}) and more generally $f_w$ to the $\phi$-image
of indecomposable weight-$w$ MZVs of depth $\geq 2$ in a $\mathbb Q$-basis.
In other words, the isomorphism $\phi$ is non-canonical starting from weight 8.
Nevertheless, the form of the all-weight result (\ref{mzvsec.38}) is unaffected by
the above choices.

The information of the all-order result (\ref{mzvsec.38}) can be encoded in the
following coaction formula without any reference to basis dependent quantities $P_w,M_w$
\cite{Drummond:2013vz},
\beq
\Delta ( F^{\mathfrak m}) _P{}^Q = \sum_{R \in S_{n-3}} (F^{\mathfrak m})_P{}^R \otimes (F^{\mathfrak dr})_R{}^Q\, .
 \label{mzvsec.40}
\eeq
Following the coaction of MZVs in \cite{Goncharov:2005sla}, the
MZVs in the second entry $F^{\mathfrak dr}$ are promoted to deRham periods
$\zeta^{\mathfrak m}_{n_1,n_2,\ldots,n_r} \rightarrow \zeta^{\mathfrak dr}_{n_1,n_2,\ldots,n_r}$
(see for instance \cite{Francislecture})
with a net effect of modding out by $\zeta_2$ since $\zeta^{\mathfrak dr}_{2} =0$.
One can straightforwardly verify (\ref{mzvsec.40}) by insertion of (\ref{mzvsec.38}) and
using the simple form (\ref{mzvsec.5}) of the deconcatenation coaction in the $f$-alphabet.

Based on $Z(1,P,n{-}1,n|1,Q,n,n{-}1) = -  \sum_{R \in S_{n-3}} S^{-1}(Q|R)_1
F_P{}^R$ with the KLT matrix $S(A|B)_1$ in (\ref{kltrec}), the 
coaction formula (\ref{mzvsec.40}) can be readily translated to
motivic and deRham versions $Z^{\mathfrak m}, Z^{\mathfrak dr}$ of
the $Z$-integrals (\ref{Zintdef}). In the first place, one arrives at the coaction 
of the $(n{-}3)! \times (n{-}3)!$ basis of $Z(1,P,n{-}1,n|1,Q,n,n{-}1)$, but
one can generalize to arbitrary $A,B \in S_n$ in
\beq
\Delta Z^{\mathfrak m}(A|B) = - \sum_{P,Q \in S_{n-3}} 
Z^{\mathfrak m}(A|1,P,n,n{-}1) S(P|Q)_1 \otimes 
Z^{\mathfrak dr}(1,Q,n{-}1,n|B)
 \label{mzvsec.41}
\eeq
by noting that both sides of the equation obey the same monodromy relations in $A$ and
IBP relations in $B$.

Note that (\ref{mzvsec.40}) and (\ref{mzvsec.41}) are special cases of more 
general coaction formulae
\beq
\Delta \bigg( \int_\gamma \omega \bigg)^{\mathfrak m} =  \sum_{j=1}^d
\bigg( \int_\gamma \omega_j \bigg)^{\mathfrak m} \otimes
\bigg( \int_{\gamma_j} \omega \bigg)^{\mathfrak dr} 
\label{mzvsec.42}
\eeq
for wider classes of integration cycles $\gamma$ and differential forms $\omega$
that were studied in the context of Feynman integrals 
\cite{Abreu:2017enx, Abreu:2017mtm, Abreu:2019wzk, Abreu:2021vhb} 
and Lauricella hypergeometric functions \cite{Brown:2019jng}. The sum over $j$ in 
(\ref{mzvsec.42}) runs over $d$-dimensional bases of twisted homologies $\{ \gamma_j\}$
and cohomologies $\{\omega_j\}$, respectively. Moreover,
these bases are understood to be chosen as orthonormal in
the sense that the zero-transcendentality part 
of $\int_{\gamma_i}\omega_j$ is given by $\delta_{ij}$. In our setting, the orthonormality 
condition is met by the Kronecker delta in the field-theory limit (\ref{ftofFPQ}) 
of $F_P{}^Q$. The superscripts ${\mathfrak m}$ and ${\mathfrak dr} $
in (\ref{mzvsec.42}) again refer to the motivic and deRham periods
with $\zeta^{\mathfrak dr}_{2} =0$ in the second entry.

\subsection{KK-like and BCJ relations within the $\alpha'$-expansion}
\label{sec:7.3.4}

We have seen in section \ref{sec:7.7} that the monodromy relations obeyed by color-ordered
open-string amplitudes deform the KK and BCJ relations of field-theory amplitudes
by trigonometric functions in $\alpha' s_{ij}$. It will now be shown that certain sectors in
the $\alpha'$-expansions of disk integrals in section \ref{sec:7.3} preserve the field-theory BCJ
relations of the SYM amplitudes they multiply. Other sectors in the $\alpha'$-expansion of 
open-superstring amplitudes will be reviewed to obey analogues of KK relations 
where the coefficients are still integers independent on $s_{ij}$.

\subsubsection{BCJ relations at all orders in $\alpha'$}
\label{sec:7.x.y}

The organization (\ref{mzvsec.38}) of the $\alpha'$-expansion of $F_P{}^Q$ can be 
used to generate solutions of the BCJ relations (\ref{BCJrelations}) or (\ref{fundBCJ}) 
at arbitrary mass dimensions. This can be seen by inserting permutations of
the string-amplitude formula (\ref{npttree}) into the monodromy relations (\ref{sepmono})
and expanding in $\alpha'$. It is crucial to note that the trigonometric factors yield
series in even zeta values (\ref{mzvsec.2}), i.e.\ exclusively the commutative
generator $f_2$ in the $f$-alphabet upon passing to motivic MZVs and taking the $\phi$-image,
\beq
\sin(\pi x)  = \pi x \exp\bigg( {-} \sum_{k=1}^\infty \frac{ \zeta_{2k} }{k} x^{2k} \bigg)\, .
\label{mzvsec.43}
\eeq
Hence, the only departures of the monodromy relations from the BCJ relations 
occur for non-zero powers of $f_2$ -- the appearance of the odd generators
$f_{2k+1}$ is unaffected by the sine functions.

In order to identify independent solutions of the BCJ relations, we impose
(the motivic version of) the monodromy relations to hold separately for the coefficient of any 
$f_2^k f_{i_1} f_{i_2}\ldots f_{i_r}$ with $k,r\in \mathbb N_0$ and $i_j \in 2\mathbb N{+}1$.
The separation of different transcendentality structures has been firstly 
applied in \cite{Broedel:2012rc} to demonstrate the KK and BCJ
relations of the tree-level matrix elements of the $\alpha'^3\zeta_3 {\Tr} \{ D^2\mathbb F^4
+\mathbb F^5\}$ operator in the superstring effective action and the $\alpha'{\Tr}\{\mathbb F^3\}$
operator of the open bosonic string. By focusing on the $f_2 \rightarrow 0$ part
of the monodromy relations, the coefficient of any $f_{i_1} f_{i_2}\ldots f_{i_r}$ in
color-ordered open-string amplitude is found to obey KK and BCJ relations,
\begin{align}
0 &= {\cal A}_{i_1,i_2,\ldots,i_r}(P \shuffle Q,n) \ , \ \ \ \ P,Q\neq \emptyset\, ,
\label{mzvsec.44} \\
0 &= \sum_{j=2}^{n-1}  (k_{p_1} \cdot k_{p_2 p_3 \ldots p_j})
{\cal A}_{i_1,i_2,\ldots,i_r}(p_2p_3\ldots p_{j}p_1 p_{j+1} \ldots p_n) \, ,
\notag
\end{align}
where ${\cal A}_{i_1,i_2,\ldots,i_r}(P)$ is a shorthand for the coefficient
of $f_{i_1} f_{i_2}\ldots f_{i_r}$ in the $\phi$-image of the motivic versions 
${\cal A}^{\mathfrak{m}}$ of superstring amplitudes ${\cal A}$, i.e.
\begin{align}
{\cal A}_{i_1,i_2,\ldots,i_r}(1,Q,n{-}1,n)
&:= \phi\big( {\cal A}^{\mathfrak{m}}(1,Q,n{-}1,n) \big) \, \big|_{f_{i_1} f_{i_2}\ldots f_{i_r}}  \label{mzvsec.45}
\\
&\phantom{:}= \sum_{R \in S_{n-3}} 
(M_{i_1} M_{i_2}\ldots M_{i_r})_Q{}^R A(1,R,n{-}1,n)\, .
\notag
\end{align}
By isolating the coefficients of $f_{i_1} f_{i_2}\ldots f_{i_r}$ in last three lines 
of (\ref{mzvsec.37}), we for instance arrive at the
simplest independent solutions to the BCJ relations at the orders of $\alpha'^{\leq 11}$ 
in table \ref{countBCJ}. The table only tracks the solutions of BCJ relations
that are realized in the $\alpha'$-expansion of superstring disk amplitudes -- a variety
of further solutions multilinear in polarization vectors can be systematically 
generated from tree-level amplitudes of bosonic or heterotic strings 
\cite{Huang:2016tag, Azevedo:2018dgo} or from building blocks of loop-level
string amplitudes \cite{GreenBZA}.
Similarly, as will be detailed in section \ref{sec:7.x.z}, the $\alpha'$-expansion of 
$Z(P|Q)$ integrals can be used to generate rational functions in $s_{ij}$
at various mass dimensions that obey BCJ relations in $Q$.

\begin{table}[h]
	\begin{center}
		\begin{tabular}{|c | c  || c  |c  |}\hline 
			$w$ &\#({\rm solutions}) &$w$ &\#({\rm solutions}) \\\hline \hline
			 0 &$A={\cal A}_{\emptyset}$   &6 &${\cal A}_{3,3}$\\ \hline
			 1 &$\times $  &7 &${\cal A}_{7}$\\ \hline
			 2 &$\times$   &8 &${\cal A}_{3,5}, \ {\cal A}_{5,3}$\\ \hline
			 3 &${\cal A}_{3}$   &9 &${\cal A}_{9}, \ {\cal A}_{3,3,3}$\\ \hline
			 4 &$\times$   &10 &${\cal A}_{3,7}, \ {\cal A}_{7,3}, \ {\cal A}_{5,5}$\\ \hline
			 5 &${\cal A}_{5}$   &11 &${\cal A}_{11},\ {\cal A}_{3,3,5},\ {\cal A}_{3,5,3},\ {\cal A}_{5,3,3}$\\ \hline
		\end{tabular}
	\end{center}
\caption{The solutions ${\cal A}_{i_1,i_2,\ldots,i_r}$ of BCJ relations at the order
of $\alpha'^{\leq 11}$ that can be read off from the coefficient of $f_{i_1}f_{i_2}\ldots f_{i_r}$ 
in the $\alpha'$-expansion of open-superstring amplitudes.}
\label{countBCJ}
\end{table}

Note that permutations ${\cal A}_{i_1,i_2,\ldots,i_r}(P)$ outside the 
$(n{-}3)!$-element basis of  ${\cal A}_{i_1,i_2,\ldots,i_r}(1,Q,n{-}1,n)$ in (\ref{mzvsec.45}) 
can be expanded via
\beq\label{againbcj.41}
{\cal A}_{i_1,i_2,\ldots,i_r}(P) = - \sum_{Q,R \in S_{n-3}} m(P|1,R,n,n{-}1) S(R|Q)_1 
{\cal A}_{i_1,i_2,\ldots,i_r}(1,Q,n{-}1,n)\,,
\eeq
i.e.\ by adapting the solutions (\ref{revbcj.41}) of BCJ relations 
to ${\cal A}_{i_1,i_2,\ldots,i_r}(P)$ in the place of SYM amplitudes.

\subsubsection{KK-like relations}
\label{KKlikesec}

Inspired by the Kleiss--Kuijf (KK) relations \eqref{KKrelation1} among tree-level
amplitudes in field theories, we shall now investigate the $\alpha'$-expansion of
disk amplitudes in (\ref{mzvsec.38}) for identities with constant coefficients. 
More precisely, we shall go beyond the coefficients (\ref{mzvsec.45}) of 
the $f_B= f_{b_1} f_{b_2}\ldots f_{b_r}$ with odd $b_j$ and identify KK-like 
relations among
\begin{align}
{\cal A}_{\ell | B}(1,Q,n{-}1,n)
&=  \phi\big( {\cal A}^{\mathfrak{m}}(1,Q,n{-}1,n) \big) \, \big|_{f_2^\ell f_{b_1} f_{b_2}\ldots f_{b_r}} \label{kklike.01} \\
&\phantom{:}= \sum_{R \in S_{n-3}} 
(P_{2\ell} M_{b_1} M_{b_2}\ldots M_{b_r})_Q{}^R A(1,R,n{-}1,n)
\notag
\end{align}
associated with arbitrary powers $\ell\geq 0$ of $f_2$, following version 3 of \cite{cdescent}. For 
instance, the well-known cyclic and reflection properties
of disk amplitudes hold separately along with each $f_2^\ell f_B$,
\begin{align}
\label{simpsyms}
\cA_{\ell | B}(1,2, \ldots,n) &= \cA_{\ell | B}(2,3, \ldots,n,1) \, , \\
\cA_{\ell | B}(1,2, \ldots,n) &=(-1)^n\cA_{\ell | B}(n,n{-}1, \ldots,2,1)\,,\notag
\end{align}
leaving at most $\frac{1}{2}(n{-}1)!$ independent permutations.
However, the coefficients (\ref{kklike.01}) of $f_2^\ell f_B$ at different values of $\ell$
obey different additional relations, so we will focus on the individual 
components. To this effect, following \cite{KKlikeFTpaper}, we will 
refer to relations of the form
\beq\label{ftkkrel}
\sum_\s c_\s {\cal A}_{\ell | B}(\s) = 0 \, ,
\eeq
with constant coefficients $c_\s \in \mathbb Q$ as  {\it KK-like} and study them separately
at each $\ell \geq 0$. For simple examples of KK-like relations, we have the permutation symmetry
${\cal A}_{1 | B}(1,2,3,4)= {\cal A}_{1 | B}(1,2,4,3)$ as well as the six-term identity~\cite{1loopbb}
\beq\label{threeperm}
{\cal A}_{1 | B}(1,2,3,4,5) + {\rm perm}(2,4,5) = 0
\eeq
universal to the coefficients of $f_2 f_B$.
These KK-like relations of ${\cal A}_{1 | B}$ clearly differ from the KK relations
(\ref{mzvsec.44}) of ${\cal A}_{0 | B}$ such as the three-term identity 
${\cal A}_{0 | B}(1,2,3,4) + {\rm cyc}(2,3,4)=0$ at four points
or the four-term identity ${\cal A}_{0 | B}(1,2\shuffle 345) =0$ at five points.

Note that the simplest examples of ${\cal A}_{1 | B}$ up to and including the
order of $\alpha'^{10}$ do not require an $f$-alphabet description and can be
equivalently obtained from the coefficient of $\zeta_2$ or products
$\zeta_2 \zeta_{2k+1}$, $\zeta_2 \zeta_{2k_1+1} \zeta_{2k_2+1}$,
$\zeta_2 \zeta_{3,5}$ in (\ref{mzvsec.33}) and (\ref{mzvsec.34}). Starting
from the order of $\alpha'^{11}$ with the MZV basis choice of \eqref{mzvsec.34}, the coefficients of $\zeta_2 \zeta_{2k+1}$
in the matrix $F$ generically receive admixtures of products $M_{i_1} M_{i_2}\ldots$
with odd $i_j$ on top of the expected $P_2 M_{2k+1}$ in (\ref{kklike.01}). As
illustrated by the coefficient
$F \, |_{ \zeta_2 \zeta_9 }=P_2 M_9 + 9 [M_3, [M_5,M_3]]$ in (\ref{mzvsec.34}), passing
to the $f$-alphabet is necessary to isolate the matrix product
$\phi(F^{\mathfrak{m}}) \, |_{ f_2 f_9 }=P_2 M_9 $ in (\ref{mzvsec.36}).
The five-point KK-like relation (\ref{threeperm}) only holds if ${\cal A}_{1|9}$ is
constructed from $P_2 M_{2k+1}$ in (\ref{kklike.01}), i.e.\ defined by
${\cal A}_{1|9} = \phi({\cal A}^{\mathfrak{m}})\, |_{f_2 f_9}$, but fails in presence
of extra terms $\sim [M_3, [M_5,M_3]]$ that would arise from 
${\cal A}\, |_{\zeta_2 \zeta_9}$.\footnote{We would like to thank Ricardo Medina 
for email correspondence on this point.} Note, however, that different MZV basis 
choices may push this issue to higher orders $\ap^{>11}$, as evidenced by the 
expansion \eqref{neww11} in which $F \, |_{ \zeta_2 \zeta_9 }=P_2 M_9$.

\subsubsection{Berends--Giele idempotents and BRST-invariant permutations}
\label{BGidempsec}

In order to write down the explicit form of KK-like amplitude relations, we
need to specify a way to generate permutations with the correct properties. As discussed in
\cite{cdescent}, the relevant permutations achieving this are related to the {\it descent algebra} of
permutations via the so-called {\it BRST-invariant permutations}\footnote{The terminology
of ``BRST-invariant permutations'' was coined in \cite{cdescent} by the analogy
of $\gamma_{1|P_1, \ldots,P_k}$ in (\ref{brstgamma})
with certain BRST invariants in the pure spinor computation of one-loop 
amplitudes. However, we are not claiming that BRST transformations act on permutations.}
$\g_{1|P_1,P_2, \ldots,P_k}$ depending on a number $k$ of  words $P_1, \ldots,P_k$
\beq\label{brstgamma}
\gamma_{1|P_1, \ldots,P_k}
= 1\big(\cE(P_1)\shuffle\cE(P_2)\shuffle \ldots \shuffle\cE(P_k)\big)\, ,
\eeq
where $\cE(P)$ is the {\it Berends--Giele
idempotent} defined
in terms of the right-action multiplication $P\circ Q$ of permutations by \cite{cdescent}
\beq\label{BGidemp}
\cE(P) := P\circ\cE_n\,,\quad|P|=n\,,
\eeq
where\footnote{In fact,
$\cE_n$ is given by the inverse
permutations of the Eulerian or Solomon idempotent \cite{solomon1968poincare,reutenauer1986theorem,loday1992serie}.}
\beq\label{kappadef}
\cE_n = \sum_{\s\in S_n}\kappa_{\s^{-1}}\s\,,\qquad
\kappa_\s = {(-1)^{d_\s} \over|\s|{|\s|-1\choose d_\s}}\, ,
\eeq
and $d_\s$ denotes the descent number of the permutation $\s$.
Moreover, they were shown to satisfy the shuffle relations
\beq\label{BGidshuffle}
\cE(R\shuffle S)=0\, , \quad R,S\neq\emptyset\,.
\eeq
This implies that the number of linearly independent BRST-invariant permutations
at $n$ points is given by
\beq\label{nogam}
\#(\g_{1|P_1, \ldots,P_k}) = {n{-} 1\stirling k}\, ,\quad \sum_{i=1}^k \len{P_i} = n{-}1\,,
\eeq
where ${p\stirling q}$ denotes the Stirling cycle numbers \cite{graham1989concrete,knuth1992two}
(traditionally called Stirling numbers of the first kind) that count the number of ways to arrange
$p$ objects into $q$ cycles.
For example,
${5\stirling q}=24, 50, 35, 10, 1$
for $q=1,2,3,4,5$. For example permutations of the above definitions,
see the~\ref{descpermsapp}.

\paragraph{KK-like amplitude relations} The KK-like relations among the component amplitudes
of \eqref{kklike.01} were observed to satisfy the following decomposition according to
the number of parts $k$ in the partitions of $n{-}1$ legs \cite{cdescent}
\begin{align}
{\cal A}_{0|B}(\g_{1|P_1, \ldots,P_k}) &= 0\, ,\quad k\neq1\, ,\notag\\
{\cal A}_{1|B}(\g_{1|P_1, \ldots,P_k}) &= 0\, ,\quad k\neq3\, ,\label{z2one}\\
{\cal A}_{\ell|B}(\g_{1|P_1, \ldots,P_k}) &= 0\, ,\quad k\neq1,3,5, \ldots,2\ell{+}1\, ,\quad \ell\ge2\, .\notag
\end{align}
In addition, it was demonstrated
in \cite{cdescent} that the even cases when $k=2m$ encode the parity and cyclicity relations
\eqref{simpsyms}. In this sense, the KK-like relations for even $k$ are equivalent to
\eqref{simpsyms}.

For example, the case $k=3$ for $n=5$ with $\g_{1|23,4,5}$ given in \eqref{fivegammaEx}
leads to
the 12-term relation after using \eqref{simpsyms}:
\begin{align}
	 &{\cal A}_{0|B}(1,2,3,4,5)
          + {\cal A}_{0|B}(1,2,3,5,4)
          + {\cal A}_{0|B}(1,2,4,3,5)
          + {\cal A}_{0|B}(1,2,4,5,3)\notag\\&
          + {\cal A}_{0|B}(1,2,5,3,4)
          + {\cal A}_{0|B}(1,2,5,4,3)
          - {\cal A}_{0|B}(1,3,2,4,5)
          - {\cal A}_{0|B}(1,3,2,5,4) \label{kklike.02}\\&
          - {\cal A}_{0|B}(1,3,4,2,5)
          - {\cal A}_{0|B}(1,3,5,2,4)
          + {\cal A}_{0|B}(1,4,2,3,5)
          - {\cal A}_{0|B}(1,4,3,2,5)=0\,,\notag
\end{align}
which can be reduced to linear combinations of the KK relations (\ref{mzvsec.44}).

In addition, given that the BRST-invariant permutations constitute a basis for permutations 
in the descent algebra\footnote{This claim
follows from the conjectural relation between the BRST-invariant permutations and the inverse of
the {\it idempotent basis} \cite{garsia1989decomposition} of the descent algebra. See
\cite{cdescent} for more details.},
any other KK-like relation can be written as a linear combination of $\g_{1|P_1, \ldots,P_k}$.
For instance, using the decomposition
\beq\label{lc5}
W_{12345}+{\rm perm}(2,4,5) =
3\g_{1|2345}
          - \half \g_{1|345,2}
          + \half \g_{1|4,235}
+ {1\over8} \g_{1|5,4,3,2} + (4\leftrightarrow5)\,,
\eeq
where a permutation $\s$ is written as $W_\s$ for typographical convenience,
it follows that the KK-like relation \eqref{threeperm} can be rewritten as
\beq
3{\cal A}_{1|B}(\g_{1|2345})
          - \half {\cal A}_{1|B}(\g_{1|345,2})
          + \half {\cal A}_{1|B}(\g_{1|4,235})
+ {1\over8} {\cal A}_{1|B}(\g_{1|5,4,3,2}) + (4\leftrightarrow5) = 0 \,,
\label{kklikeLC}
\eeq
or, equivalently after using the reflection relation \eqref{simpsyms}, $3{\cal A}_{1|B}(\g_{1|2345}) + 3{\cal A}_{1|B}(\g_{1|2354}) = 0$.
Note the crucial absence of $\g_{1|P_1, \ldots,P_k}$ with $k=3$ in the decomposition \eqref{lc5},
which provides a consistency check of \eqref{z2one}.
For another example, one can check that the following 720-term (or 360 after using \eqref{simpsyms})
linear combination vanishes,
${\cal A}_{1|B}(\g_{1|2,3,4,5,67})=0$, in agreement with the second line of \eqref{z2one} with $k=5$.

\paragraph{Basis dimensions} Using the counting \eqref{nogam} one can show that the number of
linearly independent amplitudes under the KK-like relations is given by \cite{cdescent}
\begin{align}
\label{dimbases}
\#\big({\cal A}_{0|B}(1,2, \ldots,n)\big) &={n{-}1\stirling 1}= (n{-}2)!\,,\\
\#\big({\cal A}_{1|B}(1,2, \ldots,n)\big) &= {n{-}1\stirling 3}\,,\notag\\
\#\big({\cal A}_{\ell|B}(1,2, \ldots,n)\big) &= {n{-}1\stirling 1} + {n{-}1\stirling 3}
+ \cdots +{n{-}1\stirling 2\ell{+}1}\,, \ \ \ \ \ell \geq 2 \,,\notag
\end{align}
with the implicit assumption that ${p\stirling q}=0$ for $q{>}p$.
The counting of independent permutations of the amplitudes ${\cal A}_{1|B}(1,2, \ldots,n)$ associated with $f_2 f_B$
yields ${n{-}1\stirling 3}=1,6, 35, 225,\ldots$ at $n=4,5,6,7,\ldots$ and has been
studied in \cite{KKlikeFTpaper, 1loopbb}.

A consistency check on the claim that the cyclicity and reflection symmetries \eqref{simpsyms} are encoded
in the even-$k$ BRST-invariant permutations follows from the counting \eqref{nogam} 
as $\sum_{k\;{\rm even}}^{n-1}\#\big(\g_{1|P_1, \ldots,P_k}\big)=\half(n{-}1)!$. To see this, we note
the elementary identity $\sum_{k\;\rm even}^{n-1}{n-1\stirling k} =
\half(n{-}1)!$ of Stirling cycle numbers. It furthermore follows from \eqref{dimbases} that those
coefficients ${\cal A}_{\ell | B}(1, \ldots,n)$ with $\ell \geq 2$ and $n\le 2\ell{+}3$
obey no additional KK-like relations other than the cyclicity and reflection 
symmetry \eqref{simpsyms}: the 
counting of (\ref{dimbases}) yields $\sum_{k\;\rm odd}^{n-1}{n-1\stirling k} = \frac{1}{2}(n{-}1)!$ 
independent permutations in these cases. For instance, when $\ell=2$ the number of 
linearly independent permutations of ${\cal A}_{\ell |B}(1,2,\ldots,n)$ w.r.t.\ KK-like relations 
is $\frac{1}{2}(n{-}1)!=360$ for $n=7$ but $2519=\frac{1}{2}(n{-}1)!-1$ 
for $n=8$. This last prediction has been confirmed by a brute-force
search using~\cite{FORM}. 

\subsubsection{BCJ and KK relations of $Z$-theory amplitudes}
\label{sec:7.x.z}

The BCJ and KK-like relations in specific sectors of the $\alpha'$-expansion
of string amplitudes can be traced back to analogous relations for the
disk integrals $Z(P|Q)$ in (\ref{Zintdef}). As before, the discussion hinges on
the $f$-alphabet description of the $\alpha'$-expansion and the underlying
motivic MZVs, and we employ the notation
\beq
\mathfrak{Z}(P|Q) =\phi\big(Z^{\mathfrak{m}}(P|Q) \big) \, , \ \ \ \ \ \
\mathfrak{Z}_{\times}(Q) =\phi\big(Z^{\mathfrak{m}}_{\times}(Q) \big) = \sum_{P\in S_{n-1}} \mathfrak{Z}(P,n|Q)
\label{noclutter}
\eeq
for the $\phi$-image of motivic $Z$-theory amplitudes $Z^{\mathfrak{m}}(P|Q)=Z(P|Q) 
\, |_{\zeta_{n_1,\ldots} \rightarrow \zeta^{\mathfrak{m}}_{n_1,\ldots}}$ to avoid 
cluttering.

As shown in section \ref{sec:7.7}, tree-level amplitudes of the NLSM
of Goldstone bosons can be obtained from the symmetrized versions $Z_\times(Q)$
of the $Z(P|Q)$-integrals defined in (\ref{otherDC.04}). The color-ordering
$Q$ of the NLSM amplitude (\ref{otherDC.05}) is encoded in the
Parke--Taylor integrand $\PT(Q)$ of the symmetrized integral $Z_\times(Q)$.
Hence, KK and BCJ relations of the NLSM are a simple consequence of
partial-fraction and IBP relations of Parke--Taylor integrals, see section \ref{sec:6.3.3}.

This worldsheet derivation of amplitude relations of the NLSM is actually
not tied to the low-energy limit in (\ref{otherDC.05}) since the KK and BCJ relations
of $Z(P|Q)$ w.r.t.\ the Parke--Taylor orderings $Q$ are valid at all orders in
$\alpha'$. In particular, KK and BCJ relations apply to every $\mathbb Q$-independent 
combination of MZVs in the $\alpha'$-expansion of abelian 
$Z$-integrals. After peeling off the leading power of $(\pi \alpha')^{n-2}$
exposed by the sine functions in (\ref{otherDC.08}), we expect all combinations
$f_2^\ell f_{i_1}f_{i_2}\ldots$ with $\ell \in \mathbb N_0$ and odd
letters $i_j$ to appear in the $\alpha'$-expansion of $Z_\times(Q)$ at sufficiently large
multiplicity $|Q|$.\footnote{Since 
the four-point $\alpha'$-expansion (\ref{4ptmzv.1}) is expressible in terms of
Riemann zeta values only, the onset of irreducible MZVs $\zeta_{3,5},\zeta_{3,7},
\zeta_{3,3,5},\ldots$ at higher depth is relegated to $Z_\times(Q)$ at
multiplicities $|Q|\geq 6$.}

Any combination of $f_2^\ell f_{i_1} f_{i_2}\ldots$ in the $\alpha'$-expansion of $Z_\times(Q)$ 
can be interpreted as an effective interaction among scalars with one color
degree of freedom that preserves the KK and BCJ relation of the NLSM.
By uniform transcendentality of disk integrals, the $w^{\rm th}$ subleading
order of $\alpha'$ features MZVs of weight $w$ each of which signals 
scalar interactions with $2w$ additional derivatives beyond the NLSM.
The subleading order $\sim \zeta_2(\pi \alpha')^{n-2}$ of $Z_\times(Q)$
for instance defines a four-derivative deformation of the NLSM that
preserves its amplitude relations and can also be described through the
Lagrangian in section 3.3 of \cite{Mizera:2018jbh}. 

More generally, the
$f$-alphabet images $\mathfrak{Z}_\times(Q)$ of symmetrized (motivic) disk 
integrals in (\ref{noclutter}) can be taken as generating functions of scalar
effective-field-theory amplitudes subject to KK and BCJ relations,
\beq
\mathfrak{Z}_\times(A\shuffle B,n)\,\big|_{ f_2^\ell f_{i_1} f_{i_2}\ldots } = 
\mathfrak{Z}_\times( \{ A, B\} ,n)\,\big|_{ f_2^\ell f_{i_1} f_{i_2}\ldots }  = 0
\ \ \ \ \forall \ A,B \neq \emptyset \,, \ \ \ \ \ell \geq 0 \, , \ \ \ \ i_j \in 2\mathbb N{+}1\, .
\label{Zbcj.01}
\eeq
The same type of reasoning applies to the Parke--Taylor orderings $Q$
of {\it non-abelian} $Z$-integrals and their $\phi$-images $\mathfrak{Z}(P|Q)$
in (\ref{noclutter}): each combination of MZVs in
the $\alpha'$-expansion corresponds to effective interactions of
{\it bi-colored} scalars that 
preserve the KK- and BCJ relations of bi-adjoint scalars in $Q$,
\begin{align}
\mathfrak{Z}(P|A\shuffle B,n)\,\big|_{ f_2^\ell f_{i_1} f_{i_2}\ldots } &= 
\mathfrak{Z}(P|  \{ A, B\} ,n)\,\big|_{ f_2^\ell f_{i_1} f_{i_2}\ldots }  = 0
\ \ \ \ \forall \ A,B \neq \emptyset \,, \ \ \ \ \ell \geq 0 \, , \ \ \ \ i_j \in 2\mathbb N{+}1  \, .
\label{Zbcj.02a}
\end{align}
For the amplitude relations where $P$ is varied at fixed $Q$ in turn, the
reasoning in section \ref{sec:7.x.y} implies that only the $f_2\rightarrow 0$
sector of the $\alpha'$-expansion preserves KK- and BCJ relations.
These field-theory relations of $\mathfrak{Z}(P|Q)$ at fixed $Q$ then 
hold independently for the coefficients of any $f_{i_1} f_{i_2}\ldots$ with odd $i_j$,
\begin{align}
\mathfrak{Z}(A\shuffle B,n|Q)\,\big|_{  f_{i_1} f_{i_2}\ldots } &= 
\mathfrak{Z}(  \{ A, B\} ,n|Q )\,\big|_{  f_{i_1} f_{i_2}\ldots }  = 0
\ \ \ \ \forall \ A,B \neq \emptyset \,, \ \ \ \  i_j \in 2\mathbb N{+}1
\, .
\label{Zbcj.02b}
\end{align}
For the coefficient of
$f_2^\ell f_{i_1} f_{i_2}\ldots$ in $\mathfrak{Z}(P|Q)$ at $\ell \geq 1$, we
obtain bi-colored scalar amplitudes subject to the KK-like
relations of section \ref{KKlikesec} in $P$. 

Based on a Berends--Giele recursion for the $\alpha'$-expansion of $Z$-integrals,
a proposal for the non-linear equations of motion of the underlying non-abelian
$Z$-theory can be found in \cite{BGap} and section \ref{sec:7.5}. Explicit results
up to and including the order of $\alpha'^{7}$ are publicly available from
the website \cite{gitrep}.

\subsection{String corrections from the Drinfeld associator}
\label{sec:7.4}

We shall now review a recursive all-multiplicity method to determine the polynomial
structure of the $\alpha'$-expansion of the disk integrals $F_P{}^Q$ in (\ref{Fdef}).
This methods generates all the MZVs from the Drinfeld associator (see section \ref{sec:7.2.2})
whose non-commutative variables $e_0, e_1$ are identified with specific matrices
with entries linear in $\alpha's_{ij}$. More specifically, the recursive step 
in passing from $n{-}1$ to $n$ points \cite{drinfeld},
\beq
F^{\sigma_i} = \sum_{j=1}^{(n-3)!} \big[ \Phi(e_0,e_1) \big]_{ij} \big( F^{\sigma_j} \big|_{k_{n-1}=0} \big) \, ,
\label{drinfd.01}
\eeq
is based on $(n{-}2)! \times (n{-}2)!$ matrices $e_0,e_1$ whose derivation will be described below.
The $F^{\sigma_i} $ are understood to be the $F_P{}^Q$ with $P=23\ldots n{-}2$ the
canonical ordering and $Q$ the $i^{\rm th}$ permutation $\sigma_i$ of $2,3,\ldots,n{-}2$
in lexicographical ordering. The Drinfeld associator $\Phi$ is expanded
in terms of shuffle-regularized MZVs as in (\ref{mzvsec.10}). The soft limit on the right-hand 
side of (\ref{drinfd.01}) acts recursively in the sense that 
\beq
F^{\sigma(23\ldots n{-}2)}  \big|_{k_{n-1}=0} = 
\left\{ \begin{array}{cl}
F^{\sigma(23\ldots n{-}3)} &: \ \sigma(n{-}2) = n{-}2\,, \\
0 &: \ {\rm otherwise} \, ,
\end{array} \right. 
\label{drinfd.02}
\eeq
which terminates with the three-point integral $F^\emptyset=1$.
The relevance of the Drinfeld associator for string amplitudes was firstly pointed out in
\cite{Drummond:2013vz}, among other things by relating its coaction properties with those
of the $F_P{}^Q$ in (\ref{mzvsec.40}). Nevertheless, it is an open problem to deduce
(\ref{mzvsec.40}) from the results of this section. The specific construction towards
the $\alpha'$-expansion of the $F_P{}^Q$ was given in \cite{drinfeld} and is based
on an expansion method for Selberg integrals from the mathematics literature \cite{Terasoma}.
Its description in terms of twisted deRham theory and intersection numbers of twisted
forms can be found in \cite{Kaderli:2019dny}, where $e_0,e_1$ are identified as
braid matrices.

\subsubsection{Construction of the matrices $e_0,e_1$}
\label{sec:7.4.1}

The recursion (\ref{drinfd.01}) can be derived from the deformation
\begin{align}
\hat F^{\sigma}_\nu&:= 
(2\ap)^{n-3} \! \! \! \! \! \! \! \! \! \!  \! \! \!  
\int\limits_{0 < z_{2}< z_{3} < \ldots < z_{n-2} < z_0}
\! \! \! \! \! \! \! \! \! \! \! \! \!    dz_2 \, dz_3\, \ldots \, dz_{n-2}
\prod_{1\leq p<q}^{n-1} |z_{pq}|^{-2\alpha' s_{pq}} \, \prod_{r=2}^{n-2} | z_{0r}|^{-2\alpha' s_{0r}} 
\omega^{\sigma}_\nu \, , \notag
\\
\omega^{\sigma}_\nu &= \sigma \bigg\{  
\prod_{k=2}^\nu \sum_{j=1}^{k-1} \frac{ s_{jk}}{z_{jk}} \prod_{m=\nu+1}^{n-2} \sum_{n=m+1}^{n-1} \frac{ s_{mn} }{z_{mn} } 
\bigg\}
\label{drinfd.03}
\end{align}
of the disk integrals $F^\sigma = F_{23\ldots n-2}{}^{\sigma(23\ldots n-2)}$ in
(\ref{Fdef}) by additional Mandelstam invariants $s_{0j}$ and an auxiliary
puncture $z_0 \in (0,1)$ on the disk boundary. The permutation $\sigma$ acts on the labels of 
both $s_{ij}$ and $z_{ij}$ enclosed in the curly brackets of the second line while leaving
$\sigma(1)=1$ and $\sigma(n{-}1)=n{-}1$ invariant. One can recover (\ref{drinfd.03})
from a basis of disk integrals in the $(n{+}1)$-point open-string amplitudes (\ref{npttree})
after removing the integration over $z_0 \in (0,1)$ and the associated $dz_0/z_{0j}$.

The integer $\nu=1,2,\ldots,n{-}2$ in (\ref{drinfd.03}) labels different
classes of integrands $\omega^{\sigma}_\nu$ that were related by the IBP identities
from the Koba--Nielsen factor in the undeformed case (\ref{Fdef}). By the
contributions $ | z_{0r}|^{-2\alpha' s_{0r}} $ to the Koba--Nielsen factor in
(\ref{drinfd.03}), the $n{-}2$ values of $\nu$ together with the $(n{-}3)!$
permutations $\sigma$ of $2,3,\ldots,n{-}2$ yield a total of $(n{-}2)!$ different integrals
$\hat F^{\sigma}_\nu$. The components of this $(n{-}2)!$-vector will be
ordered as $\hat F=(\hat F^{\sigma}_{n-2},\hat F^{\sigma}_{n-3},\ldots,\hat F^{\sigma}_{2},\hat F^{\sigma}_{1})$ with
lexicographic ordering for the permutations $\sigma$ indexing
the $(n{-}3)!$-component subvectors $\hat F^\sigma_{1},\ldots,\hat F^\sigma_{n-2}$.
The examples of (\ref{drinfd.03}) at $n=4$ and $5$ points are the two- and 
six-component vectors
\begin{align}
\hat F \, \big|_{n=4} &= 2\alpha' \int^{z_0}_0 dz_2 \, |z_{12}|^{-2\alpha' s_{12}}
|z_{23}|^{-2\alpha' s_{23}} |z_{02}|^{-2\alpha' s_{02}} \left( \begin{smallmatrix}  X_{12} \\ X_{23} 
\end{smallmatrix} \right) \, ,\label{drinfd.04} \\
\hat F \, \big|_{n=5} &= (2\alpha' )^2 \int^{z_0}_0 dz_3 \int^{z_3}_0 dz_2 \, 
|z_{23}|^{-2\alpha' s_{23}}
\prod_{j=2}^3
|z_{1j}|^{-2\alpha' s_{1j}}
|z_{j4}|^{-2\alpha' s_{j4}}
|z_{0j}|^{-2\alpha' s_{0j}}
  \left( \begin{smallmatrix}  X_{12} (X_{13}{+}X_{23}) \\ 
  X_{13} (X_{12}{+}X_{32}) \\ 
  X_{12}X_{34} \\
  X_{13} X_{24} \\
  (X_{23}{+}X_{24}) X_{34} \\
  (X_{32}{+}X_{34}) X_{24}
\end{smallmatrix} \right) \, , \notag
\end{align}
with the shorthand $X_{ij} = \frac{s_{ij} }{z_{ij}}$ as in section \ref{theIBPsec}.
Since the entries of the $n$-point vectors $\hat F$ form 
IBP bases, their $z_0$-derivatives are bound to yield
homogeneous Knizhnik--Zamolodchikov (KZ) equations of the 
following form
\beq
\frac{ d}{dz_0}  \hat F = \bigg( \frac{ \hat e_0}{z_0} + \frac{ \hat e_1}{1{-}z_0} \bigg) \hat F \, .
\label{drinfd.05}
\eeq
The entries of the $(n{-}2)! \times (n{-}2)!$ braid matrices $\hat e_0,\hat e_1$ are linear in
$\alpha' s_{ij}$ with $0\leq i<j\leq n{-}1$ as can be seen from the $z_0$-derivative
of the deformed Koba--Nielsen factor in (\ref{drinfd.03}). The same factors of 
$|z_{0r}|^{-2\alpha' s_{0r}}$ suppress the boundary terms $z_{n-2}=z_0$
from $\frac{ d}{dz_0} $-action on the integration limits for $z_{n-2} \in(z_{n-3},z_0)$. 
The explicit form of $\hat e_0,\hat e_1$ follows from reducing the contributions $\sum_{r=2}^{n-2} 
\frac{ s_{0r}}{z_{0r}} \omega^\sigma_\nu$ of the Koba--Nielsen derivatives 
w.r.t.\ the unintegrated variable $z_0$ to a basis of $ \omega^\sigma_\nu/z_0$ 
and $ \omega^\sigma_\nu/(1{-}z_0)$.

In fact, the recursion (\ref{drinfd.01}) only requires the kinematic limit
\beq
e_0 =  \hat e_0 \, \big|_{s_{0j}=0}  \, , \ \ \ \ \ \ 
e_1 =  \hat e_1 \, \big|_{s_{0j}=0} 
\label{drinfd.5a}
\eeq
of the braid matrices $\hat e_0,\hat e_1$ in (\ref{drinfd.05}). The four- and five-point
integrals (\ref{drinfd.04}) give rise to the following $2\times2$ and
$6\times 6$ examples:
\begin{align}
e_0 \, \big|_{n=4} &=2\alpha' \left( \begin{array}{cc} -s_{12} &s_{12} \\ 0 &0 \end{array} \right)\, , \ \ \ \ e_1 \, \big|_{n=4}  = 2\alpha' \left( \begin{array}{cc} 0 &0 \\ -s_{23} &s_{23} \end{array} \right) \, ,
\label{drinfd.06}\\
e_0  \, \big|_{n=5} &= 2\alpha'  \left( \begin{array}{cccccc}
-s_{123} &0 &s_{13}{+}s_{23} &s_{12} &s_{12} &-s_{12} \\
0 &-s_{123} &s_{13} &s_{12}{+}s_{23} &-s_{13} &s_{13} \\
0 &0 &-s_{12} &0 &s_{12} &0 \\
0 &0 &0 &-s_{13} &0 &s_{13} \\
0&0&0&0&0&0 \\
0&0&0&0&0&0
\end{array} \right) \, , \notag \\
e_1  \, \big|_{n=5}  &= 2\alpha'  \left( \begin{array}{cccccc}
0&0&0&0&0&0 \\
0&0&0&0&0&0 \\
-s_{34} &0 &s_{34} &0 &0 &0 \\
0 &-s_{24} &0 &s_{24} &0 &0 \\
-s_{34} &s_{34} &-s_{23}{-}s_{24} &-s_{34} &s_{234} &0 \\
s_{24} &-s_{24} &-s_{24} &-s_{23}{-}s_{34} &0 &s_{234} \end{array} \right)  \, .
\notag
\end{align}
The explicit form of the braid matrices $e_0,e_1$ at $n\leq 9$ points is
available in machine-readable form \cite{MZVWebsite} (where our conventions
are matched after rescaling $s_{ij} \rightarrow - 2\alpha' s_{ij}$ in the dataset
of the website), and a graphical all-multiplicity description can be found in \cite{Kaderli:2019dny}.

\subsubsection{Uniform transcendentality}
\label{sec:7.4.0}

The factorization of $\alpha'$ in the braid matrices $e_0,e_1$ as exemplified by (\ref{drinfd.06})
persists to any multiplicity~$n$. On these grounds, the $(n{-}2)!\times (n{-}2)!$ matrix 
representations $\Phi(e_0,e_1)$ of the Drinfeld associator enjoy
uniform transcendentality: the words $e_A$ in $e_0,e_1$ of length
$|A|=w$ are of the order $(\alpha')^w$ and accompanied by MZVs
$I(0;A;1)$ of transcendental weight $w$ in (\ref{mzvsec.10}). It is then easy
to show by induction that the recursion (\ref{drinfd.01}) propagates uniform
transcendentality from the $(n{-}1)$-point integrals $F^{\sigma_j} |_{k_{n-1}=0} $ 
on the right-hand side to the $n$-point integrals $F^{\sigma_i}$ on the left-hand side.

Together with the $\alpha'$-independent SYM amplitudes $A(\ldots)$ in (\ref{npttree}), 
we conclude that $n$-point open-superstring amplitudes are uniformly
transcendental. Following the discussions of
transcendentality properties in the field-theory literature such as
\cite{Fleischer:1998nb, Kotikov:2004er}, the first string-theory references
to observe and define uniform transcendentality of bases of disk integrals 
are \cite{Stieberger:2007am, nptStringII, Stieberger:2012rq}. The KLT relations (\ref{stringkltrel}) together with the uniform transcendentality
of the KLT kernel in (\ref{gravsec.6}) imply that also type II amplitudes are uniformly
transcendental. However, the $\alpha'$-dependence of the kinematic factors
$A_{(DF)^2+{\rm YM}+\phi^3}(\ldots)$ in (\ref{hetEYM.15}) obstructs uniform 
transcendentality of massless heterotic-string amplitudes (except for single-trace
gauge amplitudes with no or one graviton in (\ref{hetEYM.32}) and (\ref{hetEYM.34}) 
\cite{Stieberger:2014hba, Schlotterer:2016cxa}). 
Massless amplitudes of open or closed bosonic strings are in general non-uniformly 
transcendental for the same reason \cite{Huang:2016tag, Azevedo:2018dgo}.

The factorization of $\alpha'$ on the right-hand side of the KZ equation (\ref{drinfd.05})
can be viewed as a string-theory analogue of the so-called $\epsilon$-form of differential
equations of Feynman integrals \cite{Henn:2013pwa, Adams:2018yfj}: 
the dimensional-regularization parameter $\epsilon$ of Feynman integrals in $D_0{-}2\epsilon$
spacetime dimensions with $D_0\in \mathbb N$ serves as an expansion variable similar to
$\alpha'$ in string amplitudes. Various families of Feynman integrals admit uniformly 
transcendental bases under IBP relations 
\cite{Kotikov:1990kg, Fleischer:1998nb, Kotikov:2004er, Arkani-Hamed:2010pyv, Broedel:2018qkq}, 
also see \cite{Caola:2022ayt, Bourjaily:2022bwx} and 
\cite{Weinzierl:2022eaz, Abreu:2022mfk, Blumlein:2022zkr} for white papers and recent reviews.
In a growing number of examples, uniform transcendentality can be manifested by 
casting the differential equations of vectors $I$ of Feynman integrals into $\epsilon$-form 
$dI = \epsilon A I$, where the matrix $A$ of one forms no longer depends 
on $\epsilon$ \cite{Henn:2013pwa, Adams:2018yfj}.

\subsubsection{Regularized boundary values}
\label{sec:7.4.2}

Given the braid matrices $e_0, e_1$ derived from the KZ equation (\ref{drinfd.05})
at $s_{0j} =0$, we shall now review the origin of the recursion
(\ref{drinfd.01}) for the $\alpha'$-expansion of disk integrals.
The key idea is to use the relation (\ref{drinfeld.11}) between the
regularized boundary values $C_0,C_1$ of the solutions to a general
KZ equation. For the $(n{-}2)!$-component vector $\hat F$
in (\ref{drinfd.03}), the regularized boundary value as $z_0\rightarrow 0$ 
is given by \cite{drinfeld}
\beq
C_0 \, \big|_{s_{0j}=0} \rightarrow \big(F^\sigma \, \big|_{s_{j,n-1}=0},\underbrace{0,0,\ldots,0}_{(n{-}3)(n{-}3)!} \big)\, ,
\label{drinfd.11}
\eeq
where the $(n{-}3)(n{-}3)!$ vanishing entries stem from the subvectors with
$\nu =1,2,\ldots,n{-}3$. The $(n{-}3)!$ undeformed integrals $F^\sigma |_{s_{j,n-1}=0}$
realize the soft limit $k_{n-1} \rightarrow 0$ in (\ref{drinfd.01}) and can be
obtained from the subvector with $\nu=n{-}2$ by
rescaling of the integration variables $z_j = x_j z_0$ 
that transforms the integration domain in (\ref{drinfd.03}) to
$0<x_2<x_3<\ldots < x_{n-2}<1$. The appearance of lower-point disk
integrals from the soft limit in (\ref{drinfd.02}) is easiest to see from
the following IBP rewriting of (\ref{Fdef})
\beq
F^{\sigma(23\ldots n-2)}= 
(2\ap)^{n-3} \! \! \! \! \! \! \! \! \! \!  \! \! \!  
\int\limits_{0 < z_{2}< z_{3} < \ldots < z_{n-2} < 1}
\! \! \! \! \! \! \! \! \! \! \! \! \!    dz_2 \, dz_3\, \ldots \, dz_{n-2}
\prod_{1\leq p<q}^{n-1} |z_{pq}|^{-2\alpha' s_{pq}} 
 \sigma \bigg\{  \frac{ s_{n-2,n-1} }{ z_{n-2,n-1} }
\prod_{k=2}^{n-3} \sum_{j=1}^{k-1} \frac{ s_{jk}}{z_{jk}} 
\bigg\} \, .
\label{drinfd.12}
\eeq
If $\sigma(n{-}2)\neq n{-}2$, then the 
denominator of $\sigma \{\frac{ s_{n-2,n-1} }{ z_{n-2,n-1} }\}$ involves non-adjacent
variables $z_{\sigma(n{-}2)},z_{n-1}$ in the integration domain of (\ref{drinfd.12}).
One can set $s_{j,n-1}=0$ at the level of the integrand and reproduce the zeros on
the right-hand side of (\ref{drinfd.02}). If $\sigma(n{-}2) = n{-}2$ in turn, the Koba--Nielsen
integral over $|z_{n-2,n-1}|^{-2\alpha' s_{n-1,n-2}-1}$ results in a kinematic 
pole $s_{n-2,n-1}^{-1}$ whose residue
is obtained from setting $z_{n-2}=1$ in the integrand. This residue is given by
the $(n{-}1)$-point integral $F^{\sigma(23\ldots n-3)}$ on the right-hand side of (\ref{drinfd.02}), 
and the soft limit $s_{j,n-1}=0$ suppresses the regular terms in $s_{n-2,n-1}$ beyond the residue.

The second regularized boundary value (\ref{drinfeld.11}) obtained from the $(n{-}2)!$ integrals
in (\ref{drinfd.03}) reproduces the undeformed $n$-point disk integrals in
its first $(n{-}3)!$ components \cite{drinfeld}
\beq
C_1 \, \big|_{s_{0j}=0} \rightarrow \big(F^\sigma, \ldots \big)\, .
\label{drinfd.13}
\eeq
This can be intuitively understood from the fact that $z_0\rightarrow 1$ restores
the original integration domain $0<z_2<\ldots <z_{n-2}<1$ of $F^\sigma$,
and we are setting $s_{0j}=0$ in (\ref{drinfd.13}) to remove the deformation of the 
Koba--Nielsen factor. However,
the components in the ellipsis of (\ref{drinfd.13}) involve lower-multiplicity 
contributions from the 
difference between the regions $ z_{n-2} \in (z_{n-3},z_0)$ and $ z_{n-2} \in (z_{n-3},1)$:
for some of the components of the integrands $\omega_\nu^\sigma$ in (\ref{drinfd.03}) with
$\nu\leq n{-}3$, the difference $z_{n-2} \in (z_0,1)$ between the above regions
contributes to the $\alpha'$-expansion even though it shrinks to zero size as $z_0 \rightarrow 1$.
The detailed evaluation of these $\nu \leq n{-}3$ components of $C_1$ is subtle
and fortunately not needed to derive the recursion (\ref{drinfd.01}).

Note that the field-theory limit $[ \Phi(e_0,e_1) ]_{ij} =
\delta_{ij}+{\cal O}(\alpha'^2)$ of the associator in (\ref{drinfd.01}) together
with the three-point integral $F^\emptyset =1$ imply by
induction in $n$ that $F^{\sigma_{j}} = \delta_{j,1} + {\cal O}(\alpha'^2)$, 
i.e.\ that the $\alpha'$-expansions of all the $F^{\sigma_{j}}$ with $j\neq 1$ 
start at order $\alpha'^2$. This is one way of deriving the
orthonormal field-theory limits (\ref{ftofFPQ}) of the $F_P{}^Q$.

In summary, the relation (\ref{drinfeld.11}) between regularized boundary values 
and their representations (\ref{drinfd.11}), (\ref{drinfd.13}) for the specific solution
$\hat F$ of the KZ equation in (\ref{drinfd.03}) implies the recursion (\ref{drinfd.01}) for
$n$-point disk integrals. 

\subsubsection{Connection with twisted deRham theory and outlook}
\label{sec:7.4.3}

The $z_0$-deformed integrals (\ref{drinfd.03}) are special cases of more general Koba--Nielsen 
or Selberg integrals over the disk boundary with an arbitrary number of integrated and 
unintegrated punctures \cite{selbergref, aomoto1987gauss, Terasoma}. They obey 
KZ equations in multiple variables, and a recursion for the
braid matrices in their differential operator has been given in \cite{Mizera:2019gea}.
The discussion in the reference is tailored to specific fibration bases w.r.t.\
IBP, and the transformation matrices to the bases (\ref{drinfd.03}) in the case of 
four unintegrated punctures can be found in \cite{Kaderli:2019dny}. This is how
the all-multiplicity results for braid matrices in \cite{Mizera:2019gea} translate
into the $n$-point instances of $e_0,e_1$ in \cite{Kaderli:2019dny}.

The coaction properties (\ref{mzvsec.40}) of the $n$-point disk integrals in string amplitudes
generalize to the case of more than three unintegrated punctures at $(z_i,z_j,z_k) \rightarrow
(0,1,\infty)$, for instance to the family of
Selberg integrals (\ref{drinfd.03}) with an $(n{-}2)!$ basis of integration contours. 
The $\alpha'$-expansions of Selberg integrals with an arbitrary number of integrated 
and unintegrated punctures were investigated in \cite{Britto:2021prf}. Their coactions
in the basis choice of the reference line up with the master formula (\ref{mzvsec.42}) 
that initially arose from studies of dimensionally regulated Feynman integrals 
\cite{Abreu:2017enx, Abreu:2017mtm, Abreu:2019wzk, Abreu:2021vhb}. 
It is striking to see that the coaction formula (\ref{mzvsec.42}) manipulating 
contours $\gamma_j$ and differential forms $\omega_j$ in
twisted-(co-)homology bases is compatible
with that of the polylogarithms in the respective $\epsilon$- or $\alpha'$-expansions. 
A mathematical proof for Lauricella hypergeometric functions can be found 
in \cite{Brown:2019jng}.

Selberg integrals with arbitrary numbers of integrated and unintegrated punctures on
a disk boundary have been generalized to genus one and investigated from a multitude
of perspectives in the mathematics \cite{FelderVarchenko, Manowatanabe} and physics 
\cite{Mafra:2019ddf, Mafra:2019xms, Broedel:2019gba, Broedel:2020tmd, Kaderli:2022qeu} 
literature. These references
offer several lines of attack to expand the configuration-space integrals of one-loop
open-string amplitudes in $\alpha'$. In particular, the construction of \cite{Broedel:2019gba}
can be viewed as a direct genus-one analogue of the Drinfeld-associator 
method of this section.

\subsection{Berends--Giele recursion for disk integrals}
\label{sec:7.5}

In this section we review the construction \cite{BGap} of a Berends--Giele formula to compute the
$\ap$-expansion of $Z(P|Q)$ disk integrals \eqref{Zintdef} recursively in the length $\len{P}$, or
alternatively in the number of points of the associated disk amplitude \eqref{localFormWithZAgain}.
Given the interpretation of Berends--Giele currents as coefficients in the perturbiner solution of 
an equation of motion, this method adds support to the introduction of $Z$-theory
\cite{Carrasco:2016ldy,BGap,Carrasco:2016ygv}; the scalar theory whose
amplitudes computed by the standard Berends--Giele method \cite{BerendsME} are given by the integrals $Z(P|Q)$.

\subsubsection{Extending the field-theory limit}

The starting point behind the Berends--Giele method to evaluate disk integrals is the
assumption that the Berends--Giele method to evaluate their field-theory limit $\ap\to0$
as \cite{FTlimit}
\beq\label{ftlimBGAgain}
\lim_{\ap\to0} Z(P,n|Q,n) = \lim_{s_P\to0}s_P \phi(P|Q) \,,
\ee
can be lifted to arbitrary $\ap$ orders
\beq\label{ZfromBG}
Z(P,n|Q,n) = \lim_{s_P\to0}s_P \phi^\ap \!(P|Q)
\ee
via the introduction of an $\ap$-corrected Berends--Giele current\footnote{We adopt the notation
$\phi(P|Q)=\phi_{P|Q}$ whenever convenient.} $\phi^\ap \!(P|Q)$. In the field-theory limit case of
\eqref{ftlimBGAgain}, the Berends--Giele current $\phi(P|Q)$ is the coefficient of the perturbiner
solution $\Phi(X)$ in (\ref{doubleLie}) of the
equation of motion $\Box\Phi=[\![\Phi,\Phi]\!]$ of the bi-adjoint scalar theory as reviewed in section \ref{subsec:biadj}. Interpreting $\phi(P|Q) = \lim_{\ap\to0}\phi^\ap \!(P|Q)$, the required step to evaluate
\eqref{ZfromBG} is to obtain the $\ap$-corrections to the equation of motion of the bi-adjoint theory
and to recursively generate $\alpha'$-dependent Berends--Giele currents from its perturbiner solution
\beq\label{doubleLieAgain}
\Phi(X) := \sum_{P,Q} \phi^\ap \!(P|Q)\,t^P \otimes \tilde t^Q\, e^{k_P\cdot X},
\qquad t^P:= t^{p_1}t^{p_2} \ldots t^{p_{|P|}}
\eeq
with initial condition $\phi^\ap \!(i |j)=\delta_{ij}$ in the single-particle case.
The $\alpha'$-corrected equation of motion found in \cite{BGap} is written in the following 
compact way
\begin{align}
\frac{1}{2} \Box \Phi &=
\sum_{p=2}^{\infty}(2\ap)^{p-2}  \int^{\rm eom} \prod_{i<j}^{p} |z_{ij}|^{-2\ap
\partial_{ij}}  \label{ZEOM} \\
&\quad\times\Bigl(\,
\sum_{l=1}^{p-1}
\frac{ [\Phi_{12\ldots l} , \Phi_{p,p-1\ldots l+1}] }{(z_{12}z_{23}\ldots z_{l-1,l})
 (z_{p,p-1} z_{p-1,p-2} \ldots z_{l+2,l+1})}
+ \perm(2,3,\ldots,p{-}1)\, \Bigr)
\notag \\
&= [ \Phi_1 , \Phi_2 ] +2\ap  \int^{\rm eom} \! \! |z_{12}|^{-2\ap \partial_{12}}  |z_{23}|^{-2\ap \partial_{23}}
\bigg( \frac{[\Phi_{12},\Phi_3] }{z_{12}} +  \frac{[\Phi_{1},\Phi_{32}] }{z_{32}}  \bigg)  \notag \\
&\quad + (2\ap)^2  \int^{\rm eom}  \! \! |z_{23}|^{-2\ap \partial_{23}} \prod_{j=2}^3
 |z_{1j}|^{-2\ap \partial_{1j}} |z_{j4}|^{-2\ap \partial_{j4}}
\bigg( \frac{[\Phi_{123},\Phi_4] }{z_{12} z_{23}} {+}  \frac{[\Phi_{12},\Phi_{43}] }{z_{12} z_{43}}  
 {+}  \frac{[\Phi_{1},\Phi_{432}] }{z_{43} z_{32}}  
+(2\leftrightarrow 3)\bigg) \notag \\
&\quad  +\ldots \, , \notag
\end{align}
with $z_1=0$ and $z_{p}=1$
which is obtained directly from the local representation of the disk amplitude \eqref{stringWithTs} under
the following mappings discussed at length in \cite{BGap}. The unintegrated vertices in
\eqref{stringWithTs} are replaced as
\beq\label{VtoPhi}
\langle V_PV_QV_n\rangle \longrightarrow [\Phi_P,\Phi_Q]\,,
\eeq
where $\Phi_P$ is a shorthand for various linear combinations of $\phi^\ap \!(R|S)$ as explained 
below. The contributions spelled out at the end of (\ref{ZEOM}) descend from ${\cal A}(1,2,3)
= \langle V_1 V_2 V_3 \rangle$ as well as the four- and 
five-point amplitudes in (\ref{4ptex.3}) and (\ref{fivcex}). The
ellipsis refers to permutations of $[\Phi_{12\ldots l} , \Phi_{p\ldots l+1}]$
with $p\geq 5$ following the form of disk amplitudes \eqref{stringWithTs} at six points and beyond.

The mapping denoted by $\int^{\rm eom}$ encodes a series of rules meant to compute 
the regularized integrals over $z_2,\ldots, z_{p-1}$ appearing in \eqref{stringWithTs} after fixing
$(z_1,z_p)=(0,1)$ and expanding the Koba--Nielsen factor in a series of
$\ap$. The replacement $s_{ij}\to \p_{ij}$ in the Koba--Nielsen exponents 
will be defined in \eqref{dij} below.
The technical details involving manipulations of shuffle-regularized polylogarithms can be 
found in \cite{BGap} and lead to rational combinations of MZVs at each order
in the $\alpha'$-expansion of (\ref{ZEOM}).

Even though the origin of (\ref{ZEOM}) from a Lagrangian is unsettled, we interpret it
as the non-linear equation of motion of $Z$-theory.

\paragraph{The shorthand $\Phi_P$}
The shorthand $\Phi_P$ in the equation of motion (\ref{ZEOM}) denotes an 
expansion of several factors of $\phi^\ap \!(A|B)$ according to the following rules. First, define
\beq\label{TAB}
T^{\rm dom}_{A_1,A_2, \ldots, A_n}\otimes T^{\rm int}_{B_1, B_2, \ldots, B_n} \equiv
\phi^\ap \!(A_1|B_1) \phi^\ap \!(A_2|B_2) \cdots \phi^\ap \!(A_n|B_n)
\eeq
for arbitrary words $A_i$ and $B_j$. Next, define linear combinations
\beq\label{TAB.lin}
T^{B_1,B_2, \ldots, B_n}_{A_1,A_2, \ldots, A_n}:=
T_{A_1,A_2, \ldots, A_n}^{\rm dom}\otimes T_{\rho(B_1,B_2, \ldots, B_n)}^{\rm int} \, ,
\eeq
where the map $\rho$ on words is given in \eqref{rhomapdef}, and it is understood here to
act on the labels $i$ of the words $B_i$. It is straightforward to see that
$T^{B_1,B_2,\ldots,B_n}_{A_1,A_2, \ldots,A_n}$ satisfies the recursion
\beq\label{Trec}
T^{B_1,B_2,\ldots,B_n}_{A_1,A_2, \ldots,A_n} = T^{B_1, B_2, \ldots, B_{n-1}}_{A_1,A_2, \ldots,
A_{n-1}}\, \phi^\ap \!(A_n|B_n)
- T^{B_2, B_3, \ldots, B_{n}}_{A_1, A_2, \ldots A_{n-1}}\, \phi^\ap \!(A_n|B_1)\,,
\eeq
with initial condition $T^{B_1}_{A_1} = \phi^\ap \!(A_1 | B_1)$
which can be taken as its alternative definition. The simplest examples of \eqref{Trec} are,
\begin{align}\label{TABex}
T^{B_1, B_2}_{A_1,A_2} &= \phi^\ap \!(A_1 | B_1) \phi^\ap \!(A_2 | B_2)
- \phi^\ap \!(A_1 | B_2) \phi^\ap \!(A_2 | B_1)\,, \\
T^{B_1, B_2, B_3}_{A_1, A_2, A_3}  &=
  \phi^\ap \!(A_1 | B_1) \phi^\ap \!(A_2 | B_2) \phi^\ap \!(A_3 | B_3)
- \phi^\ap \!(A_1 | B_2) \phi^\ap \!(A_2 | B_3) \phi^\ap \!(A_3 | B_1)\cr
&{}
- \phi^\ap \!(A_1 | B_2) \phi^\ap \!(A_2 | B_1)  \phi^\ap \!(A_3 | B_3)
+ \phi^\ap \!(A_1 | B_3) \phi^\ap \!(A_2 | B_2) \phi^\ap \!(A_3 | B_1)\,,\cr
T^{B_1, B_2, B_3, B_4}_{A_1, A_2, A_3, A_4} &=
  \phi^\ap \!(A_1 | B_1) \phi^\ap \!(A_2 | B_2) \phi^\ap \!(A_3 | B_3) \phi^\ap \!(A_4 | B_4)
- \phi^\ap \!(A_1 | B_2) \phi^\ap \!(A_2 | B_1) \phi^\ap \!(A_3 | B_3)  \phi^\ap \!(A_4 | B_4)\cr
&{}
- \phi^\ap \!(A_1 | B_2) \phi^\ap \!(A_2 | B_3) \phi^\ap \!(A_3 | B_1) \phi^\ap \!(A_4 | B_4)
+ \phi^\ap \!(A_1 | B_3) \phi^\ap \!(A_2 | B_2) \phi^\ap \!(A_3 | B_1) \phi^\ap \!(A_4 | B_4)\cr
&{}
- \phi^\ap \!(A_1 | B_2) \phi^\ap \!(A_2 | B_3) \phi^\ap \!(A_3 | B_4) \phi^\ap \!(A_4 | B_1)
+ \phi^\ap \!(A_1 | B_3) \phi^\ap \!(A_2 | B_2) \phi^\ap \!(A_3 | B_4) \phi^\ap \!(A_4 | B_1)\cr
&{}
+ \phi^\ap \!(A_1 | B_3) \phi^\ap \!(A_2 | B_4) \phi^\ap \!(A_3 | B_2) \phi^\ap \!(A_4 | B_1)
- \phi^\ap \!(A_1 | B_4) \phi^\ap \!(A_2 | B_3) \phi^\ap \!(A_3 | B_2) \phi^\ap \!(A_4 | B_1)\,.\notag
\end{align}
By construction, the above satisfy the shuffle symmetries on the $B_i$ slots
\beq\label{shufT}
T^{(B_1,B_2, \ldots,B_j)\shuffle (B_{j{+}1}, \ldots,B_n)}_{A_1, A_2,\ldots,A_n} = 0\, , \ \ \ \ 
j=1,2,\ldots,n{-}1\, ,
\eeq
where the shuffle product is understood to act on the labels $i$ of $B_i$. Surprisingly, the
definition \eqref{Trec} satisfies the generalized Jacobi identities \eqref{genjac} on its $A_j$
slots; i.e.\ $T^{B_1,B_2, \ldots ,B_n}_{A_1,A_2, \ldots ,A_n}$ at fixed order of the $B_i$ slots
satisfy the same symmetries as the nested commutator $[[ \ldots[[A_1,A_2],A_3]
\ldots],A_n]$ such as $T^{B_1, B_2}_{A_2,A_1}=-T^{B_1, B_2}_{A_1,A_2}$
and $T^{B_1, B_2,B_3}_{A_1,A_2,A_3}+T^{B_1, B_2,B_3}_{A_2,A_3,A_1}
+ T^{B_1, B_2,B_3}_{A_3,A_1,A_2}=0$. 

Finally, the shorthand $\Phi_P$ for a word $P=p_1p_2 \ldots p_n$ is defined as
\beq\label{PhiPdef}
\Phi_P := T^{B_1,B_2, \ldots,B_n}_{A_{p_1},A_{p_2}, \ldots,A_{p_n}}\,,
\eeq
that is, the word $P$ captures the ordering of the labels $i$ of the words $A_i$, while the labels
$j$ of the words $B_j$ are in the canonical order. For example, $\Phi_2 = T^{B_1}_{A_2}$ as well as
\begin{align}
\Phi_{21}&=T^{B_1,B_2}_{A_2,A_1}\,,\\
\Phi_{231}&=T^{B_1,B_2,B_3}_{A_2,A_3,A_1}\,,\notag\\
\Phi_{4213}&=T^{B_1,B_2,B_3,B_4}_{A_4,A_2,A_1,A_3}\,,\notag
\end{align}
and the properties of $T^{B_1, \ldots, B_n}_{A_1,\ldots, A_n}$ readily imply
generalized Jacobi identities of $\Phi_P$ such as $\Phi_{21}=-\Phi_{12}$ and $\Phi_{123}+\Phi_{231}+\Phi_{312}=0$. The commutators of $\Phi_P$ and $\Phi_Q$ in \eqref{ZEOM} 
yield the left-to-right Dynkin bracket (\ref{ellmap}),
\beq
[\Phi_{P},\Phi_Q] = \Phi_{P \ell(Q)} \, ,
\eeq
for instance $[\Phi_1,\Phi_{32}]= \Phi_{132}-\Phi_{123} = \Phi_{231}$ or $[\Phi_{12},\Phi_{43}]
= \Phi_{1243}-\Phi_{1234}$.
In this way, the equation of motion \eqref{ZEOM} leads to a recursion for the Berends--Giele
currents $\phi^\ap \!(P|Q)$ that can be used to obtain the $\ap$-expansion of the $Z(P|Q)$
integrals using the Berends--Giele formula \eqref{ZfromBG}.

\paragraph{The equation of motion up to $\ap^3$ order}
Applying the integration rules discussed in \cite{BGap} to (\ref{ZEOM}) one 
obtains expansions such as
\begin{align}
2\ap  \int^{\rm eom} \! \! |z_{12}|^{-2\ap \partial_{12}}  |z_{23}|^{-2\ap \partial_{23}}
 \frac{1 }{z_{12}}  &= \frac{1}{\partial_{12}} \bigg( \frac{\Gamma(1-2\ap \partial_{12}) 
 \Gamma(1-2\ap \partial_{23})  }{\Gamma(1-2\ap \partial_{12}-2\ap \partial_{23}) }
- 1 \bigg) \notag \\
&= - (2\ap)^2 \zeta_2 \partial_{23} - (2\ap)^3 \zeta_3 \partial_{23}(\partial_{12}{+}\partial_{23})
 \label{ZEOMnew}  \\
 &\quad -(2\ap)^4 \zeta_4 \partial_{23}\big( \partial_{12}^2 + \tfrac{1}{4} \partial_{12}\partial_{23} 
 + \partial_{23}^2\big) + {\cal O}(\ap^5)
\notag
\end{align}
and thereby the following equation
of motion including $\ap$-corrections
\begin{align}
\half \Box \Phi &=[\Phi_{1}, \Phi_{2}]
+  \Bigl((2\ap)^2\zeta_2 \p_{12} + (2 \ap)^3\zeta_3 \p_{12}(\p_{12} + \p_{23}) \Bigr)  [\Phi_{{1}},\Phi_{{3}{2}}] \label{ZEOM3} \\
&\quad - \Bigl((2\ap)^2\zeta_2 \p_{23} +(2 \ap)^3\zeta_3 \p_{23} (\p_{12} + \p_{23}) \Bigr)  [\Phi_{{1}{2}},\Phi_{{3}}]\cr
&\quad + \Bigl( (2\ap)^2\zeta_2 +(2 \ap)^3\zeta_3 \big(\p_{21}
+ 2 \p_{31} + 2 \p_{32}
+ 2 \p_{42}  +  \p_{43}\big)  \Bigr)
[\Phi_{{1}{2}}, \Phi_{{4}{3}}] 
\cr
&\quad - \Bigl( (2\ap)^2\zeta_2 + (2 \ap)^3\zeta_3\big(2 \p_{21}
+  \p_{31} + 3 \p_{32}
 +  \p_{42} + 2 \p_{43} \big) \Bigr)
  [\Phi_{{1}{3}}, \Phi_{{4}{2}}] \cr
&\quad - 2 (2\ap)^3\zeta_3 \bigl( \p_{42} +  \p_{43} \bigr)
[\Phi_{{1}{2}{3}}, \Phi_{{4}}] 
+ (2\ap)^3\zeta_3  \bigl( 3 \p_{42} +  \p_{43} \bigr)
[\Phi_{{1}{3}{2}}, \Phi_{{4}}]   \cr
&\quad -2 (2 \ap)^3\zeta_3 \bigl( \p_{31} +  \p_{21} \bigr) [\Phi_{{1}}, \Phi_{{4}{3}{2}}] 
+ (2 \ap)^3\zeta_3  \bigl( 3 \p_{31}  +  \p_{21} \bigr) [\Phi_{{1}}, \Phi_{{4}{2}{3}}] \cr
&\quad+  (2\ap)^3\zeta_3   \Bigl(
 [\Phi_{{1}{2}}, \Phi_{{5}{3}{4}}] 
{} -  2 [\Phi_{{1}{2}}, \Phi_{{5}{4}{3}}] 
{} +  2 [\Phi_{{1}{2}{3}}, \Phi_{{5}{4}}] 
{} + 2 [\Phi_{{1}{3}}, \Phi_{{5}{2}{4}}]
{} -  [\Phi_{{1}{3}{2}}, \Phi_{{5}{4}}]\cr
&\quad \quad {} -  2 [\Phi_{{1}{3}{4}}, \Phi_{{5}{2}}]
{} -  3 [\Phi_{{1}{4}}, \Phi_{{5}{2}{3}}]
{} +  2 [\Phi_{{1}{4}}, \Phi_{{5}{3}{2}}]
{} -  2 [\Phi_{{1}{4}{2}}, \Phi_{{5}{3}}]
{} +  3 [\Phi_{{1}{4}{3}}, \Phi_{{5}{2}}]
\Bigr) + {\cal O}(\ap^4)\,,\notag
\end{align}
where
\beq\label{dij}
\p_{ij}\Phi_P := (k_{A_i}\cdot k_{A_j})\Phi_P\, .
\eeq
For a simple example of a practical calculation using these definitions, the low-energy 
expansion of disk integrals up to $\ap^2$ at any multiplicity is determined from $s_A\phi^\ap \!(A|B)$
in (\ref{ZfromBG}) as follows (here $\phi^\ap_{A|B}:= \phi^\ap \!(A|B)$, and the
initial conditions are $\phi^\ap_{i| j}=\delta_{ij}$)
\begin{align}
s_A \phi^\ap_{A|B} &=
\sum_{A_1A_2=A\atop B_1B_2=B} (\phi^\ap_{A_1|B_1} \phi^\ap_{A_2|B_2}
- \phi^\ap_{A_1|B_2} \phi^\ap_{A_2|B_1}) \\
&\quad + (2\ap)^2\zeta_2 \sum_{A_1 A_2 A_3=A\atop B_1 B_2 B_3=B}
 \Big[(k_{A_1}\cdot k_{A_2}) \bigl(
 \phi^\ap_{A_{1}|B_{1}}\phi^\ap_{A_{2}|B_{3}}\phi^\ap_{A_{3}|B_{2}}
- \phi^\ap_{A_{1}|B_{1}}\phi^\ap_{A_{2}|B_{2}}\phi^\ap_{A_{3}|B_{3}} \cr
&\quad \quad \quad \quad \quad \quad \quad\quad
\quad \quad\quad \quad \quad \quad \
+ \phi^\ap_{A_{1}|B_{3}}\phi^\ap_{A_{2}|B_{1}}\phi^\ap_{A_{3}|B_{2}}
- \phi^\ap_{A_{1}|B_{3}}\phi^\ap_{A_{2}|B_{2}}\phi^\ap_{A_{3}|B_{1}}
\bigr) \cr
&\quad \quad \quad \quad \quad \quad\quad \quad  \quad \ +  (k_{A_2}\cdot k_{A_3}) \bigl(
\phi^\ap_{A_{1}|B_{2}}\phi^\ap_{A_{2}|B_{1}}\phi^\ap_{A_{3}|B_{3}}
 - \phi^\ap_{A_{1}|B_{1}}\phi^\ap_{A_{2}|B_{2}}\phi^\ap_{A_{3}|B_{3}}\cr
&\quad \quad \quad \quad \quad \quad \quad\quad
\quad \quad\quad \quad \quad \quad \
+ \phi^\ap_{A_{1}|B_{2}}\phi^\ap_{A_{2}|B_{3}}\phi^\ap_{A_{3}|B_{1}}
 - \phi^\ap_{A_{1}|B_{3}}\phi^\ap_{A_{2}|B_{2}}\phi^\ap_{A_{3}|B_{1}} \bigr) \Bigr] \cr
&\quad +(2 \ap)^2\zeta_2 \sum_{A_1 A_2 A_3 A_4=A\atop B_1 B_2 B_3 B_4=B} 
          \Bigl[\phi^\ap_{A_{1}|B_{1}}\phi^\ap_{A_{2}|B_{2}}\phi^\ap_{A_{3}|B_{4}}\phi^\ap_{A_{4}|B_{3}}
         - \phi^\ap_{A_{1}|B_{1}}\phi^\ap_{A_{2}|B_{2}}\phi^\ap_{A_{3}|B_{3}}\phi^\ap_{A_{4}|B_{4}}\cr
&\quad \quad \quad \quad \quad \quad \quad\quad\quad \quad \
+ \phi^\ap_{A_{1}|B_{1}}\phi^\ap_{A_{2}|B_{3}}\phi^\ap_{A_{3}|B_{2}}\phi^\ap_{A_{4}|B_{4}}
         - \phi^\ap_{A_{1}|B_{1}}\phi^\ap_{A_{2}|B_{4}}\phi^\ap_{A_{3}|B_{2}}\phi^\ap_{A_{4}|B_{3}}\cr
&\quad \quad \quad \quad \quad \quad \quad\quad\quad \quad \
         + \phi^\ap_{A_{1}|B_{2}}\phi^\ap_{A_{2}|B_{1}}\phi^\ap_{A_{3}|B_{3}}\phi^\ap_{A_{4}|B_{4}}
          - \phi^\ap_{A_{1}|B_{2}}\phi^\ap_{A_{2}|B_{1}}\phi^\ap_{A_{3}|B_{4}}\phi^\ap_{A_{4}|B_{3}}\cr
&\quad \quad \quad \quad \quad \quad \quad\quad\quad \quad \
         - \phi^\ap_{A_{1}|B_{2}}\phi^\ap_{A_{2}|B_{3}}\phi^\ap_{A_{3}|B_{1}}\phi^\ap_{A_{4}|B_{4}}
         + \phi^\ap_{A_{1}|B_{2}}\phi^\ap_{A_{2}|B_{4}}\phi^\ap_{A_{3}|B_{1}}\phi^\ap_{A_{4}|B_{3}}\cr
&\quad \quad \quad \quad \quad \quad \quad\quad\quad \quad \
  - \phi^\ap_{A_{1}|B_{3}}\phi^\ap_{A_{2}|B_{1}}\phi^\ap_{A_{3}|B_{4}}\phi^\ap_{A_{4}|B_{2}}
         + \phi^\ap_{A_{1}|B_{3}}\phi^\ap_{A_{2}|B_{2}}\phi^\ap_{A_{3}|B_{4}}\phi^\ap_{A_{4}|B_{1}} \cr
&\quad \quad \quad \quad \quad \quad \quad\quad\quad \quad \
         + \phi^\ap_{A_{1}|B_{3}}\phi^\ap_{A_{2}|B_{4}}\phi^\ap_{A_{3}|B_{1}}\phi^\ap_{A_{4}|B_{2}}
         - \phi^\ap_{A_{1}|B_{3}}\phi^\ap_{A_{2}|B_{4}}\phi^\ap_{A_{3}|B_{2}}\phi^\ap_{A_{4}|B_{1}}\cr
&\quad \quad \quad \quad \quad \quad \quad\quad\quad \quad \
         + \phi^\ap_{A_{1}|B_{4}}\phi^\ap_{A_{2}|B_{1}}\phi^\ap_{A_{3}|B_{3}}\phi^\ap_{A_{4}|B_{2}}
         - \phi^\ap_{A_{1}|B_{4}}\phi^\ap_{A_{2}|B_{2}}\phi^\ap_{A_{3}|B_{3}}\phi^\ap_{A_{4}|B_{1}}\cr
&\quad \quad \quad \quad \quad \quad \quad\quad\quad \quad \
          - \phi^\ap_{A_{1}|B_{4}}\phi^\ap_{A_{2}|B_{3}}\phi^\ap_{A_{3}|B_{1}}\phi^\ap_{A_{4}|B_{2}}
         + \phi^\ap_{A_{1}|B_{4}}\phi^\ap_{A_{2}|B_{3}}\phi^\ap_{A_{3}|B_{2}}\phi^\ap_{A_{4}|B_{1}}
	 \Bigr]
	 + {\cal O}(\alpha'^3) \,.\notag
\end{align}
For example, one can show from the above recursion that
\beq\label{5ptz2}
Z(13524|32451) = - {1\over s_{13}s_{135}} + (2\ap)^2\zeta_2\Bigl(
{s_{35}\over s_{135}}
+ {s_{25}\over s_{13}}
-1
\Bigr) + {\cal O}(\alpha'^3) \, .
\eeq
From the above example, it is not hard to imagine that these calculations, despite systematic,
are long and tedious to perform by hand. A {\tt FORM} program that computes the $\ap$-expansion
of integrals of arbitrary multiplicity up to $\ap^7$ can be found in the git repository
\cite{gitrep}.

\subsubsection{Planar binary trees and $\ap$-corrections}

From the discussion above, the $\alpha'$-expansion of string disk integrals is
determined by the Berends--Giele formula (\ref{ZfromBG}) whose
currents $\phi^\ap \!(P|Q)$ are recursively generated by the
equations of motion (\ref{ZEOM}) of the non-abelian $Z$-theory \cite{BGap}.

As discussed in \cite{flas}, one can promote this setup to the theory of free Lie algebras by assuming the existence
of $\ap$-corrections to the binary-tree expansion (\ref{bMap}) as $b^\ap \!(P)$ by defining
\beq\label{phiapdef}
\phi^\ap \!(P|Q) := \langle b^\ap \!(P), Q\rangle\,,
\eeq
where $\langle A,B\rangle=\d_{A,B}$ denotes the scalar product of words defined in \eqref{AdotB}.
Using the explicit expressions of $\phi^\ap \!(P|Q)$ up to $\ap^7$ order
one can show that the Lie-polynomial form of the binary-tree expansion with $\ap$-corrections
becomes
\begin{align}\label{BMapap}
s_P b^\ap \!(P) &= \sum_{XY=P}[b^\ap \!(X),b^\ap \!(Y)] \\
&\quad + (2\ap)^2\zeta_2\sum_{XYZ=P}k_X\cdot k_Y[b^\ap \!(X),[b^\ap \!(Z),b^\ap \!(Y)]]\cr
&\quad- (2\ap)^2\zeta_2\sum_{XYZ=P}k_Y\cdot k_Z[[b^\ap \!(X),b^\ap \!(Y)],b^\ap \!(Z)]\cr
&\quad+ (2\ap)^2\zeta_2\sum_{XYZW=P}[[b^\ap \!(X),b^\ap \!(Y)],[b^\ap \!(W),b^\ap \!(Z)]]\cr
&\quad - (2\ap)^2\zeta_2\sum_{XYZW=P}[[b^\ap \!(X),b^\ap \!(Z)],[b^\ap \!(W),b^\ap \!(Y)]] +{\cal O}(\ap^3)\, .\notag
\end{align}
It is important to emphasize that the symmetries of the ``domain'' $P$ and ``integrand'' $Q$ 
are different, in particular $\phi^\ap \!(P|Q)\neq \phi^\ap \!(Q|P)$, unlike its field-theory 
version \eqref{symFTphi}. The integrand $Q$ satisfies shuffle symmetry as 
$\phi^\ap \!(P|R\shuffle S) = \langle b^\ap \!(P), R\shuffle S\rangle = 0 \ \forall \ R,S\neq \emptyset$ 
because $b^\ap \!(P)$ is a Lie polynomial, as a consequence of 
Ree's theorem 3.1 (iv) in \cite{Reutenauer}. The shuffle symmetries of the 
domain $P$ are spoiled by the monodromy properties of the disk integrals.

The Lie polynomial \eqref{BMapap} begs for a combinatorial understanding via free-Lie-algebra
methods in combination with the properties of MZVs following from the 
Drinfeld associator, whose logarithm is known to be a Lie series.

\subsection{Closed strings as single-valued open strings}
\label{sec:7.6}

In this section, we review the relation between open- and closed-string $\alpha'$-expansions
through the single-valued map of MZVs (see section \ref{sec:7.2.3}). In most parts
of this section, we shall set $\alpha'=\frac{1}{2}$ for open-string quantities and $\alpha'=2$ for 
closed strings in order to implement the rescaling of $\alpha' \rightarrow 4 \alpha'$
in (\ref{stringkltrel}) or (\ref{gravsec.21}).

On the one hand, one can already reduce (integrated) closed-string tree-level amplitudes to 
open-string computations by means of the KLT formula (\ref{stringkltrel}) or (\ref{gravsec.9}). 
On the other hand, the KLT formula does not manifest if some of the MZVs in open-string 
$\alpha'$-expansions (\ref{mzvsec.38}) cancel in between the amplitude factors and the sine
functions. As will be reviewed below, the single-valued map reduces closed-string $\alpha'$-expansions
to those of open strings while exposing all the dropouts of MZVs (including powers of
$\zeta_2$ at low weights and certain indecomposable MZVs such as $\zeta_{3,5}$).

Some of the selection rules on MZVs can already be illustrated from the
$\alpha'$-expansions (\ref{4ptmzv.1}) and (\ref{4ptmzv.3}) of four-point open-
and closed-string amplitudes. It is easy to see at the level of the exponents
of (\ref{4ptmzv.1}) and (\ref{4ptmzv.3}) that the expansions of the disk and 
sphere integrals are related by
\beq
{\rm sv} \bigg(
\frac{ \Gamma(1{-}s_{12})\Gamma(1{-}s_{23}) }{\Gamma(1{-}s_{12}{-}s_{23})}
\bigg) =
\frac{\Gamma(1{-}s_{12})\Gamma(1{-}s_{23})\Gamma(1{+}s_{12}{+}s_{23}) }{
\Gamma(1{+}s_{12})\Gamma(1{+}s_{23})\Gamma(1{-}s_{12}{-}s_{23})}\, ,
\label{svsec.1}
\eeq
where the single-valued map is applied order by order in $\alpha'$ and acts trivially on the $s_{ij}$. 
According to (\ref{mzvsec.16}), sv annihilates even zeta values $\zeta_{2k}$ while doubling 
the odd ones $\zeta_{2k+1}$. Note that we have eliminated $s_{13}={-}s_{12}{-}s_{23}$
in (\ref{4ptmzv.3}) to expose the independent variables. From the perspective of the
four-point KLT formula (\ref{gravsec.4}), the trigonometric expansion of the KLT kernel 
${\cal S}_{\alpha'}$ in terms of even zeta values via (\ref{mzvsec.43}) cancels
all the $\zeta_{2k}$ in the $\alpha'$-expansion of ${\cal A}(1,2,3,4)$ or 
$\frac{ \Gamma(1{-}s_{12})\Gamma(1{-}s_{23}) }{\Gamma(1{-}s_{12}{-}s_{23})}$  and leads 
to the relation (\ref{svsec.1}). In the rest of this section,
we will study the $n$-point generalization of this observation from several perspectives
and describe the cancellation of certain indecomposable MZVs at higher depth from
closed-string amplitudes in terms of the single-valued map.

\subsubsection{From the KLT formula to the single-valued map}
\label{sec:7.6.1}

The selection rules on MZVs in $n$-point closed-string amplitudes
were firstly identified by combining the KLT relations with the
structure (\ref{mzvsec.38}) of the open-string $\alpha'$-expansion
and exploiting conjectural properties of the $M_w, P_w$ matrices \cite{Schlotterer:2012ny}.
The construction in the reference starts from the general form (\ref{gravsec.22}) of the 
KLT relations with a symmetric choice of bases ${\cal B}_1,{\cal B}_2 \rightarrow (1,P,n{-}1,n)$
of permutations. With the expansion (\ref{npttree}) of open-string amplitudes ${\cal A}(\ldots)$ in
terms of SYM tree amplitudes $A(\ldots)$, we obtain
\begin{align}
{\cal M}^{\rm closed}_{n} &= \sum_{P,Q\in S_{n-3}} A(1,P,n{-}1,n)  G^{PQ} \tilde A(1,Q,n{-}1,n)\, ,
\label{svsec.2} \\
G^{PQ} &= \sum_{A,B \in S_{n-3}} (F^t)^{P}{}_A
m^{-1}_{\alpha'}(1,A,n{-}1,n|1,B,n{-}1,n) F_B{}^Q \, . \notag
\end{align}
The next step is to insert the $\alpha'$-expansion (\ref{mzvsec.38}) for both
$(n{-}3)!\times (n{-}3)!$ matrices of disk integrals $(F^t)^{P}{}_A$ 
and $F_B{}^Q$ as well as the observation \cite{Schlotterer:2012ny}
\beq
(\mathbb P^t) m^{-1}_{\alpha'} \mathbb P=  m^{-1} 
\, , \ \ \ \ \ \ 
 M^t_{2k+1} m^{-1}  = m^{-1} M_{2k+1} 
 \label{svsec.2a}
 \eeq
for $\mathbb P= \sum_{k=0}^\infty \zeta_2^k P_{2k}$ in order to move all the $P_w,M_w$
matrices to the right of $m^{-1}_{\alpha'}$. We have suppressed the
permutations indexing the $(n{-}3)! \times (n{-}3)!$ matrices $m^{-1}_{\alpha'}$ and
$m^{-1}$ which are simply the KLT kernels of string and field theory for
the symmetric choice of bases ${\cal B}_1,{\cal B}_2 \rightarrow (1,P,n{-}1,n)$ in 
(\ref{gravsec.22}), for instance
\beq
m_{\alpha'}^{-1}(1,2,3,4|1,2,3,4)=  
\frac{ \sin( \pi s_{12}) \sin( \pi s_{23}) }{ \pi \sin(
\pi(s_{12}{+}s_{23}))} \, , \ \ \ \
m^{-1}(1,2,3,4|1,2,3,4) = \frac{s_{12}s_{23}}{s_{12}+s_{23}}
 \label{svsec.2g}
 \eeq
at four points. By passing to motivic MZVs and applying the isomorphism $\phi$ to 
the $f$-alphabet, (\ref{svsec.2a}) leads to the following simplified form 
of (\ref{svsec.2}) \cite{Schlotterer:2012ny},
\begin{align}
\phi (G^{\mathfrak m}) &= m^{-1} \sum_{r,s=0}^\infty
\sum_{a_1,a_2,\ldots,a_r \atop{ \in 2\mathbb N+1}}
\sum_{b_1,b_2,\ldots,b_s \atop{ \in 2\mathbb N+1}}
M_{a_r}\ldots M_{a_2} M_{a_1} M_{b_1} M_{b_2}\ldots M_{b_s} 
( f_{a_1 } f_{a_2} \ldots f_{a_r} \shuffle f_{b_1} f_{b_2}\ldots f_{b_s})
\notag \\
&= m^{-1} \sum_{r=0}^\infty
\sum_{i_1,i_2,\ldots,i_r \in 2\mathbb N+1} M_{i_1} M_{i_2} \ldots M_{i_r}
\sum_{j=0}^r  ( f_{i_j} \ldots f_{i_2}  f_{i_1} \shuffle  f_{i_{j+1}} f_{i_{j+2}}\ldots f_{i_r}  )\, ,
 \label{svsec.3}
\end{align}
where the sums over words in odd $a_j,b_j$ have been rearranged 
to expose the coefficients of a given matrix product in the last line.
At this point, one can recognize the form (\ref{mzvsec.17}) of the single-valued map in
the $f$-alphabet and obtain the motivic version ${\cal M}^{\mathfrak{m},{\rm closed}}_{n} $
of the closed-string amplitude in the form \cite{Stieberger:2013wea}
\begin{align}
\phi( {\cal M}^{\mathfrak{m},{\rm closed}}_{n} )&= -  \sum_{P,Q,R \in S_{n-3}}  A(1,P,n,n{-}1) S(P|Q)_1
\notag \\
&\quad \times
\sum_{r=0}^{\infty} \sum_{i_1,\ldots,i_r \in 2\mathbb N+1}
{\rm sv} ( f_{i_1} f_{i_2} \ldots f_{i_r} ) 
( M_{i_1} M_{i_2}  \ldots M_{i_r} )_Q{}^R \tilde A(1,R,n{-}1,n)
 \label{svsec.4} \\
 &= -  \sum_{P,Q,R \in S_{n-3}}  A(1,P,n,n{-}1) S(P|Q)_1 \,
\big[ {\rm sv} \, \phi( F^{\mathfrak{m}}) \big]_Q{}^R  \tilde A(1,R,n{-}1,n)
\notag
\end{align}
upon insertion into (\ref{svsec.2}) and changing bases of left-moving
SYM amplitudes to $A(1,P,n,n{-}1)$.\footnote{This has been done via
\[
\sum_{P \in S_{n-3}} A(1,P,n{-}1,n) m^{-1}(1,P,n{-}1,n|1,Q,n{-}1,n)= -\sum_{P \in S_{n-3}} A(1,P,n,n{-}1) S(P|Q)_1\, . \]} 
In passing to the last line, we have identified the series over words
in ${\rm sv}(f_{i_j})$ as the single-valued map of $\phi( F^{\mathfrak{m}})$ 
in (\ref{mzvsec.38}), where ${\rm sv}(f_2)=0$ removes all contributions from the $P_{2k}$.

\subsubsection{Closed-string amplitudes as a field-theory double copy}
\label{sec:7.6.2}

Since the $\phi$-map retains the complete information on the
MZVs in its preimage, we can rewrite (\ref{svsec.4}) as an amplitude relation
\cite{Stieberger:2013wea}
\begin{align}
{\cal M}^{\rm closed}_{n} &=  -  \sum_{P,Q \in S_{n-3}}  A(1,P,n,n{-}1) S(P|Q)_1
\,{\rm sv} \,\tilde{\cal A}(1,Q,n{-}1,n) \, ,
\label{svsec.5a}
\end{align}
where the single-valued image of the entire open-superstring amplitude
(\ref{npttree}) can be presented in one of the following forms:
 \begin{align}
 {\rm sv} \,\tilde  {\cal A}(1,P,n{-}1,n) 
 &= \sum_{Q \in S_{n-3}} ({\rm sv} \, F)_P{}^{Q}A(1,Q,n{-}1,n)
 \label{svsec.5b} \\
 &= 
\sum_{r=0}^{\infty} \sum_{i_1,\ldots,i_r \in 2\mathbb N+1} \!
 \phi^{-1} \big[ {\rm sv} ( f_{i_1} f_{i_2}\ldots f_{i_r} ) \big] \!\sum_{Q\in S_{n-3}}\!  ( M_{i_1} M_{i_2} \ldots M_{i_r} )_P{}^Q   \tilde A(1,Q,n{-}1,n) \notag \\
 &= \sum_{Q\in S_{n-3}} \big(
 1+2\zeta_3 M_3 + 2 \zeta_5 M_5 + 2 \zeta_3^2 M_3^2 + 2 \zeta_7 M_7
 + 2 \zeta_3 \zeta_5 \{ M_3 , M_5\} + 2 \zeta_9 M_9 + \tfrac{4}{3} \zeta_3^3 M_3^3 \notag \\
 &\quad + 2 \zeta_5^2 M_5^2 + 2 \zeta_3 \zeta_7  \{ M_3 , M_7\} + 2 \zeta_{11} M_{11}
 + \zeta_3^2 \zeta_5 (M_3^2 M_5 +2 M_3 M_5 M_3 + M_5 M_3^2) \notag \\
 &\quad + 2 \big(\tfrac{1}{5} \zeta_{3,3,5} - \tfrac{4}{35} \zeta_2^3 \zeta_5
 + \tfrac{6}{25} \zeta_2^2 \zeta_7 + 9 \zeta_2 \zeta_9 \big) [ M_3,[M_3,M_5]]
 + \ldots \big)_P{}^Q   \tilde A(1,Q,n{-}1,n)
\, ,
\notag
\end{align}
with weight or $\alpha'$-orders $\geq 12$ in the ellipsis. The matrix anticommutators
$ \{ M_a , M_b\} = M_a M_b {+}M_b M_a$ along with $\zeta_3 \zeta_5$ and $\zeta_3 \zeta_7$
are the remnant of applying the single-valued map to the contributions from $\zeta_{3,5}$ and
$\zeta_{3,7}$, where the relevant terms of (\ref{mzvsec.34}) are mapped to
\begin{align}
{\rm sv} \bigg( \zeta_3 \zeta_5 M_5 M_3 + \frac{1}{5} \zeta_{3,5} [M_5,M_3] \bigg) &=  
4 \zeta_3 \zeta_5 M_5 M_3 - 2 \zeta_3 \zeta_5 [M_5,M_3] = 2 \zeta_3 \zeta_5 \{ M_3, M_5\} \, , \\
{\rm sv} \bigg( \zeta_3 \zeta_7 M_7 M_3 + \frac{1}{14} ( \zeta_{3,7} + 3 \zeta_5^2 ) [M_7,M_3] \bigg) &=  
4 \zeta_3 \zeta_7 M_7 M_3 - 2 \zeta_3 \zeta_7 [M_7,M_3] = 2 \zeta_3 \zeta_7 \{ M_3, M_7\} \, , \notag
\end{align}
see (\ref{mzvsec.15}) for ${\rm sv} \, \zeta_{3,5}$ and ${\rm sv} \, \zeta_{3,7}$. While 
$\zeta_{3,5}$, $\zeta_{3,7}$ and in fact all higher-depth MZVs with at most two odd letters
$f_{a} f_b$ in their $f$-alphabet image drop out from closed-string amplitudes, the
$\alpha'$-expansion (\ref{svsec.5b}) retains $\zeta_{3,3,5}$ as the simplest conjecturally 
irreducible higher-depth MZV in the image of the single-valued map.

Even though the derivation started out from the string-theory KLT formula
(\ref{gravsec.22}), we brought the closed-string amplitude into the form
(\ref{svsec.5a}) of a {\it field-theory} KLT formula with the $\alpha'$-independent
kernel $S(P|Q)_1$ in (\ref{kltrec}). At tree level, the closed superstring is said
to be a field-theory double copy of SYM with the single-valued open superstring.
A similar type of field-theory double copy was found for the open superstring
in (\ref{AstringP}) with the scalar $Z$-integrals in the place 
of the ${\rm sv} \,\tilde{\cal A}$: both double-copy formulae
(\ref{AstringP}) and (\ref{svsec.5a}) for open and closed superstrings
involve SYM trees as a field-theory building block and carry the entire
$\alpha'$-dependence in a string-theoretic double-copy constituent
$Z$ or ${\rm sv} \,\tilde{\cal A}$.

Permutation invariance of the field-theory
KLT formula hinges on the BCJ relations of the double-copy constituents which
are certainly satisfied for the SYM amplitudes $A(\ldots)$ on the left of the KLT matrix 
in (\ref{svsec.5a}). The single-valued open-superstring amplitudes in turn obey 
BCJ relations \cite{Stieberger:2014hba}
by the reasoning in section \ref{sec:7.x.y} -- the matrix products at each order in 
the $\alpha'$-expansion of (\ref{svsec.5b})
preserve the BCJ relations of the SYM amplitudes.

\subsubsection{Sphere integrals as single-valued disk integrals}
\label{sec:7.6.3}

The relation between closed and single-valued open superstrings as well as
the associated KLT relation (\ref{svsec.5a}) 
can be rewritten at the level of Parke--Taylor-type sphere integrals $J(P|Q)$
defined in (\ref{Jintdef}). This can be seen by inserting the single-valued map of (\ref{AstringP}),
\beq
{\rm sv} \,\widetilde{\cal A}(P)= - \sum_{Q,R \in S_{n-3}} {\rm sv} \, Z(P|1,Q,n,n{-}1) S(Q|R)_1 
\widetilde A(1,R,n{-}1,n)\, ,
 \label{svsec.7}
\eeq
into (\ref{svsec.5a}) and comparing with the
representation (\ref{gravsec.16}) of ${\cal M}^{\rm closed}_{n}$.
The coefficients of the $(n{-}3)!^2$ independent bilinears
$A(1,P,n,n{-}1)\widetilde A(1,Q,n{-}1,n)$ have to agree in both representations of 
${\cal M}^{\rm closed}_{n}$, and we conclude that \cite{Stieberger:2014hba}
\beq
{\rm sv} \, Z(P|Q) = J(P|Q)\, .
 \label{svsec.8}
\eeq
In comparing the definitions (\ref{Zintdef}) and (\ref{Jintdef})
of the disk and sphere integrals, the single-valued map is seen to 
effectively trade a disk integration over the domain $D(P)$ in
(\ref{domain}) for a sphere integration with an insertion of the
antiholomorphic Parke--Taylor factor $\overline{\PT(P)}$. This is
natural from the connection between disk orderings and Parke--Taylor
factors via Betti-deRham duality \cite{betti1, betti2}, relating simple
poles of $\overline{\PT(P)}$ in $\bar z_i{-} \bar z_j$ to inequalities 
$z_i<z_j$ characterizing the integration domain $D(P)$.

Instead of relying on the $\alpha'$-expansion (\ref{mzvsec.38}) of open-superstring 
amplitudes and the properties (\ref{svsec.2a}) of the matrices $P_{w},M_w$,
one can prove (\ref{svsec.8}) at all multiplicities and orders in $\alpha'$ via single-valued 
integration \cite{Schnetz:2013hqa}. A simple ``physicists' proof'' on the basis of the
Betti-deRham duality between $D(P)$ and $\overline{\PT(P)}$ as well as
standard transcendentality conjectures on MZVs can be found in \cite{Schlotterer:2018zce},
and the reader is referred to \cite{Brown:2018omk, Brown:2019wna} for a mathematically 
rigorous proof. Moreover, the fact that the expansion coefficients of $J(P|Q)$ are single-valued
MZVs can be explained from the study of single-valued correlation functions 
\cite{Vanhove:2018elu}.

As an important plausibility check of (\ref{svsec.8}), we note that both sides 
obey BCJ relations in both $P$ and $Q$. First, IBP
relations among Parke--Taylor factors readily imply the BCJ relations of $Z$ and $J$ in $Q$ 
and those of $J$ in $P$. Second, the sv-action $\frac{1}{\pi} \sin(\pi x) \rightarrow x$ 
on the trigonometric factors of section \ref{sec:6.5.1}
maps the monodromy relations of $Z$ in $P$ into BCJ relations, see
section \ref{sec:7.3.4} and \cite{Stieberger:2014hba}.

\subsubsection{The web of field-theory double copies for string amplitudes}
\label{sec:7.6.x}

Single-trace amplitudes ${\cal A}^{\rm het}$ of gauge multiplets in heterotic
string theories have been expressed in terms of SYM trees and the
sphere integrals $J$ in (\ref{hetEYM.32}). By the relation (\ref{svsec.8}) between
sphere and single-valued disk integrals, one can identify \cite{Stieberger:2014hba},
\beq
{\cal A}^{\rm het}(P) = {\rm sv} \, {\cal A}(P) \, ,
\label{svhet.01}
\eeq
i.e.\ single-trace amplitudes of the gauge multiplet in type I and heterotic string theories
are related by the single-valued map. Moreover, the field-theory double copy
(\ref{hetEYM.15}) together with (\ref{svsec.7}) and (\ref{svsec.8}) imply that all 
massless tree amplitudes for the heterotic
string reduce to single-valued type I amplitudes \cite{Azevedo:2018dgo},
\beq
{\cal M}^{\rm het}_n =- \sum_{P,Q \in S_{n-3}} 
A_{(DF)^2+{\rm YM}+\phi^3}(1,P,n,n{-}1) S(P|Q)_1 \,
{\rm sv} \, {\cal A}(1,Q,n{-}1,n)\, ,
\label{svhet.02}
\eeq
where the amplitudes $A_{(DF)^2+{\rm YM}+\phi^3}$ of the $(DF)^2+{\rm YM}+\phi^3$ 
field theory \cite{Johansson:2017srf} (see section \ref{sec:7.7.7}) are rational functions 
of $\alpha'$. Just like the expression (\ref{svsec.5a}) for type II amplitudes,
(\ref{svhet.02}) double-copies single-valued open superstrings with a field theory
(with $A_{(DF)^2+{\rm YM}+\phi^3}$ in the place of SYM amplitudes in case of
the heterotic string).

Similar double-copy formulae apply to bosonic strings: removing the bi-adjoint scalars
from the $(DF)^2+{\rm YM}+\phi^3$ theory leaves a simpler field theory $(DF)^2+{\rm YM}$ 
with the same massive states \cite{Johansson:2017srf}
which casts $n$-point tree-level amplitudes ${\cal A}_n^{\rm bos}$ and ${\cal M}_n^{\rm bos}$
 of open and closed bosonic strings into the compact form \cite{Azevedo:2018dgo}
\begin{align}
{\cal A}_n^{\rm bos}(R) &=
- \sum_{P,Q \in S_{n-3}} 
A_{(DF)^2+{\rm YM}}(1,P,n,n{-}1) S(P|Q)_1 
Z(R|1,Q,n{-}1,n)\, ,
\label{svhet.03}\\
{\cal M}_n^{\rm bos} &=
- \sum_{P,Q \in S_{n-3}} 
\tilde A_{(DF)^2+{\rm YM}}(1,P,n,n{-}1) S(P|Q)_1 \,
{\rm sv} \, {\cal A}^{\rm bos}(1,Q,n{-}1,n)\, .
\notag
\end{align}
Moreover, the gravity sector of heterotic-string amplitudes admits an alternative
form \cite{Azevedo:2018dgo}
\beq
{\cal M}_n^{\rm het} \, \big|_{\rm grav} =
- \sum_{P,Q \in S_{n-3}} 
\tilde A(1,P,n,n{-}1) S(P|Q)_1 \,
{\rm sv} \, {\cal A}^{\rm bos}(1,Q,n{-}1,n)\, ,
\label{svhet.04}
\eeq
where the supersymmetries arise from the opposite double-copy constituent
as compared to (\ref{svhet.02}) -- from the SYM field-theory amplitudes $\tilde A$
instead of single-valued superstring disk amplitudes.

A summary of the field-theory double-copy formulae (\ref{AstringP}), (\ref{svsec.5a}) and (\ref{svhet.01}) to (\ref{svhet.04}) for tree amplitudes in various
string theories can be found in table \ref{dcpyarray}. In all cases, the double copy refers to the
KLT formula with $\alpha'$-independent kernel $S(P|Q)_1$ and features a
field-theory building block (SYM, $(DF)^2+{\rm YM}$ or $(DF)^2+{\rm YM}+\phi^3$)
without any transcendentality. The infinite tower of massive poles characteristic to string
amplitudes occurs through the other double-copy constituent -- either
the universal basis of disk integrals $Z$ for open strings or single-valued open-string
amplitudes in case of type II, heterotic or closed bosonic strings.

\begin{table}[h]
\begin{center}
\begin{tabular}{|c || c  | c  |c  |}\hline 
string $\otimes$ QFT &SYM &$(DF)^2+{\rm YM}$ &$(DF)^2+{\rm YM}+\phi^3$ \\\hline \hline
$Z$-theory &open superstring   &open bosonic string &comp.\ open bosonic string \\ \hline
sv(open superstring)&closed superstring  &heterotic (gravity) &heterotic (gauge \& gravity) \\ \hline
sv(open bosonic string)&heterotic (gravity)   &closed bosonic string &comp.\ closed bosonic string
\\ \hline
\end{tabular}
\end{center}
\caption{Double copy constructions of tree-level amplitudes in various string theories as
presented in \cite{Azevedo:2018dgo}.}
\label{dcpyarray}
\end{table}

Note that the compactified versions of open and closed 
bosonic strings in the rightmost column of table~\ref{dcpyarray} refer to the 
geometric realization of the gauge sector of the heterotic string: the Kac--Moody
currents ${\cal J}^a(z)$ in the vertex operators of the gauge multiplet in section \ref{sec:7.7.5}
can be obtained from compactifying free bosons $\partial_z X^I(z)$ on a torus, where
$I$ labels the Cartan generators of the gauge group \cite{Gross:1984dd}.

\subsubsection{Twisted KLT relations}
\label{sec:7.6.4}

An interesting variant of the sphere integrals $J(P|Q)$ in (\ref{Jintdef})
arises in so-called chiral or twisted string theories \cite{Hohm:2013jaa, Siegel:2015axg, 
Huang:2016bdd}.
These theories are characterized by finite spectra due to a flipped level-matching
condition that can be informally identified with a relative sign flip of $\alpha'$ between 
left- and right-moving worldsheet degrees of freedom. In particular, the spectrum of 
twisted type II superstrings reduces to the associated supergravity multiplets.

At the level of the sphere integrals
in the tree-level amplitudes of twisted string theories, the sign flip between left-
and right movers applies to the antiholomorphic
part of the Koba--Nielsen factor,
\beq
\widehat J(P|Q) := \bigg({-}\frac{ \ap}{2\pi} \bigg)^{n-3}\!\!\!\!
\int \limits_{\mathbb C^{n-3}} {d^2z_1 \ d^2z_2 \ \cdots  \ d^2z_n \over {\rm vol}({\rm SL}_2(\Bbb C))}
\, \prod_{i<j}^n \bigg( \frac{ \bar z_{ij} }{ z_{ij}} \bigg)^{\frac{\ap}{2} s_{ij}}   \PT(Q) \overline{\PT(P)} \, ,
 \label{svsec.9}
\eeq
where the single-valued factors of $|z_{ij}|^{-\alpha' s_{ij}}$ in (\ref{Jintdef}) 
from the correlators of conventional strings are replaced by 
$(\frac{ \bar z_{ij} }{ z_{ij}} )^{\alpha' s_{ij}/2}$. Apart from these 
modifications of the Koba--Nielsen factors, 
the chiral correlators ${\cal K}_n$ among closed-string vertex operators 
can be freely interchanged between the type II versions of twisted and 
conventional strings \cite{Huang:2016bdd, LipinskiJusinskas:2019cej}.
Hence, the supergravity $n$-point function, computed from twisted type II strings,
takes the form of (\ref{gravsec.16}) with the modified sphere integrals 
$\widehat J(P|Q)$ in the place of $J(P|Q)$. In order to arrive at the field-theory
KLT formula (\ref{KLTrel}) for supergravity, the sphere integrals of the twisted strings
have to directly match the doubly-partial amplitudes $m(P|Q)$,
\beq
\widehat J(P|Q) = m(P|Q) \, .
 \label{svsec.10}
\eeq
However, the sphere integrals (\ref{svsec.9}) are ill-defined due of the multivalued
factors of $(\frac{ \bar z_{ij} }{ z_{ij}} )^{\alpha' s_{ij}/2}$ in the integrand. Still, one
can formally define $\widehat J(P|Q) $ by a KLT formula, where the reversal 
of $\alpha'$ along with the antiholomorphic $\bar z_{ij}$
leads to a sign-flipped version of standard $Z$-integrals
\beq
\widehat Z(P|Q) = Z(P|Q)\, \big|_{\alpha' \rightarrow - \alpha'} \, ,
 \label{svsec.11}
\eeq
namely
\beq
\widehat J(A|B) = - \sum_{P,Q \in S_{n-3}} Z(1,P,n{-}1,n|A) {\cal S}_{\alpha'}(P|Q)_1 \widehat{Z}(1,Q,n,n{-}1|B) \, .
 \label{svsec.12}
\eeq
Upon comparison with the requirement (\ref{svsec.10}),
the KLT formula for the twisted sphere integral needs to 
reproduce the $\alpha'$-independent doubly-partial amplitude,
\beq
 m(A|B) = - \sum_{P,Q \in S_{n-3}} Z(1,P,n{-}1,n|A) {\cal S}_{\alpha'}(P|Q)_1 \widehat{Z}(1,Q,n,n{-}1|B)
  \, .
 \label{other12}
\eeq
Hence, the conclusion is that a sign-flip in one of the $Z$-integrals in the
conventional KLT formula (\ref{gravsec.9}) is enough to cancel the entire tower 
of $\alpha'$-corrections. In fact, (\ref{other12}) can be deduced from the 
twisted period relations \cite{cho1995} as introduced into the physics literature
in \cite{Mizera:2017rqa}.

Another way of understanding the dropout of $\alpha'$-corrections from
(\ref{svsec.9}) is to revisit the simplification of the sphere
integrals in section \ref{sec:7.6.1}. The sign flip effectively reverses 
$M_{2k+1} \rightarrow - M_{2k+1}$ in one of the $F$-factors in the matrix $G^{PQ}$ of
sphere integrals in (\ref{svsec.2}) and turns (\ref{svsec.3}) into
\beq
\phi (G^{\mathfrak m})  \rightarrow  m^{-1} \sum_{r=0}^\infty
\sum_{i_1,i_2,\ldots,i_r \in 2\mathbb N+1} M_{i_1} M_{i_2} \ldots M_{i_r}
\sum_{j=0}^r (-1)^j 
( f_{i_j} \ldots f_{i_2}  f_{i_1} \shuffle  f_{i_{j+1}} f_{i_{j+2}}\ldots f_{i_r}  )
\, .
 \label{svsec.13}
\eeq
By the alternating signs $(-1)^j$ on the right-hand side, the coefficient of each non-trivial matrix 
product $M_{i_1} \ldots M_{i_r}$ with $r \neq 0$ cancels \cite{Huang:2016bdd}. 
Hence, the matrix $G^{PQ}$ in (\ref{svsec.2}) is mapped to the KLT kernel 
in passing to the twisted string, and we obtain supergravity amplitudes as expected.

One can also turn the logic around and impose that the expression for $G^{PQ}$
in (\ref{svsec.2}) reduces to
\beq
m^{-1}(1,P,n{-}1,n|1,Q,n{-}1,n)= \sum_{A,B \in S_{n-3}} (F^t)^{P}{}_A
m^{-1}_{\alpha'}(1,A,n{-}1,n|1,B,n{-}1,n) \widehat F_B{}^Q
 \label{svsec.14}
\eeq
with $ \widehat F_B{}^Q = F_B{}^Q\big|_{\alpha' \rightarrow - \alpha'} $ 
as in (\ref{svsec.11}). The coefficients of $f_2^{\ell=0,1,2,3,\ldots}$
or $f_{2k+1}$ in (\ref{svsec.14}) then imply the properties 
(\ref{svsec.2a}) of the matrices $P_w,M_w$.

Finally, we note an amusing variant of (\ref{other12}): instead of obtaining doubly-partial
amplitudes $m(P|Q)$ by contracting $Z(A|P)\widehat{Z}(B|Q)$
with the string-theory KLT kernel ${\cal S}_{\alpha'}(A|B)_1$, one can instead contract
the free indices $P,Q$ with the field-theory kernel $S(P|Q)_1$. In this way, one arrives
at the inverse $m_{\alpha'}$ of the {\it string-theory} KLT kernel \cite{Mizera:2016jhj}
\begin{align}
m_{\alpha'}(A|B) = - \sum_{P,Q \in S_{n-3}} Z(A|1,P,n{-}1,n) S(P|Q)_1 \widehat{Z}(B|1,Q,n,n{-}1)\, .
 \label{svsec.15}
\end{align}
Note that the heterotic and bosonic versions of twisted strings provide a
worldsheet realization of the $(DF)^2+{\rm YM}$ and $(DF)^2+{\rm YM}+\phi^3$
theories \cite{Azevedo:2019zbn}: up to the sign flip $\alpha' \rightarrow - \alpha'$
between left- and right-movers, the Parke--Taylor expansion (\ref{hetEYM.12})
of bosonic correlators takes the identical form in twisted string theories.
The massive modes in the $(DF)^2+{\rm YM}$ and $(DF)^2+{\rm YM}+\phi^3$
theories arise from asymmetric double copies of vertex operators
for an open-string tachyon and the first mass level of compactified open bosonic strings. 
Similarly, the spectrum of heterotic twisted strings contains a colorless spin-two multiplet 
from a double copy of tachyons with the first mass level of the open superstring. 

The generalizations of the correlators ${\cal K}_n, {\cal K}^{\rm bos}_n$ to massive
states also take a universal form for conventional and twisted strings
up to $\alpha' \rightarrow - \alpha'$. This was exploited in \cite{Guillen:2021mwp}
to pioneer field-theory double-copy structures in tree-level amplitudes
involving massive open- and closed-superstring states based on tools
from the heterotic twisted string.

\section{Conclusion and outlook}
\label{sec:conclu}

This work aims to give a comprehensive review of string tree-level amplitudes
in the pure spinor formalism. The manifestly spacetime supersymmetric worldsheet 
description of the pure spinor superstring reviewed in section \ref{sec:theform}
introduces massless open-string excitations in the framework of ten-dimensional
SYM, see section \ref{sec:10dSYM}. The OPEs of these superspace vertex operators
give rise to the multiparticle formalism whose rich combinatorial structure has been
presented from different perspectives in section \ref{multiSYMsec}. The multiparticle
formalism connects conformal-field-theory techniques with recursive organizations
of Feynman diagrams and led to compact formulae for $n$-point tree amplitudes
of SYM, see section \ref{SYMsec}.

The setup of the first sections is the key to find the decomposition (\ref{nptdisk}) of $n$-point
superstring disk amplitudes into a basis of color-ordered SYM trees. This is the main result of this
review whose derivation and interplay with disk integrals and their field-theory limit
is presented in section \ref{sec:diskamp}. The structure of the disk amplitude together
with its corollaries for type II superstrings and heterotic strings have profound implications
on field-theory amplitudes reviewed in section \ref{AmpRelsec} -- the color-kinematics
duality of gauge theories and Goldstone bosons as well as double-copy 
descriptions of gravity, Born--Infeld and Einstein--Yang--Mills. The last section \ref{apsec} 
is dedicated to the low-energy expansion of superstring tree-level amplitudes and
the elegant mathematical structures of the multiple zeta values therein.

From the material in this review, both the moduli-space integrand for $n$-point string
tree-level amplitudes and the $\alpha'$-expansion of the integrated expressions are
available to any desired order. We have presented the strong connectivity of both
the integrands and the integrated results with the web of double-copies among field-theory 
amplitudes and various areas of pure mathematics including combinatorics, number theory 
and algebraic geometry. The detailed control over string tree amplitudes cross-fertilizes with
ambitious questions on string dualities (say through the multiple zeta values in
multiparticle type IIB amplitudes) but also offers new connections between perturbative
string theories beyond any known duality (e.g.\ gauge amplitudes of heterotic strings
as single-valued type I amplitudes).

The diverse insights unlocked by the results on string tree-level amplitudes in this review
motivate a similar investigation of loop amplitudes, where already the last years
witnessed progress on several frontiers. We shall now give an overview of recent
loop-level developments that generalize selected aspects of this review beyond tree
level. The subsequent path through the literature is far from complete and may
quickly become outdated after the time of writing this review. The reader is
referred to \cite{Berkovits:2022fth} for an overview of loop-level amplitude computations
in the pure spinor formalism as of October 2022 and to the
white paper \cite{Berkovits:2022ivl} for a status report on a broader selection
of topics in string perturbation theory as of March 2022.

\subsection{Loop amplitudes in the pure spinor formalism}
\label{sec:9.1}

By the manifest spacetime supersymmetry of the pure spinor formalism,
it automatically incorporates a variety of cancellations in loop amplitudes
between internal bosons and fermions. In the pure spinor prescription for 
loop-level open- and closed-string amplitudes, many of these 
cancellations can be traced back
to the saturation of fermionic zero modes. The loop-amplitude prescription
in the ``minimal'' worldsheet variables of section~\ref{sec:theform} dates back
to 2004 \cite{Berkovits:2004px}, followed by its extension to ``non-minimal'' 
variables in 2005 \cite{Berkovits:2005bt}. A central ingredient in loop amplitudes
of the pure spinor superstring is a composite $b$-ghost whose explicit form
in the non-minimal variables \cite{Berkovits:2005bt} involves poles in the 
pure spinor ghosts.

Just like at tree level, the loop amplitudes computed
from these prescriptions automatically involve kinematic factors in pure spinor
superspace after integrating out the non-zero modes of the worldsheet variables.
Moreover, the pure spinor formalism is readily compatible with the chiral-splitting 
procedure \cite{DHoker:1988pdl, DHoker:1989cxq}
to express closed-string correlators at arbitrary genus as a holomorphic square
of chiral amplitudes that integrate to open-string amplitudes after specifying
boundary conditions for the endpoints. Accordingly, the subsequent
status report on explicit loop-level computations in the pure spinor formalism
refers to both open- and closed-string amplitudes unless stated otherwise.
In fact, the so-called homology invariance
of chiral amplitudes -- their single-valuedness on higher-genus surfaces under
suitable shifts of the loop momenta -- provided crucial input for
recent loop-amplitude computations in the pure spinor formalism.

The constraints from zero-mode counting facilitated the derivation of 
non-renormalization theorems in string theory \cite{Berkovits:2006vc, Berkovits:2009aw} 
and led to multiloop results on the ultraviolet structure of maximal supergravity
through a worldline version of the pure spinor formalism \cite{BjornssonWM}.
The computation of non-vanishing string loop amplitudes with the pure spinor formalism 
was initiated with the one-loop four-point amplitude in 2004 \cite{Berkovits:2004px}
and the two-loop four-point amplitude in 2005 \cite{twoloop}.
The bosonic components of the two-loop result were later on confirmed \cite{2looptwo} to 
reproduce the earlier two-loop four-point computation in the RNS formalism~\cite{DHoker:2005vch}.

The non-minimal pure spinor formalism has been used to compute one-loop
five-point amplitudes \cite{Mafra:2009wi}, the exactly normalized
four-point amplitudes at one loop \cite{1loopH} and two loops \cite{2loopH} 
as well as the low-energy limits of the three-loop four-point \cite{3loop} and
two-loop five-point \cite{Gomez:2015uha} amplitudes. In all of these cases,
the $b$-ghosts only contribute through their zero modes. However, the non-zero
modes of the $b$-ghost and the complexity of its multiparticle correlators currently 
cause a bottleneck in performing higher-order computations directly from the 
prescription. Still, consistency conditions on loop amplitudes and in particular 
the multiparticle formalism of section \ref{multiSYMsec} often allowed to 
circumvent the most daunting challenges from the $b$-ghost and led to
many recent advances on higher-point amplitudes.

The multiparticle formalism spawned simplified expressions for one-loop 
open- and closed-string amplitudes in an
integral basis at five points \cite{GreenBZA} and at six points
\cite{Mafra:2016nwr}. The latter reference also reconciles the hexagon
gauge anomaly of individual worldsheet diagrams in type I 
theories \cite{Green:1984qs, Green:1984sg} 
with BRST cohomology techniques and derives the anomaly kinematic factor \cite{anomaly}
from an explicit amplitude representation. Based on additional input from locality,
chiral splitting and the associated homology invariance, a systematic procedure
to derive one-loop correlators is described in
\cite{Mafra:2018nla, Mafra:2018pll, oneloopIII}. The resulting chiral amplitudes
enjoy a double-copy structure \cite{Mafra:2017ioj} similar to the 
KLT-type formula for the open-superstring correlator (\ref{KLTcorrel}) at genus zero.
However, the coefficients of holomorphic Eisenstein series in $(n \geq 8)$-point correlators
have so far resisted a computation from this method.

Similarly, two-loop five-point amplitudes were constructed beyond their low-energy 
limit by a confluence of BRST invariance, locality and chiral-splitting 
techniques \cite{DHoker:2020prr}. Their parity-even bosonic components
were later on verified from a first-principles computation in the RNS 
formalism \cite{DHoker:2021kks}. Finally, an exact-in-$\alpha'$
expression for the three-loop four-point amplitude was proposed in \cite{Geyer:2021oox}
based on input from the field-theory limit, ambitwistor strings and modular invariance.
It would be interesting to analyze this three-loop result from a pure spinor perspective.

The combined power of the multiparticle formalism, BRST invariance and locality has
also been used to directly propose loop integrands for ten-dimensional SYM, see
\cite{towardsOne} for one-loop integrands up to six points and \cite{towardsTwo} for
two-loop five points. The five-point results at one and two loops readily manifested 
the color-kinematics duality and induced the corresponding loop integrands for 
type II supergravity in pure spinor superspace via double copy as in \cite{Carrasco:2011mn}. 
Based on the tropical-geometry methods of \cite{Tourkine:2013rda}, these five-point 
field-theory amplitudes were independently derived from the $\alpha'\rightarrow 0$ limit 
of the corresponding string amplitudes at one loop \cite{towardsOne} and at 
two loops \cite{DHoker:2020prr}.

However, the present pure spinor methods leave open questions on the loop-level 
realization of the color-kinematics duality and double copy at $n\geq 6$ points.
The first superspace construction of one-loop six-point SYM numerators in 
\cite{towardsOne} violated certain kinematic Jacobi identities. These violations disappear\footnote{The violations of kinematic Jacobi identities in the one-loop six-point
results of \cite{towardsOne} also disappear in MHV helicity configurations upon 
dimensional reduction to $D=4$ \cite{He:2015wgf}.} 
in passing to the linearized variant of Feynman propagators \cite{He:2017spx} that typically 
arise from ambitwistor strings \cite{Geyer:2015bja}. The resulting supergravity 
integrands on linearized propagators are available in KLT- and cubic-diagram 
form \cite{He:2016mzd, He:2017spx} obtained from the forward 
limits of (\ref{KLTrel}) and (\ref{gravsec.14}). A solution of all the one-loop kinematic 
Jacobi identities on quadratic propagators was offered in
\cite{Bridges:2021ebs} by taking the field-theory limit of the corresponding
string amplitudes in different color orderings. However, the conventional cubic-diagram
double copy \cite{loopBCJ} of these color-kinematics dual SYM numerators conflicts with 
BRST invariance, so it is an open problem to construct supergravity loop integrands on 
quadratic propagators at $n\geq 6$ points.

Both the above subtleties in finding string-theory realizations
of the gravitational double copy and the non-zero mode contributions
of the $b$-ghosts kick in at the one-loop six-point level. One may speculate 
about a connection between the two kinds of challenges, for instance whether
an incorporation of the $b$-ghost into the multiparticle formalism is the
missing puzzle piece for $n$-point one-loop string amplitudes with manifest
double-copy structure in the field-theory limit. This scenario is supported by 
the role of the $b$-ghost for a kinematic algebra
and its connection with tree-level multiparticle superfields identified 
in \cite{Ben-Shahar:2021doh}. Also higher-loop string amplitudes call for further investigations 
of the $b$-ghost since its poles in the pure spinor ghosts necessitate regularization techniques
such as \cite{Aisaka:2009yp, Grassi:2009fe} at genus $g\geq 3$.

\subsection{Worldsheet integrals in loop-level string amplitudes}
\label{sec:9.2}

A central line of tree-level results in this review is driven by the Parke--Taylor bases of
disk integrals $Z$ in (\ref{Zintdef}) and sphere integrals $J$ in (\ref{Jintdef}).
First, their integration-by-parts relations and $(n{-}3)!$ bases fruitfully 
resonate with the BRST properties and BCJ relations of the accompanying 
kinematic factors in pure spinor superspace. Second, their logarithmic singularities (including 
the absence of double poles) ensure that the $\alpha'$-expansion is uniformly
transcendental and in fact realizes field-theory amplitude relations along with
infinite families of multiple zeta values. These properties of Parke--Taylor integrals
at tree level motivate the goal of constructing similar kinds of integral bases at higher genus.

At genus one, generating functions $Z_\eta$ of open-string integrals in different string theories
furnish conjectural $(n{-}1)!$-element bases with uniformly transcendental
$\alpha'$-expansions \cite{Mafra:2019ddf, Mafra:2019xms}. At suitable orders in the
bookkeeping variables $\eta_2,\eta_3,\ldots,\eta_n$, one can read off the combinations of 
theta functions for one-loop correlators of the pure spinor superstring that share the 
logarithmic singularities of the Parke--Taylor factors. More precisely, the function space 
to assemble the one-loop analogues
of the tree-level correlators ${\cal K}_n$ in (\ref{KLTcorrel}) is controlled by the loop momenta
of the chiral-splitting procedure \cite{DHoker:1988pdl, DHoker:1989cxq} and
the coefficients $g^{(k)}(z,\tau)$ of the Kronecker--Eisenstein series \cite{Broedel:2014vla},
\beq
\frac{ \theta_1'(0,\tau) \theta_1(z{+}\eta,\tau) }{\theta_1(z,\tau)\theta_1(\eta,\tau)} = \frac{1}{\eta}
+ \sum_{k=1}^{\infty} \eta^{k-1} g^{(k)}(z,\tau) \, .
\label{kroneckerEisenstein}
\eeq
After integration over the loop momenta, the moduli-space integrand of open- and 
closed-string amplitudes is expressed in terms of doubly-periodic versions $f^{(k)}(z,\tau)$ of
the Kronecker-Eisenstein coefficients $g^{(k)}(z,\tau)$ which also manifest the modular
properties in the closed-string case.

In the same way as Parke--Taylor factors share the BCJ relations of gauge-theory tree
amplitudes, the combinations of $g^{(k)}(z_{ij},\tau)$ and loop momenta
seen in one-loop correlators \cite{Mafra:2017ioj, oneloopIII} are observed to obey the same identities
as the BRST-invariant kinematic factors \cite{Mafra:2018pll}. Up to and including seven points, 
this duality between kinematics and worldsheet functions is fully established in the references
and underpins a double-copy structure in the chiral amplitudes. Starting from eight points,
the chiral amplitudes involve holomorphic Eisenstein series ${\rm G}_k(\tau)= - g^{(k)}(0,\tau)$ with $k\geq 4$, and it is an open problem to accommodate them into the duality between 
kinematics and worldsheet functions.

The loop-level analogue of the string-theory KLT relation of section \ref{sec:6.5.2} 
is uncharted terrain at the time of writing. However, the tree-level monodromy
relations whose vibrant interplay with KLT relations was illustrated in section \ref{sec:6.5.1} 
were generalized to loop level, investigated from several perspectives 
\cite{Tourkine:2016bak, Hohenegger:2017kqy, Tourkine:2019ukp, 
Casali:2019ihm, Casali:2020knc} and extended to one-loop amplitude 
relations between mixed open-and-closed-string amplitudes and pure open-string 
amplitudes \cite{Stieberger:2021daa}. It would be very interesting to relate
properties of the Kronecker--Eisenstein-type functions in the chiral correlators 
to one-loop monodromy relations as done for Parke--Taylor factors and disk orderings 
through the relations (\ref{ZKK}), (\ref{ZBCJ}) and (\ref{sepmonoZ}) of the $Z(P|Q)$ integrals.

The dependence of integrated string loop amplitudes on $\alpha'$ and the kinematic variables
is more involved than at tree level and features branch cuts in addition to the poles for
the infinite tower of massive string modes. At one loop for instance, the analytic continuations in the
external momenta required by the integral representations and compatible with the poles
and branch cuts are discussed in \cite{DHoker:1993hvl, DHoker:1993vpp, DHoker:1994gnm}. Already the $\alpha'$-expansion of one-loop string amplitudes
contains logarithms in Mandelstam invariants on top of the Laurent series in $s_{ij\ldots k}$
seen in the tree-level $\alpha'$-expansions of section \ref{apsec}. The logarithms in one-loop
four-point closed-string amplitudes due to effective tree-level interactions $D^{2k}\mathbb R^m$ 
were pioneered in \cite{Green:2008uj, DHoker:2015gmr} and computed to all orders in
$\alpha'$ in \cite{DHoker:2019blr}. Two recent lines of attack to determine the non-analytic
sector of higher-point one-loop string amplitudes are based on one-loop matrix elements of tree-level effective interactions \cite{Edison:2021ebi} and an implementation of Witten's $i\epsilon$ prescription \cite{Eberhardt:2022zay}.

A prominent motivation for the computation and low-energy expansion of string loop amplitudes
stems from their implications for string dualities. In this context, the main interest is in
the analytic contributions to the $\alpha'$-expansion which reflect new interactions
in the loop-level effective actions. For type IIB superstrings, the ${\rm SL}_2(\mathbb Z)$-invariance
w.r.t.\ the axio-dilaton field \cite{Hull:1994ys} must be realized in the coefficients of all 
independent $D^{2k}\mathbb R^m$
interactions. Perturbative string amplitudes carry important information on these modular 
invariant functions of the string coupling. Four-point string amplitudes
up to and including three loops were successfully matched with ${\rm SL}_2(\mathbb Z)$-invariant
coefficients of the $\mathbb R^4$, $D^{4}\mathbb R^4$ and $D^{6}\mathbb R^4$ 
interactions \cite{Green:1997tv, Green:1999pu, DHoker:2005jhf, Green:2005ba, 3loop, DHoker:2014oxd}.

Pure spinor methods gave rise to compact representations of loop amplitudes also beyond
four points. Together with a low-energy expansion of the worldsheet integrals, the
five-point amplitude computations at one loop \cite{GreenBZA} and two loops 
\cite{DHoker:2020tcq} made a duality analysis of $D^{2k}\mathbb R^5$ interactions 
tractable. Both references confirmed the duality properties of supermultiplet components
that violate the $U(1)$ R-symmetry of type IIB supergravity 
\cite{Boels:2012zr, Boels:2013jua, Green:2019rhz} and cannot arise
in four-point string amplitudes \cite{Green:1999qt}. Moreover, the loop-level effective actions
will involve new superinvariants starting with $D^{6}\mathbb R^5$ that 
are absent in the tree-level effective action \cite{GreenBZA, DHoker:2020tcq}. 
The classification of independent interactions and ${\rm SL}_2(\mathbb Z)$-invariant
type IIB couplings necessitates precise control over the tensor structure of multiparticle
loop amplitudes as provided by pure spinor superspace.

A frequently used strategy towards low-energy expansions of string loop amplitudes
is to first integrate over vertex insertion points prior to the complex-structure moduli $\tau$
of the genus-$g$ surface. In this way, the $\alpha'$-expansions generate infinite
families of modular invariant functions of $\tau$ w.r.t.\ ${\rm Sp}_{2g}(\mathbb Z)$ on the
worldsheet (rather than the ${\rm SL}_2(\mathbb Z)$ acting on the axio-dilaton field in
case of type IIB). These functions generalize the (single-valued) MZVs of section \ref{apsec}
to higher genus and were dubbed modular graph forms in \cite{DHoker:2015wxz, DHoker:2016mwo} 
after earlier case studies in \cite{Green:1999pv, Green:2008uj, DHoker:2015gmr}. Already at genus
one, modular graph forms stimulated interdisciplinary research
lines at the interface of string theory, algebraic geometry and number theory,
see e.g.\ \cite{Gerken:review} for an overview as of November 2020, \cite{DHoker:2022dxx} for
lecture notes and \cite{Dorigoni:2022npe} for the connection with Brown's equivariant iterated Eisenstein integrals \cite{Brown:2017qwo, Brown:2017qwo2}. The study of
higher-genus modular graph forms started with \cite{DHoker:2013fcx, DHoker:2014oxd, 
Pioline:2015qha, DHoker:2017pvk, DHoker:2018mys, Basu:2018bde} and suggests a 
generalization to modular graph tensors \cite{DHoker:2020uid}.

From a string-theory perspective, a major appeal of modular graph forms is to investigate
the loop-level generalization of the tree-level relation (\ref{svsec.8}) between closed-string 
and single-valued open-string integrals. In one-loop amplitudes of open superstrings, 
the iterated integrals over vertex-operator insertions on a cylinder- or M\"obius-strip
boundary were shown in \cite{Broedel:2014vla, Broedel:2017jdo} to yield 
elliptic MZVs \cite{Enriquez:Emzv} and elliptic polylogarithms
\cite{BrownLev}. There is a variety of evidence \cite{Zerbini:2015rss, DHoker:2015wxz, 
Broedel:2018izr, Zagier:2019eus} that modular graph forms may be viewed as single-valued 
elliptic MZVs. In particular, the closed-string counterparts \cite{Gerken:2019cxz, Gerken:2020yii}
of the conjectural $(n{-}1)!$ basis $Z_\eta$ of one-loop open-string integrals
led to an explicit all-order proposal \cite{Gerken:2020xfv} how to relate 
modular graph forms to single-valued elliptic MZVs in the 
respective $\alpha'$-expansions. On the one hand, this line of reasoning aims to
extract the more challenging configuration-space integrals over
punctured tori from the simpler iterated integrals over cylinder- and M\"obius-strip boundaries. 
On the other hand, this research direction may reveal loop-level manifestations 
of a deeper relation between closed and open strings beyond any known
string duality.

\section*{Acknowledgements}

We are grateful to Maor Ben-Shahar, Nathan Berkovits, Lorenz Eberhardt, 
Max Guillen and Callum Hunter for helpful discussions and comments on a draft version of this work.
Moreover, we would like to thank the anonymous referee for a highly diligent study of the manuscript, several important corrections and numerous valuable suggestions. CRM is
supported by a University Research Fellowship from the Royal Society.
OS is supported by the European Research Council under
ERC-STG-804286 UNISCAMP. CRM would like to thank the engineers and the construction workers
responsible for good quality of the electric poles alongside the roads in Waterloo, Canada.
Without them, this work would not have been possible in its current form. OS dedicates this work to the 18th birthday of his brother Robin Schlotterer -- finally you are old enough to have fun with pure spinors, Robin.

\appendix
\section{Gamma matrices}
\label{sec:appA}

Pure spinor calculations in ten dimensions  often involve the handling of
gamma matrices. In this appendix we review some of the most common manipulations involving
ten-dimensional gamma matrices (for a computer implementation, see \cite{ulf}).
Most of this material can also be found in \cite{guttenberg}.
In particular, the book \cite{Freedman:2012zz} contains a variety of discussions in general
dimensions and should be consulted for further reading.

\subsection{The Clifford Algebra in ${\mathbb R}^{1,9}$}
\label{sec:appA.2}

\paragraph{Lorentzian signature}
The $32\times32$ Dirac matrices $\Gamma^m$  in ten-dimensional Minkowski space ${\mathbb R}^{1,9}$
with $m=0, \ldots, 9$
satisfy the Clifford algebra
\beq\label{LorentzClifford}
\{ \Gamma^m, \Gamma^n \} = 2\eta^{mn}1_{32\times32}\, .
\eeq
The signature of the metric is the mostly plus $(-++ \cdots+)$. In the Weyl representation
of $\Gamma^m$ only the off-diagonal $16\times 16$ blocks are
non-vanishing, parameterized as
\beq\label{littlegamma}
\Gamma^m = \begin{pmatrix}
           0                        & (\gamma^m)^{\alpha\beta}\\
           (\gamma^m)_{\alpha\beta} & 0
           \end{pmatrix},
\eeq
in terms of chiral gamma $16\times16$ matrices $\g^m$ subject to
\be
\label{clifford}
\gamma^m_{\a\b} (\gamma^n)^{\b\d} + \gamma^n_{\a\b} (\gamma^m)^{\b\d}
= 2\eta^{mn}\d_\a^\d\, .
\ee

\paragraph{Numerical representation}

An explicit representation of the $16\times16$ gamma matrices \eqref{littlegamma} is given by
\begin{align}
(\gamma^0)^{\a\beta} & =\begin{pmatrix}
			 1_{8\times 8} & 0            \\
		         0 & 1_{8\times 8}
			 \end{pmatrix} \, ,
& (\gamma^0)_{\a\beta}&  = \begin{pmatrix}
			 -1_{8\times 8} & 0            \\
		        0       & -1_{8\times 8}
			 \end{pmatrix}	\, , \label{10dgammas}\\
(\gamma^i)^{\a\beta} & =\begin{pmatrix}
			 0      & \sigma^i_{a\dot{a}}            \\
			 \sigma^i_{\dot{b}b}       & 0
			 \end{pmatrix}\, , &
(\gamma^i)_{\a\beta}&  =  \begin{pmatrix}
			 0      & \sigma^i_{a\dot{a}}            \\
			 \sigma^i_{\dot{b}b}       & 0
			 \end{pmatrix}\, ,	 \notag\\
(\gamma^9)^{\a\beta} & =\begin{pmatrix}
			 1_{8\times 8} & 0            \\
			       0       & -1_{8\times 8}
			 \end{pmatrix} \, ,&
(\gamma^9)_{\a\beta}&  = \begin{pmatrix}
			 1_{8\times 8} & 0            \\
			       0       & -1_{8\times 8}
			 \end{pmatrix}	\, ,\notag
\end{align}
where $\sigma^i$ with $i=1,2,\ldots,8$ are $8\times 8$ matrices ($\mathds{1}:=1_{2\times2}$)
\begin{align}
\sigma^1_{a\dot{a}} & = \varepsilon \otimes \varepsilon \otimes \varepsilon \, ,&
\sigma^2_{a\dot{a}} & = \mathds{1} \otimes \sigma^1 \otimes \varepsilon \, ,\\
\sigma^3_{a\dot{a}} & = \mathds{1} \otimes \sigma^3 \otimes \varepsilon \, ,&
\sigma^4_{a\dot{a}} & = \sigma^1 \otimes \varepsilon \otimes \mathds{1}  \, , \notag\\
\sigma^5_{a\dot{a}} & = \sigma^3 \otimes \varepsilon \otimes \mathds{1} \, , &
\sigma^6_{a\dot{a}} & = \varepsilon \otimes \mathds{1} \otimes \sigma^1 \, ,  \notag \\
\sigma^7_{a\dot{a}} & = \varepsilon \otimes \mathds{1} \otimes \sigma^3 \, ,&
\sigma^8_{a\dot{a}} & = \mathds{1} \otimes \mathds{1} \otimes \mathds{1} \, , \notag
\end{align}
and $\sigma^i$ with $i=1,2,3$ are the Pauli matrices with $\varepsilon = i\sigma^2$
\beq\label{22m}
\sigma^1=\begin{pmatrix} 0&1\\
1&0\end{pmatrix}\,,\quad
\varepsilon = \begin{pmatrix} 0 & 1\\-1 & 0\end{pmatrix}\,,\quad
\sigma^3 = \begin{pmatrix} 1 & 0\\ 0 &-1\end{pmatrix}\,.
\eeq
While the gamma-matrix representations (\ref{10dgammas}) are tailored to
Minkowski-space $\mathbb R^{1,9}$, their Euclidean analogues can be
found in (\ref{gammarep}) below.

\paragraph{Charge conjugation and chirality matrices}
The chirality matrix in ten spacetime dimensions is given by
\begin{equation}
\label{chirality}
\Gamma = \Gamma^0\Gamma^1\ldots\Gamma^9 = \begin{pmatrix}
				   1_{16\times 16}      & 0 \\
			            0       & -1_{16\times 16}
			 \end{pmatrix}
\end{equation}
which splits a
$32$-component Dirac spinor into two $16$-component spinors of
opposite chiralities
\beq\label{deco}
\lambda = \begin{pmatrix}
               \lambda^{\alpha}\\
	       \lambda_{\alpha}
	  \end{pmatrix}
\eeq
called Weyl ($\l^\a$) and anti-Weyl ($\l_\a$). The charge conjugation matrix satisfying $C\Gamma^m
= - (\Gamma^m)^TC$ is $C=\Gamma^0$. Since it is off-diagonal, the Weyl and anti-Weyl are
inequivalent representations in ten dimensions (unlike in four).

\paragraph{Generalized Kronecker delta}

It is convenient to define the generalized Kronecker delta as
\beq\label{genKronecker}
\d^{a_1a_2 \ldots a_n}_{b_1b_2 \ldots b_n} = {1\over n!}\d^{[a_1}_{b_1}\d^{a_2}_{b_2} \cdots \d^{a_n]}_{b_n}
\eeq
which is totally antisymmetric in both sets of indices, 
e.g.\ $\d^{ab}_{mn}=\half(\d^a_m\d^b_n-\d^b_m\d^a_n)$.
Using the notation of words, in $D$ dimensions we have
\beq\label{gkcontgen}
\d^{PA}_{QA} = \frac{{D-p\choose a}}{{p+a\choose a}}\d^P_Q\,,\qquad
p:=\len{P}\, ,\; \ a:=\len{A}\,,\; \ p{+}a\le D\,.
\eeq
For example in $D=10$ we have $\d^{mna}_{pqa} = {8\over3}\d^{mn}_{pq}$.
In particular, when $D=10$ where $\d^m_m=10$, the identity \eqref{gkcontgen} gives the full contraction when $P=Q=\emptyset$ as
\beq\label{gKcont}
\d^{m_1 \ldots m_n}_{m_1 \ldots m_n} = {10\choose n}=10,45,120,210,252\quad\hbox{for }n=1,2,
3,4,5\,.
\eeq

\subsection{Fierz decompositions} Antisymmetric products of gamma matrices are defined by the
\textit{$n$-forms}
\beq\label{antig}
\g^{m_1m_2 \ldots m_n} = {1\over n!}\g^{[m_1}\g^{m_2} \ldots \g^{m_n]} \, ,
\eeq
and they lead to four possible configurations of Weyl-spinor indices, 
namely $(\g^{ \ldots})_{\a\b}$, $(\g^{ \ldots}){}^{\a\b}$, $(\g^{ \ldots})_\a{}^\b$ and
$(\g^{ \ldots})^\b{}_\a$. The mixed combinations occur at ranks $2,4,6,8,10$,
\begin{align}\label{gdu}
(\g^{m_1m_2})_\a{}^\b\, ,\quad
(\g^{m_1 \ldots m_4})_\a{}^\b\, ,\quad
(\g^{m_1 \ldots m_6})_\a{}^\b\, ,\quad
(\g^{m_1 \ldots m_8})_\a{}^\b\, ,\quad
(\g^{m_1 \ldots m_{10}})_\a{}^\b\, ,\\
(\g^{m_1m_2})^\b{}_\a\, ,\quad
(\g^{m_1 \ldots m_4})^\b{}_\a\, ,\quad
(\g^{m_1 \ldots m_6})^\b{}_\a\, ,\quad
(\g^{m_1 \ldots m_8})^\b{}_\a \, ,\quad
(\g^{m_1 \ldots m_{10}})^\b{}_\a \, ,\label{gud}
\end{align}
and are related by (anti)symmetry, i.e.\ the matrices in \eqref{gud} 
can be rewritten in terms of the matrices in \eqref{gdu}, see below.
Ranks $1,3,5,7,9$ give rise to spinor indices of alike chiralities
\begin{align}
\label{gdd}
(\g^{m_1})_{\a\b}\, ,\quad
(\g^{m_1 m_2 m_3})_{\a\b}\, ,\quad
(\g^{m_1 \ldots m_5})_{\a\b}\, ,\quad
(\g^{m_1 \ldots m_7})_{\a\b}\, ,\quad
(\g^{m_1 \ldots m_9})_{\a\b}\, ,\\
(\g^{m_1})^{\a\b},\quad
(\g^{m_1 m_2 m_3})^{\a\b}\, ,\quad
(\g^{m_1 \ldots m_5})^{\a\b}\, ,\quad
(\g^{m_1 \ldots m_7})^{\a\b}\, ,\quad
(\g^{m_1 \ldots m_9})^{\a\b}\, ,\notag
\end{align}
and their symmetry properties do not mix the two lines.

The symmetry properties of the four types of matrices with respect to their spinorial indices are
\begin{align}
\label{symgams}
\hbox{symmetric:  }
&(\g^{m_1})_{\a\b}\, ,&
&(\g^{m_1 \ldots m_4})_\a{}^\b\, ,&
&(\g^{m_1 \ldots m_5})_{\a\b}\, ,&
&(\g^{m_1 \ldots m_8})_\a{}^\b\, ,&
&(\g^{m_1 \ldots m_9})_{\a\b}\, ,\\
\hbox{antisymmetric:  }
&(\g^{m_1m_2})_\a{}^\b\, ,&
&(\g^{m_1 m_2 m_3})_{\a\b}\, ,&
&(\g^{m_1 \ldots m_6})_\a{}^\b\, ,&
&(\g^{m_1 \ldots m_7})_{\a\b}\, ,&
&(\g^{m_1 \ldots m_{10}})_\a{}^\b\, ,\notag
\end{align}
where for example $(\g^{m_1m_2})_\a{}^\b = - (\g^{m_1m_2})^\b{}_\a$, and the symmetry
properties of the matrices with all upper spinorial indices is the same as those with all lower indices, i.e.\ $(\g^{m_1})^{\a\b}=(\g^{m_1})^{\b\a}$ etc.

The Fierz decompositions of spinor bilinears reads
\begin{align}
\psi^\a\chi^\b &= {1\over16}\g_{m_1}^{\a\b}(\psi\g^{m_1}\chi)
+ {1\over96}(\g_{m_1 \ldots m_3})^{\a\b}(\psi\g^{m_1 \ldots m_3}\chi)
+ {1\over3840}(\g_{m_1 \ldots m_5})^{\a\b}(\psi\g^{m_1 \ldots m_5}\chi) \, ,\label{Fierzup} \\
\psi_\a\chi^\b &= {1\over16}\d_\a^\b(\psi\chi)
+ {1\over32}(\g_{m_1m_2})_\a{}^\b(\psi\g^{m_1 m_2}\chi)
+ {1\over384}(\g_{m_1 \ldots m_4})_\a{}^\b(\psi\g^{m_1 \ldots m_4}\chi) \, . \notag
\end{align}
For anticommuting spinors $\theta^\alpha$ and bosonic pure spinors $\lambda^{\alpha}$,
important special cases of \eqref{Fierzup} are\footnote{The first identity in
\eqref{Fierzlambdatheta} is sometimes incorrectly stated with
a coefficient $\tfrac{1}{1920}$ rather than $\tfrac{1}{3840}$. To see why the latter is correct, one
has to pay attention to the epsilon term in the trace \eqref{gamma_traces5} and the self-duality
\eqref{g5d} of $\g^{mnpqr}_{\a\b}$.}
\begin{align}
\l^{\a}\l^{\b} = \frac{1}{3840}(\l\g^{mnpqr}\l)\g_{mnpqr}^{\a\b}\,, \qquad
\t^{\a}\t^{\b} = \frac{1}{96}(\t\g^{mnp}\t)\g_{mnp}^{\a\b}\,.
\label{Fierzlambdatheta}
\end{align}

\subsection{Duality properties}
In ten-dimensional Minkowski space $\mathbb R^{1,9}$, using the convention
\beq\label{epsconv}
\e_{01 \ldots 9} = 1\,,\qquad
\e^{01 \ldots 9} = -1\,,
\eeq
with (in particular, $\e^{m_1m_2 \ldots m_{10}}\e_{m_1m_2 \ldots m_{10}} = -10!$)
\beq\label{epscont}
\e_{n_1 \ldots n_{10}}\e^{m_1 \ldots m_{10}} = -10!\d^{m_1 \ldots m_{10}}_{n_1 \ldots n_{10}}\, ,
\eeq
the antisymmetric gamma matrices ($n$-forms) are related by the duality properties
\begin{align}
(\g^{m_1 \ldots m_5})_{\a\b}&=  {1\over 5!}\e^{m_1 \ldots m_5n_1 \ldots n_5}(\g_{n_1 \ldots n_5})_{\a\b}\, ,
& (\g^{m_1 \ldots m_5})^{\a\b}&= -{1\over 5!}\e^{m_1 \ldots m_5n_1 \ldots n_5}(\g_{n_1 \ldots n_5})^{\a\b} \, ,\label{g5d}\\
(\g^{m_1 \ldots m_6})_\a{}^\b&=  {1\over 4!}\e^{m_1 \ldots m_6n_1 \ldots n_4}(\g_{n_1 \ldots n_4})_\a{}^\b\, ,
& (\g^{m_1 \ldots m_6})^\a{}_\b&= -{1\over 4!}\e^{m_1 \ldots m_6n_1 \ldots n_4}(\g_{n_1 \ldots n_4})^\a{}_\b \, ,\cr
(\g^{m_1 \ldots m_7})_{\a\b}&= -{1\over 3!}\e^{m_1 \ldots m_7n_1 \ldots n_3}(\g_{n_1 \ldots n_3})_{\a\b}\, ,
& (\g^{m_1 \ldots m_7})^{\a\b}&=  {1\over 3!}\e^{m_1 \ldots m_7n_1 \ldots n_3}(\g_{n_1 \ldots n_3})^{\a\b} \, ,\cr
(\g^{m_1 \ldots m_8})_\a{}^\b&= -{1\over 2!}\e^{m_1 \ldots m_8n_1  n_2}(\g_{n_1 n_2})_\a{}^\b\, ,
& (\g^{m_1 \ldots m_8})^\a{}_\b&=  {1\over 2!}\e^{m_1 \ldots m_8n_1  n_2}(\g_{n_1  n_2})^\a{}_\b\, ,\cr
(\g^{m_1 \ldots m_9})_{\a\b}&=   \e^{m_1 \ldots m_9n_1}(\g_{n_1})_{\a\b}\, ,
& (\g^{m_1 \ldots m_9})^{\a\b}&=  -\e^{m_1 \ldots m_9n_1}(\g_{n_1})^{\a\b} \, ,\cr
(\g^{m_1\ldots m_{10}})_\alpha{}^\beta &= \e^{m_1 \ldots m_{10}} \d^\beta_\alpha \, ,
& (\g^{m_1\ldots m_{10}})^\alpha{}_\beta &= - \e^{m_1 \ldots m_{10}}\d^\alpha_\beta \, .\notag
\end{align}
A good exercise is to check them explicitly using the $SO(1,9)$ parameterization in
\eqref{10dgammas}. In ten-dimensional Euclidean space where $\e^{12 \ldots 10} = 
\e_{12 \ldots 10}=1$, the equations above are still valid after redefining the 
Levi--Civita epsilon tensor $\epsilon \to i\epsilon$.

\subsection{Traces of gamma matrices}

In ten dimensions there are no invariant tensors with $k$ antisymmetrized 
vector indices except when $k=10$, so all the $k$-forms with even $2\leq k\leq 8$ are
traceless,
\beq\label{vantr}
(\g^{m_1m_2 \ldots m_k})^\a{}_\a = 0\,,\quad k=2,4,6,8\,.
\eeq
When $k=10$ compatibility with the duality conditions \eqref{g5d} implies
\beq\label{g10trace}
(\g^{m_1 \ldots m_{10}})_\a{}^\a = 16\e^{m_1 \ldots m_{10}}\,,\quad
(\g^{m_1 \ldots m_{10}})^\a{}_\a = -16\e^{m_1 \ldots m_{10}}\,.
\eeq
The trace relations for the $1,2,3,4$ and $5$ forms are given by \cite{partha}
\begin{align}
{\Tr}(\g^{m_1}\g_{n_1}) = (\g^{m_1})_{\a\b}(\g_{n_1})^{\b\a} &= 16\d^{m_1}_{n_1}\,,\label{gamma_traces}\\
{\Tr}(\g^{m_1m_2}\g_{n_1n_2}) = (\g^{m_1m_2})_\a{}^\b(\g_{n_1n_2})_\b{}^\a &= -16\cdot 2!\d^{m_1m_2}_{n_1n_2}\,,\notag\\
{\Tr}(\g^{m_1 \ldots m_3}\g_{n_1 \ldots n_3}) = (\g^{m_1 \ldots m_3})_{\a\b}(\g_{n_1 \ldots n_3})^{\b\a} &= -16\cdot 3!\d^{m_1 \ldots m_3}_{n_1 \ldots n_3}\,,\notag\\
{\Tr}(\g^{m_1 \ldots m_4}\g_{n_1 \ldots n_4}) =(\g^{m_1 \ldots m_4})_\a{}^\b(\g_{n_1 \ldots n_4})_\b{}^\a &= 16\cdot 4!\d^{m_1 \ldots m_4}_{n_1 \ldots n_4}\,,\notag\\
{\Tr}(\g^{m_1 \ldots m_5}\g_{n_1 \ldots n_5}) = (\g^{m_1 \ldots m_5})_{\a\b}(\g_{n_1 \ldots n_5})^{\b\a} &= 16\cdot 5!\d^{m_1 \ldots m_5}_{n_1 \ldots n_5}
+ 16\e^{m_1 \ldots m_5}{}_ {n_1 \ldots n_5}\,, \label{gamma_traces5}
\end{align}
where ${\Tr}(\g^M\g^N)=\g^M_{\a\b}(\g^N)^{\b\a}$
or ${\Tr}(\g^M\g^N)=(\g^M)_\a{}^\b(\g^N)_\b{}^\a$ depending on the lengths
of the multi-indices $M$ and $N$.
We also note the vanishing of the traces with unequal lengths of $M$ and $N$
\beq\label{uneq}
{\Tr}(\g^{m_1 \ldots m_i}\g_{n_1 \ldots n_j}) = 0\,,\quad i\neq j\, .
\eeq
These identities can be conveniently summarized using word notation as\footnote{The reversal
$\tilde Q$ in \eqref{wordtrace} is explained by noting
$[mn]=-[nm]$, $[mnp]=-[pnm]$. In general, $[P]=\begin{cases}
\phantom{-}[\tilde P] &\! \! \! : p=0,1\,\mod \, 4\\
-[\tilde P] &\! \! \! : p=2,3\,\mod \, 4
\end{cases}$.}
\beq\label{wordtrace}
\Tr\big(\g^P\g_Q\big) = 16\d^{p,q}\Big[ p!\d^P_{\tilde Q} + \d^{p,5}\epsilon^P{}_Q\Big] \, ,\quad
\len{P}=p,\;  \ \len{Q}=q\,.
\eeq

\subsection{Products of gamma matrices} The antisymmetrized products of gamma matrices 
form a basis in the space of bispinor indices,
as evidenced by the Fierz identities. In order to freely move
between upstairs and downstairs indices with the Euclidean metric, we consider the Clifford algebra
after a Wick rotation, $\{\Gamma^m,\Gamma^n\}=2\d^{mn}$. Since in the pure spinor formalism it is
convenient to consider a Weyl representation leading to off-diagonal $\g^m$ matrices as in
\eqref{littlegamma}, the Clifford algebra for the $32\times 32$ matrices
$\Gamma^m$ reduces to
\beq\label{littleclifford}
\{\g^m,\g^n\} = 2\d^{mn}
\eeq
in terms of  $16\times16$ chiral gamma matrices.
The explicit construction of such matrices can be found in \ref{u5app.1}.

We now want to convert products of gamma matrices into sums over antisymmetrized gammas in the
spinorial index basis. The starting point is the Clifford algebra \eqref{littleclifford} which implies
\beq\label{twog}
\g^m\g^n = \g^{mn} + \d^{mn}\,.
\eeq
This formula can be used iteratively when more indices are present, but the amount of generated terms
grows quickly when doing so. General formulae and strategies to handle the combinatorics exist in
the literature. For instance, a general formula for the product $\g_{m_1 \ldots m_i}\g^{n_1 \ldots
n_j}$ has been written in \cite{proeyen} using a diagrammatic method, 
while in \cite{Kuusela:2019iok} an OPE-like algorithm was presented. A nice formula was 
given in \cite{guttenberg} with the combinatorics conveniently organized as 
(note that the convention here has $1/k!$ in $[a_1 \ldots a_k]$)
\beq\label{gammaExpandGut}
\g_{a_1a_2 \ldots a_p}\g^{b_1b_2 \ldots b_q} =
\sum_{k=0}^{{\rm min}(p,q)}k!{p\choose k}{q\choose k}\d^{[b_1}_{[a_p}\d^{\phantom{[}\mkern-5mu b_2}_{\phantom{[}\mkern-5mua_{p-1}}\cdots
\d^{\phantom{[}\mkern-5mu b_k}_{\phantom{[}\mkern-5mu a_{p-k+1}}\g_{\strut a_1 \ldots a_{p-k}]}{}^{b_{k+1} \ldots b_q]}\, ,
\eeq
where all the signs in the sum (prior to antisymmetrization) are uniformly positive due to the reverse ordering chosen for some indices on
the right-hand side. This formula can be further decluttered using a notation based on words. If we adopt the convention
where a lower case letter corresponding to the word denotes the length of the word, $\len{A}:=a$,
we can rewrite \eqref{gammaExpandGut} more compactly as (here we have $1/k!$ in $[a_1 \ldots a_k]$)
\beq\label{gammaExpand}
\g_A\g^B = \sum_{{XY=A\atop ZW=B}\atop z=y} y!{a\choose y}{b\choose y}\d^{[Z}_{[\tilde
Y}\g_{\strut X]}{}^{W]}\, ,
\eeq
where $\tilde Y$ denotes the reversal of $Y$ and we note the constraint $y=z$
on the lengths of $Y$ and $Z$ due to the generalized Kronecker delta. The combinatorial coefficients
compensate the overall $1/(a!b!)$ due to the antisymmetrizations over the $A$ and $B$
indices and the normalization of the generalized Kronecker delta (\ref{genKronecker}) such that in 
the expanded result all terms have a $\pm 1$ coefficient, as follows from the iterated 
use of \eqref{twog}. For example,
\begin{align}\label{smallex}
\g_{a_1a_2}\g^{b_1b_2} &= \g_{a_1a_2}{}^{b_1b_2} + 4\d^{[b_1}_{[a_2}\g_{a_1]}{}^{b_2]} + 2\d^{b_1b_2}_{a_2a_1}\\
& =    \g_{a_1a_2}{}^{b_1b_2}
      + \d_{a_2}^{b_1} \g_{a_1}{}^{b_2}
      - \d_{a_2}^{b_2} \g_{a_1}{}^{b_1}
      + \d_{a_1}^{b_2} \g_{a_2}{}^{b_1}
      - \d_{a_1}^{b_1} \g_{a_2}{}^{b_2}
      + \d_{a_2}^{b_1} \d_{a_1}^{b_2}
      - \d_{a_2}^{b_2} \d_{a_1}^{b_1} \, .\notag
\end{align}
Another example, which when fully expanded generates $136$ terms in total, is given by
\beq\label{bigex}
\g_{a_1a_2a_3a_4a_5}\g^{b_1b_2b_3} = \g_{a_1a_2a_3a_4a_5}{}^{b_1b_2b_3}
+ 15\d^{[b_1}_{[a_5}\g_{a_1a_2a_3a_4]}{}^{b_2b_3]}
+ 60\d^{[b_1b_2}_{[a_5a_4}\g_{a_1a_2a_3]}{}^{b_3]}
+ 60\d^{b_1b_2b_3}_{[a_5a_4a_3}\g_{a_1a_2]}\,.
\eeq
The different coefficients in front of each term correspond to the numbers of terms (with $\pm 1$
coefficients) that are generated once the explicit antisymmetrization takes place (note
$136=1+15+60+60$).

Related formulas for the commutator of gamma matrices can be found in
\cite{gamma_commutators}.

\subsection{Gamma matrix identities and pure spinors}

A set of frequently used identities when manipulating pure spinor
superspace expressions is listed below (repeated indices are contracted):
\begin{align}
\label{gamIds}
\g^m_{\a(\b} \g^{m}_{\g \delta)}&= 0\,,\\
\g^{mnp}_{\a[\b}\g^{mnp}_{\g\d]} &= 0\,,\\
\g_{mnp}^{\a\b}\g^{mnp}_{\g\d}&=48(\d^\a_\g\d^\b_\d - \d^\b_\g\d^\a_\d)\,,\\
\g^{mnp}_{\a\b}\g^{mnp}_{\g\d}&=12(\g^m_{\a\d}\g^m_{\b\g} - \g^m_{\a\g}\g^m_{\b\d})\,,\\
\g^m_{\a\b}\g^m_{\d\s} &= - \half\g^m_{\a\d}\g^m_{\b\s} -
{1\over24}\g^{mnp}_{\a\d}\g^{mnp}_{\b\s}\,,\\
\g^{mnp}_{\a\b}\g^{mnp}_{\d\s} &= - 18\g^m_{\a\d}\g^m_{\b\s} +
\half\g^{mnp}_{\a\d}\g^{mnp}_{\b\s}\,,\\
\g^{mnp}_{\a\b}\g^{mnp}_{\d\s} &= -12 \g^m_{\a\b}\g^m_{\d\s} - 24\g^m_{\a\d}\g^m_{\b\s}\,,\\
(\g^{mn})_\a^{\;\; \d}(\g_{mn})_\b^{\;\; \s} &=
-8 \d_\a^\s \d_\b^\d  - 2\d_\a^\d\d_\b^\s + 4 \g^m_{\a\b}\g_m^{\d\s} \,,\\
(\g^{mnpq})_{\a}{}^\b(\g_{mnpq})_{\s}{}^\d &= 315\d_\a^\d\d_\s^\b + {21\over2}(\g^{mn})_\a{}^\d(\g_{mn})_\s{}^\b +
{1\over8}(\g^{mnpq})_\a{}^\d(\g_{mnpq})_\s{}^\b\,,\\
(\g^{mnpq})_{\a}{}^\b(\g_{mnpq})_{\s}{}^\d &= - 48 \d_\a^\b\d_\s^\d
+288 \d_\a^\d\d_\s^\b + 48 \g^m_{\a \s} \g_m^{\b \d}\,.
\end{align}
They can be derived by using that the gamma matrices form a complete basis for the spinorial
indices, see \cite{krotov} for more examples. Another important identity is the self-duality of
the five-form that yields,
\beq\label{sdual}
\g^{mnpqr}_{\a\b}\g^{mnpqr}_{\d\s} = 0\,.
\eeq
Some of the above identities are particularly useful when contracted with pure spinors. The first
identity of \eqref{gamIds}, for instance, implies
\beq\label{lagalaga}
(\l\g^m)_\a(\l\g_m)_\b = 0\,.
\eeq
To see this it suffices to contract $\g^m_{\a(\b} \g^{m}_{\g \delta)}= 0$ with $\l^\a\l^\g$ and use 
the symmetry \eqref{symgams} of the one-form $\g^m_{\b\g}=\g^m_{\g\b}$
to obtain $(\l\g^m)_\b(\l\g_m)_\d = - (\l\g^m)_\d(\l\g_m)_\b - (\l\g^m\l)\g^m_{\d\b} =-
(\l\g^m)_\d(\l\g_m)_\b=0$ based on the pure spinor constraint \eqref{espinor_puro}. 
An important corollary of (\ref{lagalaga}) is
\beq\label{laga5}
(\l\g_m)_\a (\l\g^{mnpqr}\l) = 0\, ,
\eeq
which can be proven by decomposing the five-form using \eqref{gammaExpand} as $(\l\g^{mnpqr}\l)=
(\l\g^m\g^{npqr}\l)
+ (\l\g^{npq}\l)\d^{mr}$ $
       - (\l\g^{npr}\l)\d^{mq}
       + (\l\g^{nqr}\l)\d^{mp}
       - (\l\g^{pqr}\l)\d^{mn}$ and observing that all the three-forms contracted with two pure
       spinors vanish by the antisymmetry \eqref{symgams}. We can thus
rewrite $(\l\g_m)_\a (\l\g^{mnpqr}\l) =(\l\g_m)_\a (\l\g^{m}\g^{npqr}\l)$ which vanishes
by \eqref{lagalaga}.

\section{\label{u5app}The $U(5)$ decomposition of $SO(10)$}

In this appendix we will list some of the formulae relevant for the decomposition of
various $SO(10)$ representations into their $U(5)$ components. Some useful references
are \cite{georgi} and the appendices of \cite{grassi_cov,schiappa,Aisaka:2002sd}.

\subsection{The Clifford algebra in $\mathbb{R}^{10}$}
\label{u5app.1}

For convenience, we will
consider the Wick-rotated version $SO(10)$ of the Lorentz group $SO(1,9)$.
The ten-dimensional Clifford algebra in Euclidean signature
\beq\label{cliffordalgebra}
\{\Gamma^m,\Gamma^n\} = 2\d^{mn} \, ,\qquad m,n=1,2, \ldots10
\eeq
admits a recursive construction \cite{brauerweyl} starting from the $2\times2$  representation
in terms of Pauli matrices
\beq\label{paulimat}
\sigma_1 = \begin{pmatrix}0 & 1\\ 1 & 0\end{pmatrix}\,,\quad
\sigma_2 = \begin{pmatrix}0 & -i\\ i & 0\end{pmatrix}\,,\quad
\sigma_3 = \begin{pmatrix}1 & 0\\ 0 & -1\end{pmatrix}\,,
\eeq
satisfying $\{\s_i,\s_j\}=2\d_{ij}$ for $i,j=1,2,3$. To
assemble the explicit $2^5\times 2^5$ gamma matrices $\Gamma^m$ in ten dimensions
we use the
Kronecker product of Pauli sigma matrices as follows \cite{Kaplan:2005ta}
($\mathds{1}:=1_{2\times2}$; also see (\ref{10dgammas}) for the analogous numerical 
representation of $\Gamma^m$ for Minkowski spacetime $\mathbb R^{1,9}$):
\begin{align}
\Gamma^1 &= \sigma_2\otimes \sigma_1\otimes\mathds{1}\otimes\mathds{1}\otimes\mathds{1}\, ,
& \Gamma^6 &= \sigma_2\otimes \sigma_2\otimes\mathds{1}\otimes\mathds{1}\otimes\mathds{1}\, ,\notag\\
\Gamma^2 &= \sigma_2\otimes \sigma_3\otimes \sigma_1\otimes\mathds{1}\otimes\mathds{1}\, ,
& \Gamma^7 &= \sigma_2\otimes \sigma_3\otimes\sigma_2\otimes\mathds{1}\otimes\mathds{1}\, ,\notag\\
\Gamma^3 &= \sigma_2\otimes \sigma_3\otimes\sigma_3\otimes\sigma_1\otimes\mathds{1}\, ,
& \Gamma^8 &= \sigma_2\otimes \sigma_3\otimes\sigma_3\otimes\sigma_2\otimes\mathds{1} \, ,\label{gammarep}\\
\Gamma^4 &= \sigma_2\otimes \sigma_3\otimes\sigma_3\otimes\sigma_3\otimes\sigma_1\, ,
& \Gamma^9 &= \sigma_2\otimes \sigma_3\otimes\sigma_3\otimes\sigma_3\otimes\sigma_2\, ,\notag\\
\Gamma^5 &= \sigma_2\otimes \sigma_3\otimes\sigma_3\otimes\sigma_3\otimes\sigma_3 \, ,
& \Gamma^{10} &= -\sigma_1\otimes\mathds{1}\otimes\mathds{1}\otimes\mathds{1}\otimes\mathds{1}\, .\notag
\end{align}
The properties of the Kronecker product, $(A\otimes B)(C\otimes D)=(AC\otimes BD)$ and
$(A\otimes B)^T=A^T\otimes B^T$, imply that the Clifford algebra \eqref{cliffordalgebra} is
satisfied. Moreover, the symmetry properties of the above gamma matrices are
\beq\label{gtrans}
\Gamma_m^T = \begin{cases} - \Gamma_m \, , & m=1,\ldots,5\\
+\Gamma_m \, , & m=6,\ldots,10\,.
\end{cases}
\eeq
In addition, the gamma matrices in \eqref{gammarep} are purely imaginary for $m=1, \ldots,5$  and real for
$m=6, \ldots,10$, as they are constructed with an odd or even number of $\s_2$.
This means that the representation \eqref{gammarep} is hermitian
\beq\label{gamT}
\Gamma_m^\dagger = \Gamma_m\,.
\eeq

\paragraph{Charge conjugation and chirality matrices}
Given the above symmetry property of $\Gamma_m$, the charge conjugation matrix $C$
satisfying
\beq\label{Cdef}
C\Gamma_m= - \Gamma_m^T C
\eeq
is obtained by the product of all antisymmetric $\Gamma$'s \cite{pais1962spinors}
\beq\label{Cmat}
C= \Gamma_1\Gamma_2\Gamma_3\Gamma_4\Gamma_5 = -\sigma_2\otimes\sigma_1\otimes\sigma_2\otimes\sigma_1\otimes\sigma_2\, .
\eeq
It is easy to see that $C$ is antisymmetric, off-diagonal and satisfies $C^2=1_{32\times32}$.
In addition, the chirality matrix  is given by
\beq\label{chirmat}
\Gamma_{11}= -i\Gamma_1 \ldots \Gamma_{10} = \sigma_3\otimes\mathds{1}\otimes\mathds{1}\otimes
\mathds{1}\otimes \mathds{1} = \begin{pmatrix}1_{16\times 16} & 0\\ 0 & -1_{16\times
16}\end{pmatrix}\,,
\eeq
and has the same numerical value as in the Lorentzian version \eqref{chirality}.

\subsection{\label{decsec}Vectors and Lorentz generators}
The generators $M^{mn}$ of the group $SO(10)$ are
antisymmetric $M^{mn}=-M^{nm}$ and satisfy the
Lie-algebra relations
\beq\label{lieso10}
[M^{mn},M^{pq}]= \d^{mp}M^{nq} - \d^{np}M^{mq} - \d^{mq}M^{np} + \d^{nq}M^{mp}\,.
\eeq
The vector $V^p$ and spinor $\Psi$ representations of $SO(10)$ are defined by the following
transformations\footnote{Note the consistency between \eqref{lieso10} and \eqref{vecso10} as the
transformation \eqref{lieso10} can be viewed as the two-form version of \eqref{vecso10} 
using $[M^{mn},V^r\otimes V^s]= [M^{mn},V^r]\otimes V^s + V^r\otimes[M^{mn},V^s]$ and setting $M^{pq}=V^p\otimes V^q-V^q \otimes V^p$. The sign on the right-hand side of (\ref{spinso10})
may naively appear to conflict with the fact that the Lorentz algebra (\ref{lieso10}) is obeyed
by Gamma matrices in the normalization of $M^{mn} \rightarrow - \frac{1}{2} \Gamma^{mn}$.
However, repeated Lorentz transformations lead to Gamma matrices in the opposite multiplication
order $[ M^{pq}, [M^{mn} , \Psi] ] = \frac{1}{4} (\Gamma^{mn} \Gamma^{pq}) \Psi$ such that
$[ [M^{pq}, M^{mn} ], \Psi] =  \frac{1}{4} [\Gamma^{mn}, \Gamma^{pq}] \Psi 
= - \frac{1}{2} (\delta^{p[m} \Gamma^{n]q} -\delta^{q[m} \Gamma^{n]p} )\Psi$, and
the normalization on the right-hand side of
(\ref{spinso10}) is determined by consistency with (\ref{lieso10}). 
In case of the pure spinor ghost $\Psi \rightarrow \lambda^\alpha$ and its
Lorentz current $M^{mn} \rightarrow N^{mn}$, a more detailed
version of this calculation can be found in (\ref{consN}), where
(\ref{spinso10}) is implemented through the OPE (\ref{lambda_N}).}
\begin{align}\label{vecso10}
[M^{mn},V^p] &= \d^{mp}V^n - \d^{np}V^m\, ,\\
\label{spinso10}
[M^{mn},\Psi] &= {1\over2}\Gamma^{mn}\Psi\,.
\end{align}
To decompose the vectorial representation $SO(10)\to U(5)$ we
shall split the ten components of the
$SO(10)$ vector $V^m$ with $m=1, \ldots,10$ into two vectors $v^a, v_a$ labelled by an index
$a=1, \ldots,5$ as
\beq\label{vecU5}
v^a = {1\over \sqrt{2}}\big(V^a + i V^{a+5}\big)\, ,\qquad
v_a = {1\over \sqrt{2}}\big(V^a - i V^{a+5}\big)\, ,\quad a=1, \ldots,5\,.
\eeq
Consequently, the components of tensors of $SO(10)$  are split
following the tensor products
of the vector decompositions \eqref{vecU5} with the corresponding symmetry
conditions. This implies, for example, that the generators $M^{mn}$ split as follows:
\begin{align}
\label{sumab}
m^{ab} & =  \frac{1}{2} \left( M^{ab} + iM^{a(b+5)} + iM^{(a+5)b} -M^{(a+5)(b+5)}\right) \, ,\\
m_{ab} & = \frac{1}{2} \left( M^{ab} - iM^{a(b+5)} - iM^{(a+5)b} - M^{(a+5)(b+5)}\right)\, ,\cr
m^a_b & = \frac{1}{2} \left( M^{ab} - iM^{a(b+5)} + iM^{(a+5)b} +
     M^{(a+5)(b+5)}\right)\, .\notag
\end{align}
Moreover, the trace of $m^a_b$ is given by
\beq
m = \sum_{a=1}^5 m^a_a = i\sum_{a=1}^5 M^{(a+5)a} \label{n_cor}\,.
\eeq
From the above it follows that
\begin{align}
[m^{ab},v^c] & = 0\,,
& [m_{ab},v_c] & = 0 \, ,\\
[m_{ab},v^c] & = \d_a^c v_b - \d^c_b v_a\,,
& [m^{ab},v_c] & = \d^a_c v^b - \d^b_c v^a\, ,\notag\\
[m^{a}_{b},v^c] & = -\d^c_b v^a\,,
& [m^{a}_{b},v_c] & = \d^a_c v_b\, ,\notag\\
[m,v^c] & =  -v^c\,,
& [m,v_c] & =  v_c\, .\notag
\end{align}
To derive the above commutators,
the decompositions \eqref{sumab} and \eqref{vecU5} have been used to rewrite them
in $SO(10)$ language, where \eqref{vecso10} can be applied with its outcome cast back in $U(5)$
variables using \eqref{vecU5}.

Similarly, the same strategy shows that the $SO(10)$ Lie algebra \eqref{lieso10} decomposes to
$U(5)$ as
\begin{align}
\label{so10u5}
[m_{ab},m_{cd}] &= 0\,, & [m^{ab},m^{cd}] &= 0\, ,\\
[m_{ab},m^{cd}] &= -\d_a^c m^d_b + \d_a^d m^c_b +\d_b^c m^d_a - \d_b^d m^c_a\,,
&[m_{ab},m^{c}_{d}] &= -\d^c_b m_{ad} + \d_a^c m_{bd}\, ,\notag\\
[m^{ab},m^{c}_{d}] &= -\d^a_d m^{bc} + \d^b_d m^{ac}\,,
&[m^{a}_{b},m^c_d] &= -\d^c_b m^a_d + \d^a_d m^c_b\, ,\notag\\
[m,m_{ab}] &= 2 m_{ab}\,,
& [m,m^{ab}] &= -2 m^{ab} \, ,\notag\\
[m,m^{a}_{b}] &= 0\,,
&[m,m] &= 0 \, .\notag
\end{align}
This shows that $m^a_b$ are the generators of $U(5)$ embedded in $SO(10)$. 
Moreover $m^{ab}$ and $m_{ab}$ transform as two-forms under $U(5)$, and $v_a$, 
$v^a$ transform in the defining representations ${\bf 5}$ and ${\bf \bar5}$ of $U(5)$. The trace $m$ 
is the $U(1)$ generator in the decomposition $U(5)=SU(5)\otimes U(1)$. The $U(1)$ charge 
$q_R$ of a representation $R$ is defined by $[m,R] = q_R R$ and denoted by a subscript
${\bf \bar N}_{q_R}$ for an $N$-dimensional representation of $SU(5)$. We conclude that the 
vector $V^m$ and the antisymmetric tensor $M^{mn}$ transform as follows under
the $SU(5)\otimes U(1)$ decomposition of $SO(10)$,
\begin{align}
\label{vectou5}
V^m &\to v^a \oplus v_a\, , &M^{mn} \to m^{ab}\oplus m_{ab} \oplus m^a_b\oplus m\, ,\\
{\bf 10} &\to {\bf 5}_{-1} \oplus {\bf \bar 5}_{1}\, , &{\bf 45} \to {\bf 10}_{-2}\oplus {\bf
\bar{10}}_{2}\oplus {\bf 24}_0\oplus {\bf 1}_0\, .\notag
\end{align}

\paragraph{Decomposition of the Lorentz currents OPE}
In the pure spinor formalism the $SO(10)$ to $U(5)$ decomposition must be applied to 
the OPE between the Lorentz generators for the pure spinor variables,
\begin{equation}
\label{opeApp}
N^{mn}(z)N^{pq}(w) \sim \frac{ \d^{mp}N^{nq}(w) -\d^{np}N^{mq}(w)  - \d^{mq}N^{np}(w) + \d^{nq}N^{mp}(w)}{z-w}
- 3\frac{\big(\delta^{mq}\delta^{np} - \delta^{mp}\delta^{nq}\big)}{(z-w)^2} \, .
\end{equation}
Starting from the generators in \eqref{sumab} and \eqref{n_cor}, we perform the redefinitions
\beq
\label{redefms}
n= {m\over \sqrt{5}}  \, ,\qquad 
 n^a_b= m^a_b - {1\over 5}\d^a_bm \, ,
\eeq
which identify $n^a_b$ as the traceless generator of $SU(5)$.
Using the same strategy as above leads to the
following OPEs among the $SU(5)\otimes U(1)$ decompositions
\begin{align}
\label{so10u5OPE}
n_{ab}(z)n_{cd}(w) &\sim \text{regular}\,,
& n^{ab}(z)n^{cd}(w) &\sim\text{regular}\, ,\\
n_{ab}(z)n^{cd}(w) &\sim {-\d_{[a}^c n^d_{b]} + \d_{[a}^d n^c_{b]}
- {2\over \sqrt{5}}\d^c_{[a}\d^d_{b]}n \over z-w}
- 3{\d^c_b\d^d_a - \d^c_a\d^d_b\over (z-w)^2}\,,
&n(z)n^{a}_{b}(w) &\sim \text{regular}\,,\notag\\
n^{a}_{b}(z)n^c_d(w) &\sim {-\d^c_b n^a_d + \d^a_d n^c_b\over z-w}
- 3{\d^a_d\d^c_b-{1\over5}\d^a_b\d^c_d\over(z-w)^2}\,,
& n(z)n_{ab}(w) &\sim  {2\over\sqrt{5}}{n_{ab}\over z-w} \, ,\notag\\
n^{ab}(z)n^{c}_{d}(w) &\sim {-\d^a_d n^{bc} + \d^b_d n^{ac} - {2\over5}\d^c_d n^{ab}\over z-w}\,,
& n(z)n^{ab}(w) &\sim -{2\over\sqrt{5}}{n^{ab}\over z-w}\, ,\notag\\
n_{ab}(z)n^{c}_{d}(w) &\sim {-\d^c_b n_{ad} + \d_a^c n_{bd} +{2\over5}\d^c_dn_{ab}\over z-w}\,,
&n(z)n(w) &\sim -{3\over (z-w)^2}\, .\notag
\end{align}
Redefining the $U(1)$ charge to $[n,R]= {q_R\over \sqrt{5}} R$ in view of \eqref{redefms} we see
that $(n, n^a_{b}, n^{ab}, n_{ab})$ transform as the $({\bf 1}_0, {\bf 24}_0, {\bf 10}_{-2},{\bf
\bar{10}}_{2})$
representations of $SU(5)\otimes U(1)$.

\subsection{Spinors}
To obtain the decomposition of the spinorial representation of $SO(10)$ under $SU(5)\otimes U(1)$
it will be convenient to consider the linear combinations \cite{georgi}
\begin{align}
b^a &= \frac{1}{2}\left(\Gamma^{a} + i \Gamma^{a+5}\right)\, , \quad \quad
b_a = \frac{1}{2}\left(\Gamma^{a} - i \Gamma^{a+5}\right)\, ,
\label{cran}
\end{align}
where $a=1,2,\ldots ,5$.
The Clifford algebra \eqref{cliffordalgebra} implies the fermionic oscillator algebra
\beq\label{transb}
\{b_a,b^b\}=\d_a^b\,,\qquad \{b_a,b_b\}=\{b^a,b^b\}=0\, .
\eeq
This means that the matrices $b_a$ and $b^b$ can be interpreted as
annihilation and creation operators. To exploit this interpretation we define
a vacuum $\ket{0}$ annihilated by all the $b_a$ operators, $b_a\ket{0}:=0$ (also $\bra{0}b^a:=0$) and normalized
as $\langle 0|0\rangle=1$. States are created
by acting with the creation operators $b^a$ on the vacuum, for a maximum
of $32$ states. We will also define $\bra{\psi} = \ket{\psi}^T$.
These operators also satisfy
\begin{align}\label{hermb}
b_a^\dagger &= b^a\, , &(b^a)^\dagger &= b_a\,, \\
b_a^T &= -b^a\, ,& (b^a)^T &= -b_a\, ,\notag
\end{align}
for $a=1, \ldots,5$, as can be verified from \eqref{gtrans} and \eqref{gamT}.
In this language, the  charge conjugation and chirality matrices  
in \eqref{Cmat}  and \eqref{chirmat} become
\beq\label{gam11}
C=\prod_{j=1}^5(b_j+b^j)\,,\qquad
\Gamma_{11}=\prod_{j=1}^5(b^jb_j-b_jb^j)\, .
\eeq
To connect this description with the $U(5)$ decomposition of $SO(10)$ above, we
write the generators $M^{mn}$ for the spinor representation as
\beq\label{spinM}
M^{mn} \rightarrow -{1\over2}\Gamma^{mn}=-{1\over4}[\Gamma^m,\Gamma^n]\,,
\eeq
which satisfy the commutation relations \eqref{lieso10}.
Therefore, from the expressions \eqref{sumab} and \eqref{n_cor},
the $U(5)$ Lorentz generators become
\begin{align}\label{u5ms}
m^a_b &= -{1\over2}\big(b^ab_b - b_b b^a\big)\,,
&m &= -{1\over2}\big(b^ab_a - b_a b^a\big)=-b^a b_a + {5\over2}\,,\\
m^{ab}& =-b^ab^b,
&m_{ab}&=-b_ab_b\,.\notag
\end{align}
These expressions can be verified by plugging the spinorial
representation \eqref{spinM} into the decompositions \eqref{sumab} and using the inverse of \eqref{cran}.
In this language, it is straightforward to verify the decompositions \eqref{so10u5}
using $[b_bb^a,b_db^c]=[b_bb^a,b_d]b^c + b_d[b_bb^a, b^c]$ and $[b_d,b_bb^a]=\{b_d,b_b\}b^a -
b_b\{b_d,b^a\}$.
Furthermore,
\begin{align}
\label{u5bs}
[m^a_b,b^c]&=-\d^c_b b^a\, ,
&[m^a_b,b_c]&=\d^a_c b_b\,,\\
[m^{ab},b^c] &=0\,,
&[m^{ab},b_c] &=  \d^a_c b^b -\d^b_c b^a\, ,\notag\\
[m_{ab},b^c] &= \d^c_a b_b -\d^c_b b_a \,,
&[m_{ab},b_c] &= 0\,,\notag\\
[m,b^c] &=  -b^c\,,
& [m,b_c] &= b_c\, .\notag
\end{align}
These relations identify $m^a_b$ as the generators of $U(5)$ and $m$ as the generator of $U(1)$.
Moreover \eqref{u5bs} implies that $b^c$ and $b_c$ transform in the ${\bf 5}_{-1}$ and  
${\bf\bar{5}}_{1}$ representations of $SU(5)\otimes U(1)$, respectively.

\paragraph{Pure spinors}
Recall that a ten-dimensional pure spinor was defined by Cartan as 
a bosonic Weyl spinor $\Lambda$
satisfying the equation \cite{cartan}
\beq\label{purespinorCartan}
\Lambda^T C\Gamma^m \Lambda=0\,,
\eeq
where $C$ is the $32\times 32$ charge conjugation
matrix \eqref{Cmat} and $\Gamma_{11}\Lambda = -\Lambda$. From \eqref{chirmat} we obtain $\Lambda^T = (\lambda
\;\;{} 0)$ for a $16$ dimensional bosonic spinor $\l^\a$, and this implies the familiar equation $\lambda^\a\g^m_{\a\b}\l^\b=0$.
We will now describe the pure spinor constraint using the  creation and annihilation operators
\eqref{cran}. To do this, first we will need to characterize a Weyl spinor in this language.
\begin{lemma}
The $16\oplus 16'$ states of ten-dimensional Weyl $\ket{\l}$ and anti-Weyl $\ket{\Omega}$ spinors satisfying
$\Gamma_{11}\ket{\l}={-}\ket{\l}$ and $\Gamma_{11}\ket{\Omega} = \ket{\Omega}$
are created by
\begin{align}
\label{weylu5}
\ket{\l} &= \lambda^{+}\ket{0} +\frac{1}{2}\lambda_{ab}b^b b^a\ket{0}
+\frac{1}{4!}\lambda^a \epsilon_{abcde}b^e b^d b^c b^b\ket{0}\,,\\
\ket{\Omega} &=
	       \frac{1}{5!}\omega_+\varepsilon_{abcde}b^ab^bb^cb^db^e\ket{0}
               +\frac{1}{2! 3!}\omega^{ab}\varepsilon_{abcde}b^cb^db^e\ket{0}
	       + \omega_a b^a\ket{0}\, .
	       \notag
\end{align}
These
expressions correspond to the following representation decompositions of $SO(10)\rightarrow U(5)$:
\begin{align}
\label{dimu5}
\la & \rightarrow (\lambda^+,\lambda_{ab},\lambda^a) \, ,& \omega_{\alpha} & \rightarrow (\omega_+,\omega^{ab},\omega_a) \, , \\
{\bf 16}  & \rightarrow ({\bf 1},\bar{{\bf 10}},{\bf 5})\, , &{\bf 16}'  & \rightarrow
({\bf 1}, {\bf 10},\bar{{\bf 5}})\, .\notag
\end{align}
\end{lemma}
\noindent{\it Proof.}
The chirality matrix in terms of the
creation and annihilation algebra is given by \eqref{chirmat}
such that $\Gamma_{11}\ket{0}={-}\ket{0}$ and $\{\Gamma_{11}, b^a\}=0$. 
This means that states with an even (odd)
number of creation operators acting on the vacuum have eigenvalue $-1$ ($+1$) under $\Gamma_{11}$.
This explains the expressions \eqref{weylu5}.
The number of independent components of each $U(5)$ representation in
\eqref{dimu5} follows easily from the fermionic nature of the creation operators
$b^a$ as $\#(b^{a_1} \ldots b^{a_k})={5\choose k}$.\qed

Note that the $U(5)$ components of the Weyl and anti-Weyl spinors can be extracted as
\begin{align}
\l^+ & = \langle0\ket{\l}\,, & \l_{ab} & = \bra{0}b_ab_b\ket{\l}\,, & \l^a & = {1\over4!}\epsilon^{abcde}\bra{0}b_bb_cb_db_e\ket{\l}\, , \notag\\
\om_+ & = {1\over5!}\epsilon^{abcde}\bra{0}b_ab_bb_cb_db_e\ket{\om}\,, & \om^{ab} & =
-{1\over3!}\epsilon^{abcde}\bra{0}b_cb_db_e\ket{\om}\,, & \om_a & = \bra{0}b_a\ket{\om}\,.
\label{proju5}
\end{align}
In order to obtain the number of degrees of freedom of a ten-dimensional pure spinor we will need
the following results
\beq\label{Cbs}
\bra{0}Cb^a b^b b^c b^d b^e\ket{0} = \epsilon^{abcde}\,,\qquad
Cb_a = b^aC\, ,\quad Cb^a = b_a C\,,
\eeq
which can be obtained from \eqref{chirmat} and \eqref{transb} together with the normalization
$\langle 0|0\rangle=1$. Using the Weyl spinor decomposition \eqref{weylu5} one can also show
\begin{align}
\label{temps}
\bra{0}Cb^a\ket{\lambda} &= \l^a\, ,\\
\bra{0}Cb^ab^bb^c\ket{\lambda} &= -{1\over2}\epsilon^{abcde}\l_{de}\, ,\notag\\
\bra{0}Cb^ab^bb^cb^db^e\ket{\lambda} &= \epsilon^{abcde}\l^+ \, .\notag
\end{align}
We are now ready to show

\begin{prop.} A ten-dimensional pure spinor has eleven complex degrees of freedom.
\end{prop.}
\noindent{\it Proof.}
Under the decomposition of $SO(10) \rightarrow U(5)$, the constraint \eqref{purespinorCartan}
generates two sets of independent equations (with $i=1,2,3,4,5$ each)\footnote{The 
constraint \eqref{purespinorCartan} for
both $\Gamma^1=b^1+b_1$ and $\Gamma^6=-i(b^1-b_1)$ implies
$\bra{\l}Cb_1\ket{\l}=0$ and $\bra{\l}Cb^1\ket{\l}=0$ from suitable linear combinations.}:
\begin{align}
\bra{\l}Cb^i\ket{\l} &= 0 \, , \label{satis0} \\
\bra{\l}Cb_i\ket{\l} &= 0 \,.\label{satis}
\end{align}
The transpose relation $(b^i)^T = -b_i$ in \eqref{transb} implies
$\bra{\l} =
\bra{0}\lambda_{+} +\frac{1}{2}\bra{0}b_a b_b \lambda_{ab}
+\frac{1}{24}\bra{0}b_b b_c b_d b_e \lambda^a \epsilon_{abcde}$.
So the equation \eqref{satis0}
becomes
\begin{align}
0=\bra{\l}Cb^p\ket{\l} &= \lambda^{+}\bra{0}Cb^p\ket{\lambda}
  +\frac{1}{2}\lambda_{ij}\bra{0}b_i b_j Cb^p\ket{\lambda}
  +\frac{1}{24}\lambda^i\epsilon_{ijklm}\bra{0}b_j b_k b_l b_m Cb^p\ket{\lambda}\notag\\
  &=2\lambda^{+}\lambda^a - \frac{1}{4}\epsilon^{abcde} \lambda_{bc}\lambda_{de}\,,\label{pure}
\end{align}
where we used  $b_aC=Cb^a$ from \eqref{Cbs} and \eqref{temps}.
Hence, \eqref{pure} implies that we can write the five components $\l^a$ in terms
of the others
\beq\label{solPS}
\lambda^a = \frac{1}{8\l^+}\epsilon^{abcde} \lambda_{bc}\lambda_{de}\,,
\eeq
which solves the first equation \eqref{satis0}.
Moreover, the second equation \eqref{satis} yields
\beq\label{const2}
\bra{0}Cb_b\ket{0}= 2\l^a\l_{ab}\, ,
\eeq
which is automatically satisfied when inserting the
solution \eqref{solPS} due to an over-antisymmetrization
of five-dimensional indices. Therefore the pure spinor constraint in ten dimensions removes 
only the five components \eqref{solPS} from the $16$-component Weyl spinor,
leaving a total of eleven degrees of freedom.\qed

\paragraph{Spinorial transformations in $U(5)$ language}
In the fundamentals of the pure spinor formalism it is necessary to know how the
$U(5)$ components of the pure spinor transform under the $SO(10)$ rotations. To do this
we note the interpretation \cite{georgi} $O\ket{v} = \ket{Ov}$ for an arbitrary operator $O$
that allows one to read off how the different tensor components transform under $O$.
Straightforward calculations using the operators \eqref{u5ms} imply that the right-hand side 
of the spinorial transformation \eqref{spinso10}, given by the action of 
$\half\Gamma^{mn}=-M^{mn}$, decomposes as follows
\begin{align}
m_{ab}\ket{\l} &= -\l_{ab}\ket{0}-{1\over2}\epsilon_{abcde}\l^e b^d b^c\ket{0}\,  ,\notag \\
m^{ab}\ket{\l} &= -\l^+ b^a b^b\ket{0} - {1\over2}\l_{cd}b^ab^bb^db^c\ket{0} \, ,  \label{nolast}\\
m^a_b\ket{\l} &= {1\over2}\d^a_b\ket{\l} - \l_{cb}b^a b^c\ket{0}
- {1\over3!}\l^c \epsilon_{cdefb}b^a b^fb^e b^d\ket{0}\, ,\notag\\
m\ket{\lambda} &= {5\over2}\lambda^{+}\ket{0} +\frac{1}{4}\lambda_{ab}b^b b^a\ket{0}
-\frac{1}{16}\lambda^a \epsilon_{abcde}b^e b^d b^c b^b\ket{0}\,.\notag
\end{align}
Note the factor of $\ket{\l}$ instead of $\ket{0}$ in the first term of \eqref{nolast}.

Using the projections to the $U(5)$ components \eqref{proju5} these transformations imply
\begin{align}
m_{ab}\l^+ &= -\l_{ab}\,, &
m_{ab}\l_{cd} &= -\epsilon_{abcde}\l^e\,, &
m_{ab}\l^c &= 0\,,\\
m^{ab}\l^+ &= 0\,, &
m^{ab}\l_{cd} &= -(\d^a_d\d^b_c - \d^a_c\d^b_d)\l^+\,, &
m^{ab}\l^c &= {1\over2}\epsilon^{abcde}\l_{de}\,,\notag\\
m^a_b\l^+ &= {1\over2}\d^a_b\l^+\,, &
m^a_b\l_{cd} &= -\d^a_d\l_{cb} + \d^a_c\l_{db}+{1\over2}\d^a_b\l_{cd}\,,&
m^a_b\l^c &= \d^c_b\l^a -{1\over2}\d^a_b\l^c\,,\notag\\
m\l^+ &= {5\over2}\l^+\,, &
m\l_{cd} &= {1\over2}\l_{cd}\,,&
m\l^c &= - {3\over2}\l^c\, .\notag
\end{align}
For example, $|m \l\rangle_{ab} := \bra{0}b_ab_b
m\ket{\l}=\tfrac{1}{4}\l_{fg}\bra{0}b_ab_bb^gb^f\ket{0}=\tfrac{1}{2}\l_{ab}$,
where $|m\l\rangle_{ab}$ denotes the projection of $|m\l\rangle$ into its $\mathbf{10}$ component
of $SU(5)$.

After identifying the $SU(5)\otimes U(1)$ Lorentz currents with a traceless $n^a_b$ as
\beq
\label{Nsfrom}
(n,n^a_b,n^{ab},n_{ab}) = -\bigg({m\over\sqrt{5}},m^a_b-{1\over5}\d^a_b m,m^{ab},m_{ab} \bigg)\, ,
\eeq
we arrive at the following $SO(10)\to SU(5)\otimes U(1)$ decompositions
\begin{align}
n_{ab}\lambda^+ &= \lambda_{ab} \,,
& n_{ab}\lambda_{cd} &= \epsilon_{abcde}\lambda^e\,,
& n_{ab}\lambda^c &= 0 \,,\\
n^{ab}\lambda^+ &= 0  \,,
& n^{ab}\lambda_{cd} &= -\d^{[a}_c\d^{b]}_d\l^+ \,,
& n^{ab}\lambda^c &= -\frac{1}{2}\epsilon^{abcde}\lambda_{de}\,, \notag\\
n^a_b\lambda^+ &= 0\,,
& n^a_b \l_{cd} &= \delta^a_d\l_{cb} - \delta^a_c\l_{db} -\frac{2}{5}\delta^a_b\l_{cd} \,,
& n^a_b\l^c &= -\d^c_b\l^a + {1\over5}\delta^a_b\l^c\, ,\cr
n\lambda^+ &= -\frac{\sqrt{5}}{2}\lambda^+\,,
& n\lambda_{cd} &= -\frac{1}{2\sqrt{5}}\lambda_{cd}\,,
& n\lambda^c &= \frac{3}{2\sqrt{5}}\lambda^c\,.\notag
\end{align}
These are the coefficients of the single pole in the OPE $N^{mn}\l^\a$ in (\ref{lambda_N}).

To find the $U(1)$ charges of the anti-Weyl spinor we compute
\beq
m\ket{\Omega} = -{1\over48}\omega_+\varepsilon_{abcde}b^ab^bb^cb^db^e\ket{0}
               -{1\over24}\omega^{ab}\varepsilon_{abcde}b^cb^db^e\ket{0}
	       +{3\over2}\omega_a b^a\ket{0}
\eeq
which implies, after projecting to the components via \eqref{proju5},
\begin{align}
m\omega_+ &= -{5\over2}\omega_+\,,&
m\omega^{ab} &= -{1\over2}\omega^{ab}\,,&
m\omega_a &= {3\over2}\omega_a\,.
\end{align}

\section{Combinatorics on words}%
\label{wordsapp}

In this appendix we list some of the most common maps on words used throughout this review. With
the exception of the letterification defined in \cite{genredef}, these definitions are standard
and can be found in the books \cite{lothaire1997combinatorics,Reutenauer}.

The left-to-right bracketing map $\ell (A)$, also called the Dynkin bracket, is defined recursively by
\beq\label{elldef}
\ell(123..n):=
\ell(123...n-1)n-n\ell(123...n-1)\, ,\;\;\;\ \ell(i)=i\, ,\;\;\; \ \ell(\emptyset)=0\,.
\eeq
For example,
\begin{align}
\ell(12) &= 12 - 21\, ,\\
\ell(123) &= 123 - 213 - 312 + 321\,.\notag
\end{align}
In addition, the map $\rho(A)$ is defined by
\beq\label{rhomapdef}
\rho(123 \ldots n):=\rho(123 \ldots n{-}1)n-\rho(23 \ldots n)1\,.
\eeq
For example,
\begin{align}
\rho(12) &= 12-21\,,\\
\rho(123) &= 123-213-231+321\,.\notag
\end{align}
The {\it shuffle product} $\shuffle$ is defined recursively
by
\beq\label{shuffledef}
\emptyset\shuffle P = P\shuffle\emptyset := P\, ,\qquad
iP\shuffle jQ := i(P \shuffle jQ) + j(Q \shuffle iP)\,,
\eeq
where $i$ and $j$ are letters, $P$ and $Q$ are words while
$\emptyset$ represents the empty word. For example,
\begin{align}
1 \shuffle 2 &= 12+21\, , \ \ \ \ 1 \shuffle 23 = 123 + 213 + 231 \, , \\
12\shuffle 34&=
       1234
       + 1324
       + 1342
       + 3124
       + 3142
       + 3412\, . \notag
       \end{align}
The {\it deshuffle\/} $\d(P)=X\otimes Y$ of $P$ (sometimes denoted as $P= X\shuffle Y$) 
is the sum of all pairs of words $X,Y$ such that $P$ is in the shuffle of $X$ and $Y$. An efficient
algorithm to obtain the words $X$ and $Y$ in $\d(P)=X\otimes Y$ is given by \cite{Reutenauer}
\beq\label{deltamap}
\d(a_1a_2 \ldots a_n):= \d(a_1)\d(a_2) \ldots\d(a_n) \, ,\quad
\d(a_i) := \emptyset \otimes a_i + a_i\otimes\emptyset\,,\quad
\d(\emptyset) := \emptyset\otimes\emptyset\,,
\eeq
where the product is defined as
\beq\label{compwise}
(A\otimes B)(R\otimes S):=(AR\otimes BS)\,.
\eeq
For example,
\begin{align}
\label{excupD}
\d(1) &= \emptyset\otimes 1 + 1\otimes\emptyset\,,\\
\d(12) &= \d(1)\d(2) = (\emptyset\otimes 1 +
1\otimes\emptyset)(\emptyset\otimes 2 + 2\otimes\emptyset) =
\emptyset\otimes 12  + 1\otimes 2 + 2\otimes 1 + 12\otimes\emptyset\,,\cr
\d(123) &= \delta(12)\delta(3) =
(\emptyset\otimes 12  + 1\otimes 2 + 2\otimes 1 + 12\otimes\emptyset)
(\emptyset\otimes 3 + 3\otimes\emptyset)\cr
&= \emptyset\otimes123
+ 1\otimes23
+ 2\otimes13
+ 12\otimes3
+ 3\otimes12
+ 13\otimes2
+ 23\otimes1
+ 123\otimes\emptyset\,.\notag
\end{align}
An alternative characterization of the deshuffle map is
\beq\label{deshuffle}
\d(P) = \sum_{X,Y}\langle P, X\shuffle Y\rangle\, X\otimes Y\,,
\eeq
where $\langle \cdot,\cdot\rangle$ denotes
the scalar product on words\footnote{Not to be confused with the pure spinor zero-mode measure.}
\beq\label{AdotB}
\langle A, B\rangle := \d_{A,B}\, ,\qquad
\d_{A,B}= \begin{cases}$1$\, , & \hbox{if } A=B\cr
 $0$\, , & \text{otherwise}\, . \end{cases}
\eeq
In addition, the {\it letterfication} maps a
{\it word\/} $Q$ to a {\it letter\/} $\dot q$,
\beq\label{letterif}
Q\rightarrow \dot q\,.
\eeq
The purpose of this map is to avoid deconcatenation of $\dot q$ since a letter can not be
deconcatenated. For example, suppose that the word $Q=12$
has been letterified to $\dot q = 12$ and that $P=3$. Then deconcatenating $QP$ is different
from deconcatenating $\dot qP$. For example, one gets only one term
$\sum_{XY=\dot qP}S_X T_Y = S_{\dot q}T_3 = S_{12}T_{3}$
instead of the usual two ($\sum_{XY=\dot QP}S_X T_Y = S_1T_{23} + S_{12}T_3$) 
if $Q$ is not letterified.

In the proof of \eqref{bshuffle} we used a result that was already 
implicit in the proof of the shuffle symmetry of Berends--Giele
currents in the appendix A of \cite{Gauge}:
\begin{lemma} For $A$ and $B$ non-empty words
\beq\label{shuho}
\Delta(A\shuffle B) = \emptyset \otimes (A\shuffle B) + (A\shuffle B)\otimes\emptyset +
A\otimes B + B\otimes A + \psum_{PQ=A}\psum_{RS=B} (P\shuffle R)\otimes (Q\shuffle S)\, ,
\eeq
where $\psum$ denotes deconcatenation over non-empty words.
\end{lemma}
\noindent{\it Proof.} The
deconcatenation coproduct
$\Delta(P) = \sum_{XY=P}X\otimes Y$
is a homomorphism with respect to the shuffle product, $\Delta(A\shuffle B) = \Delta(A)\shuffle \Delta(B)$
(see Proposition 1.9 of \cite{Reutenauer}).
Noting that $(P\otimes Q)\shuffle (R\otimes S) = (P\shuffle R)\otimes (Q\shuffle S)$ we get
for $A,B\neq\emptyset$
\begin{align}
\Delta(A\shuffle B) &= \Delta(A)\shuffle \Delta(B) = \sum_{PQ=A}\sum_{RS=B}(P\otimes Q)\shuffle(R\otimes S)\\
&=\emptyset \otimes (A\shuffle B) + (A\shuffle B)\otimes\emptyset +
A\otimes B + B\otimes A + \psum_{PQ=A}\psum_{RS=B} (P\shuffle R)\otimes (Q\shuffle S) \, , \notag
\end{align}
where the first four terms in the second line
come from separating off the empty words in the sums
such that the deconcatenation words in $\sum'$ are not empty.\qed

\subsection{The dual Lie polynomials}
\label{dualLiesec}

The dual Lie polynomials in $\cL^*$ are characterized by the dual basis $iQ$ satisfying
\beq\label{dualbasis}
\langle iQ,\ell(iP)\rangle = \d_{Q,P}\, ,
\eeq
where $\ell(iP)$ is the Lyndon basis of Lie polynomials when $i$ is the minimum letter of $iQ$.
Given a dual Lie polynomial $P^*$ and a Lie polynomial $P$, their expansions in their
respective bases are given by
\cite{melanconReutenauer}
\begin{align}
P^* &= \sum_Q \langle P^*, \ell(iQ)\rangle iQ\,,\label{dualExpansion}\\
P &= \sum_Q \langle P, iQ\rangle \ell(iQ)\,.\label{LieExpansion}
\end{align}
Using Ree's theorem \eqref{reestheo} it is easy to see that dual Lie polynomials are unchanged
by proper shuffles and therefore define equivalence classes $P^* + \sum R\shuffle S \sim P^*$. For
related work, see \cite{griffing1995dual} and also \cite{hadleigh}.

\section{Dynkin labels of $SO(10)$}
\label{sec:appDynkin}

In this appendix we will very briefly summarize the representation theory of $SO(10)$
in the language of Dynkin labels that was used in the main text. The practical calculations
were done using \cite{Lie}. For the mathematical background, see \cite{humphreys,fultonharris}.

An irreducible representation of the Lie algebra of $SO(10)$ is labelled by five 
indices $(a_1 a_2 a_3 a_4 a_5)$ characterizing its highest-weight vector.
For instance a scalar of $SO(10)$
is represented by $(00000)$, while a vector is the $(10000)$; see the table below for more
examples.
\begin{table}[h]
\label{tab:I}
\centering
\begin{tabular}{|c| c|}
\hline
Dynkin label & $SO(10)$ content  \\ \hline
$(00000)$ & 0-form $A$  \\ \hline
$(10000)$ & 1-form $A^m$  \\ \hline
$(01000)$ & 2-form $A^{[mn]}$  \\ \hline
$(00100)$ & 3-form $A^{[mnp]}$  \\ \hline
$(00011)$ & 4-form $A^{[mnpq]}$  \\ \hline
$(00020)\oplus(00002)$ & 5-form $A^{[mnpqr]}$  \\ \hline
$(00010)$ & anti-Weyl spinor $\psi_\a$  \\ \hline
$(00001)$ & Weyl spinor $\psi^\a$  \\ \hline
$(0000n)$ & $\l^{\a_1}\l^{a_2} \ldots \l^{\a_n}$ pure spinors \\ \hline
\end{tabular}
\end{table}
The dimension of the representation labelled by $(a_1\ldots a_5)$ is given by \cite{LieART}
\begin{align}\label{dimrep}
87091200 \, &{\rm dim}(a_1 a_2 a_3 a_4 a_5)=
(1 + a_1)
(1 + a_2)
(1 + a_3)
(1 + a_4)
(1 + a_5)
(2 + a_1 + a_2)
(2 + a_2 + a_3)\notag\\
&\times (2 + a_3 + a_4)
(2 + a_3 + a_5)
(3 + a_1 + a_2 + a_3)
(3 + a_2 + a_3 + a_4)
(3 + a_2 + a_3 + a_5)\notag\\
&\times
(3 + a_3 + a_4 + a_5)
(4 + a_1 + a_2 + a_3 + a_4)
(4 + a_1 + a_2 + a_3 + a_5)\notag\\
&\times
(4 + a_2 + a_3 + a_4 + a_5)
(5 + a_1 + a_2 + a_3 + a_4 + a_5)
(5 + a_2 + 2a_3 + a_4 + a_5)\notag\\
&\times
(6 + a_1 + a_2 + 2a_3 + a_4 + a_5)
(7 + a_1 + 2a_2 + 2a_3 + a_4 + a_5)\, .
\end{align}
For example, ${\rm dim}(00000)=1$, ${\rm dim}(10000)=10$, and ${\rm dim}(00001)=16$. From
the formula above it is easy to be convinced that these calculations are better handled by
computers, see \cite{Lie, LieART}.

Many calculations in this review require to know the decomposition of product representations. 
A common example is the familiar fact that two vectors decompose into a symmetric and 
traceless, antisymmetric and trace parts;
$V^m W^n = {1\over2}(V^m W^n + V^n W^m-{\d^{mn}\over5}V\cdot W) + {1\over2}(V^m W^n - V^m W^n) +
{1\over10}\d^{mn}V\cdot W$.
In terms of the Dynkin labels, this is represented by
\beq
(10000)\otimes(10000)=(20000)\oplus(01000)\oplus(00000) \, ,
\eeq
where $(20000)$ is the symmetric traceless and $(01000)$ is the antisymmetric part. The dimensions
match as $10\times 10= 54 + 45 + 1$.

Of special importance for us is the pure spinor representation. A single pure spinor $\l^\a$
is a Weyl spinor $(00001)$, but a product of $n$ pure spinors $\l^{\a_1}\l^{\a_2} \ldots \l^{\a_n}$
is $(0000n)$. The dimensions of the pure spinor representation $\l^n=(0000n)$ are
$16,126,672,2772,9504,28314, \ldots$ for $n=1,2,3,4,5,6, \ldots$.

\section{Pure spinor superspace correlators}
\label{PSSapp}

The result \eqref{la3th5Lie} of Lemma \ref{la3th5lemma} guarantees that any pure spinor superspace
expression with three pure spinors and five thetas
can be reduced to the unique scalar proportional to $(\l^3 \t^5)$ with coefficients carrying the
tensorial structure whose normalization is uniquely fixed by
the condition \eqref{la3th5}. Therefore one can assemble
a catalog of correlators using symmetry arguments alone.
For instance, (in contrast to the rest of this work, the antisymmetrization brackets $[m_1 m_2\ldots m_k]$ enclosing $k$ indices here include $1/k!$, e.g.\ 
$V^{[m} W^{n]}= \frac{1}{2}(V^m W^n - V^n W^m)$)
\begin{align}
\label{duo}
\langle(\l\g^{m} \t) (\l\g^{s} \t)(\l\g^{u} \t)(\t\g_{fgh}\t)\rangle &= 24\d^{msu}_{fgh} \, ,\\
\langle(\l\g_{m}\t)(\l\g_{s}\t)(\l\g^{ptu}\t)(\t\g_{fgh}\t)\rangle &=
       {288\over 7}\d^{[p}_{[m}\eta_{s][f}\d^{t}_g\d^{u]}_{h]} \, ,\notag\\
\langle(\l\g_{m}\t)(\l\g^{n rs}\t)(\l\g^{ptu}\t)(\t\g_{fgh}\t)\rangle &= {12\over
35}\e_{fghm}{}^{nprstu}+{144\over 7}\Big[
	 \d^{[n}_m\d^r_{[f}\eta^{s][p}\d^t_g\d^{u]}_{h]}
	-\d^{[p}_m\d^t_{[f}\eta^{u][n}\d^r_g\d^{s]}_{h]}
\Big]\cr
&\quad -{72\over 7}\Big[
	 \eta_{m[f}\eta^{v[p} \d^t_g\eta^{u][n}\d^r_{h]}\d^{s]}_v
	-\eta_{m[f}\eta^{v[n} \d^r_g\eta^{s][p}\d^t_{h]}\d^{u]}_v
\Big] \, ,\notag\\
\langle (\l\g^{mnpqr}\t)(\l\g_{d}\t)(\l\g_e\t)(\t\g_{fgh}\t)\rangle &=
 - {480\over 7}\big(\delta^{mnpqr}_{defgh}
 - {1\over 120}\e^{mnpqr}_{\qquad\;\; defgh}\big)\, , \notag\\
\langle (\l\g^{mnpqr}\t)(\l\g_{stu}\t)(\l\g^v\t)(\t\g_{fgh}\t)\rangle&=
{576\over 7}
\eta^{v[m}\d^n_{[s}\d^p_t \eta_{u][f}\d^q_g\d^{r]}_{h]}
-{1152\over 7}
\d^{[m}_{[s}\d^n_t\d^p_{u]}\d^q_{[f}\d^{r]}_g\d^v_{h]}  \notag\\
&\quad+{1\over 120}\e^{mnpqr}_{\qquad\;\; abcde}\left(
{576\over 7}
\eta^{v[a}\d^b_{[s}\d^c_t \eta_{u][f}\d^d_g\d^{e]}_{h]}
-{1152\over 7}
\d^{[a}_{[s}\d^b_t\d^c_{u]}\d^d_{[f}\d^{e]}_g\d^v_{h]}
\right) \, , \notag\\
\langle(\l\g^{mnp}\t)(\l\g_{qrs}\t)(\l\g_{tuv}\t)(\t\g^{ijk}\t)\rangle &=
-{1728\over 35}\Big[
\d^{[i}_a\d^{j}_{[t}\d^{k]}_{u}\d^{[m}_{v]}\d^{n}_{[q}\d^{p]}_{r}\d^a_{s]}
- \d^{[i}_a\d^{j}_{[q}\d^{k]}_{r}\d^{[m}_{s]}\d^{n}_{[t}\d^{p]}_{u}\d^a_{v]}
+ \d^{[i}_{[q}\d^{j}_{r}\eta^{k][m}\eta_{s][t}\d^{n}_{u}\d^{p]}_{v]}\notag\\
&\hskip-35pt+\d^a_{[t}\eta^{b[i}\d^j_u\eta^{k][m}\eta_{v][q}\d^n_r\eta_{s]a}\d^{p]}_b
- \d^{a}_{[q}\eta^{b[i}\d^{j}_{r}\eta^{k][m}\eta_{s][t}\d^{n}_{u}\eta_{v]a}\d^{p]}_b
- \d^{[i}_{[t}\d^{j}_{u}\eta^{k][m}\eta_{v][q}\d^{n}_{r}\d^{p]}_{s]}
\Big] \, , \notag
\end{align}
where $\d^{msu}_{fgh}$ is the antisymmetrized combination of Kronecker deltas beginning with
$\frac{1}{3!}\d^m_f\d^s_g\d^u_h$, see \eqref{genKronecker}.
To justify the first line of \eqref{duo}, note that its right-hand side is the
only tensor antisymmetric both in $[msu]$ and $[fgh]$ and which is normalized 
to yield 2880 upon contraction (because
$\d^{msu}_{msu} = 120$, see \eqref{gKcont}), therefore respecting the 
normalization \eqref{tlct}. The other identities are justified by similar means.

\paragraph{Basis of zero-mode correlators}
Using the Fierz identities \eqref{Fierzlambdatheta} appropriately, all pure spinor superspace expressions can be written
as a linear combination of the following three correlators
\cite{anomaly, stahn}
\begin{align}
\langle(\l\g^{mnpqr}\l)(\l\g^{s}\t)(\t\g_{fgh}\t)
(\t\g_{jkl}\t)\rangle &=
{1152\over 7}\d^{m}_{f}\d^n_g \d^h_j\d^p_k\d^q_{l}\d^r_s
-{2304\over 7}\d^{m}_{f} \d^n_g \d^p_{h}\d^q_{j}\d^{r}_k \d^s_{l}
\label{uind}\\
&\quad+{1\over120}\e^{mnpqr}_{\qquad\;\; abcde}\Big[
{1152\over 7}\d^{a}_{f}\d^b_g \d^h_j\d^c_k\d^d_{l}\d^e_s
-{2304\over 7}\d^{a}_{f} \d^b_g \d^c_{h}\d^d_{j}\d^{e}_k \d^s_{l}
\Big]\notag\\
&\quad+ [mnpqr][fgh][jkl]+(fgh\leftrightarrow jkl) \, ,\notag\\
\langle(\l\g^{mnpqr}\l)(\l\g^{stu}\t)(\t\g_{fgh}\t)
(\t\g_{jkl}\t)\rangle &=
{6917\over 7}\Big[
         \d^v_s\d^t_{f}\d_u^m\d_g^n\d^h_j\d^p_k\d^q_{l}\d_v^{r}
	-\d^{s}_{f}\d^t_g \d_u^m\d^n_{h}\d^p_{j}\d^q_k\d^{r}_{l}
\Big]\notag\\
&\quad+{3456\over 1125}\e^{mnpqr}_{\qquad\;\; abcde}\Big[
         \d^v_s\d^t_{f}\d_u^a\d_g^b\d^h_j\d^c_k\d^d_{l}\d_v^{e}
	-\d^{s}_{f}\d^t_g \d_u^a\d^b_{h}\d^c_{j}\d^d_k\d^{e}_{l}\Big]\notag\\
&\quad + [mnpqr][stu][fgh][jkl]+(fgh\leftrightarrow jkl)\, ,\notag\\
\langle (\lambda\gamma^{mnpqr}\lambda) (\lambda\gamma^{stuvx}\theta)
 (\theta\gamma^{fgh}\theta) (\theta\gamma^{jkl}\theta)\rangle &=
 2880\Big[{8\over7}\d^m_s \d^n_t \d^p_f \d^q_g \d^r_h \d^u_j \d^v_k \d^x_l
- {8\over7}\d^m_s \d^n_t \d^p_u \d^q_f \d^r_g \d^v_j \d^x_k \d^h_l\notag\\
&\quad + {16\over7}\d^m_s \d^n_t \d^p_u \d^q_f \d^r_j \d^v_g \d^x_k \d^h_l
- {24\over7}\d^m_s \d^n_t \d^p_f \d^q_g \d^r_j \d^u_h \d^v_k \d^x_l\Big]\notag\\
&\quad + 24\e^{mnpqr}_{\qquad \, abcde}\Big[
{8\over7}\d^a_s \d^b_t \d^c_f \d^d_g \d^e_h \d^u_j \d^v_k \d^x_l
- {8\over7}\d^a_s \d^b_t \d^c_u \d^d_f \d^e_g \d^v_j \d^x_k \d^h_l\notag\\
&\quad + {16\over7}\d^a_s \d^b_t \d^c_u \d^d_f \d^e_j \d^v_g \d^x_k \d^h_l
- {24\over7}\d^a_s \d^b_t \d^c_f \d^d_g \d^e_j \d^u_h \d^v_k \d^x_l
\Big]\notag\\
&\quad+[mnpqr][stuvx][fgh][jkl] + (fgh\leftrightarrow jkl)\, ,\notag
\end{align}
where the notation $+[i_1 \ldots i_k]\ldots$ instructs to antisymmetrize the indices
$i_1, \ldots, i_k$ including the normalization $1/k!$.
In the correlations above we obtained the epsilon terms by considering
the duality (\ref{g5d}) of the five-form gamma matrix, explaining their relative
factors of the form
$(\hbox{parity even}) + \frac{1}{120} \epsilon_{10}(\hbox{parity even})$.

Suppose one wants to compute the following pure spinor correlator
\be
\label{now}
\langle (\l\g^m \t)(\l\g^n\g^{rs}\t)(\l\g^p\g^{tu}\t)(\t\g_{fgh}\t)\rangle \, .
\ee
Using the gamma-matrix identity
$\g^m\g^{np} = \g^{mnp} + \eta^{mn}\g^p - \eta^{mp}\g^n$,
we obtain a linear combinations of correlators present in the catalog above:
\begin{align}
\label{tocompu}
\langle (\l\g^m \t)(\l\g^n\g^{rs}\t)(\l\g^p\g^{tu}\t)(\t\g_{fgh}\t)\rangle &=
\langle (\l\g^m \t)(\l\g^{nrs} \t)(\l\g^{ptu} \t)(\t\g_{fgh}\t) \rangle \\
&\quad+ 2 \langle(\l\g^{m} \t) 
\eta^{n[r}(\l\g^{s]} \t)(\l\g^{ptu} \t)(\t\g_{fgh}\t)\rangle\cr
&\quad+2 \langle(\l\g^{m} \t) 
(\l\g^{nrs} \t) \eta^{p[t}(\l\g^{u]} \t) (\t\g_{fgh}\t)\rangle\cr
&\quad+ 4 \langle(\l\g^{m} \t) 
\eta^{n[r}(\l\g^{s]} \t)
\eta^{p[t}(\l\g^{u]} \t)(\t\g_{fgh}\t)\rangle\, .\notag
\end{align}
Proceeding in this way we can quickly calculate any zero-mode correlator,
and a {\tt FORM} implementation can be found in \cite{PSS}.

\paragraph{Practicalities of pure spinor superspace component expansions}
By virtue of Fierz identities, all possible pure spinor superspace
expressions can be written in
the basis \eqref{uind} of three fundamental zero-mode correlators:
$(\l\g^{[5]}\l)(\l\g^{[n]}\t)(\t\g^{[3]}\t)(\t\g^{[3]}\t)$ with $n=1,3$ or $5$, where the notation
$\g^{[n]}$ for an integer $n$ means an antisymmetric gamma matrix with $n$ vectorial indices.
The explicit form of this basis can be found in (\ref{uind}).

However, it is often more efficient to
assemble beforehand a catalog of common correlators
and use them out of storage rather than performing the Fierz manipulations to
go to the above basis. This avoids wasteful manipulations that dramatically
simplify in the end, such as computing the simple correlator \eqref{simpleextlct} via
\beq\label{waste}
\langle(\l\g^{m} \t) (\l\g^{n} \t)(\l\g^{p} \t)(\t\g_{abc}\t)\rangle =
{1\over96} \langle (\l\g^{mrstn}\l)(\l\g^{p} \t)(\t\g_{abc}\t)(\t\g_{rst}\t)\rangle\,,
\eeq
which, as can be seen in the expression \eqref{uind}, leads to many intermediate terms.

Another approach  is to evaluate the correlators of three lambdas
and five thetas by brute-force in terms of the tensor \cite{trivedi,stahn}
\beq\label{Ttensordef}
\langle \l^{\a_1}\l^{\a_2}\l^{\a_3}\t^{\d_1}\t^{\d_2}\t^{\d_3}\t^{\d_4}\t^{\d_5} \rangle :=
T^{\a_1\a_2\a_3; \d_1\d_2\d_3\d_4\d_5} = {1\over 1792}
\g^{\a_1\d_1}_m\g^{\a_2\d_2}_n\g^{\a_3\d_3}_p \g^{\d_4\d_5}_{mnp} + [\d_1\d_2\d_3\d_4\d_5]
\eeq
where $+[\d_1 \ldots\d_5]$ instructs to antisymmetrize over the indices $\d_1, \ldots,\d_5$
including the normalization factor $1/5!$\footnote{For definiteness, the definition \eqref{Ttensordef} has 60 terms
starting with ${1\over 107520}\g^{\a_1\d_1}_m\g^{\a_2\d_2}_n\g^{\a_3\d_3}_p \g^{\d_4\d_5}_{mnp} -
\cdots$.}. Note that the right-hand side of \eqref{Ttensordef} is found to be 
symmetric in $(\a_1\a_2\a_3)$ after taking the antisymmetrizations over $\d_1, \ldots,\d_5$
and $m,n,p$ into account. In addition, it is straightforward to see that
$T^{\a\b\g;\d_1\d_2\d_3\d_4\d_5}\g^m_{\a\d_1}\g^n_{\b\d_2}\g^p_{\g\d_3}\g^{mnp}_{\d_4\d_5} = 2880$
recovers the normalization \eqref{tlct}.

Evaluating pure spinor superspace expressions using this method follows from
\be
\label{compT}
\langle \l^{\a}\l^{\b}\l^{\g}\t^{\d_1}\t^{\d_2}\t^{\d_3}
\t^{\d_4}\t^{\d_5}f_{\a\b\g\d_1\d_2\d_3\d_4\d_5}(e,\chi,k)\rangle =
T^{\a\b\g;\d_1\d_2\d_3\d_4\d_5}f_{\a\b\g\d_1\d_2\d_3\d_4\d_5}(e,\chi,k)\,,
\ee
but this usually leads to the calculation of many gamma-matrix traces, often with many free indices.
While there are closed formulae for these traces in the \ref{sec:appA.2}, doing these calculations
on demand tends to become a time-consuming task. Therefore, to avoid any spurious 
inefficiencies, the catalog approach is used in the program \cite{PSS}.
As we have seen, pure spinor superspace expressions with many external particles can
be evaluated very efficiently using multiparticle superfields in the Harnad--Shnider gauge.

\section{\label{HSapp}$\t$-expansion of SYM superfields}

A convenient gauge choice to expand the superfields of ten-dimensional
SYM in theta is the Harnad--Shnider gauge \cite{harnad},
\beq\label{HSgauge}
\theta^\alpha \Bbb A^{{\rm HS}}_\alpha = 0 \,.
\eeq
At the linearized level, the gauge $\t^\a A_\a^i =0$ has been used in
\cite{ooguri,tsimpis} to obtain the $\theta$-expansions (\ref{linTHEX}) of 
the single-particle superfields $K_i$ to arbitrary order. Since the recursive 
definition of multiparticle Berends--Giele currents ${\cal A}_\alpha^P$ in
\eqref{cAalpha} quickly generates many terms, it would be expensive to follow the recursion up to single-particle
level and then expand the multiparticle superfields using the Harnad--Shnider gauge fixing \eqref{HSgauge}. Luckily
it was shown in \cite{Gauge} that one can exploit the gauge-transformation properties of multiparticle
superfields to arrive at Berends--Giele currents satisfying the Harnad--Shnider gauge \eqref{HSgauge}.

It is easy to see that Berends--Giele currents in Lorenz gauge do not 
satisfy the condition \eqref{HSgauge}, i.e.\ $\t^\a \bA^{{\rm L}}_\a\neq0$.
The idea is to find a non-linear gauge transformation $\Bbb L$
\beq\label{toHS}
\bA^{{\rm HS}}_\a = \bA^{{\rm L}}_\a - [D_\a,\Bbb L] + [\bA^{{\rm L}}_\a , \Bbb L]
\eeq
such that $\t^\a \bA^{{\rm HS}}_\a=0$. Assuming that the superfields have been brought to
this gauge, the derivation of their $\t$-expansions proceeds in a similar way as in their
single-particle counterpart.

We start by contracting the non-linear equations
of motion \eqref{SYMeom} with $\theta^\alpha$ while assuming the 
Harnad--Shnider gauge $\t^\a \bA_\a = 0$. The result \cite{harnad}
\begin{align}
\big( {\cal D}+1\big)\Bbb A_{\b}&=(\theta\gamma^m)_\b\Bbb A_m \, , \ \ \ \ \ \ \ \ \ \
{\cal D}\Bbb A_m=(\theta\gamma_m\Bbb W)\, , \label{SYMeomt} \\
{\cal D}\Bbb W^\b&={1\over4}(\theta\gamma^{mn})^\b\Bbb F_{mn} \, , \ \ \ \
{\cal D}\Bbb F^{mn}=-(\Bbb W^{[m}\gamma^{n]}\theta)\notag
\end{align}
is most
conveniently expressed in terms of the Euler operator
\beq\label{Euler}
{\cal D} := \theta^\alpha D_\alpha = \theta^\alpha {\partial \over \partial \theta^\alpha}
\eeq
that weights the $k^{\rm th}$ order in $\theta$ by a factor of $k$.
One can therefore use (\ref{SYMeomt}) to reconstruct the entire $\theta$-expansion
of all SYM superfields from their zeroth orders $\Bbb K(\theta=0)$,
\begin{align}
[\Bbb A_{\a}]_k&={1\over k+1}(\t\gamma^m)_\alpha[\Bbb A_m]_{k-1} \, , \ \ \ \
[\Bbb A_m]_k={1\over k}(\t\gamma_m[\Bbb W]_{k-1}) \, ,\label{thetaex}  \\
[\Bbb W^\a]_k&={1\over4k}(\t\gamma^{mn})^\alpha[\Bbb F_{mn}]_{k-1} \, , \ \ \ \
[\Bbb F^{mn}]_k=-{1\over k}([\Bbb W^{[m}]_{k-1}\gamma^{n]}\t) \, ,\notag
\end{align}
where the notation $[\ldots]_k$ instructs to only keep terms of order $(\t)^k$ of
the enclosed superfields. The analogous relations for the superfields at higher mass dimensions
in (\ref{highmass}) are
\begin{align}
[\Bbb W_m^{\a}]_k&={1\over k}\bigg\{{1\over4}(\t\gamma^{pq})^\a[\Bbb F_{m\vert pq}]_{k-1} -  (\t \gamma_m)_\b
\sum_{l=0}^{k-1}\{ [\Bbb W^\beta]_l,[\Bbb W^{\a}]_{k-l-1}\}\bigg\} \, ,\label{moretheta} \\
[\Bbb F^{m\vert pq}]_k&=-{1\over k}\bigg\{([\Bbb W^{m[p}]_{k-1}\gamma^{q]}\t)
+ (\t \gamma^m)_\a \sum_{l=0}^{k-1}[ [\Bbb W^\a]_l,[\Bbb F^{ pq}]_{k-l-1}]\bigg\}\, ,\notag\\
[\Bbb W_{mn}^{\a}]_k&={1\over k}\bigg\{{1\over4}(\t\gamma^{pq})^\a[\Bbb F_{mn\vert pq}]_{k-1}-(\t\gamma_m)_\beta
\sum_{l=0}^{k-1}     \{ [\Bbb W^\beta]_l,[\Bbb W_n^{\a}]_{k-l-1}\}      \cr
&\quad - (\t\gamma_n)_\beta \sum_{l=0}^{k-1} \Big(  \{ [\Bbb W^{\beta}_{m}]_l,[\Bbb W^{\a}]_{k-l-1}\}
+ \{ [\Bbb W^\beta]_l,[\Bbb W_m^{\a}]_{k-l-1}\}
\Big)\bigg\}\,.\notag
\end{align}
Using the notation ${\cal K}_P(X,\t):= {\cal K}_P(\t)e^{k_P\cdot X}$, 
the recursions \eqref{thetaex} and \eqref{moretheta} were shown in \cite{Gauge} to yield
the following multiparticle $\theta$-expansions,
\begin{align}
{\cal A}^P_{\alpha}(\theta)&=
{1\over 2}(\theta\gamma_m)_\alpha \ce^m_P
+{1\over 3}(\theta\gamma_m)_\alpha (\theta\gamma^m{\cal X}_P)
- {1\over32}(\theta\gamma^m)^\alpha(\theta\gamma_{mnp}\theta)\cf_P^{np} \label{THEXone}\\
&\quad{}
+ {1\over60}(\t\g^m)_\alpha(\t\g_{mnp}\t)({\cal X}_P^n\g^p\t)
+ {1\over1152}(\t\g^m)_\a(\t\g_{mnp}\t)(\t\g^{p}{}_{qr}\t)\cf_{P}^{n\vert qr}\notag\\
&\quad+ \sum_{XY=P}[\cA^{X,Y}_\a]_5 + \ldots  \, ,\notag\\
{\cal A}_P^m(\t)&=
\ce_P^m
+(\t\g^m {\cal X}_P)
-{1\over8}(\t \g^{m}{}_{pq}\t) \cf_P^{pq}
+{1\over12}(\t\g^{m}{}_{np}\t)(\cX_{P}^{n}\g^{p}\t)\notag\\
&\quad{}+{1\over 192}(\t\g^{m}_{\phantom{m}nr}\t)(\t\g^{r}_{\phantom{r}pq}\t)\cf_P^{n\vert pq}
-{1\over480} (\t\g^{m}_{\phantom{m}nr}\t)(\t\g_{\phantom{r}pq}^{r}\t)(\cX^{np}_{P}\g^{q}\t)\notag\\
&\quad+ \sum_{XY=P}\Big([\cA_{X,Y}^m]_4 + [\cA_{X,Y}^m]_5\Big)  + \ldots \, , \notag\\
\cW_P^\a(\t)&=
\cX^\alpha_P
+{1\over 4}(\t \g_{mn})^{\alpha} \cf_P^{mn}
-{1\over 4}(\t \g_{mn})^{\alpha}(\cX_P^{m}\g^{n}\t)
-{1\over48} (\t\g^{\phantom{m}q}_{m})^\alpha(\t\g_{qnp}\t)\cf^{m\vert np}_P \notag\\
&\quad{}+{1\over96}  (\t\g^{\phantom{m}q}_{m})^{\alpha}(\t\g_{qnp}\t)({\cal X}^{mn}_{P}\g^{p}\t)
- {1\over 1920}(\t\g^{\phantom{m}r}_{m})^{\alpha}(\t\g^{\phantom{nr}s}_{nr}\t)(\t\g_{spq}\t)\cf^{mn\vert pq}_{P} \notag\\
&\quad+ \sum_{XY=P}\Big([\cW_{X,Y}^\a]_3 + [\cW_{X,Y}^\a]_4 + [\cW_{X,Y}^\a]_5\Big) 
 + \ldots \, , \notag\\
{\cal F}_P^{mn}(\t)&=
\cf_P^{mn}
-(\cX_P^{[m}\g^{n]}\t)+{1\over 8}(\t\g_{pq}^{\phantom{pq}[m}\t) \cf_P^{n]\vert pq}
 -{1\over12}(\t\g^{\phantom{pq}[m}_{pq}\t)(\cX_{P}^{n]p}\g^{q}\t)\notag\\
&\quad{} -{1\over192}(\t\g^{\phantom{ps}[m}_{ps}\t)\cf^{n]p\vert qr}_P(\t\g^{s}_{\phantom{s}qr}\t)
+ {1\over 480}(\t\g^{[m}_{\phantom{[m}ps}\t)(\cX^{n]pq}_{P}\g^{r}\t)(\t\g^{s}_{\phantom{s}qr}\t)\notag\\
&\quad+ \sum_{XY=P}\Big([\cF_{X,Y}^{mn}]_2  + [\cF_{X,Y}^{mn}]_3 + [\cF_{X,Y}^{mn}]_4 +
[\cF_{X,Y}^{mn}]_5\Big) + \sum_{XYZ=P} [\cF_{X,Y,Z}^{mn}]_5  + \ldots\, , \notag
\end{align}
with terms of order $\theta^{\geq 6}$ in the ellipsis. The non-linearities of the form
$\sum_{XY=P} [{\cal K}_{X,Y}]_l$ can be traced back to the quadratic
expressions in \eqref{moretheta}, e.g.
\begin{align}
[{\cal A}_\alpha^{X,Y}]_5&= 
{1\over144}(\t\g_m)_\alpha(\t\g^{mnp}\t)({\cal X}_X\g_n\t)({\cal X}_Y\g_p\t)\, , \label{appFa}\\
%
[{\cal A}_{X,Y}^m]_4 &=  {1\over24} (\t\g^{m}_{\phantom{m}np}\t) ({\cal X}_{X}\g^{n}\t) ({\cal X}_{Y}\g^{p}\t) \, , \notag\\
[{\cal W}_{X,Y}^\alpha]_3 &= -{1\over6}(\t\g_{mn})^{\alpha}(\cX_{X}\g^{m}\t)(\cX_{Y}\g^{n}\t) \, , \notag\\
[{\cal F}_{X,Y}^{mn}]_2 &=  - ({\cal X}_X\gamma^{[m}\theta)({\cal X}_Y\gamma^{n]}\theta) \, , \notag
\end{align}
and further instances as to make the complete orders $\theta^{\leq 5}$ available are spelled out in the appendix of \cite{Gauge}.
It is easy to see that these non-linear terms vanish in the single-particle case, and one
recovers the linearized expansions (\ref{linTHEX}) of \cite{tsimpis,ooguri}.

Analogous $\theta$-expansions for the superfields of higher mass dimensions start with
\begin{align}
{\cal W}_P^{m\a}(X,\t)&= e^{k_P \cdot X}\Big( \cX_P^{m\alpha} + {1\over4}(\theta\g_{np})^{\alpha}\cf^{m\vert np}_P
+ \! \! \sum_{XY=P} \! \! \big[(\cX_{X}\g^{m}\theta)\cX^{\alpha}_Y-(X\leftrightarrow Y)\big]+ \ldots \Big) 
\,,\label{THEmass}\\
{\cal F}_P^{m|pq}(X,\theta)&=e^{k_P \cdot X}  \Big(  \cf_P^{m|pq} -(\cX^{m[p}_P\gamma^{q]}\theta)
+ \! \! \sum_{XY=P} \! \! \big[(\cX_{X}\gamma^{m}\theta)\cf^{pq}_Y-(X\leftrightarrow Y)\big] + \ldots \Big) \, ,\notag
\end{align}
where the lowest two orders $\sim \theta^2, \theta^3$ in the ellipsis
along with generalizations to higher mass dimensions are spelled out in the appendix of \cite{Gauge}.

\section{\label{LorBCJapp}Redefinitions from the Lorenz gauge to the BCJ gauge}

As shown in \cite{genredef}, multiparticle superfields in the BCJ gauge can be generated by 
starting from the multiparticle superfields in
the Lorenz gauge defined recursively in \eqref{Lorenzdef}.
In contrast to the hybrid gauge discussed in section \ref{sfinbcj}, the redefinitions are more involved
and require the following iterated redefinition,
\beq\label{genRedef}
K_{[P,Q]}=L_1(\hat K_{[P,Q]})\,,
\eeq
where the operator $L_j$ is defined as the local version of the perturbiner
\eqref{Lpert}
\beq\label{Ajdef}
L_{j}(\hat K_{[P,Q]}) := \hat K_{[P,Q]}
- \frac{1}{j} \big(\hat H\otimes L_{(j+1)}(\hat K)\big)_{C([P,Q])}
- \frac{1}{j}\begin{cases} D_\alpha \hat H_{[P,Q]} &: \ K= A_\alpha\, ,\\
k_{PQ}^m \hat H_{[P,Q]} &: \ K= A^m\, , \\
0 &: \ K= W^\alpha \, ,
\end{cases}
\eeq
where we used the notation \eqref{replace} and $\hat H$ is defined below in \eqref{HhatDef}.
The action of $L_j(\hat K_{[P,Q]})$ gives rise to
$L_{(j+1)}(\hat K_{[A,B]})$ on the right-hand side with
$\len{A}{+}\len{B}<\len{P}{+}\len{Q}$. Therefore this is a iteration over the
index $j$, and it eventually stops as each step involves splitting the nested brackets in $[P,Q]$. 
The iteration built into the redefinition \eqref{genRedef} yields the infinite series of 
non-linear terms \cite{SchubertFinite} present in the finite gauge transformation of the 
corresponding perturbiner series.

The examples \eqref{examplesH} of redefinitions from the hybrid to BCJ gauge have the
following Lorenz to BCJ counterparts:
\begin{align}
A^m_{[1,2]} &= \Ahat^m_{[1,2]}\,, \label{lorbcjex}\\
A^m_{[[1,2],3]} &= \Ahat^m_{[[1,2],3]} - k^m_{123}\hat H_{[[1,2],3]}\,,\notag\\
A^m_{[[1,2],[3,4]]} &=
  \Ahat^m_{[ [ 1 , 2 ] , [ 3 , 4 ] ]}
 -  (k_{1}\cdot k_{2})   \Big(
 \hat H_{[ 1 , [ 3 , 4 ] ]} \Ahat^m_{2} 
 -  \hat H_{[ 2 , [ 3 , 4 ] ]} \Ahat^m_{1}
\Big)\notag\\
&\quad +  (k_{3}\cdot k_{4})   \Big(
 \hat H_{[ [ 1 , 2 ] , 4 ]} \Ahat^m_{3}
 -  \hat H_{[ [ 1 , 2 ] , 3 ]} \Ahat^m_{4}
\Big)
 - k^m_{1234}  \hat H_{[ [ 1 , 2 ] , [ 3 , 4 ] ]}\,,\notag\\
A^m_{[[[1,2],3],4]} &=
  \Ahat^m_{[ [ [ 1 , 2 ] , 3 ] , 4 ]}
-  (k_{1}\cdot k_{2})   \Big(
   \hat H_{[ [ 1 , 3 ] , 4 ]} \Ahat^m_{2} 
 -  \hat H_{[ [ 2 , 3 ] , 4 ]} \Ahat^m_{1} 
\Big)\notag\\
&\quad -  (k_{12}\cdot k_{3})   \Big(
  \hat H_{[ [ 1 , 2 ] , 4 ]} \Ahat^m_{3}
\Big)
 -  (k_{123}\cdot k_{4})   \Big( 
  \hat H_{[ [ 1 , 2 ] , 3 ]} \Ahat^m_{4} 
\Big)
 - k^m_{1234} \hat H_{[ [ [ 1 , 2 ] , 3 ] , 4 ]}\,, \notag\\
A^m_{[[[[1,2],3],4],5]} &=
  \Ahat^m_{[ [ [ [ 1 , 2 ] , 3 ] , 4 ] , 5 ]}
 -  (k_{1}\cdot k_{2})   \Big(
   \hat H_{[ [ 1 , 3 ] , 4 ]} \Ahat^m_{[ 2 , 5 ]} 
 +  \hat H_{[ [ 1 , 3 ] , 5 ]} \Ahat^m_{[ 2 , 4 ]}
 +  \hat H_{[ [ 1 , 4 ] , 5 ]} \Ahat^m_{[ 2 , 3 ]}\notag\\
&\qquad{} +  \hat H_{[ [ [ 1 , 3 ] , 4 ] , 5 ]} \Ahat^m_{2}
- (1\leftrightarrow2)
\Big)\notag\\
&\quad -  (k_{12}\cdot k_{3})   \Big(
   \hat H_{[ [ 1 , 2 ] , 4 ]} \Ahat^m_{[ 3 , 5 ]} 
 +  \hat H_{[ [ 1 , 2 ] , 5 ]} \Ahat^m_{[ 3 , 4 ]} 
 +  \hat H_{[ [ [ 1 , 2 ] , 4 ] , 5 ]} \Ahat^m_{3}
 - ([1,2]\leftrightarrow3)
\Big)\notag\\
&\quad -  (k_{123}\cdot k_{4})   \Big(
   \hat H_{[ [ 1 , 2 ] , 3 ]} \Ahat^m_{[ 4 , 5 ]} 
 +  \hat H_{[ [ [ 1 , 2 ] , 3 ] , 5 ]} \Ahat^m_{4} 
\Big)\notag\\
&\quad -  (k_{1234}\cdot k_{5})   \Big(
 \hat H_{[ [ [ 1 , 2 ] , 3 ] , 4 ]} \Ahat^m_{5}
\Big)
 -  \hat H_{[ [ [ [ 1 , 2 ] , 3 ] , 4 ] , 5 ]} k^m_{12345}\,,\notag\\
A^m_{[[[1,2],3],[4,5]]} &=
\Ahat^m_{[ [ [ 1 , 2 ] , 3 ] , [ 4 , 5 ] ]}
-  (k_{1}\cdot k_{2})   \Big(
\hat H_{[ 1 , [ 4 , 5 ] ]} \Ahat^m_{[ 2 , 3 ]}
 +  \hat H_{[ [ 1 , 3 ] , [ 4 , 5 ] ]} \Ahat^m_{2}
- (1\leftrightarrow2)
\Big)\notag\\
&\quad-  (k_{12}\cdot k_{3})   \Big(
  \hat H_{[ [ 1 , 2 ] , [ 4 , 5 ] ]} \Ahat^m_{3}
 -  \hat H_{[ 3 , [ 4 , 5 ] ]} \Ahat^m_{[ 1 , 2 ]}
\Big)\notag\\
&\quad -  (k_{123}\cdot k_{45})   \Big(
 \hat H_{[ [ 1 , 2 ] , 3 ]} \Ahat^m_{[ 4 , 5 ]}
\Big) \notag\\
&\quad +  (k_{4}\cdot k_{5})   \Big(
 \hat H_{[ [ [ 1 , 2 ] , 3 ] , 5 ]} \Ahat^m_{4}
 -  \hat H_{[ [ [ 1 , 2 ] , 3 ] , 4 ]} \Ahat^m_{5}
\Big)- k^m_{12345}\hat H_{[ [ [ 1 , 2 ] , 3 ] , [ 4 , 5 ] ]}\,.
\end{align}
For an example of the redefinition \eqref{genRedef} for more than one iteration of $L_j$, it is
enough to consider the
superfield $\Ahat^m_{[[12,34],56]}$. A long and tedious calculation yields \cite{genredef}
\begin{align}
A^m_{[[12,34],56]}&=\hat A^m_{[[12,34],56]}-k^m_{123456}\hat H_{[[12,34],56]}\label{genRedefExampleOnePartTwo}\\
&\quad- (k_1\cdot k_2)\Big(\hat A^m_2\hat H_{[[1,34],56]}+\hat A^m_{[2,34]}\hat H_{[1,56]}+\hat A^m_{[2,56]}\hat H_{[1,34]}\cr
&\qquad{}\qquad{}-\frac{1}{2}k^m_{234}\hat{H}_{[2,34]}\hat H_{[1,56]}
-\frac{1}{2}k^m_{256}\hat H_{[2,56]}\hat H_{[1,34]}-(1\leftrightarrow 2)\Big)\notag\\
&\quad- (k_{12}\cdot k_{34})\Big(\hat A_{34}^m\hat H_{[12,56]}-(12\leftrightarrow 34)\Big)\notag\\
&\quad- (k_{1234}\cdot k_{56})\hat A_{56}^m\hat H_{[12,34]}\notag\\
&\quad- (k_3\cdot k_4)\Big(\hat A^m_4\hat H_{[123,56]}+\hat A^m_{[12,4]}\hat H_{[3,56]}+\hat A^m_{[4,56]}\hat H_{[12,3]}\notag\\
&\qquad{}\qquad{}-\frac{1}{2}k^m_{124}\hat{H}_{[12,4]}\hat H_{[3,56]}
-\frac{1}{2}k^m_{456}\hat H_{[4,56]}\hat H_{[12,3]}-(3\leftrightarrow 4)\Big)\notag\\
&\quad- (k_5\cdot k_6)\Big(\hat A_6^m\hat H_{[[12,34],5]}-(5\leftrightarrow 6)\Big)\,.\notag
\end{align}
The factors of $1/2$ correspond to the appearance of quadratic terms of the redefining superfields
$H_{[A,B]}$
in the finite gauge transformation given by the infinite series \eqref{finGauge}.

In the above redefinitions $\hat H_{[P,Q]}$  is given by
\begin{align}
\hat H_{[A,B]}&=\hat H'_{[A,B]}
-\frac{1}{2} (\hat H\otimes\hat H)_{\tilde C([A,B])}\,,
\label{HhatDef}\\
\hat H'_{[A,B]}&=H_{[A,B]}-\frac{1}{2}\Big[\big(\hat H'_Ak^m_A
-(\hat H\otimes\hat H^m)_{\tilde C'(A)}\big) A^{B}_m-(A\leftrightarrow B)\Big]\,,
\notag\\
\hat H'_{i}&=\hat H'_{[i,j]}=0\,, \notag
\end{align}
where the $H_{[A,B]}$ are defined as they were in \eqref{genH} to \eqref{HABCdef}, and
$\hat{H}_A^m:= k_A^m\hat H_A$. Furthermore, the maps $\tilde C$ and
$\tilde C'$ in the subscripts of (\ref{HhatDef}) are variants of the 
contact-term map $C$ reviewed in section \ref{contactsec} and introduced in \cite{genredef},
\beq\label{Ctildedef}
\tilde C(i) = 0\,,\qquad \tilde C([A,B]) = [C(A),B]_r + [A,C(B)]_r \, ,
\eeq
see (\ref{contactdef}) for the definition of the $C$ map on the right-hand side,
and we use the notation
\beq\label{rbra}
[P\otimes Q,B]_r = [P,B]\otimes Q\,.
\eeq
The definitions in (\ref{HhatDef}) furthermore involve the map
\beq
\tilde C'([A,B]) = \tilde C([A,B]) - {1\over2}(k_A\cdot k_B)(A\otimes B-B\otimes A)\, .
\eeq
In this way, iterative use of \eqref{HhatDef} will reduce any $\hat H_{[A,B]}$ 
to combinations of $A^m_C$, Mandelstam invariants and the 
superfields $H_{[A,B]}$ defined in \eqref{genH} to \eqref{HABCdef}, for instance
\begin{align}
\hat H_{[[[[1,2],3],[4,5]],6]}&=H_{[[[[1,2],3],[4,5]],6]}\label{Htsixend}\\
&\quad -\frac{1}{2}H_{[[[1,2],3],[4,5]]}(k_{12345}\cdot A_6)+\frac{1}{4}H_{[[1,2],3]}(k_{123}\cdot A_{45})(k_{12345}\cdot A_6) \notag\\
&\quad -\frac{1}{2}(k_1\cdot k_2)\big( H_{[[1,3],6]} H_{[2,[4,5]]} - H_{[[2,3],6]} H_{[1,[4,5]]}\big)\notag\\
&\quad -\frac{1}{2}(k_{12}\cdot k_3)\big(  H_{[[1,2],6]} H_{[3,[4,5]]}\big)
-\frac{1}{2} (k_{123}\cdot k_{45})\big( H_{[[4,5],6]} H_{[[1,2],3]}\big)\,.\notag
\end{align}

\section{The contact-term map is nilpotent}
\label{Cnilap}

To show that the contact-term map in (\ref{contactdef}) is nilpotent\footnote{We acknowledge illuminating discussions with
Hadleigh Frost.} we will first
determine the action of the adjoint
$C^*$ on $X_1\wedge X_2\wedge X_3$, where $X_i\in\cL^*$ are dual Lie polynomials
(see~\ref{dualLiesec}). From the definition \eqref{contactAs} we know that the adjoint of
the contact-term map is the $S$ bracket. For convenience we can use
\beq\label{cstardef}
\langle X_1\wedge X_2, C(\Gamma_1)\rangle = \langle C^\star(X_1\wedge X_2),\Gamma_1\rangle
\eeq
for a Lie monomial $\Gamma_1\in\cL$ and dual words $X_1,X_2\in \cL^*$,
and $C^\star(X_1\wedge X_2)=2\{X_1,X_2\}$ is the $S$ bracket.
Recall that in \eqref{extC}
the contact-term map $C$ is extended to act on the antisymmetric product of Lie polynomials $\cL\wedge\cL$ as a graded derivation of grading $+1$
acting on Lie polynomials $\Gamma_i$ of grading $+1$,
\beq\label{CLL}
C(\Gamma_1\wedge \Gamma_2) = C(\Gamma_1)\wedge \Gamma_2 - \Gamma_1\wedge C(\Gamma_2)\,.
\eeq
To compute $C^\star(X_1\wedge X_2\wedge X_3)$ we use the definition
\beq\label{deft}
\langle C^\star(X_1\wedge X_2\wedge X_3), \Gamma_1\wedge \Gamma_2\rangle
=\langle X_1\wedge X_2\wedge X_3, C(\Gamma_1\wedge \Gamma_2)\rangle
\eeq
which exploits the fact that $C(\Gamma_1\wedge\Gamma_2)$ has grading $+3$.
Using \eqref{CLL} we get
\beq\label{needs}
\langle C^\star(X_1\wedge X_2\wedge X_3), \Gamma_1\wedge \Gamma_2\rangle =
\langle X_1\wedge X_2\wedge X_3, C(\Gamma_1)\wedge \Gamma_2\rangle
-\langle X_1\wedge X_2\wedge X_3,\Gamma_1\wedge C(\Gamma_2)\rangle \, .
\eeq
Defining 
$\langle A\otimes B, C\otimes D\rangle = \langle A,C\rangle\langle B,D\rangle$
we obtain
\beq\label{weten}
\langle A\wedge B,C\wedge D\rangle =
2\langle A,C\rangle\langle B,D\rangle - 2\langle A,D\rangle\langle B,C\rangle\, .
\eeq
To use this, we need to split the three-fold wedge product democratically into two factors:
\beq\label{democ}
X_1\wedge X_2\wedge X_3 = {1\over 3}\Big( (X_1\wedge X_2)\wedge X_3 +X_1\wedge (X_2\wedge X_3)
+(X_3\wedge X_1)\wedge X_2\Big)
\eeq
which exploits the cyclic symmetry of $X_1\wedge X_2\wedge X_3$ and the parenthesis indicates the
split. Therefore \eqref{needs} becomes
\begin{align}
\langle C^\star(X_1\wedge X_2\wedge X_3), \Gamma_1\wedge \Gamma_2\rangle &=
{1\over3}\langle (X_1\wedge X_2)\wedge X_3, C(\Gamma_1)\wedge \Gamma_2\rangle
-(\Gamma_1\leftrightarrow\Gamma_2) + {\rm cyc}(X_1,X_2,X_3)\\
&= {2\over3}\langle (X_1\wedge X_2),C(\Gamma_1)\rangle\langle X_3,\Gamma_2\rangle
-(\Gamma_1\leftrightarrow\Gamma_2) + {\rm cyc}(X_1,X_2,X_3)\cr
&= {2\over3}\langle C^\star(X_1\wedge X_2),\Gamma_1\rangle\langle X_3,\Gamma_2\rangle
-(\Gamma_1\leftrightarrow\Gamma_2) + {\rm cyc}(X_1,X_2,X_3)\cr
&= {1\over3}\langle C^\star(X_1\wedge X_2)\wedge X_3,\Gamma_1\wedge \Gamma_2\rangle
 + {\rm cyc}(X_1,X_2,X_3)\, , 
 \notag
\end{align}
where we used \eqref{weten} in the second line, $\langle X_1\wedge X_2,\Gamma_2\rangle = 0$,
and \eqref{weten} again to identify the last line. Therefore we conclude
\beq\label{Cst}
C^\star(X_1\wedge X_2 \wedge X_3) = {1\over3}C^\star(X_1\wedge X_2) \wedge X_3  + {\rm cyc}(X_1,X_2,X_3)
\eeq
which resembles the action of the (nilpotent) homology operator $\p$ of \cite{homologyOp} (see
also \cite{naryRev}).
Noting that $C^\star(X_1\wedge X_2)=2\{X_1,X_2\}\in\cL^\star$, the right-hand side is
in $\cL^*\wedge\cL^*$ and
therefore $C^\star$ can act again,
\beq\label{Cstcoh}
C^\star\circ C^\star(X_1\wedge X_2\wedge X_3) = {4\over 3}\{\{X_1,X_2\},X_3\} + {\rm
cyc}(X_1,X_2,X_3) = 0
\eeq
by virtue of the Jacobi identity of the $S$ bracket \cite{flas}. Therefore $C^\star\circ C^\star=0$ and we
conclude
\begin{prop.}
The contact-term map is nilpotent
\beq\label{Cnil}
C\circ C = 0\,.
\eeq
\end{prop.}
\noindent\textit{Proof.} Using that $C\circ C(\Gamma_1)\in \cL\wedge\cL\wedge \cL$ we get
\beq\label{Ccoho}
\langle X_1\wedge X_2\wedge X_3,C\circ C(\Gamma_1)\rangle = 
\langle C^\star(X_1\wedge X_2\wedge X_3), C(\Gamma_1)\rangle =
\langle C^\star\circ C^\star(X_1\wedge X_2\wedge X_3), \Gamma_1\rangle = 0\, .
\eeq
Therefore $C^2(\Gamma_1)=0$ for any Lie polynomial $\Gamma_1$. By induction if
$C^2(\Gamma_1) = C^2(\Gamma_2) = 0$ we also get $C^2(\Gamma_1\wedge \Gamma_2) = 0$ as
\eqref{CLL} implies
\beq
C\circ C(\Gamma_1\wedge \Gamma_2) = C^2(\Gamma_1)\wedge \Gamma_2
+ \Gamma_1\wedge C^2(\Gamma_2) \, ,
\eeq
where we used that $C(\Gamma_1)$ has grading $+2$.\qed

\section{Different representation of multiparticle vertex}
\label{Vapp}

In \cite{6ptTree, nptStringI}, an alternative representation for the
multiparticle vertex $V_{123}$
was found closely following the OPE calculation in the tree-amplitude prescription.
Its expression was denoted by $T_{123}$ and given by (note the shorthands $A_i^{k_j}:= (k_j\cdot
A_i)$ or $\g^{k^{i}}= \gamma_m k_i^m$, and repeated indices are summed)
\begin{align}
\label{T123string}
T_{123} &=
  {1 \over 6}\,\big[ (\l \g^{k^{2}}  W^1) A_{2}^{m} A_{3}^{m}
 -  (\l \g^{k^{1}}  W^2) A_{1}^{m} A_{3}^{m}
-  (\l \g^{k^{3}}  W^1) A_{2}^{m} A_{3}^{m}
+  (\l \g^{k^{3}}  W^2) A_{1}^{m} A_{3}^{m} \big]\\
&+  {1 \over 3}\big[ (\l \g^{k^{2}}  W^3) A_{1}^{m} A_{2}^{m}
-  (\l \g^{k^{1}}  W^3) A_{1}^{m} A_{2}^{m} \big]\notag\\
 &+  {1 \over 3}\big[ (\l \g^{m}  W^1) (W^{2} \g^{m}  W^3)
-  (\l \g^{m}  W^2) (W^{1} \g^{m}  W^3)
-  2 (\l \g^{m}  W^3) (W^{1} \g^{m}  W^2) \big]\notag\\
& +  {1 \over 6}\big[ (\l \g^{m}  W^1) A_{2}^{k^{3}} A_{3}^{m}
-  (\l \g^{m}  W^1) A_{2}^{k^{1}} A_{3}^{m}
+  (\l \g^{m}  W^2) A_{1}^{k^{2}} A_{3}^{m}
-  (\l \g^{m}  W^2) A_{1}^{k^{3}} A_{3}^{m}\big]\notag\\
&+  {1 \over 6}\big[ (\l \g^{m}  W^3) A_{1}^{m} A_{2}^{k^{3}}
- (\l \g^{m}  W^3) A_{1}^{k^{3}} A_{2}^{m}\big]
+  {5 \over 6}\big[ (\l \g^{m}  W^3) A_{1}^{m} A_{2}^{k^{1}}
- (\l \g^{m}  W^3) A_{1}^{k^{2}} A_{2}^{m}\big]\notag\\
& +  {1 \over 3}\big[ (\l \g^{m}  W^2) A_{1}^{m} A_{3}^{k^{2}}
- (\l \g^{m}  W^1) A_{2}^{m} A_{3}^{k^{1}}\big]
 +  {2 \over 3}\big[ (\l \g^{m}  W^2) A_{1}^{m} A_{3}^{k^{1}}
- (\l \g^{m}  W^1) A_{2}^{m} A_{3}^{k^{2}}\big]\notag\\
& +  {1 \over 2}\, V_{1} A_{2}^{k^{1}} A_{3}^{k^{1}}
+  {1 \over 3}\, V_{1} A_{2}^{k^{1}} A_{3}^{k^{2}}
+  {1 \over 6}\, V_{1} A_{2}^{k^{3}} A_{3}^{k^{1}}
-  {1 \over 6}\, V_{1} A_{2}^{m} A_{3}^{m} s_{13}
 -  {1 \over 3}\, V_{1} A_{2}^{m} A_{3}^{m} s_{12}\notag\\
& -  {1 \over 3}\, V_{2} A_{1}^{k^{2}} A_{3}^{k^{1}}
 -  {1 \over 2}\, V_{2} A_{1}^{k^{2}} A_{3}^{k^{2}}
 -  {1 \over 6}\, V_{2} A_{1}^{k^{3}} A_{3}^{k^{2}}
 +  {1 \over 6}\, V_{2} A_{1}^{m} A_{3}^{m} s_{23}
 +  {1 \over 3}\, V_{2} A_{1}^{m} A_{3}^{m} s_{12} \notag\\
& +  {1 \over 6}\, V_{3} A_{1}^{k^{2}} A_{2}^{k^{3}}
 -  {1 \over 6}\, V_{3} A_{1}^{k^{3}} A_{2}^{k^{1}}
 -  {1 \over 6}\, V_{3} A_{1}^{m} A_{2}^{m} s_{23}
 +  {1 \over 6}\, V_{3} A_{1}^{m} A_{2}^{m} s_{13} \, .\notag
\end{align}
In \cite{EOMBBs} a new representation for this object was proposed based on an analysis of the
equations of motion of a class of superfields. It was denoted by $V_{123}$ and given by
\begin{align}
\label{V123sol}
V_{123} &=
{1 \over 6}\big[ (\l \g^{k^{1}mn}  W^2) A_{1}^{m} A_{3}^{n}
-  (\l \g^{k^{2}mn}  W^1) A_{2}^{m} A_{3}^{n}\big] \notag \\
& +  {1 \over 12}\big[ (\l \g^{k^{1}mn}  W^3) A_{1}^{m} A_{2}^{n}
 +  (\l \g^{k^{2}mn}  W^3) A_{1}^{m} A_{2}^{n}
 +  (\l \g^{k^{3}mn}  W^1) A_{2}^{m} A_{3}^{n}
 -  (\l \g^{k^{3}mn}  W^2) A_{1}^{m} A_{3}^{n}\big]\notag\\
&+  {1 \over 4}\big[ (\l \g^{k^{2}}  W^1) A_{2}^{m} A_{3}^{m}
 -  (\l \g^{k^{1}}  W^2) A_{1}^{m} A_{3}^{m}
 -  (\l \g^{k^{1}}  W^3) A_{1}^{m} A_{2}^{m}
 +  (\l \g^{k^{2}}  W^3) A_{1}^{m} A_{2}^{m}\big]\notag\\
&+  {1 \over 12}\big[ (\l \g^{m}  W^3) A_{1}^{k^{3}} A_{2}^{m}
- (\l \g^{m}  W^3) A_{1}^{m} A_{2}^{k^{3}}\big]
+  {5 \over 12}\big[ (\l \g^{m}  W^2) A_{1}^{k^{2}} A_{3}^{m}
-  (\l \g^{m}  W^1) A_{2}^{k^{1}} A_{3}^{m}\big]\notag\\
&+  {1 \over 3}\big[ (\l \g^{m}  W^2) A_{1}^{m} A_{3}^{k^{2}}
-  (\l \g^{m}  W^1) A_{2}^{m} A_{3}^{k^{1}}\big]\notag\\
&+  {1 \over 2}\big[ (\l \g^{m}  W^2) A_{1}^{m} A_{3}^{k^{1}}
-  (\l \g^{m}  W^1) A_{2}^{m} A_{3}^{k^{2}}
-  (\l \g^{m}  W^3) A_{1}^{k^{2}} A_{2}^{m}
+  (\l \g^{m}  W^3) A_{1}^{m} A_{2}^{k^{1}}\big]\notag\\
&+  {1 \over 6}\big[ (\l \g^{m}  W^1) (W^{2} \g^{m}  W^3)
-  (\l \g^{m}  W^2) (W^{1} \g^{m}  W^3)
 -  2(\l \g^{m}  W^3) (W^{1} \g^{m}  W^2)\big]\notag\\
&+  {1 \over 4}\, V_{1} A_{2}^{k^{1}} A_{3}^{k^{1}}
 +  {1 \over 3}\, V_{1} A_{2}^{k^{1}} A_{3}^{k^{2}}
 -  {1 \over 12}\, V_{1} A_{2}^{k^{3}} A_{3}^{k^{1}}
 +  {1 \over 12}\, V_{1} A_{2}^{m} A_{3}^{m} s_{13}
 -  {1 \over 12}\, V_{1} A_{2}^{m} A_{3}^{m} s_{12}
 -  {1 \over 6}\, V_{1} (W^{2} \g^{k^{1}}  W^3)\notag\\
& -  {1 \over 3}\, V_{2} A_{1}^{k^{2}} A_{3}^{k^{1}}
 -  {1 \over 4}\, V_{2} A_{1}^{k^{2}} A_{3}^{k^{2}}
 +  {1 \over 12}\, V_{2} A_{1}^{k^{3}} A_{3}^{k^{2}}
 -  {1 \over 12}\, V_{2} A_{1}^{m} A_{3}^{m} s_{23}
 +  {1 \over 12}\, V_{2} A_{1}^{m} A_{3}^{m} s_{12}
 +  {1 \over 6}\, V_{2} (W^{1} \g^{k^{2}}  W^3)\notag\\
& +  {5 \over 12}\, V_{3} A_{1}^{k^{2}} A_{2}^{k^{3}}
 -  {5 \over 12}\, V_{3} A_{1}^{k^{3}} A_{2}^{k^{1}}
 -  {1 \over 6}\, V_{3} A_{1}^{m} A_{2}^{m} s_{23}
 +  {1 \over 6}\, V_{3} A_{1}^{m} A_{2}^{m} s_{13}
 +  {1 \over 3}\, V_{3} (W^{1} \g^{k^{3}}  W^2) \, .
\end{align}
Both of these building blocks satisfy the required generalized Jacobi identities:
\begin{align}
T_{123} + T_{213} &= 0 \, , & T_{123}+T_{231}+T_{312} & = 0\, , \notag \\
V_{123} + V_{213} &= 0 \, , & V_{123}+V_{231}+V_{312} & = 0\,,
\end{align}
therefore both qualify as representatives of the unintegrated vertex at multiplicity three.
A tedious calculation shows that their difference is BRST exact, namely
\beq
T_{123}-V_{123}  = Q\Sigma_{123}\, ,
\ee
where
\begin{align}
\Sigma_{123} & =
{1 \over 4}\big[ A_{2}^{k^{1}} (A_{1} \cdot A_{3})
+ A_{2}^{k^{3}} (A_{1}\cdot A_{3})
-  A_{1}^{k^{2}} (A_{2} \cdot A_{3})
- A_{1}^{k^{3}} (A_{2}\cdot A_{3})\big]\notag \\
&\quad -  {1 \over 3}\, (W^{1} \g_{m}  W^2) A_{3}^{m}
 -  {1 \over 6}\, (W^{1} \g_{m}  W^3) A_{2}^{m}
 +  {1 \over 6}\, (W^{2} \g_{m}  W^3) A_{1}^{m}\,.
\end{align}

\section{BRST-invariant permutations at low multiplicities}
\label{descpermsapp}

To help understanding the
definition of the Berends--Giele idempotent given in section~\ref{BGidempsec}, the first few permutations of \eqref{BGidemp} read as follows
\begin{align}
\cE(1) &= W_1\, ,\\
\cE(12) &= \half(W_{12}-W_{21})\,,\notag\\
\cE(123) &=
       {1\over3} \Word(1,2,3)
       - {1\over6} \Word(1,3,2)
       - {1\over6} \Word(2,1,3)
       - {1\over6} \Word(2,3,1)
       - {1\over6} \Word(3,1,2)
       + {1\over3} \Word(3,2,1)\,,\notag\\
\cE(1234) &=
        {1\over4} \Wordq(1,2,3,4)
       - {1\over12} \Wordq(1,2,4,3)
       - {1\over12} \Wordq(1,3,2,4)
       - {1\over12} \Wordq(1,3,4,2)
       - {1\over12} \Wordq(1,4,2,3)
       + {1\over12} \Wordq(1,4,3,2) \cr
&\quad       - {1\over12} \Wordq(2,1,3,4)
       + {1\over12} \Wordq(2,1,4,3)
       - {1\over12} \Wordq(2,3,1,4)
       - {1\over12} \Wordq(2,3,4,1)
       + {1\over12} \Wordq(2,4,1,3)
       + {1\over12} \Wordq(2,4,3,1)\cr
&\quad       - {1\over12} \Wordq(3,1,2,4)
       - {1\over12} \Wordq(3,1,4,2)
       + {1\over12} \Wordq(3,2,1,4)
       + {1\over12} \Wordq(3,2,4,1)
       - {1\over12} \Wordq(3,4,1,2)
       + {1\over12} \Wordq(3,4,2,1)\cr
&\quad       - {1\over12} \Wordq(4,1,2,3)
       + {1\over12} \Wordq(4,1,3,2)
       + {1\over12} \Wordq(4,2,1,3)
       + {1\over12} \Wordq(4,2,3,1)
       + {1\over12} \Wordq(4,3,1,2)
       - {1\over4} \Wordq(4,3,2,1)\, ,\notag
\end{align}
where a permutation $\s$ is written as $W_\s$ in order to avoid confusion with the rational
coefficients. Using these definitions and examples, it is easy to generate the first few
permutations of \eqref{brstgamma}. For instance, at multiplicities three and four we have
\begin{align}
\gamma_{1|2,3} &= W_{123} + W_{132}\,,\quad
\gamma_{1|23} ={1\over2}W_{123} - {1\over2}W_{132}\,,\\
\gamma_{1|2,3,4} &=
	    W_{1234}
          + W_{1243}
          + W_{1324}
          + W_{1342}
          + W_{1423}
          + W_{1432}\,,\cr
\gamma_{1|23,4} &=
	    {1\over2} W_{1234}
          + {1\over2} W_{1243}
          - {1\over2} W_{1324}
          - {1\over2} W_{1342}
          + {1\over2} W_{1423}
          - {1\over2} W_{1432}\,,
\cr
\gamma_{1|234} &=
	    {1\over3} W_{1234}
          - {1\over6} W_{1243}
          - {1\over6} W_{1324}
          - {1\over6} W_{1342}
          - {1\over6} W_{1423}
          + {1\over3} W_{1432}\,, \notag
\end{align}
where it suffices to list only the different partitions of labels as other permutations follow
from relabeling due to the
total symmetry of \eqref{brstgamma} under exchanges of any pair of words
$P_i\leftrightarrow P_j$ and the functional form of \eqref{BGidemp}.
Similarly, at multiplicity five the BRST invariant permutations are given by
\begin{align}
\label{fivegammaEx}
\gamma_{1|2,3,4,5} &=W_{1(2\shuffle3\shuffle4\shuffle5)} \, ,\\
\gamma_{1|23,4,5} &=
	    {1\over2} W_{12345}
          + {1\over2} W_{12354}
          + {1\over2} W_{12435}
          + {1\over2} W_{12453}
          + {1\over2} W_{12534}
          + {1\over2} W_{12543}\cr&
      \quad    - {1\over2} W_{13245}
          - {1\over2} W_{13254}
          - {1\over2} W_{13425}
          - {1\over2} W_{13452}
          - {1\over2} W_{13524}
          - {1\over2} W_{13542}\cr&
       \quad    + {1\over2} W_{14235}
          + {1\over2} W_{14253}
          - {1\over2} W_{14325}
          - {1\over2} W_{14352}
          + {1\over2} W_{14523}
          - {1\over2} W_{14532}\cr&
       \quad    + {1\over2} W_{15234}
          + {1\over2} W_{15243}
          - {1\over2} W_{15324}
          - {1\over2} W_{15342}
          + {1\over2} W_{15423}
          - {1\over2} W_{15432}\, ,
\cr
\gamma_{1|234,5} &=
            {1\over3} W_{12345}
          + {1\over3} W_{12354}
          - {1\over6} W_{12435}
          - {1\over6} W_{12453}
          + {1\over3} W_{12534}
          - {1\over6} W_{12543}\cr&
     \quad      - {1\over6} W_{13245}
          - {1\over6} W_{13254}
          - {1\over6} W_{13425}
          - {1\over6} W_{13452}
          - {1\over6} W_{13524}
          - {1\over6} W_{13542}\cr&
      \quad     - {1\over6} W_{14235}
          - {1\over6} W_{14253}
          + {1\over3} W_{14325}
          + {1\over3} W_{14352}
          - {1\over6} W_{14523}
          + {1\over3} W_{14532}\cr&
      \quad     + {1\over3} W_{15234}
          - {1\over6} W_{15243}
          - {1\over6} W_{15324}
          - {1\over6} W_{15342}
          - {1\over6} W_{15423}
          + {1\over3} W_{15432} \, ,
          \cr
\gamma_{1|23,45} &=
           {1\over4} W_{12345}
          - {1\over4} W_{12354}
          + {1\over4} W_{12435}
          + {1\over4} W_{12453}
          - {1\over4} W_{12534}
          - {1\over4} W_{12543}\cr&
       \quad    - {1\over4} W_{13245}
          + {1\over4} W_{13254}
          - {1\over4} W_{13425}
          - {1\over4} W_{13452}
          + {1\over4} W_{13524}
          + {1\over4} W_{13542}\cr&
    \quad       + {1\over4} W_{14235}
          + {1\over4} W_{14253}
          - {1\over4} W_{14325}
          - {1\over4} W_{14352}
          + {1\over4} W_{14523}
          - {1\over4} W_{14532}\cr&
       \quad    - {1\over4} W_{15234}
          - {1\over4} W_{15243}
          + {1\over4} W_{15324}
          + {1\over4} W_{15342}
          - {1\over4} W_{15423}
          + {1\over4} W_{15432} \, , \cr
\gamma_{1|2345} &=
	   {1\over4} W_{12345}
          - {1\over12} W_{12354}
          - {1\over12} W_{12435}
          - {1\over12} W_{12453}
          - {1\over12} W_{12534}
          + {1\over12} W_{12543}\cr&
   \quad        - {1\over12} W_{13245}
          + {1\over12} W_{13254}
          - {1\over12} W_{13425}
          - {1\over12} W_{13452}
          + {1\over12} W_{13524}
          + {1\over12} W_{13542}\cr&
      \quad     - {1\over12} W_{14235}
          - {1\over12} W_{14253}
          + {1\over12} W_{14325}
          + {1\over12} W_{14352}
          - {1\over12} W_{14523}
          + {1\over12} W_{14532}\cr&
      \quad     - {1\over12} W_{15234}
          + {1\over12} W_{15243}
          + {1\over12} W_{15324}
          + {1\over12} W_{15342}
          + {1\over12} W_{15423}
          - {1\over4} W_{15432} \, .\notag
\end{align}

\bibliography{reviewrefs}

\end{document}